\pdfoutput=1

\documentclass[12pt, oneside]{book} 
\usepackage{bibentry}
\usepackage{KUstyle}

\nobibliography*

\usepackage[hyperpageref]{backref}

\renewcommand*{\backrefalt}[4]{
      \ifcase #1 
      \or Cited on page #2.
      \else Cited on pages #2.
      \fi}

 \graphicspath{ {./img/} {./imgsplit1/}}

\usepackage{amsthm} 

\newtheorem{theorem}{Theorem}[chapter]
\newtheorem{definition}{Definition}[chapter]

\usepackage{amsmath}
\usepackage{amssymb}
\usepackage{wasysym}

\usepackage{multicol}
\usepackage{multirow}
\usepackage{booktabs}      

\usepackage[table]{xcolor}
\usepackage[export]{adjustbox}

\usepackage{tikz}
\usepackage{pgfplots}
\pgfplotsset{compat=1.18}
\usetikzlibrary{shapes.misc}

\usepackage{algorithm2e}

\usepackage{nicefrac}

\usepackage{cleveref}
\usepackage{currfile}

\usepackage{pifont}

\usepackage{caption}
\DeclareMathOperator*{\argminA}{arg\,min}

\newcommand{\trophy}{\includegraphics[page=1,height=3.3mm, trim=6mm 1.7mm 5mm 0.25mm]{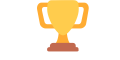}}

\newcommand{\rs}{RSs}

\newcommand{\floorkm}{\left\lfloor\frac{km}{n}\right\rfloor}

\newcommand{\our}{$_{\text{our}}$}
\newcommand{\ori}{$_{\text{ori}}$}
\newcommand{\up}{$\uparrow$}
\newcommand{\down}{$\downarrow$}

\newcommand{\ups}{\texorpdfstring{\up}{}} 
\newcommand{\dws}{\texorpdfstring{\down}{}}

\newcommand{\divori}{$_{\div{\text{-ori}}}$}
\newcommand{\divour}{$_{\div{\text{-our}}}$}
\newcommand{\mulour}{$_{\times{\text{-our}}}$}

\newcommand{\joint}{\textsc{Joint}}
\newcommand{\eff}{\textsc{Eff}}
\newcommand{\fair}{\textsc{Fair}}

\newcommand{\ind}{$_\text{ind}$}
\newcommand{\bgrp}{$_\text{b-group}$}
\newcommand{\wgrp}{$_\text{w-group}$}

\newcommand{\explainsig}{Asterisk ($^*$) denotes a statistically significant correlation ($\alpha=0.05$), after applying the Benjamini-Hochberg procedure.}

\newcommand{\nofix}{{\large $\circ$}}
\newcommand{\bbullet}{{\large $\bullet$}} 
\newcommand{\partialfix}{\raisebox{\depth}{{\tiny\LEFTcircle}}}
\newcommand{\obs}{\includegraphics[page=1,height=2.33009172mm, trim=6mm 3mm 5mm 0]{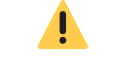}}
\newcommand{\NA}{--}

\definecolor{dark-green}{rgb}{0.0, 0.5, 0.0}
\newcommand{\cm}{{\color{dark-green}\checkmark}}
\newcommand{\xm}{{\color{red}\ding{55}}}
\newcommand{\specialcell}[2][c]{%
\begin{tabular}[#1]{@{}c@{}}#2\end{tabular}
}

\Crefname{equation}{Eq.}{Eqs.}
\Crefname{figure}{Fig.}{Figs.}
\Crefname{table}{Tab.}{Tabs.} 
\Crefname{appendix}{App.}{Apps.} 
\crefformat{section}{\S#2#1#3}
\crefformat{subsection}{\S#2#1#3}

\usepackage[most]{tcolorbox}

\theoremstyle{remark}
\newtheorem*{ex}{Example}

\usepackage{geometry}

\usepackage[all]{nowidow}

\makeatletter
\newcommand\footnoteref[1]{\protected@xdef\@thefnmark{\ref{#1}}\@footnotemark}

\ptype{PhD Thesis}
\author{Theresia Veronika Rampisela}
\title{Offline Evaluation Measures of Fairness \\
in Recommender Systems
}
\subtitle{}
\advisor{Principal supervisor: Christina Lioma}
\coadvisor{Co-supervisors: Tuukka Ruotsalo and Maria Maistro}
\date{This thesis has been submitted to the PhD School of The Faculty of Science, University of Copenhagen on 31 May 2025.}

\renewcommand\@dotsep{3.5}   

\usepackage{titlesec}
\usepackage{fancyhdr}

\fancypagestyle{plain}{         
\fancyhf{}
\fancyfoot[C]{\thepage}}%

\renewcommand{\headrulewidth}{0pt}

\newcommand\mymainpagestyle{%
\fancyhf{}      
\fancyhead[L]{\nouppercase{\footnotesize{\chaptername~ \thechapter~ | ~\leftmark\phantom{y}}} \renewcommand{\headrulewidth}{0.4pt} \headrule \renewcommand{\headrulewidth}{0pt}}
\setlength{\headheight}{25pt} 
\fancyfoot[C]{\thepage}
}

\begin{document}


\maketitle
\frontmatter 
\pagestyle{plain} 

\newpage \ \newpage 

\section*{Abstract}
\label{sec:abstract}
\addcontentsline{toc}{section}{Abstract} 

The evaluation of recommender system fairness has become increasingly important, especially with recent legislation that emphasises the development of fair and responsible artificial intelligence. 
This increasing importance has led to the emergence of various fairness evaluation measures, which quantify fairness based on different fairness definitions. 
However, many of such measures are simply proposed and further used without further analysis on their robustness. 
As a result, there is insufficient understanding and awareness of the measures' limitations. Among other issues, it is not known what kind of model outputs produce the (un)fairest score, how the measure scores are empirically distributed, and whether there are cases where the measures cannot be computed (e.g., due to division by zero). These issues cause difficulty in interpreting the measure scores and confusion on which measure(s) should be used for a specific case. 

To address the above issues, this thesis presents a series of papers that assess and overcome various theoretical, empirical, and conceptual limitations of existing recommender system fairness evaluation measures. Altogether, the papers in this thesis investigate a wide range of offline evaluation measures for different fairness notions, which are divided based on the evaluation subjects (users and items) and for different evaluation granularities (groups of subjects and individual subjects). 
Firstly, we perform theoretical and empirical analysis on a set of fairness measures, exposing flaws that limit their interpretability, expressiveness, or applicability. 
Secondly, we contribute novel evaluation approaches and measures that overcome these limitations. 
Finally, considering the measures' limitations, we recommend guidelines for the appropriate measure usage, thereby allowing for more precise selection of fairness evaluation measures in practical scenarios.

Overall, the papers in this thesis collectively contribute to advancing the state-of-the-art offline evaluation of fairness in recommender systems.

\newpage
\section*{Resumé}
\label{sec:resume}
\addcontentsline{toc}{section}{Resumé}

Evaluering af fairness (retfærdighed) i anbefalingssystemer er blevet stadig vigtigere, især i lyset af ny lovgivning, der understøtter udviklingen af fair og ansvarlig kunstig intelligens. Denne stigende opmærksomhed har ført til fremkomsten af forskellige fairness-evalueringsmetoder, som kvantificerer fairness baseret på forskellige definitioner. Imidlertid er mange af disse metoder blevet foreslået og anvendt uden en dybere analyse af deres robusthed. Dette har resulteret i en utilstrækkelig forståelse og bevidsthed om metoderne og deres begrænsninger. Blandt andet er det uklart, hvilke modeloutput der producerer de mest (un)fair scores, hvordan målingerne empirisk fordeler sig, og om der findes situationer, hvor målingerne ikke kan beregnes (f.eks. ved division med nul). Disse udfordringer vanskeliggør fortolkningen af målingernes resultater og skaber uklarhed om, hvilke metoder der bør anvendes i specifikke tilfælde.

For at adressere disse udfordringer præsenterer denne afhandling en serie af artikler, der vurderer og løser forskellige teoretiske, empiriske og konceptuelle begrænsninger ved eksisterende fairness-evalueringsmetoder i anbefalingssystemer. Artiklerne undersøger et bredt udvalg af offline evalueringsmetoder, opdelt efter evalueringssubjekter (brugere og genstande) samt på forskellige evalueringsniveauer (grupper af subjekter og individuelle subjekter). Først udføres en teoretisk og empirisk analyse af et sæt fairness-målinger, hvor der identificeres begrænsninger, der påvirker deres fortolkningsevne, udtrykskraft og anvendelighed. Derefter introduceres nye evalueringsmetoder, som overvinder disse begrænsninger. Endelig anbefales retningslinjer for korrekt anvendelse af fairness-målinger, hvilket muliggør en mere præcis udvælgelse af passende evalueringsmetoder i praktiske scenarier.

Samlet set bidrager artiklerne i denne afhandling til at fremme den nyeste udvikling inden for offline evaluering af fairness i anbefalingssystemer.

\newpage
\chapter*{List of Publications}
\addcontentsline{toc}{section}{List of Publications}
This thesis includes the following papers as chapters, listed in the order of their appearance in the thesis.\footnote{The publication status is updated as of the PhD defence date.}

\begin{enumerate}
    \item \bibentry{Rampisela2024EvaluationStudy}. \trophy \textit{ Women in RecSys Journal Paper of the Year Award 2024, Junior Category.}
    \item \bibentry{Rampisela2024CanRelevance}. \textit{Acceptance rate: 20\%.}
    \item \bibentry{Rampisela2025RelevanceGuidelines}.
    \item \bibentry{Rampisela2025JointFrontier}.  \includegraphics[height=12pt]{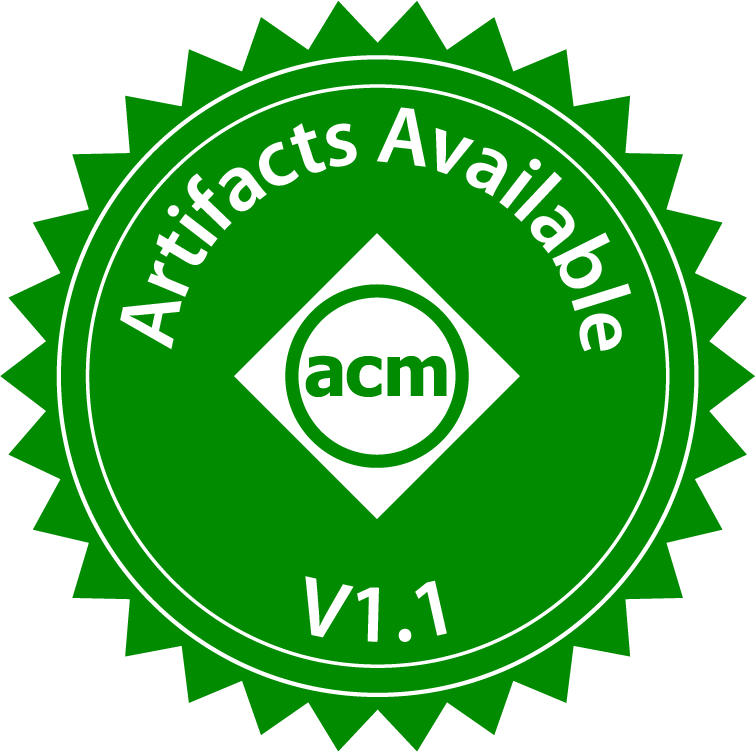} \textit{ACM Artifacts Available v1.1 badge}, \textit{acceptance rate (oral presentation): 7.5\%}.
    \item Theresia Veronika Rampisela, Maria Maistro, Tuukka Ruotsalo, and Christina Lioma. Measuring Individual User Fairness with User Similarity and Effectiveness Disparity. 2025. 10 pages. \textit{Under review.} 
    \item \bibentry{Rampisela2025StairwayFairness}. \textit{Acceptance rate: 21\%.}
\end{enumerate}


\newpage
\chapter*{Acknowledgements}
\label{sec:acks}
\addcontentsline{toc}{section}{Acknowledgements}

This PhD thesis and the entirety of my PhD journey would not have been possible without the support of many people. I would like to acknowledge them here.

First and foremost, I am extremely grateful for my PhD supervisors. From the start to the end of my studies, they have provided me with invaluable scientific and professional guidance, critical and detailed feedback, as well as many opportunities to develop my career, all of which I deeply appreciated. I have learned a lot from them. 
To Christina, thank you for consistently emphasising the importance of conducting high-quality scientific research and for showing how to explain complex concepts so that they can be easily understood. I am also thankful that you helped me understand the ``why's'' behind certain practices, in a safe environment. 
To Tuukka, I am grateful for your fresh ideas and perspectives, which have been greatly useful. Thank you for always encouraging me to look at the bigger picture. 
To Maria, thank you for your technical and mathematical expertise, as well as for your help in translating abstract ideas and thoughts into clarity. Thank you for proactively facilitating my participation in activities that are relevant to my professional growth.

I sincerely thank the Algorithms, Data \& Democracy (ADD) project for fully funding my PhD. 
Being part of the project has allowed me to meet wonderful researchers from various disciplines, providing me with an inter-disciplinary collaboration experience and broadening my perspective on the societal impacts of algorithms.

To the Information Retrieval (IR) Lab at the University of Copenhagen, thank you for the enriching presentations and discussions, as well as for the 3Cs: (paper) clinics, chats, and cakes. A special thanks goes to Vadym, Pietro, and Ervin for being a significant part of my PhD journey. Thank you for being there for me through the ups and downs! Another special thanks goes to Qiuchi and Sara, with whom I had the pleasure to share an office and hold insightful conversations.

During my studies, I had the pleasure of visiting Falk Scholer and the IR research group at RMIT University in Melbourne. This visit has given me new, invaluable insights into the diverse spectrum of IR evaluation research. 

Next, I would like to thank my parents for financially supporting my education up to my master's degree and for ensuring that I receive a good education. To my sister, thank you for being my moral support from 10,000 km away.

To July, Kim, and Morgan, thank you for letting me stay at your lovely home during my research visit in Melbourne, and for making me feel welcome in your family. 
To the Indonesian Student Association in Denmark, thank you for helping me navigate life in a foreign country. 
I am also thankful for the Mensa Foundation, which has awarded me an additional scholarship in relation to my PhD studies. 
I would also like to acknowledge Chloé Rouyer, who has provided a convenient LaTeX template that I have used for this thesis.

Finally, I thank my colleagues and friends, whom I cannot possibly name one by one, for their various forms of support and encouragement that I have received throughout this journey.

\newpage
\tableofcontents

\newpage

\mainmatter 
\mymainpagestyle{} 

\chapter{Executive Summary}
\label{chap:exsum}

\section{Introduction}
\label{s:thesis_intro}
Recommender systems are a type of Artificial Intelligence (AI) that help their users discover relevant items in the form of content, products, services, and so forth based on the users' preferences, needs, or interests \cite{AggarwalRecommenderSystems}. These systems are highly useful in both professional and daily activities, with a wide range of applications. Some common applications include recommending job positions to apply for \cite{Siting2012JobSurvey}, suggesting news articles to read \cite{Karimi2018NewsAhead}, and providing ideas for songs, videos, or movies as entertainment options \cite{Davidson2010YoutubeRecsys,Harper2015TheContext,schedl2015music}. 

Alongside the development of methods for improving recommendation quality, various ways of evaluating recommender systems have been proposed \cite{Zangerle2022EvaluationFramework,Gunawardana2022}. The purpose of evaluation is not only to assess the systems' effectiveness, but also to capture additional important aspects, such as novelty and serendipity \cite{Herlocker2004EvaluatingSystems}. This thesis focuses on one particular `beyond effectiveness' evaluation aspect: fairness. 

Ensuring fairness in AI applications, such as recommender systems, is crucial and even compulsory according to legislations such as the EU AI Act \cite{EUAIAct2024}. While the overall idea of fairness is generally to avoid disadvantaging individuals or specific groups of entities, there is no universal definition of fairness.  
Even in the specific context of recommender systems, many fairness notions arise from what different stakeholders may perceive as `fair' \cite{Wang2023ASystems,Smith2020ExploringSystems,Smith2023ScopingPerspective,Ferraro2021WhatPlatforms,Dinnissen2023AmplifyingPlatforms,Sonboli2021FairnessPerspective} (see \Cref{sss:taxonomy_fairness}). 
Recommender system stakeholders include the consumers and the item providers, which are associated with \textit{user}- and \textit{item}-side fairness, respectively \cite{Burke2018BalancedRecommendation}. User-side fairness typically concerns disparity in the users' recommendation effectiveness, which has to do with ensuring that all users are equally satisfied with their recommendations \cite{Leonhardt2018UserSystems}. 
On the other hand, item-side fairness is usually based on the exposure received by the items, which is about recommending items to users in a way that would benefit the items or their providers \cite{Ekstrand2022FairnessSystems}. For instance, recommending a product to users may increase their awareness of the product, benefiting its seller in the form of advertisement. 
Fairness can also be evaluated at different levels of granularity: for \textit{groups} of users/items and for the \textit{individual} subjects \cite{Wang2023ASystems}. Group fairness usually emphasises that different groups of users/items should receive similar treatment, or have a similar experience \cite{Wang2023ASystems,Ekstrand2022FairnessSystems}. For example, users from different age groups should get recommendations that are equally good \cite{Ekstrand2018AllEffectiveness}. In contrast, individual fairness focuses more on providing equal treatment for similar users or items \cite{Wang2022ProvidingSystems,Wu2023EquippingEmbedding}. 

The deeply contextual nature of fairness, along with the inherent limitation of not being able to label fairness for a given data instance without considering other data instances or additional information, leads to the difficulty in evaluating fairness. Most research on recommender system fairness rely on offline experiments \cite{Deldjoo2024FairnessDirections}, where one or more evaluation measures\footnote{Though some may refer to evaluation measures as \textit{metrics}, this thesis distinguishes them to avoid confusion with distance metrics, which must satisfy additional properties \cite{MetricsEncyclopaedia}.} are computed as a proxy of how fair a system is. 
There is a multitude of such measures under numerous fairness concepts, which can be categorised according to their evaluation granularity or the fairness subject (e.g., users or items), or other grouping criteria \cite{Wang2023ASystems}. 
The sheer number of recommender system fairness evaluation measures, the various fairness definitions they follow, and other factors such as how different their formulations can be, collectively make it challenging to understand their appropriate usage \cite{Majumder2021FairFairness, Richardson2021TowardsToolkits}. These factors motivate a more careful selection of these measures by determining the most suitable measure(s) for specific use cases \cite{Vassy2024Consumer-sideEvaluation}, considering their practical constraints \cite{smith2022recsysfairnessmetricsuse}.

Generally, the strengths and pitfalls of an evaluation measure can be unveiled through theoretical and empirical analyses \cite{Lima2021,Buckley2000EvaluatingStability}. Despite the existence of numerous surveys on recommender system fairness 
\cite{Verma2020FacetsRecommendation,Wang2023ASystems,Amigo2023ASystems,Smith2023ScopingPerspective,LiYunqi2023FairnessApplications,Zhao2025FairnessDiversitySurvey,Wu2023FairnessStrategies,Aalam2022EvaluationReview,Zehlike2022FairnessSystems,Pitoura2022FairnessOverview,Deldjoo2024FairnessDirections}, there is a lack of studies that theoretically dissect the fairness evaluation measures or empirically compare them. At the same time, several measures are proposed without clear analysis on their limitations. Moreover, there is no clear standard on how to assess the goodness of recommender system fairness evaluation measures (\Cref{sss:limitation_eval_measures}). This leads to several problems: 

\begin{enumerate}
    \item There is insufficient understanding of how recommender system fairness measures quantify fairness. Often, the specific kind of recommendation lists producing the (un)fairest score is unknown, and sometimes, even the theoretical measure range is unclear as it is not explicitly stated (e.g., \cite{Yang2023FARA:Optimization,Saito2022FairRanking,Jeunen2021Top-KExposure}) and it is not obvious from the mathematical formulation. 
    Suppose that a measure is bounded at 0, and that this is the fairest score. If its upper bound (the most unfair score) is unknown, it is not possible to understand whether a score of $50$ is closer to the most unfair score or to the fairest. If the upper bound is $100$, then $50$ can be deemed moderately fair. However, if the upper bound is $1000$ instead, $50$ would seem highly fair. Without knowing the measure's range or what the `ideal' output looks like, the score interpretation is highly ambiguous. This necessitates a look into the measures' \textbf{theoretical limitations}.
    
    \item Several empirical properties of recommender system fairness measures are poor\-ly understood. One such property is related to how the measure scores are distributed in their range. For example, two measures with a $[0,1]$-range score the same set of systems conflictingly \cite{Leonhardt2018UserSystems,Zhu2020FARM:APPs}: one has a compressed range of $[0,0.04]$, while the other has a wider range of $[0.6,1]$. Considering that 0 is the fairest score and 1 is the unfairest score, the systems could be seen as extremely fair based on the first measure, but moderately to highly unfair based on the second measure. The difference in the measures' observed empirical range may cause confusion in their score interpretation. 
    Knowing the measures' expressiveness is important, as a measure with a compressed empirical range could give the impression that two systems are comparably fair when they are not. Hence, such \textbf{empirical limitations} also need to be investigated. 

    \item Recommender system fairness measures may not quantify the intended aspects. For instance, individual fairness is generally understood as providing equal treatment to \textit{similar} individuals \cite{Dwork2012FairnessAwareness}, but most evaluation measures for individual fairness in recommender systems simply account for disparity or variation across all individuals, without considering their similarity. To ensure that the measures account for the intended aspects \cite{bauer2024dagstuhlevaluation}, this type of \textbf{conceptual limitations} must be addressed.
\end{enumerate}

\paragraph{Objectives.} The main objectives of this thesis are:

\begin{enumerate}
    \item To assess the theoretical, empirical, and conceptual limitations of existing recommender system fairness evaluation measures; 

    \item To design new evaluation approaches or measures that resolve the above limitations; and 

    \item To recommend the appropriate measure usage, considering their limitations.
\end{enumerate}

Overall, this thesis aims to improve state-of-the-art offline evaluation measures of fairness in recommender systems.

\paragraph{Scope.}
The scope of the papers that are part of this thesis is as follows. 
This thesis examines existing offline fairness evaluation measures in recommender systems, starting from individual item fairness (\textit{Papers 1--4} \Crefrange{chap:TORS24}{chap:WWW25}) and continues with studying user-side fairness for individuals (\textit{Paper 5}, \Cref{chap:PUF}) and for groups (\textit{Paper 6}, \Cref{chap:intersectional}). 
Individual item fairness can be evaluated by considering only item exposure (e.g., how many times it gets recommended to users) or jointly with item relevance (e.g., whether the item suits the users' needs or preferences). To avoid complexity arising from having item relevance as an additional variable, we first study measures that account only for exposure (\textit{Paper 1}, \Cref{chap:TORS24}) separately from those that account for both exposure and relevance (\textit{Papers 2--4}, \Crefrange{chap:SIGIR24}{chap:WWW25}). The fairness types evaluated and investigated in this thesis are summarised in \Cref{tab:paper_fairness_overview}.

\begin{table}
\centering
\caption{The three fairness types covered by the six papers included as chapters in this thesis (P1--P6). Fairness is categorised based on the combination of evaluation granularity (individuals vs.~groups) and evaluation subject (items vs.~users). 
Fairness for groups of items is not included as it has been extensively studied \cite{Schumacher2024PropertiesRankings,Raj2022MeasuringResults,Sakai2023VersatileRelevance}.
}
\label{tab:paper_fairness_overview}

\begin{tabular}{l|llllll}
\toprule
 Fairness type & \multicolumn{1}{c}{P1} & \multicolumn{1}{c}{P2} & \multicolumn{1}{c}{P3} & \multicolumn{1}{c}{P4} & \multicolumn{1}{c}{P5} & \multicolumn{1}{c}{P6} \\

 \midrule
Individual items & \cm & \cm & \cm & \cm &  &  \\
 \midrule
Individual users &  &  &  &  & \cm & \cm \\
 \midrule
Groups of users &  &  &  &  &  & \cm \\
 \bottomrule
\end{tabular}
\end{table}

In summary, all papers that are part of this thesis evaluate fairness in an offline setting. Altogether, the papers cover fairness for both subjects (user and item fairness), as well as for both granularity levels (group and individual fairness). Furthermore, all evaluations are based on the model output (i.e., the recommendation lists) instead of the process, e.g., based on how the users or items are represented by a model's internal mechanism \cite{Li2021LeaveUsers,Li2021TowardsNotion,Wu_Wu_Wang_Huang_Xie_2021}). For the most part, the focus is on evaluation at a fixed point in time, rather than over a time period \cite{Biega2018EquityRankings}. 

While we have covered three common types of fairness within the above-mentioned scopes, several types are excluded. 
Firstly, item-side group fairness evaluation measures are excluded as they have been extensively studied in the broader field of ranked outputs (e.g., \cite{Schumacher2024PropertiesRankings,Sakai2023VersatileRelevance,Raj2022MeasuringResults}). 
Additionally, when studying individual fairness measures, we have excluded measures of counterfactual fairness and fair representation learning for recommendation, as they are not computed based on the recommendation list (e.g., \cite{Li2021TowardsNotion,Shao2024AverageCounterfactual}). We have also excluded (item) subject fairness, as it is highly similar to item-side fairness \cite{Ekstrand2022FairnessSystems,Knees2024}. 
Finally, our studies also exclude case-specific fairness measures due to their limited applicability to the general recommendation task. For example, some measures are precisely designed only for fairness among items with little historical data \cite{Zhu2021FairnessSystems} or only for group recommendations, where the task is to recommend the same item(s) to a group of users \cite{Serbos2017FairnessRecommendations,Sacharidis2019Top-nFairness,Xiao2017Fairness-AwarePareto-Efficiency}. 

\paragraph{Chapter Outline.} The rest of this chapter is organised as follows. 
Section~\ref{s:theory} provides an overview of recommender systems, an introduction to fairness in recommender systems, as well as the theoretical background for offline evaluation measures of recommender system fairness and their limitations. Section~\ref{s:contribution} presents a detailed overview of the contributions made by the papers included as chapters in this thesis.  
Section~\ref{s:summary_future} summarises the contributions of this thesis and discusses promising directions for future research. 

\section{Theoretical Background}
\label{s:theory}

This section provides an overview of recommender systems (\Cref{ss:overview_recsys}), an introduction to fairness in recommender systems (\Cref{ss:fairness_recsys}), 
as well as the theoretical background on offline evaluation of recommender system fairness that is necessary for understanding the subsequent chapters of this thesis (\Cref{ss:offline_eval_fairness}). Cross-references to the papers included in this thesis are made when relevant.

\subsection{Overview of Recommender Systems}
\label{ss:overview_recsys}

The overview of recommender systems is explained in three parts: the general recommendation task (\Cref{sss:recommendation_task}), the basic and advanced approaches in recommendation (\Cref{sss:approaches_recommendation}), as well as how recommender systems are evaluated (\Cref{sss:eval_recsys}).

\subsubsection{The Recommendation Task}
\label{sss:recommendation_task}

The main goal of recommender systems is to help users find the most appealing items that would match their preferences, suit their needs, and so on, from a (usually large) set of items. For simplicity, this thesis refers to such items as `relevant items'. 
In this thesis, the recommendation task is to first predict the relevance score of each item to the user, then rank the items based on relevance, and finally present the list of most relevant items in decreasing relevance \cite{Maron1960RelevanceRetrieval,Cremonesi2010PerformanceTasks}.\footnote{There are also other ways of generating or presenting the recommendations, e.g., displaying them in a grid \cite{Raj2024GridLayout} or carousels \cite{FerrariDacrema2022}.} 
This is done for each user in a given set of users. Typically, the focus is on recommending items that have not been presented to the user \cite{Canamares2020}, even though re-recommendation can be useful in some cases, such as re-recommending consumables \cite{Lerche2016RemindersRecommendations} and songs \cite{manolovitz2020practical}.
 
\subsubsection{Approaches in Recommendation}
\label{sss:approaches_recommendation}

To personalise the item recommendation, recommender systems learn from the users' past interactions. These interactions exist in the form of explicit and implicit feedback, also referred to as ratings. Explicit feedback indicates whether a user likes or dislikes an item (e.g., via a like/dislike button), and to what extent (e.g., via a 5-star scale) \cite{AggarwalRecommenderSystems}. For implicit feedback, user preferences are inferred from their activities rather than explicitly expressed by the user \cite{AggarwalRecommenderSystems}. For example, clicking on or buying an item may indirectly indicate a positive preference for that item \cite{AggarwalRecommenderSystems}. 
Other information, such as user/item attributes, as well as contextual information (e.g., time and the user's location), may also be used to enhance the recommendation \cite{Adomavicius_Mobasher_Ricci_Tuzhilin_2011,AggarwalRecommenderSystems}.

Two of the most common recommendation methods are collaborative filtering methods and content-based methods \cite{AggarwalRecommenderSystems}. Collaborative filtering is based on the idea that similar users would prefer similar items \cite{Schafer2007}. Content-based methods mainly utilise item attributes/features to select other items that are similar to the ones that the users preferred in the past \cite{Pazzani2007}. This thesis mainly utilises collaborative filtering methods. Some examples of collaborative filtering methods are popularity-based models, which recommend the most popular items \cite{Ji2020PopularityBaseline}; matrix factorisation models, which characterise users and items as vectors of latent factors and predict with inner product the user ratings for the unrated items \cite{Koren2009Matrix}; and neighbourhood-based models, which average ratings from similar users or items \cite{Deshpande2004Item-basedAlgorithms,Resnick1994GroupLens:Netnews}. 
More advanced collaborative filtering methods are usually deep learning-based  \cite{He2017NeuralFiltering,Zhang2019DeepPerspectives}. Recent work has also utilised large language models as recommenders \cite{Hou2024LargeSystems,Liao2024LLaRA:Assistant}.

\subsubsection{Evaluation of Recommender Systems}
\label{sss:eval_recsys}

At a high level, recommender system evaluation protocols can be categorised into three: offline evaluation, user study, and online evaluation \cite{Zangerle2022EvaluationFramework,Gunawardana2022,Deldjoo2024FairnessDirections}. 
Offline evaluation is done on a pre-collected dataset that contains the users' past interactions (e.g., ratings \cite{Harper2015TheContext}, listening history \cite{Schedl2016TheRecommendation}), with the purpose of collecting quantitative data for predictive performance evaluation \cite{Zangerle2022EvaluationFramework}. This is contrary to the other two protocols, which require real-time user input \cite{Freyne2013}, enabling the collection of quantitative and qualitative data through eye-tracking experiments, action logs, and/or questionnaires \cite{Zangerle2022EvaluationFramework}. In user studies, a small number of participants are recruited and tasked to engage with a prototype recommender system, in a laboratory setting (controlled environment), whereas online evaluation involves collecting data from real-world users' interactions with different versions of recommender systems in production \cite{Gunawardana2022}, such as through A/B testing \cite{GomezUribe2016Netflix}. 

User studies and online evaluations are typically much more expensive than offline evaluation, in terms of planning time, experiment time, and financial cost \cite{Zangerle2022EvaluationFramework}. In contrast, the relatively low cost of conducting offline evaluations allows researchers to compare model performance on several datasets relatively quickly \cite{Herlocker2004EvaluatingSystems}. As such, offline evaluation can serve as the first evaluation phase, for example, to eliminate worse-performing recommender models from the candidate pool of models, before further testing in user studies or online evaluation \cite{Gunawardana2022}. 
This thesis focuses on offline evaluation; this evaluation protocol is described in more detail below.

Offline evaluation for recommender systems largely follows experimental practice in machine learning and information retrieval \cite{Castells2022OfflineDirections}. To conduct an offline evaluation with a pre-collected dataset, the dataset needs to be split into at least two parts: training and testing. Others split the dataset into three: training, validation, and testing \cite{Meng2020ExploringModels}. 
The training split would be input to a recommender model \cite{Herlocker2004EvaluatingSystems}, such that the model learns and infers user preferences from past interaction data. The validation split may be used to select the final model in the hyperparameter tuning process. 
The testing split serves as the ground truth for evaluating the predictions or recommendations generated by the model. 
The datasets can be split in many ways, for example, randomly or temporally  \cite{Canamares2020}. Another factor to consider is whether the data should be split per user or split globally, considering all users. Recent work recommend performing a global temporal split, as other data splitting strategies may cause the recommender model to learn from `future' interactions that would not be available yet at prediction time
\cite{JiYitong2023AEvaluation,Meng2020ExploringModels}.

Based on the testing split, model performance can be assessed through \textit{evaluation measures}. Usually, these measures are computed to assess the recommendation effectiveness, but they can also evaluate other aspects such as novelty \cite{CastellsNoveltySystems}, diversity \cite{Canamares2020}, or fairness \cite{Wang2023ASystems}; this thesis evaluates recommendation effectiveness and fairness (\Cref{sss:eval_measures_fairness}). Generally, these evaluation measures map a ranking of items to a numeric score. Traditional information retrieval measures such as Precision at cut-off $k$ (P@$k$) and Normalized Discounted Cumulative Gain at cut-off $k$ (NDCG@$k$ \cite{Jarvelin2002CumulatedTechniques}) are often used to evaluate the effectiveness of recommender systems \cite{Castells2022OfflineDirections}.\footnote{Many more effectiveness measures exist; these are just two examples, and they are not necessarily the optimal ones for all use cases.} 
The measure P@$k$ quantifies the proportion of relevant items that are within the top-$k$ recommendation list, while NDCG@$k$ compares how close a ranking is to the ideal ranking that follows a decreasing order of relevance. The choice of $k$ is typically small due to the limited size of the interface \cite{FerrariDacrema2022}, e.g., between 1 and 20 \cite{Alhijawi2023SurveyRecSys}. This means that only a few specific items will be presented to the users \cite{Cremonesi2010PerformanceTasks, Canamares2020}.

Pre-collected datasets are often incomplete \cite{Gunawardana2022,Castells2022OfflineDirections}; there may be additional items that are relevant to a user other than items that have been highly rated by the user, clicked, and so on. As such, the recommender model may struggle to learn from the limited amount of data, which could result in very low effectiveness scores \cite{Castells2022OfflineDirections}. The incompleteness issue could also affect evaluation in another way: items without ratings are usually deemed as non-relevant, resulting in an underestimation of the model's effectiveness \cite{Canamares2020}. 
Thus, the low scores do not necessarily mean that the model is ineffective \cite{Castells2022OfflineDirections}.

\subsection{Fairness in Recommender Systems}
\label{ss:fairness_recsys}

Recommender system fairness is introduced in two parts: the taxonomy of fairness followed by this thesis (\Cref{sss:taxonomy_fairness}) and strategies for fair recommendation (\Cref{sss:fair_strategies}). 
Note that fairness evaluation is covered separately in \Cref{ss:offline_eval_fairness}.

\subsubsection{Taxonomy of Fairness in Recommender Systems}
\label{sss:taxonomy_fairness}

There are many ways of viewing fairness in recommender systems. Two common ways of categorising recommender system fairness across various surveys (e.g., \cite{Li2021TutorialSystems,Wang2023ASystems,Aalam2022EvaluationReview, Amigo2023ASystems}) are by considering the fairness \textit{subject} and the partition \textit{granularity} of the subjects. 
In terms of subject, recommender system fairness is often classified into \textit{item fairness} and \textit{user fairness} \cite{Wang2023ASystems}.\footnote{Some literature also refer to item and user fairness as provider-side and consumer-side fairness, respectively \cite{Ekstrand2022FairnessSystems}.} 
Based on partition granularity, fairness is commonly divided into \textit{individual fairness} and \textit{group fairness} \cite{Amigo2023ASystems}. 

While it is not possible to provide a general definition of fairness that always holds, even after categorising fairness in terms of subject and/or granularity, the intuition behind each fairness type and its importance is provided below.

\paragraph{Item fairness.} 
Fairness for items is often defined in terms of disparity in item utility. The more equal the utility across (types of) items, the fairer the recommendation. 
Item utility has typically been associated with the \textit{exposure} provided by a recommender system to various (types of) items. The utility can also be a combination of item exposure and the system's \textit{effectiveness} of recommending items to users for whom the item would be relevant \cite{Amigo2023ASystems}.

Firstly, item exposure can be understood as how often an item gets recommended to users and at which rank position. The more often an item is recommended to users, the higher the exposure it receives. Similarly, top-ranked items can be said to receive more exposure than bottom-ranked items, as position bias causes them to be more likely seen and further actioned upon \cite{Joachims2005AccuratelyFeedback}.
As such, item fairness is important mainly for the item provider/supplier side, whose goal is for someone to discover the products, services, or content that they have provided. This discovery can lead to many actions and potential benefits \cite{Burke2017MultisidedRecommendation}. For example, on a music streaming platform, if a user gets recommended a song and listens to it, this may count towards the royalty that the artist earns.\footnote{\url{https://support.spotify.com/us/artists/article/royalties/}} 
Likewise, if a user purchases an item at an e-commerce site, the seller could earn money. Therefore, if a recommender system only recommends the same few items to everyone, many items will never be shown to users, ultimately disadvantaging the item providers.

Secondly, regarding effectiveness-based item fairness, it is important to recommend the items to the right users. In other words, exposure should be allocated proportionally to the item's relevance \cite{Biega2018EquityRankings}. For instance, a fair movie recommender system may consider that blockbuster movies would cater to the broader audience's taste and recommend these movies more frequently than movies with a niche genre, which may have fewer fans. Instead, if the system recommends niche movies to the general audience, this would increase the items' exposure and promote item fairness (considering the definition that is purely based on exposure), but at the expense of users' satisfaction. 
An example of an unfair recommendation for items would be a job recruitment platform that ranks applicants with similar qualifications far apart \cite{Deldjoo2024FairnessDirections}; here, the applicants are the `items'. The lower-ranked candidate would experience reduced visibility to recruiters, which decreases their chance of being selected to proceed in the recruitment process, jeopardising their chances of employment and thus their livelihood. 

This thesis studies item fairness in \textit{Papers 1--4} (\Crefrange{chap:TORS24}{chap:WWW25}).

\paragraph{User fairness.} Fairness for users is commonly defined in terms of user utility, e.g., the recommendation effectiveness. The most common definition revolves around the idea of having a more equal recommendation effectiveness across different (types of) users would mean that the recommendation is fairer \cite{Ekstrand2018AllEffectiveness}. It is important for all users to receive recommendations that effectively match their preferences and needs; otherwise, the users may not be satisfied with the recommendations and may abandon the platform \cite{Wang2023ASystems}. 
For instance, a job recommendation platform that constantly provides irrelevant suggestions for job positions to certain (groups of) users could be deemed unfair for these users. This is because the (groups of) users may be disadvantaged in their job searching process, as they have a harder time finding relevant jobs than other users.

Another possible definition for user fairness involves providing different treatment for different groups of users. For example, \citeauthor{Deldjoo2019RecommenderEntropy}~\cite{Deldjoo2019RecommenderEntropy} argue that paying users should receive better recommendations than non-paying users. The idea is that because paying users spend money to use the platform, their recommendation quality should be better than that of the non-paying users. 
Hence, fairness for users does not always mean that the recommendation effectiveness has to be equal for all users or user groups.

In this thesis, user fairness is covered in \textit{Papers 5--6} (\Crefrange{chap:PUF}{chap:intersectional}).

\paragraph{Individual fairness.} Individual fairness generally refers to providing equal treatment to similar individuals \cite{Dwork2012FairnessAwareness}. 
The idea is that users with similar characteristics (e.g., in their past interactions \cite{Wu2023EquippingEmbedding}) should receive similar recommendations, or that equal exposure should be given to similar items (e.g., similarity in terms of their popularity in the training data \cite{Ge2021TowardsRecommendation,pellegrini2023fairnessallinvestigatingharms} or in their learned representations \cite{Wang2022ProvidingSystems}). 
As there is no agreement on how the similarity should be measured, and that the similarity measure is often task-specific \cite{Dwork2012FairnessAwareness}, individual fairness is also often based on the idea that all individuals should be treated equally, regardless of their similarity \cite{Leonhardt2018UserSystems,Patro2020FairRec:Platforms}. Based on this notion of individual fairness, a recommender system would be highly unfair towards individual users, for example, if the recommendation effectiveness varies widely across users, meaning that some users receive high-quality recommendations and others receive low-quality recommendations. 
Similarly, a system would be highly unfair towards individual items, for instance, if some items get recommended to users much more frequently than others, while some items never get recommended at all. 

Individual fairness is investigated in each paper included in this thesis (\textit{Papers 1--6}, \Crefrange{chap:TORS24}{chap:intersectional}).

\paragraph{Group fairness.} 
Group fairness is typically associated with ensuring that the experience of different user groups or item groups is comparable \cite{Ekstrand2022FairnessSystems}. 
Under this fairness notion, similar users and items are grouped based on their attributes or characteristics. Often, these attributes (e.g., ethnicity, religion, gender) are defined in anti-discrimination laws, which prohibit discrimination based on specific group membership \cite{barocas2023fairnessMLbook}. However, users and items can also be grouped based on other attributes that do not directly relate to socio-demographic background (e.g., user activity level \cite{Li2021User-orientedRecommendation,Fu2020Fairness-AwareGraphs}, user personality traits \cite{Chen2025InvestigatingRecommendations}, or item popularity \cite{Fu2020Fairness-AwareGraphs}). The idea is that regardless of the socio-demographic background, their activity level, or their prior popularity, different groups of users/items should be able to receive similar treatment or similar utility (e.g., similar recommendation effectiveness for different user groups \cite{Ekstrand2018AllEffectiveness} or similar exposure for relevant item from different groups \cite{prost2022simpsonsparadoxrecommenderfairness}). Even if these attributes are not explicitly part of the input, they may be implicitly present \cite{FeldmanCertifying,Xu2024AModels}. Hence, ensuring that the recommendations are fair towards different groups requires more effort than omitting the use of group membership information from the input.

Unfairness may also arise from combinations of the attributes, i.e., intersectionality  \cite{Crenshaw1991MappingColor,Kearns2019AnLearning,Ekstrand2022FairnessSystems}). For example, the generated recommendations can be fair towards users when they are grouped based on only gender or only age, but extremely unfair when the intersectionality between gender and age is considered \cite{Deldjoo2025CFaiRLLM}. Other work considers fairness for different groups of users and items simultaneously \cite{Kheya2025UnmaskingFairness,Wang2024IntersectionalRecommendation}. Even though the recommendation effectiveness may be equal across different user groups, different groups of items may be recommended to each user group, which can be seen as unfair, as it may perpetuate certain existing social stereotypes  \cite{Kheya2025UnmaskingFairness}.

This thesis investigates group fairness in \textit{Paper 6} (\Cref{chap:intersectional}). 

\paragraph{Other fairness categories.} There are other ways of categorising fairness beyond item/user fairness or individual/group fairness. One such category is \textit{subject fairness}, which concerns how item subjects are treated \cite{Ekstrand2022FairnessSystems}. Subject fairness refers to having a diverse representation of individuals that the items are about, e.g., people or entities covered in news articles \cite{Ekstrand2022FairnessSystems, Knees2024}. 
In this regard, subject fairness could be seen as highly similar to item-side fairness \cite{Ekstrand2022FairnessSystems}. However, in subject fairness, the idea of exposing/recommending a diverse list of items to each user is even more important than having an overall diverse recommendation across all users, which is the typical goal of item-side fairness \cite{Ekstrand2022FairnessSystems}. Another type of fairness is \textit{counterfactual fairness}, which ensures that individuals receive the same outcome if their group membership is changed to another group \cite{Wang2023ASystems, Pearl2009CausalOverview}. 
Counterfactual fairness can also be seen as a specific type of individual fairness \cite{Kusner2017CounterfactualFairness,Wang2023ASystems}. In addition, there is also \textit{within-group fairness}, which is related to the idea of providing an equitable outcome to all individuals within a group \cite{pellegrini2023fairnessallinvestigatingharms,Pitoura2022FairnessOverview}. These are just some examples of other fairness categories; many more ways of viewing fairness exist, such as static vs.~dynamic fairness or single-sided vs.~multi-sided fairness \cite{Li2021TutorialSystems}. 

While this thesis also covers within-group fairness (\textit{Paper 6}, \Cref{chap:intersectional}), it does not cover subject fairness or counterfactual fairness, as explained in the scope of the thesis (\Cref{s:thesis_intro}).

\subsubsection{Strategies for Fair Recommendation} 
\label{sss:fair_strategies}

Improving recommendation fairness can be done in several ways. However, increasing fairness may decrease recommendation effectiveness \cite{karimi2023providertradeoffs,Zehlike2022FairGroups,Gomez2025EnhancingAwareness}. 
Hence, rather than achieving the maximum possible fairness, fairness-enhancing strategies focus on increasing fairness while maintaining an acceptable level of recommendation effectiveness. Broadly, these strategies can be categorised into data-oriented methods, ranking methods, and re-ranking methods \cite{Wang2023ASystems}.

Data-oriented methods refer to the modification or augmentation of training data, such that the recommender learns from more balanced interaction samples (e.g., in terms of user group proportions \cite{Ekstrand2018AllEffectiveness}, or user preferences \cite{Rastegarpanah2019FightingSystems}). The idea is that, with these adjusted samples, the recommender would generate more equitable recommendations towards various users or user groups. Meanwhile, fair ranking strategies are primarily in the form of recommendation models that are optimised for fairness, e.g., by incorporating fairness-related terms in the model's objective function \cite{Burke2018BalancedRecommendation,Yao2017BeyondFiltering}. As such, the model is optimised for both recommendation effectiveness and fairness. 
Lastly, fair re-ranking methods receive a recommendation list generated by a recommendation model and reorder the items in the list, with the aim of improving fairness (e.g., by pushing underrepresented items closer towards the top of the list \cite{Zehlike2017FAIR:Algorithm}). The re-ranking can be done independently per user or simultaneously across all users.

While not the main contribution of this thesis, we explore fair re-ranking methods to improve item fairness in Papers \textit{2--4} (\Crefrange{chap:SIGIR24}{chap:WWW25}).

\subsection{Offline Evaluation of Recommender System Fairness}
\label{ss:offline_eval_fairness}

This thesis focuses on evaluating recommender system fairness in an offline setting. As such, fairness evaluation is done by computing measures as a proxy of the disparity in the recommendations. 
The general idea of how fairness is quantified through the measures is explained in \Cref{sss:eval_measures_fairness}, while their limitations are covered in \Cref{sss:limitation_eval_measures}.

\subsubsection{Evaluation Measures of Fairness in Recommender Systems}
\label{sss:eval_measures_fairness}

Many survey papers have collected existing measures for quantifying fairness in recommender systems and categorised them in various ways based on their use cases \cite{Wang2023ASystems,Amigo2023ASystems,Smith2023ScopingPerspective,LiYunqi2023FairnessApplications,Wu2023FairnessStrategies,Aalam2022EvaluationReview,Zehlike2022FairnessSystems,Pitoura2022FairnessOverview,Zhao2025FairnessDiversitySurvey,Deldjoo2024FairnessDirections}. A common practice across previous work is to cluster these measures according to the evaluation subject for whom fairness is measured (i.e., users vs.~items) and the evaluation granularity, which refers to the level at which fairness is assessed (i.e., individuals vs.~groups). This clustering is aligned with the fairness taxonomy used in this thesis (\Cref{sss:taxonomy_fairness}). Note that some measures overlap across categories as they can quantify more than one fairness type \cite{Wang2023ASystems,Amigo2023ASystems}.

To illustrate the workings of existing evaluation measures for fairness in recommender systems, the typical variables in the measure equations and how they are used are summarised here, following the categorisation based on evaluation subject and evaluation granularity. The specific measures and their mathematical equations are introduced in the later chapters of this thesis; the chapters are referenced in each category below.

\paragraph{Item fairness measures.} 
Item fairness is widely defined in terms of how recommender systems allocate exposure to different (types of) items \cite{Wang2023ASystems}. As such, item fairness measures generally quantify disparity in item exposure, which can be inferred from the frequency of an item's appearance in the recommendation lists across all users. The list can either be the top-$k$ recommended items or all items. Often, item exposure is weighed based on the rank position at which the item appears. Most measures compare exposure between items either in a pairwise manner or altogether across all items. The item exposure is also compared to the number of items in the dataset and to the number of recommendation slots, i.e., the total number of items that are recommended to all users. This is because the total amount of exposure that can be distributed to the items is limited by the number of slots.

A subset of these measures also considers item relevance to the user; this thesis refers to them as `joint' measures or `relevance-aware' measures. Some measures compare item relevance to its exposure; the idea is that items with higher relevance should receive more exposure. 
Alternatively, item relevance and item exposure are combined to indicate user interaction: if a relevant item is exposed to a user, it is assumed that the user will interact with the item \cite{Saito2022FairRanking}. Unfairness is then measured as the disparity of these interactions across all items. The higher the disparity, the more unfair the recommendation.

\textit{Paper 1} (\Cref{chap:TORS24}) of this thesis presents item fairness measures that are based only on exposure (detached from relevance). \textit{Papers 2--3} (\Crefrange{chap:SIGIR24}{chap:SIGIR24_ext}) detail measures that quantify relevance-aware item fairness. 

\paragraph{User fairness measures.} 
User fairness generally refers to how fair a recommender system is to different (types of) users \cite{Wang2023ASystems}. 
Measures of this type commonly score fairness based on the difference in user utility. First, as a proxy of user utility, the recommendation for each user is evaluated for effectiveness with measures such as Precision at $k$ (P@$k$) or NDCG@$k$ (\Cref{sss:eval_recsys}). Afterwards, the effectiveness score per user is fed into another measure that aggregates the disparity in recommendation effectiveness across individuals or across groups. The aggregated score would be taken as the fairness score. The higher the disparity in recommendation effectiveness, the more unfair it is.

Some measures quantify fairness for users through the difference in the list of recommended items, regardless of the items' relevance to the users \cite{Wu2023EquippingEmbedding}. These measures first gauge the dissimilarity between the recommended items in two lists, e.g., by considering the number of overlapping items. Similar to how utility-based user fairness measures operate, the aggregated recommendation similarity across users or groups of users is regarded as the fairness score. More similar recommendations translate to a fairer score.

In this thesis, the formulations of user fairness measures are given in \textit{Papers 5--6} (\Crefrange{chap:PUF}{chap:intersectional}). \textit{Paper 5} (\Cref{chap:PUF}) focuses on both effectiveness-based and non-effectiveness-based user fairness measures, while \textit{Paper 6} (\Cref{chap:intersectional}) examines only the former.

\paragraph{Individual fairness measures.}
Individual fairness is often defined as treating similar individuals similarly \cite{Dwork2012FairnessAwareness}. There are two types of individual fairness measures: those that are similarity-based and those that are not. 
Many measures for individual fairness in recommender systems are not similarity-based; they quantify individual fairness as the disparity or variation of utility (e.g., item exposure or recommendation effectiveness) across all individuals. A handful of measures are similarity-based; they compute similarity between pairs of individuals based on the representation of the individual, e.g., the user/item embeddings \cite{Wu2023EquippingEmbedding,Wang2022ProvidingSystems}. In these measures, only pairs of individuals exceeding a similarity threshold contribute to the fairness score.

The details of individual fairness measures can be found in \textit{Papers 1--3} and \textit{Papers 5--6} (\Crefrange{chap:TORS24}{chap:SIGIR24_ext} and \Crefrange{chap:PUF}{chap:intersectional} respectively). 
\textit{Papers 1} and \textit{5} (\Cref{chap:TORS24,chap:PUF}) cover both similarity-based and non-similarity-based individual fairness measures, while \textit{Paper 2, 3} and \textit{6} investigate only the latter.

\paragraph{Group fairness measures.}
Based on the number of groups, two types of group fairness measures exist: measures that are defined only for two groups and those that can operate with two or more groups. Measures of the first type tend to directly compare utility between the two groups, e.g., taking the difference or the ratio between them. 
Some measures that can be used for two or more groups average the utility difference between group pairs, while others quantify fairness with respect to variation across groups. The group utility scores can either be weighted or used as they are (unweighted). The weights can be set arbitrarily, but they can also be determined based on the proportion of group utility (to the total utility), or proportion of the group size (to the total population) \cite{Amigo2023ASystems}.

This thesis employs group fairness measures that are explicitly defined for two or more (user) groups. These measures are overviewed in \textit{Paper 6} (\Cref{chap:intersectional}).

\subsubsection{Limitations in Existing Fairness Evaluation Measures}
\label{sss:limitation_eval_measures}

Evaluation measures for recommender system fairness may have some limitations. These limitations are either in the measure formulas or arise from the empirical use of the measure. While in most cases the measures can still be computed to evaluate fairness, these limitations may restrict the measure's applicability, interpretability, expressiveness, stability, or efficiency.
Hence, it is important to identify, analyse, and resolve these limitations when possible. 

Determining the limitations of a fairness measure is a complex problem of its own. Before the first paper in this thesis (\Cref{chap:TORS24}, \cite{Rampisela2024EvaluationStudy}), there were not many existing work that examines the possible limitations of fairness measures in recommender systems. To the best of our knowledge, the only work that touches upon the issue was \citeauthor{Raj2022MeasuringResults}~\cite{Raj2022MeasuringResults}, which mentions and addresses the issue of edge cases in group fairness measures for ranking, as part of a framework for identifying the conceptual differences between measures. 
Previous work studies the limitations of other measure types by first defining a list of properties or desiderata. However, for recommender system fairness measures, there is no standard `checklist' to follow.

Other prior work from a wide range of fields within and outside computer science have constructed a number of lists detailing properties for various types of evaluation measures: effectiveness \cite{Moffat2013SevenMetrics}, diversity \cite{Albahem2018DesirableDiversity,Amigo2018AxiomaticDiversity}, fairness in computer network \cite{jain1984quantitative}, and income inequality in the economics domain \cite{Allison1978MeasuresInequality,Bourguignon1979DecomposableMeasures}. 
Some general numerical properties, such as whether the measure score is bounded, can be readily adopted as ideal characteristics for recommender system fairness evaluation measures. However, not all properties would serve as suitable criteria. Firstly, some of these properties are not specifically tailored for fairness objectives in recommender systems, which means that they may not be particularly meaningful or relevant in the context of recommender system fairness. Secondly, some of them are not directly applicable and require modification to suit the fairness context. Thus, determining the measure limitations demands significant effort in collecting properties from existing work and adapting them for recommender system fairness evaluation measures, or even proposing new, specific criteria that suit the context.

\begin{figure}
\centering
\resizebox{0.95\textwidth}{!}{
\begin{tikzpicture}
    \node (A) at (0, 2.5) {
        \begin{tcolorbox}[colback=red!20, colframe=red!50, width=8cm, height=2cm,  valign=center, title=\textbf{Theoretical Limitations}]
            Mainly affect measure \textbf{applicability} and \textbf{interpretability}
        \end{tcolorbox}
    };

    \node (B) at (-5.5, -1) {
        \begin{tcolorbox}[colback=red!20, colframe=red!50, width=8cm, height=2cm,  valign=center,title=\textbf{Empirical Limitations}]
            Mainly affect measure \textbf{expressiveness}, \textbf{stabi\-li\-ty}, and \textbf{efficiency}
        \end{tcolorbox}
    };

    \node (C) at (5.5, -1) {
        \begin{tcolorbox}[colback=red!20, colframe=red!50, width=8cm, height=2cm,  valign=center, title=\textbf{Conceptual Limitations}]
            Mainly affect measure \textbf{applicability} and \textbf{expressiveness}
        \end{tcolorbox}
    };

    \draw[<->, line width=0.8mm] (A) -- (B);
    \draw[<->, line width=0.8mm] (A) -- (C);
    \draw[<->, line width=0.8mm] (B) -- (C);

\end{tikzpicture}}
\caption{Overview of the three types of 
limitations identified, analysed, or resolved in this thesis for existing evaluation measures of recommender system fairness.}
\label{fig:limitations}
\end{figure}
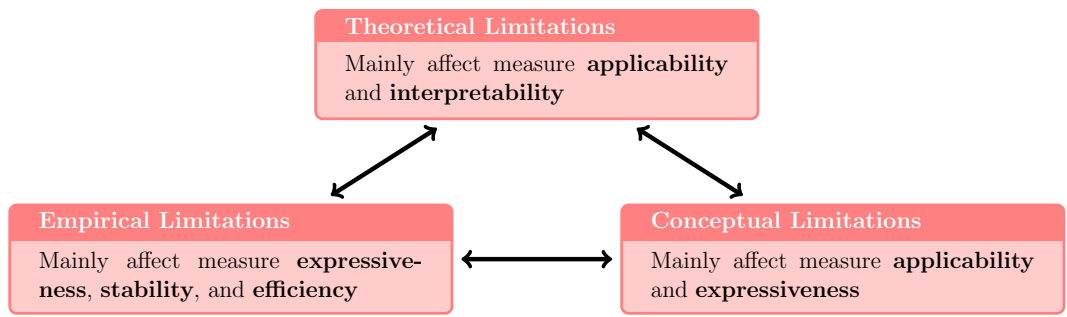

This thesis explores a series of measure limitations and groups them into three classes. The following part explains the three classes of measure limitations in terms of how they can be identified, their practical implications, and their relationships. \Cref{fig:limitations} overviews the measure limitations.

\paragraph{Theoretical limitations.}
These limitations arise directly from the measure's mathematical formulation and can be primarily identified by scrutinising the equation or testing it with different values to simulate common and extreme cases. For example, the equation may be ill-defined or fail to handle some scenarios. Consequently, for these cases, the measure cannot be computed due to, e.g., division by zero, and no score is produced. While some prior work handled these cases in their implementation, the issue with the general measure formulation itself remains, decreasing the measure's \textbf{applicability}.

In other cases, the precise range of the measure is unknown. Various reasons for this include ambiguous measure equations (e.g., details such as the logarithmic base are not stated) and the measure range is not mentioned in the original work proposing the measure. Even if the measure range is known, there are cases where the (un)fairest possible recommendation is not mapped to the endpoints of the range.\footnote{This is similar to the realisability property in effectiveness measures \cite{Moffat2013SevenMetrics}.}  
As the maximum or minimum achievable score is unknown, it is not possible to assess how close the recommendation is to the (un)fairest scenario. Thus, it is difficult to \textbf{interpret} the measure scores.

\textit{Paper 1} (\Cref{chap:TORS24}) and \textit{Paper 3} (\Cref{chap:SIGIR24_ext}) critically analyse theoretical limitations in existing individual item fairness measures for recommender systems. These papers pinpoint the specific causes and conditions resulting in the limitations. Additionally, they formulate corrections to the measures to counter the limitations or justify why the limitations are not resolvable. 

\paragraph{Empirical limitations.}
These are properties that emerge during or after the actual computation of the measure. Several measure characteristics can be tested through experiments with real and synthetic datasets, for example, \textbf{expressiveness}: Does a measure tend to always score very low or very high? How does it respond to changes in the recommendation list or item relevance? A measure with a consistently compressed empirical range has limited sensitivity, which makes it hard to distinguish the fairness of different recommendations.

On the other hand, there exist conditions that ideally should not impact the measure score or only minimally change it (e.g., ties in item relevance should not change the score considerably). If the measure score is largely affected by such conditions, its \textbf{stability} is questioned. Furthermore, there is also the question of \textbf{efficiency}: can the measure be computed in a reasonable time? A measure that requires extensive computational time may not be desirable.

All of the above concern limitations within a single measure. When comparing two or more measures, there may be concerns whether a particular measure is redundant: can the same conclusion be achieved by computing another measure instead? 
Existing literature on ranking effectiveness measures  \cite{Webber2008PrecisionRedundant} and on machine learning fairness measures \cite{Majumder2021FairFairness} highlight the \textbf{redundancy} of several measures, as computing another measure leads to similar conclusions. One way of assessing the conclusion similarity is by computing the measure agreement in ranking recommender models from the best to the worst (e.g., from the fairest to the most unfair) \cite{Sakai2014MetricsTest,Sakai2019DiversityGood}. This can be done by, for example, calculating the Kendall's $\tau$ correlation coefficient \cite{Kendall1938Correlation} between the two rankings \cite{Sakai2023VersatileRelevance}.

Each of the six papers included in this thesis covers empirical limitations. Specifically, stability-related issues are analysed in \textit{Paper 3} (\Cref{chap:SIGIR24_ext}); measure efficiency is investigated in \textit{Paper 3} (\Cref{chap:SIGIR24_ext}), \textit{Paper 4} (\Cref{chap:WWW25}), and \textit{Paper 5} (\Cref{chap:PUF}); measure expressiveness and redundancy are investigated in each paper. 

\paragraph{Conceptual limitations.} 
These limitations refer to certain important aspects pertaining to fairness that cannot be quantified by existing evaluation measures. This is because the measure may not be created for that purpose. An example of such aspects is user/item similarity. This aspect is often left out of individual fairness measures, even though individual fairness is commonly defined in terms of equality across similar individuals \cite{Dwork2012FairnessAwareness}. 
Naturally, a measure that does not account for a particular aspect in its equation can neither be used to evaluate nor expected to be sensitive to that aspect. 
This results in a need to modify existing measures or to formulate a new evaluation measure that is custom-built for the specific aspect(s), e.g., by including the missing aspect in the measure equation.

In this thesis, conceptual limitations are addressed in \textit{Papers 4--6} (\Crefrange{chap:WWW25}{chap:intersectional}). 
\textit{Paper 4} (\Cref{chap:WWW25}) proposes a new evaluation approach to jointly quantify individual item fairness and recommendation effectiveness.  
\textit{Paper 5} (\Cref{chap:PUF}) contributes a novel individual user fairness measure that considers both disparity in recommendation effectiveness and user similarity. \textit{Paper 6} (\Cref{chap:intersectional})  unifies for the first time individual fairness and group fairness evaluation for recommender system users by collecting and computing measures that are compatible with both.

\paragraph{Relationship between limitations.} 
While this thesis broadly groups the limitations into three classes, these classes are not fully separable as they are still related to each other. 
An empirical limitation, such as the expensive computational time of a measure, may be traced back to the measure equation, which relates more to theoretical limitations. Yet, it is more practical to study how computational complexity translates to the actual computational time. 
Likewise, conceptual limitations can also be inferred directly from the measure equations. However, unlike theoretical limitations, which mainly concern flaws within the equation itself, conceptual limitations stem from missing variables or components.
Lastly, about the relationship between conceptual and empirical limitations, the unique aspects quantified by a measure contribute to its expressiveness in those aspects. All in all, the three types of measure limitations should not be viewed or analysed separately but rather addressed together, as done in this thesis.

\section{Scientific Contributions}
\label{s:contribution}

This section provides a detailed summary of the papers included in the thesis. Altogether, the papers examine and resolve three categories of limitations in existing recommender system fairness evaluation measures: theoretical, empirical, and conceptual (see \Cref{s:limitations}). \Cref{tab:contrib_overview} overviews the main contributions of each paper in assessing the measure limitations, addressing the limitations by designing a new evaluation approach or correcting the measures, and recommending general usage guidelines across a collection of measures. 

\clearpage

\begin{table}[!h]
\centering
\caption{Overview of the main contributions made by the papers in this thesis based on the contribution type: assessment of \textbf{limitations} in existing recommender system fairness evaluation measures, resolution of the limitations through a new evaluation measure/approach and corrections to existing measures (\textbf{measure}), as well as \textbf{guidelines} that cover a set of fairness measures. The type of limitation spans across Theoretical (T), Empirical (E), and Conceptual (C) (\Cref{fig:limitations}).
}
\label{tab:contrib_overview}
\begin{tabular}{lccc|c|c}
\toprule
 & \multicolumn{3}{c|}{{\textbf{\specialcell{Limitation}}}}  & \multirow{2}{*}{\textbf{Measure}} & \multirow{2}{*}{\textbf{\specialcell{Guideline}}}  \\
 &\textbf{T} & \textbf{E} & \textbf{C} &  &     \\
\midrule 
Paper 1 (\Cref{chap:TORS24}) & \cm & \cm &  & \cm & \cm  \\
Paper 2 (\Cref{chap:SIGIR24})&  & \cm &  &  & \cm   \\
Paper 3 (\Cref{chap:SIGIR24_ext})& \cm & \cm &  & \cm & \cm \\
Paper 4 (\Cref{chap:WWW25})&  & \cm & \cm & \cm &   \\
Paper 5 (\Cref{chap:PUF})&  & \cm & \cm & \cm &    \\
Paper 6 (\Cref{chap:intersectional})&  & \cm & \cm &  &   \\
\bottomrule
\end{tabular}
\end{table}

\subsection{Paper 1: Evaluation Measures of Individual Item Fairness for Recommender Systems: A Critical Study}

In recent years, there has been an increasing amount of work on item-side fairness in recommender systems \cite{Wang2023ASystems}. Along with this growing research interest, various evaluation measures for this type of fairness emerge. However, the practical differences and similarities among them, as well as their properties are not well-understood yet. Motivated by the need to investigate these aspects, we survey all existing exposure-based evaluation measures for individual item fairness in recommender systems (see \Cref{sss:eval_measures_fairness}), collect eight families of such measures, and analyse each measure in depth. 

\begin{table}[htb] 
\caption{
Measures of exposure-based individual item fairness and their theoretical limitations. The table is reused from \citeauthor{Rampisela2024EvaluationStudy}~\cite{Rampisela2024EvaluationStudy}.
}
\label{tab:contrib-limitation-summary}
\resizebox{\textwidth}{!}{
\begin{tabular}{l|c||c|c|c|c|c|c|c|c|c}
\toprule
\midrule
\parbox[t]{0.9\textwidth}
{Legend\\
\bbullet: we fully resolve the limitation\\
\nofix: the limitation is unresolvable \\
\checkmark: another measure resolves the limitation} 
&\rotatebox[origin=r]{90}{Source}
&\rotatebox[origin=r]{90}{Jain \cite{jain1984quantitative}}
&\rotatebox[origin=r]{90}{QF \cite{Zhu2020FARM:APPs}}
&\rotatebox[origin=r]{90}{Ent \cite{Shannon1948ACommunication}}
&\rotatebox[origin=r]{90}{Gini \cite{Gini1912VariabilitaMutabilita}}
&\rotatebox[origin=r]{90}{Gini-w \cite{Do2021Two-sidedDominance}}
&\rotatebox[origin=r]{90}{FSat \cite{Patro2020FairRec:Platforms}}
&\rotatebox[origin=r]{90}{VoCD \cite{Wang2022ProvidingSystems}}
&\rotatebox[origin=r]{90}{II-D \cite{Wu2022JointRecommendation}}
&\rotatebox[origin=r]{90}{AI-D \cite{Wu2022JointRecommendation}}
\\
\midrule
\midrule
non-realisability: cannot reach max/min score (cause number denoted by \textit{C}) &&&&&&&&&\\
\textit{C1.} Most unfair score is only given to an impossible scenario    &us&\bbullet&\bbullet&\bbullet&\bbullet&\bbullet&\bbullet&&&\\
\textit{C2.} Fewer recommendation slots compared to number of items&us&\bbullet&\bbullet&\bbullet&\bbullet&\bbullet&\bbullet&&\nofix&\nofix\\
\textit{C3.} Number of recommendation slots is indivisible by number of items&us&\bbullet&&\bbullet&\bbullet&\nofix&&&\nofix&\nofix\\
\textit{C4.} Non-realisability due to unknown formulation of max/min score
&us&&&&&\nofix&&\nofix&\nofix&\nofix\\
\midrule 
quantity-insensitivity: ignores frequency of item recommendation   &\cite{Zhu2020FARM:APPs}&&$\checkmark$&&&&&&&\\
\midrule
undefinedness: cannot be computed (undefined value)   &us&&&\bbullet&&&&&&\\
\midrule 
always-fair: gives fairest score regardless of recommendation contents   &\cite{Patro2020FairRec:Platforms}&&&&&&\nofix&&&\\
\midrule 
item-representation-dependence: depends on how items are represented   &us&&&&&&&\nofix&&\\
\midrule 
\bottomrule
\end{tabular}
}
\end{table}

This paper contributes a novel analysis on five \textbf{theoretical limitations} of individual item fairness measures (summarised in \Cref{tab:contrib-limitation-summary}). Out of the five limitations, we are the first to identify three within these measures. For the remaining two that have been previously discovered, we provide a formal explanation for their underlying causes. Furthermore, we address three limitations by correcting the measures or suggesting the use of another measure without the same limitation. 

Two types of \textbf{measure corrections} were proposed, each addressing a specific limitation. The first method deals with the \textit{undefinedness} limitation, a product of an undefined mathematical operation. To counter this limitation, we redefine the measure equation to account for edge cases, so that the score can now be computed for those cases.  Our second method resolves the issue where measures that range between $[0,1]$ cannot actually reach 0 or 1 for various reasons (\textit{see non-realisability}, \textit{C1--C4} in \Cref{tab:contrib-limitation-summary}). Only by looking at the scores, it is not possible to understand how close the output is, to the fairest or unfairest case, which makes the scores hard to interpret. We correct this issue by normalising the measures based on the scores obtained from the (un)fairest possible recommendations. In this way, one end of the measure range is now mapped to the unfairest recommendation and the other to the fairest recommendation, easing the score interpretation. Furthermore, we verify through experiments that the corrected measures indeed resolve the limitations; we provide reasons why these limitations remain unresolved in isolated cases.

Additionally, we uncover the \textbf{empirical limitations} of the measures. We discover that some measures have limited expressiveness as they constantly score near-perfect fairness, even if the item exposure distribution is highly imbalanced. We also find that several measures may be redundant, as they tend to produce similar rankings of models.

The comprehensive investigation distinguishes this paper from prior work in two ways: Firstly, prior work studies fairness for item groups \cite{Raj2022MeasuringResults}, while we focus on fairness for individual items.  Secondly, existing work explain the high-level underlying idea behind the measures, but do not compute or compare all of them, whereas our study conducts both a theoretical and an empirical investigation.

Finally, considering the unique characteristics and challenges of each measure, we release \textbf{guidelines} to help researchers and practitioners select the appropriate measure for evaluating exposure-based individual item fairness in recommender systems.

\subsection{Paper 2: Can We Trust Recommender System Fairness Evaluation? The Role of Fairness and Relevance}

Broadly, there are two types of individual item fairness measures for recommender systems: purely exposure-based measures and `joint' measures that account for item relevance on top of item exposure, i.e., `relevance-aware' measures (\Cref{sss:eval_measures_fairness}). Our prior work (\textit{Paper 1}, \Cref{chap:TORS24}) conducts an extensive study on the exposure-based measures, but the properties and tendencies of the joint measures remain unclear as they have not been investigated. To this end, we investigate the \textbf{empirical limitations} of all individual item fairness measures that consider both item exposure and relevance.

Our investigation concentrates on two aspects: measure agreement and measure expressiveness. The purpose of studying measure agreement is to determine whether joint measures yield the same conclusion as single-aspect measures or other joint measures. We analyse measure expressiveness to understand how sensitive the measure scores are to changes in item exposure and item relevance.

This work contributes novel insights on the empirical tendencies of existing individual item fairness measures. First, we expose the fact that some joint measures are rather similar, and thus, one measure can be computed as a reasonable proxy for the other. Moreover, we discover their limited sensitivity towards changes in item relevance and exposure, even though both variables are part of the measures. We highlight several measures that consistently score very close to the fairest, impairing their interpretability, as their scores are always extremely fair. 

Considering these alarming findings, we conclude with \textbf{guidelines} for the measure usage: we advise against computing similar measures to avoid redundancy; we encourage a more careful interpretation of the measure scores; and we suggest measuring fairness and relevance separately.

\subsection{Paper 3: Relevance-aware Individual Item Fairness Measures in Recommender Systems: Limitations and Usage Guidelines}

This work extends \textit{Paper 2} (\Cref{chap:SIGIR24}) and contributes three things: an in-depth investigation on the \textbf{theoretical limitations} of existing relevance-aware individual item fairness measures; \textbf{measure corrections} that resolve a subset of the limitations; and detailed \textbf{guidelines} that specify factors to consider for measure selection (\Cref{tab:contrib-guideline}). 

\begin{table}[!ht]
    \centering
    \caption{Factors related to the measure Alignment (A), Computability (C), Interpretability (I), Expressiveness (X), Stability (S), and Efficiency (E) 
    to consider when selecting relevance-aware item fairness measures. Statements related to (A) are neutral; the rest are desirable.}
    \label{tab:contrib-guideline}
    \resizebox{\textwidth}{!}{
    \begin{tabular}{p{12cm}|*{13}{|c}}
\toprule
\midrule
\parbox[t]{0.8\textwidth}
{Legend\\
\cm: the statement applies to the measure in general \\
\obs: the statement applies to the measure, for binary relevance and single-round recommendation  \\
\xm: the statement does not apply to the measure
} 
&\rotatebox[origin=r]{90}{IAA\ori{}}
&\rotatebox[origin=r]{90}{IAA\our{}}
&\rotatebox[origin=r]{90}{IFD\divori{}}
&\rotatebox[origin=r]{90}{IFD\divour{}}
&\rotatebox[origin=r]{90}{IFD$_{\times\text{-ori}}$}
&\rotatebox[origin=r]{90}{IFD$_{\times\text{-our}}$}
&\rotatebox[origin=r]{90}{HD\ori{}}
&\rotatebox[origin=r]{90}{MME\ori{}}
&\rotatebox[origin=r]{90}{IBO/IWO\ori{}}
&\rotatebox[origin=r]{90}{IBO/IWO\our{}}
&\rotatebox[origin=r]{90}{II-F\ori{}}
&\rotatebox[origin=r]{90}{II-F\our{}}
&\rotatebox[origin=r]{90}{AI-F\ori{}}
\\
\midrule
\midrule
A1. The measure tends to agree with effectiveness measures&\cm&\cm&\xm&\xm&\xm&\xm&\cm&\xm&\xm&\xm&\cm&\cm&\xm\\
\midrule
A2. The measure tends to agree with fairness measures or disagree with effectiveness measures&\xm&\xm&\cm&\cm&\cm&\cm&\xm&\cm&\xm&\xm&\xm&\xm&\cm\\
\midrule
\midrule
C1. The measure can be computed even if the full ranking of items (not just the top-$k$) is unavailable & \xm & \xm&\xm&\xm&\cm&\cm&\cm&\cm&\cm&\cm&\cm&\cm&\cm\\
\midrule
C2. The measure can be computed even if item relevance beyond top-$k$ is unknown &\xm&\xm&\xm&\xm&\cm&\cm&\xm&\cm&\xm&\xm&\cm&\cm&\cm\\
\midrule
C3. The measure can be computed even if the total \# of relevant items per user is unknown &\xm&\xm&\xm&\xm&\cm&\cm&\xm&\cm&\xm&\xm&\xm&\xm&\xm\\
\midrule
C4. The measure can be computed even if there is an item that is not relevant to any user &\cm&\cm&\cm&\cm&\cm&\cm&\cm&\cm&\xm&\cm&\cm&\cm&\cm\\
\midrule
C5. The measure can be computed for any $k$ &\xm&\cm&\cm&\cm&\cm&\cm&\cm&\cm&\cm&\cm&\cm&\cm&\cm\\
\midrule
\midrule
I1. The measure maps the unfairest and the fairest recommendation to the endpoints of its theoretical range &\xm&\obs&\xm&\obs&\xm&\obs&\xm&\xm&\xm&\xm&\xm&\obs&\xm\\
\midrule
I2. The measure distinguishes between an item exposed at $k$ and an unexposed item at $k+1$ &\xm&\cm&\cm&\cm&\cm&\cm&\cm&\cm&\cm&\cm&\cm&\cm&\cm\\
\midrule
\midrule
X1. The measure scores are not compressed in a small interval
&\xm&\cm&\xm&\cm&\xm&\cm&\cm&\xm&\cm&\cm&\xm&\cm&\xm\\
\midrule
X2. The measure responds to changes in $k$ &\cm&\cm&\xm&\cm&\cm&\cm&\cm&\cm&\cm&\cm&\cm&\cm&\cm\\
\midrule
\midrule
S1. The measure is unaffected by item relevance beyond the top-$k$  &\xm&\xm&\xm&\xm&\cm&\cm&\xm&\cm&\xm&\xm&\xm&\xm&\xm\\
\midrule
S2. The measure is stable or becomes fairer if there are more relevant items beyond the top-$k$ &\xm&\xm&\cm&\cm&\cm&\cm&\cm&\cm&\xm&\xm&\xm&\xm&\xm\\
\midrule
S3. The measure is unaffected by ties in item relevance 
&\cm&\cm&\cm&\cm&\cm&\cm&\xm&\cm&\cm&\cm&\cm&\cm&\cm\\
\midrule
S4. The measure is unaffected by the number of users with only one relevant item &\cm&\cm&\xm&\xm&\cm&\cm&\cm&\cm&\cm&\cm&\cm&\cm&\cm\\
\midrule
\midrule
E1. The measure can be computed separately per user  &\cm&\cm&\cm&\cm&\cm&\cm&\xm&\xm&\xm&\xm&\cm&\cm&\xm\\
\midrule
E2. The measure can be computed separately per item  &\xm&\xm&\xm&\xm&\xm&\xm&\xm&\cm&\cm&\cm&\cm&\cm&\cm\\
\midrule
E3. The measure is efficient to compute ($<$60 s) for large datasets (e.g., $>1$K users and $>10$K items) &\cm&\cm&\cm&\cm&\xm&\xm&\cm&\xm&\cm&\cm&\cm&\cm&\cm\\

\bottomrule
\bottomrule
\end{tabular}
    }
\end{table}

We have previously conducted an empirical investigation on all existing relevance-aware individual item fairness measures and linked several findings on the measures' \textbf{empirical limitations} to the measure equations. Similar to \textit{Paper 1} (\Cref{chap:TORS24}), we critically analyse the measures' theoretical limitations by looking closely at their formulations. The analysis focuses on cases that would restrict the measure's applicability or interpretability. To the best of our knowledge, this is the first work that identifies the theoretical limitations of the measures, and we find five such limitations.

Most limitations are resolvable; we justify why the rest are not. To overcome the resolvable limitations, we formulate extensions of the measures in two forms. The first form is a normalised version of the measure, which correctly maps the fairest possible recommendation to 0 and the unfairest to 1. This normalisation significantly improves the score interpretability, as previously the scores were always near 0, giving an illusion of all-time maximally fair output, even for the most unfair case. 
The second form of measure extension is a redefinition of the measure, which accounts for cases that cause the measure to be incomputable (e.g., division by 0).  
We kept these modifications to a minimum, so as not to radically change the original measure design, yet the measures can now be used in a wider range of cases. 

In addition, through a series of experiments with real and synthetic datasets, this paper shows the extent of the measures' theoretical limitations in practical cases. At the same time, we confirm that our measure extensions have resolved the issues. The paper concludes with specific guidelines that connect the measure limitations and their practical implications on a wide range of evaluation scenarios. Our guidelines specify which measure should (not) be used in different cases and conditions. We hope that these guidelines will ease the selection of relevance-aware individual item fairness measures based on both their properties and their possible use cases.

\subsection{Paper 4: Joint Evaluation of Fairness and Relevance in Recommender Systems with Pareto Frontier}

\begin{figure}
    \centering
        \scalebox{.8}{
        \begin{tikzpicture}
            \tikzstyle{blue rectangle}=[fill={rgb,255: red,1; green,115; blue,178}, draw=black, shape=rectangle]
            \tikzstyle{green circle pareto}=[fill={rgb,255: red,2; green,158; blue,115}, draw=black, shape=rectangle]
            \tikzstyle{orange rectangle}=[fill={rgb,255: red,222; green,132; blue,5}, draw=black, shape=rectangle]
            \tikzstyle{pink rectangle}=[fill={rgb,255: red,204; green,120; blue,188}, draw=black, shape=rectangle]
            \tikzset{cross/.style={cross out, draw=black, minimum size=2*(#1-\pgflinewidth), inner sep=0pt, outer sep=0pt}, 
            cross/.default={5pt}}
            \begin{axis}
            [
                xmin = 0,
                ymin = 0,
                xmax = 1+0.1,
                ymax = 1+0.1,
                xlabel={\large Relevance}, 
                ylabel={\large Fairness}, 
                ticks=none,
                clip=false,
                axis lines=left,
                axis line style=thick,
                x label style={at={(axis description cs:0.5,0)},anchor=north},
                y label style={at={(axis description cs:0,.5)},anchor=south},
                legend style={at={(1.6,0.8)},
                anchor=south,legend columns=1},
                legend style={font=\small, draw=none},
                legend cell align={left}
            ]

            \addlegendimage{}
            \addlegendimage{dashed,gray!50}

            \node (0) at (0.2, 1) {};
            \node (1) at (1, 0.2) {};
            \node [style=green circle pareto] (9) at (0.5, 0.5) {};
            \node [style=orange rectangle] (bestF) at (0.2, 0.9) {};
            \node [style=blue rectangle] (bestR) at (0.65, 0.2) {};
            \draw [in=90, out=0] (0.center) to (1.center);
            \addlegendentry{Pareto Frontier (PF)};
            \draw (0.7656854249492, 0.7656854249492) 
            node[cross,red, ultra thick,
            label={[align=center, xshift=5pt, yshift=2pt]88:
            {\textbf{PF-midpoint}}\\[-2pt]{\footnotesize($\alpha=0.5$)}}] (10) {};
            \draw[dashed, color=gray!50] (9) -- (10);
            \addlegendentry{Distance to Pareto Frontier (DPFR)};
            \draw[dashed, color=gray!50] (10) -- (bestF);
            \draw[dashed, color=gray!50] (10) -- (bestR);
            \node[label={\textbf{Model C}}]  at (9) {};
            \node[label={\textbf{Model A}}]  at (bestF) {};
            \node[label={\textbf{Model B}}]  at (bestR) {};
            \end{axis}%
        \end{tikzpicture}
        }
    
    \caption{
    In this example, 
    Model A is best for fairness, 
    Model B is best for relevance, and Model C is the closest 
    to the Pareto Frontier (PF) midpoint, when relevance and fairness are equally weighted ($\alpha=0.5$). The figure has been reused from \citeauthor{Rampisela2025JointFrontier}~\cite{Rampisela2025JointFrontier} with some modifications for clarity.
    }
    \label{fig:contrib_pareto_teaser}

\end{figure}
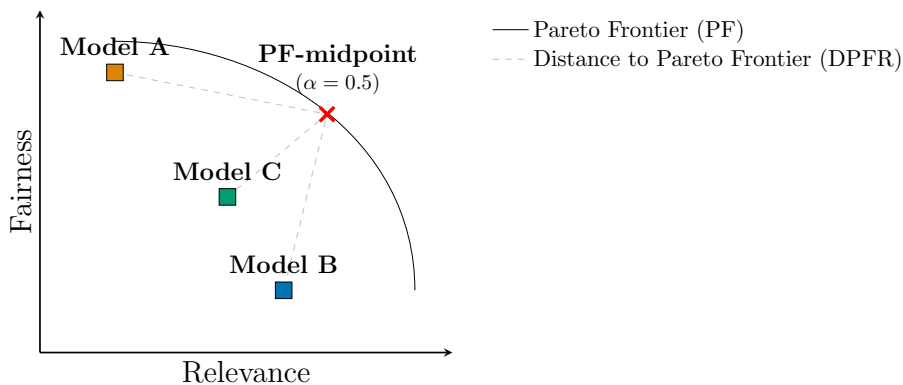

Previous work, including ours (\textit{Papers 1--3}, \Crefrange{chap:TORS24}{chap:SIGIR24_ext}), has measured recommendation fairness and its relevance either separately or together. When the two aspects are quantified separately, it is hard to select the recommender with the best balance between the two. This is because the fairest recommender may not provide the most relevant recommendations or vice versa. Yet, evaluation measures that consider both fairness for individual items and their relevance to users also do not reliably quantify the balance between both aspects. These joint measures either align exclusively with one of the two aspects, or have inconsistent alignments (\textit{Papers 2--3}, \Crefrange{chap:SIGIR24}{chap:SIGIR24_ext}). In other words, there is no existing approach that can reliably quantify both relevance and individual item fairness via a single score.

To address this \textbf{conceptual limitation}, we present a \textbf{novel evaluation approach}, \textit{Distance to Pareto Frontier} (DPFR), to find the most balanced model in terms of both fairness and relevance (see example in \Cref{fig:contrib_pareto_teaser}). 
We contribute a new algorithm that generates the Pareto Frontier based on the dataset and a selected cut-off $k$. Given these inputs, the algorithm first maximises recommendation relevance using the test set and iteratively improves fairness, until the maximally fair recommendation is formed. The final generated frontier contains a set of Pareto-optimal solutions, representing the best fairness-relevance trade-offs that are achievable based on the dataset. 

The advantages of DPFR are threefold. First, it is modular, which means that it is model-agnostic and can be computed with various pairs of existing single-aspect measures. Secondly, DPFR allows evaluation against controlled, dataset-based trade-offs of fairness and relevance that fulfill Pareto-optimality; this is important as improving recommendation fairness often comes at the cost of sacrificing its overall effectiveness \cite{karimi2023providertradeoffs,Zehlike2022FairGroups}. Thirdly, our approach is intuitive and is easily interpretable as it directly measures how far the model score is from a desired fairness-relevance balance. No existing joint evaluation approaches for recommender system fairness and relevance have all three advantages.

In our experiments with six datasets of varying characteristics, we test the compatibility of existing single-aspect measures with DPFR. We find 12 pairs of relevance and fairness measures that are compatible with DPFR. Then, we compare DPFR to existing single-aspect measures and joint measures. Our results show that neither measure type can be a reliable proxy for DPFR, as they rank models from the best to worst differently from DPFR; these are part of the measures' \textbf{empirical limitations}. 
Finally, we show that DPFR can be computed efficiently without compromising its evaluation effectiveness. 
Thus, as an alternative to existing joint evaluation approaches, we recommend computing DPFR to select the empirical best possible recommendation based on both relevance and fairness.

\subsection{Paper 5: Measuring Individual User Fairness with User Similarity and Effectiveness Disparity}

Fairness in recommender systems can be evaluated not only for items--as we have investigated in \textit{Papers 1--4} (\Crefrange{chap:TORS24}{chap:WWW25})--but also for users. In this paper, we shift our focus to fairness evaluation measures for individual users.

\begin{table}
    \caption{Overview of individual user fairness measures in recommender systems.}
    \label{tab:contrib_checklist}
    \centering
    \resizebox{0.95\textwidth}{!}{
    \begin{tabular}{llcc}
         \toprule
         Measure & Reference & Effectiveness & User Similarity \\
         \midrule
         Standard deviation & \cite{Patro2020FairRec:Platforms,Biswas2021TowardPlatforms,Wu2021TFROM:Providers,Rastegarpanah2019FightingSystems,Li2024ExplainingPerspective}   & $\checkmark$ & -\\
         Gini index & \cite{Fu2020Fairness-AwareGraphs, Leonhardt2018UserSystems} & $\checkmark$ & - \\
         Envy-based &\cite{Patro2020FairRec:Platforms, Biswas2021TowardPlatforms,Do2022OnlineSystems}& $\checkmark$ & -\\
         UF & \cite{Wu2023EquippingEmbedding} &- & $\checkmark$  \\
         \midrule
         PUF & ours & $\checkmark$ & $\checkmark$ \\
         \bottomrule
    \end{tabular}}
\end{table}

Individual fairness is commonly defined on the basis of having equal treatment for \textit{similar} individuals \cite{Dwork2012FairnessAwareness}. Yet, upon collecting all existing evaluation measures for individual user fairness (overviewed in \Cref{tab:contrib_checklist}), we discover that only one such measure \cite{Wu2023EquippingEmbedding} actually follows this definition by accounting for user similarity. This measure quantifies fairness based on the representation of items in two similar users' recommendation lists. A couple of \textbf{conceptual limitations} arise from quantifying fairness in this way. Firstly, having similar recommended items may not directly translate to having similar recommendation effectiveness: if two highly similar users get recommended identical items, they may not equally like their recommendations, as their exact preferences could differ. Secondly, even if similar users are equally satisfied with their recommendations, the measure would deem this unfair if the recommendations consist of entirely different sets of items with dissimilar representations. Thus, it is important to evaluate fairness beyond differences in item representation.

In this paper, we introduce \textit{Pairwise User unFairness} (PUF), a \textbf{novel evaluation measure} for individual user fairness that is both aligned with user utility as a key objective in recommender systems and with the individual fairness definition. PUF evaluates unfairness as the average pairwise difference of user utility, weighted by the user pair similarity. The user utility is the recommendation effectiveness score (e.g., NDCG). The idea is that if a user pair has either low similarity or a small difference in their effectiveness scores, the pair would only contribute a small increase in the PUF score, which translates to less unfairness. Meanwhile, highly similar user pairs with a large effectiveness score difference would have a higher PUF score, which means unfairer recommendations. With this measure, fairness for individual users can now be evaluated based on effectiveness disparity while also considering user similarity.

Our experiments demonstrate that PUF can reliably capture both user similarity and recommendation effectiveness, while existing measures fail to do so; these are the \textbf{empirical limitations} that we uncovered from the measures. Moreover, as no existing measures consistently yield similar conclusions to PUF, none can be used as a reliable estimate of our measure. 
By having a measure that is aligned with the individual fairness definition and user utility, we hope that our measure can serve as a tool for practitioners to meaningfully estimate performance disparity across users.

\subsection{Paper 6: Stairway to Fairness: Connecting Group and Individual Fairness}

Most work on recommender system fairness focuses exclusively on group fairness, rarely evaluating for both group and individual fairness \cite{Deldjoo2024FairnessDirections}. Furthermore, in prior studies on both fairness types, group fairness and individual fairness are evaluated with different measure families, which makes it difficult to compare them, as different measures may not have equal sensitivity or empirical ranges \cite{Rampisela2024CanRelevance,Rampisela2024EvaluationStudy,Schumacher2024PropertiesRankings}. Other literature instead evaluates fairness for different subjects or objectives, e.g., equal recommendation effectiveness across individual users and equal exposure between item groups \cite{Wu2021TFROM:Providers}, which are also not directly comparable. Consequently, the interplay between group and individual fairness is not scientifically well-understood. Analysing the relationship between the two fairness types is key to understanding how improving one fairness type can impact the other.

\begin{figure}
    \centering
   \begin{minipage}[t]{.6\columnwidth}
   \vspace{0pt}
   \includegraphics[width=\linewidth]{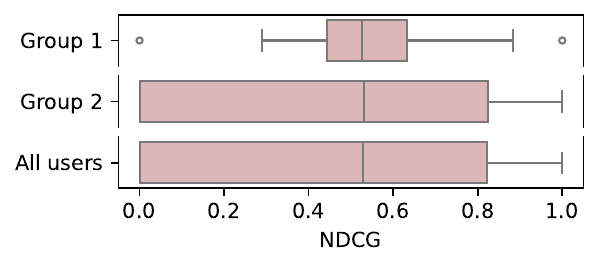}
    \end{minipage} %
   \begin{minipage}[t]{0.38\columnwidth}
   \vspace{12pt}
   \scalebox{.77}{
   \begin{tabular}{lccc}
   \toprule
      & \#user & $\uparrow$NDCG & $\downarrow$Gini \\
   \midrule
   \midrule
   Group 1  & 11 & 0.546 & \multirow{2}{*}{{\LARGE\}}\raisebox{2pt}{0.037}} \\ 
   Group 2  & 901 & 0.471& \\
   \midrule
   \midrule
   Individual & 912 & 0.472 & 0.446  \\ 
   \bottomrule
   \end{tabular}}
    \end{minipage}
   \caption{Left: NDCG score distribution of two handpicked, non-overlapping user groups from our experiment and for all individual users of the two groups. Circles denote outliers. Right: Fairness score (Gini) for groups and individuals. The lower the Gini score, the fairer. Fairness is higher between groups than across all individuals as both groups have similar average NDCG, but within-group variance is high, which means that recommendation quality varies widely across users (see the boxplots).}
    \label{fig:contrib_teaser_boxplot}
\end{figure}

We contribute the first empirical study on the relationship between user-side group and individual fairness evaluation measures. 
To allow a proper comparison of group fairness and individual fairness, we evaluate both with the same measure family. We gather all existing user-side group fairness measures and identify a subset of them that have also been used as a user-side individual fairness measure. In the given example (\Cref{fig:contrib_teaser_boxplot}), both fairness types are measured with Gini \cite{Gini1912VariabilitaMutabilita}. This example shows in practice how a recommender can be very fair towards user groups and, at the same time, much more unfair towards individual users.

In this paper, we also address a \textbf{conceptual limitation} of user-side group fairness measures. No existing user-side group fairness measure satisfies both the following criteria: 
(1) it ranges between 0 and 1, which enables a proper comparison of group and individual fairness; and 
(2) it accounts for within-group variance, which relates to individual fairness. 
To this end, we identify an extra measure that fulfills them both and compute it to evaluate user-side group and individual fairness. 

Our empirical analyses with 8 recommenders and 3 datasets reveal that group fairness scores tend to mask individual-level unfairness, regardless of how the users are grouped or the group size. 
We find that this is due to high within-group unfairness, where effectiveness scores vary widely in each user group, resulting in high utility variation across all users. Even when the group fairness measure formulation accounts for the within-group variance, the masking effect remains. 
More importantly, no group fairness measures consistently reach the same conclusion as any individual fairness measure, which is an \textbf{empirical limitation} of the measures. This necessitates the evaluation of both group and individual fairness, as the fairest model for user groups may not be the fairest for individual users.

All in all, we encourage a more comprehensive fairness evaluation by assessing both group and individual fairness. We especially highlight the importance of evaluating individual fairness on top of group fairness, as disparity across individuals goes undetected by group-level scores. Considering that most fairness-aware efforts have been centred on mitigating disparity between groups, our work motivates a push towards initiatives for improving fairness across individuals, such that all recommender system users/items can obtain a similar level of benefit.

\section{Summary and Future Work}
\label{s:summary_future}

The papers included in this thesis collectively contribute to the advancement of state-of-the-art offline evaluation approaches for recommender system fairness. In particular, this thesis contributes: 
(1) a study and guidelines on exposure-based individual item fairness measures and our proposed extensions of the measures; 
(2) an empirical study on relevance-aware individual item fairness measures and their usage guidelines;
(3) a theoretical study, empirical analyses, and guidelines on relevance-aware fairness measures, as well as our proposed measure extensions;
(4) a new joint evaluation approach for both recommendation  fairness and effectiveness;
(5) a new evaluation measure for individual user fairness; and
(6) a work that connects group fairness and individual fairness. More than 29 unique fairness measures for individual items, for individual users, or for user groups, were studied across the six papers. This number consists of both existing and new measures.

Through theoretical and empirical investigations, we expose serious flaws of existing recommender system evaluation measures in the form of theoretical, empirical, and conceptual limitations. Altogether, these flaws may severely limit the measures' interpretability, expressiveness, and applicability. On one hand, our study reveals redundancy among existing measures, meaning that not all of them must be computed to have a good overview of fairness, yet some of them conflict in their conclusions in ranking models from the fairest to the unfairest (\textit{Paper 2}, \Cref{chap:SIGIR24}). On the other hand, we show that evaluating for only one type of fairness is insufficient  (\textit{Paper 6}, \Cref{chap:intersectional}). In addition, we find that even fairness measures that operate on the same idea (e.g., parity), for the same subject (users/items), and are evaluated for the same granularity (individual/group) often differ in their theoretical and empirical properties. Practically, this means that some measures are more suitable for relative comparison among models than for a standalone evaluation of a single system. This also means that the measure scores need to be carefully interpreted as some are stricter than others. Furthermore, fairness-optimisation methods guided by these measures may require different approaches due to the measures' various alignment to effectiveness measures. Finally, we have condensed the insights from our investigations into practical guidelines for the appropriate measure usage.

While this thesis aims to provide a comprehensive coverage of offline evaluation of fairness in recommender systems and has studied a wide range of fairness measures, it is by no means exhaustive. Many questions remain unanswered, leaving gaps for further investigation. Therefore, three key potential research areas for future work on fairness in recommender systems are identified and discussed below:

\paragraph{Online and user-centred fairness evaluation.} This thesis has concentrated on a specific evaluation paradigm: offline evaluation. While this paradigm is widely employed in recommender system fairness research, it is not the sole way of evaluating fairness. There are not many studies on online evaluation for recommender system fairness (e.g., \cite{Beutel2019FairnessComparisons}), but they are worth further investigating, as they directly reflect recommender systems' performance in the real world. 
Similarly, user studies on what stakeholders perceive as fair, such as \cite{Ungruh2024PuttingRecommenders,Smith2023ScopingPerspective,Smith2024RecommendProviders, Smith2020ExploringSystems,Elahi2021BeyondSystems,Ferraro2021WhatPlatforms,Dinnissen2023AmplifyingPlatforms,Sonboli2021FairnessPerspective}, are still rather limited in coverage or participants' background diversity. This opens up avenues for large-scale online studies, as well as user studies that target a more diverse set of participants and a wider range of fairness notions. 

Even though offline evaluation is relatively faster and simpler than conducting online evaluation or user studies, their results may conflict \cite{Beel2015ComparisonSystems,Castells2022OfflineDirections,Garcin2014OfflineOnline}. This is especially crucial for fairness evaluation, as prior user studies have pointed out that some desirable fairness notions based on user perception align more with the concepts of privacy, inclusion, and transparency, rather than conventional fairness concepts in offline evaluation of recommender system fairness \cite{Smith2024RecommendProviders,Elahi2021BeyondSystems,Ferraro2021WhatPlatforms}. Although these aspects, together with fairness, are pillars of responsible artificial intelligence (AI), they are not commonly accounted for in existing fairness measures \cite{Smith2024RecommendProviders}. 
Hence, to have a more complete picture of recommender system fairness and its connection to broader societal issues, there is a need to investigate other evaluation paradigms, especially in their alignment with existing offline approaches and in relation to other pillars of responsible AI. By leveraging user studies, future work could design offline evaluation measures that cater to actual stakeholder needs and perceptions of fairness.

\paragraph{Multi-stakeholder, multi-round, and multi-attribute fairness.}
The papers included in this thesis have conducted an in-depth investigation into single-stakeholder fairness at a fixed point in time. However, in reality, recommender systems operate across multiple rounds of recommendations and involve multiple stakeholders, each of whom may be associated with multiple attributes. This calls for multi-stakeholder, multi-round, and multi-attribute fairness evaluation.

Firstly, about multi-stakeholder fairness, recommender systems are two-sided platforms that require a balance between users, items, and the possible third parties involved \cite{Ekstrand2022FairnessSystems}. As such, fairness evaluation should also consider the multi-stakeholder nature of the ecosystem to minimise unfairness towards all parties. Recent work has proposed several measures to jointly evaluate fairness across stakeholders (e.g., \cite{Wu2022JointRecommendation,Kheya2025UnmaskingFairness,Wang2024IntersectionalRecommendation}). Yet, the relationship between multi- and single-stakeholder fairness measures is currently unknown. For example, it is not known if optimising for a multi-stakeholder fairness measure would automatically improve single-stakeholder fairness scores (or vice versa).  More importantly, as there are no existing studies that `stress-test' these measures for a wide range of scenarios, it is not well-understood if these measures would work properly in all cases. 

Secondly, regarding multi-round fairness, real-world recommender systems users typically engage with the platform over a period of time. Over time and over multiple rounds of engagements, the system learns the user preferences better and iteratively provides updated recommendations that are learned from the users' previous interactions. Consequently, recommendation effectiveness may change over time. As user-side fairness is typically defined based on disparity in recommendation effectiveness, it is important to study how this type of fairness changes over time. Multi-round evaluation is also relevant for item-side fairness, which is often defined in terms of the exposure received by items. The total amount of exposure is limited directly by the number of recommendation slots. As there are more slots in multi-round recommendations, there are also more opportunities for items to appear in a user's recommendation list. Therefore, it would be interesting to study fairness considering all rounds of recommendations. 

Fairness in multiple rounds has been evaluated either separately per round \cite{Ferraro2024GenderImbalance,Yoo2024EnsuringDynamicSystems,Ferraro2021BreakRecommenders} or jointly considering all rounds \cite{Wu2022JointRecommendation,Borges2019EnhancingAutoencoders}. In \textit{Paper 1} and \textit{Paper 3} (\Cref{chap:TORS24} and \Cref{chap:SIGIR24_ext}), we analyse measures that quantify multi-round fairness for items. While we have briefly discussed their theoretical limitations under the multi-round evaluation setup, their in-depth and empirical investigations are left for future work.

Thirdly, about multi-attribute fairness, group fairness evaluation is closely tied to user/item sensitive attributes (e.g., gender, race). Our work (\textit{Paper 6}, \Cref{chap:intersectional}) studied fairness for user groups and found that fairness worsens when the groups are formed with more sensitive attributes. As such, we highlight the importance of multi-attribute fairness assessment. While our investigation is limited to treating all attributes as nominal variables, future work could explore multi-attribute fairness evaluation measures that accommodate ordinal and interval values (e.g., user age and years of work experience) in conjunction with nominal values \cite{Sakai2023VersatileRelevance,Mitchell2020DiversitySelection}.

\paragraph{Recommender system fairness and Large Language Models (LLMs).} The rise of LLMs in recent years has opened up multiple new areas of research that build upon LLMs' world knowledge \cite{petroni-etal-2019-language,zhang-etal-2023-large}, as well as their `reasoning' \cite{deepseekai2025deepseekr1incentivizingreasoningcapability,OpenAI2024} and `instruction-following' ability \cite{zhou2023instructionfollowingevaluationlargelanguage,zeng2024evaluating,ouyang2022traininglanguagemodelsfollow}. 
Beyond using LLMs as zero-shot recommenders, e.g., \cite{Hou2024LargeSystems} and \textit{Paper 6} (\Cref{chap:intersectional}), several ideas on how recommender system fairness research can benefit from LLMs are outlined below.

Recommender system fairness research can potentially exploit LLMs in at least three ways: for data bias detection, for fairness-aware approaches, and for fairness evaluation. Existing work has leveraged LLMs to detect bias in tabular datasets \cite{yu2025biasnavi}; similar tools can be developed for identifying potential biases in recommender system datasets, so that unfair biases arising from data can be managed prior to model training. 
LLMs have also been used as a fairness-aware reranker for recommendations \cite{Gao2025LLM4Rerank}. As this reranking approach is still limited to item-side fairness and for a specific fairness measure, future work could extend the approach for other fairness notions and measures. Additionally, the use of LLM-as-a-judge has gained increasing popularity; LLMs have been used to evaluate personalisation \cite{dong-etal-2024-llm}, document relevance \cite{balog2025rankersjudgesassistantsunderstanding}, and recommender system explanation \cite{Zhang2024LLMExplanation} with varying levels of success. Hence, one possible research direction is to experiment with LLMs to assess fairness. 
By doing this, LLM alignment to various fairness concepts can also be examined. 
Finally, another area that is worth exploring concerns a newly proposed fairness evaluation framework for LLM-based recommenders \cite{deldjoo2024normativeframeworkbenchmarkingconsumer,Deldjoo2025CFaiRLLM,Zhang2023IsRecommendation}. In this framework, fairness evaluation is done by comparing two recommendation lists, each generated by including and excluding a user's sensitive attributes in the prompt. Understanding how this framework aligns with other existing evaluation approaches can be crucial to assessing its validity and robustness.

Overall, the topic of fairness in recommender systems remains relevant not only in the specific field of responsible AI research, but also in the broader society. With the development of new technologies, such as LLMs, there is both a legal and practical necessity to uncover their hidden bias before further leveraging them to assess and improve fairness in other applications. Evaluating the potential harms and risks arising from these biases is the first step in mitigating and reducing unfairness.

\chapter{Evaluation Measures of Individual Item Fairness for Recommender Systems: A Critical Study}
\label{chap:TORS24}
\section*{Abstract}
Fairness is an emerging and challenging topic in recommender systems. In recent years, various ways of evaluating and therefore improving fairness have emerged. In this study, we examine existing evaluation measures of fairness in recommender systems. Specifically, we focus solely on exposure-based fairness measures of individual items that aim to quantify the disparity in how individual items are recommended to users, separate from item relevance to users. We gather all such measures and we critically analyse their theoretical properties. We identify a series of limitations in each of them, which collectively may render the affected measures hard or impossible to interpret, to compute, or to use for comparing recommendations. We resolve these limitations by redefining or correcting the affected measures, or we argue why certain limitations cannot be resolved. We further perform a comprehensive empirical analysis of both the original and our corrected versions of these fairness measures, using real-world and synthetic datasets. Our analysis provides novel insights into the relationship between measures based on different fairness concepts, and different levels of measure sensitivity and strictness. We conclude with practical suggestions of which fairness measures should be used and when. Our code is publicly available. To our knowledge, this is the first critical comparison of individual item fairness measures in recommender systems.

\section{Introduction}\label{s:introduction}

The concept of fairness in Recommender Systems (RSs) is commonly understood as treating users or items that are alike, in a similar way. It can be studied either for individual items or users (individual fairness), or for groups of items or users (group fairness). Individual fairness typically refers to similar individuals being treated similarly \cite{Biega2018EquityRankings}, where sometimes the similarity of individuals is measured through a certain metric, e.g. distance of the individual representation \cite{LiYunqi2023FairnessApplications}. 
In this work, we focus solely on fairness for individual items, and specifically on evaluation measures designed to quantify individual item fairness in RSs. While there exist comprehensive surveys on group fairness measures \cite{Raj2022MeasuringResults,Zehlike2022FairnessSystems}, to the best of our knowledge, no critical analysis of evaluation measures of individual item fairness for \rs{} has been presented. 

Individual item fairness is an important type of fairness that occurs in many RS scenarios. For example, it is helpful for new item discovery and for ensuring that recommended items come from different providers or creators. Individual item unfairness may occur due to popularity bias, which causes some items to be recommended more often than others \cite{Zhu2021Popularity-OpportunityFiltering}. Some items may not even be recommended at all. For instance, interesting content from emerging content creators could be recommended less frequently than less or equally interesting content from popular creators. 

In a traditional recommendation scenario, a model produces a top-$k$ list of recommendations across all users. This output is typically evaluated by measuring the recommendation relevance of the top-$k$ recommendations to each user. On top of that, one can measure the individual item fairness of the top-$k$ recommendations. 
On a high level, we distinguish between two definitions of individual item fairness. In the \textbf{first definition}, individual item fairness is understood as \textit{all items}\footnote{All items in the dataset or all items in the recommendations given to all users. Both definitions are used in the literature and also in our work in $\S$\ref{ss:onlyfair}.}
\textit{having equal exposure}. 
Exposure (also known as attention \cite{Biega2018EquityRankings} or coverage \cite{Wang2022ProvidingSystems}) refers to an item appearing in the top $k$ recommendations for a user. The definition of fairness above does not consider how relevant the recommended items are to the users; it only considers how uniform their exposure is. 
Given that relevance is a key aim in \rs, fairness has also been given a \textbf{second definition} as \textit{all items having an equal opportunity for exposure, where the opportunity is based on item relevance to users or similar other criteria \cite{Wu2022JointRecommendation,Diaz2020EvaluatingExposure}.}

Several evaluation measures have been used to empirically evaluate fairness for individual item fairness in \rs~\cite{Wang2023ASystems}. 
All of the measures evaluate fairness based on the recommendation list; the input is the recommendation list, and in most cases, the measure compares the exposure of items in the recommendation list and the total number of items in the dataset. Some of the measures follow the first definition of fairness, which is purely exposure-based, while other measures evaluate fairness jointly with relevance. 
Here, we focus solely on fairness measures of the first type, that is measures that evaluate only fairness without considering relevance. Even if these fairness-only measures have been used to evaluate fairness in prior work, they have not been extensively analysed for their limitations and possible consequences arising thereof. They have also not been analysed in relation to one another, as most research in individual item fairness only uses a single fairness measure. 
As a result, it is currently unclear how to interpret these measures and select which measures should be used under various circumstances. For instance, we find cases where a measure is theoretically bound between $[0,1]$ but empirically cannot reach either of the endpoints, which makes the interpretation of the measure extremely difficult. 
It is also unknown if there are theoretical limitations and cases where the measures would fail. For instance, we find that 
a measure will give the highest scores no matter the recommendation, given a certain number of users, items, or cut-off. 
We also find that a measure cannot be computed as the formulation is not well-defined for a commonly-encountered case.

The goals of this work are to address the above research gaps, by examining existing measures of individual item fairness in \rs, presenting analytical limitations and solving them (or justifying why they cannot be solved), and examining how the scores of the measures change in relation to each other in different scenarios. As such, 
we contribute the following: 
\begin{itemize}
    \item We review individual item fairness measures in \rs~($\S$\ref{s:priorwork}).
    \item We identify and analyse 5 theoretical limitations of those measures, where three of the limitations are novel 
    and two limitations have already been identified but without a formal/theoretical explanation, which we provide ($\S$\ref{s:limitations}). 
    \item We propose theoretical corrections of the measures to resolve their limitations or explain why some limitations are unresolvable ($\S$\ref{s:extensions}). Note that none of the measures is without limitations (Tab.~\ref{tab:limitation-summary}).
    \item  We present an empirical analysis of the existing measures to study the correlation among measures, investigate the relations between fairness and relevance, and compare the corrected measures against the original measures on real-world and synthetic datasets ($\S$\ref{s:exp}). 
    \item We provide insights to guide the correct use of different types of individual item fairness measures ($\S$\ref{s:discussion}).
\end{itemize}

\section{Individual Item Fairness Measures}
\label{s:priorwork}

We present our notation ($\S$\ref{ss:notation}) and the  
eight exposure-based evaluation measures of individual item fairness that we study in this work ($\S$\ref{ss:onlyfair}). 
To the best of our knowledge, these eight measures are all the measures of individual item fairness that exist in RSs.\footnote{Based on publications up to August 2022.} Several of these measures are taken from \cite{Wang2023ASystems}. 

\subsection{Notation and Examination Functions}
\label{ss:notation}

\begin{table}[tb]
\caption{Summary of the notation.}
\label{tab:notation}
\resizebox{0.98\textwidth}{!}{
\begin{tabular}{ll}
\toprule
Notation &Explanation \\
\midrule
$U=\{u_1, u_2, \dots, u_m\}$ &The set of users \\
$I = \{i_1, i_2, \dots, i_n\}$ &The set of items \\
$|U| = m$ &The unique number of users in dataset \\
$|I| = n$ &The unique number of items in dataset\\
$k$ & The cut-off threshold\\
$km$ & The total number of recommendation slots\\
$r_{u,i} \in \{0,1\}$ & The relevance of item $i$ to user $u$ \\
$z(u,i) \in \{1, 2, \dots, n\} $ & The rank position of item $i$ for user $u$\\
$z(u,i,w) \in \{1, 2, \dots, n\} $ & The rank position of item $i$ for user $u$ in round $w$\\
$R_{u}^{k}$ & The top $k$ recommendations for user $u$ \\
$R_{u,w}^{k}$ & The top $k$ recommendations for user $u$ in round $w$ \\
$R$ & The set of top $k$ unique items recommended to all users\\
$1_\mathcal{A}(x)=1$ if $x \in \mathcal{A}$, else 0 & Indicator function\\
\bottomrule
\end{tabular}}
\end{table}

\begin{table*}
\centering
\caption{Examination functions used in fairness measures.}
\label{tab:exp-weigh}
\begin{tabular}{ll>{\raggedright}p{2.5cm}p{2.5cm}<{}}
\toprule
         & Equation & Measure & Reference                        \\ 
         \midrule
uniform  & $e_{\text{unif}}(u,i)\equiv 1, \forall (u,i)$                         & Jain, QF, Ent, Gini, FSat, VoCD & \cite{jain1984quantitative,Zhu2020FARM:APPs,Shannon1948ACommunication, Gini1912VariabilitaMutabilita, Wang2022ProvidingSystems,Patro2020FairRec:Platforms} \\

DCG      & $e_{\text{DCG}}(u, i, w) = 1/\log_2 (z(u,i,w)+1)$         & Gini-w & \cite{Do2021Two-sidedDominance}   \\
RBP      & $e_{\text{RBP}}(u,i,w) = \gamma^{z(u,i,w)-1}$         & II-D, AI-D   & \cite{Wu2022JointRecommendation}                 \\
\bottomrule
\end{tabular}
\end{table*}

Given a set of users $U=\{u_1, u_2, \dots, u_m\}$, with $|U| = m$,  and 
a set of items $I = \{i_1, i_2, \dots, i_n\}$, with $|I| = n$, for each user $u \in U$ we rank all $n$ items to produce the full recommendation list. 
The list of the top $k$ recommended items to user $u$ is $R_u^{k}$, 
and the set of all top $k$ recommended items to all users is $R = \bigcup_{u \in U}{R_u^{k}}$.
If $u$ finds the recommended item $i$ relevant, we write $r_{u,i}=1$, otherwise $r_{u,i}=0$. The rank position of item $i$ in the recommendation list for user $u$ is $z(u,i)$. 
As there exist fairness measures that consider multiple rounds of recommendation, we also use the following notations for such measures: the rank position of item $i$ for user $u$ in round $w$ is $z(u,i,w)$ and $R_{u, w}^{k}$ is the list of the top $k$ recommended items to user $u$ in round $w$.
Tab.~\ref{tab:notation} summarizes our notation.

Fairness for individual items is closely linked to exposure, which is identified as the appearance of an item in the top $k$ recommendations for a user. Exposure can be quantified in several ways using different examination functions $e(\cdot)$ in various binary or graded ways. 
Examination functions are functions modelling the probability of a user seeing an item that is exposed to the user. All examination functions in this paper assume that this probability only depends on $z(u,i)$, the rank position of an item $i$ for user $u$. Tab.~\ref{tab:exp-weigh} presents the examination functions used by the individual item fairness measures that are included in this paper. These examination functions apply either no discount, logarithmic discount, or exponential-like discounting. 

The simplest examination function is uniform, $e_{\text{unif}}$, and assumes a constant weight of $1$ for each rank position \cite{jain1984quantitative,Zhu2020FARM:APPs,Shannon1948ACommunication, Gini1912VariabilitaMutabilita, Wang2022ProvidingSystems,Patro2020FairRec:Platforms}.
The two other examination functions are $e_{\text{DCG}}$ \cite{Do2021Two-sidedDominance} 
and $e_{\text{RBP}}$ \cite{Wu2022JointRecommendation}, which 
use the discount function based on Discounted Cumulative Gain (DCG)~\cite{Jarvelin2002CumulatedTechniques} and Rank-Biased Precision (RBP)~\cite{Moffat2008Rank-biasedEffectiveness} respectively.
In $e_{\text{RBP}}$, the parameter $\gamma$ is the user's patience, i.e., the probability of the user examining the next ranked item. For example, $\gamma=0.8$ in \cite{Wu2022JointRecommendation} and $\gamma=0.5$ in \cite{Diaz2020EvaluatingExposure}.

\subsection{Measures of Individual Item Fairness}
\label{ss:onlyfair}

We present the eight measures that so far have been used to quantify fairness for individual items (without considering item relevance), as well as the context in which they are used in the original work. We use the subscript $\cdot_{\text{ori}}$ to denote the original formulation of a measure as opposed to our corrected version that we present later in $\S$\ref{s:extensions}. Note that these measures are typically used with a fixed cut-off $k$, and therefore the number of recommendation slots $km$ is also fixed. 

\subsubsection[Jain's Index (Jain)]{Jain's Index (Jain) \cite{jain1984quantitative}}\label{sss:jain-ori} Jain, which was originally defined for fairness in computer networks, 
has been used in \rs~\cite{Zhu2020FARM:APPs} to measure how consistent item exposure is in relation to the number of times an item is recommended. The original work uses this measure to evaluate ``fairness of the recommendation opportunity''. Jain is the ratio between the square of the number of recommendation slots and the sum of squares of the number of times each item is recommended, where the ratio is divided by the number of items in the dataset. It is calculated as follows: 
\begin{equation}
\label{eq:jain-ori}
\text{Jain}_{\text{ori}} = 
\frac{
    \left[
        \sum\limits_{i\in I} \sum\limits_{u\in U}1_{R_{u}^{k}}(i)
    \right]^2
    }
{n \sum\limits_{i\in I} \left[\sum\limits_{u\in U}
1_{R_{u}^{k}}(i)\right]^2}
= \frac{(km)^2}{n \sum\limits_{i\in I} \left[\sum\limits_{u\in U}
1_{R_{u}^{k}}(i)\right]^2}
\end{equation}
where 
$1_{R_{u}^{k}}(i)=1$ if item $i$ is in the top $k$ recommendations for user $u$, and 0 otherwise. 
$\sum\nolimits_{u\in U}
1_{R_{u}^{k}}(i)$ counts how many times item $i$ is recommended in the top $k$ across all users. The range of Eq.~\eqref{eq:jain-ori} is $[0,1]$.\footnote{This range is based on the original paper, \cite{jain1984quantitative}} The range of Jain matters as \cite{Zhu2020FARM:APPs} analysed the absolute values of Jain, in addition to observing the difference of the scores. 
The higher the Jain score, the fairer the recommendation with respect to individual items (i.e., items are exposed consistently with respect to other items in the dataset). 
E.g., if $60\%$ of the items in the dataset are exposed equally to all users, for instance, $R_{u_1}^{3}=[i_1,i_2,i_3],\ R_{u_2}^{3}=[i_4,i_5,i_6],\ n=10$, then $\text{Jain}=0.6$. However, this interpretation does not hold and becomes less intuitive when items are not exposed equally, which is often the case. For instance, $R_{u_1}^{3}=[i_1,i_2,i_3],\ R_{u_2}^{3}=[i_1,i_2,i_4],\ R_{u_3}^{3}=[i_1,i_5,i_6],\ n=10$. In this case, $60\%$ of the items are exposed but $\text{Jain}=0.476$. In real-life, it is unlikely that items are exposed equally, which means that Jain's interpretation suffers from this limitation, more often than not. 

\subsubsection{Qualification Fairness (QF) \cite{Zhu2020FARM:APPs}} \label{sss:qf-ori}
QF is a modification of Jain that measures how many items are in the set of top $k$ recommended items $R$, divided by $n$, the total number of items in the dataset. 
The authors of the original measure explained in \cite{Zhu2020FARM:APPs} that the measure only considers whether an item in the dataset is recommended, as opposed to how many times it is recommended. 

\begin{equation}
\label{eq:qf-ori}
    \text{QF$_{\text{ori}}$} =  
    \frac{
        \left[
            \sum\limits_{i\in I} 1_{R}(i)
        \right]^2
        }
    {n \sum\limits_{i\in I}
    \left[1_{R}(i)\right]^2} 
    = \frac{
        \left[
            \sum\limits_{i\in I} 1_{R}(i)
        \right]^2
        }
    {n \sum\limits_{i\in I}
    1_{R}(i)}                                   
    = \frac{\sum\limits_{i\in I} 1_{R}(i)}{n} 
    = \frac{|R|}{n}                                 
\end{equation}

\noindent The QF range is $[0,1]$. The higher the score, the fairer the recommendation. 
Along with the relative comparison of the QF scores, the absolute values of QF scores are also taken into account in \cite{Zhu2020FARM:APPs,Mansoury2020FairMatch:Systems}, where 
a score of $1$ means that all items in the dataset are in the top $k$ at least once. 
Formally, QF (Eq.~\ref{eq:qf-ori}) is equivalent to Coverage \cite{Herlocker2004EvaluatingSystems}, a measure of diversity, which has been used to evaluate fairness too \cite{Mansoury2020FairMatch:Systems}.

\subsubsection{Entropy (Ent) \cite{Shannon1948ACommunication}} Ent measures how uniform the exposure of the recommended items is  \cite{Patro2020FairRec:Platforms,Mansoury2021ASystems,Mansoury2020FairMatch:Systems}. 
In \cite{Patro2020FairRec:Platforms}, Lorenz curves were used to detect massive differences in individual item exposures and therefore, Entropy-like measure was proposed to quantify the inequality of item exposure: 

\begin{equation}
   \text{Ent$_{\text{ori}}$} = 
        - \sum\limits_{i \in I}{p(i) \log{p(i)}} \qquad \text{and}\qquad  p(i) = \frac{\sum\limits_{u\in U}1_{R_{u}^{k}}(i)}{km}
        \label{eq:ent-ori}
\end{equation}
where $p(i)$ is the recommendation frequency of $i$, i.e., how often item $i$ is recommended in the top $k$ to any user in the dataset, divided by the available recommendation slots $km$. In~\cite{Patro2020FairRec:Platforms}, $\log$ is the log base-$n$. 
It is unclear what log base is used in \cite{Mansoury2021ASystems,Mansoury2020FairMatch:Systems}. When the log base is $b$, Ent ranges between $[0, \log_b{n}]$ while for log base-$n$, the range is $[0, 1]$. 

In the past, this measure has been used by \cite{Patro2020FairRec:Platforms, Mansoury2020FairMatch:Systems, Mansoury2021ASystems} to compare recommender models using absolute values, where a higher Ent is interpreted as having a more uniform distribution of the recommended items, and thus fairer.

\subsubsection{Gini Index (Gini) \cite{Gini1912VariabilitaMutabilita}} \label{sss:gini-ori} 
Gini is a measure of variability i.e., the mean difference from all observed quantities \cite{Ceriani2012TheGini}. It is most commonly used to measure inequality in the distribution of economic income, where the intuition is that a Gini score of 1 means that one entity receives all the income. 
Similarly in \rs, it is used to measure how much the distribution of item exposure deviates from an equal/uniform distribution \cite{Mansoury2020FairMatch:Systems, Do2022OptimizingRankings, Do2021Two-sidedDominance}. Formally, Gini is defined as follows: 

\begin{equation}
   \text{Gini$_\text{ori}$} = \frac{\sum\limits_{j=1}^n{(2j-n-1) Ex_j}}{n\sum\limits_{j=1}^n Ex_j} 
   \quad \text{and}\quad  
Ex_j = \sum\limits_{u\in U} \sum\limits_{w=1}^W 1_{R_{u,w}^{k}}(x_j)\cdot e_{(\cdot)}(u,x_j,w)
        \label{eq:gini-ori}
\end{equation}
where $x_j$ is the item with the $j$-th least amount of $Ex_j$, the total exposure received by that item across $W$ rounds of recommendations.\footnote{This is the total exposure received by each item, including items that are not recommended to any users. The scores are sorted as $Ex_1, Ex_2, \dots Ex_n$. Ties between $Ex_j$'s do not affect the final score.} $W$ is the number of rounds, and 
$1_{R_{u,w}^{k}}(i)=1$ when item $i$ is in user $u$'s top $k$ recommendation list in round $w$. 
The examination function $e_{(\cdot)}(u,x_j,w)$ used in \cite{Mansoury2020FairMatch:Systems} is uniform, $e_{\text{unif}}(u,x_j,w)\equiv 1$ and in \cite{Do2022OptimizingRankings, Do2021Two-sidedDominance} the examination function $e_{\text{DCG}}$  
(see Tab.~\ref{tab:exp-weigh}) is used.
We refer to the latter case as Gini-w. 
Gini and Gini-w's range is $[0,1]$, where 0 means that there is an equal distribution of item exposure for all items in the dataset (fairest case). 
The range is important for the interpretability of the measure. Having an interpretable range is more important for how Gini and Gini-w quantify fairness when looking at the absolute values of the measure, like in \cite{Mansoury2020FairMatch:Systems, Do2021Two-sidedDominance}, than when looking at the difference in model rankings, like in \cite{Do2022OptimizingRankings}.

\subsubsection{Fraction of Satisfied Items (FSat) \cite{Patro2020FairRec:Platforms}}\label{sss:fsat}FSat is defined in the context of maximin-shared fairness, where fairness means that each item is 
recommended at least $\frac{km}{n}$ times, as there are only $km$ slots that should ideally be distributed equally between $n$ items. However, this distribution is impossible if 
the number of slots is not divisible by the number of items in the dataset, $n \nmid km$. The requirement is relaxed to recommending each item at least $\left\lfloor \frac{km}{n}\right\rfloor$ times (the maximin share) for it to be a fair recommendation. An item is satisfied iff its exposure is more than or equal to the maximin share. $\text{FSat}$ measures the number of satisfied items divided by the total number of items: 
\begin{equation}
\label{eq:fsat-ori}
    \text{FSat}_\text{ori}=
    \frac{1}{n} 
         \sum\limits_{i \in I} 
        \delta
            \left(
            \sum\limits_{u\in U} 1_{R_u^{k}}(i) \geq \left\lfloor \frac{km}{n}\right\rfloor
            \right)
\end{equation}

\noindent where $\delta(\cdot)=1$ when the expression $\cdot$ is True and 0 otherwise. 
FSat has a range of $[0,1]$, and the higher, the fairer. The range of values matters, for example \cite{Patro2020FairRec:Platforms} has used both absolute values and difference in values to interpret FSat. 

\subsubsection{Violation of Coverage Disparity (VoCD) \cite{Wang2022ProvidingSystems}}\label{sss:vocd-ori} 
VoCD is a fairness constraint. In \cite{Wang2022ProvidingSystems}, VoCD 
is used to optimise the recommendations for fairness during the training process, but not used for evaluating the final recommendation output. We include VoCD in this work to provide insights into what transpires when VoCD is used as an evaluation measure in its current formulation. VoCD is also the only measure operating on the Lipschitz condition~\cite{Dwork2012FairnessAwareness}, which requires similar individuals to be treated similarly. 
The idea behind VoCD is that any two $\alpha$-similar recommended items should receive similar coverage. 
Two distinct items $i,i' \in R$ are $\alpha$-similar if $d(i,i')=1-sim(i,i')\leq \alpha$, 
where $d(i,i')$ is the cosine distance and $sim(i,i')$ is the cosine similarity between the embeddings\footnote{The original paper specifically defines that they use embeddings, but in principle, any representation could work.} of item $i$ and $i'$ respectively, and $\alpha$ is a parameter. 
Similar coverage means that the Coverage Disparity (CD) of those items, which is proportional to their exposure difference, must not exceed a threshold $\beta$. 
VoCD thus measures the average violation of the maximum allowed coverage disparity $\beta$ in all pairs of $\alpha$-similar recommended items:
\begin{align}
\label{eq:vocd-ori}
\begin{split}
    \text{VoCD$_\text{ori}$}
    &= \frac{1}{|A|} \sum\limits_{\forall (i, i') \in A}{\max{(CD(i,i')-\beta,0})} \\
    CD(i,i') &= 
    \left| 
    \frac{\sum\limits_{u\in U}1_{R_{u}^{k}}(i) - \sum\limits_{u\in U}1_{R_{u}^{k}}(i')
    }{
    \max\left(
        \sum\limits_{u\in U}1_{R_{u}^{k}}(i), \sum\limits_{u\in U}1_{R_{u}^{k}}(i')
        \right)} 
    \right|
\end{split}
\end{align}
\noindent where $A$ is the set of all pairs of $\alpha$-similar items and $CD(i,i')$ is the coverage disparity between item $i$ and $i'$. $\text{VoCD}=0$ means that there is no violation of coverage disparity between any pairs (i.e., fair). 
In \cite{Wang2022ProvidingSystems}, the absolute value of VoCD affects the parameter that is used to control fairness during the training process. 
VoCD is customisable  w.r.t.~fairness and similarity: a lower\footnote{This is \textit{lower} instead of \textit{higher} as the original authors \cite{Wang2022ProvidingSystems} defined the term $\alpha$-similar items based on the cosine distance between the two items being not more than $\alpha$.} 
$\alpha$ means a stricter similarity requirement and a lower $\beta$ means a stricter fairness requirement. 

\subsubsection{Individual-user-to-individual-item disparity (II-D) \cite{Wu2022JointRecommendation}} \label{sss:iid-ori} 
II-D was first defined by \cite{Diaz2020EvaluatingExposure} to quantify the mean squared difference between system exposure and random exposure in individual queries and individual items, 
where there is a distribution of rankings (stochastic rankings). II-D is a resulting component of decomposing another measure in the original work, which quantifies item fairness proportional to item relevance to users. It was redefined by \cite{Wu2022JointRecommendation} for $W$ rounds of recommendations in RSs as: 

\begin{equation}
    \label{eq:iid-ori}
     \text{II-D}_\text{ori} = \frac{1}{m} \frac{1}{n} 
     \sum\limits_{u \in U} \sum\limits_{i \in I} \left(E_{u,i}^{ } - E_{u,i}^{\sim}\right)^2
\end{equation}

\begin{equation}
    \label{eq:eui}
E_{u,i} = \frac{1}{W} \sum\limits_{w=1}^{W}1_{R_{u,w}^{k}}(i) \cdot e_{\text{RBP}}(u,i,w)
\qquad \text{and} \qquad
E_{u,i}^{\sim} = \frac{1-\gamma^{k}}{n(1-\gamma)} 
\end{equation}

\noindent where $E_{u,i}$ is the expected exposure of $i$ to $u$ as per a stochastic ranking policy,  
$E_{u,i}^{\sim}$ is the expected exposure of $i$ to $u$ based on a uniformly random distribution over all permutations of items, and $\gamma$ is an arbitrarily-set parameter for user patience. 
The examination function based on RBP (see Tab.~\ref{tab:exp-weigh}) is used in $E_{u,i}$ and the equation of $E_{u,i}^{\sim}$ is derived based on the same examination function \cite{Wu2022JointRecommendation}. The range of II-D is not well-known,\footnote{The authors of the original measure did not state the range of II-D.} but a lower value means a fairer recommendation. In \cite{Wu2022JointRecommendation}, min-max normalisation is performed on II-D post-computation, such that the range is $[0,1]$. 
This range of values matters; for example \cite{Wu2022JointRecommendation} has used the absolute values of II-D to analyse fairness and relevance trade-off, in addition to looking at the difference in model rankings based on II-D scores. 

\subsubsection{All-users-to-individual-item disparity (AI-D) \cite{Wu2022JointRecommendation}} \label{sss:aid-ori} 
AI-D computes the mean of the squared difference between system exposure and random exposure in each item. AI-D is similar to II-D in the sense that it is originally used for multiple rounds of recommendations and also a component resulting from the decomposition of another measure proposed by \cite{Wu2022JointRecommendation} that considers fairness w.r.t.~relevance. 
However, unlike II-D, AI-D is sensitive to whether an item is recommended to multiple users due to the aggregation of the difference in exposure being done per item.

\begin{equation}
    \label{eq:aid-ori}
     \text{AI-D}_\text{ori} = 
        \frac{1}{n} \sum \limits_{i \in I}
            \left(
                \frac{1}{m}\sum \limits_{u \in U} E_{u,i}^{ }
                - \frac{1}{m}\sum \limits_{u \in U} E_{u,i}^{\sim}
            \right)^2
\end{equation}
where $E_{u,i}^{ }$, $E_{u,i}^{\sim}$ are as per Eq.~\eqref{eq:eui}. The range of AI-D is not well-known, but a lower value means fairer recommendation. 
Like on II-D, a post-computation min-max normalisation is also performed by \cite{Wu2022JointRecommendation} on AI-D, resulting in a $[0,1]$-range. 
The range of values matters; for instance \cite{Wu2022JointRecommendation} has used the absolute values of AI-D to analyse fairness and relevance trade-off, on top of looking at the difference in model rankings based on AI-D scores. 

\section{Measure Limitations}
\label{s:limitations}
We identify $5$ theoretical limitations in the measures presented in $\S$\ref{s:priorwork} (summarised in Tab.~\ref{tab:limitation-summary}). 
We use the term `limitation' in the sense that regardless of the reason, a measure fails to quantify or fulfill properties that are important for evaluating fairness. 
Some of these limitations rarely occur, e.g. related to edge cases ($\S$\ref{ss:always_fair}), yet some are more likely to occur in practical scenarios ($\S$\ref{ss:nonreal} \& $\S$\ref{ss:undefinedness}). In the headings, we put the name of the affected measures in brackets. 

In practice, even if the limitation transpires by design, the design of the measure still restricts its usage under the conditions that we explain below. The identified limitations are independent of the recommender algorithm, as long as the recommender is a top-$k$ recommender, which is the most common recommendation scenario in practice. 
We accompany each measure name by  \up~or \down, denoting that the higher (\up) or the lower (\down) the score of the measure, the fairer the recommendation. 

\subsection{Limitation 1: Non-realisability}\label{ss:nonreal}  

This is a novel limitation identified by us and affects \emph{all measures}. We define non-realisability as the limitation whereby the max/min score of the evaluation measure cannot be reached at the top-$k$. 
As argued in \cite{Moffat2013SevenMetrics}, a desirable property of effectiveness measures is their realisability. While the realisability property in \cite{Moffat2013SevenMetrics} is related to the number of relevant items, non-realisability for fairness measures is related to the number of recommendation slots ($km$) and the number of items that are in the dataset ($n$); we explain this relationship below as part of the causes of this limitation. 
In practice, the non-realisability limitation makes fairness scores hard to interpret because 
when the worst or best possible fairness score varies based on the dataset ($m$, $n$) and experimental choice of threshold $k$, it is unknown whether the fairness score obtained for a model is closer to the max or min score. 
For instance, if a higher-is-fairer measure ranges in $[0, 1]$ and a model achieves a score of $0.2$, one might think that the model is not very fair. However, if the maximum achievable score for that case (e.g., if all top $k$ items are fair) is $0.22$, then the model might actually be fair, but this cannot be known 
from the score of the evaluation measure.
We identify four different causes of non-realisability.

\subsubsection{Cause 1 (\up Jain, \up QF, \up Ent, \up FSat, \down Gini, \down Gini-w)} 
\label{sss:cause-no-recommend}
Non-realisability can occur if the most unfair score is only given to an unrealistic recommendation scenario, which we explain next as it differs per measure. 
Specifically, the score can never be 0 for \up Jain, unless the number of slots is 0, i.e., $k=0$ or $m=0$, which does not make sense because $k=0$ means that no recommendation at all is outputted, and $m=0$ means that there are zero users to recommend to. The score of \up Ent can only be 0 when there are no items in the dataset ($n=0$), which also does not make sense because it means that there is nothing to recommend. For \up QF/FSat, the score can only be 0 if there are \textit{no} recommended items to \textit{any} users ($|R|=0$), which does not make sense either. 
On the other hand, the score can only be 1 for \down Gini/Gini-w when a single item is recommended at all $k$ slots for each user, which is a highly unlikely artificial outcome; to our knowledge, no reasonably performing recommender model can produce such an output. 
All of the above conditions are unrealistic. 
A consequence of the above is that fairness is overestimated by these measures. Instead of the unrealistic situations above, it is the realistically unfairest recommendation for \up Jain/QF/Ent/FSat that should be mapped to 0 and for \down Gini/Gini-w to 1. 

\subsubsection{Cause 2 (\up Jain, \up QF, \up Ent, \down Gini, \down Gini-w, \up FSat, \down II-D, \down AI-D)}\label{sss:cause2} The second cause of non-realisability that we identify is that when the number of recommendation slots $km$ is less than the number of items $n$, then the score cannot be the fairest, as some items cannot be exposed due to the limited availability of recommendation slots.  
This means that the max score (most fair) cannot be reached by the measure, even if all recommended items are fair. 
In the datasets in Tab.~ \ref{tab:dataset}, $km > n$ for $k \in \{10,20\}$, but for some very large datasets like LFM-1b \cite{Schedl2016TheRecommendation} or Four\-squ\-are NYC \cite{Yang2013ASystem}, $km < n$. 

For II-D and AI-D, we show that the score cannot be the fairest, during single or multiple rounds of recommendations. For example, given $k=1, m=2, n=3$, and considering all possible orders of recommendations for a single round of recommendation, the lowest scores for II-D and AI-D are \nicefrac{2}{9} and \nicefrac{1}{18} respectively. 
In the case of multiple recommendation rounds, the total number of slots is $kmW$, where $W$ is the number of rounds. As an example, given $k=1, m=2, W=2, n=5$, the lowest possible II-D and AI-D are 0.06 and 0.01 respectively. So, in these cases, the lowest score (that should be the fairest) is not zero, even if the recommendations are made as fair as possible (close to random exposure).

This leads to an underestimation of fairness: the fairest value of the measure cannot be reached for some datasets (regardless of recommendation quality), so the measure is not evenly robust across datasets and can underestimate fairness. 

\subsubsection{Cause 3 (\up Jain, \up Ent, \down Gini, \down Gini-w \down II-D, \down AI-D)}\label{sss:cause3} The third cause of non-realisability that we identify is that for Jain, Ent, Gini, and Gini-w, even when $km > n$, the measure still cannot reach the theoretical fairest value if the number of items is not an exact multiple of the recommendation slots, $n \nmid km$. E.g., if $km=4$, $n=3$, three slots can be filled with one unique item each, but no matter which item fills the last slot, one item will be recommended one more time than the rest. 
Likewise, the same applies for II-D and AI-D when the number of slots across all rounds ($kmW$) is not divisible by $n$. For example, when $k=m=W=2, n=3$, the minimum II-D is $0.02$ and the minimum AI-D is $0.005$. 
This limitation consequently leads to the same issue of robustness and underestimation of fairness, as described in \textit{Cause 2}.

\subsubsection{Cause 4 (\down Gini-w, \down VoCD, \down II-D, \down AI-D)}
\label{sss:cause_nonreal}
The fourth case of non-realisability that we identify is that measures cannot reach the theoretical (un)fairest value, as the exact formulation of the max/min achievable score is 
unknown,\footnote{Here, `unknown' is only related to the exact max/min formulation, as the maximum and minimum can always be computed by enumerating all possibilities, albeit being a costly process.}  
making the score hard to interpret. This happens because the most/least fair recommendation cannot be analytically determined due to parameters in the measure or item exposure being weighted by a non-uniform examination function.  
This causes the measure to have a range different from its theoretical range, i.e., $[0,1]$, which we explain next for each measure. 

\down Gini-w may not reach $0$ or $1$ due to non-uniform exposure. E.g., when $k=n=3, m=2$, the minimum and maximum \down Gini-w for all possible recommendation lists are $0.0373$ and $0.156$ respectively. As $n|km$, this is separate from \textbf{non-realisability}, \textit{Causes 1--3}.

For \down VoCD, as $CD(i,i') \in [0,1)$,  \down VoCD $\in [0,1-\beta)$. However, the score of \down VoCD depends on item similarity, making the most unfair score unreachable. Even though it is impossible to formulate an exact achievable maximum value for \down VoCD, 
we formally prove in App.~\ref{app:vocd} that $\forall \alpha \in [0,2], \beta \in [0,1)$, the maximum \down VoCD, $\text{VoCD}_{\max} \leq \frac{m-1}{m} - \beta$. This is obtained when there is only one pair of similar items which is recommended $1$ and $m$ times each. E.g., 
$R_{u_1}^{2} = [i_1, i_2],\ R_{u_2}^{2} = R_{u_3}^{2} = [i_1, i_3]$ where $\text{VoCD}_{\max}$ is $\nicefrac{2}{3}$, which happens when only $i_1$ and $i_2$ are similar. So, the most unfair score is non-realisable, as Eq.~\eqref{eq:vocd-ori} depends on item-pair similarity. 

\down II-D and \down AI-D also may not reach 0 or 1, even in the context of multiple rounds of recommendations and having enough slots for all items. For example, when $k=m=W=2, n=3$, considering all possible ways of recommending items, the minimum values of II-D and AI-D are 0.02 and 0.005 respectively, while the maximum values are 0.187 for both. We posit this to be due to the exponential-like exposure.

A summary of situations producing the theoretical most (un)fair scores in existing measures is given in Tab.~\ref{tab:situations}.

\begin{table*}[!h]
    \centering
    \caption{Situations that produce the theoretical most (un)fair score in existing individual item fairness measures.}
    \label{tab:situations}
    \begin{tabular}{l>{\raggedright}p{5.75cm}p{5.75cm}<{}}
        \toprule
         Measure&  Most Unfair & Most Fair\\
         \midrule
         Jain & no recommendation slots ($km=0$)& all items recommended the same amount\\
         QF& no items recommended ($|R|=0$)& all items exposed, no matter how many times\\
         Ent & no items in the dataset ($n=0$)& all items recommended the same amount\\
         Gini, Gini-w & a single item is recommended at all rank positions for all users  & all items recommended the same amount (or same total exposure weight)\\
         FSat& no items recommended ($|R|=0$)& all items recommended $\geq \left\lfloor\frac{km}{n} \right\rfloor$ times\\
         VoCD& not possible to deduce from the formula and description& all pairs of similar recommended items recommended similar amount with a normalised difference \\
         II-D &not possible to deduce from the formula and description& exposure distribution of recommended items matches exposure given by random distribution\\
         AI-D &not possible to deduce from the formula and description& exposure distribution of recommended items matches exposure given by random distribution, with the most possible number of unique items in top $k$\\ 
         \bottomrule
    \end{tabular}
\end{table*}

\subsection[Limitation 2: Quantity-insensitivity (QF)]{Limitation 2: Quantity-insensitivity (QF\ori)}\label{ss:quantity-insensitivity} 
This limitation is part of the design choice by the authors of the original QF measure \cite{Zhu2020FARM:APPs} based on a specific concept of fairness, which we explain next. 
Quantity-insensitivity means that the measure ignores how often an item is recommended across all users in a recommendation round. 
In economics, \cite{Allison1978MeasuresInequality} states that `sensitivity to transfer' is a basic criterion of an inequality measure. Similarly, we think that when exposure increases (or decreases) for an item, the fairness measure should be sensitive to the change. 
Meanwhile, QF\ori~ makes no distinction between items that are recommended once or more than once.
To illustrate, consider these scenarios: 1) $R_{u_1}^{2}=[i_1, i_2], R_{u_2}^{2}=[i_2, i_3],  R_{u_3}^{2}=[i_1, i_3]$ and 2) $R_{u_1}^{2}=R_{u_2}^{2}=[i_1, i_2],  R_{u_3}^{2}=[i_1, i_3]$. Assuming $n=5$,  QF\ori~ would be 0.6 for both cases, even though in the second scenario $i_1$ is recommended more times than $i_2$, which is recommended more than $i_3$.

As a result of this design choice, the score does not reflect the repeated recommendations of the same item to many users, which may indicate unfairness (e.g., popularity bias). This is a design limitation that one should be aware of when using QF.

\subsection[Limitation 3: Undefinedness (Ent)]{Limitation 3: Undefinedness (Ent\ori)} 
\label{ss:undefinedness} 
We define the limitation of undefinedness as the measure giving an undefined value. In practice, this limitation renders the measure incomputable when encountering an (edge) case.\footnote{This is reminiscent of the completeness property in \cite{Moffat2013SevenMetrics}.} 
This is not negligible, as an assessment question for the design of (fairness) measures is related to how the measure responds to edge cases \cite{Raj2022MeasuringResults}. 
For Ent\ori, the case is related to the possibility of encountering the undefined value of $\log{0}$ during computation.
Undefinedness happens when there is at least one item from the dataset that does not appear in the recommendations at all,\footnote{
$\log{p(i)}$ in Eq.~\ref{eq:ent-ori} is undefined if item $i$ is not in the recommendation list for any users ($p(i)=0$).} 
which happens often because not all items in the datasets are guaranteed to be at the top $k$. 
For example, given $R_{u_1}^{2}=R_{u_2}^{2}=[i_1, i_2]$ and $I=\{i_1, i_2, i_3\}$, the value of $p(i_3)=0$, and this entails $\log{p(i_3)}=\log{0}$. 
Such a situation is common, as it will later be seen in Tab.~\ref{tab:base-main}, and therefore we do not consider this as an edge case. When the measure is incomputable for several models, its interpretation is less meaningful. 

We exclude the case where no item is recommended to any users ($|R|=0$), as this is a trivial case and it does not make much sense to evaluate fairness when there is no item being recommended. Regardless of the triviality, we identify some measures that are incomputable under this edge case: Jain\ori, Ent\ori, Gini\ori, Gini-w\ori, and VoCD\ori. Note that we do not consider these four measures to have the undefinedness limitation just for this reason. 

\subsection{Limitation 4: Always-fair (\ups FSat)}
\label{ss:always_fair}
We define the limitation of always-fair as the measure giving the fairest score regardless of the content of the recommendation list, under a specific condition depending on the particular measure. 
This happens for \up FSat when $km<n$, as empirically discovered by \cite{Patro2020FairRec:Platforms}. The maximin share, in this case, is 0 and all items are deemed satisfied as per the definition in $\S$\ref{sss:fsat}, regardless of the actual distribution of recommended items. 
While this is partly due to the design choice of FSat which is based on the maximin share, this means that \up FSat will always be $1$, rendering the measure unsuitable for use cases where $km<n$. 
Even though this limitation has been empirically identified before, there was no formal definition of it, and that is what we do here.

\subsection{Limitation 5: Item-representation-dependence (\dws VoCD)} \label{ss:item-rep-dep}
We formally identify this limitation in this work, even though it is part of an intentional choice of the measure, as opposed to an accidental or unforeseen byproduct of the design. Item-representation-dependence means that the score of the measure varies according to how item representations are built 
(e.g., embedding, graphs). The max value of \down VoCD depends on which item pairs are similar based on how items are represented. 
Even though the dependence on item representation is part of the design choice for VoCD, the limitation of this design should be taken into consideration when one chooses a fairness measure, e.g. ensuring a same way of representing items for a fair comparison.
 
Depending on how item representations are built, there may be different pairs of similar items in the set $A$, yielding different \down VoCD scores. E.g., given $R_{u_1}^{2} = [i_1, i_2],\ R_{u_2}^{2} = [i_1, i_3]$, if $A=\{(i_1, i_2)\}$ as per one item representation, \down VoCD $= 0.5 - \beta$. However, if according to another item representation, $A=\{(i_2, i_3)\}$, then \down VoCD $=0$. 
While the former score represents a somewhat unfair RS, the latter denotes that the RS is fair. 
Note that one can use the same recommendation algorithm and the same dataset, but with different ways of representing the items, different VoCD scores may be obtained. Therefore, the limitation still holds even when the comparison is only performed within an algorithm and a dataset.

\begin{table}[tb] 
\caption{
Measures of individual item fairness and their theoretical limitations. We identify four different causes for non-realisability, denoted by \textit{C} in this table. 
}
\label{tab:limitation-summary}
\resizebox{\textwidth}{!}{
\begin{tabular}{l|c||c|c|c|c|c|c|c|c|c}
\toprule
\midrule
\parbox[t]{0.9\textwidth}
{Legend\\
\bbullet: we fully resolve the limitation\\
\nofix: the limitation is unresolvable\\
\checkmark: another measure resolves the limitation} 
&\rotatebox[origin=r]{90}{Source}
&\rotatebox[origin=r]{90}{Jain \cite{jain1984quantitative}}
&\rotatebox[origin=r]{90}{QF \cite{Zhu2020FARM:APPs}}
&\rotatebox[origin=r]{90}{Ent \cite{Shannon1948ACommunication}}
&\rotatebox[origin=r]{90}{Gini \cite{Gini1912VariabilitaMutabilita}}
&\rotatebox[origin=r]{90}{Gini-w \cite{Do2021Two-sidedDominance}}
&\rotatebox[origin=r]{90}{FSat \cite{Patro2020FairRec:Platforms}}
&\rotatebox[origin=r]{90}{VoCD \cite{Wang2022ProvidingSystems}}
&\rotatebox[origin=r]{90}{II-D \cite{Wu2022JointRecommendation}}
&\rotatebox[origin=r]{90}{AI-D \cite{Wu2022JointRecommendation}}
\\
\midrule
\midrule
non-realisability: cannot reach max/min score (cause number denoted by \textit{C}) &&&&&&&&&\\
\textit{C1.} Most unfair score is only given to an impossible scenario    &us&\bbullet&\bbullet&\bbullet&\bbullet&\bbullet&\bbullet&&&\\
\textit{C2.} Fewer recommendation slots compared to number of items&us&\bbullet&\bbullet&\bbullet&\bbullet&\bbullet&\bbullet&&\nofix&\nofix\\
\textit{C3.} Number of recommendation slots is indivisible by number of items&us&\bbullet&&\bbullet&\bbullet&\nofix&&&\nofix&\nofix\\
\textit{C4.} Non-realisability due to unknown formulation of max/min score
&us&&&&&\nofix&&\nofix&\nofix&\nofix\\
\midrule 
quantity-insensitivity: ignores frequency of item recommendation   &\cite{Zhu2020FARM:APPs}&&$\checkmark$&&&&&&&\\
\midrule
undefinedness: cannot be computed (undefined value)   &us&&&\bbullet&&&&&&\\
\midrule 
always-fair: gives fairest score regardless of recommendation contents   &\cite{Patro2020FairRec:Platforms}&&&&&&\nofix&&&\\
\midrule 
item-representation-dependence: depends on how items are represented   &us&&&&&&&\nofix&&\\
\midrule 
\bottomrule
\end{tabular}
}
\end{table}

\section{Resolving Limitations}
\label{s:extensions}

We explain how we resolve each limitation or why it is unresolvable. For the remainder of this paper, we refer to the original version of an evaluation measure $M$ as $M_{\text{ori}}$, and to our modified version of an evaluation measure $M$ as $M_{\text{our}}$. When $\cdot$\ori~or $\cdot$\our~is not specified, we refer to both the original and modified version simultaneously.

\subsection{Resolving Non-realisability (Limitation 1) and Undefinedness (Limitation 3)}
\label{ss:non_realisability}

\begin{table}[tb]
\caption{
Most (un)fair score @$k$. Note that the Most Fair @$k$ scores (except for Gini-w) remain the same as the theoretical fairest value (i.e., 0 or 1) when the number of recommendation slots is divisible by the number of items, $n \mid km$.}
\label{tab:bounds}

\centering
\begin{tabular}{lll}
\toprule
Measure &     Most Unfair @$k$ & Most Fair @$k$ \\
\midrule
   \up Jain\ori &          $\frac{k}{n}$ &        $\frac{(km)^2}{
 n\left(n\floorkm^2 
 + (km\bmod{n})\left(2 \floorkm+1\right) 
 \right)} $ \\
   \up QF\ori &            $\frac{k}{n}$ &        $\min\left(\frac{km}{n},1\right)$ \\
   \up Ent\ori &           $\log{k}$ &            Eq.~\eqref{eq:ent-max}\\ 
   \down Gini\ori &          $1-\frac{k}{n}$ &      $ \frac{(n-km \bmod n) (km \bmod n)}{kmn} $\\ 
  \down Gini-w\ori &          Eq.~\eqref{eq:gini-w-max} &      Eq.~\eqref{eq:gini-w-min} when $km\leq n$\\
   \up FSat\ori &          $\frac{k}{n}$   &      1\\
   \down VoCD\ori &          $\frac{m-1}{m} - \beta$ &      0\\
\bottomrule
\end{tabular}
\end{table}

We resolve causes 1, 2, and 3 of non-realisability via post-calculation correction of under/overestimated fairness scores based on the theoretical min and max values for a scenario with limited $km$ recommendation slots. 
Specifically, we rescale the range of the measures to the actual theoretically achievable most fair and unfair values. The rescaled measures either retain the $[0,1]$-range, or are now ranged in $[0,1]$. To rescale, we compute the measure's upper and lower bounds when possible (see Tab.~\ref{tab:bounds} and App.~\ref{app:boundsproof} for the derivations) by considering the most fair and unfair recommendation case for each measure, which we explain next.

For the measure that suffers from non-realisability due to \textit{Causes 1--2} (and quantity-insensitivity) i.e., QF, we posit that the most unfair recommendation $@k$ is when the same $k$ unique items are recommended to each of the $m$ users, resulting in the min exposure for items in $I$. 
Therefore, each of these $k$ items is recommended $m$ times. 
This is equivalent to the recommendation generated by Pop \cite{Rashid2002GettingYou} that gives the same $k$ most popular items to all users. 
We posit that the most fair case $@k$ is when $\min(km,n)$ unique items are recommended to $m$ users, i.e., the max number of items allowed by the recommendation slots is exposed.  

For measures that suffer from non-realisability, due to \textit{Causes 1--3} but not quantity-insensitivity (Jain, Ent, Gini, Gini-w, FSat), we consider the most fair recommendation to be when ($n-km \bmod n$) items are exposed $\floorkm$ times and the rest ($km \bmod n$) items are exposed $\floorkm +1$ times. The most unfair case for Jain, Ent, Gini, Gini-w, and FSat is the same as QF.

We then perform min-max normalisation (hereafter referred to as normalisation) using the most unfair/fair bounds as the min/max possible value of the measure. The general process is: let $x_{\max}$ be the max possible value of a fairness score $x$, and $x_{\min}$ be the min possible value of $x$. The normalised score of $x$, denoted by $x'$, is calculated using $x'= \frac{x-x_{\min}}{x_{\max}-x_{\min}}$. Note that this normalisation does not work when $x_{\max}=x_{\min}$ due to division by zero. This happens when the most unfair recommendation is equal to the fairest recommendation, which occurs when $k=n$. We therefore exclude this case from our corrections.

After normalisation, the measures quantify fairness of items by considering the following components: the recommendation list, the number of items in the dataset, and the most fair and most unfair recommendation. As a result, the most fair recommendation scenario and most unfair recommendation scenario are now mapped to the endpoints, instead of the unrealistic scenarios in Tab.~\ref{tab:situations}. 
Next, we explain in detail the normalisation of each measure.

For \up Jain\ori, we apply the following normalisation on Eq.~\eqref{eq:jain-ori}: as the higher the score, the fairer, we use Eq.~\eqref{eq:jain-max} as $x_{\max}$ and $x_{\min} = \frac{k}{n}$ because these are the most fair and unfair @$k$ values.  Eq.~\eqref{eq:jain-max} simplifies to $y=\frac{km}{n}$ when $km<n$, and simplifies to 1 when $n \mid km$.

\begin{equation}
\label{eq:jain-max}
     \text{Jain}_{\max} = \frac{(km)^2}{
     n\left(n\floorkm^2 
     + (km\bmod{n})\left(2 \floorkm+1\right) 
     \right)}   
\end{equation}

\begin{equation}
\label{eq:jain-norm}
    \text{Jain}_{\text{our}}  =  \frac{\text{Jain}_{\text{ori}}-\frac{k}{n}}{\text{Jain}_{\max}-\frac{k}{n}}
\end{equation}

We normalise \up QF\ori~ similarly to the above. As the higher the QF score, the fairer, we use $x_{max} = \min(y,1)$ and $x_{min} = \frac{k}{n}$. 
\begin{equation}
\label{eq:qf-norm}
\text{QF}_{\text{our}}  = 
\begin{cases}
    \frac{\text{QF}_{\text{ori}}-\frac{k}{n}}{1-\frac{k}{n}} 
    = \frac{|R|-k}{n-k} & \text{if } km \geq n \\
    \frac{\text{QF}_{\text{ori}}-\frac{k}{n}}{\frac{km}{n}-\frac{k}{n}} 
    = \frac{|R|-k}{k(m-1)} & \text{otherwise}
\end{cases}
\end{equation}

We acknowledge that by performing this normalisation on QF\ori, there is now an additional way of measuring fairness. The new score can now be interpreted as ``given that each item should be recommended at least once, how fair the recommendation is w.r.t. the most unfair and the fairest recommendation''. The most (un)fair recommendation and the normalisation depend on the number of recommendation slots, and we argue that it is better to use this number instead of using the number of items in the datasets; in practice, the number of items shown to the users is almost always limited. However, if QF is meant to be used for detecting the percentage of items that are exposed, then the QF\ori~ may be used at the expense of needing additional information on what is the best possible QF score, to know the limit of how fair the recommendation can be.
 
For \up Ent\ori, as the higher the score, the fairer, we use 
 Eq.~\eqref{eq:ent-max} as  $x_{\max}$ when $km \geq n$ or $x_{\max}=\log{km}$ otherwise, and 
 $x_{\min} = \log{k}$ for normalisation.  Eq.~\eqref{eq:ent-max} simplifies to $\log{n}$ when $n \mid km$.
\begin{align}
\label{eq:ent-max}
\text{Ent}_{\max} = 
&-
(n - km \bmod n)\left(\frac{\floorkm}{km}\log{\frac{\floorkm}{km}}\right) \\
&- (km \bmod n)\left(\frac{\floorkm+1}{km}\log{\frac{\floorkm+1}{km}}\right) 
\end{align}

However, \up Ent\ori~ still suffers from \textbf{undefinedness}, which we resolve by restricting the sum over $i$ in Eq.~\eqref{eq:ent-ori} to only recommended items:
\begin{equation}
\label{eq:ent-fair}
    \text{Ent$_{\text{def}}$} = 
    - \sum\limits_{i \in R}{p(i) \log{p(i)}} 
\end{equation}
where $p(i)$ is calculated via Eq.~\eqref{eq:ent-ori} and $p(i)>0$ because $i \in R$. Performing normalization on Eq.~\eqref{eq:ent-fair}, we obtain:
\begin{equation}
\label{eq:ent-norm}
\text{Ent}_{\text{our}}  = 
\begin{cases}
    \frac{\text{Ent}_{\text{def}}-\log{k}}{\text{Ent}_{\max}-\log{k}}& \text{if } km \geq n \\
    \frac{\text{Ent}_{\text{def}}-\log{k}}{\log{km}-\log{k}}
    = \frac{\text{Ent}_{\text{def}}-\log{k}}{\log{m}}  & \text{otherwise}
\end{cases}
\end{equation}

For \down Gini\ori, as the lower the score, the fairer, we use Eq.~\eqref{eq:gini-min} as $x_{\min}$ and $x_{\max} = 1-\frac{k}{n}$. Eq.~\eqref{eq:gini-min} simplifies to $1-y$ when $km>n$, and simplifies to 0 when $n \mid km$. 
    \begin{equation}
    \label{eq:gini-min}
     \text{Gini}_{\min} = 
     \frac{
     (n-km \bmod n) (km \bmod n)
     }
     {kmn} 
    \end{equation}
    \begin{equation}
    \label{eq:gini-norm}
    \text{Gini}_{\text{our}}  = 
    \frac{\text{Gini}_{\text{ori}}- \text{Gini}_{\min} }{1-\frac{k}{n} -  \text{Gini}_{\min} } 
\end{equation}
For \down Gini-w\ori, we only consider cases where $km \leq n$ as finding the most fair recommendation list across all users for the other cases is analytically not possible. The problem does not have a closed form solution and computing the solution requires solving a constrained optimization that considers all possible permutations of recommendations across users. 
We use Eq.~\eqref{eq:gini-w-min} as $x_{\min}$ when $km \leq n$, $x_{\min}=0$ otherwise, and Eq.~\eqref{eq:gini-w-max} as $x_{\max}$ to normalise \down Gini-w\ori.
    \begin{equation}
\label{eq:gini-w-min}
 \text{Gini-w}_{\min} =
  \frac{
    \sum\limits_{\ell=1}^k 
    \sum\limits_{j=n-\ell m+1}^{n- \ell m + m} (2j-n-1) \log_{\ell + 1}{2}}
    {mn\sum\limits_{\ell=1}^k{\log_{\ell+1}{2}}}
\end{equation}
\begin{equation}
\label{eq:gini-w-max}
 \text{Gini-w}_{\max} = 
 \frac{
 \sum\limits_{\ell=1}^k{(n-2\ell+1) \log_{\ell+1}{2}}}
    {n\sum\limits_{\ell=1}^k{\log_{\ell+1}{2}}}
\end{equation}
\begin{equation}
\label{eq:gini-w-norm}
\text{Gini-w}_{\text{our}}  = 
\begin{cases}
    \frac{
    \text{Gini-w}_{\text{ori}}- \text{Gini-w}_{\min}}
    {
    \text{Gini-w}_{\max}  -  \text{Gini-w}_{\min} } & \text{if } km \leq n \\ 
    \frac{
    \text{Gini-w}_{\text{ori}}}
    {
    \text{Gini-w}_{\max}}  & \text{otherwise}
\end{cases}
\end{equation}

For \up FSat\ori, as the higher the score, the fairer, we normalise Eq.~\eqref{eq:fsat-ori} using $x_{\max}=1$ and $x_{\min}=\frac{k}{n}$.\footnote{VoCD, II-D, and AI-D can be normalised in a similar way after computationally approximating their empirical minimum and maximum values.}  

\begin{equation}
\label{eq:fsat-norm}
\text{FSat}_{\text{our}}  = 
    \frac{\text{FSat}_{\text{ori}}-\frac{k}{n}}{1-\frac{k}{n}} 
\end{equation}

\subsection{Resolving Quantity-insensitivity (Limitation 2)}\label{ss:resolve-quantity} 
While the quantity-insensitivity limitation is due to the design choice of the QF measure ($\S$\ref{ss:quantity-insensitivity}), we reason that a solution to this limitation is needed in case one would like to use a measure that is equation-wise similar to QF\ori, but needs the measure to be sensitive to the change of exposure received by the item. Hence, the \textbf{quantity-insensitivity} limitation for \up QF\ori~ may simply be resolved by calculating \up Jain\ori~ instead, as \up QF\ori~ o\-riginates from \up Jain\ori~ (see Eq.~\ref{eq:jain-ori} and \ref{eq:qf-ori}). \up Jain\ori~ gives different weights between items that have been recommended once and more than once. Referring to the toy example given in $\S$\ref{ss:quantity-insensitivity} where \up QF\ori~ would be 0.6 for both cases, but \up Jain\ori~ $\approx 0.514$ for the first scenario and \up Jain\ori~ $=0.6$ for the second scenario. \up Jain\ori~ returns a higher fairness score in the second scenario as the distribution of the frequency count of items is more balanced.

\subsection{Unresolvable Limitations}
\label{ss:nofix} 
We resolve three out of five limitations (see Tab.~\ref{tab:limitation-summary}). The remaining limitations are unresolvable for the following reasons. 

\noindent\textbf{Non-realisability} 
due to \textit{Cause 4} cannot be resolved because no closed-form solutions have been discovered for the recommendations that produce the best and worst fairness scores for each measure. Computing the solution requires solving a constrained optimization problem that cannot be practically solved for large datasets such as \rs~data. 
While the measures could theoretically be corrected in a similar manner to the ones in $\S$\ref{ss:non_realisability}, it is only possible to do so after computing the most/least fair recommendation list, which is impractical.

\noindent\textbf{Item-representation-dependence} cannot be resolved because item representation is required by the measure to determine whether items are similar. It is avoidable if all items are considered similar to each other, but this is unrealistic. 
Moreover, to get comparable scores, all \rs~should use the same representation of items, which cannot be guaranteed. This point is not handled in the original definition of VoCD \cite{Wang2022ProvidingSystems}, where the only item representation considered is item embeddings. While it is possible to use the same external representation for item representation (e.g. similarity matrix based on tags, keyword, or other features) for different \rs, as commonly done with diversity metrics \cite{Ziegler2005ImprovingDiversification}, the fairness score may still change depending on the representation used to determine the item similarity.

\noindent\textbf{Always-fair} cannot be resolved for FSat\ori~ because attempting to resolve this would require tampering with the definition of `satisfied' based on the concept of maximin share. This definition of `satisfied' is an integral part of the measure. Replacing the maximin share criterion with another requirement would turn FSat\ori~ into a different measure. 
Therefore, as this limitation is related to the concept of measure, we are unable to recommend a solution to the limitation without changing the design of the measure.

\section{Empirical Analysis}
\label{s:exp}
We experimentally analyse the relevance and fairness of several recommenders and compare the original measures ($\S$\ref{s:priorwork}) to our corrected versions ($\S$\ref{s:extensions}) for the six datasets shown in Tab.~\ref{tab:dataset}. We present the results for Lastfm and Ml-1m in this section, and for the other datasets in App.~\ref{app:extend-result}.

\subsection[Experimental Setup]{Experimental Setup}
\label{ss:setup}

\noindent \textbf{Dataset preprocessing}. 
We use six freely available datasets (see Tab.~\ref{tab:dataset}) from~\cite{Zhao2021RecBole:Algorithms}:\footnote{\url{https://github.com/RUCAIBox/RecSysDatasets}} 
Lastfm;\footnote{\url{http://www.lastfm.com} \label{fn:lfm}} Ml-1m \cite{Harper2015TheContext}; Book-x \cite{Ziegler2005ImprovingDiversification}; Amazon-lb, Amazon-dm, and Amazon-is \cite{Ni2019JustifyingAspects}. We remove users/items with $<5$ interactions and use $80\%/10\%/10\%$ to train/validate/test, with a user-based random split for Lastfm and Book-x (timestamps are not available), and a user-based temporal split for all other datasets, i.e., the last $10\%$ of each user's interactions are in the test set. We convert ratings $\ge3$ on Ml-1m \& Amazon-* and ratings $\ge6$ on Book-x to 1. We discard the rest of the ratings. 
We choose these thresholds as the ratings are from 1--5 in Ml-1m and Amazon-*, and 0--10 in Book-x. We do not convert for Lastfm as it uses implicit feedback, so all interactions have a value of 1. For duplicate values, we keep the last interaction.  

\noindent \textbf{Recommenders}. For recommendation we use: Pop \cite{Rashid2002GettingYou} (recommends $k$ most popular items), item-based K-Nearest Neighbours (ItemKNN) \cite{Deshpande2004Item-basedAlgorithms}, 
Sparse Linear Method (SLIM) \cite{Ning2011SLIM:Systems}, 
Bayesian Personalized Ranking (BPR) \cite{RendleBPR:Feedback}, 
Neural Graph Collaborative Filtering (NGCF) \cite{Wang2019NeuralFiltering}, 
Neural Matrix Factorization (NeuMF) \cite{He2017NeuralFiltering}, and 
Variational Autoencoder with multinomial likelihood (MultiVAE) \cite{Liang2018VariationalFiltering}. 
We use training batch sizes of $4096$, 
Adam~\cite{Kingma2014Adam:Optimization} as optimizer, and 
the RecBole library ~\cite{Zhao2021RecBole:Algorithms}.
We train BPR, NGCF, NeuMF, and MultiVAE for $300$ epochs, but 
use early stopping of $10$ epochs and keep the model that produces the best NDCG@10 on the validation set. We tune hyperparameters on all models except Pop, with RecBole's hyperparameter tuning module. The hyperparameter search space and optimal hyperparameters are in App.~\ref{app:extend-exp_set_up}. 
For all recommenders, when we generate the recommendation list for a user during testing, the items in the user's train or validation set are placed at the end of the user's list to avoid re-recommending them. 

\noindent\textbf{Measures}.
We evaluate models w.r.t.~a) relevance-only measures (HR, MRR, Precision (P), Recall (R), MAP, NDCG), and b) individual item fairness measures, both the original and our corrected measures. 
All measures are computed at $k=10$, unless otherwise stated. We evaluate on the full test set of items instead of a sample of them, as doing the latter is known to yield misleading results~\cite{Krichene2020OnRecommendation}. This leads to lower performance than reported when sampling the test set. 
Lastly, 
for Ent, we use the log base-$n$. 
For VoCD we choose the values of $\alpha$ and $\beta$ such that VoCD maintains comparability with the other fairness measures: all recommended items are considered similar\footnote{
All recommended items will be treated as similar items as $sim(i,i') \geq 1-2 = -1$ and $sim(i,i')\in [-1,1]$.} ($\alpha=2$)
and thus $A$ is the set of all possible pairs of different items in the top $k$, without any tolerance for coverage disparity ($\beta=0$). We also choose this configuration to avoid reliance on similarity scores based on item embeddings. 
For II-D and AI-D, we use $\gamma=0.8$ \cite{Wu2022JointRecommendation}.

\begin{table}[tb]
\caption{Statistics of the datasets before and after our preprocessing.}
\label{tab:dataset}
\centering
\begin{tabular}{lrrrr}
\toprule
\textbf{dataset} & \multicolumn{1}{l}{\textbf{\#users}} & \multicolumn{1}{l}{\textbf{\#items}} & \multicolumn{1}{l}{\textbf{\#interactions}} & \multicolumn{1}{l}{\textbf{sparsity (\%)}} \\ 
\midrule
\multicolumn{5}{c}{\textit{original (as provided by \cite{Zhao2021RecBole:Algorithms})}}                                                   \\ \midrule
Lastfm\footnoteref{fn:lfm} & 1,892 & 17,632 & 92,834 & 99.7217\% \\
Ml-1m \cite{Harper2015TheContext} & 6,040 & 3,706 & 1,000,209 & 95.5316\% \\
Book-x \cite{Ziegler2005ImprovingDiversification} & 105,283 & 340,556 & 1,149,780 & 99.9968\% \\
Amazon-lb  \cite{Ni2019JustifyingAspects} & 416,174 & 12,120 & 574,628 & 99.9886\% \\
Amazon-dm \cite{Ni2019JustifyingAspects} & 840,372 & 456,992 & 1,584,082 & 99.9996\% \\
Amazon-is \cite{Ni2019JustifyingAspects} & 1,246,131 & 165,764 & 1,758,333 & 99.9991\% \\
\midrule
\multicolumn{5}{c}{\textit{preprocessed (by us)}}   \\ \midrule                                                                     
Lastfm & 1,859 & 2,823 & 71,355 & 98.6403\% \\
Ml-1m & 6,038 & 3,307 & 835,789 & 95.8143\% \\
Book-x & 5,639 & 7,455 & 91,385 & 99.7826\% \\
Amazon-lb & 1,644 & 791 & 16,765 & 98.7108\% \\
Amazon-dm & 11,750 & 9,462 & 116,681 & 99.8951\% \\
Amazon-is & 6,574 & 3,569 & 45,762 & 99.8050\%\\ 
\bottomrule
\end{tabular}
\end{table}

\subsection{Analysis of Relevance and Fairness}
\label{ss:performance} 
We start by studying the relevance and fairness of several recommender models. We compare a) different recommender models w.r.t. relevance and fairness scores, and b) different evaluation measures, including corrected and uncorrected measures. The goal of a) is to study whether relevance and/or fairness scores vary between models, and to obtain a ranking of models that is used in subsequent analysis ($\S$\ref{ss:corr}). The goal of b) is to study measures based on diverse concepts of fairness, and highlight their differences (and similarity, if any). 

The evaluation results are shown in Tab.~\ref{tab:base-main} for Lastfm and Ml-1m, and in App.~\ref{app:extend-performance} for the other datasets.

\begin{table*} 
\caption{Relevance \textsc{(rel)} and fairness \textsc{(fair)} scores of the recommender models on Lastfm and Ml-1m. The most relevant and most fair score per measure is in bold. $\uparrow$ means the higher the better, $\downarrow$ the lower the better. `nan' stands for `not a number'.}
\label{tab:base-main}
\resizebox{\columnwidth}{!}{
\begin{tabular}{lllrrrrrrr}
\toprule
 &  && Pop$^{*}$ & ItemKNN & SLIM & BPR & NGCF & NeuMF & MultiVAE \\
\midrule
\multirow[c]{21}{*}{Lastfm} 
& \multirow[c]{6}{*}{\textsc{rel}} 
& $\uparrow$ $\text{HR}$ & 0.236686 & 0.563206 & 0.520710 & \bfseries 0.603012 & 0.598171 & 0.571813 & 0.597633 \\
& & $\uparrow$ $\text{MRR}$ & 0.102296 & 0.320710 & 0.290459 & \bfseries 0.336577 & 0.327790 & 0.301449 & 0.326032 \\
& & $\uparrow$ $\text{P}$ & 0.029855 & 0.082948 & 0.075901 & \bfseries 0.091286 & 0.090479 & 0.082679 & 0.091070 \\
& & $\uparrow$ $\text{MAP}$ & 0.033572 & 0.129943 & 0.112303 & \bfseries 0.140626 & 0.136694 & 0.120650 & 0.136794 \\
& & $\uparrow$ $\text{R}$ & 0.078205 & 0.240326 & 0.210221 & \bfseries 0.262193 & 0.260140 & 0.238121 & 0.261565 \\
& & $\uparrow$ $\text{NDCG}$ & 0.063259 & 0.207209 & 0.183180 & \bfseries 0.223416 & 0.219215 & 0.198248 & 0.219408 \\
\cline{2-10}
& \multirow[c]{15}{*}{\textsc{fair}} 
& $\uparrow$ $\text{Jain}_{\text{ori}}$ & 0.005350 & 0.050544 & 0.029023 & 0.080549 & 0.086601 & 0.096789 & \bfseries 0.134035 \\
& & $\uparrow$ $\text{Jain}_{\text{our}}$ & 0.001824 & 0.047434 & 0.025715 & 0.077714 & 0.083822 & 0.094103 & \bfseries 0.131692 \\
& & $\uparrow$ $\text{QF}_{\text{ori}}$ & 0.009564 & 0.432519 & 0.145590 & 0.428268 & 0.399221 & 0.462983 & \bfseries 0.683316 \\
& & $\uparrow$ $\text{QF}_{\text{our}}$ & 0.006043 & 0.430501 & 0.142552 & 0.426235 & 0.397085 & 0.461074 & \bfseries 0.682190 \\
& & $\uparrow$ $\text{Ent}_{\text{ori}}$ & nan & nan & nan & nan & nan & nan & nan \\
& & $\uparrow$ $\text{Ent}_{\text{our}}$ & 0.096360 & 0.595274 & 0.439907 & 0.656154 & 0.660951 & 0.686602 & \bfseries 0.763463 \\
& & $\uparrow$ $\text{FSat}_{\text{ori}}$ & 0.009564 & 0.136734 & 0.071909 & 0.171803 & 0.178888 & 0.195537 & \bfseries 0.230960 \\
& & $\uparrow$ $\text{FSat}_{\text{our}}$ & 0.006043 & 0.133665 & 0.068610 & 0.168859 & 0.175969 & 0.192677 & \bfseries 0.228226 \\
& & $\downarrow$ $\text{Gini}_{\text{ori}}$ & 0.995080 & 0.908416 & 0.968311 & 0.884752 & 0.885149 & 0.864875 & \bfseries 0.780612 \\
& & $\downarrow$ $\text{Gini}_{\text{our}}$ & 0.998564 & 0.908251 & 0.970668 & 0.883591 & 0.884004 & 0.862877 & \bfseries 0.775067 \\
& & $\downarrow$ $\text{Gini-w}_{\text{ori}}$ & 0.995984 & 0.915773 & 0.972331 & 0.895154 & 0.896870 & 0.879730 & \bfseries 0.793962 \\
& & $\downarrow$ $\text{Gini-w}_{\text{our}}$ & 0.998747 & 0.918313 & 0.975028 & 0.897637 & 0.899358 & 0.882170 & \bfseries 0.796164 \\
& & $\downarrow$ $\text{VoCD}_{\text{ori}}$ & 0.669135 & 0.609061 & 0.704415 & 0.640846 & 0.656609 & 0.641465 & \bfseries 0.598510 \\
& & $\downarrow$ $\text{II-D}_{\text{ori}}$ & \bfseries 0.000970 & \bfseries 0.000970 & \bfseries 0.000970 & \bfseries 0.000970 & \bfseries 0.000970 & \bfseries 0.000970 & \bfseries 0.000970 \\
& & $\downarrow$ $\text{AI-D}_{\text{ori}}$ & 0.000668 & 0.000061 & 0.000114 & 0.000037 & 0.000034 & 0.000032 & \bfseries 0.000019 \\

\cline{1-10}
\multirow[c]{21}{*}{Ml-1m} 
& \multirow[c]{6}{*}{\textsc{rel}} 
& $\uparrow$ $\text{HR}$ & 0.273435 & 0.336204 & 0.343326 & \bfseries 0.348791 & 0.337198 & 0.324114 & 0.331070 \\
& & $\uparrow$ $\text{MRR}$ & 0.112995 & 0.136303 & 0.139593 & \bfseries 0.142428 & 0.140765 & 0.133319 & 0.130503 \\
& & $\uparrow$ $\text{P}$ & 0.045661 & 0.054190 & 0.053445 & \bfseries 0.055167 & 0.055018 & 0.052302 & 0.051308 \\
& & $\uparrow$ $\text{MAP}$ & 0.025940 & 0.036132 & 0.036830 & \bfseries 0.038389 & 0.037460 & 0.033568 & 0.035284 \\
& & $\uparrow$ $\text{R}$ & 0.040195 & 0.064935 & 0.071923 & \bfseries 0.073647 & 0.067244 & 0.060861 & 0.068755 \\
& & $\uparrow$ $\text{NDCG}$ & 0.056219 & 0.073897 & 0.075873 & \bfseries 0.078110 & 0.076032 & 0.070162 & 0.072263 \\
\cline{2-10}
& \multirow[c]{15}{*}{\textsc{fair}} 
& $\uparrow$ $\text{Jain}_{\text{ori}}$ & 0.006867 & 0.023143 & 0.045463 & \bfseries 0.068296 & 0.057974 & 0.059396 & 0.065298 \\
& & $\uparrow$ $\text{Jain}_{\text{our}}$ & 0.003857 & 0.020192 & 0.042593 & \bfseries 0.065508 & 0.055148 & 0.056576 & 0.062499 \\
& & $\uparrow$ $\text{QF}_{\text{ori}}$ & 0.040822 & 0.188388 & 0.236166 & 0.444209 & 0.306018 & 0.384336 & \bfseries 0.485939 \\
& & $\uparrow$ $\text{QF}_{\text{our}}$ & 0.037913 & 0.185927 & 0.233849 & 0.442524 & 0.303913 & 0.382469 & \bfseries 0.484380 \\
& & $\uparrow$ $\text{Ent}_{\text{ori}}$ & nan & nan & nan & nan & nan & nan & nan \\
& & $\uparrow$ $\text{Ent}_{\text{our}}$ & 0.187196 & 0.435739 & 0.552900 & 0.650556 & 0.606893 & 0.623249 & \bfseries 0.652672 \\
& & $\uparrow$ $\text{FSat}_{\text{ori}}$ & 0.020865 & 0.069549 & 0.114605 & 0.164500 & 0.147263 & 0.153311 & \bfseries 0.167523 \\
& & $\uparrow$ $\text{FSat}_{\text{our}}$ & 0.017895 & 0.066727 & 0.111920 & 0.161965 & 0.144677 & 0.150743 & \bfseries 0.164998 \\
& & $\downarrow$ $\text{Gini}_{\text{ori}}$ & 0.993229 & 0.970583 & 0.942655 & 0.893080 & 0.919816 & 0.908995 & \bfseries 0.888977 \\
& & $\downarrow$ $\text{Gini}_{\text{our}}$ & 0.996201 & 0.973246 & 0.944935 & 0.894681 & 0.921783 & 0.910813 & \bfseries 0.890521 \\
& & $\downarrow$ $\text{Gini-w}_{\text{ori}}$ & 0.994377 & 0.973088 & 0.948836 & 0.902403 & 0.927651 & 0.918188 & \bfseries 0.894798 \\
& & $\downarrow$ $\text{Gini-w}_{\text{our}}$ & 0.996731 & 0.975391 & 0.951082 & 0.904539 & 0.929846 & 0.920361 & \bfseries 0.896916 \\
& & $\downarrow$ $\text{VoCD}_{\text{ori}}$ & 0.789888 & 0.733042 & 0.737530 & 0.705762 & 0.724919 & 0.712774 & \bfseries 0.699980 \\
& & $\downarrow$ $\text{II-D}_{\text{ori}}$ & \bfseries 0.000828 & \bfseries 0.000828 & \bfseries 0.000828 & \bfseries 0.000828 & \bfseries 0.000828 & \bfseries 0.000828 & \bfseries 0.000828 \\
& & $\downarrow$ $\text{AI-D}_{\text{ori}}$ & 0.000381 & 0.000096 & 0.000049 & 0.000032 & 0.000038 & 0.000038 & \bfseries 0.000030 \\
\bottomrule
\multicolumn{10}{l}{\multirow{3}{*}{
\parbox[t]{1.35\textwidth}{*The scores of our \textsc{fair} measures for Pop are not 0 or 1, because in our experiment set-up, items from users' train or validation splits are excluded from the top $k$ recommendation list.}}}
\end{tabular}%
}
\end{table*}

\subsubsection{Discussion of recommendation models in Tab.~\ref{tab:base-main}} 

BPR has the highest relevance scores, while MultiVAE generally has the highest fairness scores. Between BPR and MultiVAE, we observe that the relevance scores are higher in BPR but the fairness scores are higher for MultiVAE. 
E.g., in Lastfm, NDCG $=0.223$ for BPR and NDCG $=0.219$ for MultiVAE. Meanwhile, the higher-is-better fairness scores range between $[0.078, 0.656]$ for BPR and $[0.132, 0.763]$ for MultiVAE. 
This is not always observed for all models. E.g., for ItemKNN and SLIM, better fairness tends to be accompanied by better relevance. Furthermore, other discrepancies exist: the relevance of ItemKNN and MultiVAE is on par, but their fairness scores e.g., the scores of \up Jain of ItemKNN, are only half of those achieved by MultiVAE. 
Generally, the recommenders agree in relative ordering of scores, but some models have higher scores for .\our~than .\ori~and vice-versa, e.g., for Lastfm, \down Gini\ori$>$\down Gini\our~ for ItemKNN but \down Gini\ori$<$\down Gini\our~ for SLIM. Overall, we observe that: 

\begin{itemize}
    \item A recommender model that is the best in terms of relevance may also be relatively fair.
    \item The recommender models mostly have a similar ordering of the scores of fairness measures: if a fairness measure has a higher value than another in a recommender model X, it is the same for recommender model Y.
    \item Some models achieve relatively similar relevance scores, but with a huge disparity between their fairness scores.
\end{itemize}

\subsubsection{Discussion of fairness evaluation measures in Tab.~\ref{tab:base-main}} 
For the higher-is-better measures, Jain, QF, FSat, and Gini, the scores of the original measures and our measures are similar. Both \up Jain and \up QF should range in $[0,1]$, but \up Jain is very close to 0 i.e., ($\sim$0.1 or less), and \up QF scores are $\sim$0.7 (in Lastfm) and $\sim$0.5 (in Ml-1m). 
Similarly, the scores of \down II-D and \down AI-D are also very close to 0 while \down Gini scores are closer to 1. 
While these are due to the different underlying fairness ideas between the measures, the big differences in scores may cause confusion, e.g. that a recommendation is very unfair based on \up Jain or \down Gini, or moderately fair based on \up QF. For Lastfm and Ml-1m, we also see that the absolute scores for the same recommender, e.g., MultiVAE, follow the same order from the lowest to the highest: \up Jain, \up FSat, \up QF, and \up Ent. 
This indicates that \up Jain tends to give lower scores (more unfair) than the other measures. We observe similar trends for \down II-D and \down AI-D, which tend to give lower scores (more fair) compared to other lower-is-better measures.
The weighted \down Gini-w is also more strict than the unweighted \down Gini as \down Gini-w tends to give more unfair scores than \down Gini. We study further the strictness of these measures in $\S$\ref{ss:insert}.

We also observe that scores of \down AI-D are hardly distinguishable. They differ only in the fourth or more decimal point for both Lastfm and Ml-1m. However, differences in other measures can be seen in the first or second decimal point. The small scores of AI-D may be due to the measure quantifying the disparity between item exposure and random exposure, which is very little when we have a large number of items. This finding suggests that when computing \down AI-D, care should be taken to avoid rounding errors and failure to distinguish the scores due to the floating-point format.

For all datasets and models, the original Ent cannot be calculated because it returns NaN due to zero division errors. This happens because there are items in the dataset that are not recommended. 
Our corrected version of this measure ($\S$\ref{s:extensions}) does not suffer from this problem.

For the same dataset and $k$, regardless of the recommender, the II-D scores are always the same/constant. Due to the fixed amount of slots $km$, within a single recommendation round, the exposure values $E_{u,i} \in \{1, \gamma, \dots, \gamma^k, 0\}$ (see Eq.~\ref{eq:eui}) for all user-item pairs, and the number of user-item pair having a specific exposure value $E_{u,i}$ is always $m$. Both of these properties lead to constant II-D scores, as II-D is calculated by taking a mean squared difference between each $E_{u,i}$ value and a constant value based on random expected exposure. When considering cases with multiple rounds of recommendations, the score of II-D may not remain constant anymore, as the user-item exposure values are aggregated across recommendation rounds, resulting in the possibility of $E_{u,i}$ having linear combinations of values from the set above. We have also illustrated in $\S$\ref{sss:cause_nonreal} that II-D is not constant in multiple recommendation rounds.

Overall, we observe that: 
\begin{itemize}
    \item The different fairness measures have different ranges in these experiments, even if theoretically they have the same range. 
    \item The original Ent is always incomputable in the experiments, and our corrected Ent resolves the issue. 
    \item II-D\ori~ remains constant for the same dataset, rendering this measure notably less meaningful under this single-round experimental set-up. 
    \item Both \down II-D\ori~ and \down AI-D\ori~ have minuscule values, indicating near-perfect fairness even if this contradicts other fairness scores.   
\end{itemize}

\subsection{Correlation between Measures}
\label{ss:corr}

When comparing different recommender models, sometimes the ranking of the models (e.g., from the most to least fair score) is more concerning than the absolute values of the measures that we have seen in Tab.~\ref{tab:base-main}. Motivated by this, we analyse the measures' correlation in order to study the agreement of model rankings based on different measures of relevance and fairness. We compare the following things: 
1) the agreement between measures of the same type (relevance or fairness); 
2) the agreement between measures of different types;
3) the agreement between the original measures and our corrections to the original measures; 
4) the agreement between measures across different datasets. 
By performing this analysis, we also gain insights into how measures that capture different fairness concepts (dis)agree with one another. 

\begin{figure}
\centering
    \includegraphics[width=\textwidth]{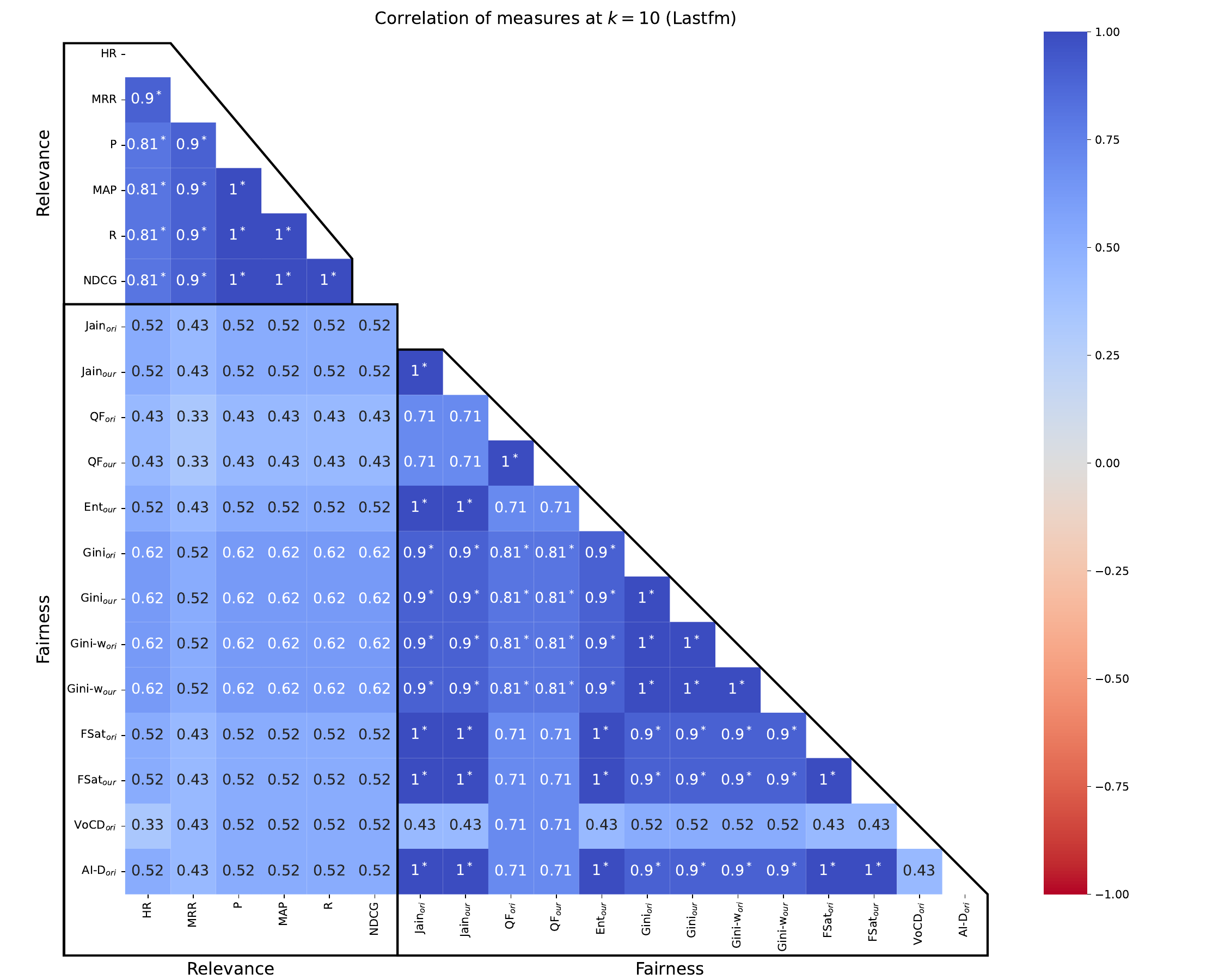}

\caption{Correlation (Kendall's $\tau$) between relevance and fairness measures for Lastfm. \explainsig}
\label{fig:corr-lastfm}
\end{figure}

\begin{figure}
\centering
    \includegraphics[width=\textwidth]{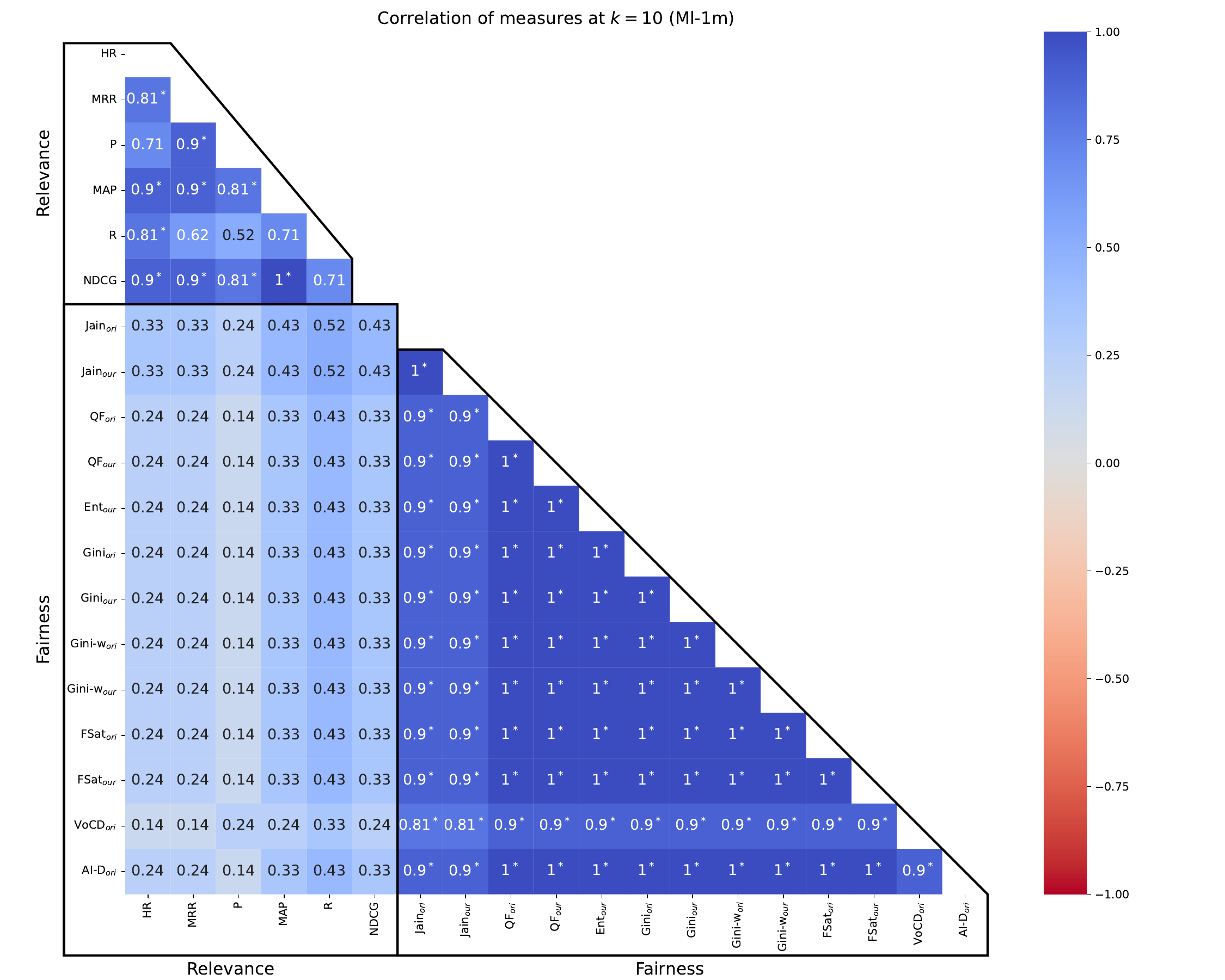}
    \caption{Correlation (Kendall's $\tau$) between relevance and fairness measures for Ml-1m. \explainsig}
\label{fig:corr-ml1m}
\end{figure}

We use Kendall's $\tau$ between measures to compute ranking agreement. Fig.~\ref{fig:corr-lastfm}--\ref{fig:corr-ml1m} show the Kendall's $\tau$ values between relevance measures and fairness measures for Lastfm and Ml-1m (see App.~\ref{app:corr} for the other data\-sets). The computation is as follows: for each dataset, we rank the models based on the most relevant or most fair scores. 
We omit Ent\ori~as it produces NaN in Tab.~\ref{tab:base-main} and we also omit II-D as the scores for one dataset are the same across models. 
We compute the correlation significance and correct errors arising from multiple testings for a dataset, using the Benjamini-Hochberg (BH) procedure that is based on false discovery rate  \cite{Benjamini1995ControllingTesting}. Upon correction, some correlations are still significant; these are indicated by an asterisk ($^*$) in Fig.~\ref{fig:corr-lastfm}--\ref{fig:corr-ml1m}.\footnote{We also use two more conservative procedures separately to correct the errors: Bonferroni and Holm \cite{Holm1979AProcedure}. Upon correction we obtain no significant results for any tests across the six datasets.}

We first analyse the correlation among measures of the same type. The relevance measures are highly correlated with each other: $[0.81,1]$ for Lastfm and $[0.52,1]$ for Ml-1m, as expected \cite{Webber2008PrecisionRedundant}. The fairness measures are also strongly correlated with each other: $[0.71,1]$ for Lastfm and $[0.81,1]$ for Ml-1m, except VoCD\ori~ for Lastfm, $[0.43,0.71]$. This is expected from how the measures treat items when computing fairness; VoCD\ori~ only considers items in the recommendation list, while the remaining fairness measures consider all items in the dataset. 
For Ml-1m, all the computed correlations between fairness measures are significant after applying the BH procedure. On the other hand, after applying the same procedure for Lastfm, neither of QF nor VoCD, has significant correlations with the rest of the fairness measures, except for QF and Gini/Gini-w. 
It is also reasonable for QF to not correlate significantly with most of the measures, as it is the only measure insensitive to the difference in the number of times an item is recommended. 

Interestingly, even though in Tab.~\ref{tab:base-main} the scores of \up Jain, \up QF, \up Ent, and \up FSat occupy different parts of their range, these measures are highly correlated. The same goes for \down Gini and \down AI-D. This shows that even measures based on different concepts of fairness are still capable of producing similar rankings of models. 
Nevertheless, the absolute scores of the measures can be misinterpreted due to the measures occupying different parts of their range.

Our corrected fairness measures are always perfectly correlated with the original fairness measures (1 in both datasets). This is expected because our corrected versions are obtained by normalization, which does not change the relative order of the models.

Regarding the correlations between measures of different types, we see different trends between relevance and fairness measures for Lastfm and Ml-1m. In Lastfm, we see moderate correlations between fairness and relevance measures, $[0.33,0.62]$, but these are lower for Ml-1m $[0.14,0.52]$. These findings are expected as the fairness measures do not consider relevance. None of these correlations are significant after applying the BH procedure. Yet, some correlations between fairness and relevance measures are significant for Book-x and Amazon-is (App.~\ref{app:corr}). 

\subsection{Max/min Achievable Fairness}
\label{ss:maxmin}

\begin{figure*}
    \centering
    \includegraphics[width=.95\textwidth]{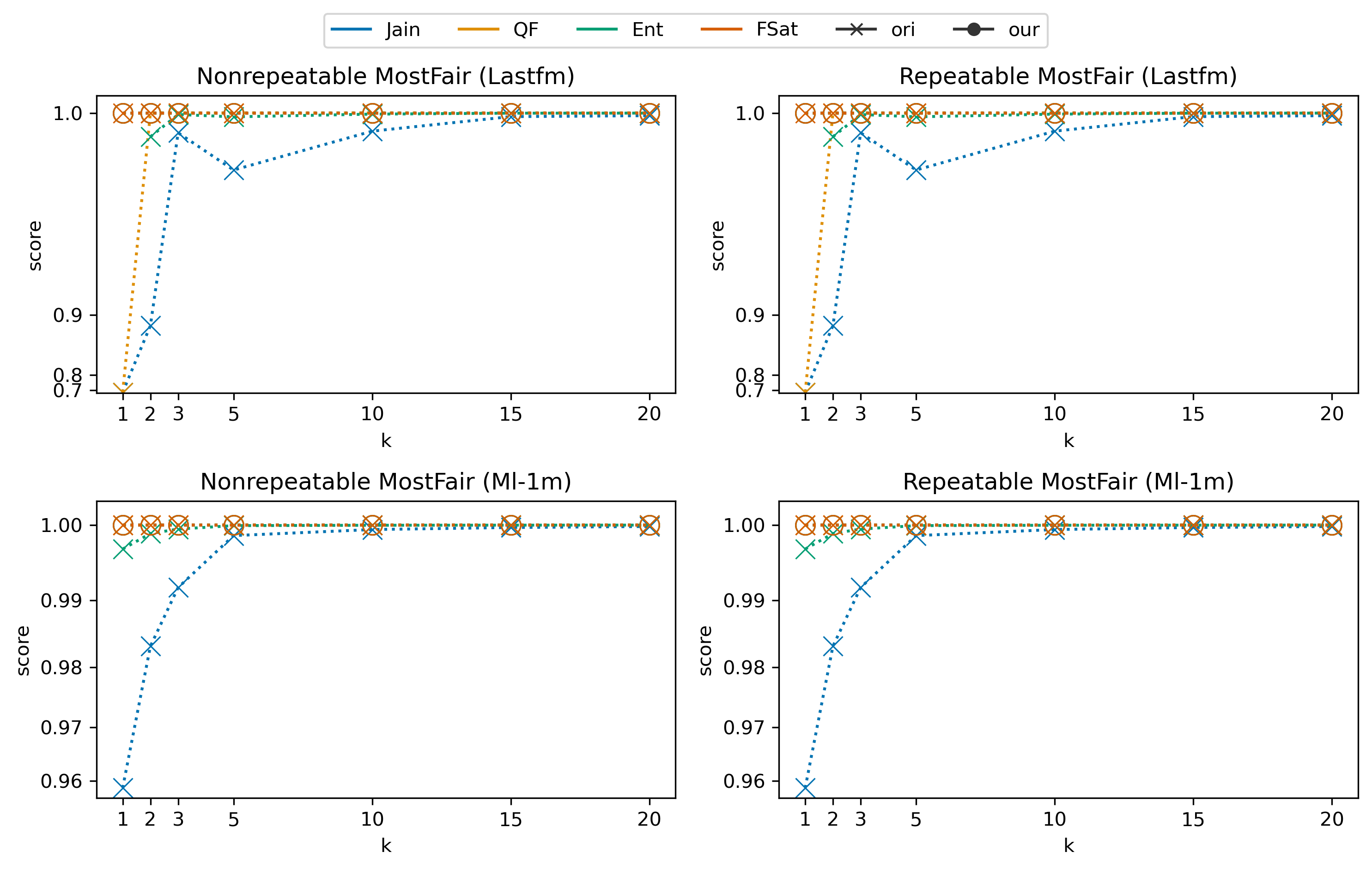}
    \caption{Most fair scores with varying $k$ for higher-is-fairer fairness measures for Lastfm and Ml-1m. All scores from the corrected measures (denoted by `our') measures overlap with each other.}
    \label{fig:mostfair_higher_better}
    \centering
    \includegraphics[width=.95\textwidth]{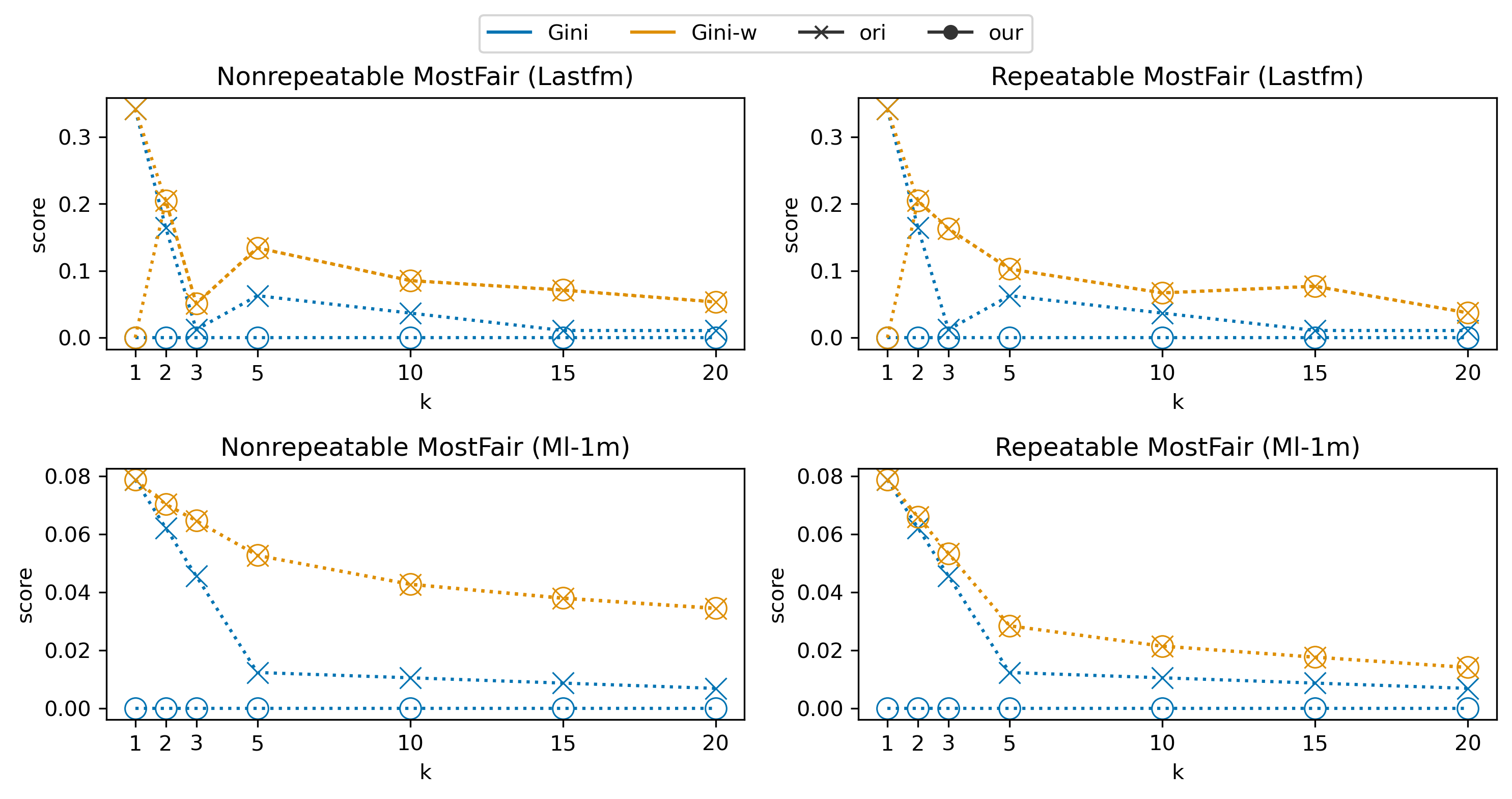}
    \caption{Most fair scores with varying $k$ for lower-is-fairer fairness measures for Lastfm and Ml-1m.}
    \label{fig:mostfair_lower_better}
\end{figure*}

\begin{figure*}
    \centering
    \includegraphics[width=0.85\textwidth]{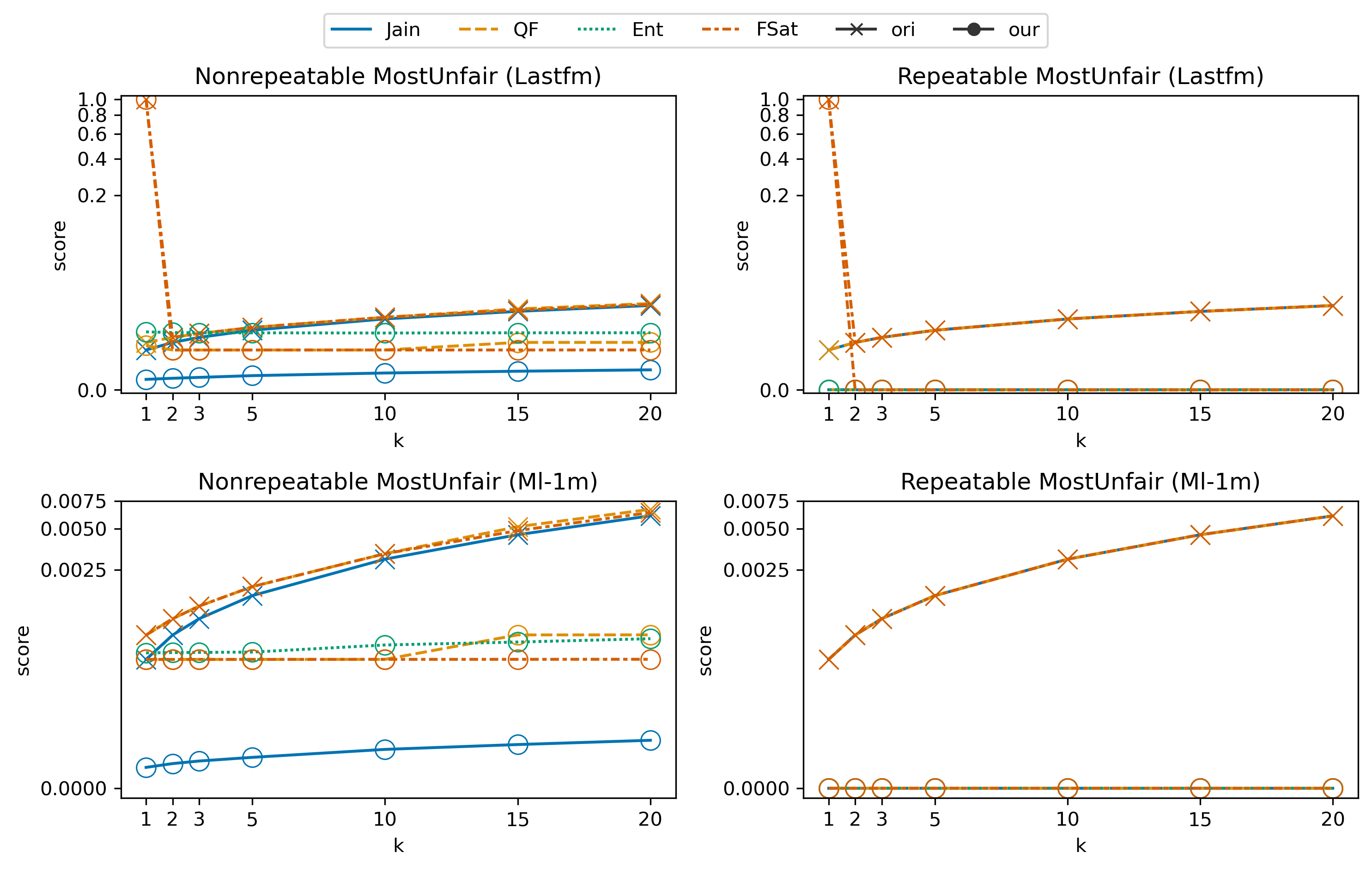}
    \caption{Most unfair scores with varying $k$ for higher-is-fairer fairness measures for Lastfm and Ml-1m. On Repeatable MostUnfair, all scores from the corrected measures (denoted by `our') overlap with each other for the shown values of $k>1$ for Lastfm and for all shown values of $k$ for Ml-1m.}
    \label{fig:mostunfair_higher_better}
    
    \centering
    \includegraphics[width=0.85\textwidth]{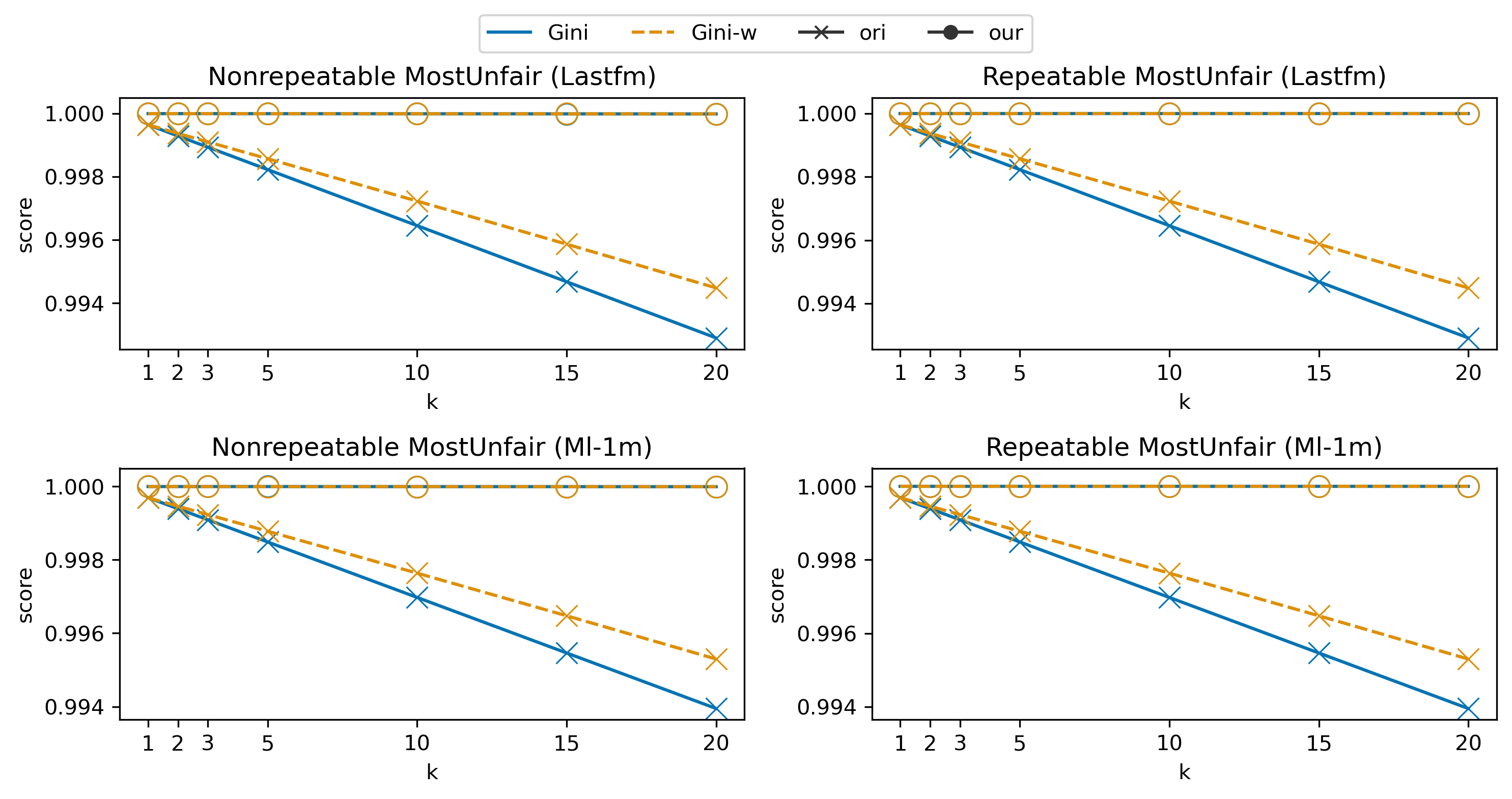}
    \caption{Most unfair scores with varying $k$ for lower-is-fairer fairness measures for Lastfm and Ml-1m. On Repeatable MostUnfair, all scores from the corrected measures (denoted by `our') overlap with each other for all shown values of $k$.}
    \label{fig:mostunfair_lower_better}
\end{figure*}

The aim of this experiment is to quantify the extent to which the fairness measures can achieve their theoretical maximum and minimum fairness value (0 or 1) for different datasets and different $k \in \{1,2,3,5,10,15,20\}$. This relates to the \textbf{non-realisability} limitation (\textit{Causes 1--3}). We experiment solely with the fairness measures for which we have resolved this limitation, namely Jain, QF, Ent, Gini, Gini-w, and FSat. 
We primarily compare the original (uncorrected) versions of these measures against the corrected ones. 
We use two settings: repeatable recommendation, where items in the train/val split can be re-recommended to users following practical cases in industry settings; and nonrepeatable recommendation, 
which is the typical setting for evaluating recommender systems in academic work. 
For each setting, we devise two recommenders: MostFair and MostUnfair. Repeatable MostFair aims to recommend each item in the dataset the same amount of times. However, this is impossible if $n \nmid km$ and in this case some items are recommended $\floorkm$ times while others $\floorkm+1$ times. For Nonrepeatable MostFair, for each user we generate a list of \textit{recommendable items}, defined as items in $I$ that have not appeared in their corresponding train/val split. For one user at a time, we then recommend the least popular $k$ recommendable items based on the current recommendation lists of all users.
Repeatable MostUnfair recommends the same $k$ items to each user. Nonrepeatable MostUnfair does the same, but if any of those $k$ items is a non-recommendable item, the non-recommendable item is replaced by a recommendable item. The results of this experiment for Lastfm and Ml-1m are presented in Fig.~\ref{fig:mostfair_higher_better}--\ref{fig:mostunfair_lower_better} and for the remaining datasets in App.~\ref{app:maxmin}. We discuss the findings below.

\noindent\textbf{Theoretical maximum fairness}. For both nonrepeatable and repeatable settings, all original measures fail to achieve their theoretical maximum fairness values due to the \textbf{non-realisability} limitation (\textit{Causes 2--3}). The scores of the original measures get closer to the theoretical maximum fairness values as $k$ increases. 
However, these scores are still not equal to the theoretical maximum fairness value. In the original measures, having more slots due to larger $k$ does not guarantee that the scores would be higher as well. E.g., the score of \up Jain\ori~ in Fig.~\ref{fig:mostfair_higher_better} is higher at $k=3$ compared to $k=5$, because of the changing values of $km \bmod n$ for different values of $k$. 
Our corrected versions always reach their theoretical maximum fairness values for both repeatability settings, except for Gini-w\our. This behaviour is due to the unresolvable non-realisability (\textit{Cause 4)} limitation for Gini-w. However, Gini-w\our~ can still reach the theoretical most fair value when $k=1$ for Lastfm (Fig.~\ref{fig:mostfair_lower_better}), while the original version fails to do so.

\noindent\textbf{Theoretical minimum fairness}. 
All original measures fail to reach the theoretical minimum values for all experimented values of $k$ for all settings due to \textbf{non-realisability} limitation (\textit{Cause 1}). This happens less frequently in our measures (Fig.~\ref{fig:mostunfair_higher_better}). Our measures successfully achieve the theoretical minimum fair values under the repeatable settings, 
except for FSat in Lastfm (Fig.~\ref{fig:mostunfair_higher_better}). This is because when $k=1$, there are not enough slots for the items and due to the \textbf{always-fair} limitation that is unresolvable ($\S$\ref{ss:nofix}), the score for \up FSat is 1, which is not the theoretical minimum fair value. Additionally, the scores of the original measures diverge from the theoretical minimum fairness value with larger $k$. This happens to our measures only in the nonrepeatable setting because the normalization is done by assuming that any item can be recommended to any users. This assumption is not true in the nonrepeatable settings because some items cannot be re-recommended to some users. 
However, differently from the original measures, the scores of our measures diverge less as $k$ increases. 

Overall, while the difference in scores between the original measures and our versions is not large, our measures quantify the actual most (un)fair situations more accurately than the original measures. The difference between the original measures and our versions in the most unfair recommendation under the repeatable setting is $\frac{k}{n}-0 = \frac{k}{n}$ for Jain, QF, Gini, and FSat; and $\log{k}$ for Ent (see Tab.~\ref{tab:bounds}). However, the difference would be greater in item-poor domains where $n$ is small and therefore possibly close to $k$, e.g. insurance \cite{BorgBruun2022LearningDomain}. 
The scores of the original measures also change with $k$. This makes their interpretation harder because the distance between the original scores and the theoretical maximum/minimum fair score also changes without an intuitive pattern for different values of $k$, as seen in the \up Jain\ori~ scores in Fig.~\ref{fig:mostfair_higher_better} which can increase or decrease as $k$ increases. Furthermore, the original measures suffer particularly for low $k$ values, which are the most important rank positions in real-life \rs. The scores of our measures rarely change with different values of $k$.

\subsection{Sliding Window: Relevance and Fairness at Different Rank Positions}
\label{ss:sliding}

This experiment studies how relevance and fairness scores of all measures vary at decreasing rank positions. The experiment aims to observe 1) the change in relevance scores, if any, as items should ideally be placed in the ranks according to decreasing order of true relevance; and 2) whether and how the fairness scores change across different rank positions. Due to bias in recommenders, popular items tend to be given more exposure. Thus, we expect the relevance scores to decrease and the fairness scores to become more fair at decreasing rank positions.
We study how the above changes may generally differ between relevance measures and fairness measures, as well as between different fairness measures, including the ones with different fairness notions. 

\begin{figure*}
    \centering
    \includegraphics[width=\textwidth]{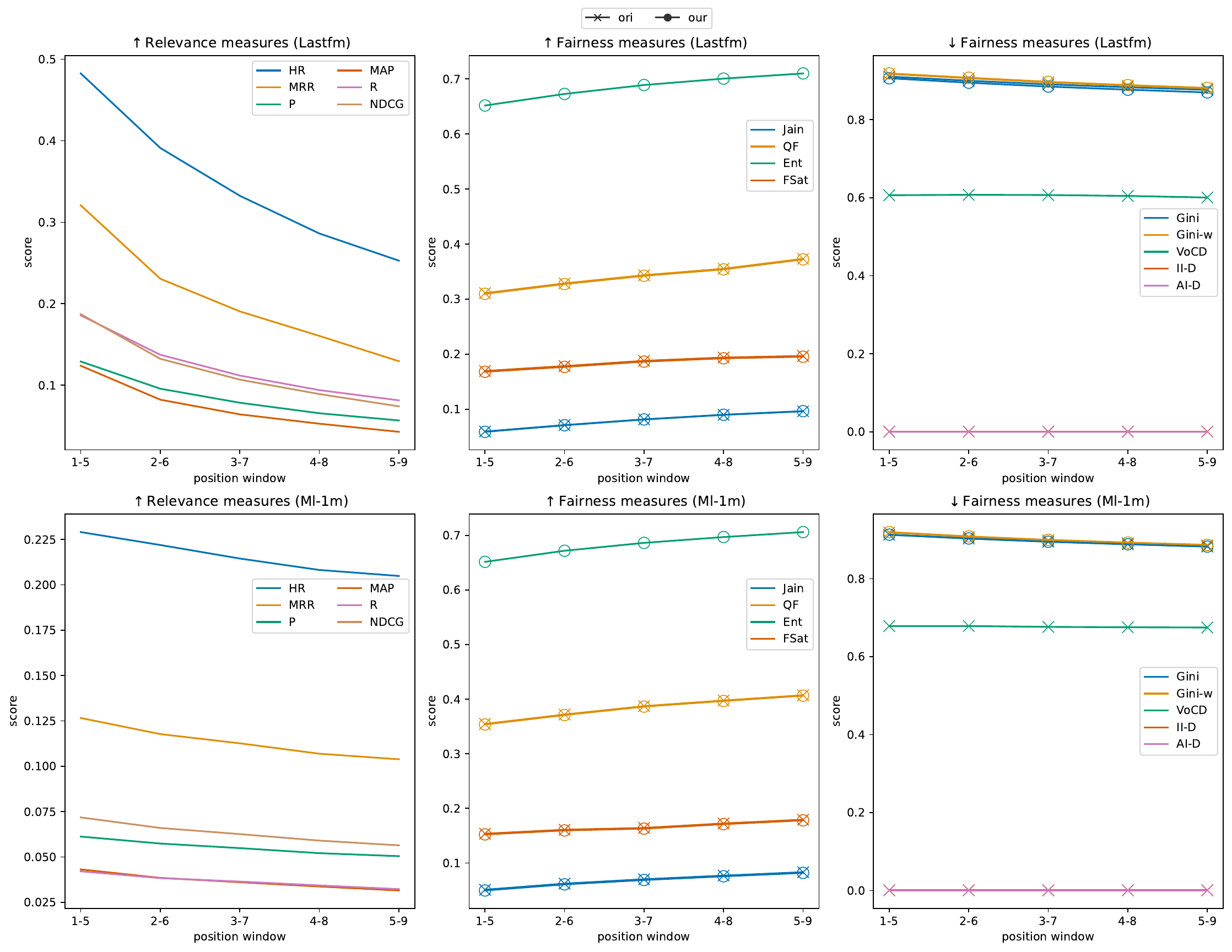}
    \caption{Sliding window evaluation for BPR model, on Lastfm and Ml-1m. Each row of figures is for one dataset, each column is for the different groups of measures (relevance, higher-is-better fairness, lower-is-better fairness measures). II-D and AI-D lines overlap.}
    \label{fig:sliding-1}
\end{figure*}

We conduct this experiment as follows. We use the runs from the BPR model, which is the best in our experiments. Given one run, we compute the measures for different sliding windows of rank positions in rankings 1--5, 2--6, and so on until 5--9. 
We reorder the recommended items such that items that were previously recommended at the top positions are now at the bottom positions when we change the window according to decreasing rank. 
The results for Lastfm and Ml-1m are presented in Fig.~\ref{fig:sliding-1} and for the rest of the datasets in App.~\ref{app:sliding}. 

The following observations from Fig.~\ref{fig:sliding-1} apply to both the original fairness measures and our corrected versions of these measures unless otherwise stated. All relevance scores decrease as rank decreases. 
The drop of relevance scores for Ml-1m ($[0.04,0.23] \rightarrow [0.03, 0.20]$) is less extreme than in Lastfm ($[0.12,0.48] \rightarrow [0.04, 0.25]$). This is partly because the test set of Lastfm has at most five relevant items per user, while on average, Ml-1m has many more. While relevance scores decrease, 
fairness measures show that fairness slightly increases down the rank, except for \down VoCD\ori. 
The range of higher-is-better fairness measures increases from $[0.06,0.65]\rightarrow[0.10,0.71]$ for Lastfm and $[0.05,0.65]\rightarrow[0.08,0.71]$ for Ml-1m. The range of \down Gini and \down Gini-w, decreases from $[0.91,0.92]\rightarrow[0.87,0.88]$ for Lastfm and $[0.91,0.92]\rightarrow[0.88,0.89]$ for Ml-1m. 
\down VoCD\ori~ seems invariant to changes in the position window ($0.61 \rightarrow 0.60$ for Lastfm and $0.68\rightarrow0.67$ for Ml-1m). This may be because VoCD\ori~is the only measure that considers fairness exclusively for recommended items, and the recommended items differ a little in terms of the number of times they are recommended as rank decreases.
\down AI-D has even smaller changes in scores as the values are already minuscule in the first place, while \down II-D is always constantly small for a dataset. 
The small values, compared to other measures, are due to these measures quantifying fairness using different concepts from other measures, i.e. comparing exposure to random exposure (also observed and explained in $\S$\ref{ss:performance}). 
The ranges of all fairness measures are roughly the same across datasets, but the range of relevance measures varies across datasets. This also holds for the datasets in App.~\ref{app:sliding}. This may be due to the distribution of the recommended items being similar across datasets, and the distribution of the number of relevant items differing across datasets, as explained above for Lastfm and Ml-1m.

Fairness measures are also somewhat invariant to changes in relevance. This is anticipated as the equations of fairness measures are independent of relevance values.

\subsection{Measure Strictness and Sensitivity through Artificial Insertion of Items}
\label{ss:insert} 
We have observed in $\S$\ref*{ss:performance} that different fairness measures vary in their strictness of quantifying fairness (e.g., some measures give scores close to the most fair values, and the opposite for others). It is however unknown how sensitive fairness measures are, given the change of the number of times an item is exposed in the recommendation list across all users. Therefore, the goal of this experiment is to study the strictness and sensitivity of the measures, and compare these aspects between measures of similar and different fairness concepts. Knowing the strictness and sensitivity of the measures matters as this affects how we interpret the scores of the measures. For example, if one uses a measure that tends to produce scores close to the most fair value, they must be aware that the score may not reflect fairness accurately.

As such, we devise an experiment to specifically study how the relevance measures, existing fairness measures and our corrected fairness measures scores change when we artificially control the fraction of jointly least exposed and relevant items in the recommendation list.  
We start with an initial recommendation list. We define a \textit{least exposed (LE) item} as an item in the dataset with the least exposure, based on the current recommendation list.\footnote{This fairness concept is closely tied to all measures in this work, except for VoCD which concerns only items in the recommendation list, as opposed to in the dataset (Tab.~\ref{tab:situations}).} An LE item in this experiment is therefore an item that has not appeared in the current recommendation list. We define a \textit{relevant item} as per the labels of relevance. 

From the initial recommendation list, we insert jointly LE and relevant items, one item at a time. 
We create a synthetic dataset with $m=1000$ users and $n=10000$ items. The number of items is exactly the number of recommendation slots $km$ for a cut-off $k=10$. 
We artificially generate a ranking of top $k$ as follows. The artificial insertion of jointly LE and relevant items begins with the recommendation of the same 10 items $i_1, i_2, \dots, i_{10}$ to all users. These items are irrelevant to each user except $u_1$, as we keep the recommendation list for $u_1$ the same throughout the experiment. This is because we want to keep the number of items exactly $km$ where theoretically each item could be recommended exactly once and if we have to completely replace all $m$ users' recommendation lists, we would need to have more than $km$ items. We expect the relevance measures to give scores close to zero on this initial recommendation list as only $u_1$ has relevant items. We expect the fairness measures to give scores that are equal to or close to the theoretical most unfair scores.\footnote{\label{fn:close_to}We say ``close to'' due to the \textbf{non-realisability} (\textit{Cause 4}) limitation in some measures.}

Let $P$ be the fraction of items in the $k$ that are artificially inserted by us. We vary from $P=0$, the original recommendation where we have not inserted any items artificially, to $P=1$ where all items in the $k$ are jointly LE and relevant items that are artificially inserted by us. We increase $P$ in steps of $\nicefrac{1}{k}$. 
From the bottom of a user's recommendation list, we replace one item at a time with a known jointly LE and relevant item, until we end with a recommendation list of different $km$ items across all users, that are all relevant only to the user to whom that item is recommended. 
At the end of the insertion process, each user is recommended exactly 10 relevant items, and those items are also fair w.r.t.~the entire recommendation list for all users, considering all items in the dataset; item fairness is not defined w.r.t.~a specific user. 
We expect the relevance measures to give scores of 1 on the final recommendation list and fairness measures to give scores that are (close to) the fairest scores.\footnoteref{fn:close_to}

\begin{figure}
    \centering
    \includegraphics[width=\textwidth]{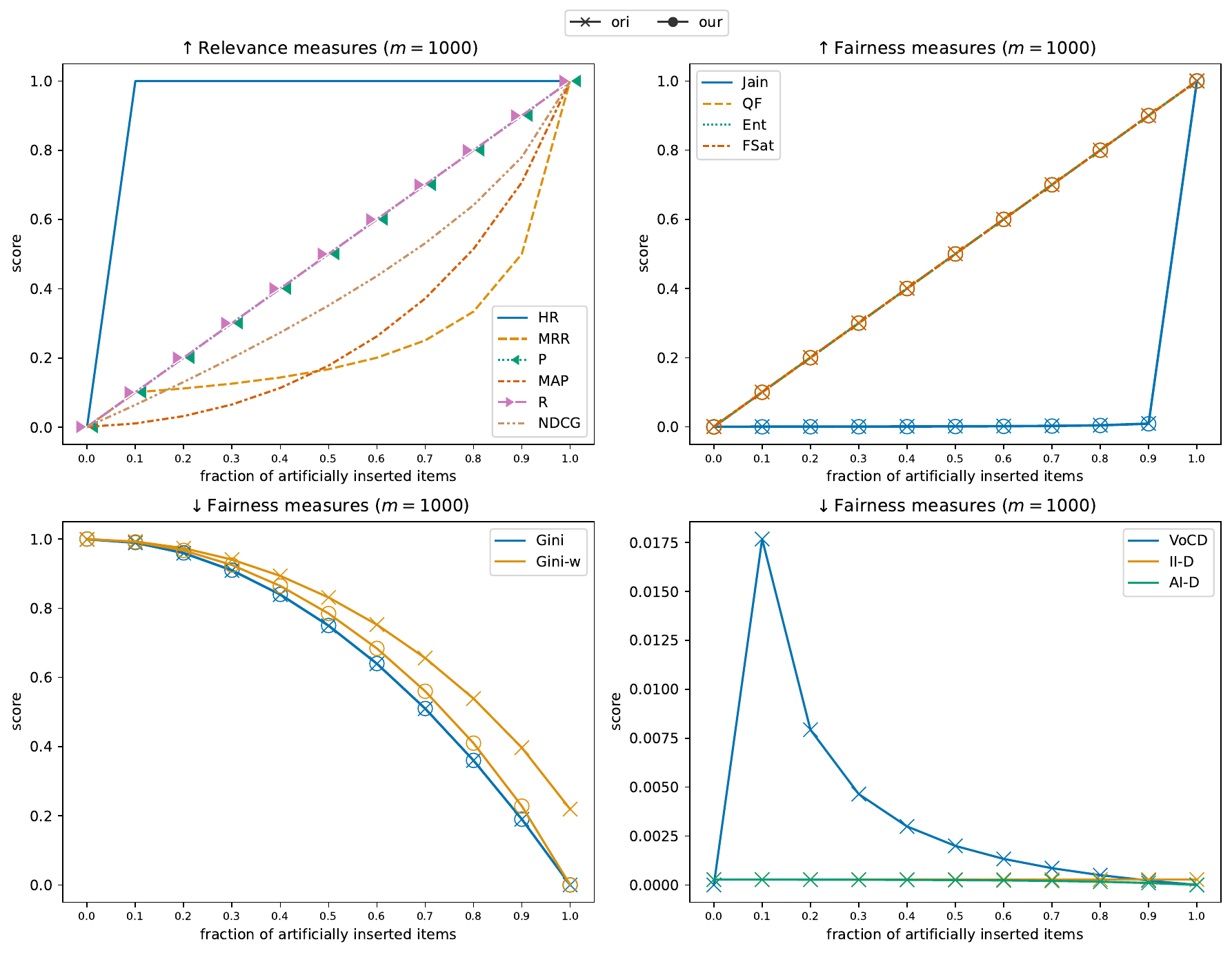}
    \caption{Results for jointly LE and relevant item insertion. All measures are at $k=10$. QF\ori~and FSat\ori~overlap. QF\our, FSat\our, and Ent\our~also overlap.}
    \label{fig:artificial-fair}
\end{figure}

The results of this experiment are presented in Fig.~\ref{fig:artificial-fair}. We see that all relevance measures increase as we add more relevant items.\footnote{The relevance measures do not start from 0 when $P=0$, as there is one user with ten relevant items.} 
The following observations apply to both the original fairness measures and to our corrected versions of these measures, unless otherwise specified. All fairness measures, except VoCD\ori~ and II-D\ori, indicate more fairness as we increase $P$, but with varying sensitivity, explained next. \up Jain is one of the strictest fairness measures. Even when the proportion of LE items is 0.9 (item $i_1$ is recommended to all users, but the rest of the recommendation lists are filled with different items), the \up Jain score is still close to 0, which translates to unfair while \up QF, \up Ent, and \up FSat are 0.9, which is close to the fairest score of 1. 
The scores of QF\ori~are exactly the same as FSat\ori, because all items in the recommendation list are recommended once, which is also the maximin share (defined in $\S$\ref{ss:onlyfair}). 
\up QF\our, \up Ent\our, and \up FSat\our~also give identical scores. This is expected as the increase in the scores is constant and proportional to the fraction of artificially inserted LE items, yet this is interesting as these three measures are based on three different fairness notions (QF being insensitive to the number of times an item is recommended, and FSat being based on maximin-shared fairness). 
 
Meanwhile, the increase of fairness in \down Gini and \down Gini-w follows a non-linear trend, with \down Gini-w being stricter than \down Gini. The non-linear trend is also expected as Gini and Gini-w are based on the Lorenz curve, a graphical representation of the cumulative proportion of exposure to the cumulative proportion of items. We also see that Gini-w\our~ is able to reach the theoretical most fair when the entire recommendation list consists of artificially inserted LE items, while Gini-w\ori~ fails. 
\down VoCD\ori~ is insensitive to the insertion of items as it only considers fairness for recommended items. The number of times these items are recommended across all users does not differ much in this set-up, therefore \down VoCD\ori~ returns scores that are close to the fairest.
Most notably, \down II-D\ori~ and \down AI-D\ori~ are very close to 0 (on the scale of $10^{-3}$ or even smaller) even when the same $k$ items are recommended to all users. \down II-D \ori~ remains constant, while \down AI-D\ori~ is rather insensitive to the addition of LE and relevant items. The small scores are due to the measures quantifying fairness according to the closeness of item exposure with random exposure, while other measures have no such comparisons. Therefore, for these measures, the scales are not very meaningful, even though for AI-D, the scores still indicate improvement as we insert more LE items. 
    
We see similar trends with $m \in \{100, 500\}$, but the change of the scores is most stable with $m=1000$. As we increase the number of users (and items), the range of VoCD, II-D, and AI-D scores also becomes more compressed. In contrast, the range of the other measures remains similar. We also observe a similar but opposite trend of results when we artificially insert known irrelevant and multiple copies of items already in the recommendation list. Both of these results are in App.~\ref{app:insert}.

Overall, the artificial insertion experiment indicates that several measures respond linearly to the insertion of LE items 
i.e., \up QF, \up FSat, and \up Ent, while the rest do so non-linearly. 
This can affect the interpretation of these scores, as we observe that it is generally harder to achieve a high fairness score in some measures. In some other measures, it is also easier to improve fairness when starting from a relatively fair situation, but much harder when starting from a completely unfair situation. 

\section{Related Work}\label{s:prevwork}

Prior work \cite{Do2021Two-sidedDominance, Wang2022ProvidingSystems,Zhu2020FARM:APPs,Wu2022JointRecommendation,Patro2020FairRec:Platforms,Mansoury2020FairMatch:Systems,Mansoury2021ASystems} proposes exposure-based individual item fairness measures but does not provide a comprehensive analysis of the limitations of the measures. Our work differs from this because we extensively analyse individual item fairness measures specifically for \rs, identify novel and previously-known limitations in them, and address these limitations. Meanwhile, several other work uses individual item fairness measures that also takes into account the relevance of the item to users  
\cite{Morik2020ControllingLearning-to-Rank, Borges2019EnhancingAutoencoders,Saito2022FairRanking,Wu2022JointRecommendation,Zhu2021FairnessSystems}. The investigation of these measures (of fairness and relevance) is reserved for our future work.

\citeauthor{Amigo2023ASystems}~\cite{Amigo2023ASystems} overview fairness measures in \rs~ and characterise them according to five dimensions. 
They focus on generalising the measures into  broad categories and studying the relationship between multi-stakeholder fairness, whereas we analyse each individual item fairness measure both theoretically and empirically. 
We also focus on the relationship among individual item fairness measures. 

\citeauthor{Raj2022MeasuringResults}~\cite{Raj2022MeasuringResults} analyse fairness measures for provider-side group fairness in ranked outputs. They include one measure for individual item fairness, II-D (referred to as EED in~\cite{Raj2022MeasuringResults}, which was originally proposed by \cite{Diaz2020EvaluatingExposure}), 
but only analyse it as a group fairness measure. 
They list some questions to assess the design of fairness measures, which we exploit in this work, e.g., the examination function used (Tab.~\ref{tab:exp-weigh}), the ideal/fair criteria (Tab.~\ref{tab:situations}), and if the measure incorporates relevance ($\S$\ref{ss:onlyfair}). 
Both our work and theirs identify the limitations in the measures e.g., edge cases where zero values cause undefinedness in the measure computation ($\S$\ref{ss:undefinedness}). 
They propose to use a small constant to avoid computing $\log{0}$ in group fairness metrics, but we do not use the same approach as using a small constant can introduce noise in the measure computation. 

\citeauthor{Majumder2021FairFairness}~\cite{Majumder2021FairFairness} examine classification measures for both individual and group fairness through empirical analysis and cluster the measures based on correlation values. They analyse the correlation among fairness measures but do not investigate the 
limitations of those measures. They find disagreements between the measures when labelling a model as fair or unfair. We did not do this mapping, as this may lead to loss of valuable information regarding the range and the actual values of the measures. 
However, we show that disagreements also exist between several individual item fairness measures in \rs.

All previous work finds that several fairness measures 
are highly correlated, and are insensitive to changes in data  
\cite{Amigo2023ASystems,Majumder2021FairFairness,Raj2022MeasuringResults}. Our analysis in $\S$\ref{s:exp} confirms these findings in a different experimental set-up and sheds light onto additional limitations which have not been reported or corrected previously. 
Next, we provide practical guidelines for choosing among individual item fairness measures.

\section{Discussion}
\label{s:discussion}
\subsection{Summary of Theoretical Corrections}
In this paper, we critically analyse individual item fairness measures in \rs~w.r.t.~their limitations. We point out a total of five theoretical limitations in the measures, and identify that each measure suffers from two or more limitations. Some limitations are due to intentional design choices of the measure, while the remaining limitations go against some pre-defined desirable properties of (fairness) evaluation measures. 

We posit that evaluation measures of fairness should have two important utilities: 
1) to assess systems/models in isolation (i.e., evaluating `how fair' a single system is, with one endpoint being the most unfair and the other being the most fair); 
2) to compare different \rs~and make a decision about whether and to what extent system A is more fair than system B, based on the measure.\footnote{Note that utility 1) is for assessment purposes, not a goal for model development; it may not be possible for relevant recommendations to be maximally fair at the same time.} 
A measure should ideally be usable for both use cases. At the present stage, none of the individual item fairness measures is suitable for the first use case, hence the need to modify the current measures for it.\footnote{This is unlike relevance measures, where there is still room for choice, as several measures have reachable endpoints (e.g., NDCG, RR@$k$) \cite{Moffat2013SevenMetrics}.} 

Note that having limitation(s) does not mean that the measures are completely unusable. Some measures can still be used despite having limitations, as long as one is aware of these limitations. For instance, in the case of measures that empirically cannot reach the endpoints of $[0,1]$, it is still possible to use the measure to compare fairness between two or more systems. 
Yet, evaluating fairness for a single system using such a measure is a challenge, as having a score of, for example, 0.6 does not always mean that there is still 40\% room for improvement. This point should be kept in mind, especially given the common practice of interpreting a single score in comparison to the known range of that measure. 

We also provide theoretical solutions to address the three resolvable limitations, and we argue why the remaining limitations cannot be resolved. The first set of solutions guarantees that the measures range in a bounded interval, e.g., $[0, 1]$, where both the theoretical minimum and maximum scores are achievable, with one endpoint corresponding to the most unfair recommendation list, and the other to the fairest recommendation list. This set of solutions considers the number of recommendation slots and the number of items in the dataset, and at the same time assures that the measures are well-defined for both common and edge cases.
Our second solution ensures that the affected measure is sensitive to the change of exposure received by an item, thereby fulfilling a desired property of fairness measure. 

\subsection{Summary of Empirical Findings} 
Extensive empirical experiments were conducted to compute relevance and fairness scores for both the original measures and for our corrected versions of these measures. 
The experiments utilised six datasets and seven recommendation models, including state-of-the-art models and well-established baseline models. 
Even though the models are all trained to optimise for relevance, we discover that the fairest model is not necessarily the worst in terms of relevance scores. This was unexpected as the fairness measures used in this work are detached from relevance. However, we also see the common observation where several models have higher relevance scores, but exhibit lower fairness. 

Our results empirically show that relevance measures and fairness measures have different ranges, which makes the interpretation of the fairness measures difficult. Further, the range of fairness measures is incomparable between different measures, and for some measures, this range is also not lower/upper-bounded empirically. Other noticeable observations include some fairness measures that tend to score much lower/higher compared to other measures, as well as uncorrected measures that are incomputable or produce constant values, given any recommendation list based on the same dataset. While the actual scores of the measures may differ, we found that most fairness measures have a high and significant agreement in ranking the recommenders from the most to least fair. The strong agreement is observed between the corrected and uncorrected measures, and even between some measures that are based on different fairness concepts.
Altogether, our results show that:
\begin{itemize}
    \item Despite limitations in quantifying the extent of fairness, the measures agreed in the ordering of models according to their individual item fairness. 
    \item Some original measures do not reflect the absolute quantity or differences in fairness.
    \item The corrected measures are required to reliably use the measures for scenarios that may contain commonly occurring and edge cases.
\end{itemize}

\subsection{Guidelines of the Appropriate Use of the Fairness Measures}
Next, we summarise guidelines on using these fairness measures based on the above theoretical and empirical findings. 

\noindent\textbf{Use original fairness measures only to evaluate relative fairness}. 
\textit{Relative fairness} refers to comparing the relative ordering of fairness scores. 
The original fairness measures suffer from theoretical limitations that limit their usage in settings where data distributions or recommendation scenarios do not fulfil the theoretical premises of the original measures. Moreover, the original measures may be more difficult to interpret as their range and scaling do not always match the intuitive expectations of being between 0 and 1. While the original measures as proposed outside recommendation can be used as their ranges are known and can be easily interpreted, we advise using them to evaluate only relative fairness. 

\noindent\textbf{Use our corrected fairness measures to evaluate absolute fairness}.
\textit{Absolute fairness} refers to measuring how close a model's recommendation is to the most (un)fair recommendation scenario. To evaluate absolute fairness, we recommend using our corrected measures for Jain, QF, Ent, Gini, and FSat. Our fairness measures are always perfectly correlated with the original measures, thus providing results that align with the original measures. Further, our measures are well-defined and have better interpretability w.r.t.~how the minimum/maximum scores correspond to the most unfair/fair recommendation scenario, as shown in $\S$\ref{ss:maxmin}. Our fairness measures are highly correlated with each other, but because they operate on different scales, one should not deduce that a model is (un)fair based on the absolute fairness measurement scores. 

Note that both FSat\ori~ and FSat\our~ should never be used when $km<n$ due to the unresolvable \textbf{always-fair} limitation, as the score will always be perfectly fair regardless of the recommendation. \down Gini-w\our~ should preferably be used when one has equal or more items than recommendation slots ($km\leq n$), as the measure works ideally in that setting: a score of 0 means the recommendation is perfectly fair, while a score of 1 means the unfairest possible recommendation. For the remaining cases, which commonly happens in many public recommendation datasets for any cut-off $k$ (Tab.~\ref{tab:dataset}), even if the most unfair recommendation entails a score of 1, the most fair recommendation is not mapped to a score of 0 in Gini-w\our. Yet, this is still better than Gini-w\ori~ which maps unrealistic scenarios to 0 and 1. For Ent, we recommend using our correction, since our correction avoids the \textbf{undefinedness} limitation, and would produce the same score as Ent\ori~ when all items are recommended.  

We discourage using the rest of the measures due to their tendency to have scores that are not representative of fairness, e.g. scale mismatch between II-D/AI-D and the rest of the measures. Additionally, II-D should not be used for single-round recommendations as the scores are always constant.

\section{Conclusions}

We have presented a novel investigation into the theoretical and empirical limitations of current evaluation measures of individual item fairness in recommender systems. 
We have further amended these measures to correct their limitations or have argued why some limitations are impossible to resolve. Extensive experiments on real-life and synthetic data reveal novel insights on how individual item fairness measures should and should not be used. 

In the present work, we solely concentrated on measures that quantify individual item fairness independently of recommendation performance. We reserve the analysis of fairness measures that are tied to relevance for our future work. Future work should investigate whether measures that aim to simultaneously quantify both recommendation performance, or relevance, and fairness suffer from similar limitations and empirical behaviours than the measures studied here. Other future work could further explore the relationship between individual item fairness and item group fairness \cite{Wu2022JointRecommendation}, or fairness between users and items \cite{Amigo2023ASystems}. 
All measures studied here also assume that exposure is the key factor for fairness, while there might be other factors to consider for fairness, e.g. speed of the recommendation or the wait time from when an item is introduced to a system until it gets recommended. 

The empirical studies could also be extended to account for the behaviour and performance of the measures using additional datasets. However, while we cannot exclude the possibility that the experiments on other settings, domains, or datasets could lead to new insights, it is unlikely that they would affect our conclusions and guidelines on the appropriate use of the studied measures. Future work could also utilize our corrected measures to optimize recommendation models for fairness to reveal whether the corrected measures could improve recommendation models as opposed to only measuring the performance of existing models. 

\section*{Code Availability}
Our source code (in Python 3.10) is publicly available on \url{https://github.com/theresiavr/individual-item-fairness-measures-recsys} and usable under the MIT License, with proper attribution to this work and possibly other related work. Other restrictions regarding the usability of the code may apply for the RecBole library \cite{Zhao2021RecBole:Algorithms}.

\section*{Acknowledgements}
The work is supported by the Algorithms, Data, and Democracy project (ADD-project), funded by Villum Foundation and Velux Foundation, as well as the Academy of Finland. 
We also thank the anonymous reviewers who have provided insightful comments and suggestions to improve earlier versions of the manuscript.

\section{Appendix}

\subsection{Mathematical Workings for Bounds in Tab.~\ref{tab:bounds}}
\label{app:boundsproof}

We provide the derivation of the obtained min/max achievable value of Jain, QF, Ent, Gini, FSat, and VoCD. The min/max achievable values are obtained from the most (un)fair recommendation scenarios described in $\S$\ref{ss:non_realisability}.

\subsubsection{Jain's Index}
The most unfair case for \up Jain produces Jain$_{\min}$ and the most fair case for \up Jain produces Jain$_{\max}$.
\begin{equation*}
\text{Jain}_{\min} = \frac{(km)^2}{n \sum\limits_{i\in I} \left[\sum\limits_{u\in U}
1_{R_{u}^{k}}(i)\right]^2} =  \frac{(km)^2}{n(km^2)} = \frac{k}{n}
\end{equation*}
\begin{align*}
\text{Jain}_{\max} 
&= \frac{(km)^2}{n \sum\limits_{i\in I} \left[\sum\limits_{u\in U}
1_{R_{u}^{k}}(i)\right]^2} \\
&=  \frac{(km)^2}{
n\left((n-km \bmod n)\floorkm^2 + (km \bmod n)\left(\floorkm+1\right)^2\right)} \\
&= \frac{(km)^2}{
n\left(n\floorkm^2-(km\bmod{n})\floorkm^2 + (km\bmod{n})\left(\floorkm^2+2\floorkm+1\right)\right)
} \\
&= \frac{(km)^2}{
         n\left(n\floorkm^2 
         + (km\bmod{n})\left(2 \floorkm+1\right) 
         \right)}   
\end{align*}

\subsubsection{Qualification Fairness}
The most unfair case for \up QF produces QF$_{\min}$ and the most fair case for \up QF produces QF$_{\max}$.
\begin{align*}
    \text{QF}_{\min} 
    = \frac{(k\cdot1)^2}{n(k \cdot 1^2)} = \frac{k}{n}
\end{align*}
When there are not enough recommendation slots for all items, $km < n$:
\begin{align*}
    \text{QF}_{\max}
    = \frac{(km\cdot 1)^2}{nkm(1^2)} = \frac{km}{n}
\end{align*}
When $km \geq n$, all items can be recommended, hence  QF$_{\max}=\frac{n}{n}=1$.

\subsubsection{Entropy}
The most unfair case for \up Ent produces Ent$_{\min}$ and the most fair case for \up Ent produces Ent$_{\max}$. The first term in Ent$_{\max}$ comes from $n-km\bmod{n}$ that are each recommended $\floorkm$ times and the second comes from $km\bmod{n}$ items that are each recommended $\floorkm + 1$ times.
\begin{equation*}
\text{Ent}_{\min} 
     = -k \cdot \frac{1}{k} \log{\frac{1}{k}} 
     = - \log{k^{-1}}          
     = \log{k}
\end{equation*}
\begin{align*}
\text{Ent}_{\max} = 
&-
(n - km \bmod n)\left(\frac{\floorkm}{km}\log{\frac{\floorkm}{km}}\right) \\
&- (km \bmod n)\left(\frac{\floorkm+1}{km}\log{\frac{\floorkm+1}{km}}\right) 
\end{align*}

\subsubsection{Gini Index}
The most unfair case for \down Gini produces Gini$_{\max}$ and the most fair case for \down Gini produces Gini$_{\min}$.

\begin{align*}
    \text{Gini}_{\max}  
        =& \frac{\sum\limits_{j=n-k+1}^n(2j-n-1)m}{nk(m)} \\
        =& \frac{\sum\limits_{n-k+1}^n{2j} - n\sum\limits_{n-k+1}^n{1} - \sum\limits_{n-k+1}^n{1}}{nk} \\
        =& \frac{2\sum\limits_{n-k+1}^n{j} - n(n-(n-k+1)+1) - (n-(n-k+1)+1)}{nk} \\
        =& \frac{2\frac{(n-k+1+n)(n-(n-k+1)+1)}{2}-nk-k}{nk} \\
        =& \frac{(2n-k+1)(k)-nk-k}{nk} \\
        =& \frac{(k)(2n-km+1-n-1)}{nk} \\
        =& \frac{n-k}{n} \\
        =& 1- \frac{k}{n}
\end{align*}

To derive Gini$_{\min}$, we use the pairwise difference formula of Gini:

\begin{equation*}
    \text{Gini} = \frac{\sum\limits_{(i,i')}{
    \left|
    \sum\limits_{u\in U}{1_{R_{u}^{k}}(i)}
    -\sum\limits_{u\in U}{1_{R_{u}^{k}}(i')}
    \right|
    }}
    {2n^2{\bar{x}}}
\end{equation*}

where $\bar{x}$ is the average number of times an item is recommended, for the most fair case, calculated as follows:

\begin{equation*}
\bar{x} = \frac{(n - km \bmod n)\floorkm + (km \bmod n)(\floorkm+1)}{n}
= \frac{n\floorkm +km \bmod n }{n}
\end{equation*}

We simplify the numerator and denominator separately, for clarity in the proof. 
To simplify the numerator, in the most fair case, there are $2(n-km \bmod n)(km \bmod n)$ pairs of items with an absolute difference of $\left(\floorkm +1\right) - \floorkm=1$, which is the difference of the number of times the items are recommended. The rest of the pairs have 0 differences. Hence, the numerator is $2(n-km \bmod n)(km \bmod n)$. Putting everything together:
\begin{align*}
    \text{Gini}_{\min} 
    &=  \frac{2(n-km \bmod n)(km \bmod n)}{
    2n^2\left(\frac{n\floorkm +km \bmod n }{n}\right)
    }\\
    &= \frac{(n-km \bmod n)(km \bmod n)}{
    n\left(n\floorkm +km \bmod n\right)
    } \\
    &= \frac{(n-km \bmod n)(km \bmod n)}{
    n\left(n\floorkm +km -n\floorkm\right)
    } \\
    &= \frac{(n-km \bmod n)(km \bmod n)}{
    kmn
    } 
\end{align*}

The most unfair case for \down Gini-w produces Gini-w$_{\max}$ and the most fair case for \down Gini produces Gini-w$_{\min}$. 

To obtain Gini-w$_{\max}$, we derive the numerator and denominator separately using Eq.~\ref{eq:gini-ori}. We first compute the numerator of Gini-w$_{\max}$, considering that the items with the least to the most exposure are as follows: the first $n-k$ items are with zero exposure (as they are not present in the top $k$), one item is exposed $m$ times at position $k$, one item is exposed $m$ times at position $k-1$, and so on, until one last item that is exposed $m$ times at the top of the recommendation list. The exposure received by those items respectively are $0, m\log_{k+1}{2}, m\log_{k}{2}, \dots, m\log_{2}{2}$. Therefore, the numerator of Gini-w$_{\max}$ can be written as $m \sum\limits_{\ell=1}^k{(n-2\ell+1) \log_{\ell+1}{2}}$. Meanwhile, the denominator of Gini-w$_{\max}$ is simply $n$ times the total exposure received by the items: $mn \sum\limits_{\ell=1}^k{\log_{\ell+1}{2}}$. Putting the numerator and denominator together:
\begin{equation*}
 \text{Gini-w}_{\max} = 
 \frac{
 m\sum\limits_{\ell=1}^k{(n-2\ell+1) \log_{\ell+1}{2}}}
    {mn\sum\limits_{\ell=1}^k{\log_{\ell+1}{2}}}
=  \frac{
 \sum\limits_{\ell=1}^k{(n-2\ell+1) \log_{\ell+1}{2}}}
    {n\sum\limits_{\ell=1}^k{\log_{\ell+1}{2}}}
\end{equation*}

To obtain Gini-w$_{\min}$, we also derive the numerator and denominator separately using Eq.~\ref{eq:gini-ori}. Note that for Gini-w$_{\min}$ we only consider cases where $km\leq n$ due to the unresolvable limitation of non-realisability, \textit{Cause 4} ($\S$\ref{ss:nofix}). First, we explain how to obtain the numerator. With the restriction of $km\leq n$, to make the recommendation the fairest, the $km$ items that are recommended must be unique, leaving $n-km$ items exposed. Thus, the items with the least to the most exposure are as follows: the first $n-km$ items receive zero exposure, the next $m$ items will be recommended once each at position $k$, another set of $m$ items each at position $k-1$, and so on until the last set of $m$ items that will each be recommended at the top of the recommendation list. The numerator of  Gini-w$_{\min}$ can then be calculated as follows:

\begin{align*}
    &\sum\limits_{j=(n-km)+1}^{n-km+m} (2j-n-1) \log_{k + 1}{2}
    \\ 
    & +
    \sum\limits_{j=(n-km+m)+1}^{n-km+2m} (2j-n-1) \log_{k}{2} \\
    & +
    \dots \\
    & +
    \sum\limits_{j=(n-m)+1}^{n} (2j-n-1) \log_{2}{2} \\
    &=\sum\limits_{\ell=1}^k 
    \sum\limits_{j=n-\ell m+1}^{n- \ell m + m} (2j-n-1) \log_{\ell + 1}{2}
\end{align*}

As for the denominator, it is obtained the same way as in Gini-w$_{\max}$, resulting in  $mn \sum\limits_{\ell=1}^k{\log_{\ell+1}{2}}$ as the total exposure received by the items remains the same for the same cut-off $k$ and the number of user $m$. Putting the numerator and denominator together:

\begin{equation*}
 \text{Gini-w}_{\min} =
  \frac{
    \sum\limits_{\ell=1}^k 
    \sum\limits_{j=n-\ell m+1}^{n- \ell m + m} (2j-n-1) \log_{\ell + 1}{2}}
    {mn\sum\limits_{\ell=1}^k{\log_{\ell+1}{2}}}
\end{equation*}

\subsubsection{Fraction of Satisfied Items}
The most unfair case for \up FSat produces FSat$_{\min}$ and the most fair case for \up FSat produces FSat$_{\max}$. Note that $k \leq n$ thus $1 \geq \frac{k}{n} \Leftrightarrow m \geq \frac{km}{n}\geq\floorkm$.
\begin{align*}
    \text{FSat}_{\min} 
    =  \frac{1}{n}
    \left(k \cdot \delta\left(
    m \geq \floorkm
    \right)\right) 
    = \frac{k}{n}
\end{align*}
\begin{align*}
    \text{FSat}_{\max} 
    &=  \frac{1}{n} 
    (n-km \bmod n)\cdot \delta\left(\floorkm\geq\floorkm\right) \\
    &\quad +
    \frac{1}{n}(km \bmod n)\cdot \delta\left(\floorkm+1\geq\floorkm\right)
    \\
    &=\frac{1}{n}[ 
    (n-km \bmod n) + km \bmod n]
    =\frac{n}{n}
    =1
\end{align*}

\subsubsection{Proofs for the maximum value of VoCD}
\label{app:vocd}
First, we prove in Theorem~\ref{th:max-vocd} that  when there is only one pair of similar items, the maximum VoCD value can be obtained when the two items are recommended $1$ time and $m$ times each. We then use Theorem~\ref{th:max-vocd} to show that when there are more than one pair of similar items, the maximum VoCD value does not increase (Theorem~\ref{th:vocd-not-increase}).

\begin{theorem}
\label{th:max-vocd}
If there is only one pair of $(i,i')\in A$, $\text{VoCD}_{\max}$ is obtained when  $\sum\limits_{u\in U}1_{R_{u}^{k}}(i)=1$ and $\sum\limits_{u\in U}1_{R_{u}^{k}}(i')=m$
\end{theorem}
\begin{proof}
We prove by contradiction: if an item $i$ is recommended, $1 \leq \sum\limits_{u\in U}1_{R_{u}^{k}}(i) \leq m$. 
Suppose $\exists x_1, x_2$ where $1<x_1<x_2<m$ such that 
$\frac{m-1}{m} < \frac{x_2-x_1}{x_2} \Leftrightarrow
\frac{-1}{m} < \frac{-x_1}{x_2} \Leftrightarrow
\frac{1}{m} >\frac{x_1}{x_2} \Leftrightarrow
m  < \frac{x_2}{x_1}$. 
However, $x_2 < m \Leftrightarrow \frac{x_2}{x_1} < \frac{m}{x_1} < m$, which contradicts the previous inequality, so it must be $x_1 = 1$ and $x_2 = m$.
\end{proof}
\begin{theorem}
\label{th:vocd-not-increase}
The max VoCD score does not increase with $|A|$.
\end{theorem}
\begin{proof}
Case 1: for each item pair in $A$ that is disjoint from the other item pairs, e.g., $(i_1,i_2)$ and $(i_3,i_4)$, by Theorem~\ref{th:max-vocd}, the maximum score for each pair and hence the average score of those pairs, is still $\frac{m-1}{m} - \beta$. 
Case 2: suppose the pairs are not disjoint, e.g., $(i_1, i_3), (i_2, i_3)$, and $i_1, i_2, i_3$ are recommended $x_1, x_2, x_3$ times respectively, 
where $1\leq x_1\leq x_2 \leq x_3 \leq m$. We show that for Case 2, the maximum value does not increase.

Note that $x_3 \leq m \Leftrightarrow 
-\frac{1}{x_3} \leq -\frac{1}{m} \Leftrightarrow
-\frac{x_1}{2x_3} \leq -\frac{x_1}{2m} \leq -\frac{1}{2m}$ and $-\frac{x_2}{2x_3} \leq -\frac{1}{2m}$. Thus,
$\frac{1}{2}
    \left(
        \frac{x_3-x_1}{x_3}+\frac{x_3-x_2}{x_3}
    \right)
    = 1 - \frac{x_1}{2x_3}-\frac{x_2}{2x_3} \leq 1-\frac{1}{m}$.
\end{proof}

\subsection{Extended Results of Experiments}
\label{app:extend-result}

\subsubsection{Experimental set-up}
\label{app:extend-exp_set_up}

We first report the hyperparameter search space for each recommender (Tab.~\ref{tab:searchhyper}) and the optimal hyperparameters for each model and each dataset (Tab.~\ref{tab:besthyper}).

\begin{table}
    \centering
    \caption{Hyperparameter search space for each recommender.}
    \label{tab:searchhyper}
   \resizebox{\columnwidth}{!}{\begin{tabular}{ll}
\toprule
 & Hyperparameter search space \\
 \midrule
ItemKNN & k: {[}10, 20, 30, 40, 50, 60, 70, 80, 90, 100, 150, 200, 250, 300, 400, 500, 1000{]} \\
 & shrink: {[}0.0, 0.5, 1.0{]} \\
 \midrule
\multirow{2}{*}{SLIM} & alpha: {[}0.2, 0.5, 0.8, 1.0{]} \\
 & l1 ratio: {[}0.01, 0.02, 0.05, 0.1, 0.5{]} \\
 \midrule
\multirow{2}{*}{BPR} & embedding size: {[}16, 32, 64, 128, 256, 512, 1024, 2048, 4096{]} \\
 & lr: {[}5e-5, 1e-4, 5e-4, 1e-3, 5e-3, 1e-2, 5e-2{]} \\
 \midrule
\multirow{4}{*}{NGCF} & dropout prob: {[}0.1, 0.2{]} \\
 & embedding size: {[}64, 128, 256, 512{]} \\
 & hidden size: {[}{[}64, 64, 64{]}, {[}128, 128, 128{]}, {[}256, 256, 256{]}{]} \\
 & lr: {[}5e-4, 1e-3, 5e-3{]} \\
 \midrule
\multirow{3}{*}{NeuMF} & dropout prob: {[}0.1, 0.2{]} \\
 & hidden size: {[}{[}128, 64{]}, {[}128, 64, 32{]}, {[}64, 32, 16{]}, {[}32, 16, 8{]}{]} \\
 & lr: {[}5e-4, 1e-3, 5e-3, 1e-2{]} \\
 \midrule
\multirow{4}{*}{MultiVAE} & dropout prob: {[}0.1, 0.2, 0.5{]} \\
 & hidden size: {[}{[}100{]}, {[}300{]}, {[}600{]}{]} \\
 & latent dimension: {[}64, 128, 256{]} \\
 & lr: {[}5e-4, 1e-3, 5e-3, 1e-2{]} \\
 \bottomrule
\end{tabular}}
\end{table}

\begin{table}
    \centering
    \caption{Optimal hyperparameters for each recommender and each dataset.}
    \label{tab:besthyper}
    \resizebox{\columnwidth}{!}{\begin{tabular}{lllllll}
\toprule
{} &                                    ItemKNN &                                             SLIM &                                                  BPR &                                                                                                     NGCF &                                                                           NeuMF &                                                                                     MultiVAE \\
\midrule
Amazon-lb &    \parbox[t]{1.5cm}{k: 20, \\shrink: 1.0} &  \parbox[t]{1.8cm}{alpha: 0.2, \\l1 ratio: 0.01} &  \parbox[t]{3cm}{embedding size: 4096, \\lr: 0.0001} &   \parbox[t]{3.5cm}{dropout prob: 0.2, \\embedding size: 256, \\hidden size: [128,128,128], \\lr: 0.005} &   \parbox[t]{3.2cm}{dropout prob: 0.2, \\hidden size: [128,64,32], \\lr: 0.005} &   \parbox[t]{2.5cm}{dropout prob: 0.5, \\hidden size: [600], \\latent dim: 128, \\lr: 0.005} \\
\midrule
Lastfm    &   \parbox[t]{1.5cm}{k: 300, \\shrink: 1.0} &  \parbox[t]{1.8cm}{alpha: 0.2, \\l1 ratio: 0.01} &  \parbox[t]{3cm}{embedding size: 2048, \\lr: 0.0005} &   \parbox[t]{3.5cm}{dropout prob: 0.2, \\embedding size: 512, \\hidden size: [256,256,256], \\lr: 0.001} &     \parbox[t]{3.2cm}{dropout prob: 0.2, \\hidden size: [32,16,8], \\lr: 0.001} &   \parbox[t]{2.5cm}{dropout prob: 0.5, \\hidden size: [600], \\latent dim: 64, \\lr: 0.0005} \\
\midrule
Ml-1m     &  \parbox[t]{1.5cm}{k: 1000, \\shrink: 1.0} &  \parbox[t]{1.8cm}{alpha: 0.2, \\l1 ratio: 0.01} &  \parbox[t]{3cm}{embedding size: 2048, \\lr: 0.0001} &  \parbox[t]{3.5cm}{dropout prob: 0.1, \\embedding size: 256, \\hidden size: [256,256,256], \\lr: 0.0005} &   \parbox[t]{3.2cm}{dropout prob: 0.2, \\hidden size: [64,32,16], \\lr: 0.0005} &     \parbox[t]{2.5cm}{dropout prob: 0.1, \\hidden size: [600], \\latent dim: 64, \\lr: 0.01} \\
\midrule
Book-x    &    \parbox[t]{1.5cm}{k: 20, \\shrink: 1.0} &  \parbox[t]{1.8cm}{alpha: 0.2, \\l1 ratio: 0.01} &  \parbox[t]{3cm}{embedding size: 4096, \\lr: 0.0001} &  \parbox[t]{3.5cm}{dropout prob: 0.1, \\embedding size: 512, \\hidden size: [256,256,256], \\lr: 0.0005} &     \parbox[t]{3.2cm}{dropout prob: 0.1, \\hidden size: [32,16,8], \\lr: 0.001} &  \parbox[t]{2.5cm}{dropout prob: 0.5, \\hidden size: [600], \\latent dim: 128, \\lr: 0.0005} \\
\midrule
Amazon-is &   \parbox[t]{1.5cm}{k: 250, \\shrink: 1.0} &  \parbox[t]{1.8cm}{alpha: 0.2, \\l1 ratio: 0.01} &  \parbox[t]{3cm}{embedding size: 4096, \\lr: 0.0001} &   \parbox[t]{3.5cm}{dropout prob: 0.1, \\embedding size: 512, \\hidden size: [256,256,256], \\lr: 0.001} &   \parbox[t]{3.2cm}{dropout prob: 0.2, \\hidden size: [128,64,32], \\lr: 0.001} &   \parbox[t]{2.5cm}{dropout prob: 0.5, \\hidden size: [600], \\latent dim: 128, \\lr: 0.005} \\
\midrule
Amazon-dm &    \parbox[t]{1.5cm}{k: 10, \\shrink: 1.0} &  \parbox[t]{1.8cm}{alpha: 0.2, \\l1 ratio: 0.01} &  \parbox[t]{3cm}{embedding size: 4096, \\lr: 0.0001} &   \parbox[t]{3.5cm}{dropout prob: 0.2, \\embedding size: 512, \\hidden size: [256,256,256], \\lr: 0.001} &  \parbox[t]{3.2cm}{dropout prob: 0.2, \\hidden size: [128,64,32], \\lr: 0.0005} &    \parbox[t]{2.5cm}{dropout prob: 0.2, \\hidden size: [600], \\latent dim: 256, \\lr: 0.01} \\
\bottomrule
\end{tabular}
}
\end{table}

\subsubsection{Analysis of relevance and fairness}
\label{app:extend-performance}

We present the performance scores of the recommender systems on the Amazon-* and Book-x datasets in Tab.~\ref{tab:base-extra1} \& \ref{tab:base-extra2}. The scores of the original version of Ent cannot be calculated due to zero divisions error for the same reasons explained in $\S$\ref{ss:performance}. The constant scores of II-D have also been explained in the same section. The best relevance and fairness scores are bolded.

BPR generally performs the best in relevance, with the exception of Amazon-lb where NeuMF is best, while ItemKNN gives the best fairness scores. Other trends observed on Amazon-* and Book-x are similar to that on Lastfm and Ml-1m.

\subsubsection{Correlation between measures}
\label{app:corr}

\begin{figure*}[p]
\centering
    \includegraphics[width=\textwidth]{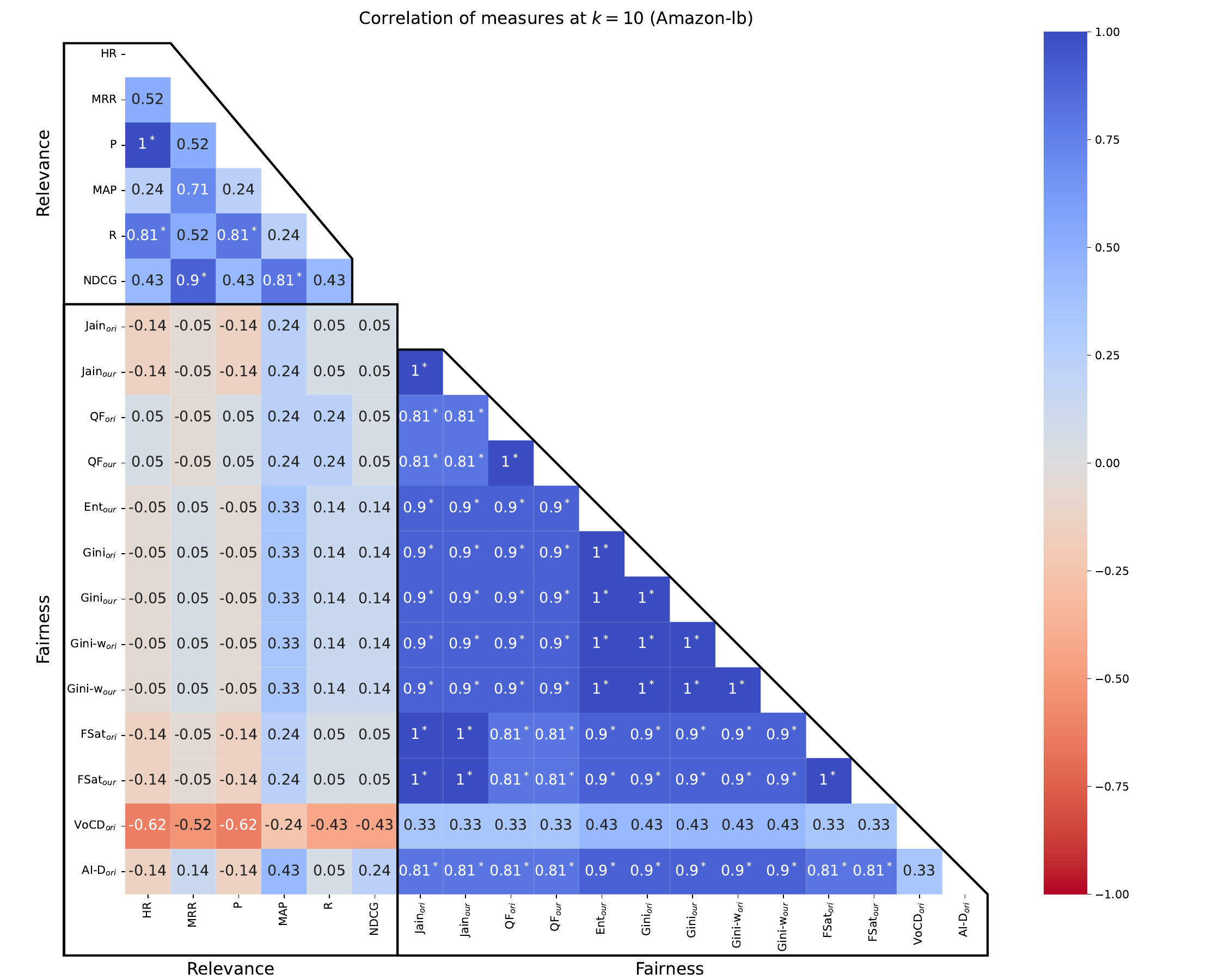}
\caption{Correlation (Kendall's $\tau$) between relevance and fairness measures for Amazon-lb. \explainsig}
\label{fig:corr-lb}
\end{figure*}

\begin{figure*}[p]
\centering
    \includegraphics[width=\textwidth]{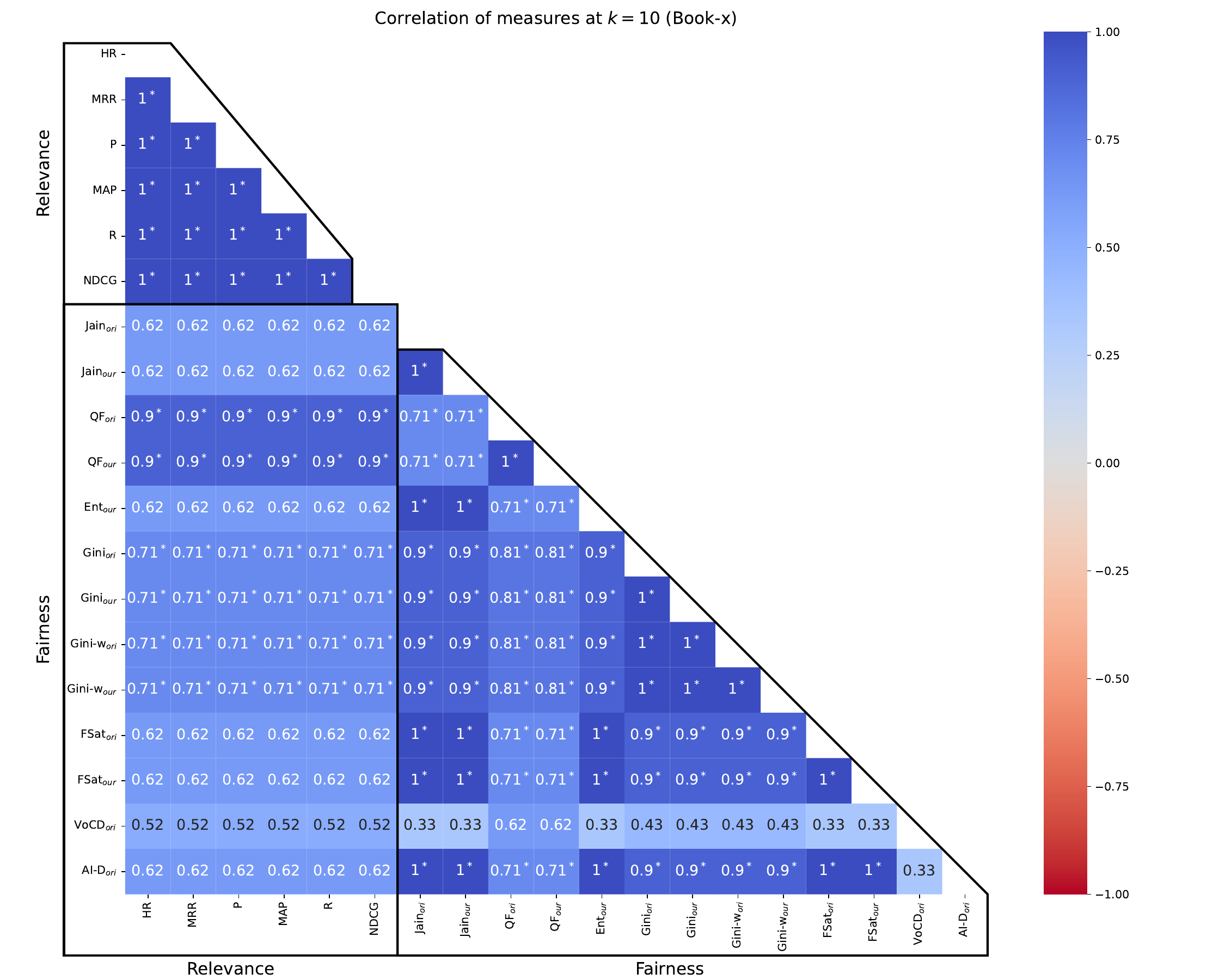}
\caption{Correlation (Kendall's $\tau$) between relevance and fairness measures for Book-x. \explainsig}
\label{fig:corr-bookx}
\end{figure*}

\begin{figure*}[p]
\centering
    \includegraphics[width=\textwidth]{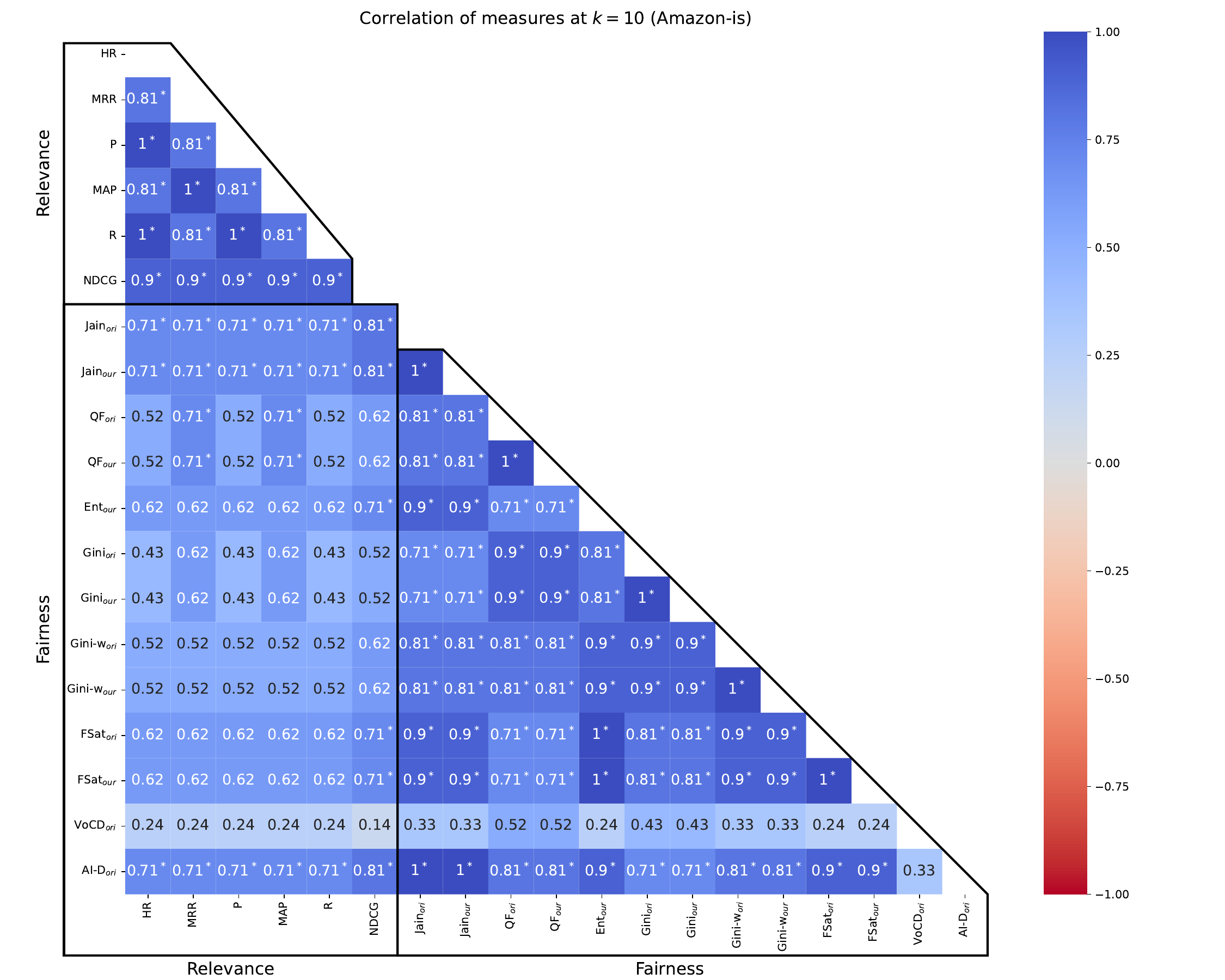}
\caption{Correlation (Kendall's $\tau$) between relevance and fairness measures for Amazon-is. \explainsig}
\label{fig:corr-is}
\end{figure*}

\begin{figure*}[p]
\centering
    \includegraphics[width=\textwidth]{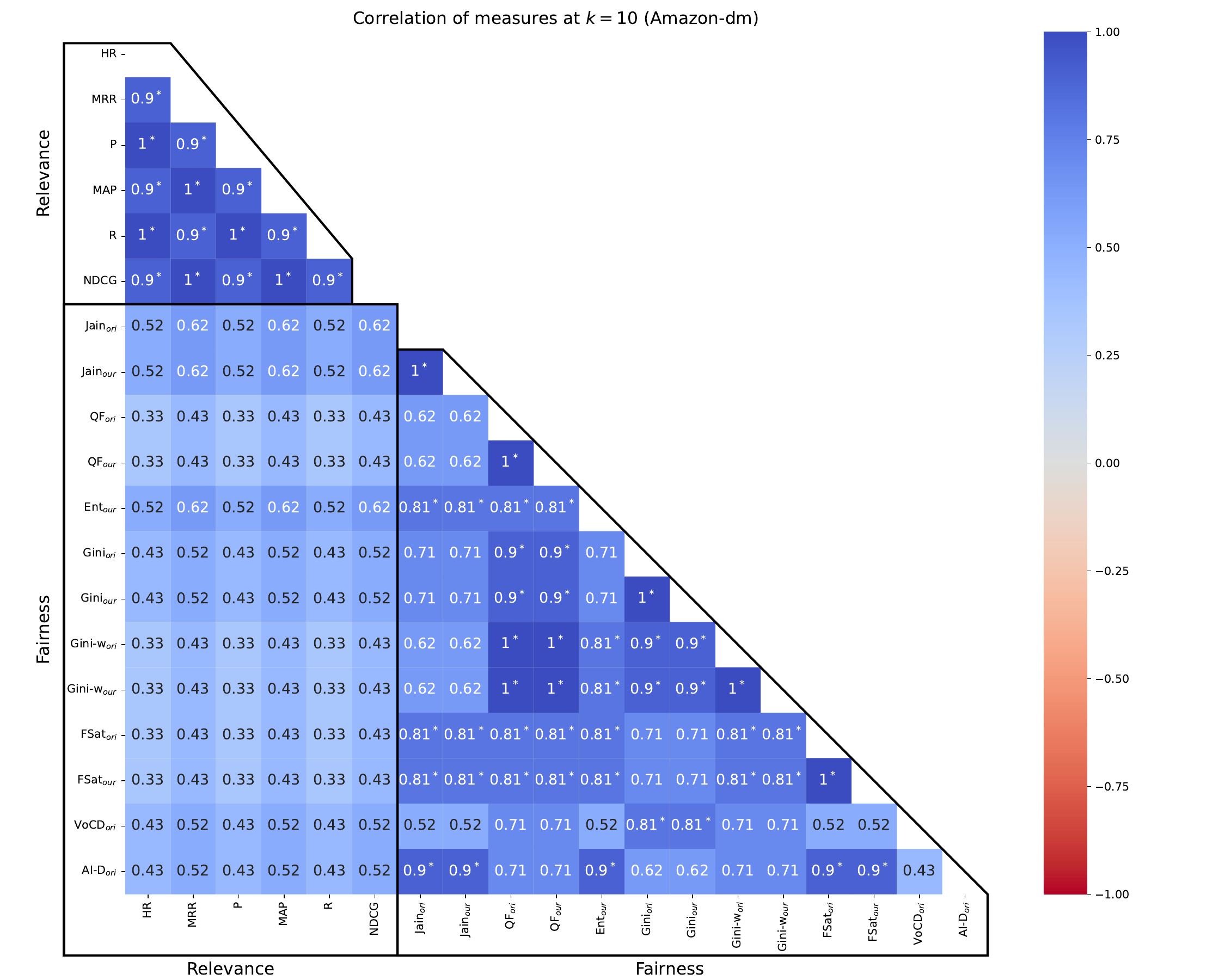}
\caption{Correlation (Kendall's $\tau$) between relevance and fairness measures for Amazon-dm. \explainsig}
\label{fig:corr-dm}
\end{figure*}

We show the Kendall's Tau values between relevance measures and fairness measures for the Amazon-* and Book-x datasets in Fig.~\ref{fig:corr-lb}--\ref{fig:corr-dm}.

\subsubsection{Max/min achievable fairness}
\label{app:maxmin}
\begin{figure}[p]
    \centering
    \includegraphics[width=\textwidth]{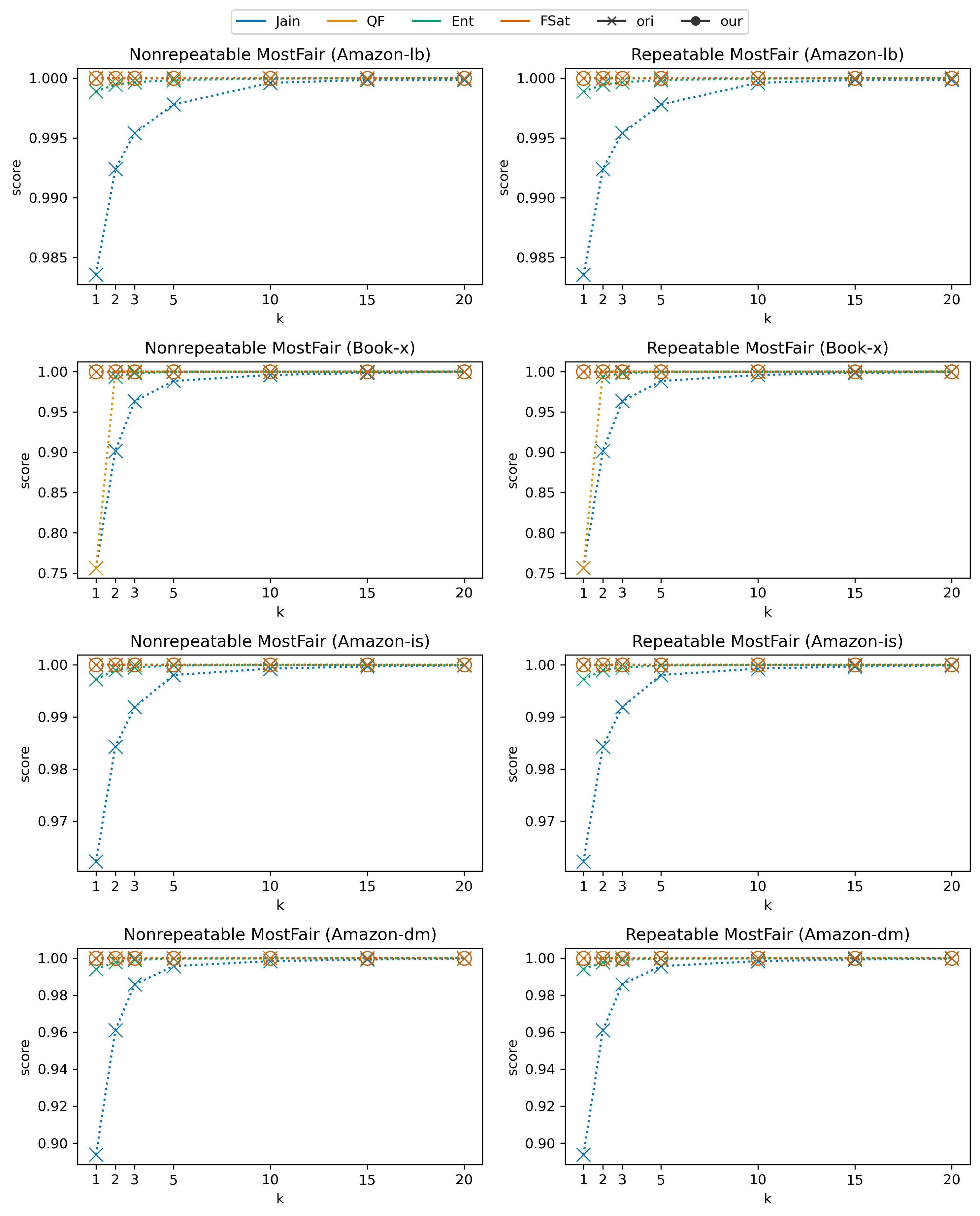}
    \caption{Most fair scores with varying $k$ for higher-is-fairer fairness measures on Amazon-* and Book-x.}
    \label{fig:app_mostfair_higher_better}
\end{figure}

\begin{figure}[p]
    \centering
    \includegraphics[width=\textwidth]{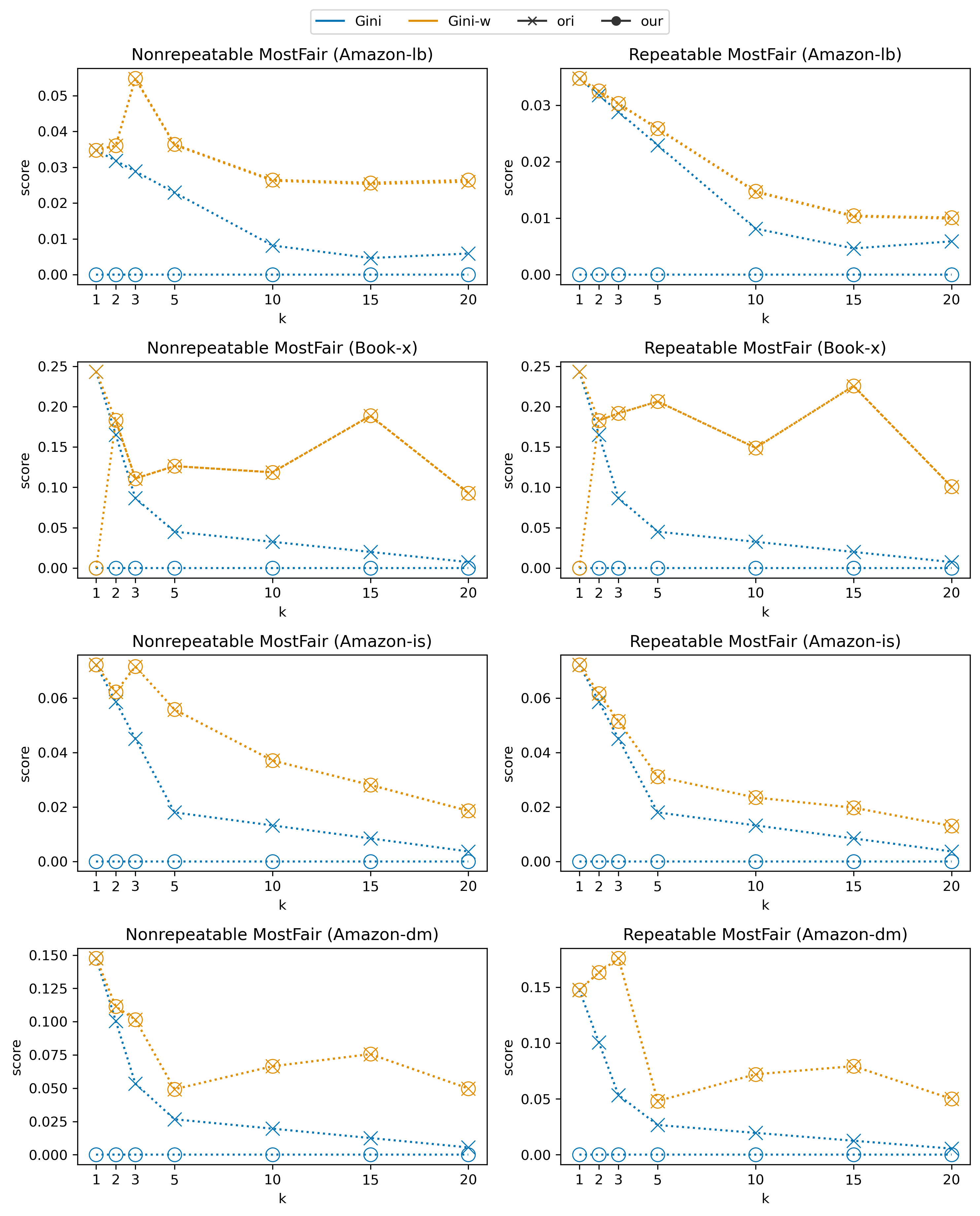}
    \caption{Most fair scores with varying $k$ for lower-is-fairer fairness measures on Amazon-* and Book-x.}
    \label{fig:app_mostfair_lower_better}
\end{figure}

\begin{figure}[p]
    \centering
    \includegraphics[width=0.95\textwidth]{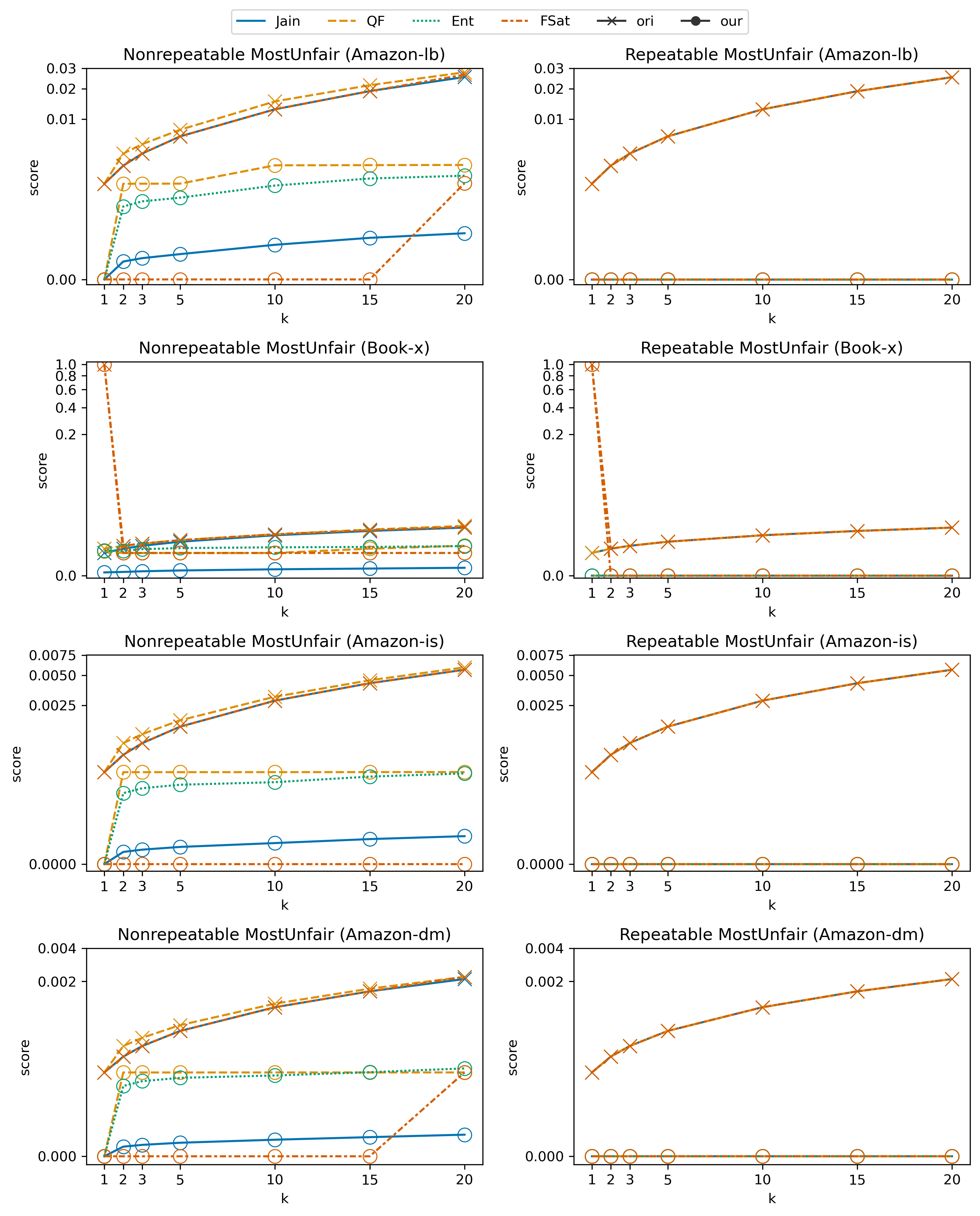}
    \caption{Most unfair scores with varying $k$ for higher-is-fairer fairness measures on Amazon-* and Book-x.}
    \label{fig:app_mostunfair_higher_better}
\end{figure}

\begin{figure}[p]
    \centering
    \includegraphics[width=0.95\textwidth]{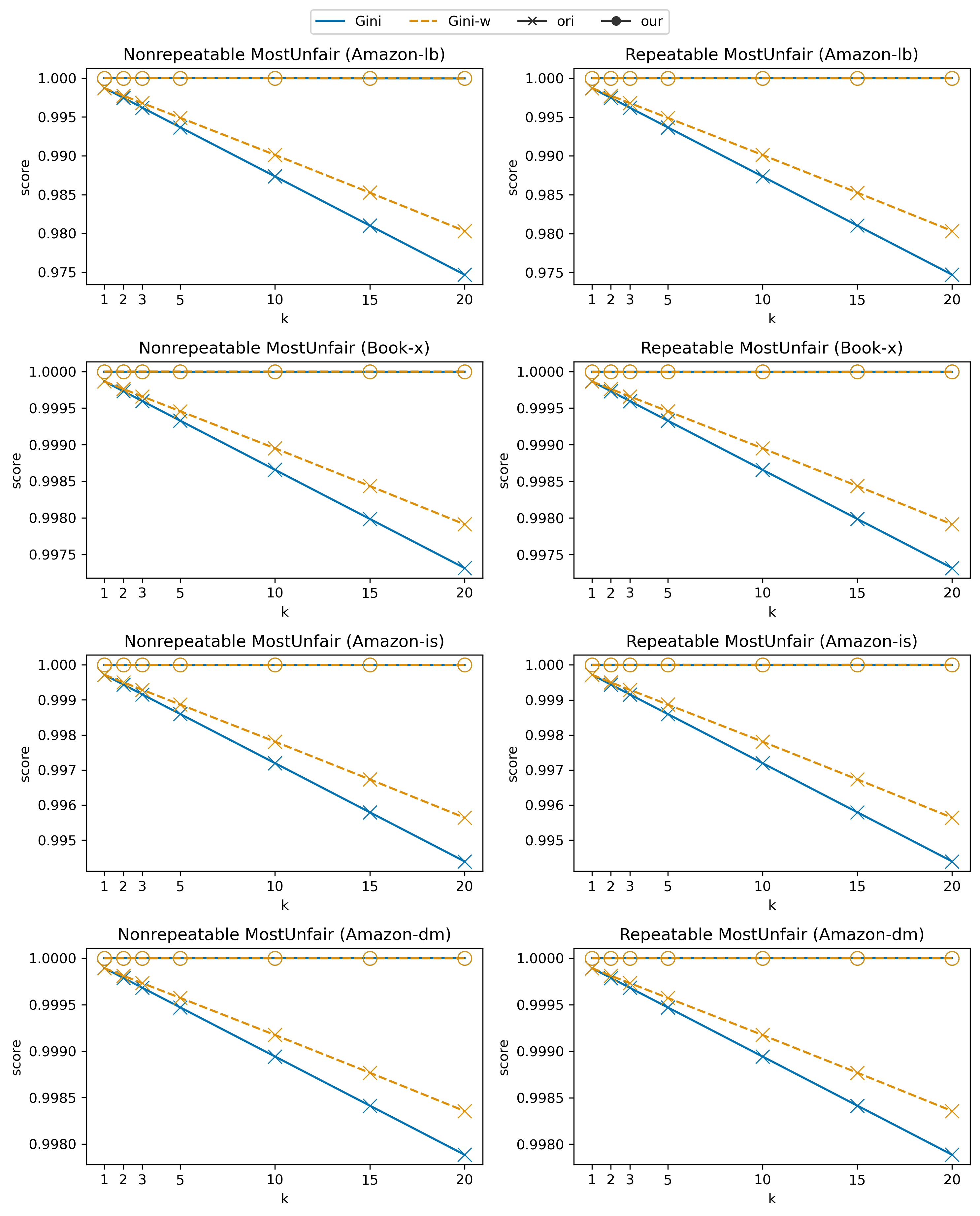}
    \caption{Most unfair scores with varying $k$ for lower-is-fairer fairness measures on Amazon-* and Book-x.}
    \label{fig:app_mostunfair_lower_better}
\end{figure}

The results of the max/min achievable fairness experiment for Amazon-* and Book-x are in Fig.~\ref{fig:app_mostfair_higher_better}--\ref{fig:app_mostunfair_lower_better}.

\subsubsection{Sliding window: relevance and fairness at different rank positions}
\label{app:sliding}
\begin{figure}[p]
    \centering
    \includegraphics[width=\textwidth]{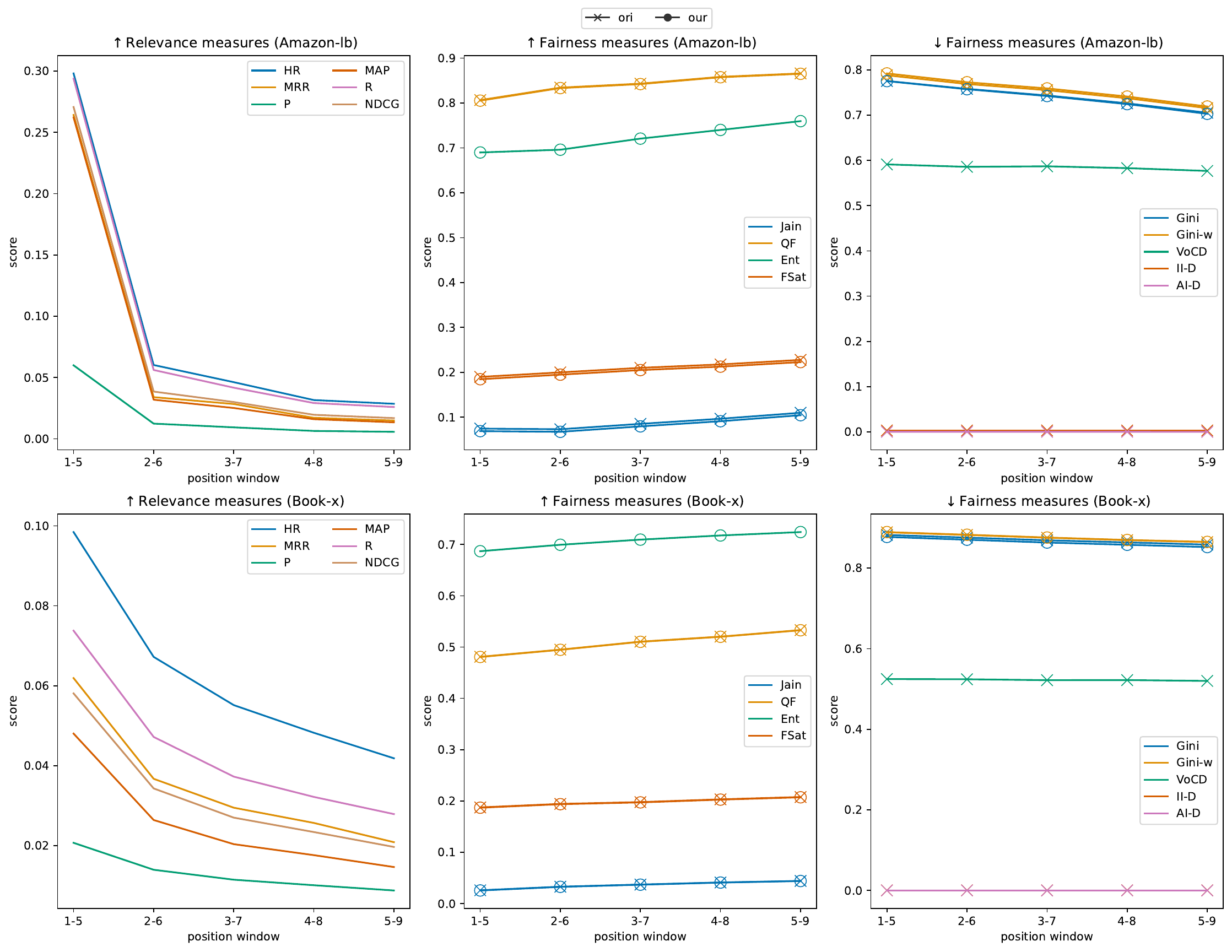}
    \caption{Sliding window evaluation for BPR model, on Amazon-lb and Book-x. Each row is for one dataset, each column is for the different groups of measures (relevance, higher-is-better fairness, lower-is-better fairness measures). II-D and AI-D lines overlap.}
    \label{fig:app_sliding_nonfairrel_amazonlb-book-x}
\end{figure}

\begin{figure}[p]
    \centering
    \includegraphics[width=\textwidth]{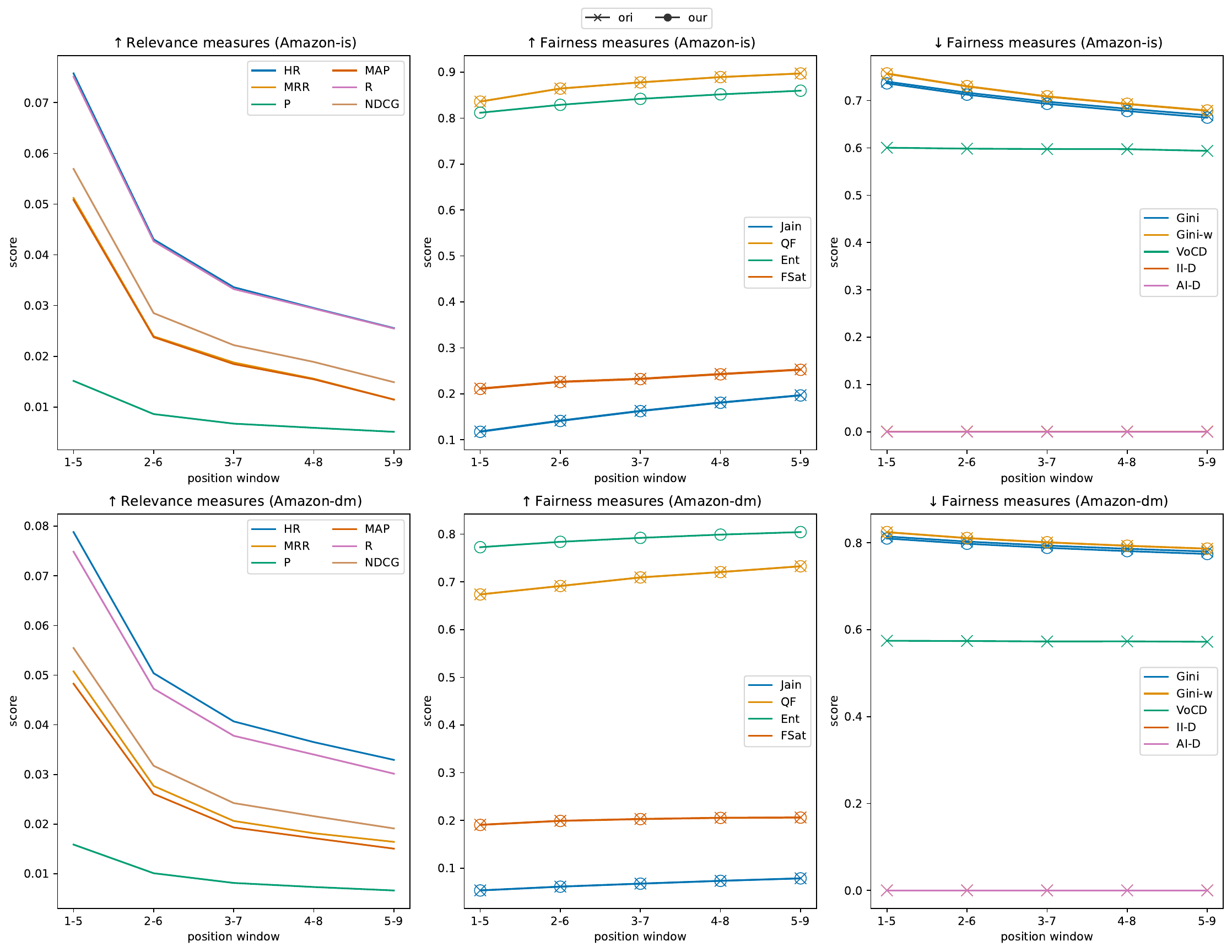}
    \caption{Sliding window evaluation for BPR model, on Amazon-is and Amazon-dm. Each row is for one dataset, each column is for the different groups of measures (relevance, higher-is-better fairness, lower-is-better fairness measures). II-D and AI-D lines overlap.}
    \label{fig:app_sliding_nonfairrel_amazonisdm}
\end{figure}

The results of the sliding window experiment for Amazon-* and Book-x are in Fig.~\ref{fig:app_sliding_nonfairrel_amazonlb-book-x}--\ref{fig:app_sliding_nonfairrel_amazonisdm}.

\subsubsection{Measure strictness and sensitivity through artificial insertion of items}
\label{app:insert}
\begin{figure}[p]
    \centering
    \includegraphics[width=\textwidth]{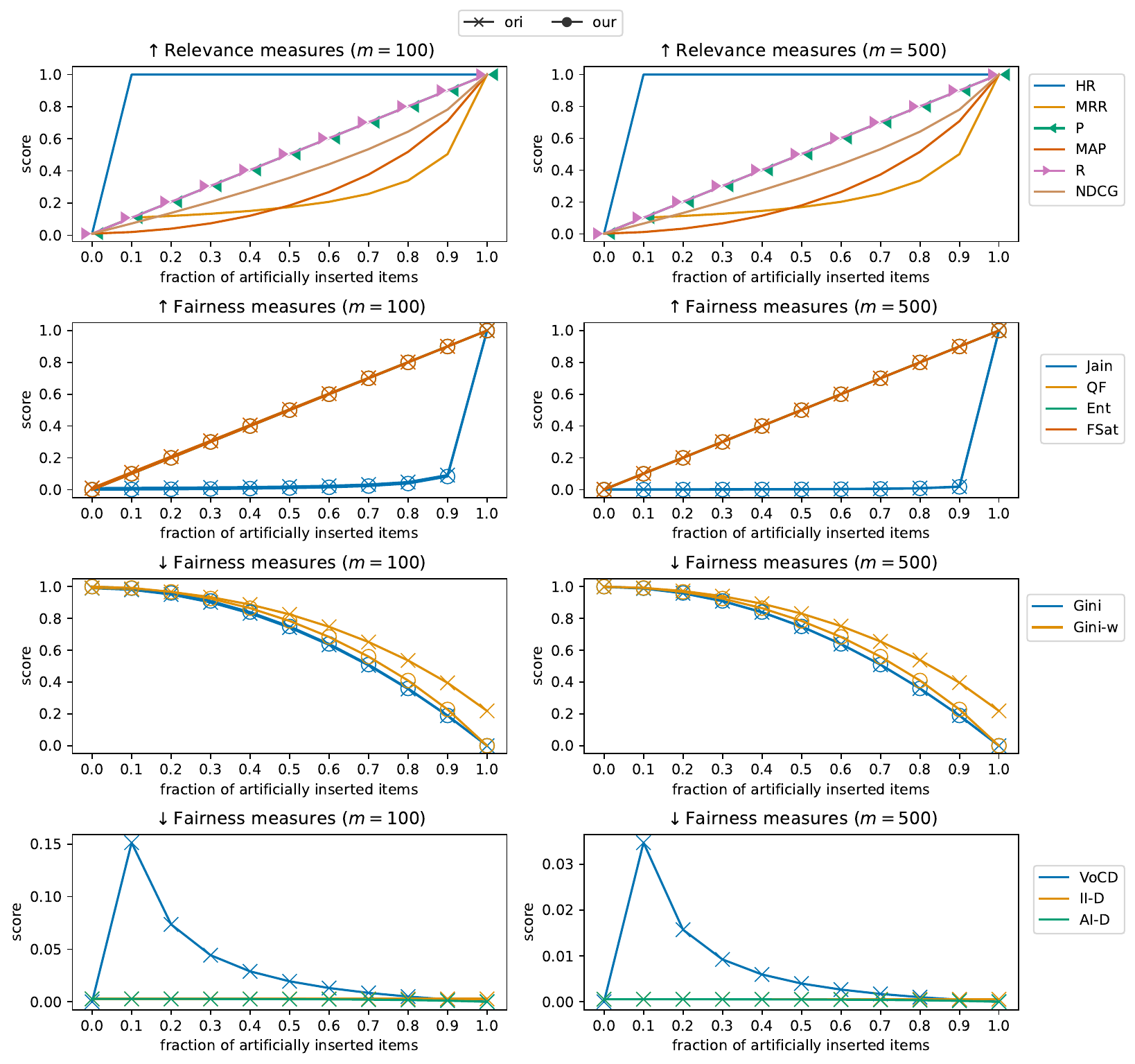}
    \caption{Results for jointly LE and relevant item insertion for $m\in\{100,500\}$. All measures are calculated at $k=10$. QF\ori~and FSat\ori~overlap. QF\our, FSat\our, and Ent\our~also overlap.}
    \label{fig:artificial-fair-extend}
\end{figure}

We present in Fig.~\ref{fig:artificial-fair-extend} the extended results of artificially inserting least exposed (LE) and relevant items for $m=\{100,500\}$ for relevance measures and fairness measures. The changes in scores are less stable compared to $m=1000$, but the general trends are the same.

\begin{figure*}[p]
    \centering
    \includegraphics[width=\textwidth]{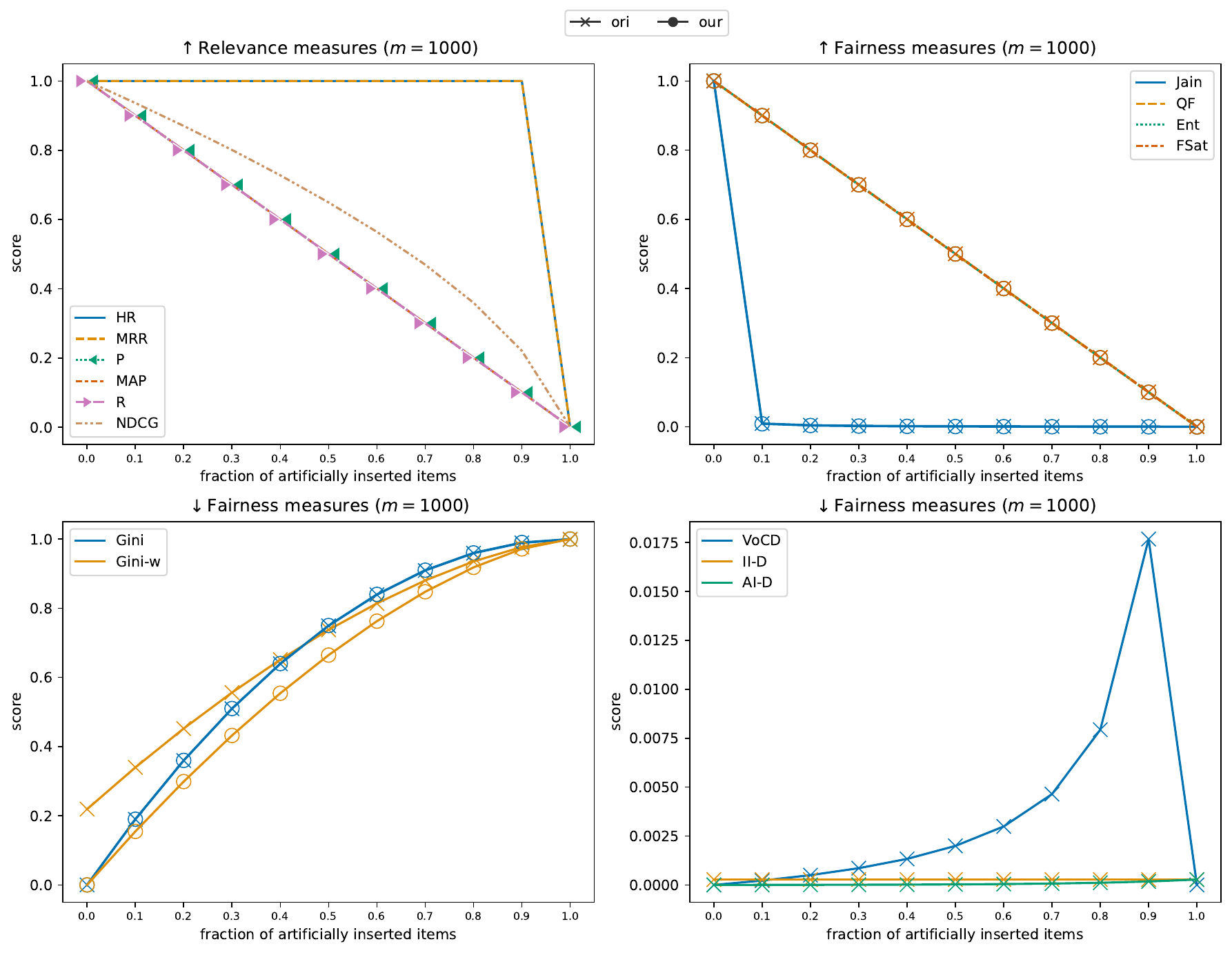}
    \caption{Results for jointly most exposed (ME) and irrelevant item insertion. All measures are calculated at $k=10$. QF\ori~and FSat\ori~overlap. QF\our, FSat\our, and Ent\our~also overlap.}
    \label{fig:artificial-unfair}
\end{figure*}

We also experiment with the artificial insertion of multiple copies of items that are already in the recommendation list and the insertion of irrelevant items, using a similar methodology. We refer to the insertion of multiple item copies as inserting the \textit{most exposed (ME) items}, as in this experiment we aim to maximise exposure of as few items as possible. This is done by iteratively inserting a copy of several items that currently have the most exposure, one copy at a time. 
We swap the starting and ending recommendation list of the artificial insertion of LE and relevant items such that at the end of the experiments, only $k$ unique items are in the recommendation list. These $k$ items will get the most exposure, while the rest of the items in the dataset get zero exposure. The item replacement is still done from the bottom of the recommendation list. 
In Fig.~\ref{fig:artificial-unfair}, we see that the trends of the measures are similar, but the opposite to that of the artificial insertion of LE and relevant items.

\begin{table*}[tb]
\caption{Relevance \textsc{(rel)} and fairness \textsc{(fair)} scores of the recommender models for Amazon-lb and Book-x. The most relevant and most fair score per measure is in bold. $\uparrow$ means the higher the better, $\downarrow$ the lower the better. `nan' stands for `not a number'.}
\label{tab:base-extra1}
\resizebox{\columnwidth}{!}{
\begin{tabular}{lllrrrrrrr}
\toprule
 &  && Pop$^{*}$  & ItemKNN & SLIM & BPR & NGCF & NeuMF & MultiVAE \\
\midrule
\multirow[c]{21}{*}{Amazon-lb} 
& \multirow[c]{6}{*}{\textsc{rel}} 
& $\uparrow$ $\text{HR}$ & 0.257908 & 0.309611 & 0.313869 & 0.320560 & 0.305961 & \bfseries 0.325426 & 0.321776 \\
& & $\uparrow$ $\text{MRR}$ & 0.222141 & 0.265047 & \bfseries 0.270739 & 0.267188 & 0.263003 & 0.267511 & 0.264885 \\
& & $\uparrow$ $\text{P}$ & 0.025912 & 0.031022 & 0.031752 & 0.032360 & 0.030718 & \bfseries 0.032968 & 0.032664 \\
& & $\uparrow$ $\text{MAP}$ & 0.221568 & 0.264694 & \bfseries 0.268219 & 0.265084 & 0.261340 & 0.263527 & 0.261339 \\
& & $\uparrow$ $\text{R}$ & 0.252251 & 0.308698 & 0.306767 & 0.315237 & 0.302616 & \bfseries 0.318375 & 0.314852 \\
& & $\uparrow$ $\text{NDCG}$ & 0.228580 & 0.275196 & \bfseries 0.278262 & 0.277727 & 0.271526 & 0.277666 & 0.275024 \\
\cline{2-10}
& \multirow[c]{15}{*}{\textsc{fair}} 
 & $\uparrow$ $\text{Jain}_{\text{ori}}$ & 0.017661 & 0.188400 & 0.050735 & 0.134711 & \bfseries 0.193167 & 0.060450 & 0.028521 \\
 && $\uparrow$ $\text{Jain}_{\text{our}}$ & 0.005085 & 0.178079 & 0.038596 & 0.123681 & \bfseries 0.182909 & 0.048439 & 0.016089 \\
 && $\uparrow$ $\text{QF}_{\text{ori}}$ & 0.024020 & \bfseries 0.978508 & 0.379267 & 0.921618 & 0.967130 & 0.790139 & 0.386852 \\
 && $\uparrow$ $\text{QF}_{\text{our}}$ & 0.011524 & \bfseries 0.978233 & 0.371319 & 0.920615 & 0.966709 & 0.787452 & 0.379001 \\
 && $\uparrow$ $\text{Ent}_{\text{ori}}$ & nan & nan & nan & nan & nan & nan & nan \\
 && $\uparrow$ $\text{Ent}_{\text{our}}$ & 0.094992 & \bfseries 0.822309 & 0.460701 & 0.731135 & 0.812621 & 0.564546 & 0.310875 \\
 && $\uparrow$ $\text{FSat}_{\text{ori}}$ & 0.022756 & 0.285714 & 0.134008 & 0.212389 & \bfseries 0.294564 & 0.158028 & 0.074589 \\
 && $\uparrow$ $\text{FSat}_{\text{our}}$ & 0.010243 & 0.276569 & 0.122919 & 0.202305 & \bfseries 0.285531 & 0.147247 & 0.062740 \\
& & $\downarrow$ $\text{Gini}_{\text{ori}}$ & 0.983735 & \bfseries 0.580725 & 0.915792 & 0.710528 & 0.606176 & 0.841397 & 0.955542 \\
& & $\downarrow$ $\text{Gini}_{\text{our}}$ & 0.996300 & \bfseries 0.584731 & 0.926915 & 0.717292 & 0.610723 & 0.850940 & 0.967509 \\
& & $\downarrow$ $\text{Gini-w}_{\text{ori}}$ & 0.986481 & \bfseries 0.605205 & 0.914661 & 0.731609 & 0.628605 & 0.861775 & 0.959218 \\
& & $\downarrow$ $\text{Gini-w}_{\text{our}}$ & 0.996317 & \bfseries 0.611240 & 0.923782 & 0.738904 & 0.634873 & 0.870368 & 0.968782 \\
& & $\downarrow$ $\text{VoCD}_{\text{ori}}$ & \bfseries 0.509398 & 0.560071 & 0.654539 & 0.617237 & 0.590049 & 0.645770 & 0.661114 \\
& & $\downarrow$ $\text{II-D}_{\text{ori}}$ & \bfseries 0.003439 & \bfseries 0.003439 & \bfseries 0.003439 & \bfseries 0.003439 & \bfseries 0.003439 & \bfseries 0.003439 & \bfseries 0.003439 \\
& & $\downarrow$ $\text{AI-D}_{\text{ori}}$ & 0.002443 & \bfseries 0.000158 & 0.000610 & 0.000280 & 0.000165 & 0.000728 & 0.001385 \\

\cline{1-10}
\multirow[c]{21}{*}{Book-x} 
& \multirow[c]{6}{*}{\textsc{rel}} 
& $\uparrow$ $\text{HR}$ & 0.034581 & 0.115269 & 0.059762 & \bfseries 0.130342 & 0.103564 & 0.085299 & 0.089910 \\
& & $\uparrow$ $\text{MRR}$ & 0.012664 & 0.063709 & 0.034535 & \bfseries 0.066125 & 0.041422 & 0.037474 & 0.039687 \\
& & $\uparrow$ $\text{P}$ & 0.003476 & 0.012999 & 0.006349 & \bfseries 0.014276 & 0.010977 & 0.008973 & 0.009576 \\
& & $\uparrow$ $\text{MAP}$ & 0.008713 & 0.047170 & 0.023412 & \bfseries 0.050778 & 0.032601 & 0.029294 & 0.031962 \\
& & $\uparrow$ $\text{R}$ & 0.022909 & 0.082607 & 0.037784 & \bfseries 0.097505 & 0.078578 & 0.064687 & 0.069501 \\
& & $\uparrow$ $\text{NDCG}$ & 0.013074 & 0.059732 & 0.029559 & \bfseries 0.065820 & 0.045847 & 0.039839 & 0.042925 \\

\cline{2-10}
& \multirow[c]{15}{*}{\textsc{fair}} 
& $\uparrow$ $\text{Jain}_{\text{ori}}$ & 0.001420 & \bfseries 0.375894 & 0.002297 & 0.036191 & 0.020391 & 0.030978 & 0.050532 \\
& & $\uparrow$ $\text{Jain}_{\text{our}}$ & 0.000080 & \bfseries 0.376670 & 0.000961 & 0.035046 & 0.019157 & 0.029804 & 0.049469 \\
& & $\uparrow$ $\text{QF}_{\text{ori}}$ & 0.002414 & \bfseries 0.899799 & 0.019584 & 0.671630 & 0.662777 & 0.519920 & 0.587928 \\
& & $\uparrow$ $\text{QF}_{\text{our}}$ & 0.001075 & \bfseries 0.899664 & 0.018267 & 0.671189 & 0.662324 & 0.519275 & 0.587374 \\
& & $\uparrow$ $\text{Ent}_{\text{ori}}$ & nan & nan & nan & nan & nan & nan & nan \\
& & $\uparrow$ $\text{Ent}_{\text{our}}$ & 0.014182 & \bfseries 0.916559 & 0.156894 & 0.688733 & 0.623813 & 0.646908 & 0.716425 \\
& & $\uparrow$ $\text{FSat}_{\text{ori}}$ & 0.002012 & \bfseries 0.392622 & 0.017170 & 0.158820 & 0.120724 & 0.145942 & 0.188062 \\
& & $\uparrow$ $\text{FSat}_{\text{our}}$ & 0.000672 & \bfseries 0.391807 & 0.015850 & 0.157690 & 0.119543 & 0.144795 & 0.186971 \\

& & $\downarrow$ $\text{Gini}_{\text{ori}}$ & 0.998615 & \bfseries 0.548876 & 0.996714 & 0.853353 & 0.877575 & 0.894874 & 0.848862 \\
& & $\downarrow$ $\text{Gini}_{\text{our}}$ & 0.999955 & \bfseries 0.534458 & 0.997988 & 0.849604 & 0.874675 & 0.892579 & 0.844955 \\
& & $\downarrow$ $\text{Gini-w}_{\text{ori}}$ & 0.998906 & \bfseries 0.582320 & 0.996386 & 0.863144 & 0.889082 & 0.904719 & 0.860833 \\
& & $\downarrow$ $\text{Gini-w}_{\text{our}}$ & 0.999954 & \bfseries 0.582930 & 0.997431 & 0.864049 & 0.890015 & 0.905668 & 0.861736 \\
& & $\downarrow$ $\text{VoCD}_{\text{ori}}$ & 0.663365 & 0.564473 & 0.720722 & 0.570963 & \bfseries 0.537809 & 0.600639 & 0.611070 \\
& & $\downarrow$ $\text{II-D}_{\text{ori}}$ & \bfseries 0.000368 & \bfseries 0.000368 & \bfseries 0.000368 & \bfseries 0.000368 & \bfseries 0.000368 & \bfseries 0.000368 & \bfseries 0.000368 \\
& & $\downarrow$ $\text{AI-D}_{\text{ori}}$ & 0.000350 & \bfseries 0.000001 & 0.000176 & 0.000013 & 0.000027 & 0.000015 & 0.000009 \\

\bottomrule
\multicolumn{10}{l}{\multirow{3}{*}{
\parbox[t]{1.4\textwidth}{*The scores of our \textsc{fair} measures for Pop are not 0 or 1, because in our experiment set-up, items from users' train or validation splits are excluded from the top $k$ recommendation list.}}}
\end{tabular}}
\end{table*}

\begin{table*}[tb]
\caption{Relevance \textsc{(rel)} and fairness \textsc{(fair)} scores of the recommender models for Amazon-is and Amazon-dm. The most relevant and most fair score per measure is in bold. $\uparrow$ means the higher the better, $\downarrow$ the lower the better. `nan' stands for `not a number'.}
\label{tab:base-extra2}
\resizebox{\columnwidth}{!}{
\begin{tabular}{lllrrrrrrr}
\toprule
 &  && Pop$^{*}$  & ItemKNN & SLIM & BPR & NGCF & NeuMF & MultiVAE \\
\midrule
\multirow[c]{21}{*}{Amazon-is} 
& \multirow[c]{6}{*}{\textsc{rel}} 
 & $\uparrow$ $\text{HR}$ & 0.031792 & 0.084271 & 0.028902 & \bfseries 0.100548 & 0.095224 & 0.077122 & 0.081838 \\
& & $\uparrow$ $\text{MRR}$ & 0.011671 & 0.047550 & 0.023075 & \bfseries 0.054403 & 0.048366 & 0.042206 & 0.041294 \\
& & $\uparrow$ $\text{P}$ & 0.003179 & 0.008427 & 0.002890 & \bfseries 0.010055 & 0.009522 & 0.007712 & 0.008184 \\
& & $\uparrow$ $\text{MAP}$ & 0.011593 & 0.047313 & 0.022880 & \bfseries 0.054002 & 0.048093 & 0.041905 & 0.041036 \\
& & $\uparrow$ $\text{R}$ & 0.031462 & 0.083916 & 0.028572 & \bfseries 0.099825 & 0.094717 & 0.076387 & 0.080993 \\
& & $\uparrow$ $\text{NDCG}$ & 0.016224 & 0.055902 & 0.024254 & \bfseries 0.064820 & 0.059064 & 0.050052 & 0.050436 \\

\cline{2-10}
& \multirow[c]{15}{*}{\textsc{fair}} 
 & $\uparrow$ $\text{Jain}_{\text{ori}}$ & 0.002925 & \bfseries 0.493178 & 0.002958 & 0.167723 & 0.101140 & 0.061059 & 0.093401 \\
& & $\uparrow$ $\text{Jain}_{\text{our}}$ & 0.000124 & \bfseries 0.492108 & 0.000156 & 0.165503 & 0.098685 & 0.058463 & 0.090919 \\
& & $\uparrow$ $\text{QF}_{\text{ori}}$ & 0.004483 & \bfseries 0.989633 & 0.014570 & 0.956851 & 0.869151 & 0.919866 & 0.856823 \\
& & $\uparrow$ $\text{QF}_{\text{our}}$ & 0.001686 & \bfseries 0.989604 & 0.011801 & 0.956729 & 0.868783 & 0.919640 & 0.856420 \\
& & $\uparrow$ $\text{Ent}_{\text{ori}}$ & nan & nan & nan & nan & nan & nan & nan \\
& & $\uparrow$ $\text{Ent}_{\text{our}}$ & 0.012261 & \bfseries 0.939484 & 0.024132 & 0.832303 & 0.746126 & 0.741374 & 0.766085 \\
& & $\uparrow$ $\text{FSat}_{\text{ori}}$ & 0.003642 & \bfseries 0.373494 & 0.008686 & 0.223032 & 0.171477 & 0.170636 & 0.204819 \\
& & $\uparrow$ $\text{FSat}_{\text{our}}$ & 0.000843 & \bfseries 0.371734 & 0.005901 & 0.220849 & 0.169149 & 0.168306 & 0.202585 \\
& & $\downarrow$ $\text{Gini}_{\text{ori}}$ & 0.997136 & \bfseries 0.445874 & 0.997066 & 0.679338 & 0.798275 & 0.763940 & 0.769841 \\
& & $\downarrow$ $\text{Gini}_{\text{our}}$ & 0.999937 & \bfseries 0.439697 & 0.999866 & 0.676963 & 0.797837 & 0.762944 & 0.768941 \\
& & $\downarrow$ $\text{Gini-w}_{\text{ori}}$ & 0.997738 & \bfseries 0.476640 & 0.997518 & 0.703146 & 0.818197 & 0.784878 & 0.784863 \\
& & $\downarrow$ $\text{Gini-w}_{\text{our}}$ & 0.999926 & \bfseries 0.477685 & 0.999705 & 0.704688 & 0.819991 & 0.786599 & 0.786584 \\
& & $\downarrow$ $\text{VoCD}_{\text{ori}}$ & 0.592760 & \bfseries 0.523208 & 0.787225 & 0.615085 & 0.640122 & 0.616732 & 0.648353 \\
& & $\downarrow$ $\text{II-D}_{\text{ori}}$ & \bfseries 0.000768 & \bfseries 0.000768 & \bfseries 0.000768 & \bfseries 0.000768 & \bfseries 0.000768 & \bfseries 0.000768 & \bfseries 0.000768 \\
& & $\downarrow$ $\text{AI-D}_{\text{ori}}$ & 0.000738 & \bfseries 0.000002 & 0.000705 & 0.000011 & 0.000019 & 0.000036 & 0.000020 \\
\cline{1-10}
\multirow[c]{21}{*}{Amazon-dm} 
& \multirow[c]{6}{*}{\textsc{rel}} 
& $\uparrow$ $\text{HR}$ & 0.022809 & 0.087660 & 0.005702 & \bfseries 0.108596 & 0.093787 & 0.073872 & 0.079660 \\
& & $\uparrow$ $\text{MRR}$ & 0.009252 & 0.048030 & 0.004607 & \bfseries 0.054654 & 0.043576 & 0.033947 & 0.036754 \\
& & $\uparrow$ $\text{P}$ & 0.002289 & 0.008928 & 0.000570 & \bfseries 0.011004 & 0.009489 & 0.007489 & 0.008051 \\
& & $\uparrow$ $\text{MAP}$ & 0.008463 & 0.044812 & 0.004106 & \bfseries 0.051930 & 0.041533 & 0.032263 & 0.034956 \\
& & $\uparrow$ $\text{R}$ & 0.020528 & 0.081430 & 0.005123 & \bfseries 0.102600 & 0.089059 & 0.069847 & 0.075436 \\
& & $\uparrow$ $\text{NDCG}$ & 0.011454 & 0.054259 & 0.004447 & \bfseries 0.064525 & 0.053219 & 0.041538 & 0.044960 \\
\cline{2-10}
& \multirow[c]{15}{*}{\textsc{fair}} 
 & $\uparrow$ $\text{Jain}_{\text{ori}}$ & 0.001092 & \bfseries 0.358718 & 0.001080 & 0.068624 & 0.078606 & 0.038114 & 0.121317 \\
& & $\uparrow$ $\text{Jain}_{\text{our}}$ & 0.000036 & \bfseries 0.358605 & 0.000023 & 0.067745 & 0.077754 & 0.037155 & 0.120578 \\
& & $\uparrow$ $\text{QF}_{\text{ori}}$ & 0.001902 & \bfseries 0.957620 & 0.003593 & 0.843585 & 0.720672 & 0.843691 & 0.879624 \\
& & $\uparrow$ $\text{QF}_{\text{our}}$ & 0.000846 & \bfseries 0.957575 & 0.002539 & 0.843419 & 0.720377 & 0.843525 & 0.879496 \\
& & $\uparrow$ $\text{Ent}_{\text{ori}}$ & nan & nan & nan & nan & nan & nan & nan \\
& & $\uparrow$ $\text{Ent}_{\text{our}}$ & 0.008686 & \bfseries 0.940649 & 0.008735 & 0.779117 & 0.768888 & 0.764866 & 0.840448 \\
& & $\uparrow$ $\text{FSat}_{\text{ori}}$ & 0.001480 & \bfseries 0.404143 & 0.002219 & 0.187804 & 0.199535 & 0.197738 & 0.246354 \\
& & $\uparrow$ $\text{FSat}_{\text{our}}$ & 0.000423 & \bfseries 0.403512 & 0.001164 & 0.186945 & 0.198688 & 0.196890 & 0.245556 \\

& & $\downarrow$ $\text{Gini}_{\text{ori}}$ & 0.998925 & \bfseries 0.462579 & 0.998932 & 0.778288 & 0.814128 & 0.771164 & 0.702994 \\
& & $\downarrow$ $\text{Gini}_{\text{our}}$ & 0.999982 & \bfseries 0.452327 & 0.999989 & 0.774693 & 0.811289 & 0.767418 & 0.697811 \\
& & $\downarrow$ $\text{Gini-w}_{\text{ori}}$ & 0.999154 & \bfseries 0.493364 & 0.999132 & 0.791188 & 0.828961 & 0.788203 & 0.718667 \\
& & $\downarrow$ $\text{Gini-w}_{\text{our}}$ & 0.999979 & \bfseries 0.493772 & 0.999957 & 0.791841 & 0.829646 & 0.788854 & 0.719261 \\
& & $\downarrow$ $\text{VoCD}_{\text{ori}}$ & 0.668615 & \bfseries 0.532741 & 0.796493 & 0.615043 & 0.643785 & 0.611526 & 0.616569 \\
& & $\downarrow$ $\text{II-D}_{\text{ori}}$ & \bfseries 0.000290 & \bfseries 0.000290 & \bfseries 0.000290 & \bfseries 0.000290 & \bfseries 0.000290 & \bfseries 0.000290 & \bfseries 0.000290 \\
& & $\downarrow$ $\text{AI-D}_{\text{ori}}$ & 0.000282 & \bfseries 0.000000 & 0.000278 & 0.000004 & 0.000004 & 0.000008 & 0.000002 \\

\bottomrule
\multicolumn{10}{l}{\multirow{3}{*}{
\parbox[t]{1.4\textwidth}{*The scores of our \textsc{fair} measures for Pop are not 0 or 1, because in our experiment set-up, items from users' train or validation splits are excluded from the top $k$ recommendation list.}}}
\end{tabular}%
}
\end{table*}



\chapter{Can We Trust Recommender System Fairness Evaluation? The Role of Fairness and Relevance}
\label{chap:SIGIR24}

\section*{Abstract}
Relevance and fairness are two major objectives of recommender systems (RSs). 
Recent work proposes measures of RS fairness that are either independent from relevance (fairness-only) or conditioned on relevance (joint measures). 
While fairness-only measures have been studied extensively, we look into whether joint measures can be trusted. 
We collect all joint evaluation measures of RS relevance and fairness, and ask: How much do they agree with each other? To what extent do they agree with relevance/fairness measures? How sensitive are they to changes in rank position, or to increasingly fair and relevant recommendations? 
We empirically study for the first time the behaviour of these measures across 4 real-world datasets and 4 recommenders. We find that most of these measures: 
i) correlate weakly with one another and even contradict each other at times; 
ii) are less sensitive to rank position changes than relevance- and fairness-only measures, meaning that they are less granular than traditional RS measures; and 
iii) tend to compress scores at the low end of their range, meaning that they are not very expressive. 
We counter the above limitations with a set of guidelines on the appropriate usage of such measures, i.e., they should be used with caution due to their tendency to contradict each other and of having a very small empirical range. 

\section{Introduction}
Recent increased focus on fairness in recommender systems (RSs) has led to studies on how to evaluate different notions of fairness in RS. A recent survey \cite{Wang2023ASystems} shows that prior work on fairness evaluation in RS mainly focuses on group fairness (e.g., \cite{Raj2022MeasuringResults,Amigo2023ASystems,Zehlike2022FairnessSystems}), but less so on individual fairness. 
Individual fairness is commonly understood as treating similar individuals similarly \cite{Dwork2012FairnessAwareness}. 
Unlike group fairness evaluation, evaluating individual fairness does not require information on sensitive attributes (e.g., gender, age) to identify protected groups \cite{Lazovich2022MeasuringMetrics}. Such information is often unavailable due to privacy and legal issues. Further, intersectionality between different group characteristics complicates group fairness \cite{Crenshaw1991MappingColor,Ekstrand2022FairnessSystems}. Individual fairness is known to lead into group fairness, but not vice versa \cite{Bower2021IndividuallyRanking}. 
Overall, individual fairness gives a broader view by assessing distribution across all individuals in the population \cite{Lazovich2022MeasuringMetrics}. For all these reasons, we focus on individual fairness, particularly individual item fairness, which is typically broadly defined w.r.t.~exposure received by items, i.e., how uniform the exposure distribution between items is \cite{Rampisela2024EvaluationStudy}. Yet, fairness beyond exposure also matters, i.e., the exposure should be proportional to item relevance \cite{Patro2022FairDirections,Smith2023ScopingPerspective,Biega2018EquityRankings}. 

Individual item fairness is measured by measures that (i) are detached from relevance (fairness-only measures, defined by exposure); or (ii) are conditioned on relevance (joint measures considering exposure w.r.t.~relevance). Measures of type (i) have been extensively analysed \cite{Rampisela2024EvaluationStudy}, but to our knowledge, this is not the case for measures of type (ii). The growing number of measures of type (ii) necessitates a thorough look into their usage in RS evaluation. 

We present a comprehensive study into the empirical properties of all joint measures of individual item fairness and relevance, motivated by the question of how much can we practically trust these measures, particularly: 
\textbf{RQ1.} To what extent do the joint measures agree with existing relevance- and fairness-only measures? 
\textbf{RQ2.} To what extent do the joint measures agree with each other?
\textbf{RQ3.} How sensitive are the joint measures across decreasing rank positions? and 
\textbf{RQ4.} How sensitive are the joint measures given increasingly fair and relevant recommendations?

We identify some alarming limitations in the measures, and we reflect on their best usage in practice. This is the first in-depth study on individual item fairness measures that consider relevance in RS.

\section{Individual Item Fairness \& Relevance}
\label{sigir_s:priorwork}

We present the notation ($\S$\ref{sigir_ss:notation}) and all existing joint evaluation measures of individual item fairness and relevance ($\S$\ref{sigir_ss:fairrel}). 

\subsection{Notation and Definitions}
\label{sigir_ss:notation}

Given a set of $n$ items, $I = \{i_1, i_2, \dots, i_n\}$, and a set of $m$ users, $U=\{u_1, u_2, \dots, u_m\}$, an ordered 
list of the $n$ items is created for each $u \in U$. This list is created in each recommendation round $w$, where $w \in \{1, 2, \dots, W\}$; a round means an occurrence in which a user receives a list of recommendations. 
If an item $i$ is \emph{relevant} to user $u$, we write $r_{u,i}=1$, otherwise $r_{u,i}=0$. Relevance can also be denoted as real values, $r_{u,i}\in[0,1]$. The list of user $u$'s top $k$ recommended items in round $w$ is $L_{u, w}$ and the rank position of item $i$ in user $u$'s recommendation list is $z(u,i,w)$. 
For cases with only one recommendation round, user $u$'s list of top $k$ recommended items is $L_u$ and the rank position of item $i$ for user $u$ is $z(u,i)$. 

While several different definitions of fairness exist, the definitions commonly used in prior work on individual item fairness are closely linked to item exposure 
\cite{Rampisela2024EvaluationStudy,Amigo2023ASystems,Zehlike2022FairnessSystems}. An item is \emph{exposed} when it is recommended at the top $k$ to a user. 
The probability of a user seeing an item exposed to them can be modelled using various \emph{examination functions}, $e(\cdot)$. Examination functions typically assume that the viewing probability depends only on the position $z(u,i,w)$ or $z(u,i)$. This is a common choice across all measures in $\S$\ref{sigir_ss:fairrel}. 

The examination functions used in this work are shown in Tab.~\ref{sigir_tab:exp-weigh}: the linear examination function, $e_{\text{li}}$ and its min-max normalised version $\tilde{e}_{\text{li}}$ apply a linear discount to each rank position up to $k$ \cite{Borges2019EnhancingAutoencoders}.
Meanwhile, discounts based on Discounted Cumulative Gain (DCG)~\cite{Jarvelin2002CumulatedTechniques} and Rank-Biased Precision (RBP)~\cite{Moffat2008Rank-biasedEffectiveness} are used in $e_{\text{DCG}}$  \cite{Singh2019PolicyRanking,Oosterhuis2021ComputationallyFairness} 
and $e_{\text{RBP}}$ \cite{Wu2022JointRecommendation,Jeunen2021Top-KExposure} respectively. The parameter $\gamma$ in $e_{\text{RBP}}$ is the user's patience, i.e., the probability of the user examining the next ranked item. The user patience parameter is commonly set at 
e.g., $\gamma \in \{0.8, 0.9\}$  \cite{Wu2022JointRecommendation, Jeunen2021Top-KExposure}.
In the inverse examination function $e_{\text{inv}}$, the inverse of the rank position is used as a discount factor \cite{Saito2022FairRanking}. Overall, we use three types of examination functions (linear, discounted, and inverse), which assume that item exposure diminishes with decreasing ranking either linearly, or with an increasing penalty that is either proportional to the rank decrease or the inverse of the rank. Generally, the most punishing is the inverse function, and the least punishing is the DCG-based discount function.

\begin{table}[tb]
\caption{Examination functions used in this work. $\tilde{e}$ is the min-max normalised examination function.}
\label{sigir_tab:exp-weigh}
\resizebox{\columnwidth}{!}{
\begin{tabular}{llll}
\toprule
         & equation & measure & reference                        \\ 
         \midrule
linear   & \parbox[t]{5cm}{$e_{\text{li}}(u, i, w) = k+1-z(u,i,w)$ \\
    $\tilde{e}_{\text{li}}(u, i, w) = \frac{ e_{\text{li}}(u, i, w)-1}{k-1} = \frac{k-z(u,i,w)}{k-1}$}                                                   & IAA   &  \cite{Borges2019EnhancingAutoencoders}                        \\
DCG      & $e_{\text{DCG}}(u, i, w) = 1/\log_2 (z(u,i,w)+1)$         & IFD & \cite{Singh2019PolicyRanking, Oosterhuis2021ComputationallyFairness}   \\
RBP      & $e_{\text{RBP}}(u,i,w) = \gamma^{z(u,i,w)-1}$         & HD, II-F, AI-F   & \cite{Wu2022JointRecommendation, Jeunen2021Top-KExposure}                 \\
inverse$^*$  & $e_{\text{inv}}(p) = 1/p$                                & MME, IBO/IWO & \cite{Saito2022FairRanking}        \\
\bottomrule
\end{tabular}
}
\\
{\footnotesize $^*$In the published version, the inverse examination function has been written as $e_{\text{inv}}(u,i) = 1/z(u,i)$. The correct formulation of the function depends directly on the recommendation slot at position $p$.}
\end{table}

\subsection{Joint Measures of Fairness and Relevance}
\label{sigir_ss:fairrel}

We present measures that evaluate fairness considering relevance (\textsc{Fair+Rel} or joint measures henceforth). 
To our knowledge, we include all \textsc{Fair+Rel} measures for RSs published up to October 2023. Each measure uses an exposure function, which is linked to the fairness of item distribution in the recommendation and, therefore, measures item fairness jointly with relevance. 
We use \up~for measures where the higher the score, the fairest the recommendation, and vice versa for \down. 
All measures--except HD--are defined 
for multiple recommendation rounds or stochastic rankings, where a distribution over rankings is considered \cite{Biega2018EquityRankings}.

\subsubsection{Inequity of Amortized Attention (IAA) \cite{Biega2018EquityRankings}}
\label{sigir_ss:iaa-ori} 
\down IAA\footnote{This measure is called IAA in \cite{Raj2022MeasuringResults} and L1-norm in \cite{Wang2023ASystems}.} measures fairness as the aggregated difference between item exposure and its relevance in a series of rankings that have been generated by a stochastic process \cite{Biega2018EquityRankings}. 
The intuition behind IAA is that for a sequence of rankings to be fair, 
items should be allocated exposure proportional to their relevance to the user. 
The item position is a proxy of its exposure level. 
IAA was modified in~\cite{Borges2019EnhancingAutoencoders} to account for multiple recommendation rounds (stochastic rankings): 
\begin{equation}
    \label{sigir_eq:iaa-ori-rel}
    \text{IAA} = \frac{1}{m}\sum\limits_{u \in U}\text{IAA}(u) 
    \end{equation}%
    \begin{equation}
    \label{sigir_eq:iaa-ori-u}
    \text{IAA}(u) = \frac{1}{n}\frac{1}{W}
        \sum_{i \in I}
        \left|
        \sum_{w=1}^W 1_{L_{u,w}}(i) \cdot \tilde{e_{.}}(u,i,w)
        -  
        \tilde{r}(u,i,w) 
        \right|
\end{equation}

In~\cite{Borges2019EnhancingAutoencoders}, 
 $\tilde{e}_{(\cdot)}(u,i,w)$ is the min-max normalised linear examination function $\tilde{e}_{\text{li}}(u, i,\allowbreak  w)$ (see Tab.~\ref{sigir_tab:exp-weigh}) 
and $\tilde{r}(u,i,w) \in [0,1]$ is the min-max normalised relevance 
value of item $i$ for user $u$ in round $w$, $r_{u,i,w}$.\footnote{Note that the normalised exposure value for a recommended item at $k$ is zero.}
Both the min and max relevance values are taken from the values associated with all items for each user per round, i.e., $\min_{i \in I} r_{u,i,w}$. 
This value is the aggregated relevance over all items for user $u$ in round $w$; the higher the relevance, the closer to 1. The higher the relevance value differs from item exposure, the more unfair. The range of IAA is $[0,1]$. 

\subsubsection{Individual Fairness Disparity (IFD) \cite{Singh2019PolicyRanking,Oosterhuis2021ComputationallyFairness}} \down IFD is the average pairwise difference of the combined value of item exposure and item merit. Merit is defined as a function of relevance.\footnote{We use the item relevance value as the item merit, as per \cite{Singh2019PolicyRanking}.} 
Similar to IAA, IFD follows the principle of allocating exposure to an item based on its relevance. 
While IAA computes the difference between the exposure and relevance of each item, IFD computes the disparity of exposure allocation between item pairs. 
Based on how exposure and merit are combined, two variations of IFD exist: IFD$_{\div}$, where item exposure is divided by item relevance \cite{Singh2019PolicyRanking}, 
and IFD$_{\times}$, where the division is replaced by multiplication \cite{Oosterhuis2021ComputationallyFairness}. The term IFD$_{(\cdot)}$ or IFD refers to the measure in general. 
The two versions slightly differ in the pairwise difference computation, the formation of set of item pairs, and the exposure weighting scheme.\footnote{Exposure is weighed proportional to $e_{\text{DCG}}$ in \cite{Singh2019PolicyRanking}; to simplify, we use $e_{\text{DCG}}$ directly.} 
Both IFD versions have been used to measure fairness in ranking \cite{Singh2019PolicyRanking, Oosterhuis2021ComputationallyFairness, Yang2023FARA:Optimization, Yang2023Marginal-Certainty-AwareAlgorithm}. 
\begin{equation}
\label{sigir_eq:IFD}
\text{IFD}_{(\cdot)} = \frac{1}{m} \sum\limits_{u \in U}\text{IFD}_{(\cdot)}(u) 
\end{equation}
\begin{equation}
\label{sigir_eq:IFD-u-div}
\text{IFD}_{\div}(u) = 
\frac{1}{|H_u|} \sum_{(i,i')\in H_u} \max{ \left\{0, 
J_{\div}(u,i) - J_{\div}(u,i')
\right\}
} 
\end{equation}
\begin{equation}
\label{sigir_eq:IFD-u-mult}
\text{IFD}_{\times}(u) = 
\frac{1}{n(n-1)} \sum_{i \in I} \sum_{i' \in I\setminus{i}}
\left[
    J_{\times}(u,i) - J_{\times}(u,i')
\right]^2
\end{equation}
\begin{equation}
    J_{\div}(u,i) = \frac{1}{W} \sum\limits_{w=1}^{W} e_{\text{DCG}}(u,i,w)/r_{u,i,w}
\end{equation}
\begin{equation}\label{sigir_eq:ifd_j_x}
    J_{\times}(u,i) = \frac{1}{W} \sum\limits_{w=1}^{W} 
    r_{u,i,w} \cdot 1_{L_{u,w}}(i) \cdot e_{\text{DCG}}(u,i,w) 
\end{equation}
 $J_{(\cdot)}(u,i)$ is the function combining the expected exposure and relevance of item $i$ for user $u$ and $H_u = \{(i,i') \in I\ |\ r_{u,i} \geq r_{u,i'} > 0\}$.
The range of IFD$_{\div}$ is $[0,\infty)$ and it is 0 only when the exposure received by each relevant item is exactly proportional to its relevance \cite{Singh2019PolicyRanking}. The range of IFD$_{\times}$ is $[0,\infty)$ based on empirical results \cite{Yang2023FARA:Optimization}. 

\subsubsection{Hellinger Distance (HD) \cite{Jeunen2021Top-KExposure}}  
\down HD has been used as a measure of individual item fairness in top $k$ contextual bandits, by quantifying the difference between the relevance- and click-distributions of the top $k$ items sorted according to (ground truth) relevance \cite{Jeunen2021Top-KExposure}. The click probability is based on user patience, system-allocated item exposure, and item relevance.
A recommendation is fair based on HD when the click probability of an item is proportional to the relevance probability of that item. To compute the relevance and click distributions, a list of top $k$ items is created for each user by sorting items based on their (ground truth) relevance; this list is the reference list used in the next step. Another list of items is created based on system prediction and used to get the click probability. For each item in the reference list, we compute its click probability based on its order in the second list. Next, the relevance probabilities of items at the same position in the reference list are aggregated across users and similarly for the click probabilities. For each rank position, two aggregated values are obtained: relevance and click. The aggregated values are the inputs to the distance metric (Eq.~\eqref{sigir_eq:HD}).
\begin{equation}\label{sigir_eq:HD}
    \text{HD} = \frac{1}{\sqrt{2}} \sqrt{
     \sum_{p=1}^k \left(\sqrt{q_p'} - \sqrt{c_p'}\right)^2}
\end{equation}
\begin{equation}
    q_p' = \frac{1}{m} \sum_{u \in U} \sum_{i \in I} 
    \delta\left(z^*(u,i) = p\right) \cdot r'_{u,i}
\end{equation}
\begin{equation}
     c_p' = \frac{1}{m} \sum_{u \in U} \frac{c^*_{u,p}}{\sum_{\ell=1}^{k}c^*_{u,\ell}}
\end{equation}
\begin{equation}
    c^*_{u,p} = \sum_{i \in I} 
    \delta\left(z^*(u,i) = p\right) \cdot c^{full}_{u,i}
\end{equation}
\begin{equation}
    c^{full}_{u,i} = c'_{u,p}\ \text{if } \exists p : z(u,i)=p\, \text{, otherwise } 0
\end{equation}
\begin{equation}
    c_{u,p} = \sum_{i \in L_u} 
    \delta\left(z(u,i) = p\right) \cdot r_{u,i} \cdot \gamma \ e_{\text{RBP}}(u,i) \cdot s_{u,p}
\end{equation}
\begin{equation}
    s_{u,p} = \prod_{1 \le j < p} 1 - \sum_{i \in L_u} \delta(z(u,i) = j) \cdot r_{u,i} 
\end{equation}

\noindent where $q_p'$ and $c_p'$ are the normalised relevance and click probability of the item at position $j$ respectively, where click depends on both relevance and exposure. The position of item $i$ based on ground-truth relevance is $z^*(u,i)$. The click probability of user $u$ for item at position $p$, $c_{u,p}$ depends on $s_{u,p}$, the probability that items before position $p$ were irrelevant to the user, and the user patience $\gamma \ e_{\text{RBP}}(u,i)$. 
$r'_{u,i} = r_{u,i}/\sum_{i \in I} r_{u,i}$
is the user-wise normalised relevance value of item $i$ to user $u$, and 
$c'_{u,p} = c_{u,p}/\sum^{k}_{p=1} c_{u,p}$ 
is the user-wise normalised click probability. The value of $\delta(\cdot) = 1$ when the expression $\cdot$ is True and 0 otherwise. HD ranges between $[0,\infty)$. 

\subsubsection{Mean Max Envy (MME) \cite{Saito2022FairRanking}}\label{sigir_sss:mme} 
\down MME uses the concept of envy-freeness, where a recommendation is fair when each item is not disadvantaged by its own exposure allocation compared to being allocated the exposure of any other item. In other words, MME computes unfairness as the disadvantage suffered by the item, if the exposure allocation of an item is swapped with another item. The disadvantage is computed based on an impact score that uses exposure and relevance: given full recommendation lists (size $n$) across all users, 
we swap each item $i$ with another item $i'$ and compute the impact score before and after the swap for all rank positions and users. If the score of the item $i$ before the swap is greater or equal to its score after the swap, we have envy-freeness for item $i$ w.r.t.~item $i'$.
MME thus computes the average maximum difference of impact 
imposed if item $i$ is replaced with another item $i'$. 
E.g., let $L_{u_1} = [i_1, i_2],\  L_{u_2} = [i_1, i_3]$ and let us swap item $i_3$ with $i_1$. Item $i_3$ will be exposed to both users at the top position, like $i_1$ did, and then impact is recomputed. 
MME is computed as follows:

\begin{equation}
    \label{sigir_eq:mme-ori}
    \text{MME} = \frac{1}{n} \sum_{i \in I} 
    \left\{
        \max_{i' \in I} Imp_i(i') - Imp_i(i)  
    \right\}
\end{equation}
\begin{equation}
        \label{sigir_eq:imp}
        Imp_i(i') = \sum\limits_{u \in U} \sum\limits_{p=1}^k r_{u,i} \cdot e_{\text{inv}}(p) \cdot X_{u,i',p} 
\end{equation}
\begin{equation}
     X_{u,i',p} = \frac{1}{W} \frac{1}{m} \sum\limits_{w=1}^W 1_{L_{u,w}}(i') \cdot \delta(z(u,i',w)=p)  
\end{equation}
where 
$Imp_i(i')$ is the impact when we allocate the exposure of item $i'$ to item $i$, 
$X_{u,i',p}$ is the probability that item $i'$ is recommended to user $u$ at position $p$ in $W$ rounds of recommendations, 
and $e_{\text{inv}}(p)$ is the exposure weight of item at position $p$, where the weight is discounted with the inverse examination function (see Tab.~\ref{sigir_tab:exp-weigh}).\footnote{This has been changed from the original text ``$e_{\text{inv}}(u,i)$ is the exposure weight of item $i'$ to user $u$'', as $e_{\text{inv}}$ depends directly on $p$.} 
MME ranges within $[0,\infty)$. 

\subsubsection{Item Better-Off (IBO) \& Item Worse-Off (IWO) \cite{Saito2022FairRanking}} \up IBO and \down IWO use the principle of \textit{dominance over uniform ranking}, where fairness means each item 
has a better impact (as defined in MME) under the current ranking policy, than if it were under the uniform random ranking policy $\pi_{unif}$, which samples all possible permutations of items uniformly at random. 
IBO/IWO measures the percentage of items for which our current ranking policy increases/decreases impact by at least 10\% compared to $\pi_{unif}$\footnote{In \cite{Saito2022FairRanking}, 10\% is hard coded, but this can be a variable. We also use 10\%.}:
\begin{equation}
    \label{sigir_eq:ibo-our}
     \text{IBO} = \frac{100}{|I^-|} 
     \sum_{i \in I^-} 
   \delta\left(Imp_i(i) \geq 1.1 \cdot Imp_{i}^{unif}\right)
\end{equation}
\begin{equation}
    \label{sigir_eq:iwo-our}
     \text{IWO} = 
     \frac{100}{|I^-|} 
     \sum_{i \in I^-} 
    \delta\left(Imp_i(i) \leq 0.9 \cdot Imp_{i}^{unif}\right) 
\end{equation}
\begin{equation}
\label{sigir_eq:imp-unif}
    Imp^{unif}_{i} =  \frac{1}{m}\frac{1}{n} \sum \limits_{p=1}^{k} \frac{1}{p} \cdot \sum \limits_{u \in U} r_{u,i} 
\end{equation}

\noindent where $I^{-}$ is the set of items with at least one user that finds the item relevant. This ensures that the set of items that cause $\delta(\cdot)=1$ in IBO is disjoint from that in IWO.\footnote{We exclude items with no relevant users, as 
for these items $Imp_i(i)=Imp^{unif}_i=0$, causing the same items being considered `better-off' and `worse-off' at the same time.} 
$Imp_i(i)$ is as per Eq.~\eqref{sigir_eq:imp} and $Imp^{unif}_i$ is the impact if item $i$ is exposed according to $\pi_{unif}$ using $e_{\text{inv}}(p)$ as examination function (see Tab.~\ref{sigir_tab:exp-weigh}).\footnote{This has been changed from the original text ``$e_{\text{inv}}(u,i)$'', as the function depends directly on position $p$.} 
Note that the above definitions are modifications to the formulation of \cite{Saito2022FairRanking} to avoid computational issues that result in 
division by zero (undefinedness limitation \cite{Rampisela2024EvaluationStudy}).\footnote{We move the divisor $Imp^{unif}_i$ from the left-hand side to the right-hand side.} 
IBO/IWO ranges between $[0,100]$.

\subsubsection{Individual-user-to-individual-item fairness (II-F) \cite{Wu2022JointRecommendation}} 
\down II-F was first defined by \cite{Diaz2020EvaluatingExposure} to quantify unfairness as the disparity between system exposure and target exposure in individual queries and individual items. II-F was redefined by \cite{Wu2022JointRecommendation} for RSs as: 
\begin{equation}
    \label{sigir_eq:iif-ori}
     \text{II-F} = \frac{1}{m} \frac{1}{n} 
     \sum\limits_{u \in U} \sum\limits_{i \in I} \left(E_{u,i}^{ } - E_{u,i}^*\right)^2 
\end{equation}
\begin{equation}
    \label{sigir_eq:eui}
E_{u,i} = \frac{1}{W} \sum\limits_{w=1}^{W}1_{L_{u,w}}(i) \cdot e_{\text{RBP}}(u,i,w)
\end{equation}
\begin{equation}
\label{sigir_eq:eui-star}
E_{u,i}^{*} = \frac{r_{u,i}}{|R_u^*|} \cdot\frac{1-\gamma^{|R_u^*|}}{1-\gamma}  \, \text{if } |R_u^*|>0\, \text{, otherwise }0
\end{equation}
\noindent where $E_{u,i}$ is the expected exposure of $i$ to $u$ as per a stochastic ranking policy. 
$E_{u,i}^{*}$ is the expected exposure of $i$ to $u$ as per an ideal stochastic ranking policy, which assumes that relevant items get equal expected exposure~\cite{Diaz2020EvaluatingExposure}. 
Thus, the recommendation is fair based on II-F if the system exposure 
matches the exposure allocated to items under an ideal ranking policy.
The examination function based on RBP (see Tab.~\ref{sigir_tab:exp-weigh}) is used in $E_{u,i}$ and the equation of $E_{u,i}^{*}$ is derived based on the same examination function \cite{Wu2022JointRecommendation}. 
$|R_u^*|$ is the number of relevant items for user $u$. II-F ranges between $[0,1]$.

\subsubsection{All-users-to-individual-item fairness (AI-F) \cite{Wu2022JointRecommendation}} \down AI-F evaluates how much \rs~ under/overexpose an item to all users as the mean deviation of overall system exposure over target exposure:
\begin{equation}
    \label{sigir_eq:aif-ori}
     \text{AI-F} = 
        \frac{1}{n} \sum \limits_{i \in I}
            \left(
                \frac{1}{m}\sum \limits_{u \in U} E_{u,i}^{ }
                - \frac{1}{m}\sum \limits_{u \in U} E_{u,i}^{*}
            \right)^2
\end{equation}
where $E_{u,i}^{ }$, $E_{u,i}^{*}$ are as per Eq.~\eqref{sigir_eq:eui}--\eqref{sigir_eq:eui-star}. 
Similar to II-F, AI-F also quantifies fairness based on how close the system exposure is to the target exposure. In II-F, this disparity is computed individually between each user-item pair, while in AI-F item exposure is first aggregated across users prior to computing the difference in exposure. Due to this aggregation,  
AI-F would have a better fairness score than II-F when there is a greater number of unique items in the recommendation, as opposed to having the same few items exposed to all users. The range of AI-F is $[0,1]$. 

\section{Experimental Setup}
\label{sigir_ss:setup}

We study the above \textsc{Fair+Rel} measures across different recommenders and datasets. 
Our general experimental setup follows, and we provide the description of the experiments in $\S$\ref{sigir_s:exp}.\footnote{Our code: \href{https://github.com/theresiavr/can-we-trust-recsys-fairness-evaluation}{https://github.com/theresiavr/can-we-trust-recsys-fairness-evaluation}.}

\noindent \textbf{Datasets.} We use four real-world datasets of varying sizes and domains: 
Lastfm (music) \cite{Cantador20112nd2011}, 
Amazon Luxury Beauty, i.e., Amazon-lb (e-commerce) \cite{Ni2019JustifyingAspects},  
QK-video (videos) \cite{Yuan2022Tenrec:Systems}, and
ML-10M (movies) \cite{Harper2015TheContext}. QK-video is as provided by  \cite{Yuan2022Tenrec:Systems}, and the rest are as provided by \cite{Zhao2021RecBole:Algorithms}. For QK-video, we use only the `sharing' interactions.

\begin{table}[tb]
\caption{Statistics of 
the preprocessed datasets.}
\label{sigir_tab:dataset}
\centering
\resizebox{0.85\columnwidth}{!}{
\begin{tabular}{lrrrr}
\toprule
dataset & \multicolumn{1}{l}{\#users ($m$)} & \multicolumn{1}{l}{\#items ($n$)} & \multicolumn{1}{l}{\#interactions} & \multicolumn{1}{l}{sparsity (\%)} \\ 
\midrule                                                           
Lastfm \cite{Cantador20112nd2011} & 1,859 & 2,823 & 71,355 & 98.64\% \\
Amazon-lb \cite{Ni2019JustifyingAspects} & 1,054 & 791 & 12,397 & 98.51\% \\
QK-video \cite{Yuan2022Tenrec:Systems} & 4,656 & 6,423 & 51,777 & 99.83\% \\ 
ML-10M \cite{Harper2015TheContext} & 49,378 & 9,821 & 5,362,685 & 98.89\% \\
\bottomrule
\end{tabular}
}
\end{table}

\noindent \textbf{Preprocessing.} We keep only users and items with at least 5 interactions (5-core filtering). When there are duplicate interactions, we keep the most recent one. Ratings equal/above 3 are converted to 1, and the rest are discarded for Amazon-lb and ML-10M, as their ratings range between $[1,5]$ and $[0.5, 5]$ respectively. No conversions are done for Lastfm and QK-video as they only have implicit feedback. Tab.~\ref{sigir_tab:dataset} presents statistics of the preprocessed datasets. 

\noindent \textbf{Data splits.} Global temporal splits \cite{Meng2020ExploringModels} with a ratio of 6:2:2 form the train/val/test sets from the preprocessed datasets for Amazon-lb and ML-10M. Global random splits with the same ratio are used for Lastfm and QK-video as they have no timestamps. 
Only users with at least 5 interactions in the train set are kept in all splits. 

\noindent \textbf{Recommenders.} We use four well-known top $k$ recommenders: 
item-based K-Nearest Neighbour (ItemKNN) \cite{Deshpande2004Item-basedAlgorithms}, Bayesian Personalised Ranking (BPR), \cite{RendleBPR:Feedback}, Variational Autoencoder with multinomial likelihood (MultiVAE) \cite{Liang2018VariationalFiltering}, and Neigh\-bour\-hood-enriched Contrastive Learning (NCL) \cite{Lin2022ImprovingLearning}. 
We train BPR, MultiVAE, and NCL using RecBole \cite{Zhao2021RecBole:Algorithms} for 300 epochs with early stopping. The configuration with the best NDCG@10 during validation is taken as the final model.\footnote{The hyperparameter search space and best values are in the code repository.} 
During testing, all unobserved items are selected as candidates for recommendation and each user's train/val items are excluded from their own recommendations. 

\noindent \textbf{Fair re-ranker.} 
As the models are not directly optimised for fairness, we use a re-ranker to obtain fairer recommendations. 
The top $k'$ items are re-ranked to provide exposure to items that were outside the top $k$, where $k'$ is ideally larger than the cut-off $k=10$. In RS datasets, normally there are very few relevant items per user, so $k'$ should not be too big (e.g., 100). We choose $k'=25$ for all datasets and models. The re-ranking is done per user with COMBMNZ (CM) \cite{Lee1997AnalysesCombination} as a robust rank fusion method.\footnote{Other re-rankers exist but do not suit  
our setup, e.g., \cite{Wang2022ProvidingSystems} requires computing item similarity, but true similarity is challenging to obtain \cite{Dwork2012FairnessAwareness,Tsepenekas2023ComparingDistributions}.} 
CM fuses two lists of scores, one based on relevance and one based on fairness, to create a new ranking for each user. 
The relevance-based score is the min-max normalised predicted relevance score. The fairness-based score is first obtained from the coverage score of each top $k'$ items based on their appearance in the top $k$. Then, we compute 1 minus the normalised coverage to allocate higher score for items with lower exposure, thus increasing fairness. 
The combined scores are sorted to generate the final fused ranking of relevance and fairness. 

\noindent\textbf{Measures}. 
Recommendation models are evaluated using all the joint measures of relevance and fairness (\textsc{Fair+Rel}) presented in $\S$\ref{sigir_ss:fairrel}.\footnote{For IAA, the ground truth relevance is used to compute the relevance score. For HD, $\gamma=0.9$ as per 
\cite{Jeunen2021Top-KExposure}. For II-F and AI-F, $\gamma=0.8$  as per 
\cite{Wu2022JointRecommendation}. Note that 
IBO/IWO are normalised to [0,1] for consistency with the other measures.} 
As comparison to the joint measures, we evaluate relevance only (\textsc{Rel}) with: Hit Rate (HR), MRR, Precision (P), Recall (R), MAP, and NDCG. We also evaluate fairness only (\textsc{Fair}) with:\footnote{We use the modified versions of these measures as per \cite{Rampisela2024EvaluationStudy}.} 
Jain Index (Jain) \cite{jain1984quantitative,Zhu2020FARM:APPs}, Qualification Fairness (QF) \cite{Zhu2020FARM:APPs}, Entropy (Ent) \cite{Patro2020FairRec:Platforms,Shannon1948ACommunication},
Fraction of Satisfied Items (FSat) \cite{Patro2020FairRec:Platforms}, and Gini Index (Gini) \cite{Gini1912VariabilitaMutabilita,Mansoury2020FairMatch:Systems}. Unless otherwise stated, all measures are computed at $k=10$.

\section{Empirical Analysis}
\label{sigir_s:exp}
We present the evaluation results of all \textsc{Fair+Rel}, \textsc{Rel}, and \textsc{Fair} measures, in $\S$\ref{sigir_ss:performance}. 
We study their correlation in $\S$\ref{sigir_ss:corr}, their sensitivity across different top-$k$ positions in $\S$\ref{sigir_ss:sliding} and across increasing levels of relevance and fairness in $\S$\ref{sigir_ss:insert}.

\subsection{Evaluation Results of All Measures}
\label{sigir_ss:performance}

Tab.~\ref{sigir_tab:base-rerank-all} shows the scores of all \textsc{Fair+Rel}, \textsc{Rel}, and \textsc{Fair} measures, per dataset and recommender/re-ranking. $\uparrow$ means the higher the score, the better, and vice versa for $\downarrow$. Overall we observe the following. 

\noindent \textbf{Best model agreement.} We aim to study whether the measures agree on the same best model. We note two main trends. First, for all datasets, the best model based on \textsc{Rel} measures is always different from the one based on \textsc{Fair} measures, except for QF in Amazon-lb. This means that the fairest model is not necessarily the best in terms of relevance. 
Second, while all \textsc{Rel} measures agree on the same best model per dataset (except MRR and MAP for Amazon-lb) and all the \textsc{Fair} measures always agree on the same best model (except QF), the \textsc{Fair+Rel} measures disagree on the best model. Occasionally, some \textsc{Fair+Rel} measures agree with another more often (e.g., IBO with IWO, or IAA with HD and II-F, or MME with AI-F and sometimes IFD), but there is no overall consistency. The agreement between some joint measures may be due to their similar formulations: both IBO/IWO are the fractions of items with an impact score greater/lower than a threshold; MME/AI-F aggregate exposure across users prior to computing the exposure difference, while IAA/HD/II-F do not; and MME/IFD are pairwise measures. 

\noindent \textbf{Range of scores.} We identify three issues on the score range of the \textsc{Fair+Rel} measures: (1) extremely small scales for several joint measures; (2) scale mismatch between single-aspect measures and joint measures; and (3) scale mismatch between joint measures. 
About (1), for all datasets and models, several \down \textsc{Fair+Rel} scores are extremely small ($\leq10^{-3}$), and these scores do not allow to distinguish across models per dataset. 
For example, IFD$_{\times}$ is always close to 0 across all datasets, as the term Eq.~\eqref{sigir_eq:ifd_j_x} is often 0 due to the low number of relevant items per user.\footnote{For all four datasets, the median number of relevant items per user is at most 46.} For MME and II-F/AI-F, Eq.~\eqref{sigir_eq:imp} and Eq.~\eqref{sigir_eq:eui-star} often result in 0 for the same reason as IFD$_{\times}$. 
About (2), while the above \textsc{Fair+Rel} scores differ in the fourth or later decimal point, the differences in the \textsc{Rel} and \textsc{Fair} scores are in the second decimal point or before. E.g., the NDCG (\textsc{Rel} score) of MultiVAE-CM and NCL for Lastfm differs by $\sim$0.16 and their Jain (\textsc{Fair} score) differs by $\sim$0.14. These examples imply non-negligible differences, but the joint scores of IAA/IFD$_{\times}$/MME/II-F/AI-F only differ by $\leq 10^{-3}$, which may seem negligible.\footnote{We use NDCG and Jain as they tend to be stricter$^*$ than some other single-aspect measures used in this work. $^*$Note that this footnote has been changed from the original text ``... as they are more sensitive to changes than HR and QF'' for clarity.} 
These inconsistencies in the difference of magnitude make the scores hard to understand. 
About (3), we see large gaps in the score range of all joint measures, e.g., between IWO, HD, and AI-F, despite all of them being lower-is-better measures. E.g., in ML-10M, \down IWO $\approx 1$ (very unfair) based on its theoretical $[0,1]$-range, \down HD is about a quarter of the \down IWO score (somewhat fair), while \down AI-F $\approx 0$ (extremely fair). This discrepancy causes confusion in score interpretation. 

Finally, we group all \textsc{Fair+Rel} measures into 3 clusters: 
(i) IAA/HD/II-F, which align more with \textsc{Rel} measures; 
(ii) IFD/MME/AI-F, which align more with \textsc{Fair} measures; and
(iii) IBO/IWO, which do not consistently align with any single-aspect measure. Within the same cluster, especially in (i), measures often have large differences in their score ranges (up to $\Delta\approx0.7$).

\begin{table*}
\centering
\caption{Relevance (\textsc{Rel}), fairness (\textsc{Fair}), and joint \textsc{Fair+Rel} scores at \(k=10\) without and with re-ranking the top \(k'=25\) items using COMBMNZ (CM). Bold marks the most relevant/fair score per measure. The score 0.000 does not mean the scores are exactly 0; this is due to the measures having small scores (\(<10^{-3}\) and rounding to 3 d.p.
}
\label{sigir_tab:base-rerank-all}
\resizebox{0.75\linewidth}{!}{

\begin{tabular}[t]{lll*{2}{r}|*{2}{r}|*{2}{r}|*{2}{r}}
\toprule
 &  & model & \multicolumn{2}{c|}{ItemKNN} & \multicolumn{2}{c|}{BPR} & \multicolumn{2}{c|}{MultiVAE} & \multicolumn{2}{c}{NCL} \\ 
\midrule
 &  & re-ranker & - & CM & - & CM & - & CM & - & CM \\
\midrule
\multirow[c]{21}{*}{\rotatebox[origin=c]{90}{Lastfm}} & \multirow[c]{6}{*}{\rotatebox[origin=c]{90}{\textsc{Rel}}} & $\uparrow$ $\text{HR}$ & 0.765 & 0.581 & 0.773 & 0.587 & 0.778 & 0.523 & \bfseries 0.793 & 0.571 \\
 &  & $\uparrow$ $\text{MRR}$ & 0.484 & 0.270 & 0.492 & 0.280 & 0.476 & 0.232 & \bfseries 0.503 & 0.260 \\
 &  & $\uparrow$ $\text{P}$ & 0.172 & 0.089 & 0.178 & 0.092 & 0.176 & 0.076 & \bfseries 0.184 & 0.087 \\
 &  & $\uparrow$ $\text{MAP}$ & 0.137 & 0.053 & 0.141 & 0.058 & 0.138 & 0.045 & \bfseries 0.148 & 0.050 \\
 &  & $\uparrow$ $\text{R}$ & 0.218 & 0.114 & 0.224 & 0.119 & 0.224 & 0.098 & \bfseries 0.234 & 0.110 \\
 &  & $\uparrow$ $\text{NDCG}$ & 0.245 & 0.119 & 0.252 & 0.126 & 0.247 & 0.102 & \bfseries 0.261 & 0.115 \\
\cline{2-11}
 & \multirow[c]{5}{*}{\rotatebox[origin=c]{90}{\textsc{Fair}}} & $\uparrow$ $\text{Jain}$ & 0.042 & 0.094 & 0.058 & 0.140 & 0.097 & \bfseries 0.222 & 0.082 & 0.215 \\
 &  & $\uparrow$ $\text{QF}$ & 0.474 & \bfseries 0.679 & 0.362 & 0.528 & 0.517 & 0.678 & 0.453 & 0.657 \\
 &  & $\uparrow$ $\text{Ent}$ & 0.589 & 0.735 & 0.610 & 0.740 & 0.707 & \bfseries 0.826 & 0.671 & 0.810 \\
 &  & $\uparrow$ $\text{FSat}$ & 0.129 & 0.216 & 0.147 & 0.228 & 0.202 & \bfseries 0.321 & 0.178 & 0.286 \\
 &  & $\downarrow$ $\text{Gini}$ & 0.904 & 0.790 & 0.910 & 0.818 & 0.839 & \bfseries 0.696 & 0.872 & 0.728 \\
\cline{2-11}
 & \multirow[c]{9}{*}{\rotatebox[origin=c]{90}{\textsc{Fair+Rel}}} & $\uparrow$ $\text{IBO}$ & 0.209 & 0.256 & 0.208 & 0.253 & 0.261 & 0.278 & 0.242 & \bfseries 0.292 \\
 &  & $\downarrow$ $\text{IWO}$ & 0.791 & 0.744 & 0.792 & 0.747 & 0.739 & 0.722 & 0.758 & \bfseries 0.708 \\
 &  & $\downarrow$ $\text{IAA}$ & 0.004 & 0.004 & 0.004 & 0.004 & 0.004 & 0.004 & \bfseries 0.004 & 0.004 \\
 &  & $\downarrow$ $\text{IFD}_{\div}$ & 0.074 & 0.053 & 0.075 & 0.054 & 0.073 & \bfseries 0.049 & 0.076 & 0.052 \\
 &  & $\downarrow$ $\text{IFD}_{\times}$ & 0.000 & 0.000 & 0.000 & 0.000 & 0.000 & \bfseries 0.000 & 0.000 & 0.000 \\
 &  & $\downarrow$ $\text{HD}$ & 0.099 & 0.177 & 0.104 & 0.174 & 0.095 & 0.203 & \bfseries 0.092 & 0.177 \\
 &  & $\downarrow$ $\text{MME}$ & 0.001 & 0.001 & 0.001 & 0.001 & 0.001 & 0.001 & 0.001 & \bfseries 0.001 \\
 &  & $\downarrow$ $\text{II-F}$ & 0.001 & 0.002 & 0.001 & 0.002 & 0.001 & 0.002 & \bfseries 0.001 & 0.002 \\
 &  & $\downarrow$ $\text{AI-F}$ & 0.000 & 0.000 & 0.000 & 0.000 & 0.000 & 0.000 & 0.000 & \bfseries 0.000 \\
\cline{1-11}
\multirow[c]{21}{*}{\rotatebox[origin=c]{90}{Amazon-lb}} & \multirow[c]{6}{*}{\rotatebox[origin=c]{90}{\textsc{Rel}}} & $\uparrow$ $\text{HR}$ & \bfseries 0.046 & 0.016 & 0.011 & 0.021 & 0.039 & 0.014 & 0.034 & 0.011 \\
 &  & $\uparrow$ $\text{MRR}$ & 0.020 & 0.011 & 0.003 & 0.007 & \bfseries 0.023 & 0.004 & 0.022 & 0.003 \\
 &  & $\uparrow$ $\text{P}$ & \bfseries 0.005 & 0.002 & 0.001 & 0.002 & 0.004 & 0.002 & 0.004 & 0.001 \\
 &  & $\uparrow$ $\text{MAP}$ & 0.006 & 0.004 & 0.002 & 0.004 & 0.006 & 0.003 & \bfseries 0.006 & 0.001 \\
 &  & $\uparrow$ $\text{R}$ & \bfseries 0.013 & 0.005 & 0.005 & 0.010 & 0.010 & 0.008 & 0.012 & 0.003 \\
 &  & $\uparrow$ $\text{NDCG}$ & \bfseries 0.011 & 0.005 & 0.003 & 0.006 & 0.010 & 0.004 & 0.011 & 0.002 \\
\cline{2-11}
 & \multirow[c]{5}{*}{\rotatebox[origin=c]{90}{\textsc{Fair}}} & $\uparrow$ $\text{Jain}$ & 0.271 & \bfseries 0.431 & 0.223 & 0.359 & 0.035 & 0.097 & 0.026 & 0.080 \\
 &  & $\uparrow$ $\text{QF}$ & \bfseries 0.650 & 0.612 & 0.549 & 0.594 & 0.222 & 0.286 & 0.229 & 0.310 \\
 &  & $\uparrow$ $\text{Ent}$ & 0.802 & \bfseries 0.839 & 0.747 & 0.809 & 0.418 & 0.558 & 0.371 & 0.534 \\
 &  & $\uparrow$ $\text{FSat}$ & 0.370 & \bfseries 0.438 & 0.314 & 0.376 & 0.114 & 0.152 & 0.091 & 0.138 \\
 &  & $\downarrow$ $\text{Gini}$ & 0.665 & \bfseries 0.598 & 0.747 & 0.660 & 0.949 & 0.899 & 0.959 & 0.910 \\
\cline{2-11}
 & \multirow[c]{9}{*}{\rotatebox[origin=c]{90}{\textsc{Fair+Rel}}} & $\uparrow$ $\text{IBO}$ & \bfseries 0.062 & 0.029 & 0.019 & 0.038 & 0.029 & 0.029 & 0.038 & 0.024 \\
 &  & $\downarrow$ $\text{IWO}$ & \bfseries 0.938 & 0.971 & 0.981 & 0.962 & 0.971 & 0.971 & 0.962 & 0.976 \\
 &  & $\downarrow$ $\text{IAA}$ & \bfseries 0.011 & 0.011 & 0.011 & 0.011 & 0.011 & 0.011 & 0.011 & 0.011 \\
 &  & $\downarrow$ $\text{IFD}_{\div}$ & 0.005 & 0.003 & 0.003 & \bfseries 0.002 & 0.005 & 0.002 & 0.005 & 0.003 \\
 &  & $\downarrow$ $\text{IFD}_{\times}$ & 0.000 & 0.000 & 0.000 & 0.000 & 0.000 & 0.000 & 0.000 & \bfseries 0.000 \\
 &  & $\downarrow$ $\text{HD}$ & \bfseries 0.580 & 0.630 & 0.661 & 0.626 & 0.597 & 0.653 & 0.598 & 0.667 \\
 &  & $\downarrow$ $\text{MME}$ & 0.001 & \bfseries 0.001 & 0.001 & 0.001 & 0.003 & 0.001 & 0.004 & 0.001 \\
 &  & $\downarrow$ $\text{II-F}$ & \bfseries 0.006 & 0.006 & 0.006 & 0.006 & 0.006 & 0.006 & 0.006 & 0.006 \\
 &  & $\downarrow$ $\text{AI-F}$ & 0.000 & \bfseries 0.000 & 0.000 & 0.000 & 0.001 & 0.000 & 0.002 & 0.000 \\
\bottomrule
\end{tabular}
}

\end{table*}

\begin{table*}
\centering
\caption*{Table \ref{sigir_tab:base-rerank-all} (continued): (\textsc{Rel}), fairness (\textsc{Fair}), and joint \textsc{Fair+Rel} scores at \(k=10\) without and with re-ranking the top \(k'=25\) items using COMBMNZ (CM). Bold marks the most relevant/fair score per measure. The score 0.000 does not mean the scores are exactly 0; this is due to the measures having small scores (\(<10^{-3}\) and rounding to 3 d.p.
}
\resizebox{0.75\linewidth}{!}{
\begin{tabular}[t]{lll*{2}{r}|*{2}{r}|*{2}{r}|*{2}{r}}
\toprule
 &  & model & \multicolumn{2}{c|}{ItemKNN} & \multicolumn{2}{c|}{BPR} & \multicolumn{2}{c|}{MultiVAE} & \multicolumn{2}{c}{NCL} \\ 
\midrule
 &  & re-ranker & - & CM & - & CM & - & CM & - & CM \\
\midrule
\multirow[c]{21}{*}{\rotatebox[origin=c]{90}{QK-video}} & \multirow[c]{6}{*}{\rotatebox[origin=c]{90}{\textsc{Rel}}} & $\uparrow$ $\text{HR}$ & 0.040 & 0.047 & 0.099 & 0.045 & 0.109 & 0.061 & \bfseries 0.130 & 0.077 \\
 &  & $\uparrow$ $\text{MRR}$ & 0.013 & 0.013 & 0.039 & 0.015 & 0.039 & 0.021 & \bfseries 0.048 & 0.024 \\
 &  & $\uparrow$ $\text{P}$ & 0.004 & 0.005 & 0.011 & 0.005 & 0.012 & 0.006 & \bfseries 0.014 & 0.008 \\
 &  & $\uparrow$ $\text{MAP}$ & 0.005 & 0.005 & 0.017 & 0.006 & 0.018 & 0.009 & \bfseries 0.022 & 0.010 \\
 &  & $\uparrow$ $\text{R}$ & 0.014 & 0.019 & 0.043 & 0.019 & 0.051 & 0.027 & \bfseries 0.061 & 0.033 \\
 &  & $\uparrow$ $\text{NDCG}$ & 0.009 & 0.010 & 0.029 & 0.011 & 0.031 & 0.016 & \bfseries 0.038 & 0.019 \\
\cline{2-11}
 & \multirow[c]{5}{*}{\rotatebox[origin=c]{90}{\textsc{Fair}}} & $\uparrow$ $\text{Jain}$ & 0.483 & \bfseries 0.589 & 0.081 & 0.379 & 0.012 & 0.032 & 0.020 & 0.071 \\
 &  & $\uparrow$ $\text{QF}$ & \bfseries 0.901 & 0.790 & 0.625 & 0.823 & 0.100 & 0.163 & 0.201 & 0.365 \\
 &  & $\uparrow$ $\text{Ent}$ & 0.933 & \bfseries 0.937 & 0.755 & 0.903 & 0.420 & 0.547 & 0.507 & 0.674 \\
 &  & $\uparrow$ $\text{FSat}$ & 0.443 & \bfseries 0.547 & 0.212 & 0.382 & 0.052 & 0.090 & 0.077 & 0.150 \\
 &  & $\downarrow$ $\text{Gini}$ & 0.472 & \bfseries 0.442 & 0.807 & 0.570 & 0.982 & 0.959 & 0.966 & 0.902 \\
\cline{2-11}
 & \multirow[c]{9}{*}{\rotatebox[origin=c]{90}{\textsc{Fair+Rel}}} & $\uparrow$ $\text{IBO}$ & 0.033 & 0.038 & \bfseries 0.054 & 0.036 & 0.031 & 0.036 & 0.043 & \bfseries 0.054 \\
 &  & $\downarrow$ $\text{IWO}$ & 0.967 & 0.962 & \bfseries 0.946 & 0.964 & 0.969 & 0.964 & 0.957 & \bfseries 0.946 \\
 &  & $\downarrow$ $\text{IAA}$ & 0.001 & 0.001 & 0.001 & 0.001 & 0.001 & 0.001 & \bfseries 0.001 & 0.001 \\
 &  & $\downarrow$ $\text{IFD}_{\div}$ & 0.009 & \bfseries 0.007 & 0.014 & 0.008 & 0.014 & 0.009 & 0.015 & 0.010 \\
 &  & $\downarrow$ $\text{IFD}_{\times}$ & 0.000 & \bfseries 0.000 & 0.000 & 0.000 & 0.000 & 0.000 & 0.000 & 0.000 \\
 &  & $\downarrow$ $\text{HD}$ & 0.576 & 0.560 & 0.490 & 0.565 & 0.478 & 0.535 & \bfseries 0.457 & 0.519 \\
 &  & $\downarrow$ $\text{MME}$ & 0.000 & \bfseries 0.000 & 0.000 & 0.000 & 0.000 & 0.000 & 0.000 & 0.000 \\
 &  & $\downarrow$ $\text{II-F}$ & 0.001 & 0.001 & 0.001 & 0.001 & 0.001 & 0.001 & \bfseries 0.001 & 0.001 \\
 &  & $\downarrow$ $\text{AI-F}$ & 0.000 & \bfseries 0.000 & 0.000 & 0.000 & 0.000 & 0.000 & 0.000 & 0.000 \\
\cline{1-11}
\multirow[c]{21}{*}{\rotatebox[origin=c]{90}{ML-10M}} & \multirow[c]{6}{*}{\rotatebox[origin=c]{90}{\textsc{Rel}}} & $\uparrow$ $\text{HR}$ & 0.487 & 0.443 & 0.512 & 0.386 & 0.417 & 0.387 & \bfseries 0.521 & 0.402 \\
 &  & $\uparrow$ $\text{MRR}$ & 0.282 & 0.225 & 0.299 & 0.185 & 0.237 & 0.191 & \bfseries 0.302 & 0.203 \\
 &  & $\uparrow$ $\text{P}$ & 0.137 & 0.105 & 0.146 & 0.088 & 0.107 & 0.096 & \bfseries 0.154 & 0.094 \\
 &  & $\uparrow$ $\text{MAP}$ & 0.089 & 0.060 & 0.095 & 0.047 & 0.067 & 0.054 & \bfseries 0.101 & 0.052 \\
 &  & $\uparrow$ $\text{R}$ & 0.022 & 0.018 & 0.025 & 0.012 & 0.020 & 0.016 & \bfseries 0.026 & 0.013 \\
 &  & $\uparrow$ $\text{NDCG}$ & 0.150 & 0.113 & 0.160 & 0.092 & 0.119 & 0.100 & \bfseries 0.167 & 0.100 \\
\cline{2-11}
 & \multirow[c]{5}{*}{\rotatebox[origin=c]{90}{\textsc{Fair}}} & $\uparrow$ $\text{Jain}$ & 0.011 & 0.027 & 0.037 & \bfseries 0.115 & 0.003 & 0.006 & 0.024 & 0.069 \\
 &  & $\uparrow$ $\text{QF}^{*}$ & 0.044 & 0.068 & 0.145 & \bfseries 0.216 & 0.014 & 0.025 & 0.086 & 0.132 \\
 &  & $\uparrow$ $\text{Ent}$ & 0.407 & 0.514 & 0.596 & \bfseries 0.716 & 0.238 & 0.324 & 0.519 & 0.638 \\
 &  & $\uparrow$ $\text{FSat}^{*}$ & 0.044 & 0.068 & 0.145 & \bfseries 0.216 & 0.014 & 0.025 & 0.086 & 0.132 \\
 &  & $\downarrow$ $\text{Gini}$ & 0.987 & 0.971 & 0.945 & \bfseries 0.879 & 0.997 & 0.993 & 0.969 & 0.930 \\
\cline{2-11}
 & \multirow[c]{9}{*}{\rotatebox[origin=c]{90}{\textsc{Fair+Rel}}} & $\uparrow$ $\text{IBO}$ & 0.031 & 0.046 & 0.069 & \bfseries 0.091 & 0.012 & 0.018 & 0.054 & 0.074 \\
 &  & $\downarrow$ $\text{IWO}$ & 0.969 & 0.954 & 0.931 & \bfseries 0.909 & 0.988 & 0.982 & 0.946 & 0.926 \\
 &  & $\downarrow$ $\text{IAA}$ & 0.008 & 0.009 & 0.008 & 0.009 & 0.009 & 0.009 & \bfseries 0.008 & 0.009 \\
 &  & $\downarrow$ $\text{IFD}_{\div}$ & 0.018 & 0.012 & 0.019 & 0.011 & 0.016 & \bfseries 0.010 & 0.020 & 0.012 \\
 &  & $\downarrow$ $\text{IFD}_{\times}$ & 0.000 & 0.000 & 0.000 & \bfseries 0.000 & 0.000 & 0.000 & 0.000 & 0.000 \\
 &  & $\downarrow$ $\text{HD}$ & 0.221 & 0.255 & 0.226 & 0.262 & 0.265 & 0.273 & \bfseries 0.218 & 0.257 \\
 &  & $\downarrow$ $\text{MME}$ & 0.001 & 0.001 & 0.001 & \bfseries 0.001 & 0.003 & 0.001 & 0.001 & 0.001 \\
 &  & $\downarrow$ $\text{II-F}$ & 0.000 & 0.000 & 0.000 & 0.000 & 0.000 & 0.000 & \bfseries 0.000 & 0.000 \\
 &  & $\downarrow$ $\text{AI-F}$ & 0.000 & 0.000 & 0.000 & \bfseries 0.000 & 0.000 & 0.000 & 0.000 & 0.000 \\
\bottomrule
\end{tabular}
}
\\
\raggedright
{\scriptsize *For ML-10M, QF $\equiv$ FSat, as QF is computed based on the 
\% of recommended items from all items, 
which in this case is equivalent to FSat. 
}
\end{table*}

\subsection{Correlation between Measures (RQ1 \& RQ2)}
\label{sigir_ss:corr}

\begin{figure*}
    \centering
    \includegraphics[width=0.98\textwidth, trim=0.25cm 0.6cm 0.3cm 0.3cm clip=True]{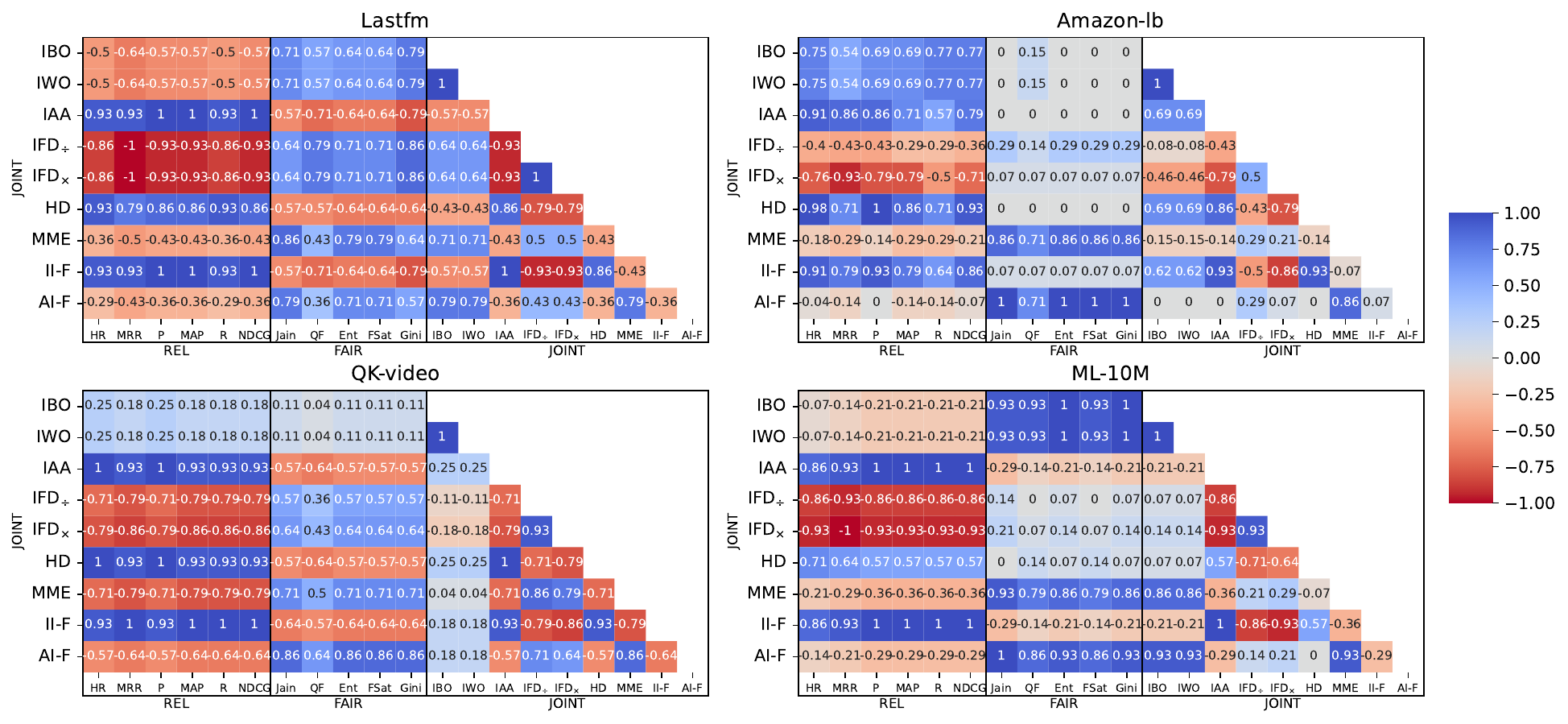}
    \caption{Kendall's $\tau$ correlation between joint \textsc{Fair+Rel} measures, \textsc{Rel}, and \textsc{Fair} measures.}
    \label{sigir_fig:corr-grid}
\end{figure*}

We compute Kendall's $\tau$ correlation between the orderings of the recommenders produced by the scores of each measure, to study how much the \textsc{Fair+Rel} measures agree among themselves but also with \textsc{Rel}-only and \textsc{Fair}-only measures, when ranking recommenders 
(Fig.~\ref{sigir_fig:corr-grid}). As \textsc{Rel} and \textsc{Fair} measures do not always correlate strongly with each other \cite{Rampisela2024EvaluationStudy}, we do not expect \textsc{Fair+Rel} measures to correlate strongly with either \textsc{Rel} or \textsc{Fair} measures.

\noindent\textbf{RQ1. Agreement of joint and single-aspect measures.} Overall, there is no consistent correlation between \textsc{Rel} and \textsc{Fair+Rel} measures. IBO/IWO's correlations vary wildly ( $\tau \in [-0.64, 0.77]$); 
IAA, HD, and II-F have moderate-to-strong positive correlations ($\tau \in [0.57, 1]$); 
IFD and MME have weak-to-strong negative correlations ($\tau \in [-1, -0.29]$ for IFD and $\tau \in [-0.79, -0.14]$ for MME); and AI-F has non-positive correlations ($\tau \in [-0.64,0]$). 

The correlations between \textsc{Fair} and \textsc{Fair+Rel} measures are inconsistent. The correlations of IBO/IWO vary largely again, albeit less than with \textsc{Rel} measures. IAA/HD/II-F have two distinct trends across groups of datasets: they have negative moderate-to-strong correlations ($\tau \in [-0.79,-0.57]$) for Lastfm and QK-video, but weak correlations for Amazon-lb and ML-10M ($\tau \in [-0.29,0.14]$). Similarly, IFD has high correlations for Lastfm and QK-video (except with QF for QK-video), $\tau \in [0.57, 0.86]$, and weak or zero correlations for the other datasets ($\tau\in [0, 0.29]$). Conversely, MME and AI-F have strong correlations except with QF for Lastfm ($\tau \in [0.5,1]$). 

Note that \textsc{Fair+Rel} measures strongly agreeing with \textsc{Rel} measures do not always strongly disagree with \textsc{Fair} measures, and vice versa. E.g., IAA/HD/II-F strongly correlates with \textsc{Rel} measures for Amazon-lb, but they correlate weakly with \textsc{Fair} measures. 

\noindent\textbf{RQ2. Agreement between joint measures.} Overall we find that the three clusters of joint measures identified in $\S$\ref{sigir_ss:performance} show strong positive correlations between measures inside the same cluster and strong negative correlations between measures from different clusters. E.g., IBO always perfectly correlates with IWO, due to their similar formulation. 
IAA, HD, and II-F agree strongly with one another, $\tau \in [0.57,1]$. IFD$_{\div}$ correlates highly with IFD$_{\times}$, $\tau \in [0.5,1]$, as their formulations are similar. MME always agrees strongly with AI-F, $\tau \in [0.79,0.93]$. IFD sometimes has moderate-to-strong correlations with MME and AI-F, $\tau \in [0.43,0.86]$ for Lastfm and QK-video, but the correlations are weaker for Amazon-lb and ML-10M, $\tau \in [0.07,0.29]$. In contrast, IAA/HD/II-F strongly disagrees with IFD, $\tau \in [-0.93,-0.5]$ except for IFD$_{\div}$ in Amazon-lb ($\tau=-0.43$).\footnote{The published version reported `` $\tau \in [-0.71,-0.5]$''; the lower bound has been corrected and the conclusion remains the same.}

Based on the above, we conclude that: IBO/IWO has inconsistent relationships with single-aspect and joint measures; IAA/HD/II-F do not align with fairness; and IFD/MME/AI-F highly disagree with relevance (even if IFD sometimes disagrees with \textsc{Fair} measures too). Among the joint measures, IBO/IWO weakly correlate with the single-aspect measures for QK-video, and similarly with IFD$_{\div}$ for Amazon-lb, but this is not consistent. We thus argue that no joint measures reliably account for both relevance and fairness.

\subsection{Measure Sensitivity at Different Ranks (RQ3)
}
\label{sigir_ss:sliding}

\begin{figure*}
    \centering
    \includegraphics[width=\textwidth, trim=0cm 0cm 0cm 0.2cm, clip=True]{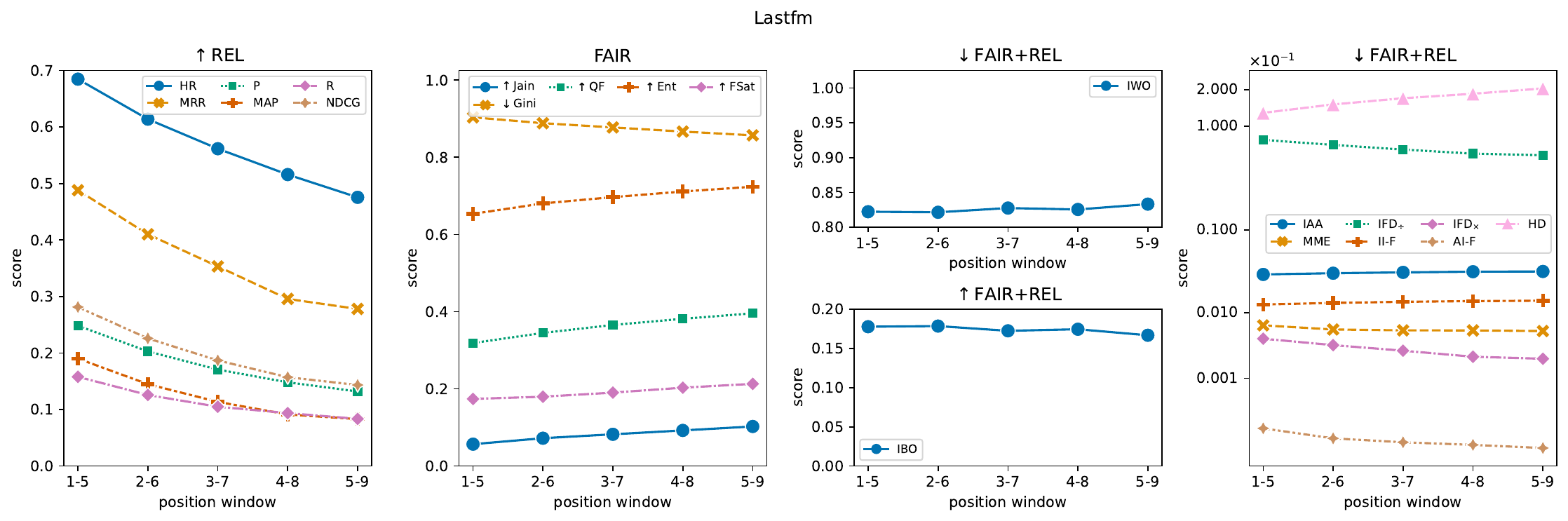}
    \includegraphics[width=\textwidth]{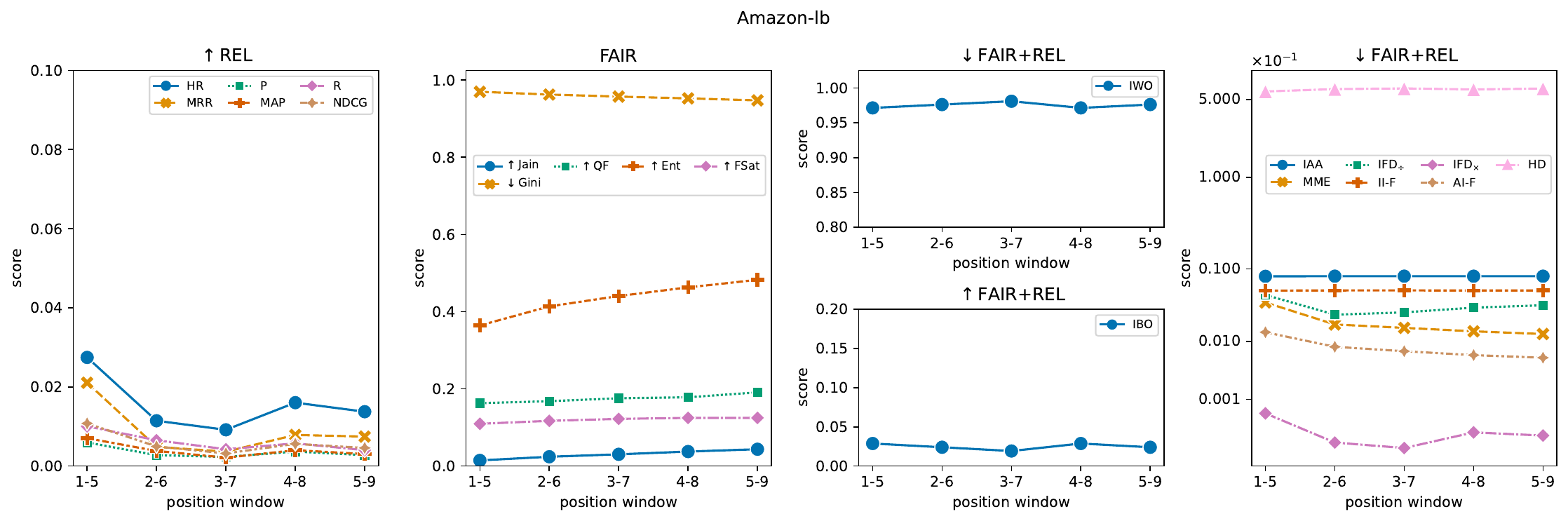}
    \includegraphics[width=\textwidth, trim=0cm 0.5cm 0cm 0cm]{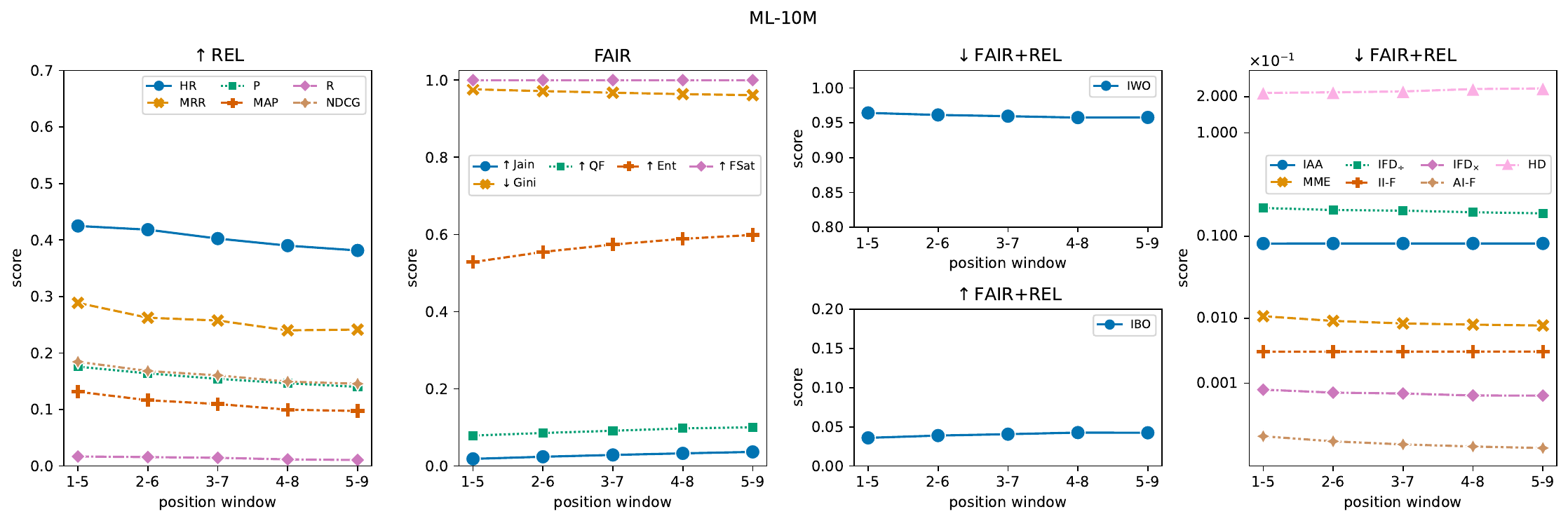}
    \caption{Sliding window evaluation (\(k=5\)) of NCL for Lastfm, Amazon-lb, and ML-10M. The last column is in exponential scale.}
    \label{sigir_fig:sliding}
\end{figure*}

We now study how sensitive the joint measures are at decreasing rank positions, compared to \textsc{Rel} and \textsc{Fair} measures. When moving down the rank, \textsc{Rel} scores are known to decrease while \textsc{Fair} scores are known to improve \cite{Rampisela2024EvaluationStudy}. For this analysis, we use only the runs of the non-reranked NCL model as it generally has the best \textsc{Rel} scores. We compute all measures at $k=5$ for each sliding window, where the windows consist of items at decreasing rank positions: 1--5, 2--6, $\dots$, 5--9. Fig.~\ref{sigir_fig:sliding} shows the results for Lastfm, Amazon-lb, and ML-10M, which represent the overall trends in all our datasets; results for QK-video are shown in the appendix (in our code repository). 

We find that, as expected, as we move down the rank, \textsc{Rel} overall decreases and \textsc{Fair} improves. However, the joint measures are notably less sensitive to changes in rank position. 
Changes with decreasing rank position in the single-aspect scores are up two magnitudes greater than in the joint measures, and the latter do not reflect these differences to the same scale. We posit that the insensitivity is due to the effect of changing relevance being masked by that of fairness and vice versa. This masking makes the scores hard to interpret. Further, the very small scores of \down IAA, IFD$_{\times}$, MME, II-F, and AI-F 
imply extremely fair recommendations (we explain the reasons for this in $\S$\ref{sigir_ss:performance}), even if \textsc{Rel} and \textsc{Fair} scores are low. Thus, these joint measures do not account well for relevance and fairness simultaneously. Last, we note that as we move down the rank, IAA/HD/II-F worsen, IFD/MME/AI-F improve, and IBO/IWO are inconsistent across datasets. This follows the three groups of joint measures discussed above. 

\subsection{Artificial Insertion of Items (RQ4)}
\label{sigir_ss:insert}

Lastly, we study how sensitive the joint measures are to different proportions of relevant items and item fairness in the ranking. Assessing this sensitivity is important as it affects score interpretation; if a joint measure is unresponsive to significant changes in relevance and fairness distribution, its score may not reflect both the fairness and relevance of the recommendations accurately. 

We start with a recommendation list having the worst \textsc{Rel} and \textsc{Fair} scores, and gradually insert more relevant and fair items to it (we explain `fair items' below). 
We observe how the joint measures respond to these changes, compared to the \textsc{Rel} and \textsc{Fair} measures.

We cannot use real-life datasets for this analysis, so we build a synthetic dataset with $m=1000$ and $n=10000$, and artificially generate rankings of items per user, as per \cite{Rampisela2024EvaluationStudy}. 
The initial ranking contains the same $k=10$ items for all users, to whom these items are irrelevant, except for one user.\footnote{This is to keep the number of items exactly $km$.} 
In each iteration, an item from the bottom of each user's top $k$ ranking is replaced by a relevant item having less exposure (hence more fair). The final ranking thus contains $km$ unique items across all users, where each item is relevant only to the user that receives that item in the top $k$. We expect all measures to initially score the worst possible, and then gradually improve as more relevant and fair items enter the ranks. 

\begin{figure}
    \centering
    \includegraphics[width=0.75\columnwidth, trim=1.5cm 0.5cm 1.5cm 0cm]{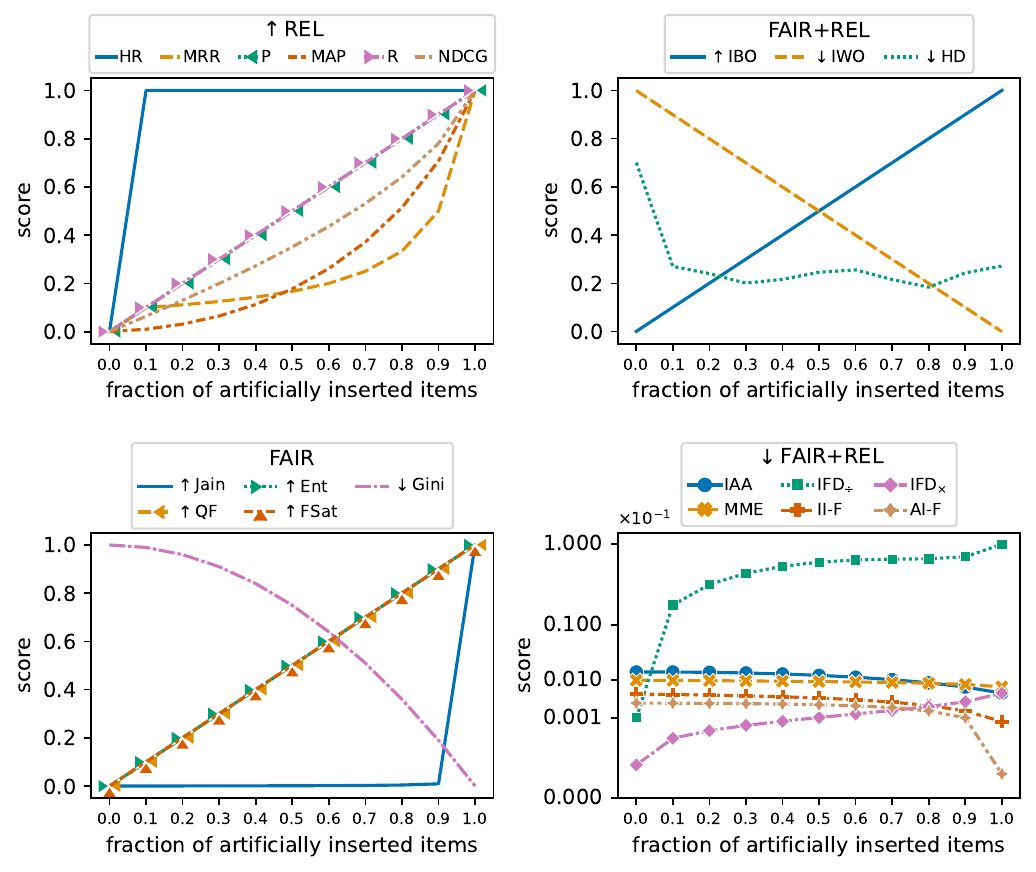}
    \caption{Artificial insertion of items with \(m=1000\) (users).}
    \label{sigir_fig:insert}
\end{figure}

Fig.~\ref{sigir_fig:insert} shows the results of this analysis. Overall, we see that most joint measures are not very sensitive to changes in \textsc{Rel} and \textsc{Fair} scores, i.e., they may vary, but negligibly. This verifies the scale mismatch between most joint measures and the single-aspect measures observed in $\S$\ref{sigir_ss:performance} \& $\S$\ref{sigir_ss:sliding}. 
While the overall change is negligible for most measures, a common observation between the joint measures is that their scores become (slightly) better as \textsc{Rel} and \textsc{Fair} scores improve. An exception to this is IFD. This is because IFD measures fairness based on the pairwise difference in the combined value of exposure and relevance. Thus, when the relevant items start to be moved to the top $k$, the gap between the exposure weight of relevant items in and outside the top $k$ increases, and so does unfairness. Among joint measures that (slightly) improve with more insertion, there are also differences. 
IBO/IWO improve linearly; as both measures are percentages of items, the change is proportional to the amount of inserted items. HD also improves, but its improvement fluctuates due to randomness introduced by the unstable sort in the computation, as per the original implementation in \cite{Jeunen2021Top-KExposure}. \down IAA/IFD$_{\times}$/MME/II-F/AI-F improve non-linearly. However, their scores are extremely close to 0, i.e., on the scale of $10^{-3}$ or less. The lower bound of the measure is 0, hence these small scores indicate that the recommendation is close to the fairest, even at the start of the process where the \textsc{Rel} and \textsc{Fair} scores are the worst in the entire progression. These joint measures are also rather insensitive to changes in \textsc{Rel} and \textsc{Fair} scores. Here, their score range is (0, 0.0015), while the range of \textsc{Rel}, \textsc{Fair}, and IBO/IWO scores is [0,1]. 

\section{Related Work}\label{sigir_s:prevwork}

\noindent\textbf{Fairness evaluation in RSs}. Among prior work on fairness evaluation in RSs \cite{Wang2023ASystems,Amigo2023ASystems,Zehlike2022FairnessSystems,Raj2022MeasuringResults, Rampisela2024EvaluationStudy}, our study is close to
\citeauthor{Amigo2023ASystems}~\cite{Amigo2023ASystems}, who study RS relevance and fairness for groups/individuals and between
items and users. Yet, the focus of our work, i.e., individual item fairness, is not covered in \cite{Amigo2023ASystems}. \citeauthor{Raj2022MeasuringResults}~\cite{Raj2022MeasuringResults} overview evaluation measures for item group fairness. Their study includes the IAA measure as a group fairness measure (whereas we focus on individual item fairness). Lastly, 
\citeauthor{Rampisela2024EvaluationStudy}~\cite{Rampisela2024EvaluationStudy} survey individual item fairness measures that are exclusively linked with fairness and identify the limitations within them, while we focus on  measures that jointly account for both fairness and relevance. 

\noindent\textbf{Joint measures of relevance and fairness}. Outside the strict domain of individual item fairness for RS, there exist other measures that quantify relevance and fairness jointly: \citeauthor{Gao2022FAIR:Evaluation}~\cite{Gao2022FAIR:Evaluation} present a measure combining KL-divergence and IDCG to jointly quantify relevance and group fairness in IR \cite{Gao2022FAIR:Evaluation}. In \cite{Xu2023P-MMF:System}, utility and provider fairness in RSs are simultaneously evaluated with a weighted sum between relevance and fairness. Another approach used in \cite{Garcia-Soriano2021Maxmin-FairConstraints} to evaluate individual fairness in ranking is to compare item position based on ground truth relevance against its position in system-produced rankings. None of the joint measures in our work is a combination of two single-aspect measures as in \cite{Gao2022FAIR:Evaluation} or in the form of weighted sum as in \cite{Xu2023P-MMF:System}. The measure in \cite{Garcia-Soriano2021Maxmin-FairConstraints} is similar to HD \cite{Jeunen2021Top-KExposure}. However, we do not use it in our work because it was not defined for RS fairness, and considerable modifications and assumptions are required prior to using it to evaluate RS fairness.

\section{Appropriate Usage of Joint Measures}
\label{sigir_s:discussion}

We find that joint measures of relevance and fairness (1) tend to align differently with single-aspect measures; 
(2) most of them consistently score almost perfect fairness, even when recommendations are highly irrelevant and unfair based on single-aspect measures; and (3) are rather unresponsive to changes in the recommendation relevance and fairness, especially compared to single-aspect measures. 
Next, we suggest how to best use these joint measures.

\noindent
\textbf{Avoid using similar joint measures}.
In $\S$\ref{sigir_ss:corr} \& $\S$\ref{sigir_ss:sliding} we find three groups of similar joint measures: (i) IAA/HD/II-F, (ii) IFD/MME/AI-F, and (iii) IBO/IWO. Only one measure per group should be used. Yet, considering that typically recommendations are evaluated with \textsc{Rel} measures, we discourage using measures in (i), as they are highly aligned with \textsc{Rel} measures. Measures in (ii) correlate strongly with \textsc{Fair} measures, and can be viable options, and likewise for measures in (iii) that do not consistently correlate with single-aspect measures. However, we argue that measures in (iii) are more useful than those in (ii). Measures in (ii) can be replaced by \textsc{Fair} measures, which are faster to compute and do not need complete relevance judgements, while still achieving highly similar conclusions.

\noindent
\textbf{Be aware of the unintuitive or inconsistent behaviour, insensitivity, and computational complexity of the measures}. 
We recommend that practitioners be aware of the measure limitations in groups (ii) and (iii). Specifically, both IFD versions worsen with higher percentages of jointly relevant and fair recommendations, while the opposite should happen in a healthy measure. 
IFD$_{\div}$ is unaffected by different cut-off $k$ values, as its original formulation only considers full rankings, so it should not be used when different $k$ matters. MME is costly to compute as it is a pairwise measure ($\sim$30 mins for the larger datasets), and the same applies to \down IFD$_{\times}$, albeit to a lesser extent. Further, IFD$_{\times}$/MME/AI-F tend to have extremely small scores, which are therefore hard to interpret and discriminate across runs. They are also rather insensitive to changes reflected by single-aspect measures, meaning that their overall expressiveness is limited. IBO/IWO is sensitive in this aspect. Considering the limitations and the redundancy between measures, IBO/IWO seem to be the most viable measure out of the existing ones, but it is the least consistent between all other measures, due to varying alignments for different datasets, so it should be interpreted cautiously. 

\noindent
\textbf{Avoid score misinterpretation in measures with small empirical scales.}
Due to the small empirical scales of \down IAA/IFD/MME/II-F/AI-F, 
their scores tend not to represent fairness, or relevance and fairness jointly, i.e., scoring very close to 0 even if the recommendation is very irrelevant and unfair based on \textsc{Rel} and \textsc{Fair} measures. Moreover, different systems can have very similar scores which do not translate to similar performance. 
E.g., two models differing in scores by only 0.001 can be interpreted as performing the same, even though the measure has a small empirical range and is not very sensitive to begin with. This issue can be fixed via apriori/posthoc normalisation based on experimental values of the measures \cite{Wu2022JointRecommendation}.  

\noindent\textbf{Measure fairness separately from relevance}. 
As most joint measures (IAA/IFD\ /MME/II-F/AI-F) are difficult to interpret because their scores tend to be compressed in a very low range, and are also rather insensitive to changes in fairness and relevance, we recommend measuring individual item fairness and relevance separately. Otherwise, the joint scores can be close to the theoretical fairest value even if \textsc{Rel} and \textsc{Fair} scores are low ($\S$\ref{sigir_ss:insert}). 
Overall, the above joint measures have unreliable scores, are not as sensitive as the \textsc{Fair} measures, and are subject to more under/overestimation of fairness than \textsc{Fair} measures which have more consistent empirical range. The remaining joint measures are not reliable either: IBO/IWO aligns inconsistently to the single-aspect measures, while HD is almost always consistent with \textsc{Rel} measures and thus does not add another dimension of fairness measurement. It is also unstable due to sorting of items with identical relevance level. 

Overall, the joint measures cannot be compared easily as they have different scales, and they quantify two aspects that are hard to combine due to mismatching scales. The measures tend to correlate highly with either \textsc{Rel} or \textsc{Fair} measures, instead of having a good balance between them. As such, optimising for a joint measure directly may not result in a simultaneously optimal recommendation based on \textsc{Rel} and \textsc{Fair} scores. Another obstacle in measuring fairness is the need to consider user-item relevance in the entire dataset (not just the recommended items), which can be an issue with extremely sparse datasets. It is thus inherently difficult to devise a measure that can jointly quantify relevance and fairness.

\section{Conclusions and Future Work}

We presented a novel empirical study on the properties of all evaluation measures that jointly account for individual item fairness and relevance in recommender systems. We found that out of 9 joint measures, 3 align with traditional relevance-only measures, 4 agree more with fairness-only measures, and the rest behave inconsistently. We also found that only a few joint measures are sensitive to a simultaneous decrease in relevance and increased fairness in the recommendation. Most surprisingly, nearly all joint measures are almost unresponsive to increases in relevance and fairness. Even worse, the majority tend to compress scores at the low end of their range, giving the illusion of an extremely fair recommendation, even when the relevance- and fairness-only scores are close to the theoretical worst value. Based on these findings, we formulated recommendations on the appropriate usage of these measures. 

Future work includes improving the design of joint measures by addressing or mitigating the limitations of the current measures outlined, to have a single score that reflects recommendation relevance and fairness more accurately and in a more balanced way. The individual fairness and relevance measures can also be optimised jointly with multi-objective approach, to obtain both fair and relevant recommendations. Future work can also investigate whether the findings hold when the models are directly optimised for fairness, or when different family of models are used.

\section*{Acknowledgements}
Funded by Algorithms, Data \& Democracy (Villum \& Velux funds).

\chapter{Relevance-aware Individual Item Fairness Measures in
Re\-com\-men\-der Systems: Limitations and Usage Guidelines}
\label{chap:SIGIR24_ext}

\section*{Abstract}
Recommender Systems (RSs) aim to provide relevant items to users, with a recent emphasis on improving recommendation fairness. Quantifying fairness of the recommended items can be done with two types of evaluation measures: measures that are purely based on item exposure (\textit{exposure-based}) and measures that account for both item exposure and item relevance (\textit{relevance-aware}). While exposure-based measures have been thoroughly analysed, relevance-aware measures have not been examined in such detail yet. We gather all existing relevance-aware individual item fairness measures for RSs and study their theoretical properties. We find that all measures suffer from one or more limitations, which may cause issues in their computation, interpretability, or expressiveness. To address this, we correct the affected measures or explain why a limitation is unresolvable. Further, we empirically investigate the extent of the limitations on the measures and compare the original measures to our reformulations under common and extreme evaluation scenarios across real-world and synthetic data. Our experiments show that our reformulated measures successfully resolve the issues in the original measures. 
We conclude by providing practical guidelines on how to select measures for a range of use cases.

\section{Introduction}
\label{ext_s:intro}

Fairness has become an increasingly important evaluation aspect in Recommender Systems (RSs). 
RS fairness can be evaluated on an individual level, where the aim is to ensure that similar individuals receive similar treatments \cite{Dwork2012FairnessAwareness}, or on a group level, which aims to provide similar treatment to both dominant and protected groups (e.g., gender minority) \cite{Deldjoo2024FairnessDirections,Ekstrand2022FairnessSystems}. In this work, we focus on individual fairness, specifically for items.

Individual item fairness in RSs is based on item exposure, i.e., how often an item is recommended to users. 
There are two broad ways of defining fairness for individual items. One definition focuses solely only the exposure distribution of items \cite{Rampisela2024EvaluationStudy}, where the more uniform the exposure distribution across all items, the fairer. Another definition considers both item exposure and item relevance to users, where fairness means that items of similar relevance should be recommended a comparable proportion of times 
\cite{Patro2022FairDirections,Smith2023ScopingPerspective,Biega2018EquityRankings}. Each fairness notion has its own advantages: the exposure-only definition ensures that each item gets a chance to be seen by users, whereas the definition based on exposure and relevance maintains user utility, which is affected by user-item relevance.

Following these two broad definitions, two types of evaluation measures for RS individual item fairness exist: \textit{exposure-only fairness measures}, which do not consider relevance, and \textit{relevance-aware fairness measures}, which consider item exposure w.r.t.~relevance. While exposure-only measures have been thoroughly examined with respect to their limitations \cite{Rampisela2024EvaluationStudy}, to the best of our knowledge, the same has not been done for relevance-aware measures. Given that relevance-aware measures are commonly used in RS fairness evaluation \cite{Wang2023ASystems}, they warrant a closer inspection in terms of their potential strengths and weaknesses (e.g., whether they can be easily interpreted, whether there are cases for which they cannot be used), as well as how they relate to exposure-only fairness measures and to common relevance-only measures, like NDCG.

To this end, in our prior work \cite{Rampisela2024CanRelevance} we empirically studied relevance-aware individual item fairness measures for RSs. In this paper, we extend \citeauthor{Rampisela2024CanRelevance}~\cite{Rampisela2024CanRelevance} as follows:
\begin{enumerate}
    \item We identify five theoretical limitations of those measures and analyse the extent of the limitations in various evaluation scenarios. To the best of our knowledge, we are the first to identify these limitations in the measures.
    \item We amend the measures to resolve the limitations or justify why some limitations cannot be resolved.
    \item We empirically show that our corrected measures overcome the limitations of the original measures in evaluating both real-world and synthetic data. 
    \item We recommend detailed guidelines on measure usage for various practical evaluation scenarios.
\end{enumerate}

\section{Evaluation Measures of Fairness for Individual Items}
\label{ext_s:notation_measures}
We present our notation (\Cref{ext_ss:notation}) and review existing individual item fairness measures: exposure-based measures (\Cref{ext_ss:onlyexp}) and relevance-aware measures (\Cref{ext_ss:exprel}). We also review exposure-based measures as we compare them to relevance-aware measures in our experiments (\Cref{ext_s:setup}).

\subsection{Notation and Examination Functions}\label{ext_ss:notation}

Given a set of $n$ items, $I = \{i_1, i_2, \dots, i_n\}$ and a set of $m$ users, $U = \{u_1, u_2, \dots, u_m\}$, for each user $u \in U$ we rank the $n$ items to obtain the full recommendation list $L_u^n$.\footnote{For simplicity, we assume that all $n$ items are always ranked.} 
We denote the list of user $u$'s top-$k$ recommended items as $L_u^{k}$; for brevity, we write this as $L_u$. If item $i$ is relevant to $u$, we write $r_{u,i}=1$, otherwise $r_{u,i}=0$. The set of all items that are relevant for user $u$ is $R_u^*$. 
Item $i$'s rank position in user $u$'s recommendation list is $z(u,i)$. Some fairness measures are defined for multiple recommendation rounds. In this case, we denote the item $i$'s rank position for user $u$ in round $w$ as $z(u,i,w)$ and item $i$'s relevance to $u$ in round $w$ as $r_{u,i,w}$. We summarise our notation in \Cref{ext_tab:notation}.

Item fairness in RSs is centred on the concept of exposure, which refers to the item appearance in a user's top-$k$ recommendation. Item exposure can be quantified (weighed) using various examination functions $e(\cdot)$, which model the probability of a user viewing an item. All examination functions studied in this work, presented in \Cref{ext_tab:exp-weigh}, assume that the probability only depends on the rank position $p$, or the item position for the user ($z(u,i)$ or $z(u,i,w)$). These functions either apply linear, logarithmic, exponential, or inverse discounting. In general, the exposure weight decreases at the highest rate with the inverse discount and decreases the slowest with the penalty based on Discounted Cumulative Gain (DCG).

The linear examination function, $e_\text{li}$ and its normalised version $\tilde{e}_\text{li}$ discount rank positions up to $k$ linearly \cite{Borges2019EnhancingAutoencoders}. Meanwhile, the examination functions $e_\text{DCG}$ and $e_\text{RBP}$ apply discounts based on DCG \cite{Jarvelin2002CumulatedTechniques} and Rank-Biased Precision (RBP) \cite{Webber2008PrecisionRedundant} respectively. The function $e_\text{RBP}$ has a user patience parameter $\gamma$, i.e., the probability of a user viewing the next item in the rank. The patience parameter has been set at, e.g., $\gamma \in \{0.8,0.9\}$ in previous literature \cite{Wu2022JointRecommendation,Jeunen2021Top-KExposure}. In $e_\text{inv}$, the inverse examination function, the discount factor is based on the inverse of the rank position \cite{Saito2022FairRanking}. 

\begin{table}[tb]
    \caption{Summary of the notation}
    \label{ext_tab:notation}
    \begin{tabular}{ll}\toprule
        Notation &Explanation \\\midrule
        $U=\{u_1, u_2, \dots, u_m\}$ &The set of users \\
        $I = \{i_1, i_2, \dots, i_n\}$ &The set of items \\
        $|U| = m$ &The number of unique users in dataset \\
        $|I| = n$ &The number of unique items in dataset\\
        $p$ & The rank position\\
        $k$ & The cut-off threshold\\
        $W$ & The number of recommendation rounds\\
        $r_{u,i} \in \{0,1\}$ & The relevance of item $i$ to user $u$ \\
        $r_{u,i,w} \in \{0,1\}$ & The relevance of item $i$ to user $u$ in round $w$\\
        $R_{u}^*$ & The set of relevant items for user $u$ \\
        $z(u,i) \in \{1, 2, \dots, n\} $ & The rank position of item $i$ for user $u$\\
        $z(u,i,w) \in \{1, 2, \dots, n\} $ & The rank position of item $i$ for user $u$ in round $w$\\
        $L_{u}^{k} = L_{u}$ & The top-$k$ recommendations for user $u$ \\
        $L_{u,w}$ & The top-$k$ recommendations for user $u$ in round $w$ \\
        $L_{u}^{n}$ & The full recommendation list for user $u$ \\
        $1_\mathcal{A}(x)=1$ if $x \in \mathcal{A}$, else 0 & Indicator function\\
        \bottomrule
    \end{tabular}
\end{table}

\begin{table}
\centering
\caption{Examination functions used by measures studied in this work. We denote as $\tilde{e}$ the min-max normalised examination function.}
\label{ext_tab:exp-weigh}
\begin{tabular}{llll}
\toprule
         & equation & measure & reference                        \\ 
\midrule
linear   & \parbox[t]{5cm}{$e_{\text{li}}(u, i, w) = k+1-z(u,i,w)$ \\
        $\tilde{e}_{\text{li}}(u, i, w) = \frac{ e_{\text{li}}(u, i, w)-1}{k-1} = \frac{k-z(u,i,w)}{k-1}$} & IAA   &  \cite{Borges2019EnhancingAutoencoders}                        \\ 
DCG      & $e_{\text{DCG}}(u, i, w) = 1/\log_2 (z(u,i,w)+1)$ & IFD & \cite{Singh2019PolicyRanking, Oosterhuis2021ComputationallyFairness}   \\
RBP      & $e_{\text{RBP}}(u,i,w) = \gamma^{z(u,i,w)-1}$    & HD, II-F, AI-F   & \cite{Wu2022JointRecommendation, Jeunen2021Top-KExposure}  \\
inverse$^*$  & $e_{\text{inv}}(p) = 1/p$                 & MME, IBO/IWO & \cite{Saito2022FairRanking} \\
\bottomrule
\end{tabular}
\\
{\footnotesize $^*$The inverse examination function directly depends on the rank position $p$.}
\end{table}

\subsection{Exposure-based Fairness Measures}
\label{ext_ss:onlyexp}

Exposure-based individual item fairness measures use the item exposure distribution, without considering item relevance to users. In the top-$k$ recommendation scenario, where each user receives a list of $k$ unique items, instead of evaluating the recommendation lists across all users for their relevance, e.g., with Precision@$k$, exposure-based measures evaluate the
recommendations for individual item fairness based on item exposure. Generally, the recommendation is deemed fairer when the exposure is more uniformly distributed across items. 

Multiple exposure-based measures have been used to quantify individual item fairness in RSs. All of these measures evaluate fairness based on item exposure in the top-$k$ recommendation lists across all users and compare the aggregated item exposure to the total number of items in the dataset. Some measures are adapted from other domains, e.g., the Gini Index was originally developed to quantify income inequality \cite{Gini1912VariabilitaMutabilita}, while the Jain Index was developed to quantify fairness in computer networks \cite{jain1984quantitative}; others were specifically defined for fairness in recommendation, e.g., QF \cite{Zhu2020FARM:APPs}. For a more detailed review of these measures, we refer the reader to \citeauthor{Rampisela2024EvaluationStudy}~\cite{Rampisela2024EvaluationStudy}.

\subsection{Relevance-aware Fairness Measures}
\label{ext_ss:exprel}
We present seven measures that have been used to evaluate fairness considering relevance (\textsc{Joint} measures henceforth). To our knowledge, these are all the \textsc{Joint} measures for RSs published up to 1 July 2024. Each measure employs an examination function (\Cref{ext_tab:exp-weigh}), which quantifies item exposure, and therefore the measure evaluates recommendation fairness of item exposure jointly with item relevance.

We present these measures next. We precede a measure name with \up\ (\down) to indicate that a higher (lower) measure score means a fairer recommendation. The subscript $\cdot_\text{ori}$ denotes the original measure formulation, as opposed to our reformulation that is presented in \Cref{ext_s:resolve}.
Note that all \textsc{Joint} measures, except Hellinger Distance (HD), are defined for multiple rounds of recommendations or stochastic rankings, where the evaluation is performed on a distribution over rankings \cite{Biega2018EquityRankings}.

\subsubsection{Inequity of Amortized Attention (IAA) \cite{Biega2018EquityRankings}}
\label{ext_ss:iaa-ori} 
\down IAA\footnote{The measure is named IAA in \citeauthor{Raj2022MeasuringResults}~\cite{Raj2022MeasuringResults} and L1-norm in \citeauthor{Wang2023ASystems}~\cite{Wang2023ASystems}.} quantifies the aggregated difference between item exposure and item relevance to a user, in a stochastically-generated series of rankings \cite{Biega2018EquityRankings}. 
Intuitively, a sequence of rankings is fair based on IAA when the item exposure is proportional to its relevance to the user. The rank position of the item is a proxy of its exposure. To measure fairness in multiple recommendation rounds, IAA was modified in \citeauthor{Borges2019EnhancingAutoencoders}~\cite{Borges2019EnhancingAutoencoders}: 
\begin{equation}
    \label{ext_eq:iaa-ori}
    \text{IAA}_\text{ori} = \frac{1}{m}\sum\limits_{u \in U}\text{IAA}_\text{ori}(u) 
    \end{equation}%
    \begin{equation}
    \label{ext_eq:iaa-ori-u}
    \text{IAA}_\text{ori}(u) = \frac{1}{n}\frac{1}{W}
        \sum_{i \in I}
        \left|
        \sum_{w=1}^W 1_{L_{u,w}}(i) \cdot \tilde{e_{.}}(u,i,w)
        -  
        \tilde{r}(u,i,w) 
        \right|
\end{equation}

\noindent where $\tilde{e}_{(\cdot)}(u,i,w)$ is the min-max normalised linear examination function $\tilde{e}_{\text{li}}(u, i, w)$ (see Tab.~\ref{ext_tab:exp-weigh}) 
and $\tilde{r}(u,i,w) \in [0,1]$ is the min-max normalised item $i$ relevance to user $u$ in round $w$, $r_{u,i,w}$ \cite{Borges2019EnhancingAutoencoders}. 

The min/max relevance values are based on each item's relevance to a user $u$ in a particular round $w$, i.e., $\min_{i \in I} r_{u,i,w}$. The higher the relevance, the closer the value is to 1. The IAA score is more unfair with a higher difference between item relevance and its exposure. IAA ranges between $[0,1]$. 

\subsubsection{Individual Fairness Disparity (IFD) \cite{Singh2019PolicyRanking,Oosterhuis2021ComputationallyFairness}} 
\down IFD averages the pairwise difference of the combined item exposure and item merit, where merit can be defined as a function of item relevance.\footnote{The item relevance value is taken as the item merit, as per \citeauthor{Singh2019PolicyRanking}~\cite{Singh2019PolicyRanking}.} 
Similar to IAA, IFD is also based on the concept of allocating item exposure considering its relevance. 
However, while IAA is based on the disparity between the exposure of each item and its relevance, IFD measures the difference in exposure allocation between item pairs. 
There are two IFD variations, based on how exposure is combined with relevance: 
IFD$_{\div}$, which divides item exposure by its relevance \cite{Singh2019PolicyRanking}, 
and IFD$_{\times}$, which multiplies them \cite{Oosterhuis2021ComputationallyFairness}. Further, the two variants slightly differ in how they compute the pairwise difference, how they form the set of item pairs, and how they weigh item exposure.\footnote{While exposure is weighed proportional to $e_{\text{DCG}}$ in \citeauthor{Singh2019PolicyRanking}~\cite{Singh2019PolicyRanking}, we directly employ $e_{\text{DCG}}$ for simplicity.}  
Fairness in ranking has been evaluated using both versions \cite{Singh2019PolicyRanking, Oosterhuis2021ComputationallyFairness, Yang2023FARA:Optimization, Yang2023Marginal-Certainty-AwareAlgorithm}. We write IFD$_{(\cdot)}$ or IFD when referring to the measure in general. 
\begin{equation}
\label{ext_eq:IFD}
\text{IFD}_{(\cdot)} = \frac{1}{m} \sum\limits_{u \in U}\text{IFD}_{(\cdot)}(u) 
\end{equation}
\begin{equation}
\label{ext_eq:IFD-u-div}
\text{IFD}_{\div \text{-ori}}(u) = 
\frac{1}{|H_u|} \sum_{(i,i')\in H_u} \max{ \left\{0, 
J_{\div}(u,i) - J_{\div}(u,i')
\right\}
} 
\end{equation}
\begin{equation}
\label{ext_eq:IFD-u-mult}
\text{IFD}_{\times\text{-ori}}(u) = 
\frac{1}{n(n-1)} \sum_{i \in I} \sum_{i' \in I\setminus{i}}
\left[
    J_{\times}(u,i) - J_{\times}(u,i')
\right]^2
\end{equation}
\begin{equation}\label{ext_eq:ifd_j_div}
    J_{\div\text{-ori}}(u,i) = \frac{1}{W} \sum\limits_{w=1}^{W} \frac{e_{\text{DCG}}(u,i,w)}{r_{u,i,w}}
\end{equation}
\begin{equation}\label{ext_eq:ifd_j_x}
    J_{\times\text{-ori}}(u,i) = \frac{1}{W} \sum\limits_{w=1}^{W} 
    r_{u,i,w} \cdot 1_{L_{u,w}}(i) \cdot e_{\text{DCG}}(u,i,w) 
\end{equation}
\noindent where the $J_{(\cdot)}(u,i)$ function combines item $i$'s 
exposure and relevance for user $u$ and $H_u = \{(i,i') \in I\ |\ r_{u,i} \geq r_{u,i'} > 0\}$. 
IFD$_{\div}$ ranges between $[0,\infty)$ and it is 0 when the exposure of each relevant item is exactly proportional to its relevance \cite{Singh2019PolicyRanking}. Based on empirical results, IFD$_{\times}$ ranges between $[0,\infty)$  \cite{Yang2023FARA:Optimization}. 

\subsubsection{Hellinger Distance (HD) \cite{Jeunen2021Top-KExposure}}  

\down HD has been used to quantify the fairness of individual items for top-$k$ contextual bandits by taking the difference between the relevance- and interaction-distributions of the top-$k$ items with the highest (ground truth) relevance \cite{Jeunen2021Top-KExposure}. The interaction probability depends on item exposure, item relevance, and user patience (i.e., the probability of a user viewing the next item in the ranking). 
HD deems a recommendation fair when an item's interaction probability (\Cref{ext_eq:c_p_prime}) is proportional to its relevance probability (\Cref{ext_eq:q_p_prime}). To obtain the relevance and interaction probabilities, for each user, $k$ items with the highest (ground truth) relevance are selected to construct a reference list used in another step. The interaction probability of each item in the reference list depends on its rank position in another item list, which is based on model predictions. In the reference list, for each item at the same position (but for different users), the relevance probabilities are aggregated across users. We therefore obtain two aggregated values for each rank position: relevance and interaction. These aggregated values are then inputted to the distance metric (Eq.~\eqref{ext_eq:HD}).
\begin{equation}\label{ext_eq:HD}
    \text{HD}_\text{ori} = \frac{1}{\sqrt{2}} \sqrt{
     \sum_{p=1}^k \left(\sqrt{q_p'} - \sqrt{c_p'}\right)^2}
\end{equation}
\begin{equation}\label{ext_eq:q_p_prime}
    q_p' = \frac{1}{m} \sum_{u \in U} \sum_{i \in I} 
    \delta\left(z^*(u,i) = p\right) \cdot r'_{u,i}
\end{equation}
\begin{equation}\label{ext_eq:c_p_prime}
     c_p' = \frac{1}{m} \sum_{u \in U} \frac{c^*_{u,p}}{\sum_{\ell=1}^{k}c^*_{u,\ell}}
\end{equation}
\begin{equation}
    c^*_{u,p} = \sum_{i \in I} 
    \delta\left(z^*(u,i) = p\right) \cdot c^{full}_{u,i}
\end{equation}
\begin{equation}
    c^{full}_{u,i} = c'_{u,p}\, \text{if } \exists p : z(u,i)=p\, \text{, otherwise } 0
\end{equation}
\begin{equation}
    c_{u,p} = \sum_{i \in L_u} 
    \delta\left(z(u,i) = p\right) \cdot r_{u,i} \cdot \gamma \ e_{\text{RBP}}(u,i) \cdot s_{u,p}
\end{equation}
\begin{equation}
    s_{u,p} = \prod_{1 \le j < p} 1 - \sum_{i \in L_u} \delta(z(u,i) = j) \cdot r_{u,i} 
\end{equation}
\noindent where $q_p'$ and $c_p'$ are the normalised relevance and interaction probability of the item at position $p$ respectively, and interaction is based on both relevance and exposure. Item $i$'s position sorted based on ground-truth relevance is $z^*(u,i)$. User $u$'s probability of interacting with an item at position $p$, $c_{u,p}$, is based on three variables: $s_{u,p}$, the probability that all items before position $p$ are irrelevant for $u$, and the user patience $\gamma \ e_{\text{RBP}}(u,i)$. 
The user-wise normalised relevance value of item $i$ to user $u$ is $r'_{u,i} = r_{u,i}/\sum_{i \in I} r_{u,i}$ and the user-wise normalised interaction probability is $c'_{u,p} = c_{u,p}/\sum^{k}_{p=1} c_{u,p}$.
The value of $\delta(\cdot) = 1$ if the expression $\cdot$ is true, otherwise it is 0. The range of HD is between 0 and 1 \cite{HellingerMathematics}. 

\subsubsection{Mean Max Envy (MME) \cite{Saito2022FairRanking}}\label{ext_sss:mme} 
\down MME is based on the envy-freeness concept, which deems a recommendation fair if no item is disadvantaged by its own exposure allocation, compared to having any other item's exposure allocation. At a high level, MME can be said to quantify unfairness as the disadvantage suffered by an item if its exposure allocation is swapped with that of another item. 
The disadvantage is measured based on an impact score, which depends on item exposure and relevance: given the full recommendation lists, $L_u^n$ (size $n$) across all users, each item $i$ is first swapped with another item $i'$. For each item, the pre- and post-swap impact scores are computed. If item $i$'s pre-swap score is not less than its post-swap score, we have envy-freeness for item $i$ w.r.t.~item $i'$. MME averages the maximum difference in the impact imposed if item $i$'s exposure allocation is swapped with that of another item $i'$. MME is defined as follows:
\begin{equation}
    \label{ext_eq:mme-ori}
    \text{MME}_\text{ori} = \frac{1}{n} \sum_{i \in I} 
    \left\{
        \max_{i' \in I} Imp_i(i') - Imp_i(i)  
    \right\}
\end{equation}
\begin{equation}
        \label{ext_eq:imp}
        Imp_i(i') = \sum\limits_{u \in U} \sum\limits_{p=1}^k r_{u,i} \cdot e_{\text{inv}}(p) \cdot X_{u,i',p} 
\end{equation}
\begin{equation}
     X_{u,i',p} = \frac{1}{W} \frac{1}{m} \sum\limits_{w=1}^W 1_{L_{u,w}}(i') \cdot \delta(z(u,i',w)=p)  
\end{equation}
where $Imp_i(i')$ is the impact of item $i$ when its exposure allocation is swapped with that of item $i'$, $X_{u,i',p}$ is the probability of recommending item $i'$ to user $u$ at position $p$ across $W$ recommendation rounds, 
and $e_{\text{inv}}(p)$ is the exposure weight of item at position $p$, where the weight is discounted with the inverse examination function (see \Cref{ext_tab:exp-weigh}). The range of MME is $[0,\infty)$. 

\subsubsection{Item Better-Off (IBO) \& Item Worse-Off (IWO) \cite{Saito2022FairRanking}} \up IBO and \down IWO are based on the principle of \textit{dominance over uniform ranking}, where fairness is achieved when every item has a higher impact (as defined in MME \Cref{ext_sss:mme}) under the current ranking policy, than if it were under the uniform random ranking policy $\pi_{unif}$, which randomly samples all possible item permutations. IBO/IWO is the percentage of items for which the current ranking policy improves/worsens impact by at least 10\% compared to $\pi_{unif}$\footnote{The threshold is 10\% in \citeauthor{Saito2022FairRanking}~\cite{Saito2022FairRanking} and in our work, but this can be a variable.}:
\begin{equation}
\label{ext_eq:ibo-ori}
 \text{IBO}_\text{ori} = \frac{100}{n} 
 \sum_{i \in I} 
\delta 
\left(
     \frac{Imp_{i}(i)}{Imp^{unif}_i} \geq 1.1
\right)
\end{equation}
\begin{equation}
\label{ext_eq:iwo-ori}
 \text{IWO}_\text{ori} = 
 \frac{100}{n} 
 \sum_{i \in I} 
\delta 
\left(
     \frac{Imp_{i}(i)}{Imp^{unif}_i} \leq 0.9
\right)
\end{equation}
\begin{equation}
\label{ext_eq:imp-unif}
    Imp^{unif}_{i} =  \frac{1}{m}\frac{1}{n} \sum \limits_{p=1}^{k} \frac{1}{p} \cdot \sum \limits_{u \in U} r_{u,i} 
\end{equation}
\noindent $Imp_i(i)$ is as per Eq.~\eqref{ext_eq:imp} and $Imp^{unif}_i$ is the impact  item $i$ has under the ranking policy $\pi_{unif}$, with $e_{\text{inv}}(p)$ as the examination function (see Tab.~\ref{ext_tab:exp-weigh}). The range of IBO/IWO is $[0,100]$.

\subsubsection{Individual-user-to-Individual-item Fairness (II-F) \cite{Wu2022JointRecommendation}} 
\down II-F was first proposed by \citeauthor{Diaz2020EvaluatingExposure}~\cite{Diaz2020EvaluatingExposure} as an unfairness measure in ranking, where unfairness is the difference between system exposure and target exposure of individual queries and individual items. II-F was reformulated for RSs \cite{Wu2022JointRecommendation} as: 
\begin{equation}
    \label{ext_eq:iif-ori}
     \text{II-F}_\text{ori} = \frac{1}{m} \frac{1}{n} 
     \sum\limits_{u \in U} \sum\limits_{i \in I} \left(E_{u,i}^{ } - E_{u,i}^*\right)^2 
\end{equation}
\begin{equation}
    \label{ext_eq:eui}
E_{u,i} = \frac{1}{W} \sum\limits_{w=1}^{W}1_{L_{u,w}}(i) \cdot e_{\text{RBP}}(u,i,w)
\end{equation}
\begin{equation}
\label{ext_eq:eui-star}
E_{u,i}^{*} = \frac{r_{u,i}}{|R_u^*|} \cdot\frac{1-\gamma^{|R_u^*|}}{1-\gamma}  \, \text{if } |R_u^*|>0\, \text{, otherwise }0
\end{equation}
\noindent where $E_{u,i}$ is the system exposure, i.e., the expected exposure of $i$ to $u$ under a stochastic ranking policy. $E_{u,i}^{*}$ is the target exposure, i.e., the expected exposure of $i$ to $u$ under an ideal stochastic ranking policy, where relevant items receive equal expected exposure~\cite{Diaz2020EvaluatingExposure}. As such, II-F deems a recommendation fair when the system-allocated item exposure is equal to the exposure allocated to items by the ideal stochastic ranking policy. 
The RBP-based examination function (see \Cref{ext_tab:exp-weigh}) is used in $E_{u,i}$ and the formulation of $E_{u,i}^{*}$ also depends on this examination function \cite{Wu2022JointRecommendation}. $|R_u^*|$ is the number of relevant items for user $u$. II-F ranges between $[0,1]$.

\subsubsection{All-users-to-Individual-item fairness (AI-F) \cite{Wu2022JointRecommendation}} 
\down AI-F quantifies how much RSs~ under/overexpose an item to all users by averaging the deviation of the overall system exposure from a target exposure:
\begin{equation}
    \label{ext_eq:aif-ori}
     \text{AI-F}_\text{ori} = 
        \frac{1}{n} \sum \limits_{i \in I}
            \left(
                \frac{1}{m}\sum \limits_{u \in U} E_{u,i}^{ }
                - \frac{1}{m}\sum \limits_{u \in U} E_{u,i}^{*}
            \right)^2
\end{equation}
where the system exposure $E_{u,i}^{ }$ and the target exposure $E_{u,i}^{*}$ are given in \Crefrange{ext_eq:eui}{ext_eq:eui-star} (the same as the ones for II-F). 
As with II-F, AI-F also evaluates fairness based on the disparity between the system exposure and the target exposure. While this difference is computed per user-item pair in II-F, AI-F computes the exposure difference after summing item exposure across users. Consequently, given a larger percentage of unique items in the recommendation (as opposed to recommending the same few items to all users), \down AI-F would give a fairer score than \down II-F. AI-F has a range of $[0,1]$.

\section{Measure Limitations}
\label{ext_s:limitation}

The measures in \Cref{ext_ss:exprel} suffer from theoretical limitations. Following \citeauthor{Rampisela2024EvaluationStudy}~\cite{Rampisela2024EvaluationStudy}, the term ``limitation'' refers to a measure failing to quantify or fulfill criteria that are important in fairness evaluation. Regardless of the reason (e.g., deliberate design choice or unintentional flaw), the limitation restricts the usage of the measure. Several limitations may render the measure completely unusable (\Cref{ext_ss:undefinedness}, \Cref{ext_ss:top_k_insensitivity}), but others can still be used with an awareness of the limitation and how it affects the measure or its interpretation (\Cref{ext_ss:nonreal}, \Cref{ext_ss:non_localisation}, \Cref{ext_ss:zero_exposure}). Here, we consider measure usage in the case of top-$k$ recommendations, as one of the most popular recommendation scenarios.

Altogether, we identify five limitations affecting the measures, summarised in \Cref{ext_tab:limitation-summary}. In each heading below, we put in brackets the names of the measures affected by the limitation. 

\begin{table}[tb] 
\caption{
Measures of relevance-aware individual item fairness and their theoretical limitations. We have identified two causes of non-realisability, denoted by \textit{C} in this table. 
Source refers to where the limitation is first identified. Empty cells mean that the measure does not suffer from the limitation. 
}
\label{ext_tab:limitation-summary}

\resizebox{\textwidth}{!}{
\begin{tabular}{l|c||c|c|c|c|c|c|c|c}
\toprule
\midrule
\parbox[t]{0.6\textwidth}
{Legend\\
\bbullet: we fully resolve the limitation\\
\partialfix: we partially resolve the limitation for single-round and binary relevance cases \\
\nofix: the limitation is unresolvable (\Cref{ext_ss:nofix})
} 
&\rotatebox[origin=r]{90}{Source}
&\rotatebox[origin=r]{90}{IAA \cite{Borges2019EnhancingAutoencoders}}
&\rotatebox[origin=r]{90}{IFD$_{\div}$ \cite{Singh2019PolicyRanking}}
&\rotatebox[origin=r]{90}{IFD$_{\times}$ \cite{Oosterhuis2021ComputationallyFairness}}
&\rotatebox[origin=r]{90}{HD \cite{Jeunen2021Top-KExposure}}
&\rotatebox[origin=r]{90}{MME \cite{Saito2022FairRanking}}
&\rotatebox[origin=r]{90}{IBO/IWO \cite{Saito2022FairRanking}}
&\rotatebox[origin=r]{90}{II-F \cite{Wu2022JointRecommendation}}
&\rotatebox[origin=r]{90}{AI-F \cite{Wu2022JointRecommendation}}
\\
\midrule
\midrule
Non-realisability: cannot reach max/min score (cause number denoted by \textit{C}) &&&&&&&&&\\
\textit{C1.} Non-realisability due to unknown formulation of max/min score &\cite{Rampisela2024EvaluationStudy}&&&&\nofix&\nofix&\nofix&&\nofix \\
\textit{C2.} Non-uniform exposure weight&us&\partialfix&\partialfix&\partialfix&&&&\partialfix& \\
\midrule
Non-localisation: requires item relevance information beyond cut-off $k$ &\cite{Moffat2013SevenMetrics}&\nofix&\nofix&&\nofix&&\nofix&\nofix&\nofix\\
\midrule 
Undefinedness: cannot be computed (undefined value)   
&\cite{Rampisela2024EvaluationStudy}&\bbullet&&&&&\bbullet&&\\

\midrule 
Zero-exposure: gives zero exposure weight to exposed item   &us&\bbullet&&&&&&&\\
\midrule 
Top-$k$-insensitivity: gives nonzero exposure weight to unexposed item $k$  &us&&\bbullet&&&&&&\\
\bottomrule
\bottomrule
\end{tabular}
}
\end{table}

\subsection{Limitation 1: Non-realisability}
\label{ext_ss:nonreal}
All \textsc{Joint} measures are affected by this limitation, which is defined as the measure theoretically failing to reach the max/min score, given all possible item rankings across users \cite{Rampisela2024EvaluationStudy}. The non-realisability limitation causes difficulty in interpreting the fairness score, as the best or worst possible fairness score changes depending on the dataset distribution, (e.g., how many relevant items a user has), or experimental choices (e.g., the cut-off $k$). Thus, how far the fairness score is from the max or min score (based on its range) cannot be known. For example, if a lower-is-fairer measure has a [0,1]-range and a model scores 0.2, one might think that the model is not perfectly fair and that fairness can still be improved. Yet, if a score of 0.2 is already the minimum that can be achieved given any combination of values, it would mean that the recommendation cannot be made fairer; none of the \textsc{Joint} measures scores can show this and their scores may therefore be misleading. In other words, this means that the achievable measure range may not actually be [0,1] as per the theoretical range, but only a subrange of it.

This limitation is first defined in \citeauthor{Rampisela2024EvaluationStudy}~\cite{Rampisela2024EvaluationStudy} in relation to the number of recommendation slots and the number of items in the dataset. Here, we show that the same limitation occurs in relation to the item relevance labels and exposure weights. 
Four causes of non-realisability were previously identified in \citeauthor{Rampisela2024EvaluationStudy}~\cite{Rampisela2024EvaluationStudy}.
We also identify a new cause of non-realisability in the \textsc{Joint} measures. Next, we explain the two causes of non-realisability, one that has been identified in \citeauthor{Rampisela2024EvaluationStudy}~\cite{Rampisela2024EvaluationStudy} and the new cause that we identified.

\subsubsection{Cause 1 (\down HD, \down MME, \up IBO/\down IWO, \down AI-F)} 
\label{ext_sss:cause4}

The first cause of non-realisability is due to the exact formulation of the max/min achievable score being unknown \cite{Rampisela2024EvaluationStudy}: given a set of user-item relevance labels, how do we order items in the recommendation list to produce the maximum or minimum fairness score? In other words, the (un)fairest recommendation cannot be determined solely through theoretical analysis due to the complexity of the measure (e.g., score aggregation per item, various permutations of user-item relevance values and their combination with item exposure weight). While one can always enumerate all permutations to get the achievable max/min scores, it is a highly costly process; thus, in the examples below, we use toy examples with a small number of users and items. Even if our hypothetical examples here are only with small values, we argue that a fairness measure should still work properly given any number of users or items. Further, in real-life settings, there are recommendation use cases with only a few items, e.g., insurance product recommendation \cite{BorgBruun2022LearningDomain}. Evaluating fairness in such cases would need a fairness measure that works properly for a small number of items.
Note that all measures affected by this limitation except HD are defined for multiple rounds of recommendations, but since MME, IBO, and IWO have only been computed under cases equivalent to a single round of recommendation (i.e., a deterministic ranking policy), we only provide examples for a single recommendation round for these measures.

\begin{ex}[HD]
 In principle, \down HD ranges between 0 and 1, but for $k=m=2$ and $n=3$, its maximum score is only 0.707, which is achieved when all top-$k$ items are irrelevant. This maximum score is obtained after iterating through all permutations of the recommendation list and item relevance values while assuming binary relevance $r_{u,i} \in \{0,1\}$. Given the score of 0.707, when comparing it to the theoretical range of [0,1], one may think that the recommendation is rather unfair, but not severely unfair, even if this is the most unfair score that is empirically achievable given the experimental choice and dataset composition. Hence, the non-realisability may hinder an accurate representation of the fairness level.
\end{ex}

\begin{ex}[MME, IBO, IWO] 
The most fair \down MME, $\text{MME}=0$, means that recommendations are envy-free, which we explain in \Cref{ext_sss:mme}. For $k>1$, an envy-free recommendation is not always guaranteed 
\cite{Saito2022FairRanking}, so $\text{MME}=0$ may be unreachable. For example, when $k=n=3$, $m=2$, and when each item is relevant to all users, the minimum achievable \down MME is $\frac{1}{18}$ under a deterministic ranking policy. 

While the upper bound (fairest score) of \down MME is unclear, the fairest scores for \up IBO ($\text{IBO}=1$) and \down IWO ($\text{IWO}=0$), are not always achievable. For example, under a deterministic ranking policy, when $k=n=3$, $m=2$, and for any combinations of relevance values for all users and items, the maximum achievable \down MME for any permutations of recommendation lists across all users is $\frac{7}{18}$, the fairest \up IBO is $\frac{2}{3}$, and the fairest \down IWO is  $\frac{1}{3}$. 

Both \up IBO/\down IWO compare the impact, based on exposure and relevance, produced by a recommendation list to the thresholded impact, based on the impact produced by recommendation lists with uniform exposure. The thresholded impact depends on a multiplying constant which is set arbitrarily and a priori (1.1 for \up IBO and 0.9 for \down IWO in Eq.~\eqref{ext_eq:ibo-ori}--\eqref{ext_eq:iwo-ori}). 
It is thus possible that the optimal distribution of impact in our recommendation list is sometimes less than the thresholded impact (for \up IBO) or more than the thresholded impact (for \down IWO) due to the dataset composition. Intuitively, not all items may have a better impact than the thresholded impact, which causes \up IBO $<1$, and not all items would have a worse impact than the thresholded impact, which causes \down IWO $>0$. So, comparing impacts as above may cause the fairest score to be unreachable.
\end{ex}

\begin{ex}[AI-F] 
\down AI-F may also not reach the unfairest score of 1. For example, given $k=m=2, n=3$, for all permutations of items in a possible recommendation list and user-item binary relevance values, its maximum score is only 0.88. This score is obtained when the same $k$ items are recommended to all users, and both items are irrelevant to each user. Even if the setup is extended to multi-round settings, the maximum AI-F would remain 0.88, obtained from recommending the same $k$ items to all users in all rounds. This is because the target exposure $E_{u,i}^*$ (\Cref{ext_eq:eui-star}) is independent of the number of rounds, while the system exposure $E_{u,i}$ producing the unfairest AI-F in multi-round settings is the same as that of the single-round, as it produces the maximum aggregated difference with the target exposure. 
\end{ex}

\subsubsection{Cause 2 (\down IAA, \down IFD$_{\div}$, \down IFD$_{\times}$, \down II-F)} We identify a second cause of non-realisability, which is due to the items having non-uniform exposure. 
All measures affected by this limitation compare and combine item exposure and relevance through some functions (e.g., subtraction, division, multiplication). When the exposure weight and relevance do not change at the same rate down the ranks, aggregating these scores at the user/item level, e.g., through computing their differences, will not be exactly zero. As such, the resulting fairness score will also have some residuals due to this discrepancy. As a result, a lower-is-fairer score that should have been 0 (the fairest possible) may not be achieved, leading to the misinterpretation that the recommendation is not completely fair, even if it is already the fairest possible, given the dataset composition.

\begin{ex}[IAA]
For \down IAA, in an ideal effectiveness-based ranking, relevance and exposure weight may decrease at different rates down the rank. This causes the fairest score to be unreachable, as the minimum score cannot be achieved if the normalised relevance value does not decrease at the same rate as the examination function down the ranks. 
For example, for $k=2, n=4$, assume a user's full (untruncated) recommendation list with normalised relevance values of $[1,0.8,0,0]$, and the list of normalised exposure weights given by the linear examination function $[1,0,0,0]$. The two exposed items are with the highest relevance among the 4 items, but $\text{IAA}_{\text{ori}}(u)=\frac{0.8}{4} = 0.2 > 0$ even though the recommendations are made as fair as possible. The issue persists when extending this to a multi-round scenario ($W>1$). In the previous example, the fairest $\text{IAA}(u)$ remains 0.2 for any $W$, as the recommendation list is already ordered by decreasing relevance, assuming the relevance $r_{u,i,w}$ for a specific user-item pair stays constant throughout different rounds. Similarly, the most unfair score cannot be achieved if the normalised relevance value does not decrease at the same rate as the exposure weight. In this case, the maximum IAA is only 0.7 instead of 1.\footnote{Obtained by having a recommendation list with non-decreasing normalised relevance values, i.e, $[0, 0, 0.8, 1]$.}
\end{ex}

\begin{ex}[IFD]
\down IFD$_{\div}$ and \down IFD$_{\times}$ may neither reach the fairest score of 0 nor the unfairest score of 1. Under a binary relevance scenario, for $k=n=10$, considering all possible combinations of recommendation list and item relevance values, the fairest \down IFD$_{\div}(u)$ is 0.003, while the unfairest is 0.158.\footnote{Excluding the trivial case where the user only has one relevant item out of all items in the dataset.} Similarly, the fairest \down IFD$_{\times}(u)$ can only be 0.0167 and the unfairest is at most 0.265. For non-binary relevance, suppose $r_{u,i} \in \{0,0.5,1\}$, \down IFD$_{\div}(u)$ can only be 0 when there are exactly two items with nonzero relevance. Otherwise, its minimum is 0.006, and its maximum is 0.428. Under the same setting, \down IFD$_{\times}(u)$ is bounded between 0.004 and 0.265. For similar reasons, these issues would also extend to multi-round scenarios, even though the lower/upper bounds may change (i.e., it may be possible to obtain an (un)fairer score than with single rounds).
\end{ex}

\begin{ex}[II-F and AI-F]
The unfairest and fairest \down II-F score may not be achievable either. For instance, given $k=m=2, n=3$, for all possible recommendation lists and item relevance values (considering only binary relevance), the maximum \down II-F is only 0.88 (obtained through recommending all relevant items at the top-$k$), while its minimum is 0.007 (obtained through recommending all irrelevant items at the top-$k$). We also enumerate all possibilities for two and three recommendation rounds. The minimum II-F can be 0 across two rounds, but it is still not exactly 0, (i.e., 0.0007) if there are three rounds. However, the maximum II-F will still be 0.88 in either case, for the same reason as we explain for AI-F in \Cref{ext_sss:cause4}. Hence, even in multi-round scenarios, the theoretical max/min II-F score may not be achievable.
\end{ex}

\subsection{Limitation 2: Non-localisation (\dws IAA, \dws IFD\texorpdfstring{$_{\div}$}{}, \dws HD, \ups IBO/\dws IWO, \dws II-F, \dws AI-F)}
\label{ext_ss:non_localisation}
The non-localisation limitation is based on the localisation property of effectiveness measures in Information Retrieval (IR) \cite{Moffat2013SevenMetrics}. The localisation property is defined as ``A score at depth $k$ can be computed based solely on knowledge of the documents that appear in the top-$k$'' \cite{Moffat2013SevenMetrics}.
We redefine the property as a measure limitation in RSs, where the measure requires item relevance information (e.g., whether the item is relevant to the user, the total number of relevant items) beyond the top-$k$. While this may be a deliberate design choice, it is still a limitation, as having additional relevant items beyond the top-$k$ may cause the measure score to appear fairer than if there are no additional relevant items. In practice, this may lead to overestimation of fairness, and instability in the evaluation results. Since most RS datasets are highly sparse, with only a few relevant items per user, their fairness evaluation can also be affected by this limitation. 

We explain the limitation through an example. Suppose that given $k=2, n=5$, a user's full recommendation list $L_u^n$ has the normalised relevance values of $[1,1,?,?,?]$, where a `$?$' represents unknown relevance, and the list of normalised exposure weights given by the linear examination function is $[1,0,0,0,0]$. The relevance of an item may be unknown due to, for example, the item not being shown to the user or the item being shown to the user but not rated/interacted with. We refer to items with unknown relevance as unobserved items henceforth. 
Following the convention in IR evaluation \cite{Moffat2008Rank-biasedEffectiveness}, unobserved items are usually regarded conservatively as irrelevant. In this case, \down IAA$(u)=0.2$. However, if all unobserved items are actually relevant instead, \down IAA$(u)$ would be 0.8, and thus the difference from the actual score is 0.6.\footnote{This difference is the same as `residual' in \citeauthor{Moffat2008Rank-biasedEffectiveness}~\cite{Moffat2008Rank-biasedEffectiveness}.} Considering that IAA ranges in $[0,1]$, the large difference is worrying, as a score of 0.8 could be interpreted as highly unfair, while 0.2 is rather fair. 
Even if there is only one unobserved item, the difference in IAA based on the two ways of treating the unobserved items would be 0.2, a non-negligible difference.  

Using the same example as the one used for IAA, we show that IFD$_\div(u)$ also suffers from the same limitation. When all unobserved items are regarded as irrelevant, \down IFD$_\div(u)=0.092$. The opposite treatment will result in a higher \down IFD$_\div(u)$, i.e., 0.114. Worse, having only one additional relevant item can cause an even higher difference in the IFD$_\div(u)$ scores. If the additional relevant item is at position $n$, e.g., $[1,1,0,0,1]$, \down IFD$_\div(u)=0.136$.

IBO, IWO, II-F, AI-F, and HD also suffer from non-localisation. Item relevance information of all user-item pairs is required to compute \Cref{ext_eq:imp-unif} for IBO/IWO and to compute \Cref{ext_eq:eui-star} for II-F/AI-F. On the other hand, HD technically requires only the top-$k$ items with the highest ground truth relevance per user (\Cref{ext_eq:c_p_prime}). However, obtaining this list of items requires (partially) sorting the ground truth relevance of all items, regardless of their position in the predicted ranking.

\subsection{Limitation 3: Undefinedness (IAA, IBO/IWO)} 
\label{ext_ss:undefinedness}

The undefinedness limitation is defined as the measure being incomputable, e.g., the measure giving an undefined value when encountering a specific case (but not necessarily an edge case) \cite{Rampisela2024EvaluationStudy}. In both IAA and IBO/IWO, this happens due to division by zero. Consequently, fairness cannot be evaluated as the measure does not yield any scores. 

For IAA, when $k=1$, the denominator of the linear examination function $\tilde{e}_{\text{li}}(u,i,w)$ is $0$. The choice of $k=1$ can be seen on mobile phone applications, where there is limited screen space and thus only the top recommendation is presented to users \cite{Biega2018EquityRankings,Christakopoulou2018QR:Recommendation} or in conversational RSs \cite{Liu2020TowardsDialogs}. For IBO/IWO, a division by zero occurs when $Imp^{unif}_i=0$. This happens when item $i$ is not relevant for any users (see \Cref{ext_eq:imp-unif}).

Note that IFD$_{\div}$ has been stated to have the zero denominator problem when encountering irrelevant items with $r_{u,i}=0$ (\Cref{ext_eq:ifd_j_div}) \cite{Oosterhuis2021ComputationallyFairness,Yang2023FARA:Optimization}, but this is actually false. The original measure \cite{Singh2019PolicyRanking} only computes fairness based on relevant items. Hence, IFD$_{\div}$ does not suffer from the undefinedness limitation. 

We exclude the trivial, unrealistic cases where there are no users ($m=0$) or no items ($n=0$) for this limitation, as evaluating fairness (or any other dimensions) under these scenarios does not make sense. In addition, we identify the following other cases that cause a measure to be undefined. 
IFD$_{\times}$ is undefined when $n<2$, as it is a pairwise measure. IFD$_{\div}$ is undefined if there is no relevant item for a user (based on the ground truth) as the divisor in \Cref{ext_eq:IFD-u-div} will be zero.\footnote{This is related to the completeness property \cite{Moffat2013SevenMetrics}.} Similarly, HD requires at least one relevant item per user, otherwise, the divisor in \Cref{ext_eq:c_p_prime} will be zero. However, we do not consider these measures to suffer from the undefinedness limitation for the following reasons. IFD$_\times$ does not suffer from undefinedness for $n<2$ as evaluating item fairness when there is only one item does not make sense as only one item can be recommended to any users. IFD$_{\div}$ and HD also do not suffer from undefinedness for users with zero relevant items, as such users are generally not included in the evaluation, which is also the case when evaluating effectiveness for RSs.

\subsection{Limitation 4: Zero-exposure (\dws IAA)} 
\label{ext_ss:zero_exposure}
We define the zero-exposure limitation as the measure's examination function giving zero exposure weight to an item that is recommended at the top-$k$. While this may arguably be an intentional design choice, it is unintuitive and confusing as there is no difference between the exposure weight assigned to an exposed item and an unexposed item. 

For \down IAA, in \Cref{ext_eq:iaa-ori},  $\tilde{e}_{\text{li}}(u,i,w)= 0$ for an (exposed) item at position $k$. The zero value is due to min-max normalisation that re-scales all exposure weights of the top-$k$ recommended items within $[0,1]$. However, an item at $k$ is still exposed; weighing the item with a nonzero exposure value makes more sense, to distinguish its weight from the unexposed item at position $k+1$. 
In practice, assigning zero exposure weight for an item at position $k$ may result in an underestimation of fairness if the item is relevant and vice versa. For example, if an item at $k=10$ has a relevance value of 1, having a positive exposure weight (e.g., 0.1) at that position would result in a 0.9 difference between the item exposure weight and the item relevance. If the exposure weight is 0 instead, the difference would be larger (i.e., 1). In this case, the \down IAA score is higher (more unfair) than it should be.

\subsection{Limitation 5: Top-\texorpdfstring{$k$}{}-insensitivity (\dws IFD\texorpdfstring{$_{\div}$)}{}}
\label{ext_ss:top_k_insensitivity}
We define the limitation of top-$k$-insensitivity as the measure's inability to respond to the change in cut-off $k$. This means that given a set of recommendations across users, the fairness score would remain the same given any values of $k$. A measure that cannot reflect a change in fairness level given a different $k$ is not ideal, as recommendations are commonly evaluated for both effectiveness and fairness at different cut-offs. This limitation can also be seen as the inverse of the zero-exposure limitation, in the sense that a nonzero exposure weight is assigned to an unexposed item.

IFD$_{\div}$ suffers from this limitation as \Cref{ext_eq:ifd_j_div} does not incorporate the function $1_{L_{u,w}}(i)$, which indicates whether item $i$ is recommended at top-$k$. 
Consequently, items at position $k+1$ and beyond are assigned nonzero exposure weights, which is unintuitive. For example, given $n=3$, suppose a user $u$'s full recommendation list, $L_u^n$ has the following relevance values $[0, 1, 1]$. The exposure weights for the three items based on $e_{\text{DCG}}$ are $[1, 0.631, 0.500]$. In this case, given any $k \in \{1,2,3\}$, \down IFD$_{\div}=0.033$. Yet, we may obtain a different IFD$_{\div}$ score if we assign zero exposure weights to the unexposed items. Suppose $k=2$, the exposure weights should have been $[1, 0.631, \underline{0}]$ instead and \down IFD$_{\div}=0.158$, which is higher than the first score. With $k=1$, \down IFD$_{\div}=0$, which is lower than the initial score. Hence, this limitation may lead to the under/overestimation of fairness, as the \down IFD$_{\div}$ score is lower/higher than it should be.

\section{Resolving Limitations}
\label{ext_s:resolve}

We show how to resolve some of the measure limitations, either fully or partially, and explain why some remain unresolvable (\Cref{ext_tab:summary_resolve}). To distinguish between the original version of an evaluation measure $M$ and our modification of it, we use $M_\text{ori}$ and $M_\text{our}$, respectively. When referring to the measure in general, we omit the  $\cdot_\text{ori}$ or $\cdot_\text{our}$ subscripts.

 \subsection{Resolving Non-realisability (Limitation 1) and Top-\texorpdfstring{$k$}{}-insensitivity (Limitation 5)}
\label{ext_ss:resolve_nonreal}

We resolve the \textbf{top-$k$-insensitivity} limitation for the only measure that suffers from it, IFD$_\div$, and we partially resolve \textbf{non-realisability} due to Cause 2 for all measures that suffer from the limitation under Cause 2 (i.e., IAA, IFD, and II-F). We resolve top-$k$-insensitivity by employing the top-$k$ indicator function in the measure. 
We partially resolve non-realisability by normalising the measure scores based on their theoretically achievable minimum and maximum values. Specifically, we min-max normalise the score per user, using the ideal and worst recommendation list (in terms of fairness), similar to the idea of normalising DCG \cite{Jarvelin2002CumulatedTechniques}.\footnote{As the normalisation is done per user, prior to averaging across all users, correcting through posthoc normalisation on the final score of the original measures is not possible.} 
As a result of the normalisation, the scores 0 and 1 refer to the fairest and unfairest possible recommendations, respectively, and the (un)fairest recommendation is unique for each measure. We provide the intuition on obtaining the (un)fairest recommendation list in this subsection, and the mathematical workings in the appendix. In some cases, it is difficult to identify the (un)fairest recommendation directly from the measure equations, so we simulate all possible rankings for various data distributions (with a small number of items) and experimental choices (e.g., $k$). We then compute the measure and identify the rankings that produce the (un)fairest scores. 
\Cref{ext_tab:example_fair_unfair} summarises the strategies to achieve the (un)fairest recommendations per measure.

We resolve the non-realisability limitation (Cause 2) only for single-round cases (or multiple rounds with a deterministic ranking policy) and for binary relevance scenarios, as the aggregation across rounds and non-binary relevance introduce extra complexity in formulating the (un)fairest recommendation list. While this may seem like a constrained setting, both scenarios are popular in RS evaluation \cite{Canamares2020,Wang2023ASystems}. Furthermore, we also exclude the highly unlikely scenario where the relevance value of all $n$ items are the same for a user $u$ ($r_{u,1} = r_{u,2} = \cdots = r_{u,n}$).\footnote{In this case, the particular user $u$ may either be excluded from the computation or the user score may be set to 0 for lower-is-fairer measures and to 1 for higher-is-better measures.}

\begin{table*}[]
\caption{Summary of ranking strategies to achieve the (un)fairest scores of the \textsc{Joint} measures.}
\label{ext_tab:example_fair_unfair}
\centering
\resizebox{0.95\linewidth}{!}{
\begin{tabular}{l|*{2}{p{6cm}}}
\toprule
Measure                            & Fairest & Unfairest \\
\midrule
IAA\our$(u)$ &  As many relevant items as possible at the top-$k$  &  As few relevant items as possible at the top-$k$  \\
\midrule
II-F\our$(u)$  &   As many relevant items as possible at the top-$k$       &    As few relevant items as possible at the top-$k$       \\
\midrule
IFD$_{\div\text{-our}}(u)$    & As few relevant items as possible at the top-$k$        & Half of the relevant items at the top; the other half unexposed          \\
\midrule
IFD$_{\times\text{-our}}(u)$  &   As few relevant items as possible at the top-$k$        &   $^*$As many relevant items as possible at the top-$k$         \\
\bottomrule
\end{tabular}}
\begin{minipage}{0.95\linewidth}
{\footnotesize *Works for most cases, but not all. See \Cref{ext_sss:resolve_ifd_mul}.}
\end{minipage}
\end{table*}

\subsubsection{Inequity of Amortized Attention (IAA)} To resolve the non-realisability limitation in IAA, we compute IAA$_{\min}(u)$, i.e., the fairest possible \down IAA for user $u$, and IAA$_{\max}(u)$, the unfairest possible \down IAA for $u$. The fairest IAA$(u)$ is obtained from a recommendation list that is based on monotonically decreasing item relevance for the user, while the unfairest IAA$(u)$ is based on a list with monotonically increasing item relevance. To minimise IAA$(u)$, we have to minimise the absolute difference between each item's relevance score and its exposure weight. Intuitively, all relevant items ($r_{u,i}=1$, for binary relevance case) should be placed at the top of the list, as item exposure weights at the top-$k$ are nonzero, and the irrelevant items ($r_{u,i}=0$) should be recommended beyond the top-$k$, where the exposure weights are zero. In contrast, maximising IAA$(u)$ requires doing the opposite, i.e., placing the irrelevant items at the top and the relevant items at the bottom. 

We reformulate IAA as follows:
\begin{equation}\label{ext_eq:iaa-our}
    \text{IAA}_{\text{our}} = \frac{1}{m} \sum\limits_{u \in U}\text{IAA}_{\text{our}}(u)
\end{equation}

\begin{equation}\label{ext_eq:iaa-u-our}
    \text{IAA}_{\text{our}}(u) = \frac{\text{IAA}_\text{ori}(u)-\text{IAA}_{\min}(u)}{\text{IAA}_{\max}(u) - \text{IAA}_{\min}(u)}
\end{equation}

\noindent where all IAA$_{(.)}(u)$ in \Cref{ext_eq:iaa-u-our} are computed with the corrected examination function, which we explain later in \Cref{ext_ss:resolve_undef}.

\subsubsection{Individual Fairness Disparity with Division (IFD\texorpdfstring{$_\div$}{})} IFD$_{\div\text{-ori}}$ suffers from \textbf{top-$k$-insensitivity} as it does not employ the function that indicates whether an item $i$ is in the top-$k$ recommendations for user $u$ in round $w$, $1_{L_{u,w}}(i)$. As such, we resolve this limitation by incorporating the indicator function in the $J_{\div\text{-ori}}(u,i)$ function, which combines item exposure and relevance. We reformulate $J_{\div\text{-ori}}(u,i)$  as follows:
\begin{equation}\label{ext_eq:ifd_j_div_our}
    J_{\div\text{-our}}(u,i) = \frac{1}{W} \sum\limits_{w=1}^{W} \frac{1_{L_{u,w}}(i) \cdot e_{\text{DCG}}(u,i,w)}{r_{u,i,w}}
\end{equation}

The above modification zeros out the combined item exposure and item relevance values for items that are ranked beyond the top-$k$. Hence, given a different cut-off $k$, IFD$_{\div\text{-our}}$ (\Cref{ext_eq:ifd-our}) can more accurately quantify the disparity between the combined exposure-relevance scores of items in and out of the top-$k$.

As IFD$_{\div\text{-ori}}$ also suffers from non-realisability, we explain how to normalise its score per user. To obtain the fairest \down IFD$_{\div}(u)$, i.e., IFD$_{\div\text{-}\min}(u)$, we place all relevant items at the bottom of the list, as IFD$_{\div}$ takes the pairwise difference of the exposure weight of only the relevant items (after resolving the top-$k$-insensitivity limitation). Thus, when the relevant items are at the bottom of the list, their exposure weights are (closer to) 0, resulting in the least \down IFD$_{\div}(u)$ score. 

Obtaining the recommendation list with the unfairest \down IFD$_{\div}(u)$, i.e., IFD$_{\div\text{-}\max}(u)$, is not as straightforward. For the base case where there are two relevant items for a user ($|R_u^*|=2$), the maximum difference can be obtained when placing one relevant item at the top, and the other one at the bottom.\footnote{There may be multiple possible recommendation lists that result in the same maximum IFD$_{\div}(u)$ score, this is just one possibility.} Generalising to an even number of $|R_u^*|$, following the same idea, placing half of the relevant items at the top and the rest at the bottom should achieve the maximum IFD$_{\div}(u)$. For example, if the dataset has $n=8$ items, and 4 are relevant to $u$, the items should be placed such that the item relevance is ordered in the following way: $[1,1,0,0,0,0,1,1]$.  

If $|R_u^*|$ is odd, $|R_u^*|$-1 of the items can be recommended with the same strategy for an even number of $|R_u^*|$. The positioning of the remaining item can be closer to either the top or the bottom, e.g., $[1,1,\underline{1},0,0,0,1,1]$ or $[1,1,0,0,0,\underline{1},1,1]$. Through a simulation with various numbers of items, $n\in \{10, 15, 20\}$, cut-off $k \in \{1, 3, 5, 10\}$, and an odd number of relevant items, $|R_u^*| \in \{3,5,\dots,k-1\}$, we find that placing the item either way always produces the unfairest IFD$_{\div}(u)$.

We reformulate IFD$_{\div}(u)$ as follows: 
\begin{equation}\label{ext_eq:ifd-our}
    \text{IFD}_{(\cdot)\text{-our}} = \frac{1}{m} \sum\limits_{u \in U}\text{IFD}_{(\cdot)\text{-our}}(u)
\end{equation}

\begin{equation} \label{ext_eq:IFD_div_our}
    \text{IFD}_{\div\text{-our}}(u) = \frac{\text{IFD}_{\div\text{-ori}}(u)-\text{IFD}_{\div\text{-}\min}(u)}{\text{IFD}_{\div\text{-}\max}(u) - \text{IFD}_{\div\text{-}\min}(u)}
\end{equation}

\noindent where all IFD$_{\div\text{-}(\cdot)}(u)$ in \Cref{ext_eq:IFD_div_our} are computed with our reformulation of the function combining item exposure and relevance, $J_{\div\text{-our}}$ (\Cref{ext_eq:ifd_j_div_our}). Note that this normalisation only applies to users with at least two relevant items ($|R_u^*|\geq 2$). As IFD$_\div(u)$ takes the pairwise difference between the combined exposure-relevance value of a user's relevant items, IFD$_{\div\text{-ori}}(u)$ is always zero when a user only has one relevant item. One way to treat such users is to assign IFD$_{\div\text{-our}}(u)=0$ (the fairest score). Alternatively, they can also be excluded from the evaluation. As their scores would always be zero, these users would pull the score of \down IFD$_{\div\text{-our}}$ down, making the score seem fairer than it should be.

\subsubsection{Individual Fairness Disparity with Multiplication (IFD\texorpdfstring{$_\times$}{})} \label{ext_sss:resolve_ifd_mul}
For IFD$_{\times\text{-ori}}$, we obtain the fairest \down IFD$_{\times\text{-}\min}(u)$ through the same recommendation list that produces IFD$_{\div\text{-}\min}(u)$, i.e., all relevant items are at the bottom. As items at the bottom have less exposure weight than the ones at the top, this ranking minimises the pairwise difference of the item scores, which are the item exposure weight times its relevance. 

Formulating the item ranking with the unfairest IFD$_{\times\text{-ori}}$, i.e., \down IFD$_{\times\text{-}\max}(u)$ is not feasible through analysing the measure equation. To find the ranking that produces IFD$_{\times\text{-}\max}(u)$, we simulate rankings with various numbers of items, $n\in \{10, 15, 20\}$, cut-off $k \in \{1, 3, 5, 10\}$, and relevant items, $|R_u^*| \in \{2,3,\dots,k-1\}$. For each combination of the experimental choices, we permute all possible rankings, compute IFD$_{\times\text{-ori}}$ for each ranking, and identify which rankings produce the unfairest score. 
We find that the items can be ranked in at least one of these two ways: 
(i) all relevant items at the top; or 
(ii) $a$ items at the top, and $|R_u^*|-a$ items at the bottom. 
Our simulation shows that strategy (i) produces the unfairest score for small cut-offs $k \in \{1, 3\}$. 
As $|R_u^*|$ increases, we observe that strategy (i) fails and strategy (ii) succeeds in achieving the unfairest score; this happens at different amounts of $|R_u^*|$, given a different number of items $n$.\footnote{We find that with a larger $n$, strategy (i) starts failing at a larger $|R_u^*|$.} 
For example, given $n=20$ and $k=10$, strategy (i) works for $|R_u^*| \in \{2, 3, \dots, 6\}$, and for $|R_u^*|\in \{7, \dots, n-k\}$ onwards, the unfairest recommendation list is in the form of (ii), i.e., placing 6 relevant items at the top, and $|R_u^*|-6$ items at the bottom. For $|R_u^*|\in \{n-k+1, \dots, n-1\}$, i.e., the general form is $|R_u^*|-(n-k) = |R_u^*|-10$ items at the top, and the rest at the bottom. 
There are some exceptions, where this increasing pattern does not hold. For example, given $k=n=10$, for $|R_u^*| \in \{6,7\}$, strategy (ii) works, i.e., placing 5 items at the top and 1 item at the bottom for $|R_u^*|=6$; we denote this as 5--1. For $|R_u^*|=7$, the list is in the form of 5--2, but for $|R_u^*|=8$, the list is in the form of 4--4 instead, which deviates from the pattern. However, this deviation is unexpected as the evaluation scenario is highly unlikely (i.e., the cut-off is the same as the number of items), and in any case, the scores of the 4--4 and 5--3 lists only differ by $\sim$0.0006.

To find IFD$_{\times\text{-}\max}(u)$ (or a good estimate of it), we precompute IFD$_{\times\text{-ori}}(u)$ with the two strategies for a given $k, m \text{, and } |R_u^*|$, and take the maximum as IFD$_{\times\text{-}\max}(u)$. 
We then compute IFD$_{\times\text{-our}}$, which resolves the non-realisability limitation in the measure, by averaging IFD$_{\times\text{-our}}(u)$ across all users: 
\begin{equation}
    \text{IFD}_{\times\text{-our}}(u) = \frac{\text{IFD}_{\times\text{-ori}}(u)-\text{IFD}_{\times\text{-}\min}(u)}{\text{IFD}_{\times\text{-}\max}(u) - \text{IFD}_{\times\text{-}\min}(u)}
\end{equation}

\subsubsection{Individual-user-to-Individual-item Fairness (II-F)} To resolve the non-realisability in II-F, we first rewrite II-F (\Cref{ext_eq:iif-ori}) as follows, to separate the II-F score per user:
\begin{equation}
     \text{II-F}_{(\cdot)} = \frac{1}{m} \sum\limits_{u \in U} \text{II-F}_{(\cdot)}(u)
\end{equation}
 \begin{equation}\label{ext_eq:iif_u_ori}
    \text{II-F}_{\text{ori}}(u) =  \frac{1}{n} \sum\limits_{i \in I} \left(E_{u,i}^{ } - E_{u,i}^*\right)^2 
 \end{equation}
We then min-max normalise II-F$(u)$:
\begin{equation}\label{ext_eq:iif-u-our}
    \text{II-F}_{\text{our}}(u) = \frac{\text{II-F}_\text{ori}(u)-\text{II-F}_{\min}(u)}{\text{II-F}_{\max}(u) - \text{II-F}_{\min}(u)}
\end{equation}
To obtain II-F$_{\min}(u)$, i.e., the fairest \down II-F$(u)$, all relevant items should be ranked at the top, while to obtain II-F$_{\max}(u)$, i.e., the unfairest \down II-F$(u)$, all relevant items should be ranked at the bottom. This is because minimising the sum of squares in \Cref{ext_eq:iif_u_ori} requires $E_{u,i}$, i.e., the exposure weight of item $i$ in user $u$'s recommendation list to be as close as possible to $E_{u,i}^*$, which is computed based on item relevance. Therefore, relevant items (with $r_{u,i}=1$) should be placed at the top, so that they are subtracted with nonzero exposure weights, and irrelevant items ($r_{u,i}=0$) should be ranked below, such that the difference between exposure weight and  $E_{u,i}^*$ is minimised. Following a similar line of reasoning, the opposite should be done to maximise the sum of squares. We also verify through simulations with $\gamma=0.8$, $n\in \{10, 15, 20\}$, cut-off $k \in \{1, 3, 5, 10\}$, and relevant items, $|R_u^*| \in \{2,3,\dots,k-1\}$ that these ranking strategies always produce the min/max II-F$(u)$, or close to that (i.e., with a negligible difference of <$10^{-3}$). II-F\our{}, which is computed with the min-max normalised II-F$(u)$ (\Cref{ext_eq:iif-u-our}), resolves the non-realisability limitation in II-F\ori. 

\subsection{Resolving Undefinedness (Limitation 2) and Zero-ex\-po\-su\-re (Limitation 4)} 
\label{ext_ss:resolve_undef}

We resolve \textbf{undefinedness} for all measures that suffer from it (i.e., IAA, IBO, and IWO) and we resolve \textbf{zero-exposure} for IAA, which is the only measure that has it.

For IAA$_\text{ori}$, the cause of \textbf{undefinedness} and \textbf{zero-exposure} is the erroneous min-max normalisation in the linear examination function (see Tab.~\ref{ext_tab:exp-weigh}). We correct the normalisation by substituting $k$ with $k+1$ in the linear examination function (Tab.~\ref{ext_tab:exp-weigh}), which results in:
\begin{equation}
\label{ext_eq:our-norm-e-li}
    \tilde{e}_{\text{our-li}}(u, i, w) = \frac{ e_{\text{li}}(u, i, w)-1}{k} = \frac{k+1-z(u,i,w)}{k}
\end{equation}
This modification ensures that items at rank $\leq k$ receive nonzero exposure and items at $k+1$ onwards receive zero exposure. 
Accordingly, the IAA corrected for the non-realisability limitation  (IAA$_\text{our}$, \Cref{ext_eq:iaa-our}) needs to be computed with this corrected examination function. 

In the following example, we show how using the corrected examination function may result in a different score than the original examination function. For $k=2, n=4$, assume a user's recommendation list with normalised relevance values of $[0.8,1, 0, 0]$. The normalised exposure weights given by the original examination functions are $[1,\underline{0},0,0]$, while the ones given by the corrected linear examination function $\tilde{e}_{\text{our-li}}(u, i, w)$ are $[1,\underline{0.5},0,0]$. The score of IAA$_\text{our}(u)=0.2$ with the former, whereas IAA$_\text{our}(u)=0.133$ with the latter. 

For IBO\ori/IWO\ori, to avoid \textbf{undefinedness} due to division by zero, we move the divisor $Imp_i^{unif}$ in the left-hand side of the inequalities in \Cref{ext_eq:ibo-ori} \&  \Cref{ext_eq:iwo-ori} to the right-hand side. We reformulate Eq.~\eqref{ext_eq:ibo-ori} \&  \eqref{ext_eq:iwo-ori} as follows: 
  \begin{equation}
    \label{ext_eq:ibo-our}
     \text{IBO}_\text{our} = \frac{100}{|I^-|} 
     \sum_{i \in I^-} 
   \delta\left(Imp_i(i) \geq 1.1 \cdot Imp_i^{unif}\right)
\end{equation}
\begin{equation}
    \label{ext_eq:iwo-our}
     \text{IWO}_\text{our} = 
     \frac{100}{|I^-|} 
     \sum_{i \in I^-} 
    \delta\left(Imp_i(i) \leq 0.9 \cdot Imp_i^{unif}\right) 
\end{equation}  
where $I^{-}$ is the set of items with at least one user that finds the item relevant. This ensures that the set of items that cause $\delta(\cdot)=1$ in IBO\our~ is disjoint from that in IWO\our. We exclude items which no users would find relevant, as for these items $Imp_i(i)=Imp^{unif}_i=0$, causing the same items to be considered `better-off' and `worse-off' at the same time. Otherwise, this results in problems in the interpretability of the scores as some items may be double counted in IBO\ori~and IWO\ori.

\begin{table}[]
    \centering
    \caption{Summary of how we fully or partially resolve the limitations. Dash `--' means that the limitation does not apply.}
    \label{ext_tab:summary_resolve}
    \resizebox{\linewidth}{!}{
    \begin{tabular}{l|>{\raggedright}*{4}{p{3cm}}<{}}
        \toprule
        Measure    & Non-realisability & Undefinedness & Zero-exposure & Top-$k$-insensitivity \\
        \midrule
        IAA & Min-max normalising score per user& Redefining the linear examination function $e_{\text{li}}$& Redefining the linear examination function $e_{\text{li}}$& \NA   \\
        \midrule
        IFD$_\div$ & \textit{idem} & \NA& \NA & Adding top-$k$ indicator function \\
        \midrule
        IFD$_\times$ & \textit{idem} & \NA & \NA & \NA\\
        \midrule
        II-F & \textit{idem} & \NA& \NA& \NA\\
        \midrule
        IBO/IWO & \NA& Excluding items that no users will find relevant & \NA& \NA\\
        \bottomrule
    \end{tabular}}
\end{table}

\subsection{Unresolvable Limitations}
\label{ext_ss:nofix}

We have resolved some measure limitations and summarised these in \Cref{ext_tab:summary_resolve}. We explain why the rest are unresolvable.

\noindent\textbf{Non-realisability} (Cause 1) cannot be resolved as it is unknown what kind of recommendation list produces the (un)fairest score \cite{Rampisela2024EvaluationStudy}. Without this formulation of the min/max score, the measures cannot be normalised to be between 0 and 1. 
The measures that suffer from non-realisability under Cause 1 are more complex than the ones under Cause 2; the former aggregate scores across items and tend to have more variables than the latter. As such, no explicit solution or mathematical formulation of the measures' upper and/or lower bounds has been determined. Furthermore, computationally solving for the bounds (e.g., through constrained optimization) is also impractical, considering the large size of a typical RS dataset. Technically, if the theoretically achievable best and worst fairness scores are known, the non-realisability in HD, MME, IBO/IWO, and AI-F can be resolved in a similar manner to the measures in \Cref{ext_ss:resolve_nonreal}.

\noindent\textbf{Non-localisation} cannot be resolved as the measures require item relevance information beyond the top-$k$ by their definitions. In previous work, to avoid having unknown item relevance information beyond the top-$k$, the model-predicted relevance is used instead of the relevance based on the test set (e.g., IAA \cite{Biega2018EquityRankings}, HD \cite{Jeunen2021Top-KExposure}). Consequently, the fairness score depends on how well the model estimates relevance, and will also be incomparable to that produced with other models that may have different estimations of relevance. When comparing multiple models, even if the estimated relevance from one model is used to compute the fairness score for all models, this raises the question of which model should be used as a reference. 
Alternatively, one can also report the possible measure range, in a similar way to RBP residuals \cite{Moffat2008Rank-biasedEffectiveness}. However, computing the \joint{} measures range may be more complex than the RBP range. Unlike the RBP range, which narrows when given more document relevance labels, the \joint{} measure range may widen with fewer missing labels beyond the top-$k$, due to, e.g., pairwise comparison (\Cref{ext_ss:non_localisation}). Thus, reporting the range of the \joint{} measures can be challenging.

For IFD$_{\div}$, II-F, and AI-F, non-localisation can technically be resolved by modifying the measure partially. For example, if we have a fusion of IFD$_{\div}$ and IFD$_{\times}$, where the exposure-relevance aggregation is still done through the division operation, but for all item-pairs, instead of only pairs of relevant items, then this modification would result in a measure that does not suffer from non-localisation. For II-F and AI-F, the `target exposure' (\Cref{ext_eq:eui-star}) can be changed to another target that does not involve the number of relevant items for a user. However, all these approaches are substantial changes beyond simple corrections, which may fundamentally alter the behaviour or the concept behind the measure.

\section{Experimental Setup}
\label{ext_s:setup}

We study the original \textsc{Joint} measures (\Cref{ext_ss:exprel}) and compare them to our corrected versions (\Cref{ext_s:resolve}), which resolve some limitations of the original measures (see \Cref{ext_tab:limitation-summary} for the summary of limitations and \Cref{ext_tab:summary_resolve} for the summary of the resolutions). We also investigate the robustness of the original measures by analysing the extent of the limitations that they suffer from. Here, we provide our general experimental setup, and we further describe our experiments in \Cref{ext_s:experiment}.\footnote{Our code: \href{https://github.com/theresiavr/relevance-aware-item-fairness-measures-recsys}{https://github.com/theresiavr/relevance-aware-item-fairness-measures-recsys}.}

\subsubsection*{Datasets} We use four real-world datasets with varying sizes from several domains: music (Lastfm) \cite{Cantador20112nd2011}, e-commerce
(Amazon Luxury Beauty, i.e., Amazon-lb) \cite{Ni2019JustifyingAspects},  
videos (QK-video) \cite{Yuan2022Tenrec:Systems}, and
movies (ML-10M) \cite{Harper2015TheContext}. All of the datasets are as provided by \citeauthor{Zhao2021RecBole:Algorithms}~\cite{Zhao2021RecBole:Algorithms}, except for QK-video, which is obtained from \cite{Yuan2022Tenrec:Systems}. QK-video has several interaction types; we only use the `sharing' interactions.

\begin{table}[tb]
\caption{Statistics of 
the preprocessed datasets.}
\label{ext_tab:dataset}
\centering
\begin{tabular}{lrrrr}
\toprule
dataset & \multicolumn{1}{l}{\#users ($m$)} & \multicolumn{1}{l}{\#items ($n$)} & \multicolumn{1}{l}{\#interactions} & \multicolumn{1}{l}{sparsity (\%)} \\ 
\midrule                                                           
Lastfm \cite{Cantador20112nd2011} & 1,859 & 2,823 & 71,355 & 98.64 \\
Amazon-lb \cite{Ni2019JustifyingAspects} & 1,054 & 791 & 12,397 & 98.51 \\
QK-video \cite{Yuan2022Tenrec:Systems} & 4,656 & 6,423 & 51,777 & 99.83 \\ 
ML-10M \cite{Harper2015TheContext} & 49,378 & 9,821 & 5,362,685 & 98.89 \\
\bottomrule
\end{tabular}
\end{table}

\subsubsection*{Preprocessing} We retain users and items with at least 5 interactions (5-core filtering). To handle duplicate interactions, we keep the most recent one. For Amazon-lb and ML-10M, we convert ratings equal/above 3 to 1 and discard the rest, as their ratings range between $[1,5]$ and $[0.5, 5]$, respectively. Lastfm and QK-video are not converted as they only have implicit feedback. Tab.~\ref{ext_tab:dataset} presents statistics of the preprocessed datasets. 

\subsubsection*{Data Splits} To form the train/val/test sets, we apply global temporal splits with a ratio of 6:2:2 on the preprocessed Amazon-lb and ML-10M \cite{Meng2020ExploringModels}. As Lastfm and QK-video have no timestamps, we use global random splits with the same ratio. From all splits, we remove users with less than 5 interactions in the train set. 

\subsubsection*{Recommenders} We rank items with four well-known top-$k$ recommenders: 
item-based K-Nearest Neighbour (ItemKNN) \cite{Deshpande2004Item-basedAlgorithms}, Bayesian Personalised Ranking (BPR) \cite{RendleBPR:Feedback}, Variational Autoencoder with multinomial likelihood (MultiVAE) \cite{Liang2018VariationalFiltering}, and Neighbourhood-enriched Contrastive Learning (NCL) \cite{Lin2022ImprovingLearning}. 
 BPR, MultiVAE, and NCL are trained for 300 epochs with early stopping using RecBole \cite{Zhao2021RecBole:Algorithms}. The final model is the model with the best Normalised Discounted Cumulative Gain at 10 (NDCG@10) on the validation split.\footnote{We provide the hyperparameter search space and the best values in the code repository.} To generate recommendations during testing, we exclude each user's train/val items, and rank the rest of the items.

\subsubsection*{Fair Re-rankers}  
The recommenders above are not trained to optimise for fairness, so the items are re-ranked to make the recommendations fairer. We re-rank the top-$k'$ items to increase the exposure of items that are originally ranked beyond the top-$k$. We set $k'$ to be greater than the cut-off $k=10$. However, to maintain recommendation relevance, $k'$ should not be overly big (e.g., 100) as a typical RS dataset has very few relevant items per user. For all datasets and recommenders, we set $k'=25$. We re-rank items per user with COMBMNZ (CM) \cite{Lee1997AnalysesCombination}, a robust rank fusion method.\footnote{Other re-rankers exist but do not fit our setup, e.g., item similarity computation is needed \cite{Wang2022ProvidingSystems}, but true similarity is hard to obtain \cite{Dwork2012FairnessAwareness,Tsepenekas2023ComparingDistributions}.} 
CM creates a new ranking for each user by fusing two lists of scores: one based on relevance and another based on exposure. We min-max normalise the predicted relevance score to get the relevance-based score. We create the fairness-based scores from the coverage of each top-$k'$ item, i.e., based on how many times the item appears in the top-$k$. We then compute 1 minus the normalised coverage to assign a higher score for less exposed items, thus increasing fairness. After the two lists of scores are fused, they are sorted to generate the final ranking that considers both item relevance and item fairness. 

\subsubsection*{Measures} 
We evaluate the models with all relevance-aware fairness measures (\textsc{Joint} measures) presented in \Cref{ext_ss:exprel} and their corrected versions if they exist (\Cref{ext_s:resolve}). For IAA, we use the ground truth relevance as item relevance. For IFD$_{\div\text{-our}}$, we define IFD$_{\div\text{-our}}(u)=0$ if $u$ only has 1 relevant item in the test split. We set $\gamma=0.9$ for HD as per 
\cite{Jeunen2021Top-KExposure} while $\gamma=0.8$ for II-F and AI-F, following 
\cite{Wu2022JointRecommendation}. We rescale IBO/IWO into [0,1] to be consistent with other measures' scale. 
For comparison to the \textsc{Joint} measures, we evaluate recommendation effectiveness (\textsc{Eff}) with: Hit Rate (HR), Mean Reciprocal Rank (MRR), Precision (P), Recall (R), Mean Average Precision (MAP), and NDCG. We also evaluate fairness (\textsc{Fair}) with measures that consider only item exposure (without factoring in item relevance):
Jain Index (Jain) \cite{jain1984quantitative,Zhu2020FARM:APPs}, Qualification Fairness (QF) \cite{Zhu2020FARM:APPs}, Entropy (Ent) \cite{Patro2020FairRec:Platforms,Shannon1948ACommunication},
Fraction of Satisfied Items (FSat) \cite{Patro2020FairRec:Platforms}, and Gini Index (Gini) \cite{Gini1912VariabilitaMutabilita,Mansoury2020FairMatch:Systems}. We compute the corrected \textsc{Fair} measures as per \citeauthor{Rampisela2024EvaluationStudy}~\cite{Rampisela2024EvaluationStudy}. We compute all measures at $k=10$, unless stated otherwise.

\section{Empirical Analysis}
\label{ext_s:experiment}

We evaluate recommendation performance with all \textsc{Eff}, \textsc{Fair}, and \textsc{Joint} measures (\Cref{ext_ss:performance}) and study their correlations (\Cref{ext_ss:corr}). We compare the achievable scores of the original \textsc{Joint} measures and our corrections (\Cref{ext_ss:exp_maxmin}). We analyse their sensitivity to various cut-offs (\Cref{ext_ss:k_sensitive}), varying number of missing relevance labels (\Cref{ext_ss:exp_nonlocal}), and different recommendation effectiveness and exposure distribution (\Cref{ext_ss:insert}).

\subsection{Evaluation Results of All Measures}
\label{ext_ss:performance}

\begin{table}
    \centering
    \caption{ 
    Effectiveness (\textsc{Eff}), fairness (\textsc{Fair}), and \textsc{Joint} scores at $k=10$ without and with re-ranking the top 25 items using COMBMNZ (CM). Bold indicates the most effective/fair score per measure. The score 0.000 does not mean that the scores are exactly 0; this is because the scores are small ($<10^{-3}$) and rounded to 3 d.p. The term `nan' stands for ``not a number''.
    }
    \label{ext_tab:base-rerank-all}
    \resizebox{0.6\textwidth}{!}{
    \begin{tabular}[t]{lll*{2}{r}|*{2}{r}|*{2}{r}|*{2}{r}}
    \toprule
     &  & model & \multicolumn{2}{c|}{ItemKNN} & \multicolumn{2}{c|}{BPR} & \multicolumn{2}{c|}{MultiVAE} & \multicolumn{2}{c}{NCL} \\ 
    \midrule
     &  & re-ranker & - & CM & - & CM & - & CM & - & CM \\
    \midrule
    \multirow[c]{26}{*}{\rotatebox[origin=c]{90}{Lastfm}} & \multirow[c]{6}{*}{\rotatebox[origin=c]{90}{\textsc{Eff}}} & $\uparrow$ $\text{HR}$ & 0.765 & 0.581 & 0.773 & 0.587 & 0.778 & 0.523 & \bfseries 0.793 & 0.571 \\
     &  & $\uparrow$ $\text{MRR}$ & 0.484 & 0.270 & 0.492 & 0.280 & 0.476 & 0.232 & \bfseries 0.503 & 0.260 \\
     &  & $\uparrow$ $\text{P}$ & 0.172 & 0.089 & 0.178 & 0.092 & 0.176 & 0.076 & \bfseries 0.184 & 0.087 \\
     &  & $\uparrow$ $\text{MAP}$ & 0.137 & 0.053 & 0.141 & 0.058 & 0.138 & 0.045 & \bfseries 0.148 & 0.050 \\
     &  & $\uparrow$ $\text{R}$ & 0.218 & 0.114 & 0.224 & 0.119 & 0.224 & 0.098 & \bfseries 0.234 & 0.110 \\
     &  & $\uparrow$ $\text{NDCG}$ & 0.245 & 0.119 & 0.252 & 0.126 & 0.247 & 0.102 & \bfseries 0.261 & 0.115 \\
    \cline{2-11}
     & \multirow[c]{5}{*}{\rotatebox[origin=c]{90}{\textsc{Fair}}} & $\uparrow$ $\text{Jain}$ & 0.042 & 0.094 & 0.058 & 0.140 & 0.097 & \bfseries 0.222 & 0.082 & 0.215 \\
     &  & $\uparrow$ $\text{QF}$ & 0.474 & \bfseries 0.679 & 0.362 & 0.528 & 0.517 & 0.678 & 0.453 & 0.657 \\
     &  & $\uparrow$ $\text{Ent}$ & 0.589 & 0.735 & 0.610 & 0.740 & 0.707 & \bfseries 0.826 & 0.671 & 0.810 \\
     &  & $\uparrow$ $\text{FSat}$ & 0.129 & 0.216 & 0.147 & 0.228 & 0.202 & \bfseries 0.321 & 0.178 & 0.286 \\
     &  & $\downarrow$ $\text{Gini}$ & 0.904 & 0.790 & 0.910 & 0.818 & 0.839 & \bfseries 0.696 & 0.872 & 0.728 \\
    \cline{2-11}
     & \multirow[c]{15}{*}{\rotatebox[origin=c]{90}{\textsc{Joint}}} & $\uparrow$ $\text{IBO}_{\text{ori}}$ & nan & nan & nan & nan & nan & nan & nan & nan \\
     &  & $\uparrow$ $\text{IBO}_{\text{our}}$ & 0.209 & 0.256 & 0.208 & 0.253 & 0.261 & 0.278 & 0.242 & \bfseries 0.292 \\
     &  & $\downarrow$ $\text{IWO}_{\text{ori}}$ & nan & nan & nan & nan & nan & nan & nan & nan \\
     &  & $\downarrow$ $\text{IWO}_{\text{our}}$ & 0.791 & 0.744 & 0.792 & 0.747 & 0.739 & 0.722 & 0.758 & \bfseries 0.708 \\
     &  & $\downarrow$ $\text{IAA}_{\text{ori}}$ &\bfseries  0.004 & \bfseries 0.004 & \bfseries 0.004 & \bfseries 0.004 & \bfseries 0.004 & \bfseries 0.004 & \bfseries 0.004 & \bfseries 0.004 \\ 
     &  & $\downarrow$ $\text{IAA}_{\text{our}}$ & 0.764 & 0.887 & 0.757 & 0.881 & 0.762 & 0.903 & \bfseries 0.749 & 0.894 \\
     &  & $\downarrow$ $\text{IFD}_{\div\text{-ori}}$ & 0.074 & 0.053 & 0.075 & 0.054 & 0.073 & \bfseries 0.049 & 0.076 & 0.052 \\
     &  & $\downarrow$ $\text{IFD}_{\div\text{-our}}$ & 0.442 & 0.241 & 0.453 & 0.250 & 0.445 & \bfseries 0.205 & 0.464 & 0.232 \\
     &  & $\downarrow$ $\text{IFD}_{\times\text{-ori}}$ & \bfseries 0.000 & \bfseries 0.000 & \bfseries 0.000 & \bfseries 0.000 & \bfseries 0.000 & \bfseries 0.000 & \bfseries 0.000 & \bfseries0.000 \\
     &  & $\downarrow$ $\text{IFD}_{\times\text{-our}}$ & 0.275 & 0.126 & 0.282 & 0.134 & 0.273 & \bfseries 0.107 & 0.290 & 0.122 \\
     &  & $\downarrow$ $\text{HD}_{\text{ori}}$ & 0.099 & 0.177 & 0.104 & 0.174 & 0.095 & 0.203 & \bfseries 0.092 & 0.177 \\
     &  & $\downarrow$ $\text{MME}_{\text{ori}}$ & \bfseries 0.001 & \bfseries 0.001 & \bfseries0.001 & \bfseries0.001 & \bfseries0.001 & \bfseries0.001 & \bfseries0.001 & \bfseries 0.001 \\
     &  & $\downarrow$ $\text{II-F}_{\text{ori}}$ & \bfseries0.001 & 0.002 & \bfseries0.001 & 0.002 & \bfseries0.001 & 0.002 & \bfseries 0.001 & 0.002 \\
     &  & $\downarrow$ $\text{II-F}_{\text{our}}$ & 0.755 & 0.884 & 0.748 & 0.878 & 0.753 & 0.901 & \bfseries 0.739 & 0.890 \\
     &  & $\downarrow$ $\text{AI-F}_{\text{ori}}$ & \bfseries0.000 & \bfseries0.000 & \bfseries0.000 & \bfseries0.000 & \bfseries0.000 & \bfseries0.000 & \bfseries0.000 & \bfseries 0.000 \\
    \cline{1-11}
    \multirow[c]{26}{*}{\rotatebox[origin=c]{90}{Amazon-lb}} & \multirow[c]{6}{*}{\rotatebox[origin=c]{90}{\textsc{Eff}}} & $\uparrow$ $\text{HR}$ & \bfseries 0.046 & 0.016 & 0.011 & 0.021 & 0.039 & 0.014 & 0.034 & 0.011 \\
     &  & $\uparrow$ $\text{MRR}$ & 0.020 & 0.011 & 0.003 & 0.007 & \bfseries 0.023 & 0.004 & 0.022 & 0.003 \\
     &  & $\uparrow$ $\text{P}$ & \bfseries 0.005 & 0.002 & 0.001 & 0.002 & 0.004 & 0.002 & 0.004 & 0.001 \\
     &  & $\uparrow$ $\text{MAP}$ &  \bfseries0.006 & 0.004 & 0.002 & 0.004 &  \bfseries 0.006 & 0.003 & \bfseries 0.006 & 0.001 \\
     &  & $\uparrow$ $\text{R}$ & \bfseries 0.013 & 0.005 & 0.005 & 0.010 & 0.010 & 0.008 & 0.012 & 0.003 \\
     &  & $\uparrow$ $\text{NDCG}$ & \bfseries 0.011 & 0.005 & 0.003 & 0.006 & 0.010 & 0.004 & \bfseries 0.011 & 0.002 \\
    \cline{2-11}
     & \multirow[c]{5}{*}{\rotatebox[origin=c]{90}{\textsc{Fair}}} & $\uparrow$ $\text{Jain}$ & 0.271 & \bfseries 0.431 & 0.223 & 0.359 & 0.035 & 0.097 & 0.026 & 0.080 \\
     &  & $\uparrow$ $\text{QF}$ & \bfseries 0.650 & 0.612 & 0.549 & 0.594 & 0.222 & 0.286 & 0.229 & 0.310 \\
     &  & $\uparrow$ $\text{Ent}$ & 0.802 & \bfseries 0.839 & 0.747 & 0.809 & 0.418 & 0.558 & 0.371 & 0.534 \\
     &  & $\uparrow$ $\text{FSat}$ & 0.370 & \bfseries 0.438 & 0.314 & 0.376 & 0.114 & 0.152 & 0.091 & 0.138 \\
     &  & $\downarrow$ $\text{Gini}$ & 0.665 & \bfseries 0.598 & 0.747 & 0.660 & 0.949 & 0.899 & 0.959 & 0.910 \\
    \cline{2-11}
     & \multirow[c]{15}{*}{\rotatebox[origin=c]{90}{\textsc{Joint}}} & $\uparrow$ $\text{IBO}_{\text{ori}}$ & nan & nan & nan & nan & nan & nan & nan & nan \\
     &  & $\uparrow$ $\text{IBO}_{\text{our}}$ & \bfseries 0.062 & 0.029 & 0.019 & 0.038 & 0.029 & 0.029 & 0.038 & 0.024 \\
     &  & $\downarrow$ $\text{IWO}_{\text{ori}}$ & nan & nan & nan & nan & nan & nan & nan & nan \\
     &  & $\downarrow$ $\text{IWO}_{\text{our}}$ & \bfseries 0.938 & 0.971 & 0.981 & 0.962 & 0.971 & 0.971 & 0.962 & 0.976 \\
     &  & $\downarrow$ $\text{IAA}_{\text{ori}}$ & \bfseries 0.011 & \bfseries 0.011 & \bfseries 0.011 & \bfseries 0.011 & \bfseries 0.011 & \bfseries 0.011 & \bfseries 0.011 & \bfseries 0.011 \\
     &  & $\downarrow$ $\text{IAA}_{\text{our}}$ & \bfseries 0.988 & 0.995 & 0.996 & 0.994 & 0.990 & 0.994 & 0.989 & 0.998 \\
     &  & $\downarrow$ $\text{IFD}_{\div\text{-ori}}$ & 0.005 & 0.003 & 0.003 & \bfseries 0.002 & 0.005 & \bfseries 0.002 & 0.005 & 0.003 \\
     &  & $\downarrow$ $\text{IFD}_{\div\text{-our}}$ & 0.016 & 0.006 & \bfseries 0.001 & 0.005 & 0.018 & \bfseries 0.001 & 0.016 & 0.004 \\
     &  & $\downarrow$ $\text{IFD}_{\times\text{-ori}}$ & \bfseries 0.000 & \bfseries 0.000 & \bfseries 0.000 & \bfseries 0.000 & \bfseries 0.000 & \bfseries 0.000 & \bfseries 0.000 & \bfseries 0.000 \\
     &  & $\downarrow$ $\text{IFD}_{\times\text{-our}}$ & 0.010 & 0.006 & 0.002 & 0.005 & 0.012 & 0.003 & 0.012 & \bfseries 0.001 \\
     &  & $\downarrow$ $\text{HD}_{\text{ori}}$ & \bfseries 0.580 & 0.630 & 0.661 & 0.626 & 0.597 & 0.653 & 0.598 & 0.667 \\
     &  & $\downarrow$ $\text{MME}_{\text{ori}}$ & \bfseries0.001 & \bfseries 0.001 & \bfseries0.001 & \bfseries0.001 & 0.003 & \bfseries 0.001 & 0.004 & \bfseries 0.001 \\
     &  & $\downarrow$ $\text{II-F}_{\text{ori}}$ & \bfseries 0.006 & \bfseries 0.006 &\bfseries 0.006 &\bfseries 0.006 &\bfseries 0.006 &\bfseries 0.006 &\bfseries 0.006 &\bfseries 0.006 \\
     &  & $\downarrow$ $\text{II-F}_{\text{our}}$ & \bfseries 0.989 & 0.995 & 0.997 & 0.995 & 0.990 & 0.995 & 0.990 & 0.998 \\
     &  & $\downarrow$ $\text{AI-F}_{\text{ori}}$ & \bfseries 0.000 & \bfseries 0.000 & \bfseries 0.000 & \bfseries 0.000 & 0.001 & \bfseries 0.000 & 0.002 & \bfseries 0.000 \\
    \bottomrule
    \end{tabular}
    }
    \end{table}

\begin{table}
\centering
    \caption*{Table \ref{ext_tab:base-rerank-all} (continued): 
    Effectiveness (\textsc{Eff}), fairness (\textsc{Fair}), and \textsc{Joint} scores at $k=10$ without and with re-ranking the top 25 items using COMBMNZ (CM). Bold indicates the most effective/fair score per measure. The score 0.000 does not mean that the scores are exactly 0; this is because the scores are small ($<10^{-3}$) and rounded to 3 d.p. The term `nan' stands for ``not a number''.
    }
    \resizebox{0.58\textwidth}{!}{
    \begin{tabular}{lll*{2}{r}|*{2}{r}|*{2}{r}|*{2}{r}}
    \toprule
     &  & model & \multicolumn{2}{c|}{ItemKNN} & \multicolumn{2}{c|}{BPR} & \multicolumn{2}{c|}{MultiVAE} & \multicolumn{2}{c}{NCL} \\ 
    \midrule
     &  & re-ranker & - & CM & - & CM & - & CM & - & CM \\
    \midrule
    \multirow[c]{26}{*}{\rotatebox[origin=c]{90}{QK-video}} & \multirow[c]{6}{*}{\rotatebox[origin=c]{90}{\textsc{Eff}}} & $\uparrow$ $\text{HR}$ & 0.040 & 0.047 & 0.099 & 0.045 & 0.109 & 0.061 & \bfseries 0.130 & 0.077 \\
     &  & $\uparrow$ $\text{MRR}$ & 0.013 & 0.013 & 0.039 & 0.015 & 0.039 & 0.021 & \bfseries 0.048 & 0.024 \\
     &  & $\uparrow$ $\text{P}$ & 0.004 & 0.005 & 0.011 & 0.005 & 0.012 & 0.006 & \bfseries 0.014 & 0.008 \\
     &  & $\uparrow$ $\text{MAP}$ & 0.005 & 0.005 & 0.017 & 0.006 & 0.018 & 0.009 & \bfseries 0.022 & 0.010 \\
     &  & $\uparrow$ $\text{R}$ & 0.014 & 0.019 & 0.043 & 0.019 & 0.051 & 0.027 & \bfseries 0.061 & 0.033 \\
     &  & $\uparrow$ $\text{NDCG}$ & 0.009 & 0.010 & 0.029 & 0.011 & 0.031 & 0.016 & \bfseries 0.038 & 0.019 \\
    \cline{2-11}
     & \multirow[c]{5}{*}{\rotatebox[origin=c]{90}{\textsc{Fair}}} & $\uparrow$ $\text{Jain}$ & 0.483 & \bfseries 0.589 & 0.081 & 0.379 & 0.012 & 0.032 & 0.020 & 0.071 \\
     &  & $\uparrow$ $\text{QF}$ & \bfseries 0.901 & 0.790 & 0.625 & 0.823 & 0.100 & 0.163 & 0.201 & 0.365 \\
     &  & $\uparrow$ $\text{Ent}$ & 0.933 & \bfseries 0.937 & 0.755 & 0.903 & 0.420 & 0.547 & 0.507 & 0.674 \\
     &  & $\uparrow$ $\text{FSat}$ & 0.443 & \bfseries 0.547 & 0.212 & 0.382 & 0.052 & 0.090 & 0.077 & 0.150 \\
     &  & $\downarrow$ $\text{Gini}$ & 0.472 & \bfseries 0.442 & 0.807 & 0.570 & 0.982 & 0.959 & 0.966 & 0.902 \\
    \cline{2-11}
     & \multirow[c]{15}{*}{\rotatebox[origin=c]{90}{\textsc{Joint}}} & $\uparrow$ $\text{IBO}_{\text{ori}}$ & nan & nan & nan & nan & nan & nan & nan & nan \\
     &  & $\uparrow$ $\text{IBO}_{\text{our}}$ & 0.033 & 0.038 & \bfseries 0.054 & 0.036 & 0.031 & 0.036 & 0.043 & \bfseries 0.054 \\
     &  & $\downarrow$ $\text{IWO}_{\text{ori}}$ & nan & nan & nan & nan & nan & nan & nan & nan \\
     &  & $\downarrow$ $\text{IWO}_{\text{our}}$ & 0.967 & 0.962 & \bfseries 0.946 & 0.964 & 0.969 & 0.964 & 0.957 & \bfseries 0.946 \\
     &  & $\downarrow$ $\text{IAA}_{\text{ori}}$ & \bfseries 0.001 &\bfseries  0.001 &\bfseries  0.001 &\bfseries  0.001 &\bfseries  0.001 &\bfseries  0.001 & \bfseries 0.001 & \bfseries 0.001 \\
     &  & $\downarrow$ $\text{IAA}_{\text{our}}$ & 0.991 & 0.989 & 0.970 & 0.989 & 0.966 & 0.983 & \bfseries 0.959 & 0.980 \\
     &  & $\downarrow$ $\text{IFD}_{\div\text{-ori}}$ & 0.009 & \bfseries 0.007 & 0.014 & 0.008 & 0.014 & 0.009 & 0.015 & 0.010 \\
     &  & $\downarrow$ $\text{IFD}_{\div\text{-our}}$ & \bfseries 0.015 & \bfseries 0.015 & 0.040 & 0.016 & 0.040 & 0.022 & 0.049 & 0.026 \\
     &  & $\downarrow$ $\text{IFD}_{\times\text{-ori}}$ & \bfseries 0.000 & \bfseries 0.000 & \bfseries 0.000 & \bfseries 0.000 & \bfseries 0.000 & \bfseries 0.000 & \bfseries 0.000 & \bfseries 0.000 \\
     &  & $\downarrow$ $\text{IFD}_{\times\text{-our}}$ & \bfseries 0.007 & \bfseries 0.007 & 0.023 & 0.008 & 0.023 & 0.012 & 0.028 & 0.014 \\
     &  & $\downarrow$ $\text{HD}_{\text{ori}}$ & 0.576 & 0.560 & 0.490 & 0.565 & 0.478 & 0.535 & \bfseries 0.457 & 0.519 \\
     &  & $\downarrow$ $\text{MME}_{\text{ori}}$ & \bfseries 0.000 & \bfseries 0.000 & \bfseries0.000 & \bfseries0.000 & \bfseries0.000 & \bfseries0.000 & \bfseries0.000 & \bfseries0.000 \\
     &  & $\downarrow$ $\text{II-F}_{\text{ori}}$ & \bfseries0.001 & \bfseries0.001 & \bfseries0.001 & \bfseries0.001 & \bfseries0.001 & \bfseries0.001 & \bfseries 0.001 & \bfseries0.001 \\
     &  & $\downarrow$ $\text{II-F}_{\text{our}}$ & 0.992 & 0.990 & 0.973 & 0.990 & 0.970 & 0.985 & \bfseries 0.964 & 0.983 \\
     &  & $\downarrow$ $\text{AI-F}_{\text{ori}}$ & \bfseries0.000 & \bfseries 0.000 & \bfseries0.000 & \bfseries0.000 & \bfseries 0.000 & \bfseries 0.000 & \bfseries 0.000 & \bfseries 0.000 \\
    \cline{1-11}
    \multirow[c]{26}{*}{\rotatebox[origin=c]{90}{ML-10M}} & \multirow[c]{6}{*}{\rotatebox[origin=c]{90}{\textsc{Eff}}} & $\uparrow$ $\text{HR}$ & 0.487 & 0.443 & 0.512 & 0.386 & 0.417 & 0.387 & \bfseries 0.521 & 0.402 \\
     &  & $\uparrow$ $\text{MRR}$ & 0.282 & 0.225 & 0.299 & 0.185 & 0.237 & 0.191 & \bfseries 0.302 & 0.203 \\
     &  & $\uparrow$ $\text{P}$ & 0.137 & 0.105 & 0.146 & 0.088 & 0.107 & 0.096 & \bfseries 0.154 & 0.094 \\
     &  & $\uparrow$ $\text{MAP}$ & 0.089 & 0.060 & 0.095 & 0.047 & 0.067 & 0.054 & \bfseries 0.101 & 0.052 \\
     &  & $\uparrow$ $\text{R}$ & 0.022 & 0.018 & 0.025 & 0.012 & 0.020 & 0.016 & \bfseries 0.026 & 0.013 \\
     &  & $\uparrow$ $\text{NDCG}$ & 0.150 & 0.113 & 0.160 & 0.092 & 0.119 & 0.100 & \bfseries 0.167 & 0.100 \\
    \cline{2-11}
     & \multirow[c]{5}{*}{\rotatebox[origin=c]{90}{\textsc{Fair}}} & $\uparrow$ $\text{Jain}$ & 0.011 & 0.027 & 0.037 & \bfseries 0.115 & 0.003 & 0.006 & 0.024 & 0.069 \\
     &  & $\uparrow$ $\text{QF}^*$ & 0.044 & 0.068 & 0.145 & \bfseries 0.216 & 0.014 & 0.025 & 0.086 & 0.132 \\
     &  & $\uparrow$ $\text{Ent}$ & 0.407 & 0.514 & 0.596 & \bfseries 0.716 & 0.238 & 0.324 & 0.519 & 0.638 \\
     &  & $\uparrow$ $\text{FSat}^*$ & 0.044 & 0.068 & 0.145 & \bfseries 0.216 & 0.014 & 0.025 & 0.086 & 0.132 \\
     &  & $\downarrow$ $\text{Gini}$ & 0.987 & 0.971 & 0.945 & \bfseries 0.879 & 0.997 & 0.993 & 0.969 & 0.930 \\
    \cline{2-11}
     & \multirow[c]{15}{*}{\rotatebox[origin=c]{90}{\textsc{Joint}}} & $\uparrow$ $\text{IBO}_{\text{ori}}$ & nan & nan & nan & nan & nan & nan & nan & nan \\
     &  & $\uparrow$ $\text{IBO}_{\text{our}}$ & 0.031 & 0.046 & 0.069 & \bfseries 0.091 & 0.012 & 0.018 & 0.054 & 0.074 \\
     &  & $\downarrow$ $\text{IWO}_{\text{ori}}$ & nan & nan & nan & nan & nan & nan & nan & nan \\
     &  & $\downarrow$ $\text{IWO}_{\text{our}}$ & 0.969 & 0.954 & 0.931 & \bfseries 0.909 & 0.988 & 0.982 & 0.946 & 0.926 \\
     &  & $\downarrow$ $\text{IAA}_{\text{ori}}$ & \bfseries 0.008 & 0.009 & \bfseries 0.008 & 0.009 & 0.009 & 0.009 & \bfseries 0.008 & 0.009 \\
     &  & $\downarrow$ $\text{IAA}_{\text{our}}$ & 0.850 & 0.887 & 0.840 & 0.910 & 0.880 & 0.899 & \bfseries 0.830 & 0.902 \\
     &  & $\downarrow$ $\text{IFD}_{\div\text{-ori}}$ & 0.018 & 0.012 & 0.019 & 0.011 & 0.016 & \bfseries 0.010 & 0.020 & 0.012 \\
     &  & $\downarrow$ $\text{IFD}_{\div\text{-our}}$ & 0.164 & 0.126 & 0.176 & \bfseries 0.102 & 0.132 & 0.110 & 0.185 & 0.109 \\
     &  & $\downarrow$ $\text{IFD}_{\times\text{-ori}}$ & \bfseries 0.000 & \bfseries 0.000 & \bfseries 0.000 & \bfseries 0.000 & \bfseries 0.000 & \bfseries 0.000 & \bfseries 0.000 & \bfseries 0.000 \\
     &  & $\downarrow$ $\text{IFD}_{\times\text{-our}}$ & 0.164 & 0.121 & 0.175 & \bfseries 0.097 & 0.131 & 0.105 & 0.181 & 0.107 \\
     &  & $\downarrow$ $\text{HD}_{\text{ori}}$ & 0.221 & 0.255 & 0.226 & 0.262 & 0.265 & 0.273 & \bfseries 0.218 & 0.257 \\
     &  & $\downarrow$ $\text{MME}_{\text{ori}}$ & \bfseries 0.001 & \bfseries 0.001 & \bfseries 0.001 & \bfseries 0.001 & 0.003 & \bfseries 0.001 & \bfseries 0.001 & \bfseries 0.001 \\
     &  & $\downarrow$ $\text{II-F}_{\text{ori}}$ & \bfseries 0.000 & \bfseries 0.000 & \bfseries 0.000 & \bfseries 0.000 & \bfseries 0.000 & \bfseries 0.000 & \bfseries 0.000 & \bfseries 0.000 \\
     &  & $\downarrow$ $\text{II-F}_{\text{our}}$ & 0.846 & 0.885 & 0.836 & 0.908 & 0.878 & 0.898 & \bfseries 0.828 & 0.900 \\
     &  & $\downarrow$ $\text{AI-F}_{\text{ori}}$ & \bfseries 0.000 & \bfseries 0.000 & \bfseries 0.000 & \bfseries 0.000 & \bfseries 0.000 & \bfseries 0.000 & \bfseries 0.000 & \bfseries 0.000 \\
    \bottomrule
    \end{tabular}
    } 
    \\
    \raggedright
    {\footnotesize *QF $\equiv$ FSat for ML-10M, as QF is computed based on the 
    percentage of recommended items from all items, 
    which in this case is equivalent to FSat. 
    }
\end{table}

We compute all \textsc{Eff} and \textsc{Fair} measures (i.e., single-aspect measures), as well as both the original and corrected \textsc{Joint} measures. We present all scores per dataset and model in \Cref{ext_tab:base-rerank-all}. We denote higher-is-better measures with $\uparrow$ and vice versa with $\downarrow$. 

Often, the purpose of evaluation is to identify the best model in terms of effectiveness, fairness, or other aspects, so that one can choose which model to deploy. However, these aspects can be measured by one or more measures, and they may or may not reach the same conclusion regarding the best model. Here, we study if the measures agree on the best model and we further analyse the measure agreement when they rank recommender models in \Cref{ext_ss:corr}. We also analyse the score range as an indicator of the measure expressiveness. 

\subsubsection*{Best model agreement} 
Do the measures agree on the same best model? We answer this question by comparing the agreement on the best model: 
(i) between \textsc{Eff} and \textsc{Fair} measures; 
(ii) among the \textsc{Joint} measures; as well as 
(iii) between the original and corrected \textsc{Joint} measures. 

Generally, all \eff{} measures agree on the best model across all four datasets, and likewise for the \fair{} measures. In contrast, the \joint{} measures tend to disagree on the best model more often than the single-aspect measures. 

For each dataset, we observe the following. 
First, the best model according to \textsc{Eff} measures for each dataset always differs from the one given by \textsc{Fair} measures, except for QF in Amazon-lb. This means that the most effective model is generally not the fairest. 
Second, even though the best model per dataset is always the same for all \textsc{Eff} measures (except MRR and MAP for Amazon-lb) and likewise for \textsc{Fair} measures (except QF), this is not the case for the \textsc{Joint} measures. Some \textsc{Joint} measures always have the same best model per dataset (e.g., IBO\our~ with IWO\our, IAA with HD and II-F, or MME with AI-F), but the trend is less consistent for IFD, which only agrees with MME/AI-F occasionally for some datasets/measure versions. The \textsc{Joint} measures that have the same best models tend to have similar formulations: both IBO/IWO are the percentage of items with an impact score higher/lower than a threshold; MME/AI-F aggregate exposure across users prior to taking the difference in exposure, while IAA/HD/II-F do not; and MME/IFD are pairwise measures. 
Third, the original IAA/II-F/IFD$_\times$ always agree with their corrected versions.\footnote{As this cannot be seen from the rounded scores in \Cref{ext_tab:base-rerank-all}, we verify this from the pre-rounded scores.} 
Yet, the original IFD$_\div$ mostly disagrees with its correction. This disagreement may be because IFD$_{\div\text{-ori}}$ assigns nonzero exposure weight for items beyond top-$k$, while IFD$_{\div\text{-our}}$ does not. In IAA/II-F/IFD$_\times$ the exposure weight assignment does not differ significantly between the original and corrected version.

In summary, we can cluster all \textsc{Joint} measures into three groups based on their general alignment to single-aspect measures: 
(i) \textsc{Eff}-aligned measures, IAA/HD/II-F 
(ii) \textsc{Fair}-aligned measures, IFD/MME/AI-F; and 
(iii) measures without consistent alignment, IBO/IWO. 
The \textsc{Joint} measure alignment to single-aspect measure can somewhat be explained through their formulation. For example, the fairest ranking based on IAA and II-F is the one that places all relevant items at the top, which aligns with \textsc{Eff} measures. As for AI-F, the measure scores fairer when there are more unique items exposed to the users, which aligns with \textsc{Fair} measures. We also find that most original \joint{} measures have the same best models as the corrected versions, except for IFD$_\div$. Therefore, computing IFD\divori{} may lead to a different conclusion from IFD\divour{}. 
We revisit the measure agreement in \Cref{ext_ss:corr}.

\subsubsection*{Score range} 
 We find three issues with the range of the original \textsc{Joint} measure scores: (i) consistently small scores for several \textsc{Joint} measures; (ii) scale mismatch between single-aspect measures and \textsc{Joint} measures; and (iii) scale mismatch between the original \textsc{Joint} measures. 
 
Regarding (i), the extremely low scores of the original \down \textsc{Joint} measure scores ($\leq10^{-3}$) for all datasets and models, make it hard to distinguish the performance of models for a specific dataset. For instance, IFD$_{\times\text{-ori}}$ always scores near 0 for all datasets, as \Cref{ext_eq:ifd_j_x} is often 0 due to the low number of relevant items per user.\footnote{The median number of relevant items per user in a dataset is at most 46, across all four datasets} For the same reason, MME\ori~ and II-F\ori/AI-F\ori, \Cref{ext_eq:imp} and \Cref{ext_eq:eui-star} often result in 0. These low scores may also be due to the \textbf{non-realisability} limitation, as the measures cannot reach the unfairest score of 1. 
Our versions of IFD$_{\times}$ and II-F do not suffer from this issue, as they have been normalised based on the achievable max/min score. Therefore, distinguishing the IFD\mulour{} and II-F\our{} scores across models for a dataset is easier than doing so for the original measures. 

Regarding (ii), i.e., the scale mismatch between single-aspect measures and \joint{} measures, while the above \textsc{Joint} scores differ in the fourth decimal point or later, the \textsc{Eff} and \textsc{Fair} scores differ in the second decimal point or before. For instance, the NDCG (\textsc{Eff} score) of MultiVAE-CM and NCL for Lastfm differs by $\sim$0.16 and their Jain (\textsc{Fair} score) differs by $\sim$0.14. These differences are non-negligible, but the joint scores of IAA\ori/IFD$_{\times\text{-ori}}$/MME\ori/II-F\ori/AI-F\ori~ differ only by $\leq 10^{-3}$, which may seem negligible.\footnote{NDCG and Jain are used in the comparison as they tend to be stricter$^*$ than some other single-aspect measures used in this work.} As the difference of magnitude is inconsistent between the single-aspect and \textsc{Joint} measures, the scores are harder to interpret. On the other hand, the differences in the scores of IAA\our/IFD$_{\times\text{-our}}$/II-F\our~ per dataset can either be seen in the first or second decimal point, except for IFD$_{\times\text{-our}}$, which differs in the third decimal point for Amazon-lb and QK-video. 
Hence, our version of IAA/IFD/II-F is more expressive than the original version, as it better matches the change in the single-aspect measures.

Regarding (iii), i.e., the scale mismatch between the original \joint{} measures, we observe considerable gaps in the range of all \textsc{Joint} measure scores, e.g., between IWO\our, HD\ori, and AI-F\ori, despite all three being lower-is-better measures. For instance, in ML-10M, \down IWO\our~ $\approx 1$ (very unfair) based on its theoretical $[0,1]$-range, \down HD\ori~ is roughly a quarter of the \down IWO\our~ score (somewhat fair), and \down AI-F\ori~ $\approx 0$ (extremely fair). This discrepancy causes confusion in score interpretation. Between some of the original and the reformulated measures, we also find huge differences. For example, across all models for ML-10M: \down IAA\ori~ $\approx 0.010$ (very fair), while \down IAA\our~ $\approx 0.9$ (very unfair). Similarly, \down II-F\ori~ $\approx 0.000$ (very fair), while \down II-F\our~ $\approx 0.9$ (very unfair). Similar gaps can also be observed for the same measure pairs in the other datasets. Thus, evaluation with IAA\ori~ or II-F\ori~ scores may result in a misleading conclusion of having extremely fair models, even if the models are actually very unfair. In contrast, evaluating with IAA\our{} or II-F\our{} leads to a more accurate conclusion.

\subsubsection*{Undefined scores} The scores of IBO\ori/IWO\ori~ are undefined for all datasets and all models. Due to division by zero, the computation results in `nan', which stands for `not a number'. The measures are incomputable as some items are not relevant to any users, which causes \Cref{ext_eq:imp-unif} to be 0, and this is a manifestation of the \textbf{undefinedness} limitation (\Cref{ext_ss:undefinedness}). Our reformulation of both measures (\Cref{ext_ss:resolve_undef}) resolves this issue.

\subsection{Measure Correlations}
\label{ext_ss:corr}

\begin{figure}
    \centering
    \includegraphics[width=0.98\textwidth, trim=0.25cm 0.6cm 0.3cm 0.3cm clip=True]{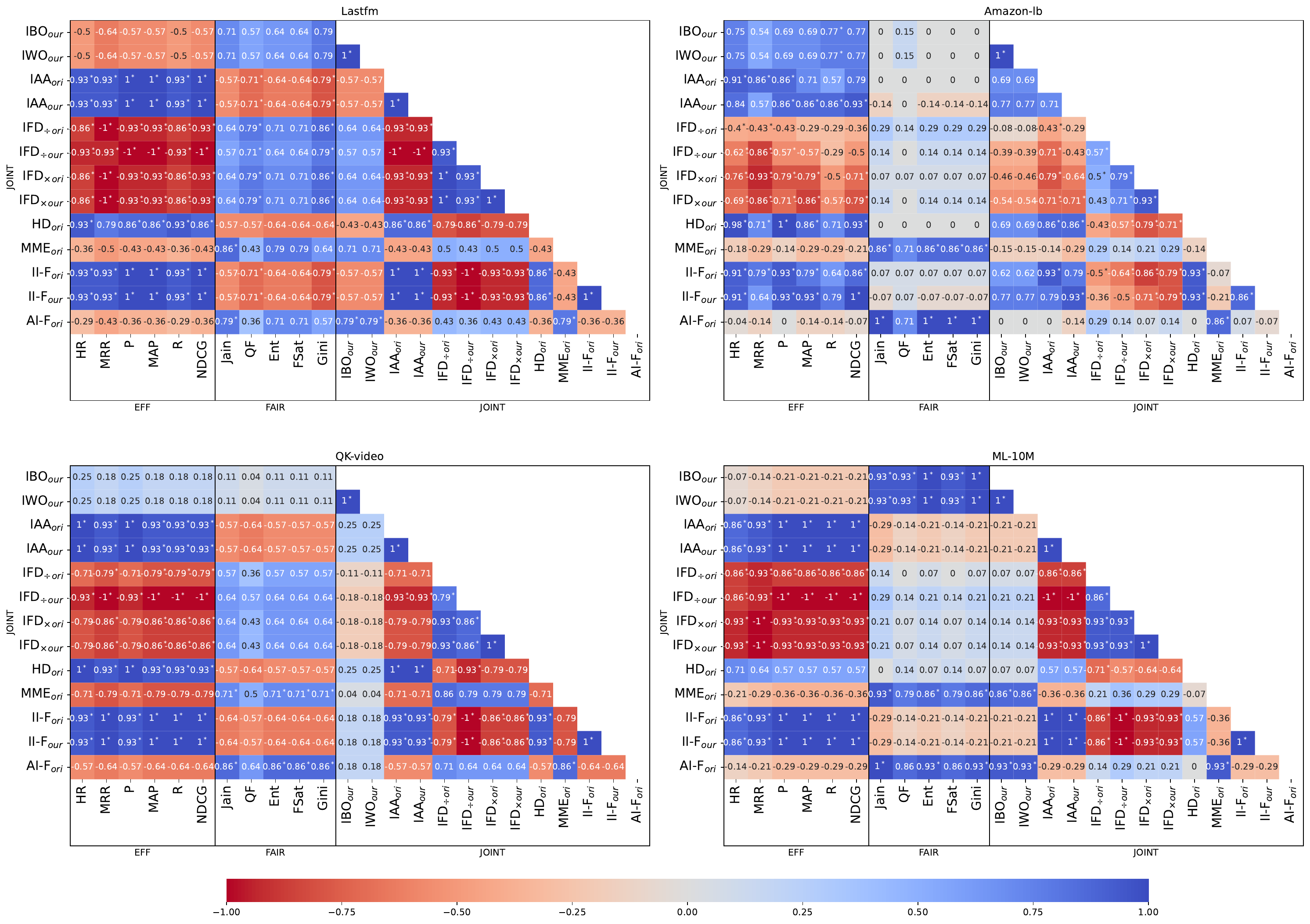}
    \caption{Kendall's $\tau$ correlation between \textsc{Joint}, \textsc{Eff}, and \textsc{Fair} measures. Asterisk  ($^*$) denotes a statistically significant correlation ($\alpha=0.05$) after applying Bonferroni's correction.}
    \label{ext_fig:corr-grid}
\end{figure}

When comparing evaluation measures, it is important to understand the extent to which they agree in ranking recommender models based on the measure scores. 
To this end, we compute Kendall's $\tau$ correlation as a measure of agreement between the orderings of the recommenders based on the measure scores computed in \Cref{ext_ss:performance}. 
We also report the statistical significance of the correlations ($\alpha=0.05$) and apply Bonferroni's correction to account for multiple testing errors within a dataset.  
We present the result in \Cref{ext_fig:corr-grid}. 
We study three types of agreement: (i) between \textsc{Joint} measures and single-aspect measures; (ii) among \textsc{Joint} measures in general; and (iii) between each original \textsc{Joint} measure and its corrected version.

\subsubsection*{(i) Agreement of joint and single-aspect measures}  \textsc{Eff} measures are known to not correlate strongly with \textsc{Fair} measures \cite{Rampisela2024EvaluationStudy}, so we do not expect the \textsc{Joint} measures to correlate strongly with either \textsc{Eff} or \textsc{Fair} measures, as they should account for both aspects. 
Generally, we find that \textsc{Joint} measures do not correlate consistently with \textsc{Eff} measures. 
IBO/IWO's correlations largely vary ($\tau \in [-0.64, 0.77]$); 
IAA, HD, and II-F have positive correlations that are either moderate or strong ($\tau \in [0.57, 1]$); 
IFD and MME have negative correlations that range from weak to strong ($\tau \in [-1, -0.29]$ for IFD and $\tau \in [-0.79, -0.14]$ for MME); and AI-F has non-positive correlations ($\tau \in [-0.64,0]$). 

We also find that the \textsc{Joint} measures do not correlate consistently with \textsc{Fair} measures either. Again, the correlations of IBO/IWO have large variance, though less than with \textsc{Eff} measures. 
IAA/HD/II-F shows two distinct trends across datasets: they have negative correlations that are moderate to strong ($\tau \in [-0.79,-0.57]$) for Lastfm and QK-video,  but correlate weakly for Amazon-lb and ML-10M ($\tau \in [-0.29,0.14]$). The correlations of IFD also vary across datasets. On one hand, IFD strongly correlates with \textsc{Fair} measures for Lastfm and QK-video (except with QF for QK-video), $\tau \in [0.57, 0.86]$, and on the other hand, IFD weakly correlates to \textsc{Fair} measures for the other datasets ($\tau\in [0, 0.29]$). For all datasets, MME and AI-F strongly correlate with \textsc{Fair} measures, except for QF in Lastfm ($\tau \in [0.5,1]$). 

In summary, even though we expect the \joint{} measures to have weak correlations with both \eff{} and \fair{} measures as the \joint{} measures consider both item relevance and exposure, this is only true for IBO/IWO. Note that \textsc{Joint} measures with strong agreement to \textsc{Eff} measures do not always strongly disagree with \textsc{Fair} measures, and vice versa. For instance, IAA/HD/II-F strongly agree with \textsc{Eff} measures for Amazon-lb, but their correlations with \textsc{Fair} measures are weak.

\subsubsection*{(ii) Agreement among \textsc{Joint} measures.} 
We expect the agreement among \joint{} measures to follow the groupings found in \Cref{ext_ss:performance}. 
Overall, the three \textsc{Joint} measure groups identified in \Cref{ext_ss:performance} indeed have strong positive correlations between measures of the same group and strong negative correlations between measures from different groups. For instance, IBO consistently correlates perfectly with IWO, as their formulation is similar. 
IAA, HD, and II-F strongly agree with each other, $\tau \in [0.57,1]$. IFD$_{\div}$ correlates moderately or highly with IFD$_{\times}$, $\tau \in [0.43,1]$ due to their similar formulations. MME and AI-F always correlate highly, $\tau \in [0.79,0.93]$. IFD occasionally has moderate-to-strong correlations with MME and AI-F, $\tau \in [0.36,0.86]$ for Lastfm and QK-video, but the correlations are weaker for Amazon-lb and ML-10M, $\tau \in [0.07,0.36]$. Conversely, IAA/HD/II-F have moderate-to-strong disagreement with IFD, $\tau \in [-1,-0.36]$ except for IFD$_{\div}$ in Amazon-lb ($\tau=-0.29$).

From the findings of (i) and (ii), we conclude that: IBO/IWO has inconsistent relationships with single-aspect and \textsc{Joint} measures; IAA/HD/II-F strongly disagrees with \textsc{Fair} measures; and IFD/MME/AI-F strongly disagrees with \textsc{Eff} measures (even if IFD occasionally disagrees with \textsc{Fair} measures as well). Among the \textsc{Joint} measures, IBO/IWO has weak correlations with the single-aspect measures for QK-video, and similarly with IFD$_{\div}$ for Amazon-lb, but the trend is inconsistent. Thus, we argue that no \textsc{Joint} measures reliably account for both recommendation effectiveness and fairness, where fairness is quantified purely based on exposure. 

\subsubsection*{(iii) Agreement between original and corrected \textsc{Joint} measures} 
We expect each original \textsc{Joint} measure to strongly correlate with its respective corrected version, as most (but not all) corrections are linear transformations. As the corrected measures min-max normalises the original  \joint{} score per user prior to averaging across users, the original measures may not perfectly correlate to their corrections. In general, the original \textsc{Joint} measures strongly agree with the corrected ones ($\tau \in [0.57,1]$), and the correlations tend to be weaker in Amazon-lb. We cannot compare IBO\ori/IWO\ori~with IBO\our/IWO\our~ as the original version returns undefined scores for all datasets and models (\Cref{ext_ss:performance}).
IAA\ori, IFD$_{\times\text{-ori}}$, and II-F\ori~perfectly correlate with their respective corrected versions for all datasets other than Amazon-lb, where $\tau \in [0.71, 0.93]$. IFD$_{\div\text{-ori}}$ have weaker (but still strong) correlations to IFD$_{\div\text{-our}}$ ($\tau \in [0.57, 0.93]$). We posit that this is because correcting the top-$k$-insensitivity of IFD$_{\div\text{-ori}}$ results in a non-linear transformation of the combined exposure-relevance score in \Cref{ext_eq:ifd_j_div}. 

In conclusion, while the original measures strongly agree with the corrected ones, they do not always give identical rank orderings of the models. This means that in some cases, by computing the original measure, we may reach different conclusions on the relative performances of the model as opposed to computing the corrected version.

\subsection{Achievable Fairness Scores (Related to Non-realisabili\-ty)}
\label{ext_ss:exp_maxmin}

\begin{figure}
    \centering
    \includegraphics[width=\textwidth, trim=0cm 0.5cm 0cm 0cm]{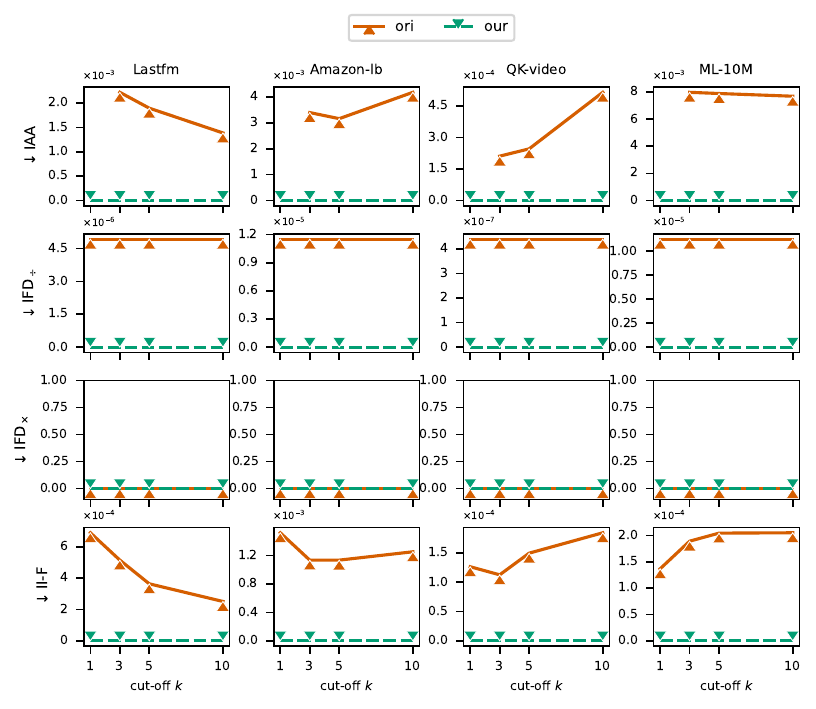}
    \caption{The fairest achievable \down \textsc{Joint} measure scores for varying $k$, both the original (ori) and corrected (our) versions. IAA\ori~is incomputable at $k=1$ due to the undefinedness limitation. Both IFD$_\times$ versions overlap.}
    \label{ext_fig:mostfair}
\end{figure}

\begin{figure}
    \centering
    \includegraphics[width=\textwidth,trim=0cm 0.5cm 0cm 0cm]{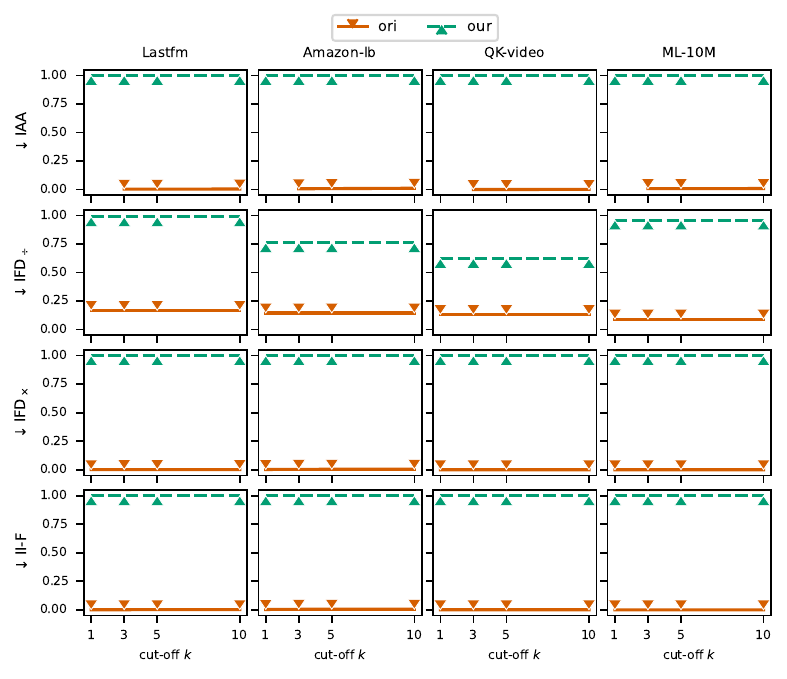}
    \caption{The unfairest achievable \down \textsc{Joint} measure scores for varying $k$, both the original (ori) and corrected (our) versions. IAA\ori~is incomputable at $k=1$ due to the undefinedness limitation.}
    
    \label{ext_fig:mostunfair}
\end{figure}

Given the (un)fairest possible ranking, to what extent can the \textsc{Joint} measures achieve their theoretical fairest or unfairest score (0 or 1) for different datasets and different cut-off $k$? 
Here, we investigate: 
(i) the severity of the \textbf{non-realisability} limitation (\textit{Cause 4}) in the original \textsc{Joint} measures; and 
(ii) the extent to which the corrected measures resolve the limitation. 
As such, we experiment only with the \textsc{Joint} measures for which we have resolved the limitation, i.e., IAA, IFD$_\div$, IFD$_\times$, and II-F, and evaluate recommendations at various cut-offs, $k\in\{1,3,5,10\}$. To generate the recommendations, we employ three strategies: \texttt{top} (rank all relevant items at the top), \texttt{bottom} (rank all relevant items at the bottom), \texttt{half-half} (rank half of the relevant items at the top and rank the rest at the bottom). Each strategy corresponds to either the fairest or unfairest recommendation as per \Cref{ext_tab:example_fair_unfair}, which differs per measure. Specifically, \texttt{top} corresponds to the fairest ranking for IAA and II-F and the unfairest for IFD$_\times$,\footnote{For IFD$_\times$, we experimented with the two strategies described in \Cref{ext_sss:resolve_ifd_mul} and found that the \texttt{top} strategy results in an unfairer score than the other.} \texttt{bottom} corresponds to the fairest ranking for IFD and the unfairest for IAA and II-F, and \texttt{half-half} corresponds to the unfairest ranking for IFD$_{\div\text{-our}}$. For every measure, we employ two strategies, that each correspond to the measure's fairest and unfairest ranking. We show the results in \Cref{ext_fig:mostfair,ext_fig:mostunfair}.


\subsubsection*{Fairest achievable score (\Cref{ext_fig:mostfair})} All of our corrected versions of the measures in this experiment always successfully reach the theoretical fairest score (i.e., 0) for all datasets and choices of $k$. On the other hand, the original \textsc{Joint} measures of IAA, IFD$_\div$, and II-F always fail to reach the fairest score (i.e., 0) due to the non-realisability limitation. While the scores may seem to be close to 0, they may cause confusion as they are not exactly 0, even if the recommendation cannot be fairer than that. Only IFD$_{\times\text{-ori}}$ manages to achieve 0, and this is because the number of relevant items per user ($|R_u^*|$) is much lower than $n-k$, which is the number of unexposed items per user. If $|R_u^*|>n-k$ instead, like the example in \Cref{ext_ss:nonreal}, IFD$_{\times\text{-ori}}$ may fail to reach 0, as an item that is both relevant and recommended at the top-$k$ would contribute a nonzero score. 
Among IAA, IFD$_{\div}$, and II-F, the measure with the highest difference between the original and the corrected scores is IAA, followed by II-F and IFD$_\div$. For IAA and II-F, this difference changes across various $k$, but there is no consistent trend across datasets. Consequently, the scores are more unpredictable and harder to interpret as the lower bound changes with different $k$. IFD$_{\div\text{-ori}}$ remains constant for different $k$ as it suffers from the \textbf{top-$k$-insensitivity} limitation, though in this case, achieving the same score for various $k$ is arguably the more desired case as the fairest achievable score should be 0, no matter the $k$. Further, IAA\ori~ is incomputable for $k=1$ due to the \textbf{undefinedness} limitation, as setting $k=1$ causes division by zero in the original, normalised linear examination function (\Cref{ext_tab:exp-weigh}). IAA\our~ does not suffer from this issue.

\subsubsection*{Unfairest achievable score (\Cref{ext_fig:mostunfair})} All original \textsc{Joint} measures in this experiment never reach the theoretical unfairest score (i.e., 1) for all combinations of dataset and $k$. Even more worrying, the measures' achievable unfairest scores are all less than 0.25, which is very far from 1, and much closer to the theoretical fairest score (i.e., 0) instead. 
In contrast, all of our corrected measures always achieve 1, except for IFD$_\div$. \down IFD$_{\div\text{-our}}$ cannot reach 1 because the datasets have some users with only one relevant item in the test set. In our experiment setup, we set \down IFD$_{\div\text{-our}}(u)=0$ for all users with only one relevant item, as this measure computes the pairwise difference of the combined exposure-relevance score from the relevant items. If these users are excluded from the evaluation, the score of IFD$_{\div\text{-our}}$ would be 1.  For different $k$ values, the scores of both the original and our versions of the measures do not vary, which is a desirable property, as the unfairest score should be constant for different cut-offs.\footnote{We verify this from the actual measure scores, and not just through the visualisation in \Cref{ext_fig:mostunfair}.} 
Other than issues related to the non-realisability of the original measures, we also observe that IAA\ori~ cannot be computed at $k=1$ due to the \textbf{undefinedness} limitation, but this issue is resolved in IAA\our.

Based on the above results, we find that the empirical ranges of the original version of IAA, IFD, and II-F are constrained to a small part of their theoretical range. As their unfairest achievable scores are extremely close to the theoretical fairest score, any scores given by the measures would be deceptively fair, leading to challenges in interpreting the scores. 
We show that all of our corrected measures resolve the compressed range issue. The scores of the corrected measures, except for IFD$_{\div\text{-our}}$, would also be easier to interpret, as 0 is mapped to the fairest possible recommendation, and 1 is mapped to the unfairest possible recommendation, unlike IAA\ori, IFD$_{\times\text{-ori}}$, and II-F\ori, which has scores other than 0 or 1 mapped to the (un)fairest scores. Therefore, the corrected measures can successfully reach the theoretical max/min score, except for IFD\divour, which fails to achieve the theoretical max score unless all users with only one relevant item are excluded from the evaluation. 

\subsection{Sensitivity to Cut-off \texorpdfstring{$k$}{} (Related to Top-\texorpdfstring{$k$}{}-insensitivity)}
\label{ext_ss:k_sensitive}

\begin{figure}
    \centering
    \includegraphics[width=\linewidth, trim=0cm 0.5cm 0cm 0cm]{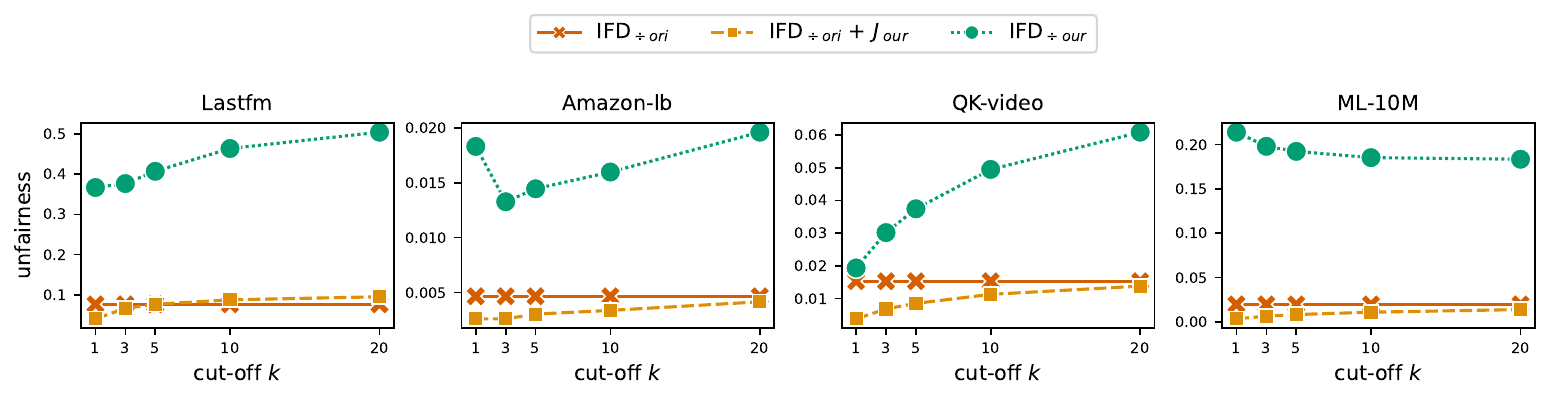}
    \caption{The original and corrected versions of IFD$_\div$ across different cut-offs on the runs from the NCL model.}
    \label{ext_fig:top_k_sensitivity}
\end{figure}

This experiment aims: (i) to show how IFD$_{\div\text{-ori}}$, which suffers from the \textbf{top-$k$-insensitivity} limitation, is unaffected by the choice of $k$; and (ii) to investigate whether the corrected version, IFD$_{\div\text{-our}}$ resolves this issue. In this experiment, we evaluate the runs from the NCL model at $k \in \{1, 3, 5, 10, 20\}$. We choose the NCL runs as they generally perform the best in terms of effectiveness. 
We compute the following measures: IFD$_{\div\text{-ori}}$, IFD$_{\div\text{-our}}$, and IFD$_{\div\text{-ori}} + J$\our, i.e., IFD$_{\div\text{-ori}}$ computed with $J_{\div{\text{-our}}}$ (\Cref{ext_eq:ifd_j_div_our}). The latter is IFD$_{\div\text{-ori}}$ that resolves only top-$k$-insensitivity, as opposed to IFD$_{\div\text{-our}}$, which resolves both top-$k$-insensitivity and non-realisability.

We present the results in \Cref{ext_fig:top_k_sensitivity}. For all datasets, IFD$_{\div\text{-ori}}$ is invariant to the choice of cut-off, while the two corrected versions vary with different $k$. The score changes are even visible 
in the non-rescaled corrected version (IFD$_{\div\text{-ori}} + J$\our), albeit to a smaller extent compared to those in IFD$_{\div\text{-our}}$. Among the corrected versions, there are differences in both the score magnitude and the score change, given different $k$. The scores of of IFD$_{\div\text{-our}}$ is always higher, which means that evaluating with IFD$_{\div\text{-ori}} + J$\our~ would lead to an underestimation of the recommendation fairness. 
For IFD$_{\div\text{-ori}} + J$\our, unfairness generally increases with a larger $k$. The same trend applies to IFD$_{\div\text{-our}}$ as well, for Lastfm and QK-video. However, as $k$ increases, \down IFD$_{\div\text{-our}}$ decreases for ML-10M. For QK-video, \down IFD\divour{} first decreases, then increases. This means that for Amazon-lb and ML-10M, the two corrected IFD$_{\div}$ versions show different behaviour with increasing $k$, e.g., recommending the top-20 items is fairer than only recommending the top-1 item for IFD\divour, but the opposite applies for IFD\divori{} + $J$\our. 

We posit that the findings for \down IFD$_{\div\text{-our}}$ differ across datasets due to the disparity in the effectiveness scores across datasets (\Cref{ext_tab:base-rerank-all}). The effectiveness disparity across datasets means that there are more relevant items at the top-$k$ in one dataset than the other, and the difference in the rank positions of the relevant items affects the \down IFD$_{\div\text{-our}}$ scores. In addition, while IFD\divori{} + $J$\our{} behaves monotonically, IFD\divour{} does not. 
This is because IFD\divour{} averages the min-max normalised IFD\divori{} + $J$\our{} score per user instead of directly min-max normalising the final IFD\divori{} + $J$\our{}. Therefore, IFD\divour{} is not expected to show the same monotonic trend as IFD\divori{} + $J$\our{}.

To conclude, we find that IFD$_{\div\text{-ori}}$ is insensitive to changes in the cut-off, and the corrected measure resolves this issue. Our results also show that while simply reformulating the exposure-relevance combination function $J_\div$ is sufficient to resolve the top-$k$-insensitivity limitation, leaving the non-realisability limitation unresolved would lead to differing conclusions that may affect the perceived recommendation fairness (based on the score magnitude). The difference in the conclusion also concerns the choice of $k$ that would result in the fairest recommendation, which may be important in deciding how many items to recommend to each user.

\subsection{Effect of Relevant Items Beyond the Top-\texorpdfstring{$k$}{} (Related to Non-localisation)}
\label{ext_ss:exp_nonlocal}

In RS datasets, there are typically a lot of items whose relevance to a specific user is unknown, as the item has not been shown to the user, or the user did not interact with it when it was shown. We refer to such items as `unobserved items'. During evaluation, these items are usually considered irrelevant, while they can actually be relevant. Additionally, some items could also be misrated, e.g., they are rated lower than intended due to misclick. In practice, items that are rated lower than a threshold may be discarded during preprocessing. Consequently, these items are regarded as irrelevant, even though the opposite could be true. Here, we investigate whether treating unobserved items or potentially misrated items as either relevant or irrelevant affects the \joint{} measure scores. 

IAA, IFD$_\div$, HD, IBO, IWO, II-F, and AI-F suffer from the \textbf{non-localisation} limitation (\Cref{ext_tab:limitation-summary}), i.e., they require item relevance information beyond the top-$k$. While we do not resolve this limitation, we study how additional relevant items beyond the top-$k$ affect the \textsc{Joint} measure scores. Quantifying the effect of additional relevant items is important to understand the \textsc{Joint} measure robustness against changing the number of relevant items for a user, even if these extra items are not recommended at the top-$k$. For example, if there are actually more relevant items, how much does the initial score underestimate or overestimate the recommendation fairness?

To answer the above questions, we use the recommendations generated by the NCL model, which overall has the best \textsc{Eff} scores across all datasets. We simulate the additional relevant items by treating an unobserved item or a low-rated item as `relevant', one item per user at a time. We do this iteratively until each user has ten additional relevant items. Only the items ranked beyond the top-$k$ are considered, i.e., items at position $k+1, k+2, \dots, n-1, n$. We employ two strategies to choose the 10 items: the first 10 irrelevant items, starting from position $k+1, k+2, \dots$ (\texttt{top}) and the last 10 irrelevant items, starting from position $n, n-1, \dots$ (\texttt{bottom}). We compute at $k=10$ the measures that suffer from non-localisation: IWO, IAA, IFD$_\div$, HD, II-F, and AI-F. IBO also suffers from non-localisation, but we do not compute it, as it correlates perfectly with IWO (\Cref{ext_ss:corr}). We compute both the original measures and their reformulations, if the latter exists. Even if the reformulated measures do not aim to resolve non-localisation, the compressed empirical range of the original measures (\Cref{ext_ss:exp_maxmin}) may mislead the result interpretations. For example, if the original measure scores only vary a little given additional relevant items, it would seem that the additional relevant items do not considerably affect the scores. 

\begin{figure}
    \centering
    \includegraphics[height=0.62\paperheight, trim=1.5cm 0.5cm 1.5cm 0cm]{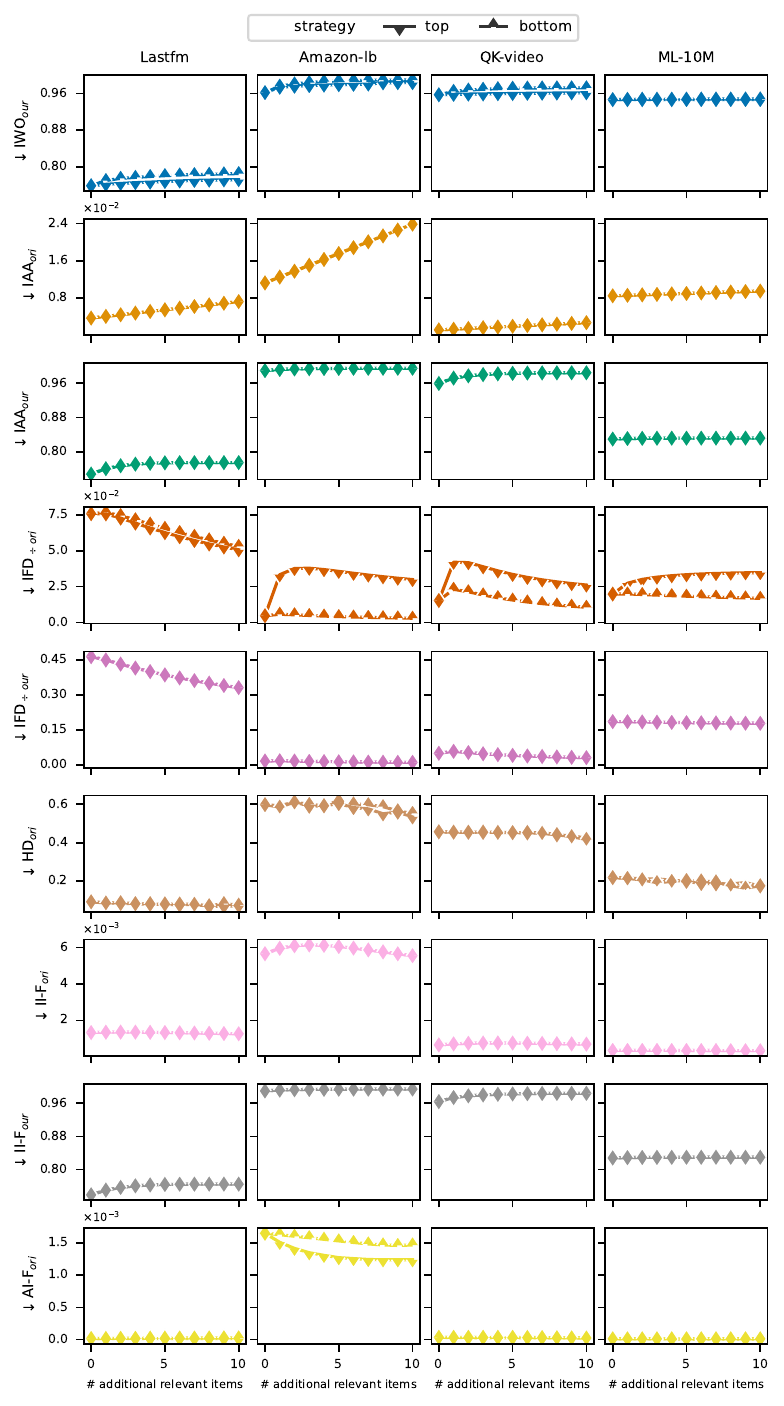}
    \caption{\textsc{Joint} measure scores for different numbers of artificially added relevant items per user. 
    We treat unobserved items or low-rated items as `relevant', starting from position $k+1$ downwards (\texttt{top}) or from position $n$ upwards (\texttt{bottom}).
    }
    \label{ext_fig:nonlocal}
\end{figure}

We present the experiment results in \Cref{ext_fig:nonlocal} and compare the effect of additional relevant items across datasets, measures, and item selection strategy. IWO\ori~ scores are unavailable for all datasets due to the \textbf{undefinedness} limitation.

\subsubsection*{Effect across datasets}  
The measure scores for ML-10M are generally more stable when adding the number of relevant items, compared to the scores for other datasets. 
This may be due to ML-10M having more relevant items per user (with a median of 46) than the other datasets, which have a median of 8 or less. As most RS datasets are sparse, each user may not have many relevant items. Thus, the measure scores would be more susceptible to the effect of additional relevant items.

\subsubsection*{Effect across measures} The measures respond differently to additional relevant items: (i) II-F\ori~ and AI-F\ori~ seem to be mostly stable (their maximum change is less than $10^{-3}$);
(ii) IWO\our, IAA\ori, IAA\our, and II-F\our~ overall become more unfair; and (iii) IFD\divori, IFD\divour, and HD\ori~ generally become fairer. 

About (i), the apparent stability of II-F\ori~ and AI-F\ori~ may be due to their compressed empirical range, as II-F\our, which has a more reasonable empirical range, exhibits a different trend. For all datasets, the range of II-F\ori{} across varying the number of additional relevant items is one or two magnitudes smaller than that of II-F\our. 

About (ii), we explain why IWO\our, IAA, and II-F\our~ become more unfair. Each relevant item beyond the top-$k$ has a zero exposure weight (as it is unexposed) and a relevance score of 1 (as it is relevant). For \down IAA, adding more relevant items beyond the top-$k$ results in a higher difference between the item exposure weight and item relevance value. For \down II-F, changing the item relevance from 0 to 1 would increase the item target exposure (\Cref{ext_eq:eui-star}). As a result, the difference between the item exposure and target exposure increases. For \down IWO, when there are more users for whom an item $i$ is relevant, $Imp^{unif}_{i}$ (\Cref{ext_eq:imp-unif}) increases. $Imp^{unif}_{i}$ is the upper bound of the inequality in \Cref{ext_eq:iwo-our}. When this upper bound increases, the inequality can be satisfied by more items, resulting in a more unfair IWO. 

About (iii), \down IFD$_\div$ becomes fairer as having a relevant item beyond the top-$k$ lowers the average pairwise difference between the combined item exposure-relevance score per user. Next, we explain why \down HD\ori~ also decreases. HD\ori~ becomes fairer when the difference between the normalised item relevance at $p$ ($q_p'$, \Cref{ext_eq:q_p_prime}) and normalised item interaction probability at $p$ ($c_p'$, \Cref{ext_eq:c_p_prime}) becomes smaller, for $p\leq k$. As such, the value of $c_p'$ is unaffected by additional relevant items. However, $q_p'$ decreases, as the user-wise normalised relevance, $r_{u,i}' = r_{u,i}/\sum_{i \in I} r_{u,i}$, decreases given additional relevant items per user. Consequently, \down HD becomes fairer. 

More interestingly, for some measures, we note that having fewer additional relevant items causes higher fluctuations than having a greater amount of additional relevant items. For example, having 1--3 additional relevant items per user causes a higher score deviation from the score for no additional items, compared to the deviation caused by 8--10 additional items per user. This is evident for IFD\divori, II-F\ori, and HD\ori, particularly for Amazon-lb. For \down IFD\divori, this is because the measure takes the mean pairwise difference of the combined exposure-relevance value and the exposure weight decreases down the rank positions. Thus, the mean pairwise difference decreases with a greater number of additional relevant items beyond the top-$k$. 
For \down HD\ori, this may be due to both the treatment of items with identical relevance by the unstable sorting method and the same reason for increased fairness explained in the previous paragraph (i.e., the normalised item relevance decreases, while the normalised item interaction probability remains constant). 
For \down II-F\ori, given additional relevant items, the target exposure (\Cref{ext_eq:eui-star}) decreases and the system exposure (\Cref{ext_eq:eui}) remains constant as it is unaffected by item relevance. Hence, the difference between the target and system exposure becomes smaller, resulting in a lower final score.

Given the high likelihood of having many unobserved items in RS datasets, the knowledge of which measures under/overestimate fairness and how the measure is affected by the number of additional relevant items may be useful in choosing which \textsc{Joint} evaluation measure to use. Measures that underestimate fairness (IFD\divori, IFD\divour, and HD\ori) may be more desirable to avoid overclaiming, for example. In our experiments, we also find that these measures tend to become fairer with a greater amount of relevant items, and thus can serve as an `upper bound'. Meanwhile, measures that overestimate fairness (IWO\our, IAA, and II-F\our) tend to amplify the overestimation with a greater amount of relevant items.

\subsubsection*{Effect across item selection strategy} How the additional relevant items are selected with the \texttt{top} and \texttt{bottom} strategy do not matter for IAA, IFD\divour, and II-F, but matters for IWO\our, IFD\divori, HD\ori, and AI-F\ori. Between the affected measures, there are also differences: the \texttt{bottom} strategy overestimates fairness more than the \texttt{top} strategy for IWO\our, whereas the opposite applies for IFD\divori. The trends of HD\ori{} follow that of IFD\divori, albeit to a lesser extent. As additional relevant items may exist close to the top-$k$, close to the bottom, or somewhere in between, measures that do not have large differences between the \texttt{top} and \texttt{bottom} item selection strategies are more robust than the other measures, and thus more desirable.

All in all, the trends across measures, datasets, and the placement of additional relevant items vary. 
First, the \joint{} measure scores are generally less stable against additional relevant items, when there are fewer relevant items per user in the original dataset. Therefore, the non-localisation limitation may be concerning for datasets with few relevant items per user. 
Second, the \joint{} measures respond in three different ways to a greater amount of relevant items: 
(i) II-F\ori/AI-F\ori{} are deceptively stable due to their constricted empirical range; 
(ii) IWO\our/IAA/II-F\our{} become more unfair; and 
(iii) IFD$_\div$/HD\ori{} become fairer. 
We note that this measure grouping differs from the one in \Crefrange{ext_ss:performance}{ext_ss:corr}; the alignment of \joint{} measures to \textsc{Eff} and \textsc{Fair} measures has little to do with how the measures are affected by additional relevance items. For IFD\divori/HD\ori/II-F\ori, fewer additional relevant items also mean worse over/underestimation of fairness. This second set of results could be useful in choosing suitable measures, e.g., if many relevant items are expected to be missing, measures that become fairer with a greater number of additional relevant items would be more appropriate. 
Third, IAA/IFD\divour/II-F are stable against the placement of the additional relevant items, and the rest are not. Considering the second and third sets of results, IFD\divour{} is the most robust against the non-localisation limitation, while IWO is the least robust.

\subsection{Sensitivity to Varying Recommendation Effectiveness and Item Exposure Distribution}
\label{ext_ss:insert}

\begin{figure}
    \centering
    \includegraphics[width=0.6\textwidth]{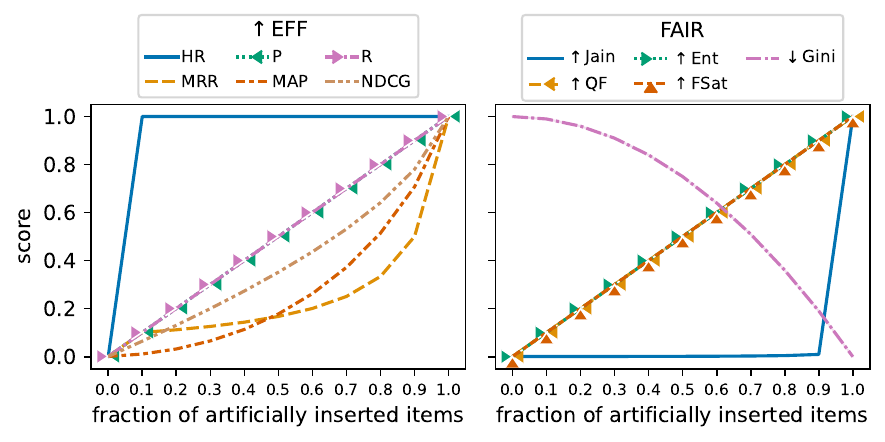}
    \includegraphics[width=0.98\textwidth]{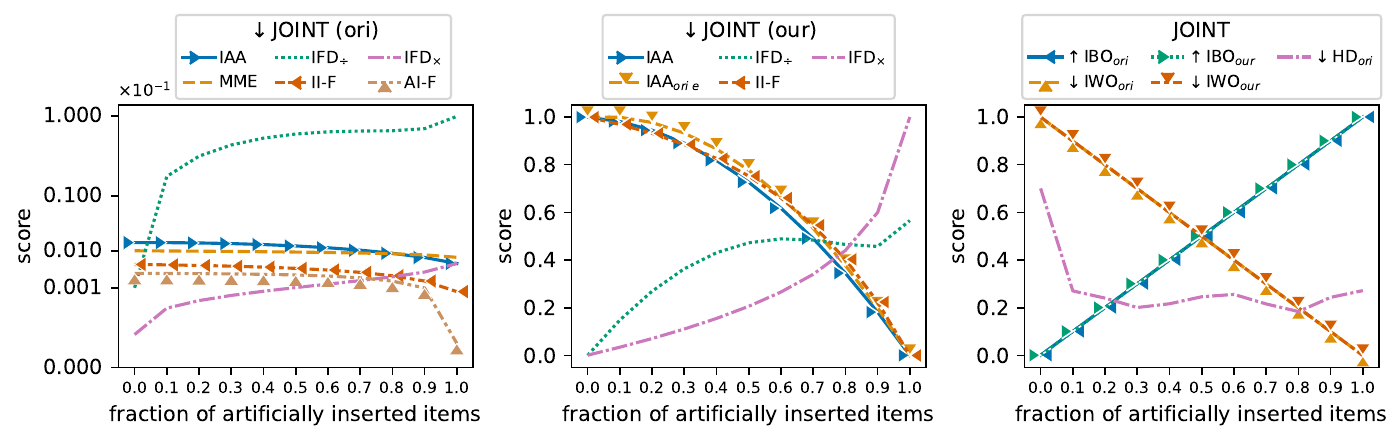}
    \caption{Artificial insertion of items with $m=1000$ (users). IAA$_\text{ori e}$ is IAA$_\text{our}$ computed with the original examination function $e_\text{li}$.     
    }
    \label{ext_fig:insert}
\end{figure}

How sensitive are the \joint{} measures towards varying proportions of relevant items and item exposure distribution in the recommendation? We assess this sensitivity as it affects score interpretation. If significant changes in item relevance and item exposure distribution do not affect a \joint{} measure, its score may not be an accurate indicator of recommendation fairness, specifically fairness notions that consider both item relevance and item exposure. We study the following: (i) the sensitivity of the original \joint{} measures compared to our corrected versions and (ii) how the inserted items affect \joint{} measures.

To investigate the above, we start with a ranking with the worst \eff{} and \fair{} scores, and replace a fraction of the ranking with items that are both more relevant and less exposed, one item per user at a time. The real-life datasets do not suit this analysis, so we generate a synthetic dataset with $m=1000$ users and $n=10{,}000$ items. We then generate artificial rankings per user, following \cite{Rampisela2024EvaluationStudy}. The first ranking has the same $k=10$ items for all users. These items are irrelevant to all users, except for one.\footnote{This is to keep the number of items exactly $km$ \cite{Rampisela2024EvaluationStudy}.} 
For each user, we replace the item at position $k$ with a relevant item that has less exposure, which improves both \eff{} and \fair. 
As IAA and IFD require the full recommendation list of each user, the remaining relevant items are ranked at the bottom of the full ranking (e.g., starting from position $n$ backwards). 
We do this iteratively for positions $k-1, k-2, \dots, 1$. Hence, the final rankings across all users altogether have $km$ unique items, where each item is relevant only to the user who has it at the top-$k$. 

We compute all \eff, \fair, and \joint{} measures at $k=10$. As a result of the replacements, all \eff{} and \fair{} measures are expected to score (close to) the worst possible at the start and improve gradually as more items are inserted into the original rankings. Based on the strategies to obtain the most (un)fair scores for IAA, IFD, and II-F (\Cref{ext_tab:example_fair_unfair}), IAA and II-F are expected to start from the unfairest score and end with the fairest score, while both IFD should start from the fairest score, and IFD$_\times$ should end with the unfairest score. 
We show the analysis results in \Cref{ext_fig:insert}.

\subsubsection*{Sensitivity of the original \joint{} measures vs.~ours}
Generally, most original \textsc{Joint} measures (\down IAA/IFD$_\times$/MME/II-F/AI-F) are largely insensitive to changes in \textsc{Eff} and \textsc{Fair} scores, i.e., their scores vary only a little. 
Further, the scores are always very close to 0, i.e., the scores range between (0, 0.0015). Since all of these measures are bounded below by 0, these small scores suggest that the recommendation is almost perfectly fair, even when both the \textsc{Eff} and \textsc{Fair} scores are the worst possible. This finding is consistent with the compressed empirical range observed for real-life datasets (\Cref{ext_ss:performance} \& \Cref{ext_ss:exp_maxmin}). In contrast, the empirical ranges of IBO/IWO/HD and all of our corrected \joint{} measures are closer to the theoretical range of $[0,1]$. The corrected version of \down IAA/IFD/II-F changes as expected, i.e., starting from 0 (or 1), and gradually reaching 1 (or 0). \down IFD\divour{} does not reach 1 as the final ranking does not correspond to the most unfair ranking for IFD\divour{}. 
Hence, our corrected measures are more responsive to changes in the \eff{} and \fair{} scores, compared to the original ones.

\subsubsection*{The trends of \joint{} measures} All \joint{} measures, both the originals and ours, improve (slightly) as the \eff{} and \fair{} scores become better, except for IFD. This is because IFD evaluates fairness by taking the pairwise difference of the items' exposure-relevance combined values; thus, moving the relevant items from the bottom to the top-$k$ increases the disparity between the relevant items' exposure at and outside the top-$k$. The trends also differ among the \joint{} measures that become (slightly) fairer with more inserted items: IBO/IWO improves proportionally to the amount of inserted items, as both are the fractions of items with better/worse impact scores than a threshold; HD improves in an irregular manner as the unstable sorting\footnote{as per the original implementation in \citeauthor{Jeunen2021Top-KExposure}~\cite{Jeunen2021Top-KExposure}} of items based on their ground truth relevance introduces randomness. IAA/IFD/MME/II-F/AI-F improves non-linearly. The trend of IAA and II-F closely resembles that of Gini, while the trend of IFD\mulour{} is close to 1-Gini. 

Here, we also see a manifestation of the \textbf{zero-exposure} limitation in IAA. We compare IAA\our{} to  IAA$_\text{ori e}$, i.e., IAA\our{} computed with the uncorrected linear examination function $e_{\text{li}}$. 
The uncorrected version has the same value for 0.0 and 0.1 fractions of artificially inserted items in the recommendation. The item at position 10 (corresponding to 0.1 fraction of inserted items in the recommendation) is regarded as unexposed by IAA$_\text{ori e}$, so replacing it with a relevant item does not make a difference. In this case, we cannot directly compare IAA\our{} to IAA\ori{} in \Cref{ext_fig:insert}, as the IAA\ori{} scores are very small and presented in an exponential axis, which makes it hard to see the score difference between 0.0 and 0.1 fractions of artificially inserted items.

In conclusion, the original versions of IAA/IFD/MME/II-F/AI-F are insensitive to increasing \eff~and \fair~scores, while our corrections to IAA/IFD/II-F reflect the change reasonably. For some original measures (MME/AI-F), the compressed empirical range issue persists, but this issue is not present for HD\ori/IBO/IWO. 
These findings are also helpful in better understanding the characteristics of MME, AI-F, HD, and IBO/IWO, which suffer from the \textbf{non-realisability} limitation but have not been corrected for it, i.e., it is not possible to derive a general formulation of the most (un)fair ranking for these measures. This analysis sheds light on the kind of ranking that produces a relatively fairer score than other rankings and thus can be used to work towards a general formulation of the most (un)fair ranking.

\section{Guidelines for Measure Usage}
\label{ext_s:guideline}

Based on the theoretical analysis (\Cref{ext_s:limitation}) and the empirical findings (\Cref{ext_s:experiment}), we formulate detailed guidelines on the appropriate usage of the relevance-aware measures (summarised in \Cref{ext_tab:consideration-summary}). 

\begin{table}[!ht]
    \centering
    \caption{Factors related to the measure Alignment (A), Computability (C), Interpretability (I), Expressiveness (X), Stability (S), and Efficiency (E) 
    to consider when selecting relevance-aware item fairness measures. Statements related to (A) are neutral; the rest are desirable.}
    \label{ext_tab:consideration-summary}
    \resizebox{\textwidth}{!}{
    
    }
\end{table}

\subsubsection*{Alignment} Alignment refers to the measure's tendency to agree with another measure, e.g., in having the same best model or ranking of models. We find that some relevance-aware fairness measures align more with effectiveness measures, some agree more with exposure-based measures ($\S$\ref{ext_ss:corr}), and the rest have inconsistent alignment to the single-aspect measures. Based on their alignment, the measures can be grouped into three: 
(i) \eff-aligned measures (IAA/HD/II-F);
(ii) \fair-aligned measures (IFD/MME/AI-F), which either strongly agree with \fair{} measures or disagree with \eff{} measures; 
and
(iii) IBO/IWO.\footnote{Some measures in group (ii) agree with the \fair{} measures for only 2/4 datasets. Other measures in this group  disagree with the \eff{} measures for only 2/4 datasets.} 
For some measures, their alignment with the single-aspect measures can be explained based on what kind of ranking results in the most (un)fair scores (\Cref{ext_tab:example_fair_unfair}). For example, IAA and II-F align with \eff{} measures as placing all relevant items at the top will result in the most effective recommendations and the fairest IAA/II-F. In contrast, IFD disagrees with \eff{} measures, as ranking all relevant items at the bottom produces the fairest IFD, but the worst recommendation effectiveness.

Our suggestions are as follows:

\begin{itemize}
    \item Avoid using relevance-aware measures that are highly similar to each other in their alignment to single-aspect measures. Instead, compute one measure per group to represent a set of highly similar measures; other measures in the same group do not need to be computed. 
    \item Prioritise evaluating with non-\eff-aligned measures (i.e., IFD/MME/AI-F/\ IBO/IWO). Typically, recommendations are first evaluated with \eff{} measures, so quantifying fairness with \eff-aligned measures is not expected to add a new perspective, as the conclusion would mostly align with the \eff{} measures. 
\end{itemize}

\subsubsection*{Computability} Computability is the extent to which the measure can be computed, considering that some information required for the computation may not be available, or that some dataset composition or experimental choices would render the measure incomputable. Given five computability-related conditions (C1--C5), only IFD$_\times$ and MME can be computed for all cases. The second `best' measures in terms of computability are II-F and AI-F, which can be computed for 4 out of 5 computability-related cases. They cannot be computed if the total number of relevant items per user is unknown. 
The most restricted measure in terms of computability is IAA\ori, which can be computed only if 4/5 conditions are fulfilled. As such, we advise the following:

\begin{itemize}
    \item The most flexible choices are IFD$_\times$/MME, especially if the total number of relevant items per user is unknown. This may often be the case, e.g., in online evaluation. Otherwise, II-F/AI-F can also be computed.
    \item Avoid using the original version of IAA as there are too many requirements.
    \item Compute the corrected version of IAA instead, as it can be computed in more cases than the original.
\end{itemize}

\subsubsection*{Interpretability} Measure interpretability refers to how easily we can understand how fair the recommendation is, simply by looking at the measure scores. When a measure has a theoretical range of [0,1], normally we would expect that one endpoint is mapped to the unfairest recommendation, and another to the fairest recommendation. Unfortunately, this is not the case for any of the original measures, as we have analysed in \Cref{ext_s:limitation}. When we do not know the empirical achievable range of the measures, it is harder to understand the fairness level, and thus measure interpretability is lower. We have corrected IAA, IFD$_\div$, IFD$_\times$, and II-F, such that a score of 0 means the fairest possible recommendation, and 1 means the unfairest possible recommendation in binary relevance and single-round recommendation scenarios. 

Another factor that affects measure interpretability, albeit to a smaller extent, is how item exposure is weighed. IAA\ori{} treats the item at position $k$ as unexposed, which can confuse the interpretation of the scores, e.g., when the item relevance at position $k$ is changed. Our version of IAA does not suffer from the same issue.

Based on the above, we suggest the following:

\begin{itemize}
    \item For binary relevance and single-round recommendations, prioritise computing IAA\our, IFD\our, or II-F\our, as they are more interpretable than the rest.
    \item Always compute IAA\our{} over IAA\ori{}, as the latter quantifies exposure in a faulty way.
    \item The rest of the original measures can technically still be used to compare which recommendation is fairer, but not for quantifying how close the recommendation is to the (un)fairest possible recommendation.
\end{itemize}

\subsubsection*{Expressiveness} Expressiveness refers to how much the measure scores can change in quantifying fairness for various distributions of item exposure and item relevance. An expressive measure is desirable, as its score can be easily distinguishable for different rankings. We find that most original relevance-aware measures (IAA/IFD/MME/II-F/AI-F), especially IFD$_{\times}$/MME/AI-F are not expressive enough, as their scores are almost always close to the fairest even when the items are mostly irrelevant and unfair based on the single-aspect measures of \eff{} and \fair{}. In other words, their scores tend to be compressed to the low end of their range, misleading us into thinking that the recommendation is extremely fair, even when both \eff{} and \fair{} scores are almost the worst possible. 

Further, they are less responsive to changes in the item exposure and relevance distribution than the single-aspect measures. 
Consequently, their scores are harder to distinguish across runs and need to be computed at higher precision (i.e., more decimal places) so that they can be distinguished. The scores can also be easily misinterpreted. For example, two models with a difference of only 0.001 in the scores can be interpreted to have similar performance, even if the measure has a small empirical range and is not very sensitive to begin with. 
Correcting IAA/IFD/II-F resolves the issues in the original measures. For other measures, the issue can be resolved by normalising the scores based on observed values of the measures \cite{Wu2022JointRecommendation}.  

We also look into the measure expressiveness in terms of the choice of the cut-off $k$. IFD\divori{} is the only measure that is insensitive to the change in $k$ due to the \textbf{top-$k$-insensitivity} limitation, while other measures, including IFD\divour, do not have this issue.

Thus, concerning the measure expressiveness, we encourage to:
\begin{itemize}
    \item Compute HD\ori/IBO/IWO or IAA\our/IFD\our/II-F\our, as they are more expressive than the others.
    \item Avoid computing the original IFD$_\times$/MME/AI-F, as they are extremely inexpressive.
    \item Be aware of potential misinterpretation in measures with small empirical scales (\down IAA\ori/IFD\ori/MME\ori/II-F\ori/AI-F\ori), as they often score very close to 0. 
    \item Refrain from computing IFD\divori, unless when evaluating on the full/untruncated rankings.
\end{itemize}

\subsubsection*{Stability} Measure stability refers to the measure being unaffected by certain conditions that should not change the measure score. The conditions are: having additional relevant items beyond the top-$k$ (S1--S2), having ties in the items' ground truth relevance (S3), and having users with only one relevant item (S4). For S1--S2, in common RS datasets, most users only have a few relevant items each. Hence, RS fairness evaluation may suffer from fluctuations if there are more relevant items than expected, even if these relevant items are not in the top-$k$. For S2, given more relevant items beyond the top-$k$, we desire measures that remain constant or become fairer to avoid overestimating fairness. In practice, we may not know the number of additional relevant items. If the measure becomes unfairer with more relevant items beyond the top-$k$, fairness is overestimated based on the existing item relevance. Thus, it is better to have a score that can guarantee that the score will not become more unfair in case of additional relevant items beyond the top-$k$. For S3, sorting items based on their ground truth relevance to compute HD introduces some noise depending on the sorting algorithm, especially when there are many items with the same relevance. For S4, IFD$_\div$ automatically gives the fairest score (0) when a user only has 1 relevant item, pulling the score down when there is a large percentage of such users. 
Out of all stability-related conditions, S4 is the easiest to handle, as users with only 1 relevant item can be excluded from the computation, as IFD$_\div$ is a pairwise measure. Overall, only IFD$_\times$ and MME are stable against the four stability-related conditions. 

Considering the above, we make the following suggestions:

\begin{itemize}
    \item Compute IFD$_\times$ and MME, as they are the most stable and the remaining measures are only stable against 2/4 stability-related conditions.
    \item HD should ideally not be used for cases where the items' ground truth relevance values include a lot of ties.
    \item When computing IFD$_\div$, calculate the number of users with a single relevant item. If too many such users exist, we suggest excluding them from evaluation and reporting the number of users with a single relevant item.
\end{itemize}

\subsubsection*{Efficiency} Is the measure computationally expensive? Can some parts of the computation be potentially parallelised? Here, we discuss the computational efficiency of the measures and the potential to compute them in a more efficient manner. Most relevance-aware measures are relatively cheap to compute (under 60 seconds per model, per dataset, based on the measure computation time in \citeauthor{Rampisela2025JointFrontier}~\cite{Rampisela2025JointFrontier}), except for MME and IFD$_{\times}$.\footnote{The computation time of IFD is based on our experiments \Cref{ext_ss:exp_maxmin}.} 
Both MME and IFD$_{\times}$ are computationally expensive as they are pairwise measures. Computing MME can take up to 30 minutes for larger datasets, while IFD$_{\times}$ is slightly faster ($\sim$12 mins). However, we note that the computation of both measures can also be parallelised, as the scores can first be computed either per user or item prior to the aggregation. Therefore, in terms of computational efficiency, our suggestions are as follows:

\begin{itemize}
    \item Avoid computing MME and IFD$_{\times}$ for datasets with many items.
    \item Consider parallelising the measure computation to save time.
\end{itemize}

\subsubsection*{Final suggestions} 
Overall, considering the alignment, computability, interpretability, expressiveness, stability, and computational efficiency of the relevance-aware individual item fairness measures, we provide the following suggestions:

\begin{itemize}
    \item Among all relevance-aware individual item fairness measures, computing IFD\mulour{} is one of the better choices, as it exhibits the most desirable characteristics in terms of its alignment to single-aspect measures, computability, interpretability, expressiveness, and stability. Even though it is also the second most computationally expensive measure, and can take more than 10 minutes to compute for larger datasets, the computation time can potentially be improved with parallelisation.

    \item IBO\our/IWO\our{} are good alternatives, as they align with neither \eff{} nor \fair{} measures, suggesting that they quantify a different evaluation dimension from the single-aspect measures. IBO\our/IWO\our{} also do not suffer from expressiveness or efficiency issues.
    
    \item Compute the corrected measures whenever possible, as they have more desirable properties overall than the original measures.

\end{itemize}

\section{Related Work}
\label{ext_s:literature}

\subsubsection*{Fairness evaluation in ranking and RSs} 
There exist many surveys on fairness in ranking or specifically in RSs \cite{Wang2023ASystems,Amigo2023ASystems,Zehlike2022FairnessSystems,Aalam2022EvaluationReview,Pitoura2022FairnessOverview,LiYunqi2023FairnessApplications,Smith2023ScopingPerspective,Wu2023FairnessStrategies,Deldjoo2024FairnessDirections}. Most of them provide a general overview of how fairness is quantified based on various dimensions, e.g., stakeholders and granularity, yet none has provided an in-depth analysis on the existing fairness measures. Aside from the survey papers, there is other prior work that are closer to ours. \citeauthor{Cherumanal2021EvaluatingRetrieval}~\cite{Cherumanal2021EvaluatingRetrieval} analysed the agreement between relevance, fairness, and diversity measures in the argument retrieval task. 
Additionally, previous work has investigated both the theoretical and empirical properties of fairness measures in ranking \cite{Schumacher2024PropertiesRankings,Raj2022MeasuringResults,Rampisela2024EvaluationStudy}. Similar to our work, previous work highlights the limitations of existing measures in various practical settings and provides a better understanding of the distinctive ways in which the measures quantify fairness.
However, the granularity of the measures differs:  \citeauthor{Schumacher2024PropertiesRankings}~\cite{Schumacher2024PropertiesRankings} and \citeauthor{Raj2022MeasuringResults}~\cite{Raj2022MeasuringResults} focus on evaluation measures for group fairness, whereas we focus on individual fairness measures. \citeauthor{Rampisela2024EvaluationStudy}~\cite{Rampisela2024EvaluationStudy} also study individual item fairness, but they examine only exposure-based fairness measures that are detached from relevance, while we investigate measures that jointly account for both item exposure and item relevance.

\subsubsection*{Group fairness and individual fairness}
Based on the granularity, RS fairness can either be evaluated at the group or individual level, although most existing work focuses on the former \cite{Deldjoo2024FairnessDirections}. Group fairness is defined as providing similar treatment to both dominant and protected groups (e.g., gender minority) \cite{Deldjoo2024FairnessDirections,Ekstrand2022FairnessSystems}, while individual fairness is defined as ensuring that similar individuals receive similar treatment \cite{Dwork2012FairnessAwareness}. Evaluating individual fairness has several benefits over group fairness. Firstly, individual fairness evaluation does not need sensitive demographic information (e.g., gender, race), which is often used for group fairness evaluation \cite{Lazovich2022MeasuringMetrics}, but may not be readily available due to privacy or legal issues. Secondly, evaluating individual fairness provides a more comprehensive overview by considering the distribution across all individuals \cite{Lazovich2022MeasuringMetrics}, rather than only considering a few groups, which can be too simplistic as the grouping may not accurately represent the individuals' whole identity \cite{Ekstrand2022FairnessSystems,Deldjoo2024FairnessDirections}.

\subsubsection*{Accounting for relevance and exposure fairness jointly}
Other than using an evaluation measure \textit{per se}, jointly quantifying item relevance and exposure can also be done through other approaches. \citeauthor{Cherumanal2021EvaluatingRetrieval}~\cite{Cherumanal2021EvaluatingRetrieval} compute the harmonic mean of relevance and exposure-based fairness measures to rank the systems based on both aspects, while \citeauthor{Rampisela2025JointFrontier}~\cite{Rampisela2025JointFrontier} compare the model scores to the Pareto Frontier, which considers the trade-off between relevance and fairness. 
Beyond individual item fairness, there are other evaluation measures that jointly quantify relevance and fairness. \citeauthor{Gao2022FAIR:Evaluation}~\cite{Gao2022FAIR:Evaluation} introduce a measure combining KL-divergence and IDCG to jointly evaluate relevance and group fairness in IR. \citeauthor{Xu2023P-MMF:System}~\cite{Xu2023P-MMF:System} measure utility and provider fairness in RSs jointly using a weighted sum of the two aspects. The relevance-aware measures in our work differ from the two previous measures \cite{Gao2022FAIR:Evaluation,Xu2023P-MMF:System} as none of them combine two single-aspect measures \cite{Gao2022FAIR:Evaluation, Rampisela2025JointFrontier,Cherumanal2021EvaluatingRetrieval} or apply a weighted sum of the two aspects \cite{Xu2023P-MMF:System}. In contrast, \citeauthor{Garcia-Soriano2021Maxmin-FairConstraints}~\cite{Garcia-Soriano2021Maxmin-FairConstraints} quantifies fairness and relevance for individual fairness in ranking similarly to HD \cite{Jeunen2021Top-KExposure}, as both compare item positions based on ground truth relevance to their positions based on predicted rankings. However, we do not use the measure in \citeauthor{Garcia-Soriano2021Maxmin-FairConstraints}~\cite{Garcia-Soriano2021Maxmin-FairConstraints} because it requires significant modifications to customise it for RS fairness.

\section{Conclusion}
\label{ext_s:conclusion}

We have conducted a novel theoretical and empirical investigation of existing relevance-aware individual item fairness measures in Recommender Systems (RSs), which account for both item exposure and item relevance. We identified theoretical measure limitations and studied the extent of the limitations under common and extreme evaluation settings. Further, we addressed these limitations by redefining measures where possible, or explaining why some limitations are unresolvable. Our comprehensive empirical analysis using real-life and synthetic data led to practical guidelines for selecting the appropriate measures for various use cases.

In this work, we focused primarily on deterministic/single rankings of items with binary relevance and, more briefly discussed measure limitations under non-binary relevance and multi-round scenarios at a conceptual level. Future work should conduct a deeper investigation, both theoretically and empirically, into relevance-aware fairness evaluation given a distribution of rankings (i.e., stochastic rankings or multi-round recommendations), to align with distributionally-informed evaluation in RSs \cite{Ekstrand2023Distributionally-InformedEvaluation}, and for graded relevance. Other future work could also explore the dynamics between individual and group fairness evaluation measures, as well as multistakeholder evaluation, which accounts for user and item fairness jointly \cite{Wang2024IntersectionalRecommendation,Burke2025De-centeringSystems}.

\section*{Acknowledgements}
The work is supported by the Algorithms, Data, and Democracy project (ADD-project), funded by the Villum Foundation and Velux Foundation. 
Pietro Tropeano contributed to a sketch of the proof. 

\section{Appendix}
\subsection{Mathematical workings for ranking strategies to achieve the (un)fairest \textsc{Joint} scores}
\label{ext_app:math}
We provide the derivation of the fairest ranking for the original IAA$(u)$, IFD$_\div(u)$ (computed with $J_{\div\text{-our}}$), IFD$_\times(u)$, and II-F$(u)$. For IAA$(u)$, we also show the workings behind the unfairest ranking.

\subsubsection{Inequity of Amortized Attention (IAA)}
We show that ranking relevant items at the top-$k$ (i.e., ranking items in non-increasing relevance) produces the fairest \down IAA$(u)$. 
Let $a_p^{\text{IAA}}$ be the exposure of item at position $p$, where item exposure in IAA is based on the product of the top-$k$ indicator function and the examination function, i.e., $1_{L_u}(i) \cdot \tilde{e}_{li\text{-our}}(u,i)$; for brevity, we denote this as $a_p$.
Let $r_{u,i} \in \{0,1\}$ be the relevance of item $i$ to user $u$; for brevity, we denote this as $r_i$. 
Formally, 
\begin{theorem} 
\label{ext_theo:iaa-fair}
Given any two items $i$ and $i'$, where  $z(u,i)=p$, $z(u,i')=p'$, $r_i > r_{i'}$,  ranking the item with higher relevance ($i$) closer to the top than the item with lower relevance ($i'$) would result in a fairer (or equal) \down IAA$(u)$ than doing vice versa:
\begin{equation*}
    r_i > r_{i'} 
    \wedge
    p < p'
    \Rightarrow
    |a_p-r_{i}| + |a_{p'}-r_{i'}| 
    \leq   
    |a_p-r_{i'}| + |a_{p'}-r_{i}| 
\end{equation*}
\end{theorem}

\begin{proof}
We prove \Cref{ext_theo:iaa-fair} by contradiction. Assume \Cref{ext_theo:iaa-fair} is false, i.e.:
\begin{align*}
  \neg \left( 
    r_i > r_{i'} \wedge p < p' \Rightarrow |a_p-r_{i}| + |a_{p'}-r_{i'}| \leq   |a_p-r_{i'}| + |a_{p'}-r_{i}|
    \right) \\
   \neg \left( 
     \neg (r_i > r_{i'} \wedge p < p') \vee (|a_p-r_{i}| + |a_{p'}-r_{i'}|  \leq  |a_p-r_{i'}| + |a_{p'}-r_{i}|)
    \right) \\
   (r_i > r_{i'} \wedge p < p' ) \wedge (|a_p-r_{i}| + |a_{p'}-r_{i'}|  >  |a_p-r_{i'}| + |a_{p'}-r_{i}|)
\end{align*}

As $r_i > r_{i'}$ and we deal with binary relevance, $r_i=1$ and $r_{i'}=0$. Substituting these values to the LHS of the inequality, we get:  
$|a_p-r_{i}| + |a_{p'}-r_{i'}| 
= |a_p-1| + |a_{p'}|
= |a_p-1| + a_{p'} 
$ (as $a_{p'} \geq 0$).
Substituting $r_i=1$ and $r_{i'}=0$ to the RHS of the inequality, we get: 
$
|a_p-r_{i'}| + |a_{p'}-r_{i}|
= |a_p| + |a_{p'}-1|
= a_p + |a_{p'}-1|
$ (as $a_{p} \geq 0$). 

Putting the LHS and RHS of the inequality together, we have:
\begin{align*} 
|a_p-1| + a_{p'} & > a_p + |a_{p'}-1| \\
|a_p-1| - |a_{p'}-1| & > a_p -  a_{p'}  \\
|a_p-1| - |a_{p'}-1| & > 0 \ \qquad \qquad \qquad
                   &(\because p<p' 
                    \Leftrightarrow a_p \geq a_{p'}  
                    \Leftrightarrow a_p  -a_{p'} \geq 0 
                     &)\\
(1-a_p) - (1-a_{p'}) &> 0 \qquad \qquad \qquad &(\because a_p \leq 1 \Leftrightarrow a_p -1 \leq 0&) \\
 a_{p'} &> a_p \qquad \qquad \qquad &(\text{Contradiction with } p<p' 
                    \Leftrightarrow a_p \geq a_{p'}&)
\end{align*}
Hence, \Cref{ext_theo:iaa-fair} holds.
\end{proof}

Next, we show that ranking relevant items at the bottom (i.e., ranking items in non-decreasing relevance) produces the unfairest \down IAA$(u)$. 

\begin{theorem} 
\label{ext_theo:iaa-unfair}
Given any two items $i$ and $i'$, where  $z(u,i)=p$, $z(u,i')=p'$, $r_{i} < r_{i'}$,  ranking the item with lower relevance $(i)$ closer to the top than the item with higher relevance $(i')$ would result in an unfairer (or equal) \down IAA$(u)$ than doing vice versa:
\begin{equation*}
    r_{i} < r_{i'} 
    \wedge
     p < p'
    \Rightarrow
    |a_{p}-r_{i}| + |a_{p'}-r_{i'}| 
    \geq   
    |a_{p'}-r_{i}| + |a_{p}-r_{i'}| 
\end{equation*}
\end{theorem}

\begin{proof} 
We prove \Cref{ext_theo:iaa-unfair} by contradiction. Assume \Cref{ext_theo:iaa-unfair} is false, i.e.:
\begin{align*}
    \neg \left(r_{i} < r_{i'} \wedge p < p' \Rightarrow |a_{p}-r_{i}| + |a_{p'}-r_{i'}|  \geq |a_{p'}-r_{i}| + |a_{p} r_{i'}| \right) \\
    \neg \left(\neg (r_{i} < r_{i'} \wedge p < p') \vee (|a_{p}-r_{i}| + |a_{p'}-r_{i'}|  \geq |a_{p'}-r_{i}| + |a_{p} r_{i'}| )\right) \\
    (r_{i} < r_{i'} \wedge p < p') \wedge (|a_{p}-r_{i}| + |a_{p'}-r_{i'}| < |a_{p'}-r_{i}| + |a_{p} - r_{i'}|)
\end{align*}

As $r_i < r_{i'}$ and we deal with binary relevance, $r_i=0$ and $r_{i'}=1$. Substituting these values to the left-hand side (LHS) of the inequality, we get:  
$|a_p-r_{i}| + |a_{p'}-r_{i'}| 
= |a_p| + |a_{p'}-1|
= a_p + |a_{p'}-1|
$ (as $a_{p} \geq 0$).
Substituting $r_i=0$ and $r_{i'}=1$ to the right-hand side (RHS) of the inequality, we obtain: 
$
|a_p-r_{i'}| + |a_{p'}-r_{i}|
= |a_p-1| + |a_{p'}|
= |a_p-1| + a_{p'}
$ (as $a_{p'} \geq 0$).
Putting the LHS and RHS of the inequality together, we have:
\begin{align*} 
a_p + |a_{p'}-1| &< |a_p-1| + a_{p'} \\
a_p -  a_{p'} &< |a_p-1| - |a_{p'}-1| \\
0 &< |a_p-1| - |a_{p'}-1| \ \qquad
                   &(\because p<p' 
                    \Leftrightarrow a_p \geq a_{p'}  
                    \Leftrightarrow a_p  -a_{p'} \geq 0 
                     &)\\
0 &< (1-a_p) - (1-a_{p'}) \qquad &(\because a_p \leq 1 \Leftrightarrow a_p -1 \leq 0&) \\
 a_{p} &< a_{p'} \qquad &(\text{Contradiction with } p<p' 
                    \Leftrightarrow a_p \geq a_{p'}&)
\end{align*}
Hence, \Cref{ext_theo:iaa-unfair} holds.
\end{proof}

\subsubsection{Individual Fairness Disparity with Division (IFD\texorpdfstring{$_\div$}{})}
We show that ranking relevant items at the bottom (i.e., ranking items in non-decreasing relevance) produces the fairest \down IFD$_\div(u)$. We denote as $a_p^\text{IFD$_\div$}$ the exposure of item at position $p$, where item exposure in IFD$_\div$ is based on $J_{\div\text{-our}}$ (\Cref{ext_eq:ifd_j_div_our}). For brevity, we denote as $a_p$ the item exposure.

\begin{theorem} 
\label{ext_theo:ifd-div-fair}
Given any two relevant items $i$ and $i'$, where  $z(u,i)=p$, $z(u,i')=p'$, $r_{i} = r_{i'} = 1$,  ranking the items closer to the bottom (beyond the top-$k$) would result in a fairer (or equal) \down IFD$_\div(u)$ than doing vice versa. 
Formally, 
\begin{equation}
\label{ext_eq:ifd-theorem}
    \max\left\{0,\frac{a_{p}}{r_{i}}-\frac{a_{p'}}{r_{i'}}\right\} 
    +     \max\left\{0,\frac{a_{p'}}{r_{i'}}-\frac{a_{p}}{r_{i}}\right\}
\end{equation}
is:
(1) lower for  $k < p < p'$ than for $p<k<p'$ (ranking 1 item at the top-$k$ and another one beyond the top-$k$); and 
(2) lower for $k < p < p'$ than for $p<p'<k$ (ranking both items at the top-$k$).

\begin{proof}
    We prove (1) and (2) separately. 
    To prove (1), we start from the premise $k < p < p'$. Ranking the two relevant items beyond the top-$k$ results in  $a_p=a_{p'}=0$. As such, \Cref{ext_eq:ifd-theorem} is 0. 
    Meanwhile, for $p<k<p'$, we have $a_p>0$ and $a_{p'}=0$. As $a_p> a_{p'}$, the expression $\frac{a_{p}}{r_{i}}-\frac{a_{p'}}{r_{i}} = a_p >0$, while $\frac{a_{p'}}{r_{i'}}-\frac{a_{p}}{r_{i}}= -a_{p} <0$. 
    \Cref{ext_eq:ifd-theorem} evaluates to $\max\{0,a_p\} + \max\left\{0,-a_{p}\right\} = a_p > 0$. Therefore, (1) holds.

    To prove (2), we use the results of proving (1), i.e., when $k < p < p'$, \Cref{ext_eq:ifd-theorem} $=0$. For $p<p'<k$, we have $a_p > a_{p'}>0$. The expression $\frac{a_{p}}{r_{i}}-\frac{a_{p'}}{r_{i}} = a_p - a_{p'}>0$, while $\frac{a_{p'}}{r_{i'}}-\frac{a_{p}}{r_{i}}=  a_{p'} - a_{p} <0$.
    \Cref{ext_eq:ifd-theorem} evaluates to $\max\{0,a_p - a_{p'}\} + \max\left\{0,a_{p'} - a_{p} \right\} = a_p - a_{p'} + 0 > 0$. Thus, (2) holds.
    
As both (1) and (2) hold, \Cref{ext_theo:ifd-div-fair} holds.
\end{proof}
\end{theorem}

\subsubsection{Individual Fairness Disparity with Multiplication (IFD\texorpdfstring{$_\times$}{})}
We show that ranking relevant items at the bottom (i.e., ranking items in non-decreasing relevance) produces the fairest \down IFD$_\times(u)$. 
We denote as $a_p^\text{IFD$_\times$}$ the exposure of item at position $p$, where item exposure in IFD$_\times$ is based on $J_{\times\text{-ori}}$ (\Cref{ext_eq:ifd_j_x}). For brevity, we denote as $a_p$ the item exposure.

\begin{theorem} 
\label{ext_theo:ifd-mul-fair}
Given any two items $i$ and $i'$, where  $z(u,i)=p$, $z(u,i')=p'$, $r_{i} < r_{i'}$,  ranking the item with lower relevance $(i)$ closer to the top than the item with higher relevance $(i')$ would result in a fairer (or equal) \down IFD$_\times(u)$ than doing vice versa:
\begin{equation*}
    r_{i} < r_{i'} 
    \wedge
     p < p'
    \Rightarrow
    (a_{p}r_{i} - a_{p'}r_{i'})^2 
    \leq   
    (a_{p'}r_{i} - a_{p}r_{i})^2 
\end{equation*}
\end{theorem}
\begin{proof}
    We prove \Cref{ext_theo:ifd-mul-fair} by contradiction. Assume \Cref{ext_theo:ifd-mul-fair} is false, i.e.:
    \begin{align*}
        \neg \left(
            r_{i} < r_{i'} \wedge p < p' \Rightarrow (a_{p}r_{i} - a_{p'}r_{i'})^2 \leq (a_{p'}r_{i} - a_{p}r_{i})^2 
            \right) \\
        \neg \left(
            \neg \left(r_{i} < r_{i'} \wedge p < p' \right) 
            \vee \left[(a_{p}r_{i} - a_{p'}r_{i'})^2 \leq (a_{p'}r_{i} - a_{p}r_{i})^2 \right] 
            \right) \\
            (r_{i} < r_{i'} \wedge p < p' )
            \wedge \left[(a_{p}r_{i} - a_{p'}r_{i'})^2 > (a_{p'}r_{i} - a_{p}r_{i})^2 \right]
    \end{align*}
    As $r_i < r_{i'}$ and we deal with binary relevance, $r_i=0$ and $r_{i'}=1$. Substituting these values to the inequality, we get $a_{p'}^2 > a_p^2$. There are three possibilities of how the rank positions $p, p'$ relate to the cut-off $k$:

    \begin{enumerate}
        \item Both rank positions are beyond $k$, $k<p<p'$. 
        In this case, $a_p=a_{p'}=0$ as items at position $p$ and $p'$ are unexposed. As $a_{p'}^2 = a_p^2$, this contradicts $a_{p'}^2 > a_p^2$.  
        
        \item Both rank positions are at the top-$k$, $p<p'<k$.
         Here, $a_p > a_{p'} > 0 \Leftrightarrow a_p^2 > a_{p'}^2$, which contradicts $a_{p'}^2 > a_p^2$.
        
        \item One rank position is at the top-$k$, and another one beyond top-$k$, $p<k<p'$. 
        Here, $a_p > 0 = a_{p'} \Leftrightarrow a_p^2 > a_{p'}^2 $, which contradicts $a_{p'}^2 > a_p^2$.
    \end{enumerate}
    Therefore, \Cref{ext_theo:ifd-mul-fair} holds via proof by contradiction.
\end{proof}

\subsubsection{Individual-user-to-Individual-item Fairness (II-F)}
We show that ranking relevant items at the top-$k$ (i.e., ranking items in non-increasing relevance) produces the fairest \down II-F$(u)$. We denote as $a_p^\text{II-F}$ the exposure of item at position $p$, where item exposure in II-F is $E_{u,i}$  (\Cref{ext_eq:eui}). 
For brevity, we denote as $a_p$ the item exposure and as $t_i$ the target exposure of item $i$, $E_{u,i}^*$ (\Cref{ext_eq:eui-star}).

\begin{theorem} 
\label{ext_theo:iif-fair}
Given any two items $i$ and $i'$, where  $z(u,i)=p$, $z(u,i')=p'$, $r_i > r_{i'}$,  ranking the item with higher relevance ($i$) closer to the top than the item with lower relevance ($i'$) would result in a fairer (or equal) \down II-F$(u)$ than doing vice versa:
\begin{equation*}
    r_{i} > r_{i'} \wedge p < p' 
    \Rightarrow
    (a_{p}-t_{i})^2 + (a_{p'}-t_{i'})^2 
    \leq   
    (a_{p'}-t_{i})^2 + (a_{p}-t_{i'})^2 
\end{equation*}
\end{theorem}

\begin{proof}
    We prove \Cref{ext_theo:iif-fair} by contradiction. Assume \Cref{ext_theo:iif-fair} is false, i.e.:
    \begin{align*}
        \neg \left(
            r_{i} > r_{i'} \wedge p < p' 
                \Rightarrow
                (a_{p}-t_{i})^2 + (a_{p'}-t_{i'})^2 
                \leq   
                (a_{p'}-t_{i})^2 + (a_{p}-t_{i'})^2
            \right) \\
        \neg \left(
            \neg \left(r_{i} > r_{i'} \wedge p < p'  \right) 
            \vee 
            \left[(a_{p}-t_{i})^2 + (a_{p'}-t_{i'})^2 
                \leq   
                (a_{p'}-t_{i})^2 + (a_{p}-t_{i'})^2 \right] 
            \right) \\
            (r_{i} > r_{i'} \wedge p < p' )
            \wedge 
                \left[(a_{p}-t_{i})^2 + (a_{p'}-t_{i'})^2 
                >   
                (a_{p'}-t_{i})^2 + (a_{p}-t_{i'})^2
                \right]
    \end{align*}
    As $r_i > r_{i'}$ and we deal with binary relevance, $r_i=1$ and $r_{i'}=0$, which means the target exposures of item $i$ and $i'$ are $t_i>t_{i'}=0$. 
    Substituting these values to the inequality, we get:
    \begin{align*} 
    (a_{p}-t_{i})^2 + a_{p'}^2 &>  (a_{p'}-t_{i})^2 + a_{p}^2 \\
    (a_{p}-t_{i})^2 - (a_{p'}-t_{i})^2  &>   a_{p}^2 - a_{p'}^2 \\
   ( a_{p}^2-2a_pt_i +t_{i}^2)  - (a_{p'}^2-2a_{p'}t_i +t_{i}^2)  &>  a_{p}^2 - a_{p'}^2 \\
   (-2a_pt_i)  - (-2a_{p'}t_i)  &> 0 \\
     2a_{p'}t_i  &> 2a_pt_i \\
     a_{p'}  &> a_p
    \end{align*}
However, as $p<p' \Leftrightarrow a_p \geq a_{p'}$, there is a contradiction. Hence, \Cref{ext_theo:iif-fair} holds. 
\end{proof}

\chapter{Joint Evaluation of Fairness and Relevance in Recommender Systems with Pareto Frontier}
\label{chap:WWW25}

\section*{Abstract}
Fairness and relevance are two important aspects of recommender systems (RSs). Typically, they are evaluated either (i) separately by individual measures of fairness and relevance, or (ii) jointly using a single measure that accounts for fairness with respect to relevance. 
However, approach (i) often does not provide a reliable joint estimate of the goodness of the models, as it has two different best models: one for fairness and another for relevance. 
Approach (ii) is also problematic because these measures tend to be ad-hoc and do not relate well to traditional relevance measures, like NDCG. 
Motivated by this, we present a new approach for jointly evaluating fairness and relevance in RSs: Distance to Pareto Frontier (DPFR). 
Given some user-item interaction data, we compute their Pareto frontier for a pair of existing relevance and fairness measures, and then use the distance from the frontier as a measure of the jointly achievable fairness and relevance. Our approach is modular and intuitive as it can be computed with existing measures. 
Experiments with 4 RS models, 3 re-ranking strategies, and 6 datasets show that existing metrics have inconsistent associations with our Pareto-optimal solution, making DPFR a more robust and theoretically well-founded joint measure for assessing fairness and relevance.
Our code: \href{https://github.com/theresiavr/DPFR-recsys-evaluation}{github.com/theresiavr/DPFR-recsys-evaluation}.

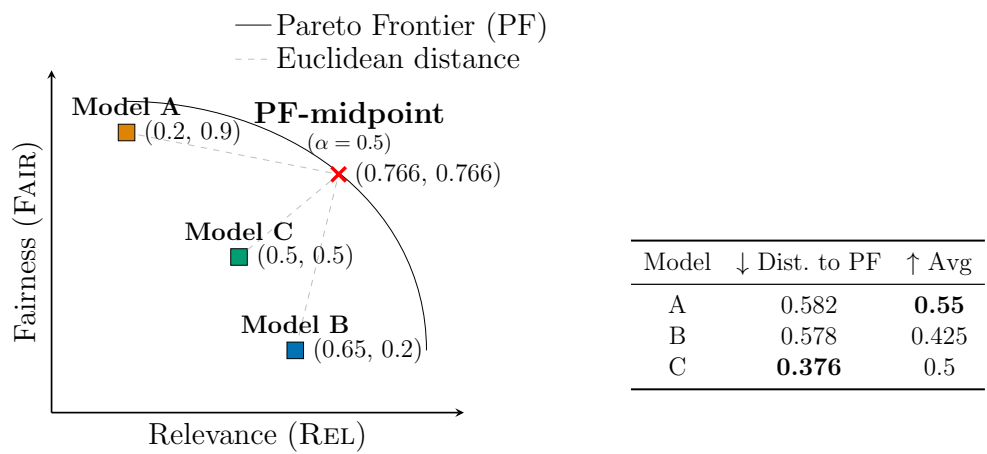
\begin{figure}
    \centering
    \begin{minipage}{.6\columnwidth}
        \centering
        \scalebox{.8}{
        \begin{tikzpicture}
            \tikzstyle{blue rectangle}=[fill={rgb,255: red,1; green,115; blue,178}, draw=black, shape=rectangle]
            \tikzstyle{green circle pareto}=[fill={rgb,255: red,2; green,158; blue,115}, draw=black, shape=rectangle]
            \tikzstyle{orange rectangle}=[fill={rgb,255: red,222; green,132; blue,5}, draw=black, shape=rectangle]
            \tikzstyle{pink rectangle}=[fill={rgb,255: red,204; green,120; blue,188}, draw=black, shape=rectangle]
            \tikzset{cross/.style={cross out, draw=black, minimum size=2*(#1-\pgflinewidth), inner sep=0pt, outer sep=0pt}, 
            cross/.default={5pt}}
            \begin{axis}
            [
                xmin = 0,
                ymin = 0,
                xmax = 1+0.1,
                ymax = 1+0.1,
                xlabel={\large Relevance (\textsc{Rel})}, 
                ylabel={\large Fairness (\textsc{Fair})}, 
                ticks=none,
                clip=false,
                axis lines=left,
                axis line style=thick,
                x label style={at={(axis description cs:0.5,0)},anchor=north},
                y label style={at={(axis description cs:0,.5)},anchor=south},
                legend style={at={(0.83,0.97)}, anchor=south,legend columns=1},
                legend style={font=\small, draw=none},
                legend cell align={left}
            ]

            \addlegendimage{}
            \addlegendimage{dashed,gray!50}

            \node (0) at (0.2, 1) {};
            \node (1) at (1, 0.2) {};
            \node [style=green circle pareto] (9) at (0.5, 0.5) {};
            \node [style=orange rectangle] (bestF) at (0.2, 0.9) {};
            \node [style=blue rectangle] (bestR) at (0.65, 0.2) {};
            \draw [in=90, out=0] (0.center) to (1.center);
            \addlegendentry{\large Pareto Frontier (PF)};
            \draw (0.7656854249492, 0.7656854249492) 
            node[cross,red, ultra thick,
            label={[align=center]0:{\normalsize(0.766, 0.766)}}, 
            label={[align=center, xshift=5pt, yshift=2pt]88:{\large\textbf{PF-midpoint}}\\[-2pt]{\footnotesize($\alpha=0.5$)}}] (10) {};
            \draw[dashed, color=gray!50] (9) -- (10);
            \addlegendentry{\large Euclidean distance};
            \draw[dashed, color=gray!50] (10) -- (bestF);
            \draw[dashed, color=gray!50] (10) -- (bestR);
            \node[label={\textbf{Model C}}, label={0:{(0.5, 0.5)}}]  at (9) {};
            \node[label={\textbf{Model A}}, label={0:{(0.2, 0.9)}}]  at (bestF) {};
            \node[label={\textbf{Model B}}, label={0:{(0.65, 0.2)}}]  at (bestR) {};
            \end{axis}%
        \end{tikzpicture}
        }
    \end{minipage}%
    \begin{minipage}[t][][b]{0.4\columnwidth}

     \scalebox{.8}{
        \begin{tabular}{ccc}
            \toprule
            Model & $\downarrow$ Dist.~to PF & $\uparrow$ Avg\\
            \midrule
            A & 0.582 & \textbf{0.55} \\
            B & 0.578 & 0.425 \\
            C & \textbf{0.376} & 0.5 \\
            \bottomrule
        \end{tabular}}
    \end{minipage}
    
    \caption{
    $(x, y)$ denotes the pair of relevance and fairness score. 
    Example: 
    Model A is best for fairness, 
    Model B is best for relevance, and Model C is the closest 
    to the Pareto Frontier (PF) midpoint, when relevance and fairness are equally weighted ($\alpha=0.5$). 
    Averaging relevance and fairness (Avg) leads to falsely concluding that Model A is best for both aspects. Note that distance to PF also beats other existing measures of fairness and relevance (see $\S$\ref{pareto_ss:corr}).
    }
    \label{fig:pareto_teaser}

\end{figure}

\section{Introduction
}

Relevance and fairness are important aspects of recommender systems (RSs). Relevance is typically evaluated using common ranking measures (e.g., NDCG), while various fairness measures for RSs exist \citep{Wang2023ASystems,Amigo2023ASystems}. 
Some fairness measures integrate relevance, so that they evaluate fairness w.r.t.~relevance. 
The problem with these joint measures is that they tend to be ad-hoc, unstable, and they do not account very well for both aspects simultaneously \citep{Rampisela2024CanRelevance}. 
Another way of evaluating relevance and fairness is to use a different measure for each aspect. However, this does not always provide a reliable joint estimate of the goodness of the models, as it may have two different best models: one for fairness and another for relevance.   
This can be avoided by aggregating the scores of the two measures into a single score, or by aggregating the resulting model rankings into one using rank fusion. These approaches are also problematic because: 
(i) the scores of the two measures may 
have different distributions and different scales, making them hard to combine; 
(ii) the two measures may not even be computed with the same input, making their combination hard to interpret (relevance scores are computed for individual users and then averaged, while fairness measures for individual items are typically based on individual item recommendation frequency); and 
(iii) the resulting scores are less understandable as it is unknown how close the models are to an ideal balance of fairness and relevance, e.g., an acceptable trade-off between fairness and relevance scores. 

To address the above limitations, we contribute an approach that builds on the set of all Pareto-optimal solutions \citep{Censor1977ParetoProblems}. Our approach addresses issue (i) and (ii) above by avoiding direct combination of measures. We directly address (iii) by computing the distance of the model scores to a desired fairness-relevance balance. 
Our approach uses Pareto-optimality, a popular concept in multi-objective optimization problems across domains, including RSs \citep{Ribeiro2015MultiobjectiveSystems}. 
A recommendation is Pareto-optimal if there are no other possible recommendations with the same \textsc{Rel} score that achieve better fairness.\footnote{The opposite is also true, but in RS scenario the \textsc{Rel} score is usually the primary objective, 
not the \textsc{Fair} score.} 
In other words, given Pareto-optimal solutions, we cannot get other recommendations that empirically perform better, 
unless relevance is sacrificed. 
In our approach, we combine existing \textsc{Fair} measures and \textsc{Rel} measures as follows. We build a Pareto Frontier (PF) that first maximises relevance, finds the best fairness achievable under the relevance constraint, and then jointly quantifies fairness and relevance as the distance from an optimal solution, see \Cref{fig:pareto_teaser}. 

Our approach, \emph{Distance to PF of Fairness and Relevance} (DPFR) has several strengths. First, DPFR is \emph{modular}; it can be used with well-known existing measures of relevance and fairness. DPFR is also \emph{tractable} as one can control the weight ($\alpha$) of fairness w.r.t.~relevance. As the resulting score is the distance to the scores of a traditional relevance measure and a well-known fairness measure, DPFR is also \emph{intuitive} in its interpretation. Most importantly, DPFR is a principled way of jointly evaluating relevance and fairness based on an empirical best solution that uses Pareto-optimality. Experiments with different RS models, re-ranking approaches and datasets show that there exists a noticeable gap between using current measures of relevance and fairness and our Pareto-optimal joint evaluation of relevance and fairness. This gap is bigger in larger datasets and when using rank-based relevance measures (i.e., MAP, NDCG), as opposed to set-based relevance measures (i.e., Precision, Recall). 

In this work, we focus on \textbf{individual item fairness}. This type of fairness is commonly defined as all items having equal exposure, where exposure typically refers to the frequency of item appearance in the recommendation list across all users~\citep{Patro2020FairRec:Platforms, Mansoury2020FairMatch:Systems, Rampisela2024EvaluationStudy}. Individual item fairness is important in ensuring that each item/product in the system has a chance to be recommended to any user \citep{Lazovich2022MeasuringMetrics}.

\section{Related Work}\label{s:prev_work}

Evaluating fairness and relevance together is a type of multi-aspect evaluation. 
However, none of the existing multi-aspect evaluation methods \citep{Maistro2021PrincipledRankings, Lioma2017EvaluationLists,Palotti2018MM:Engines} can be used in this case as 
these methods require separate labels that are unavailable in RS scenarios. 
Specifically, it is not possible to label an item as `fair', because item fairness depends on other recommended items. The same item can be a fair recommendation in one ranking, but unfair in another. In RSs, fairness is typically defined as treating users or items without discrimination \citep{Biega2018EquityRankings}. This is often quantified as the opportunity for 
having equal relevance (for users) or exposure (for items) 
\citep{Biega2018EquityRankings, Wang2022ProvidingSystems}, computed either individually or for
groups of items/users \citep{Raj2022MeasuringResults, Zehlike2022FairnessSystems}. 

The problem of jointly evaluating RS relevance and fairness is further aggravated by the fact that improved fairness is often achieved at the expense of relevance to users \citep{Mehrotra2018TowardsSystems}. 
We posit that this trade-off makes multi-objective optimization a suitable solution. Pareto optimality is a well-known objective for such optimization, 
and it has been previously used in RS but only to recommend items to users \citep{Ribeiro2015MultiobjectiveSystems,  Zheng2022AOptimization, Ge2022TowardLearning, Xu2023P-MMF:System}. 
Because the true PF is often unknown due to the problem complexity \citep{Laszczyk2019SurveyMeasures,Audet2020PerformanceOptimization}, prior work has used the model's training loss w.r.t.~two different aspects \citep{Lin2019ARecommendation} or scores from different models \citep{Nia2022RethinkingNetworks,Paparella2023Post-hocRecommendation} to generate the PF. 
Our work differs from this in terms of both the purpose of using Pareto-optimal solutions, and the nature of the PF. Specifically, we exploit Pareto-optimality through PF as a robust \textit{evaluation} method, instead of as a recommendation method. 
In addition, our generated PF is based on the ground truth (i.e., the test set), a common RS evaluation approach, instead of the recommender models' empirical performance, which may not be optimal. Thus, our PF is also model-agnostic, as opposed to the PF in \citep{Xu2023P-MMF:System}. 
Our approach differs also from FAIR~\citep{Gao2022FAIR:Evaluation} since the PF considers the empirically achievable optimal solution based on the dataset, while FAIR compares against the desired fairness distribution which might not be achievable. Lastly, \citep{Paparella2023Post-hocRecommendation} selects the optimal solution based on its distance to the utopia point (the theoretical ideal scores), whereas the utopia point may not be realistic due to dataset or measure characteristics~\citep{Rampisela2024EvaluationStudy,Moffat2013SevenMetrics}. Since our PF is generated based on test data, any of its solutions is empirically achievable.

\section{Distance to Pareto Frontier (DPFR)}
\label{s:our_method}

We present definitions 
($\S$\ref{ss:motivation}),  
and then explain DPFR in different steps: 
given a \textsc{Fair} and a \textsc{Rel} measure, how to 
generate PF based on the ground truth data in the test set ($\S$\ref{ss:generation}); how to choose a reference point in the PF based on $\alpha$ (e.g., the midpoint for $\alpha=0.5$) ($\S$\ref{ss:pareto-for-eval}); and how to compute the distance of the \textsc{Fair} and \textsc{Rel} scores to the reference point with a distance measure $d$ ($\S$\ref{ss:pareto-for-eval}). 
Additionally, we present a computationally efficient adaptation of DPFR ($\S$\ref{ss:compute-eff}). 

\subsection{Definitions}
\label{ss:motivation}

We adapt the Pareto-optimality definition \citep{vanVeldhuizen1999MultiobjectiveInnovations}: the multi-objective problem is finding the optimum \textsc{Fair} score $s_f$, and \textsc{Rel} score, $s_r$ from a list of possible recommendations across all users. We define the tuple $s = (s_r, s_f) \in S$, where $S$ is the Cartesian product of all possible \textsc{Rel} and \textsc{Fair} scores. The relation $\geq_A$ ($>_A$) means `better or equal to' (`better to') according to an aspect $A \in \{\textsc{Rel}, \textsc{Fair}\}$. 

\begin{definition}[Pareto Dominance]
A tuple $s = (s_r, s_f)$ dominates $s'=(s_r', s_f')$ iff $s$ is partially better than $s'$, i.e., $s_r \geq_\textsc{Rel} s'_r$ and $s_f \geq_\textsc{Fair} s'_f$, in addition to $s_r >_\textsc{Rel} s'_r$ or $s_f >_\textsc{Fair} s'_f$. 
\end{definition}

\begin{definition}[Pareto Optimality]
A solution (recommendation list) that has \textsc{Rel} and \textsc{Fair} scores of $x = (x_r, x_f) \in S$ is Pareto-optimal iff there is no other solution with $x'=(x_r', x_f') \in S$ that dominates $x$.
\end{definition}

\begin{definition}[Pareto Frontier]
The set of all Pareto-optimal tuples.
\end{definition}

\subsection{Pareto Frontier generation}
\label{ss:generation}
Given user-item preference data (e.g., test set), the aim is to explore the empirical, maximum feasible fairness towards individual items, such that the recommendation satisfies Pareto-optimality w.r.t.~fairness scores across all items and an average relevance score across users, e.g., MAP@$10=0.9$.\footnote{This is how \textsc{Fair} and \textsc{Rel} measures are usually computed.} This is done to measure how far a model performance is, from these Pareto-optimal solutions.
Enumerating all possible recommendations for users and items to find the complete set of Pareto-optimal solutions is computationally infeasible, and there is no analytical solution either. Instead, we contribute an algorithm that iteratively builds upon a maximally relevant initial recommendation list. Our algorithm iteratively finds Pareto-optimal recommendations by prioritising relevance over fairness, as recommendations are usually optimised for relevance (with or without fairness). This prioritisation is known as lexicographic optimization \citep{Ryu2018Multi-objectiveWeight}. 
We call our algorithm {\sc Oracle2Fair} (full technical description in App.~\ref{app:algo}). 
Our algorithm generates the PF of fairness and relevance in two steps: 
 \textbf{(1) initialisation} of the recommendations with an \textit{Oracle} (App.\ref{app:algo}, Algorithm~\ref{alg:oracle}).  The \textit{Oracle} generates a recommendation with the highest empirical score for relevance, based on user interactions that are part of the test set. This step is followed by \textbf{(2) replacements} to make the recommendations as \textit{Fair} as possible; at the end of this algorithm, the \textsc{Fair} scores should reach the empirically fairest score while maintaining as much relevance as possible. 
Throughout the PF generation, items in a user's train/val split are not recommended to the same user. Henceforth, \emph{relevant items} refers to the items in a user's test split.

\noindent \textbf{(1) Initialisation}. 
The Oracle recommends at most $k=10$ relevant items, from the $n$ items in the dataset, to each of the $m$ users in the test split, one user at a time. 
The recommendation begins with users having exactly $k$ items in the test split; only these items can be recommended to those users to gain the maximum relevance. Recommendations to other users are made maximally relevant and fair as follows: if a user has $>k$ relevant items, we pick $k$ items with the least exposure among them. Item exposure is computed based on what has been recommended to other users who already have exactly $k$ items. Note that this process is not trivial (see App.~\ref{app:algo}, 
Algo~\ref{alg:oracle}, ll.~\ref{ln:startgtk}--\ref{ln:endgtk}). If a user has $<k$ relevant items, we recommend those items at the top (to maximise top-weighted \textsc{Rel} measures) and fill the rest of their recommendation slots with the least exposed items in the dataset (Algo~\ref{alg:oracle}, ll.~\ref{ln:startltk}--\ref{ln:endltk}). This least-exposure prioritisation strategy ensures that the solutions are Pareto-optimal. 

\noindent \textbf{(2) Replacements}. 
The algorithm iteratively replaces the recommended items to achieve maximum fairness, such that each replacement results in a fairer recommendation than the previous. 
We compute the \textsc{Fair} and \textsc{Rel} measures after each replacement as follows. The most popular item, which is recommended most often, is replaced with one of these item types, 
in decreasing order of priority: an unexposed item, then the least popular item in the recommendation; this increases fairness from the previous recommendations. We do this one item and one user at a time, starting with the users that have the most popular item at the bottom of their recommendation list, to ensure that the decrease in relevance is minimum as the replacement item is mostly not relevant to that user. Nonetheless, the \textsc{Oracle2Fair} prioritises replacing the recommendations of users for whom the replacement item is relevant (if any). As fairness increases and relevance decreases/stays the same from the previous recommendation, the new recommendation is also Pareto-optimal. 
We continue the replacement until the maximum times any item is recommended is $\left\lceil km/n \right\rceil$, i.e., the upper bound of how many times an item can be recommended, if all items in the dataset must appear in the recommendation as uniformly as possible. We explicitly used $\left\lceil km/n \right\rceil$ as a stopping condition for \textsc{Oracle2Fair}. 
To ensure maximum \textsc{Rel} scores (especially in top-weighted measures), each time a replacement takes place, we rerank the recommendations based on descending relevance.

The resulting (\textsc{Rel, Fair}) scores reflecting Pareto-optimal recommendations from this process make up the PF. If there are duplicates in the \textsc{Rel} value, we keep the best \textsc{Fair} score for a single value of \textsc{Rel}. While it cannot be reasonably verified that the resulting PF matches the theoretical PF, this is a close way to build the full PF, as opposed to building the PF from trained models scores ($\S$\ref{s:prev_work}).

\subsection{Distance computation} 
\label{ss:pareto-for-eval}

For each pair of \textsc{Fair} and \textsc{Rel} measures, we find a reference point using a tunable parameter $\alpha \in [0,1]$; $\alpha=0$ means only relevance is accounted for, and $\alpha=1$ means only fairness is accounted for. Next, we explain how to compute the reference point. We first use the following equation to find the length of a subset $T$ of the PF: 
$
    lenPF(T) = \sum_{t=1}^{|T|-1} d_E(x^t, x^{t+1})
$.
 Given that $P$ is the set of all Pareto-optimal solutions, $x^t = (x_r^{t}, x_f^{t})$ is the pair of Pareto-optimal solutions $(x_r,x_f)$ with the $t$-th highest $x_r$ in $P$, and $d_E$ is the Euclidean distance. The overall PF length is $lenPF(P)$ or simply $lenPF$. 
 
 The reference point is $s_{\alpha} = x^{t'}$, where $t'$ is computed as follows:
 
\begin{equation*}
    t' = \argminA_{j \in \left[1,\dots,|P|-1\right]}\left|lenPF(T^j)-\ \alpha \cdot lenPF \right|
\end{equation*}

$T^{j}$ is a subset of $P$ containing the $j$ highest $x_r$ scores. So, the reference point is a point in the PF whose cumulative traversal distance is closest to the $\alpha$-weighted PF distance travelled from the first point in the PF. The reference point $s_\alpha$ is how far the PF is traversed, from the pair with the best \textsc{Rel} score to the one with the best \textsc{Fair} score, multiplied by $\alpha$. As the PFs 
may have different density of points along the frontiers, the reference point is not computed based on a percentile (e.g., median) to avoid bias towards the denser part. 
Next, the distance between each model's ($x_r, x_f$) scores and the reference point $s_\alpha$ is computed with a distance measure $d$ that accommodates 2d-vectors. The model with the closest distance is the best model in terms of both relevance and fairness, given the weight $\alpha$. 
We call this \emph{Distance to Pareto frontier of Fairness and Relevance} (DPFR). 

\subsection{Efficient computation of Pareto Frontier}
\label{ss:compute-eff}

Generating the PF as in $\S$\ref{ss:generation} is costly. An efficient alternative is to compute a subset of the PF. We pick a fixed amount of Pareto-optimal solutions to compute, $p$ (e.g., 10). However, to reliably approximate the PF, these solutions should be spread according to the PF distribution, as opposed to e.g., only computing the first $p$ points of the PF. The spread of the points is important, as the reference point in DPFR is computed based on the overall estimated PF. 
In the estimated PF, the first point corresponds to the measure scores of the initial recommendation given by the Oracle, and the rest are spread evenly throughout the PF generation. 
To select at which point of the \textsc{Oracle2Fair} algorithm the measures should be computed, we first estimate the total number of replacements needed by examining the distribution of recommended items frequency. This is done by getting the individual frequency count of all items in the recommendation, and subtracting the ideal upper bound of item count $\lceil km/n \rceil$ ($\S$\ref{ss:generation}) from each count. 
The number of expected replacements is the sum of the difference between the item frequency count and the ideal upper bound of item count in $\S$\ref{ss:generation}. Items with recommendation frequency counts less than the upper bound are excluded. With the estimated total number of replacement $numRep$, we compute the measures every $numRep\ \text{div} (p-1)$ replacements done by \textsc{Oracle2Fair}, such that the measures are computed a total of $p-1$ times during the replacement process + 1 time before the replacement starts. These $p$ points are spread evenly in terms of distance in the PF, which is important as DPFR is a distance-based measurement. 

\section{Experimental Setup}
\label{s:experiments}

We study how our joint evaluation approach, DPFR, compares to existing single- and multi-aspect evaluation measures of relevance and fairness. 
Next, we present our experimental setup. 
The experiments are run in a cluster of CPUs and GPUs (e.g., Intel(R) Xeon(R) Silver 4214R CPU @ 2.40GHz, AMD EPYC 7413 and 7443, Titan X/Xp/V, Titan RTX, Quadro RTX 6000, A40, A100, and H100).

\noindent \textbf{Datasets.} We use six real-world datasets of various sizes and domains: 
e-commerce (A\-ma\-zon Luxury Beauty, i.e., Amazon-lb \citep{Ni2019JustifyingAspects}),
movies (ML-10M and ML-20M \citep{Harper2015TheContext}), 
music (Lastfm \citep{Cantador20112nd2011},  
videos (QK-video \citep{Yuan2022Tenrec:Systems}), 
and jokes (Jester \citep{Goldberg2001Eigentaste:Algorithm}). 
The datasets are as provided by \citep{Zhao2021RecBole:Algorithms}, except for QK-video, which we obtain from \citep{Yuan2022Tenrec:Systems}. 
Statistics of the datasets are in \Cref{pareto_tab:stats} and extended statistics are in App.~\ref{app:stat}.

\begin{table}[tb]
\caption{Statistics of the preprocessed datasets.}

\resizebox{0.98\columnwidth}{!}{
\begin{tabular}{lrrrr}
\toprule
\textbf{dataset} & \multicolumn{1}{l}{\textbf{\#users ($m$)}} & \multicolumn{1}{l}{\textbf{\#items ($n$)}} & \multicolumn{1}{l}{\textbf{\#interactions}} & \multicolumn{1}{l}{\textbf{sparsity (\%)}} \\ 
\midrule                                                                   
Lastfm \citep{Cantador20112nd2011}& 1,859 & 2,823 & 71,355 & 98.64\% \\
Amazon-lb \citep{Ni2019JustifyingAspects} & 1,054 & 791 & 12,397 & 98.51\% \\
QK-video \citep{Yuan2022Tenrec:Systems} & 4,656 & 6,423 & 51,777 & 99.83\% \\
Jester \citep{Goldberg2001Eigentaste:Algorithm} & 63,724 & 100 & 2,150,060 & 66.26\% \\ 
ML-10M \citep{Harper2015TheContext} & 49,378 & 9,821 & 5,362,685 & 98.89\% \\
ML-20M \citep{Harper2015TheContext} & 89,917 & 16,404 & 10,588,141 & 99.28\% \\
\bottomrule
\end{tabular}
}
\label{pareto_tab:stats}
\end{table}

\noindent \textbf{Preprocessing.} We remove duplicate interactions (we keep the most recent). We keep only users and items with $\geq5$ interactions. 
We convert ratings equal/above the following threshold to $1$ and discard the rest: for Amazon-lb and ML-*, the threshold is 3, as their ratings range between $[1,5]$ and $[0.5, 5]$ resp.; the threshold for Jester is $0$, as the ratings range in $[-10, 10]$. Lastfm and QK-video have no ratings. QK-video has several interaction types; we use the `sharing' interactions. 
For Jester, we remove users with $>80$ interactions, to provide a large enough number of item candidates for recommendation during testing.\footnote{\label{fn:jester}Some users in Jester have interacted with almost all 100 items. If a user has 80 items in the train/val set, there would only be 20 candidate items to recommend during test, which makes it easier to achieve higher relevance. 
} 

\noindent \textbf{Data splits.} To obtain the train/val/test sets for Amazon-lb and ML-*, we use global temporal splits \citep{Meng2020ExploringModels} with a ratio of 
6:2:2 on the preprocessed datasets. Global random splits with the same ratio are used for the other datasets that have no timestamps. 
From all splits, we remove users with $<5$ interactions in the train set. 

\noindent\textbf{Measures}.
We measure relevance (\textsc{Rel}) with Hit Rate (HR), MRR, Precision (P), Recall (R), MAP, and NDCG. We measure individual item fairness (\textsc{Fair}), as per  \citep{Rampisela2024EvaluationStudy}, with Jain Index (Jain) \citep{jain1984quantitative, Zhu2020FARM:APPs}, Qualification Fairness (QF) \citep{Zhu2020FARM:APPs}, Gini Index (Gini) \citep{Gini1912VariabilitaMutabilita, Mansoury2020FairMatch:Systems}, Fraction of Satisfied Items (FSat) \citep{Patro2020FairRec:Platforms}, and Entropy (Ent) \citep{Shannon1948ACommunication, Patro2020FairRec:Platforms}.\footnote{
These are set-based measures, but we do not expect the conclusions to differ for rank-based measures (App.~\ref{app:discussion}).
} 
We also use joint measures (\textsc{Fair+Rel}) as per \citep{Rampisela2024CanRelevance}: 
Item Better-Off (IBO) \citep{Saito2022FairRanking},\footnote{
The measure Item Worse-Off is not used as its formulation is highly similar to IBO.}
Mean Max Envy (MME) \citep{Saito2022FairRanking},
Inequity of Amortized Attention (IAA) \citep{Biega2018EquityRankings, Borges2019EnhancingAutoencoders}, Individual-user-to-individual-item fairness (II-F) \citep{Diaz2020EvaluatingExposure, Wu2022JointRecommendation}, and 
All-users-to-individual-item fairness (AI-F) \citep{Diaz2020EvaluatingExposure}. 
We denote by $\uparrow$/$\downarrow$ measures where higher/lower is better.
DPFR is computed with Euclidean distance and $\alpha=0.5$ (PF midpoint) for all datasets, so the midpoint is based on the empirically achievable scores, per dataset and measure pairs. 
For all runs, we use $k=10$. 

\noindent\textbf{Recommenders}. We use 4 
common collaborative filtering-based recommenders: ItemKNN \citep{Deshpande2004Item-basedAlgorithms},
BPR \citep{RendleBPR:Feedback},
MultiVAE \citep{Liang2018VariationalFiltering}, and 
NCL \citep{Lin2022ImprovingLearning}, with RecBole \citep{Zhao2021RecBole:Algorithms} and tune their hyperparameters. We train for 300 epochs with early stopping, and keep the configuration with the best NDCG@10 during validation. Each user's train/val items are excluded from their recommendations during testing. 

\noindent\textbf{Fair Re-rankers.} 
To have fairer recommendations, we reorder the top $k'$ items that are pre-optimised for relevance. Ideally $k'>k$ to allow exposing items that are not in the top $k$. As there are few relevant items per user in RS data,\footnote{The median number of relevant items per user across all datasets is 2--53, see App.~\ref{app:stat}.}
$k'$ should not be too big (e.g., 100). So, we re-rank the top $k'=25$ items per dataset and model using three methods: GS, CM, and BC (explained below). 
We re-rank separately per user for CM and BC, or altogether for GS, for all $k'm$ recommended items, where $m$ is the number of users. 
Other fair ranking methods exist but cannot be used as they apply to group or two-sided fairness only (e.g., \citep{Zehlike2017FAIR:Algorithm, Zehlike2020ReducingApproach, Patro2020FairRec:Platforms}), or to stochastic rankings only (e.g., \citep{Wu2022JointRecommendation, Oosterhuis2021ComputationallyFairness}), or do not scale to larger datasets (e.g., \citep{Biega2018EquityRankings, Saito2022FairRanking}).
    
   \noindent \textit{1. Greedy Substitution (GS) \citep{Wang2022ProvidingSystems}} is a re-ranker for individual item fairness. We modify the GS algorithm, to replace the most popular items with the least popular ones, both considering how many times an item is at the top $k'$ recommendations for all users (App.~\ref{app:gs}). As such, items can be swapped across users. To determine which items are most popular (i.e., to be replaced) and least popular (replacement items), the parameter $\beta=0.05$ is used. We pair these two item types, and for each pair, we calculate the loss of (predicted) relevance if the items are swapped. We then replace up to 25\% of the initial recommendations, starting from item pairs with the least loss. 
    
    \noindent \textit{2. COMBMNZ (CM) \citep{Lee1997AnalysesCombination}} is a common rank fusion method. Two rankings are fused for each user: one based on the (min-max) normalised predicted relevance score and another based on the coverage of each top $k'$ item (to approximate fairness). We calculate item coverage only based on their appearance in the top $k$ across all users and min-max normalise the score across all users. 
    As favouring items with higher coverage would boost unfairness, we generate the ranking using $1$ minus the normalised coverage. CM uses a multiplier based on the item appearance count in the two rankings above; this count is also only based on the top $k$. 
    The resulting ranking is a fused ranking of fairness and relevance.
    
    \noindent \textit{3. Borda Count (BC)} is a common rank fusion method. For each user, we combine the original recommendation list and the rankings based on increasing item coverage, as in CM. Unlike CM, BC uses points. Higher points are given to items placed at the top. The result is a fused ranking of fairness and relevance. 

\section{Experimental Results}
\label{ss:result-pareto}
We now present the evaluation scores of 16 runs (4 recommenders $\times$ 3 re-rankers, including no reranking) ($\S$\ref{sss:results-diversity}). The relevance and fairness scores of these runs are the input to our DPFR approach. 
Not all combinations of evaluation measures are suited for PF. We explain this in $\S$\ref{sss:no-fit}. We present the generated PF ($\S$\ref{sss:result-pf}) and compare existing measures to DPFR ($\S$\ref{pareto_ss:corr}). We compare the results of efficient DPFR to other joint evaluation approaches ($\S$\ref{ss:compare-eff}).

\subsection{Groundwork Runs} \label{sss:results-diversity}

The scores of \textsc{Rel}, \textsc{Fair}, and \textsc{Fair+Rel} measures for our 16 runs are shown in the appendix (\Cref{tab:base-rerank-all-1}--\ref{tab:base-rerank-all-2}). 
Two main findings emerge from \Cref{tab:base-rerank-all-1}--\ref{tab:base-rerank-all-2}. First, for all six datasets, \textbf{none of the best models according to \textsc{Rel} are also the best according to \textsc{Fair} measures}. This is similar to our toy example (\Cref{fig:pareto_teaser}), where one model ranks highest for fairness and another for relevance. 
Second, \textbf{the five \textsc{Fair+Rel} measures have no unanimous agreement on the best model}. 
IBO has a different best model from the others in 4/6 times, but sometimes agrees with one or more \textsc{Fair} measures. 
MME and AI-F agree on the best model 5/6 times, and sometimes agree on the best model with \textsc{Fair} measures.
The best model according to IAA and II-F is always the same, and 4/6 times the same as the best model based on the \textsc{Rel} measures. 
The overall picture is inconclusive, with some \textsc{Fair+Rel} measures aligning more with \textsc{Fair} measures, and others aligning more with \textsc{Rel} measures. 

\begin{figure}[htbp]
    \centering
    \includegraphics[width=0.75\columnwidth]{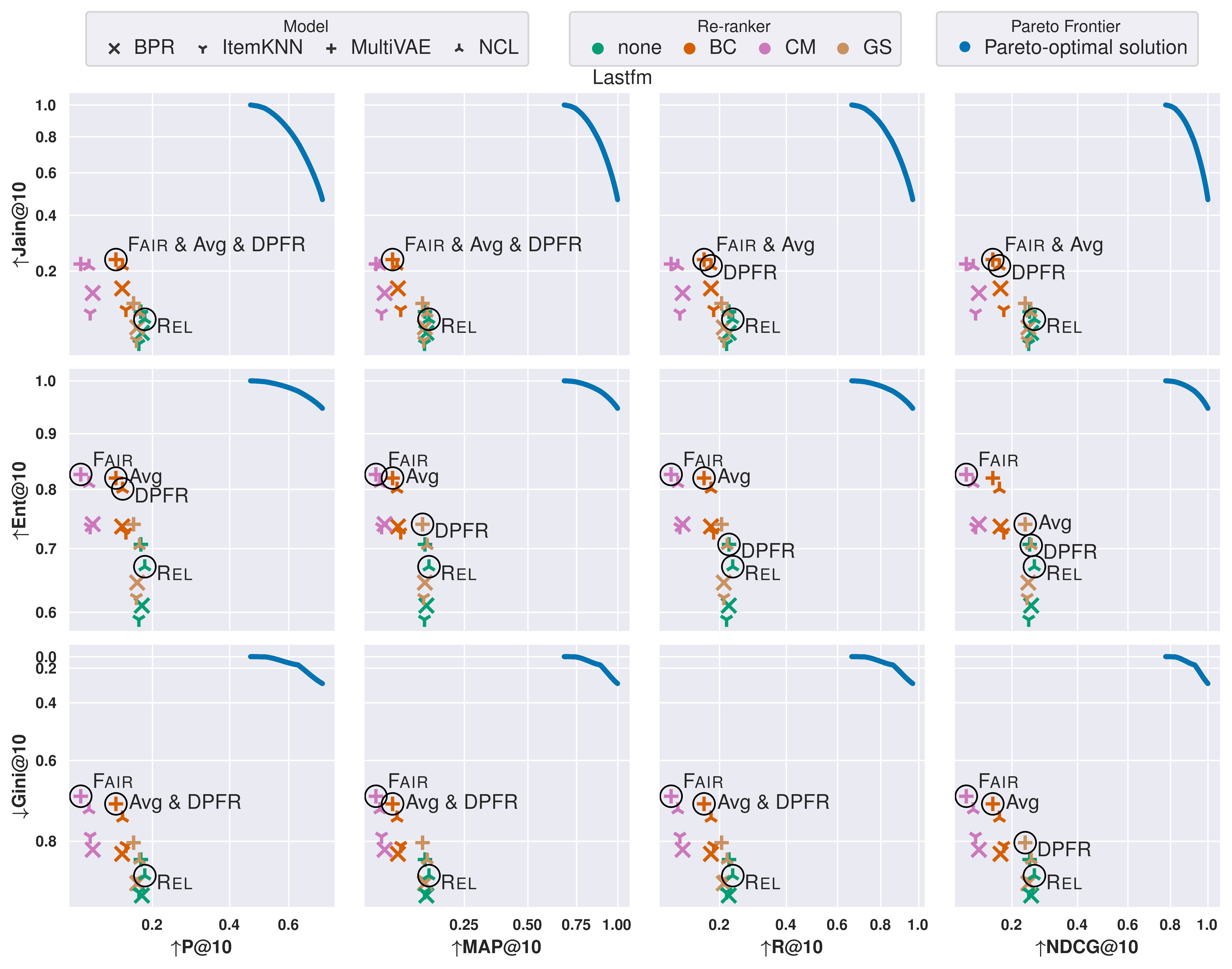}
    \includegraphics[width=0.75\columnwidth]{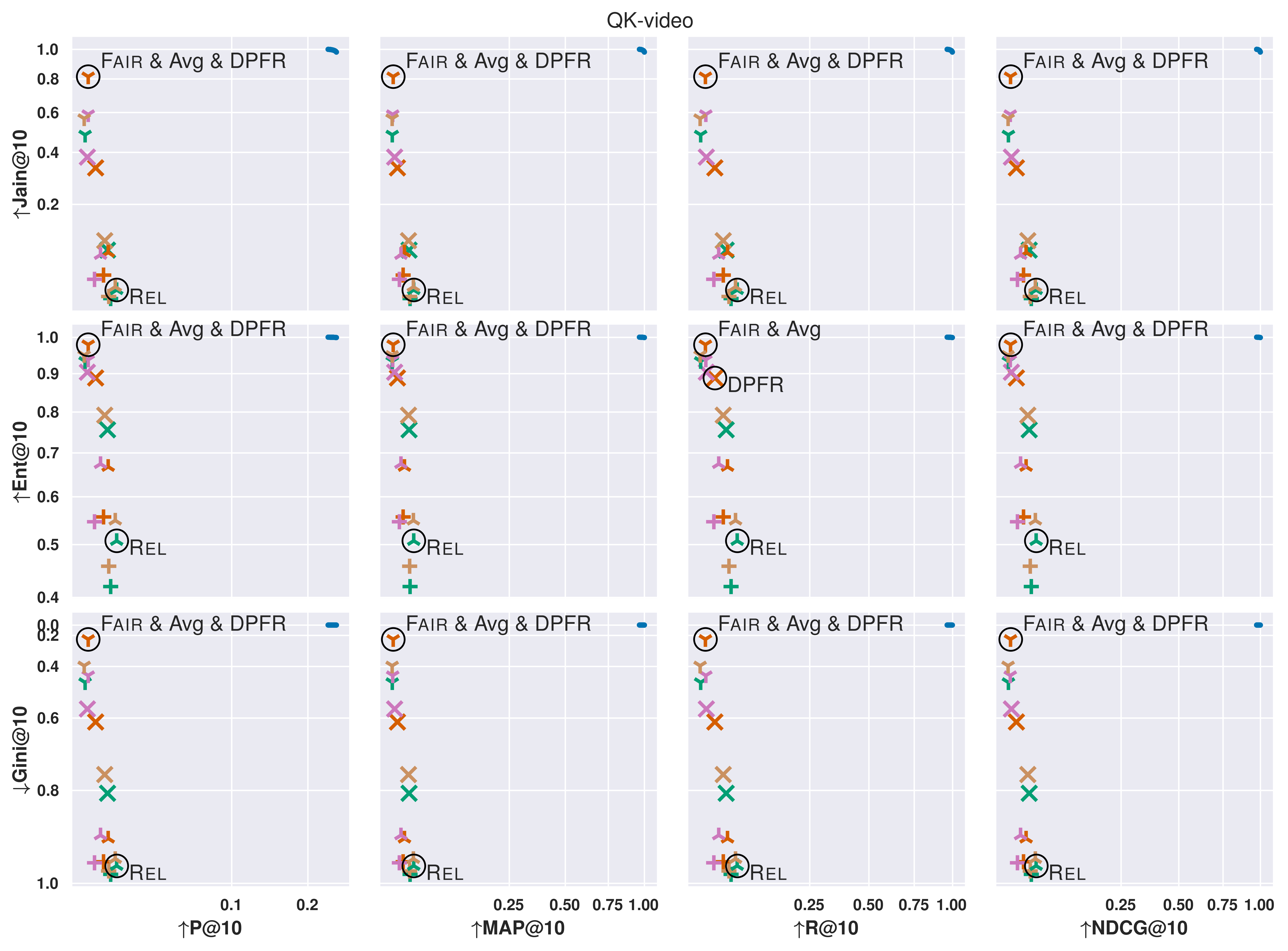}
    \caption{Pareto Frontier of fairness and relevance (in blue) and recommender scores for Lastfm and QK-video on exponential-like scales. 
    \textsc{Rel, Fair}, Avg (mean of \textsc{Rel, Fair}), and DPFR are the best model per evaluation approach.}
    \label{fig:pareto-pair-plot}
\end{figure}

\subsection{Measure Compatibility with DPFR}
\label{sss:no-fit}

 Which pairs of \textsc{Rel} and \textsc{Fair} measures are suitable to generate the PF? We answer this based on the PF slope. The slope is calculated using the two endpoints of the PF, i.e., the start and end of the  \textsc{Oracle2Fair} algorithm. A slope of zero means the \textsc{Rel} scores of the PF vary, but the \textsc{Fair} scores do not. As we compute the PF for multiple measures simultaneously, we expect a zero gradient for cases where the initial recommendation according to a \textsc{Fair} measure is already the fairest, even if other \textsc{Fair} scores are not. An undefined gradient value occurs when the initial recommendation is already the fairest and the most relevant according to a pair of \textsc{Fair} and \textsc{Rel} measures. Thus, we posit that a PF with a gradient value other than zero or undefined makes the corresponding pair of measures fit for PF generation (it allows for trade-offs in both aspects). 
The \textsc{Rel}-\textsc{Fair} measure pairs that are fit for DPFR based on their gradient are: \{P, MAP, R, NDCG\} $\times$ \{Jain, Ent, Gini\} (App.~\ref{app:actual-scores}, \Cref{tab:app-gradient}). Only results from these pairs are shown henceforth. Next, we explain what causes an undefined or zero gradient. 

\noindent \textbf{Causes of zero/undefined gradient.}
Generating the PF requires a ranking of items. Any score that is based on a single relevant item, e.g., HR and MRR, is by design not suitable. 
Out of the \textsc{Fair} measures, QF and FSat sometimes behave inconsistently depending on the dataset properties, as follows. 
A dataset with relatively few relevant items can already be made maximally fair at the start of the PF generation, as QF quantifies fairness with ignorance to frequency of exposure; the score does not change as long as the same set of items appears in the top $k$ recommendations of all users, no matter how many times each. 
When all items in the dataset already occur in the initial recommendations of our Oracle, nothing can be done to improve QF. 
For FSat, in few cases, the score is already maximum at the start of the PF generation. A maximum FSat score is achieved when all items in the dataset have at least the maximum possible exposure, if the available recommendation slots are shared equally across all items.\footnote{$\left\lfloor km/n\right\rfloor$ times (the total number of recommendation slots across users divided by the number of items).} 
In principle, QF and FSat can still be used for DPFR when the initial recommendation by Oracle is not the maximum yet. Otherwise, the interpretation would be less meaningful in joint evaluation, as there is no trade-off between different aspects.

\subsection{The Generated PF} \label{sss:result-pf}

\Cref{fig:pareto-pair-plot} shows the PF plots of the pairs of \textsc{Fair} and \textsc{Rel} measures that are suited for DPFR, only for Lastfm and QK-video, which  are representative of the overall trends in all our datasets (see \Cref{fig:app-pairplot} in the Appendix). The scores plotted are those computed in $\S$\ref{ss:generation}. The corresponding scores of our recommendation models are in App.~\ref{app:actual-scores}. 
We see that, as the recommendations are made fairer, the generated PF for all datasets is a series of monotonic scores of \textsc{Fair}, specifically monotonic increasing \textsc{Fair} scores (except $\downarrow$Gini), and the remaining measures are monotonic decreasing. 
The monotonic property  theoretically and empirically holds for the \textsc{Fair} measures, as we replace an item with the most exposure by another item with the least exposure, thereby making the recommendation fairer. 
Note that some users do not have exactly $k$ items in the test set, so the perfect relevance score cannot be reached for Precision@$k$ and Recall@$k$ \citep{Moffat2013SevenMetrics}. NDCG and MAP are implemented with normalisation\footnote{Only the first $\min{(|R^*_u|, k)}$ items in a user $u$'s recommendations are considered, where $R^*_u$ is the set of relevant items for user $u$.} so that they can still achieve a score of 1 in this situation. 

The datasets which were randomly split as they have no timestamps (QK-video, Jester) have relatively short, compact PF. This happens because the random split results in a uniform distribution of items in each split, which means that items in the test split are quite diverse (64--100\% of all unique items in the dataset). Considering that the \textsc{Oracle2Fair} algorithm starts by recommending items in the test split and stops when the recommendation reaches the fairest, there is not much room for change in \textsc{Fair} scores, as the initial recommendation is already rather fair. Additionally, there are not many relevant items per user in these datasets (i.e., the median for both datasets is 6 or less); random non-relevant items were chosen to make up for the remaining recommendations.\footnote{The randomly-split Lastfm does not have a short PF because on average it has more relevant items per user compared to QK-video and Jester (see App.~\ref{app:stat}).} 
Thus, the PF generation decreases relevance only marginally in 2/6 cases. 
Correspondingly, we find that in QK-video and Jester, there exist Pareto-optimal recommendations, that are close to maximally fair and maximally relevant, with the exception of P@10. These can be seen in the measure pairs of \{MAP, R, NDCG\} $\times$ \{Jain, Ent, Gini\}, where the PF is close to the coordinates of $(1,1)$, or $(1,0)$ for 
the measure pairs with Gini. Thus, in theory, a fair recommendation does not necessarily have to sacrifice relevance.

\subsection{Agreement between Measures} \label{pareto_ss:corr}

We study the agreement between DPFR and other evaluation approaches in ranking our 16 runs from best to worst. Low agreement means that the other approaches have few ties to the Pareto-optimal solutions that DPFR uses, and vice versa. 
We compare DPFR to (a) existing \textsc{Rel} and \textsc{Fair} measures, (b) existing joint \textsc{Fair+Rel} measures ($\S$\ref{s:experiments}), and also (c) the average (arithmetic mean) of \textsc{Fair} and \textsc{Rel} scores from the selected measure pairs that are used to generate the PFs. 
To compute the average for a measure where lower values are better (i.e., Gini), we compute $1-$the Gini score instead.

\subsubsection{Comparison of existing measures to DPFR}
We find that for all datasets and all measure pairs, \textbf{the best model as per  DPFR is always different from the best model as per \textsc{Rel} measures}. Moreover, \textbf{half the time, the best model as per DPFR is different from the best model as per a \textsc{Fair} measure}. 
Existing \textsc{Fair+Rel} measures tend to have the same best model as either \textsc{Fair} or \textsc{Rel} measures (73.3\% of the time), instead of having a more balanced evaluation of both aspects. These findings are expected as existing joint evaluation measures use relevance in their formulation differently than the \textsc{Rel} measures. 
Overall, the best model found with DPFR is less skewed towards relevance or fairness.

\subsubsection{Correlation of measures}
\begin{figure}[htbp]
    \centering   \includegraphics[width=0.405\columnwidth, trim = 2.5mm 2.5mm 2mm 0mm, clip,]{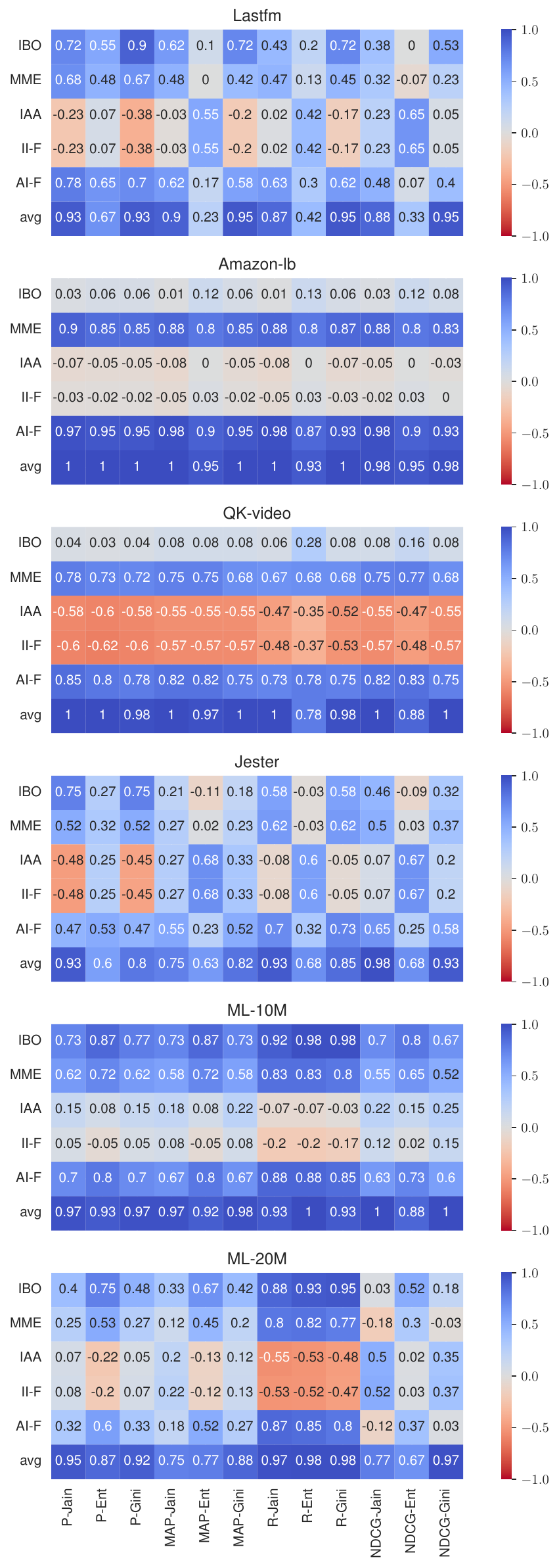}
    \caption{ 
    Kendall's $\tau$ correlation heatmap between the rank ordering of existing joint evaluation measures (including the average of \textsc{Fair} and \textsc{Rel} scores, avg), and DPFR.}
    \label{fig:corrplot}
\end{figure}

For each dataset, we compute the Kendall's\footnote{Ties are handled, unlike in Spearman's $\rho$.} $\tau$ \citep{Kendall1945TheProblems} correlations between the ranking given by DPFR and by the joint evaluation baselines (see \Cref{fig:corrplot}). 
Rankings are considered equivalent if $\tau \geq 0.9$ \citep{Maistro2021PrincipledRankings, Voorhees2001EvaluationDocuments}. 
We see similar agreement trends in datasets where recommenders have higher \textsc{Rel} scores (Lastfm and Jester) or lower (Amazon-lb and QK-video). 

Overall, most times DPFR orders models differently ($\tau<0.9$) than all \textsc{Fair+Rel} measures except AI-F. We see similar trends between IAA and II-F, and between MME and AI-F. IBO can have similar trends as MME (except for Amazon-lb and QK-video). For all datasets, IAA and II-F have overall either weak or negative $\tau$ with DPFR (e.g., $[-0.2, 0.25]$ for ML-10M and $[-0.62, -0.35]$ for QK-video). A notable exception is DPFR with \{MAP, R, NDCG\} $\times$ {Ent} for Lastfm and Jester, where we see moderate correlations, $\tau \in [0.42,0.68]$.

Ent differs from this trend because DPFR with \{MAP, R, NDCG\} $\times$ \{Jain, Gini\} has PF gradients of greater magnitude. This only affects Lastfm and Jester (they have higher \textsc{Rel} scores than the other datasets). DPFR with P has different patterns from other \textsc{Rel} measures: the raw DPFR scores of pairs involving P are lower on average, as the scores from Oracle do not start from $1$, but much less, and therefore closer to the models' scores (\Cref{fig:pareto-pair-plot}). 
Meanwhile, IBO has varying $\tau$ across datasets: a huge range of $\tau$, i.e., $[0.00,0.9]$ for Lastfm, weak correlations $[0.01, 0.13]$ for Amazon-lb, and moderate to strong correlations $[0.67, 0.98]$ for ML-10M. These variations might be because IBO is based on the number of items satisfying a certain criterion, rather than an average of scores across users and/or items, i.e., how other \textsc{Fair+Rel} measures are defined. 

Among the joint measures, AI-F correlates the strongest with DPFR, as both AI-F and DPFR, indirectly or directly, consider 
the recommendation frequency of each item and compare it with that of other items. 
However, the rank orderings given by AI-F are not equivalent to DPFR, as $\tau < 0.9$ for 5/6 datasets (excl. Amazon-lb). 
For the same measure-pair and between datasets, the $\tau$ of AI-F and DPFR also varies a lot. E.g., $\tau=0.07$ for NDCG-Ent for Lastfm, but $\tau=0.9$ 
for Amazon-lb. We thereby do not recommend using any of the \textsc{Fair+Rel} measures (none correlates with Pareto optimality). 

Taking the mean of \textsc{Fair} and \textsc{Rel} scores (avg) at a glance seems to correlate highly with DPFR. However, while it gives equivalent rankings ($\tau\geq 0.9$) in some cases (e.g., for Amazon-lb, most of ML-10M and QK-video, and half of ML-20M), it only does so for (1) datasets with lower \textsc{Rel} scores (Amazon-lb, QK-video), i.e., in cases where all models perform poorly, we have low variance in \textsc{Rel}, which leads to fairness dominating both avg and DPFR; (2) datasets with low variance in \textsc{Fair} scores (ML-*).  
In such cases, quantifying the evaluation jointly is challenging as one aspect dominates over the other. 
In the other datasets, the rank ordering given by the average is inconsistent: sometimes $\tau \geq 0.9$ for one dataset, but not for the others. This inconsistency between datasets holds for all measure pairs, except for P-Jain and NDCG-Gini. Due to these inconsistencies, we discourage using the arithmetic mean. 

Overall, our correlation analysis shows that existing joint \textsc{Fair+Rel} evaluation measures cannot be used as a reliable proxy for DPFR. 

\subsubsection{Best model disagreement} 

We take a closer look at how DPFR relates to computing averages, as they are similar approaches in terms of combining scores from a measure pair. As comparing the raw scores of DPFR and the average is invalid, we instead count the disagreement between the best model based on DPFR and the mean of \textsc{Fair} and \textsc{Rel} scores  (\Cref{tab:best_model_disagreement}). 
The aim is to study whether one would come to the same conclusion regarding the best model, using the two different joint evaluation approaches.

Among the 12 measure pairs that are fit for DPFR, we find that \textbf{the best model according to DPFR is not always the same according to the average of \textsc{Fair} and \textsc{Rel} scores of the same pair}; in one case the disagreement is up to 58.33\% (for Lastfm). The disagreement is generally much higher in the more complex rank-based measures (0--83.33\%) compared to simpler set-based \textsc{Rel} measures (0--50.00\%). 
Therefore, there are many cases where the mean of \textsc{Fair} and \textsc{Rel} scores is not the best case, especially for Lastfm and Jester where \textsc{Rel} scores are higher and vary more. 
In these two datasets, more often than not, DPFR leads to different conclusions than a simple average. Yet, sometimes the average agrees with DPFR in the best model: for QK-video, disagreement is low (8.33\%), and there is a perfect agreement on the best model for Amazon-lb; we posit that these are due to equally poor and low variance in the \textsc{Rel} scores. This is in line with our correlation analysis. 
As there is a huge range of variability across datasets (0--58.33\%), we do not recommend using a simple average to get the same result as DPFR, as it is unreliable and inconsistent. 
Generally, averaging fails to reach the same conclusion as DPFR almost half the time for the rank-based \textsc{Rel} measures,\footnote{Note that the published version of the work does not contain the phrase ``for the rank-based \textsc{Rel} measures''. This phrase has been added for clarity.} 
especially when the \textsc{Rel} and \textsc{Fair} scores vary highly.

\begin{table}[tbp]
\centering
\caption{The percentage of best model disagreement when taking the mean of \textsc{Fair} and \textsc{Rel} scores as opposed to using DPFR, separated by the \textsc{Rel} measure type.  P@$k$ and R@$k$ are set-based, NDCG and MAP are rank-based. We only consider the 12 measure pairs with a nonzero, defined gradient ($\S$\ref{sss:no-fit}).}
\label{tab:best_model_disagreement}
\begin{tabular}{lrrr}
\toprule
{} &                    Set-based & Rank-based &    All \\
\hline
\midrule
Lastfm    &          50.00 &      66.67 &  58.33 \\
Amazon-lb &          0.00 &       0.00 &   0.00 \\
QK-video  &          16.67 &       0.00 &   8.33 \\
Jester    &          16.67 &      83.33 &  50.00 \\
ML-10M    &           0.00 &      66.67 &  33.33 \\
ML-20M       &       0.00 &       50.00 &  25.00 \\
\hline
\midrule
All datasets &      13.89 &       44.44 &  29.17 \\
\bottomrule
\end{tabular}
\end{table}

\subsection{Efficient DPFR} 
\label{ss:compare-eff}

\subsubsection{The efficiency of the PF generation} 
We study the efficiency of DPFR by comparing the PF, an estimated version of PF on a subset of points, and the \textsc{Fair+Rel} measures. The estimated version of PF uses 3--12 points as per $\S$\ref{ss:compute-eff}. 
We compute the amount of points in the estimated PF as \% of those in the PF, and the resulting computation times. One point in the PF translates to one round of computing all \textsc{Fair} and \textsc{Rel} measures, so fewer points mean faster. For brevity, we report the estimated PF with only 3, 6, and 12 points in \Cref{tab:big-est-time} 
(see App.~\ref{app:actual-scores}, \Cref{tab:corr_est} for extended results).

\begin{table}[tbp]
    \centering
    \caption{Efficiency and effectiveness of PF, estimated PF (Est.~PF), and \textsc{Fair+Rel} measures: \% of data points in the Est.~PF (\% pts), computation time. Dist is the average distance between midpoints in the Est.~PF and PF over 12 measure pairs. Minimum agreement (Min $\tau$) is the Kendall's $\tau$ correlation between DPFR with PF and Est.~PF. 
    Both PFs compute 11 \textsc{Rel} and \textsc{Fair} measures simultaneously. 
    The times for other evaluation measures are averaged (Avg/model) and summed (All models) over 16 model combinations.}
    \label{tab:big-est-time}
\resizebox{\linewidth}{!}{
     \begin{tabular}{lllrrrrrr}
\toprule
& & {} &  Lastfm &  Amazon-lb &  QK-video &  Jester &  ML-10M &  ML-20M \\
\midrule
\multicolumn{2}{c}{\#pts}& PF& 4882 & 847 & 499 & 16202 & 2781 & 3783 \\
\cline{1-9}
\multicolumn{2}{c}{\multirow[c]{3}{*}{\%pts}} & Est.~PF (12 pts) & 0.25 &      1.42 &     2.40 &   0.07 &   0.43 &   0.32 \\
& & Est.~PF (6 pts) &  0.12 &      0.71 &     1.20 &   0.04 &   0.22 &   0.16 \\
& & Est.~PF (3 pts) & 0.06 &      0.35 &     0.60 &   0.02 &   0.11 &   0.08\\
\hline
\midrule
\multirow[c]{14}{*}{\rotatebox[origin=c]{90}{$\downarrow$ Computation time (mins.)}}& & PF           &   19.18 &       0.56 &     10.49 &  847.42 &   28.99 &   75.77 \\
& & Est.~PF (12 pts) &    2.02 &       0.19 &      4.23 &  552.16 &    1.90 &    2.60 \\
& & Est.~PF (6 pts)  &    2.00 &       0.19 &      4.12 &  551.72 &    1.84 &    2.54 \\
& & Est.~PF (3 pts)  &    2.01 &      0.19 &     4.07 &  552.26 &   1.82 &   2.52\\
\cline{2-9}
& \multirow[c]{5}{*}{\rotatebox[origin=c]{90}{Avg/model}} & IBO    &    <0.3s &     <0.3s &      0.01 &    0.01 &   <0.3s &     0.01  \\
& & MME    &    2.04 &       0.03 &     19.51 &    0.09 &   15.25 &    89.13 \\
& & IAA    &    <0.3s &     <0.3s &      0.01 &    0.02 &    0.01 &     0.02 \\
& & II-F   &    <0.3s &     <0.3s &      0.01 &    0.01 &    0.01 &     0.02 \\
& & AI-F   &    <0.3s &     <0.3s &      0.01 &    0.01 &   <0.3s &     0.01 \\
\cline{2-9}
& \multirow[c]{5}{*}{\rotatebox[origin=c]{90}{All models}} & IBO  &    0.02 &     <0.3s &      0.10 &    0.12 &    0.06 &     0.15 \\
& & MME  &   32.63 &       0.49 &    312.14 &    1.38 &  244.03 &  1426.10 \\
& & IAA  &    0.03 &     <0.3s &      0.10 &    0.36 &    0.13 &     0.30  \\
& & II-F &    0.04 &     <0.3s &      0.16 &    0.14 &     0.1 &     0.25 \\
& & AI-F &   0.03 &     <0.3s &      0.11 &    0.13 &    0.07 &     0.17 \\
\hline
\midrule
\multicolumn{2}{c}{\multirow[c]{3}{*}{$\uparrow$ 
Min $\tau$}} & Est.~PF (12 pts)& 0.95 &       1.00 &      1.00 &    0.98 &    0.98 &    0.97\\
\multicolumn{2}{c}{} &  Est.~PF (6 pts)&  0.90 &       0.97 &      1.00 &    0.98 &    0.95 &    0.92\\
\multicolumn{2}{c}{} &  Est.~PF (3 pts)&  0.78 &       0.98 &      1.00 &    1.00 &    0.97 &    0.75\\
\midrule
\multicolumn{2}{c}{\multirow[c]{3}{*}{$\downarrow$ Dist.}} & Est.~PF (12 pts)& 0.01 &      0.02 &     0.00 &   0.00 &   0.01 &   0.01 \\
\multicolumn{2}{c}{} &  Est.~PF (6 pts)& 0.03 &      0.05 &     0.00 &   0.00 &   0.02 &   0.02 \\
\multicolumn{2}{c}{} &  Est.~PF (3 pts)& 0.03 &      0.05 &     0.00 &   0.00 &   0.03 &   0.05 \\
\bottomrule
\end{tabular}
}
\end{table}

The PF (\Cref{fig:pareto-pair-plot}) has hundreds to tens of thousands of points each, while the estimated PF only contains 0.02--2.40\% of the points, which means reduced computational complexity for the PF generation. In terms of actual computation time (\Cref{tab:big-est-time}), computing the PF with \textsc{Oracle2Fair} take 0.56--75.77 mins to compute, but only 0.19--4.07 mins for the PFs estimated with 3 points for all datasets except Jester. For Jester, it takes $\sim$14 hours and the estimation only takes $\sim$9 hours. 
However, this is expected for Jester as it has 62K users in the test split, as opposed to the 3.5K or fewer in the other datasets (i.e., see \Cref{tab:data_split} in App.~\ref{app:stat}). 
While computing the estimated PFs is on average slower than computing the joint measures IBO, IAA, II-F, and AI-F, it is expected as the (estimated) PFs compute 11 measures simultaneously. Yet, in most cases (except Amazon-lb and Jester), the estimated PFs are still faster to compute than the time to compute MME for one model per dataset, let alone to compute MME for all models. For ML-20M, computing the estimated PF is even up to 35 times faster than computing MME of one model.

\subsubsection{The effectiveness of efficient DPFR} 
To what extent is the DPFR from the efficiently generated PF (estimated PF) a reasonable proxy for fairness-relevance joint evaluation using the PF, in terms of giving a similar ordering of models? 
We compare the DPFR from the PF and estimated PF using Kendall's $\tau$ correlations. Further, as DPFR is computed based on a $\alpha$-based reference point lying on the PF, to quantify possible accuracy loss of the estimated PF, in \Cref{tab:big-est-time} we show the error of the midpoint estimation. This error is the Euclidean distance between the reference point in the PF and estimated PF (i.e., the midpoint in our case), as per \citep{Wang2016AEngineering}.

We first analyse the error of the midpoint estimation. For the 12 measure pairs and 6 datasets, the midpoint coordinates on average do not move much: the distance is 0.00--0.05, even when the PF is only estimated with 3 points. 
Ergo, the correlations between the rank ordering of models given by the DPFR of PF and its estimation, are still equivalent ($\tau \geq 0.9$) when estimated with 6 or 12 points \citep{Maistro2021PrincipledRankings,Voorhees2001EvaluationDocuments}. Even the 3-point estimation maintains high agreement ($\tau \in [0.75,1]$), with only 5 cases having $\tau<0.9$ across 6 datasets and 12 measure-pairs. 
Therefore, it is possible to only compute a small number of points in the PF, e.g., 6 or 12 points, and still make a reliable PF estimation, evidenced by the small shift of the PF midpoint and the rank ordering of the models remaining equivalent ($\tau \geq 0.9$), if not identical ($\tau=1$) for all measure pairs and datasets. 

\section{Discussion and Conclusions}
Recommendation evaluation has long used measures that quantify only relevance, but has recently shifted to include fairness. However, there exists no de-facto way to robustly quantify these two aspects. We propose a novel approach (DPFR) that uses fairness and relevance measures under a joint evaluation scheme for RSs. DPFR computes the empirical best possible recommendation, jointly accounting for a given pair of relevance and fairness measures, in a principled way according to Pareto-optimality. DPFR is modular, tractable, and intuitively understandable. It can be used with existing measures for relevance and fairness, and allows for different trade-offs of relevance and fairness. We empirically show that existing evaluation measures of fairness w.r.t.~relevance \citep{Biega2018EquityRankings,Diaz2020EvaluatingExposure,Wu2022JointRecommendation, Saito2022FairRanking, Borges2019EnhancingAutoencoders} behave inconsistently: they disagree with optimal solutions based on DPFR computed on more robust and well-understood measures of relevance, such as NDCG, and fairness, such as Gini. 
We uncover some weaknesses of these measures, but more research is warranted to study their behaviour. Admittedly, existing joint measures are not originally defined to be aligned with existing relevance and fairness measures \citep{jain1984quantitative,Zhu2020FARM:APPs,Gini1912VariabilitaMutabilita,Mansoury2021ASystems,Mansoury2020FairMatch:Systems,Patro2020FairRec:Platforms}, so it is not surprising that they have different results from DPFR. However, existing measures show varying performance also from each other and from well-understood relevance and fairness measures. Thus, DPFR can be a viable alternative for robust, interpretable, and provenly optimal evaluation in offline scenarios. We also show that DPFR can be computed fast while reaching equivalent conclusions. 
Overall, DPFR demonstrates distinct benefits in mitigating false conclusions by up to 58\% compared to basic aggregation methods like averaging.\footnote{The published version concluded ``by up to 50\%'', but the performance of DPFR is better than previously reported.} 
Surprisingly, simple averaging aligns more with our Pareto-optimal based DPFR, than existing joint measures. 
We recommend combining either MAP-Ent or NDCG-Ent: the conclusions are distinguishable from simply averaging, or taking the best model 
based on fairness or relevance measures.

\section*{Acknowledgements}
The work is supported by the Algorithms, Data, and Democracy project (ADD-project), funded by Villum Foundation and Velux Foundation. Qiuchi Li contributed to the idea of computing the reference point. We thank the DIKU IR Lab and the anonymous reviewers, who have provided helpful feedback to improve earlier versions of the manuscript.

\section{Appendix}

\subsection{Extended Dataset Statistics}\label{app:stat}
\Cref{tab:data_split} presents the statistics of each dataset split. For several datasets (e.g., Amazon-lb and ML-*), the number of users in the test split is significantly less than the number of users in the train split. \Cref{tab:rel_item_stat} presents the statistics of items in the test split, per user.

\begin{table}[htbp]
    \centering
        \caption{Number of [users, items, and interactions] in the train, validation, and test split after preprocessing. 
    } 
    \label{tab:data_split}
\resizebox{\columnwidth}{!}{
\begin{tabular}{llll}
\toprule
{} &                   train &                     val &                    test \\
\midrule
Lastfm    &     [1842, 2821, 42758] &     [1831, 2448, 14248] &     [1836, 2476, 14237] \\
Amazon-lb &       [1054, 552, 8860] &        [470, 204, 1811] &        [437, 209, 1726] \\
QK-video  &     [4656, 6245, 34345] &      [3470, 4095, 8726] &      [3514, 4101, 8706] \\
Jester    &   [63724, 100, 1294511] &    [62137, 100, 427623] &    [62167, 100, 427926] \\
ML-10M    &  [49378, 6838, 4944064] &    [2695, 7828, 296914] &    [1523, 7880, 121707] \\
ML-20M    &  [89917, 8719, 9882504] &    [4987, 10742, 472243] &    [2178, 13935, 233394] \\
\bottomrule
\end{tabular}
}
\end{table}

\begin{table}[htbp]
    \caption{Statistics of items in the test split, per user, i.e., the number of relevant items per user.}
    \label{tab:rel_item_stat}
    \centering
\begin{tabular}{lrrrr}
\toprule
{} &    mean &  min &  median &   max \\
\midrule
Lastfm    &    7.75 &    1 &       8 &    19 \\
Amazon-lb &    3.95 &    1 &       3 &    16 \\
QK-video  &    2.48 &    1 &       2 &    16 \\
Jester    &    6.88 &    1 &       6 &    29 \\
ML-10M    &   79.91 &    1 &      46 &  1632 \\
ML-20M    &  107.16 &    1 &      53 &  2266 \\
\bottomrule
\end{tabular}
\end{table}

\subsection{Algorithms for Generating Pareto Frontier}\label{app:algo}
We present the pseudocodes of the algorithms for generating the Pareto Frontier: the Oracle (Algorithm \ref{alg:oracle}) and  \textsc{Oracle2Fair} (Algorithm \ref{alg:oracle2fair}). We provide the worst-case time complexity analysis for both algorithms and discuss a possible edge case. 
We denote as $k$ the recommendation cut-off, as $m$ the number of users, as $n$ the number of items, as $H$ the maximum number of items in a user's interaction history ($H_u$) across all users, and as $R$ the maximum number of items in a user's test split ($R_u^*$) across all users. We use the binary logarithm ($\log_2{}$). For brevity, we omit lines with $O(1)$, and we state the reasoning for each line (L).

\SetKwComment{Comment}{/* }{ */}
\RestyleAlgo{ruled}
\LinesNumbered

\begin{algorithm*}[htbp]
{\scriptsize
\caption{Oracle \\ Create recommendations with the highest relevance}\label{alg:oracle}

\KwData{ \\
$I$: all items in the dataset; \\
$H_u$: items in train-val split for each user $u \in U$;  \\
$R_u^*$: items in test split (relevant items) for each user $u \in U$; \\
$k$: number of recommended items
}
\KwResult{\\
$rec$: most relevant recommendation\\\\
$result$: a list of relevance and fairness scores\\
$itemNotInRec$: items that are not in the recommendation}

\Comment{Handle users with exactly $|R_u^*|=k$}
\lForEach{$u \in U$ where $|R_u^*| = k$}{{$rec[u] \gets R_u^*$}}

\Comment{Handle users with $|R_u^*| > k$}

\For{$K=k+1$ \KwTo $max(|R_u^*|)$}{\label{ln:startgtk}
    $userWithK \gets$ get users where $|R_u^*|=K$ \\
    \ForEach{$u \in userWithK$}{
        $takenItem[u] \gets {R_u^* \cap rec}$ \\
        $weight[u] \gets sum(countInRec(takenItem[u]))$ \\
        }
    $sortUserWithK \gets$ sort $userWithK$ by the least weight \\
    $tempRec[u] \gets R_u^* \setminus takenItem[u]$ \\
    keep only max $k$ items in $tempRec[u]$ \\

    \ForEach{$u \in sortUserWithK$}{
        $rec[u] \gets tempRec[u]$ \\
        $numItemToAdd \gets k - |tempRec[u]|$ \\
        sort $takenItem[u]$ by the least item count \\
        $rec[u].append(takenItem[u][:numItemToAdd])$  
    }
}  \label{ln:endgtk}

\Comment{Handle users with $|R_u^*| < k$}
$remainUser \gets$ get users where $|R_u^*|<k$ \\ \label{ln:startltk}
\lForEach{$u \in remainUser$}{$rec[u] \gets |R_u^*|$} 
$itemNotInRec \gets I \setminus rec$ \\
\ForEach{$u \in remainUser$}{
    \While{$|rec[u]|<k$ and $itemNotInRec \neq \emptyset$}{
        \For{$item \in itemNotInRec$}{
         \If{$item \notin H_u$}{
         $rec[u].append(item)$ \\
         $itemNotInRec \gets itemNotInRec\setminus \{item\}$
         }
        }    
    }

    \If{$|rec[u]|<k$}{
        \While{$|rec[u]|<k$}{

        $candItem \gets$ least popular item in $rec$ that is not in $H_u \cup R_u^* \cup rec[u]$ \\
        $rec[u].append(candItem)$
    }
    }
} \label{ln:endltk}
$result \gets calculateScores(rec)$ 
}
\end{algorithm*}

\begin{algorithm*}[htbp]
{\footnotesize
\caption{\textsc{Oracle2Fair} \\After recommending maximally relevant items, iteratively change the recommendation list to increase fairness until maximum fairness is reached}\label{alg:oracle2fair}
\KwData{$H_u, R_u^*,I, k$}
\KwResult{\\
$rec$: most fair possible recommendation;\\
$result$: a list of relevance and fairness scores}

$rec, result, itemNotInRec \gets Oracle(I, H_u, R_u^*)$

\Comment{Get the most popular item in the recommendations \& its frequency count}
 $newMostPop \gets mostPop \gets getMostPopItem(rec)$  \\ \label{ln:updatecountbegin}
 $newCntPop \gets cntPop \gets cnt(mostPop, rec)$ \\

$uWithMostPop \gets$ all users with $mostPop \in rec[u]$ \\
sort $uWithMostPop$ by largest index of $mostPop$ in $rec[u]$ \\ \label{ln:updatecountend}
\For{$i \in itemNotInRec$}{
\lIf{$cntPop=1$}{break}    
\If{$newMostPop \neq mostPop$}{
    $mostPop \gets newMostPop$ \\ \label{ln:mostpopstart}
    update $uWithMostPop$ following $mostPop$ \\ \label{ln:mostpopend}
}
\If{$newMostPop = mostPop$}{
    $candU \gets$ all $u$ in $uWithMostPop$ where $i \notin H_u$  \\
    \If{$\exists u \in candU$ with $i \in R_u^*$ \label{ln:recommendstart}}{
        recommend $i$ to the top $u$ from $candU$ with $i \in R_u^*$
    }
    \lElse{recommend $i$ to the top $u$ from $candU$}
    reorder $rec[u]$ so all relevant items are at the top \\
    $result.append(calculateScores(rec))$ \\ \label{ln:recommendend}
    $itemNotInRec \gets itemNotInRec \setminus \{i\}$ \\
    $newMostPop \gets getMostPopItem(rec)$ \\
    $newCntPop \gets cnt(mostPop, rec)$ 
    }
  }
\Else{
    do lines \ref{ln:updatecountbegin}--\ref{ln:updatecountend} \\ \label{ln:dolines}
    $i \gets leastPop \gets getLeastPopItem(rec)$ \\ \label{ln:getleastpop}
    $m, n \gets |U|, |I|$ \\
    \While{$cntPop > \lceil km/n \rceil$}{
    \lIf{$newMostPop \neq mostPop$}{do lines \ref{ln:mostpopstart}--\ref{ln:mostpopend}}
    \If{$newMostPop = mostPop$}{
    $candU \gets$ all $u$ in $uWithMostPop$ where $i \notin H_u \cup rec[u]$  \\
    do lines \ref{ln:recommendstart}--\ref{ln:recommendend} \\
    do lines \ref{ln:dolines}--\ref{ln:getleastpop}
    }
    }
}
}
\end{algorithm*}

\subsubsection{Time complexity of the Oracle (\Cref{alg:oracle})}
\label{app:analyse-oracle}

The line-by-line time complexity analysis of Oracle (\Cref{alg:oracle}) is as follows:

\begin{itemize}
    \item L1: $O(km)$, creating list of size $k$ for $m$ users
    \item L2--17: $O(R^2m + Rm \log{m} + Rmk\log{k})$, resulting from at most $R$ iterations of:
    \begin{itemize}
        \item L3: $O(m)$, linear search on list of size $m$
        \item L4--7: $O(mR)$, at most $m$ times looking up at most $R$ values
        \item L8: $O(m \log{m})$, sorting a list of size $m$
        \item L11--16: $O(mk \log{k})$, at most $m$ times sorting list of size $k$
    \end{itemize}
    
    \item  L18: $O(m)$, linear search on list of size $m$
    \item  L19: $O(m)$, $m$ assignment operations
    \item  L21--36: $O(k^2m^2)$, resulting from at most $m$ iterations of: 
    \begin{itemize}
        \item L22--29: $O(kH)$, at most $k$ times of linear search on list of size $H$
        \item L30--L35: $O(k^2m)$, $k$ times of counting in a list of size $km$.
    
        In most cases, the number of users $m >> H$, so the time complexity of this block is dominated by $O(k^2m)$.
    \end{itemize}
    \item L37: $O(km)$ computing relevance/fairness measures for $m$ users based on recommendation list of size $k$

\end{itemize}

Overall, the time complexity of Oracle is dominated by that of L2 and L21--36. Hence, the time complexity of Oracle is $O(R^2m + Rm \log{m} + Rmk \log{k} + k^2m^2)$.

\subsubsection{Time complexity of \textsc{Oracle2Fair} (\Cref{alg:oracle2fair})}

The line-by-line time complexity analysis of \textsc{Oracle2Fair} (\Cref{alg:oracle2fair}) is as follows:

\begin{itemize}
    \item L1: $O(R^2m + Rm \log{m} + Rmk \log{k} + k^2m^2)$ (derived in \Cref{app:analyse-oracle})
    \item L2--4: $O(km)$, $m$ times of counting and linear search on list of size $k$
    \item L5: $O(m \log{m})$, sorting a list of size $m$ (e.g., with Tim Sort)
    \item L6--24: $O(max(Hmn, kmn))$, resulting from at most $n$ iterations of: 
    \begin{itemize}
        \item L10: $O(km)$, $m$ times of linear search on list of size $k$
        \item L13: $O(Hm)$, $m$ times linear search on list of size at most $H$
        \item L14: $O(km)$
        \item L18: $O(k \log{k})$, sorting a list of size $k$
        \item L21--L22: $O(km)$; the term $O (k \log{k})$ is dominated by $O(km)$, as typically the cut-off $k$ is much smaller than the number of users $m$ (and hence $\log{k} < m$).
    \end{itemize}

     \item L26: $O(km + m \log{m})$, combining L2--4 and L5
     \item L27: $O(km)$, $m$ times counting on list of size $k$
     \item L29--34: $O(max(Hkm^2,k^2m^2)+km^2\log{m})$, resulting from at most $km$ iterations (derived below) of:
     \begin{itemize}
         \item L30: $O(km)$, from L10
         \item L32: $O(Hm)$, $m$ times linear search on list of size at most $H$
         \item L33: $O(km)$, from L14 and L18
         \item L34: $O(km+m \log{m})$, from L26--27
     \end{itemize}
\end{itemize}

We estimate the worst-case complexity of the number of iterations of the code block in L29--34, by assuming that the initial recommendation is the unfairest. The unfairest recommendation happens when the same $k$ unique items are recommended to all $m$ users. 
This code block aims to reduce the max frequency count of all items to $\left\lceil km/n \right\rceil$. Hence, the number of iterations is $\sum^{k} (m-\left\lceil km/n \right\rceil) \leq k(m-\frac{km}{n}-1) = km-\frac{k^2m}{n}-k \leq km$.

All in all, the time complexity of \textsc{Oracle2Fair} is dominated by that of L1, L6--24, and L29--34. Hence, the overall time complexity of \textsc{Oracle2Fair} is $O(R^2m + Rm \log{m} + Rmk \log{k} +  k^2m^2 + km^2\log{m} + kmn )$ if $k\geq H$, or $O(R^2m + Rm \log{m} + Rmk \log{k} +  Hkm^2 + km^2\log{m} + Hmn )$ otherwise. To simplify the time complexity, further assumptions need to be made for one or more variables.

\subsubsection{Possible edge cases}
There might be edge cases, for example, datasets where the train/val set of each user contains almost all items in the dataset (each user has rated/clicked most items in the dataset). In this case, if we do not want to re-recommend items in the train/val set to users, some items may have item counts more than
$\left\lceil km/n \right\rceil$ at the end of the process, and \Cref{alg:oracle2fair} might not halt. However, such datasets are rare in recommender systems.

\subsection{Modifications to the GS Algorithm}\label{app:gs}
The original GS algorithm \citep{Wang2022ProvidingSystems} increases individual item fairness within clusters of similar items. The item similarity is determined based on the item embedding. As our experiments and the \textsc{Fair} measures do not deal with the additional constraint of item similarity, we consider all items as similar. Therefore, we only have a single cluster of items.

On top of that, we also modify GS to increase computational efficiency. In the original GS algorithm, for each pair of candidate items for replacement $i$ and candidate items to be replaced $i'$, the algorithm finds all users that have $i$ in the original recommendation list. The algorithm then computes the loss in relevance (computed using predicted relevance value) if item $i$ is replaced by $i'$. Until this point, our modified algorithm does the same. The difference is that we save each $i, i', u$, and the loss associated, while the original algorithm only saves the information for the one user $u^*$, whose recommendation list will suffer the least loss when we replace $i$ with $i'$. The original GS then proceeds to make the replacement, update the pool of candidate items for replacement and to be replaced, and go through the entire process again. Initially, we found that with the GS algorithm, around 20\% of the initial recommendations are replaced during the process, meaning that for Amazon-lb, there are at least $437 \times 10 \times 0.2 \geq 800$ iterations of the process (\Cref{tab:data_split}). The number of iterations is much bigger for ML-10M, which has more than three times the  recommendation slots as Amazon-lb, and hence it is very costly to use the GS algorithm as is. 

Our modified GS utilises the saved information earlier. After going through all pairs of $(i,i')$, we sort the saved list from the smallest to the largest loss, and (attempt to) perform the replacement using the first $P$ pairs, where $P$ is 25\% of the number of recommendation slots. During the replacement process, if the item that is supposed to be replaced no longer exists in the user's recommendation list, we simply skip the replacement.

\subsection{Extended Results}\label{app:actual-scores}
\begin{figure*}[p]
    \centering
    \includegraphics[width=0.48\columnwidth]{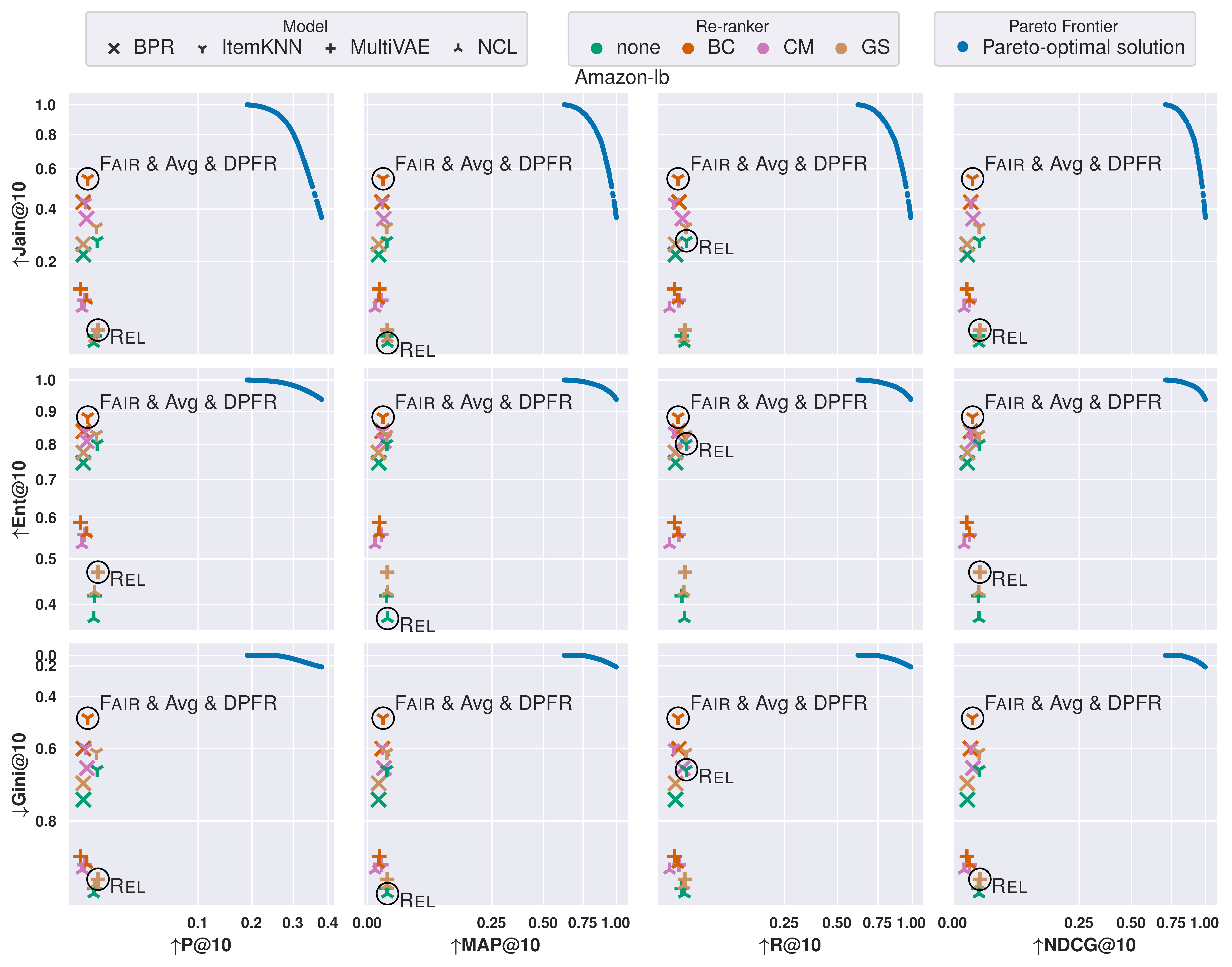}
    \includegraphics[width=0.48\columnwidth]{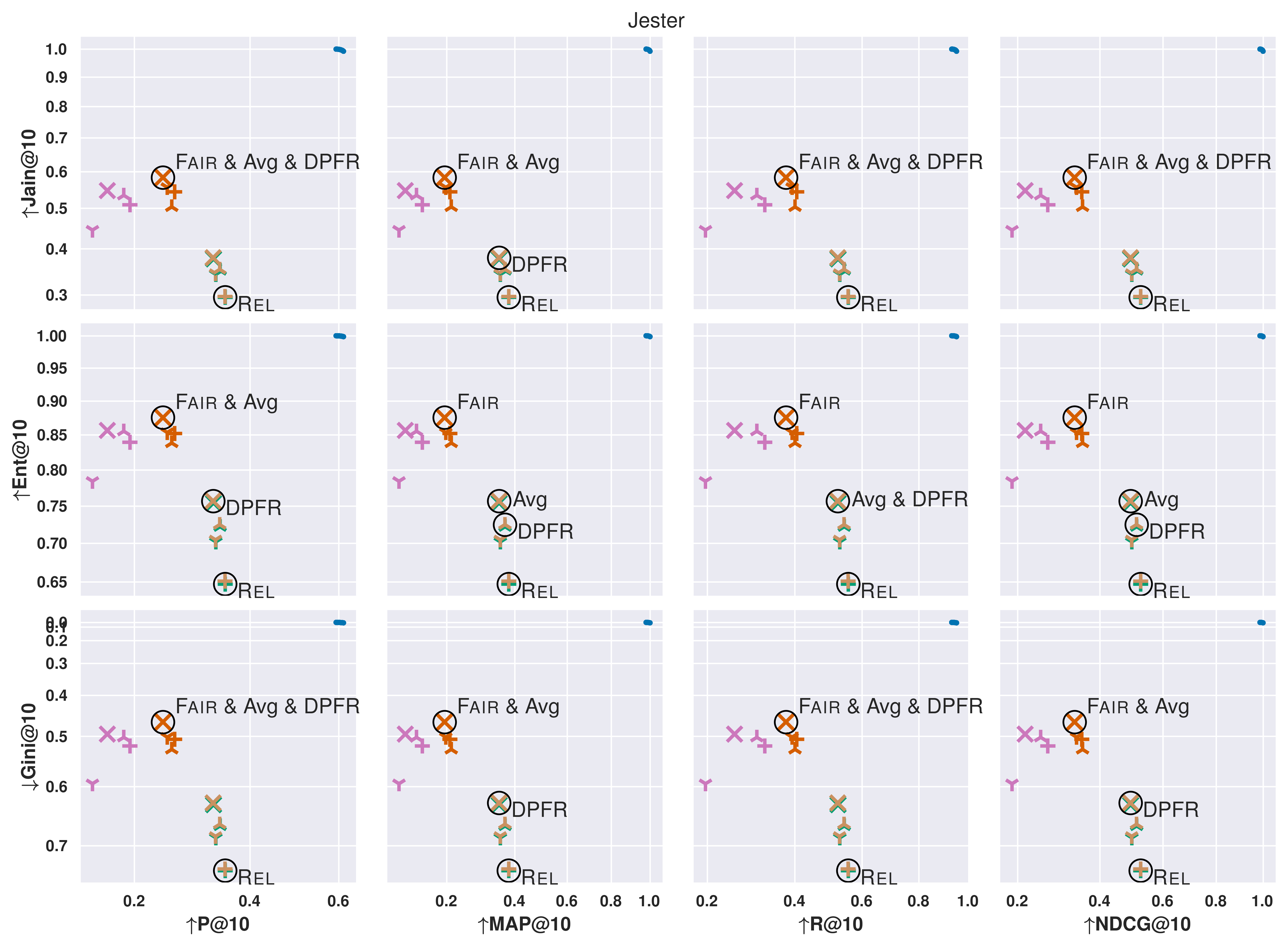}
    \includegraphics[width=0.48\columnwidth]{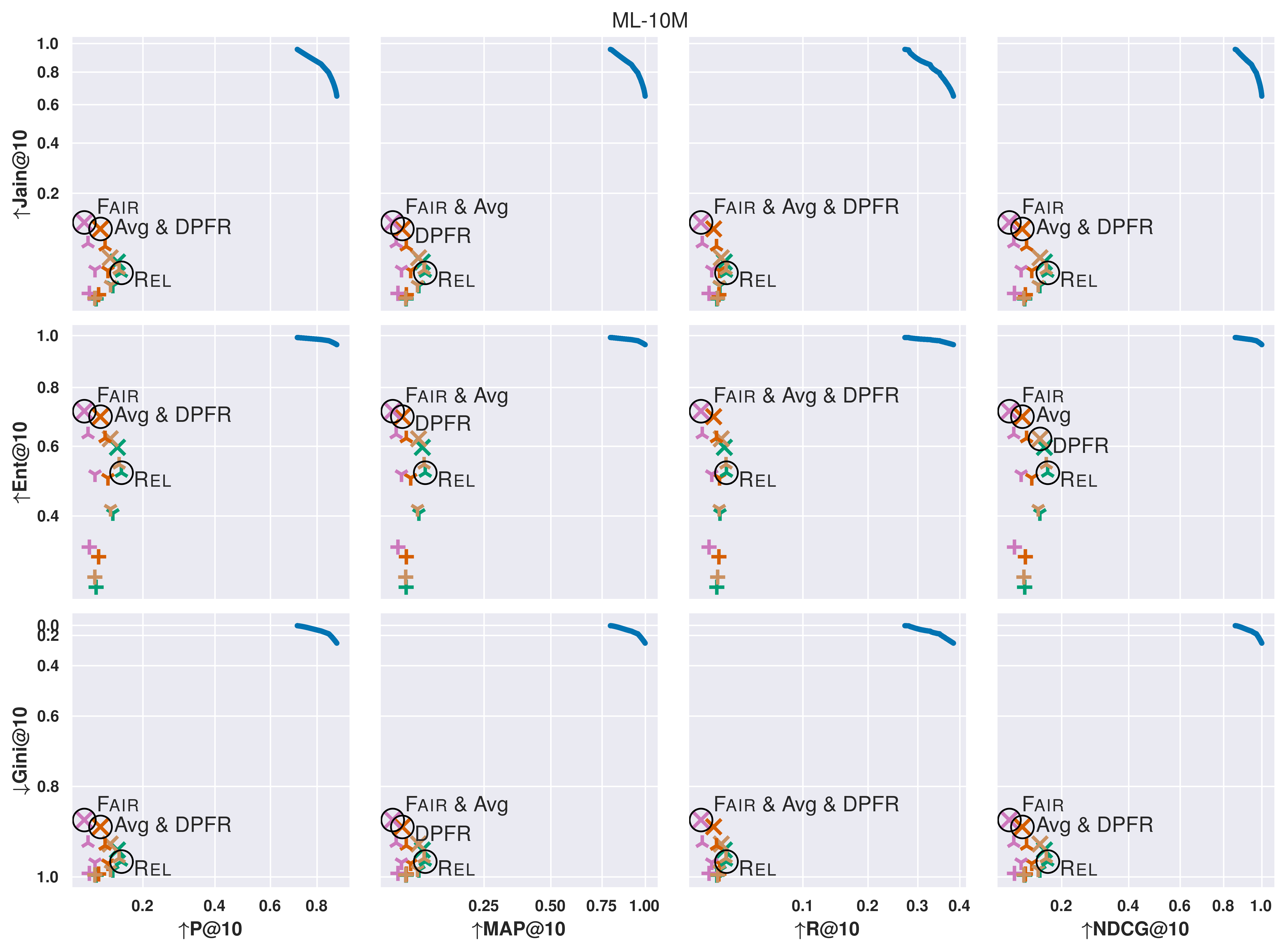}
    \includegraphics[width=0.48\columnwidth]{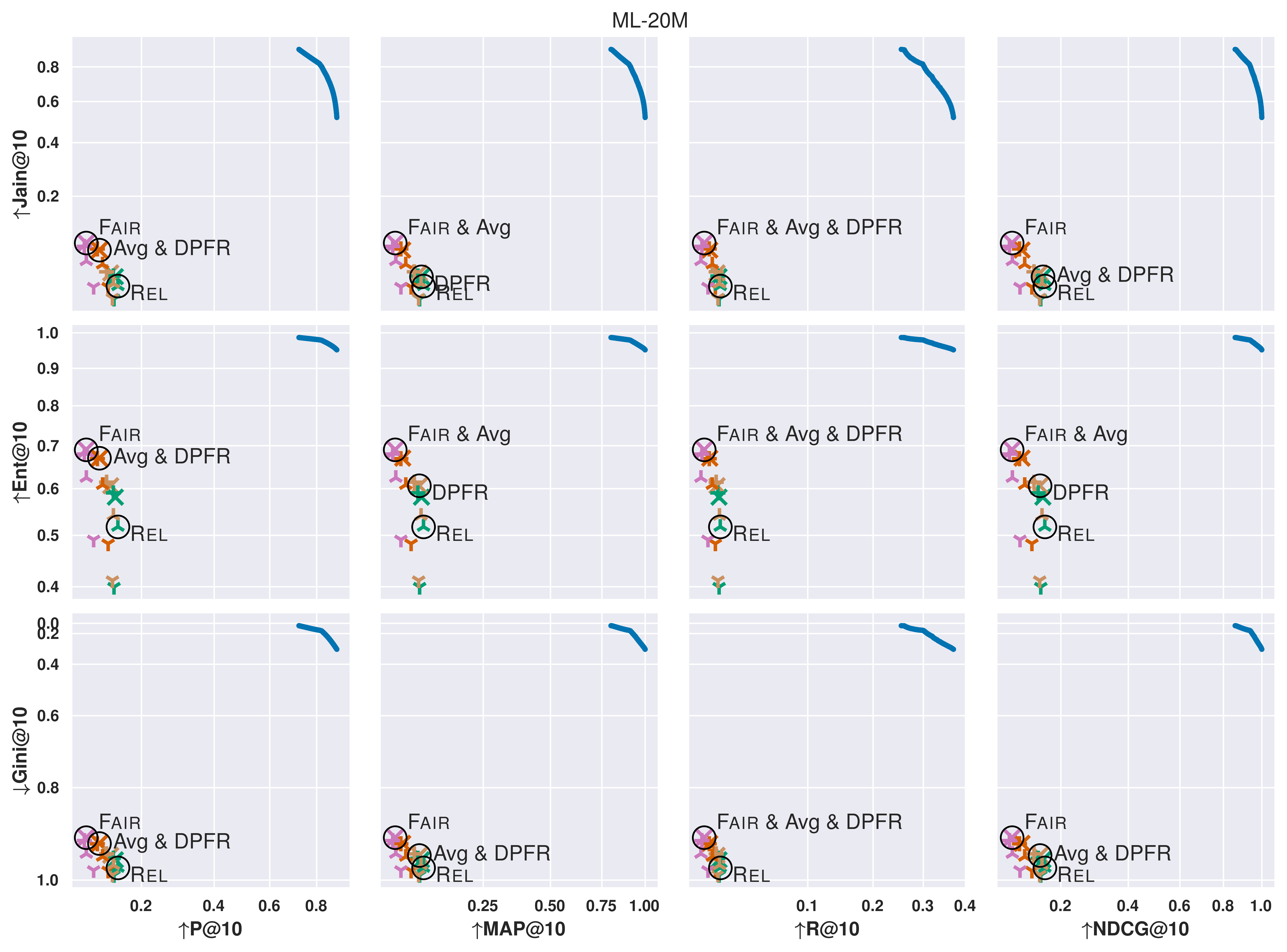}
    
    \caption{Pareto Frontier of fairness and relevance (in blue), together with recommender model scores for Amazon-lb, Jester, and ML-*. \textsc{Fair} measures are on the $y$-axis and \textsc{Rel} measures are on the $x$-axis. We implement exponential-like scales to enhance the visibility of the model plots. The \textsc{Rel, Fair}, Avg, and DPFR denote the best model based on each evaluation approach.}
    \label{fig:app-pairplot}

\end{figure*}

We present the actual scores of the recommender models in \Cref{tab:base-rerank-all-1}--\ref{tab:base-rerank-all-2}. In \Cref{tab:app-gradient}, we present the gradient values of the PF, used in determining which pair of measures are suitable for DPFR. In \Cref{fig:app-pairplot} we present the Pareto Frontier (PF) of fairness and relevance together with recommender model scores in \Cref{tab:base-rerank-all-1}--\ref{tab:base-rerank-all-2} for Amazon-lb, Jester, and ML-*. In \Cref{tab:dpfr-scores,tab:dpfr-scores2} we present the DPFR scores for all datasets. 
In \Cref{tab:corr_est} we present the Kendall's $\tau$ correlation scores of the DPFR from estimated PF and the PF.

\subsection{Further Discussions}
\label{app:discussion}

\subsubsection{The impact of replacing frequently recommended items}
In this work, we replace frequently recommended items in two cases: during the Pareto Frontier generation (PF) with the {\sc Oracle2Fair} algorithm (\Cref{ss:generation}) and as part of the fair rerankers (\Cref{s:experiments}). In both cases, none of the replacements significantly affect the overall recommendation performance:

\noindent\textbf{1. Replacement in the {\sc Oracle2Fair} algorithm.} The {\sc Oracle2Fair} algorithm is used to generate the recommendation lists whose scores make up the PF (not the scores from the model-generated recommendation lists, that we compare to a point in the PF). The replacements are done on lists separate from the model's recommendation lists. Therefore, the replacements in the {\sc Oracle2Fair} algorithm do not affect the recommendation performance based on relevance and fairness measures. 

\noindent\textbf{2. Replacement in the fair rerankers.} We look at recommendation performance based on NDCG (relevance) and Gini (fairness). In all six datasets, when comparing the best non-reranked model to its reranked versions (e.g., for Lastfm, it is NCL vs NCL-BC, NCL-CM, and NCL-GS), the decrease in NDCG is not more than 0.26, and the decrease is even below 0.15 in all datasets excluding Jester, (\Cref{app:actual-scores}, \Cref{tab:base-rerank-all-1,tab:base-rerank-all-2}). For Jester, the 0.26 decrease in NDCG is exchanged for an improvement in Gini by 0.218. Hence, we believe that the impact of item replacement is reasonable.

\subsubsection{Using rank-based fairness measures}

Suppose we have two \textsc{Fair} measures, e.g., set-based Gini (Gini) and rank-based Gini (Gini-w). In \citep{Rampisela2024EvaluationStudy}, the absolute scores of Gini and Gini-w do not differ considerably and the two measures correlate strongly with Kendall's $\tau \geq 0.9$. Hence, generating the PF with rank-based fairness measures such as Gini-w is not expected to result in significantly different conclusions from the set-based version.

\begin{table*}[p]
\centering
\caption{Relevance (\textsc{Rel}), fairness (\textsc{Fair}), and joint fairness and relevance (\textsc{Fair+Rel}) scores at $k=10$ of the recommender models for Lastfm, Amazon-lb, and QK-video, without and with re-ranking the the top $k'=25$ items using Borda Count (BC), COMBMNZ (CM), and Greedy Substitution (GS). The most relevant or most fair score per measure is in bold. $\uparrow$ means the higher the better, $\downarrow$ the lower the better.}
\label{tab:base-rerank-all-1}
\resizebox{\textwidth}{!}{
\begin{tabular}{lll*{4}{r}|*{4}{r}|*{4}{r}|*{4}{r}}
\toprule
 &  & model & \multicolumn{4}{c|}{ItemKNN} & \multicolumn{4}{c|}{BPR} & \multicolumn{4}{c|}{MultiVAE} & \multicolumn{4}{c}{NCL} \\ 
\midrule
 &  & reranking & - & BC & CM & GS & - & BC & CM & GS & - & BC & CM & GS & - & BC & CM & GS \\
\midrule
\multirow[c]{16}{*}{\rotatebox[origin=r]{90}{Lastfm}} & \multirow[c]{6}{*}{\rotatebox[origin=r]{90}{\textsc{Rel}}} & $\uparrow$ $\text{HR}$ & 0.765 & 0.742 & 0.581 & 0.750 & 0.773 & 0.729 & 0.587 & 0.751 & 0.778 & 0.693 & 0.523 & 0.734 & \bfseries 0.793 & 0.726 & 0.571 & 0.765 \\
 &  & $\uparrow$ $\text{MRR}$ & 0.484 & 0.333 & 0.270 & 0.481 & 0.492 & 0.323 & 0.280 & 0.488 & 0.476 & 0.285 & 0.232 & 0.470 & \bfseries 0.503 & 0.311 & 0.260 & 0.499 \\
 &  & $\uparrow$ $\text{P}$ & 0.172 & 0.147 & 0.089 & 0.167 & 0.178 & 0.140 & 0.092 & 0.169 & 0.176 & 0.129 & 0.076 & 0.161 & \bfseries 0.184 & 0.141 & 0.087 & 0.173 \\
 &  & $\uparrow$ $\text{MAP}$ & 0.137 & 0.085 & 0.053 & 0.135 & 0.141 & 0.080 & 0.058 & 0.138 & 0.138 & 0.070 & 0.045 & 0.132 & \bfseries 0.148 & 0.079 & 0.050 & 0.144 \\
 &  & $\uparrow$ $\text{R}$ & 0.218 & 0.186 & 0.114 & 0.211 & 0.224 & 0.180 & 0.119 & 0.211 & 0.224 & 0.163 & 0.098 & 0.205 & \bfseries 0.234 & 0.180 & 0.110 & 0.220 \\
 &  & $\uparrow$ $\text{NDCG}$ & 0.245 & 0.181 & 0.119 & 0.241 & 0.252 & 0.173 & 0.126 & 0.244 & 0.247 & 0.155 & 0.102 & 0.235 & \bfseries 0.261 & 0.170 & 0.115 & 0.252 \\
\cline{2-19}
 & \multirow[c]{5}{*}{\rotatebox[origin=r]{90}{\textsc{Fair}}} & $\uparrow$ $\text{Jain}$ & 0.042 & 0.101 & 0.094 & 0.046 & 0.058 & 0.151 & 0.140 & 0.067 & 0.097 & \bfseries 0.236 & 0.222 & 0.115 & 0.082 & 0.216 & 0.215 & 0.095 \\
 &  & $\uparrow$ $\text{QF}$ & 0.474 & 0.642 & \bfseries 0.679 & 0.533 & 0.362 & 0.491 & 0.528 & 0.402 & 0.517 & 0.658 & 0.678 & 0.554 & 0.453 & 0.622 & 0.657 & 0.502 \\
 &  & $\uparrow$ $\text{Ent}$ & 0.589 & 0.727 & 0.735 & 0.622 & 0.610 & 0.736 & 0.740 & 0.646 & 0.707 & 0.820 & \bfseries 0.826 & 0.740 & 0.671 & 0.801 & 0.810 & 0.705 \\
 &  & $\uparrow$ $\text{FSat}$ & 0.129 & 0.197 & 0.216 & 0.152 & 0.147 & 0.211 & 0.228 & 0.177 & 0.202 & 0.293 & \bfseries 0.321 & 0.249 & 0.178 & 0.269 & 0.286 & 0.221 \\
 &  & $\downarrow$ $\text{Gini}$ & 0.904 & 0.810 & 0.790 & 0.879 & 0.910 & 0.827 & 0.818 & 0.887 & 0.839 & 0.715 & \bfseries 0.696 & 0.803 & 0.872 & 0.748 & 0.728 & 0.840 \\
\cline{2-19}
 & \multirow[c]{5}{*}{\rotatebox[origin=r]{90}{\textsc{Fair+Rel}}} & $\uparrow$ $\text{IBO}$ & 0.209 & 0.270 & 0.256 & 0.227 & 0.208 & 0.263 & 0.253 & 0.228 & 0.261 & \bfseries 0.314 & 0.278 & 0.281 & 0.242 & 0.308 & 0.292 & 0.265 \\
 &  & $\downarrow$ $\text{MME}$ & 0.001 & 0.001 & 0.001 & 0.001 & 0.001 & 0.001 & 0.001 & 0.001 & 0.001 & 0.000 & 0.001 & 0.001 & 0.001 & \bfseries 0.000 & 0.001 & 0.001 \\
 &  & $\downarrow$ $\text{IAA}$ & 0.004 & 0.004 & 0.004 & 0.004 & 0.004 & 0.004 & 0.004 & 0.004 & 0.004 & 0.004 & 0.004 & 0.004 & \bfseries 0.004 & 0.004 & 0.004 & 0.004 \\
 &  & $\downarrow$ $\text{II-F}$ & 0.001 & 0.001 & 0.002 & 0.001 & 0.001 & 0.001 & 0.002 & 0.001 & 0.001 & 0.001 & 0.002 & 0.001 & \bfseries 0.001 & 0.001 & 0.002 & 0.001 \\
 &  & $\downarrow$ $\text{AI-F}$ & 0.000 & 0.000 & 0.000 & 0.000 & 0.000 & 0.000 & 0.000 & 0.000 & 0.000 & 0.000 & 0.000 & 0.000 & 0.000 & \bfseries 0.000 & 0.000 & 0.000 \\
\cline{1-19}
\multirow[c]{16}{*}{\rotatebox[origin=r]{90}{Amazon-lb}} & \multirow[c]{6}{*}{\rotatebox[origin=r]{90}{\textsc{Rel}}} & $\uparrow$ $\text{HR}$ & \bfseries 0.046 & 0.021 & 0.016 & 0.043 & 0.011 & 0.014 & 0.021 & 0.011 & 0.039 & 0.007 & 0.014 & \bfseries 0.046 & 0.034 & 0.021 & 0.011 & 0.034 \\
 &  & $\uparrow$ $\text{MRR}$ & 0.020 & 0.011 & 0.011 & 0.020 & 0.003 & 0.005 & 0.007 & 0.003 & 0.023 & 0.003 & 0.004 & \bfseries 0.024 & 0.022 & 0.006 & 0.003 & 0.022 \\
 &  & $\uparrow$ $\text{P}$ & 0.005 & 0.002 & 0.002 & 0.005 & 0.001 & 0.001 & 0.002 & 0.001 & 0.004 & 0.001 & 0.002 & \bfseries 0.005 & 0.004 & 0.002 & 0.001 & 0.004 \\
 &  & $\uparrow$ $\text{MAP}$ & 0.006 & 0.004 & 0.004 & 0.006 & 0.002 & 0.003 & 0.004 & 0.002 & 0.006 & 0.002 & 0.003 & 0.006 & \bfseries 0.006 & 0.002 & 0.001 & 0.006 \\
 &  & $\uparrow$ $\text{R}$ & \bfseries 0.013 & 0.007 & 0.005 & 0.013 & 0.005 & 0.008 & 0.010 & 0.005 & 0.010 & 0.005 & 0.008 & 0.012 & 0.012 & 0.007 & 0.003 & 0.011 \\
 &  & $\uparrow$ $\text{NDCG}$ & 0.011 & 0.006 & 0.005 & 0.011 & 0.003 & 0.005 & 0.006 & 0.003 & 0.010 & 0.003 & 0.004 & \bfseries 0.011 & 0.011 & 0.004 & 0.002 & 0.011 \\
\cline{2-19}
 & \multirow[c]{5}{*}{\rotatebox[origin=r]{90}{\textsc{Fair}}} & $\uparrow$ $\text{Jain}$ & 0.271 & \bfseries 0.547 & 0.431 & 0.324 & 0.223 & 0.432 & 0.359 & 0.259 & 0.035 & 0.123 & 0.097 & 0.043 & 0.026 & 0.098 & 0.080 & 0.031 \\
 &  & $\uparrow$ $\text{QF}$ & 0.650 & \bfseries 0.679 & 0.612 & 0.663 & 0.549 & 0.630 & 0.594 & 0.571 & 0.222 & 0.294 & 0.286 & 0.254 & 0.229 & 0.315 & 0.310 & 0.265 \\
 &  & $\uparrow$ $\text{Ent}$ & 0.802 & \bfseries 0.882 & 0.839 & 0.829 & 0.747 & 0.839 & 0.809 & 0.776 & 0.418 & 0.587 & 0.558 & 0.469 & 0.371 & 0.560 & 0.534 & 0.426 \\
 &  & $\uparrow$ $\text{FSat}$ & 0.370 & \bfseries 0.538 & 0.438 & 0.435 & 0.314 & 0.410 & 0.376 & 0.364 & 0.114 & 0.159 & 0.152 & 0.138 & 0.091 & 0.146 & 0.138 & 0.115 \\
 &  & $\downarrow$ $\text{Gini}$ & 0.665 & \bfseries 0.492 & 0.598 & 0.613 & 0.747 & 0.601 & 0.660 & 0.703 & 0.949 & 0.882 & 0.899 & 0.930 & 0.959 & 0.898 & 0.910 & 0.943 \\
\cline{2-19}
 & \multirow[c]{5}{*}{\rotatebox[origin=r]{90}{\textsc{Fair+Rel}}} & $\uparrow$ $\text{IBO}$ & 0.062 & 0.038 & 0.029 & \bfseries 0.067 & 0.019 & 0.029 & 0.038 & 0.019 & 0.029 & 0.019 & 0.029 & 0.033 & 0.038 & 0.033 & 0.024 & 0.033 \\
 &  & $\downarrow$ $\text{MME}$ & 0.001 & 0.001 & 0.001 & 0.001 & 0.001 & \bfseries 0.001 & 0.001 & 0.001 & 0.003 & 0.001 & 0.001 & 0.003 & 0.004 & 0.001 & 0.001 & 0.004 \\
 &  & $\downarrow$ $\text{IAA}$ & 0.011 & 0.011 & 0.011 & \bfseries 0.011 & 0.011 & 0.011 & 0.011 & 0.011 & 0.011 & 0.011 & 0.011 & 0.011 & 0.011 & 0.011 & 0.011 & 0.011 \\
 &  & $\downarrow$ $\text{II-F}$ & 0.006 & 0.006 & 0.006 & \bfseries 0.006 & 0.006 & 0.006 & 0.006 & 0.006 & 0.006 & 0.006 & 0.006 & 0.006 & 0.006 & 0.006 & 0.006 & 0.006 \\
 &  & $\downarrow$ $\text{AI-F}$ & 0.000 & \bfseries 0.000 & 0.000 & 0.000 & 0.000 & 0.000 & 0.000 & 0.000 & 0.001 & 0.000 & 0.000 & 0.001 & 0.002 & 0.000 & 0.000 & 0.002 \\
\cline{1-19}
\multirow[c]{16}{*}{\rotatebox[origin=r]{90}{QK-video}} & \multirow[c]{6}{*}{\rotatebox[origin=r]{90}{\textsc{Rel}}} & $\uparrow$ $\text{HR}$ & 0.040 & 0.046 & 0.047 & 0.038 & 0.099 & 0.063 & 0.045 & 0.089 & 0.109 & 0.089 & 0.061 & 0.103 & \bfseries 0.130 & 0.102 & 0.077 & 0.124 \\
 &  & $\uparrow$ $\text{MRR}$ & 0.013 & 0.014 & 0.013 & 0.013 & 0.039 & 0.018 & 0.015 & 0.038 & 0.039 & 0.028 & 0.021 & 0.038 & \bfseries 0.048 & 0.030 & 0.024 & 0.047 \\
 &  & $\uparrow$ $\text{P}$ & 0.004 & 0.005 & 0.005 & 0.004 & 0.011 & 0.007 & 0.005 & 0.010 & 0.012 & 0.009 & 0.006 & 0.011 & \bfseries 0.014 & 0.011 & 0.008 & 0.013 \\
 &  & $\uparrow$ $\text{MAP}$ & 0.005 & 0.005 & 0.005 & 0.005 & 0.017 & 0.008 & 0.006 & 0.016 & 0.018 & 0.012 & 0.009 & 0.017 & \bfseries 0.022 & 0.013 & 0.010 & 0.021 \\
 &  & $\uparrow$ $\text{R}$ & 0.014 & 0.018 & 0.019 & 0.014 & 0.043 & 0.028 & 0.019 & 0.039 & 0.051 & 0.039 & 0.027 & 0.047 & \bfseries 0.061 & 0.045 & 0.033 & 0.058 \\
 &  & $\uparrow$ $\text{NDCG}$ & 0.009 & 0.011 & 0.010 & 0.009 & 0.029 & 0.015 & 0.011 & 0.027 & 0.031 & 0.022 & 0.016 & 0.030 & \bfseries 0.038 & 0.025 & 0.019 & 0.037 \\
\cline{2-19}
 & \multirow[c]{5}{*}{\rotatebox[origin=r]{90}{\textsc{Fair}}} & $\uparrow$ $\text{Jain}$ & 0.483 & \bfseries 0.815 & 0.589 & 0.567 & 0.081 & 0.333 & 0.379 & 0.101 & 0.012 & 0.038 & 0.032 & 0.014 & 0.020 & 0.076 & 0.071 & 0.023 \\
 &  & $\uparrow$ $\text{QF}$ & 0.901 & \bfseries 0.956 & 0.790 & 0.924 & 0.625 & 0.809 & 0.823 & 0.678 & 0.100 & 0.155 & 0.163 & 0.127 & 0.201 & 0.331 & 0.365 & 0.253 \\
 &  & $\uparrow$ $\text{Ent}$ & 0.933 & \bfseries 0.979 & 0.937 & 0.950 & 0.755 & 0.888 & 0.903 & 0.792 & 0.420 & 0.557 & 0.547 & 0.458 & 0.507 & 0.667 & 0.674 & 0.549 \\
 &  & $\uparrow$ $\text{FSat}$ & 0.443 & \bfseries 0.659 & 0.547 & 0.522 & 0.212 & 0.346 & 0.382 & 0.259 & 0.052 & 0.089 & 0.090 & 0.070 & 0.077 & 0.140 & 0.150 & 0.104 \\
 &  & $\downarrow$ $\text{Gini}$ & 0.472 & \bfseries 0.235 & 0.442 & 0.397 & 0.807 & 0.613 & 0.570 & 0.761 & 0.982 & 0.957 & 0.959 & 0.976 & 0.966 & 0.909 & 0.902 & 0.952 \\
\cline{2-19}
 & \multirow[c]{5}{*}{\rotatebox[origin=r]{90}{\textsc{Fair+Rel}}} & $\uparrow$ $\text{IBO}$ & 0.033 & 0.038 & 0.038 & 0.035 & 0.054 & 0.050 & 0.036 & 0.052 & 0.031 & 0.042 & 0.036 & 0.033 & 0.043 & \bfseries 0.060 & 0.054 & 0.047 \\
 &  & $\downarrow$ $\text{MME}$ & 0.000 & \bfseries 0.000 & 0.000 & 0.000 & 0.000 & 0.000 & 0.000 & 0.000 & 0.000 & 0.000 & 0.000 & 0.000 & 0.000 & 0.000 & 0.000 & 0.000 \\
 &  & $\downarrow$ $\text{IAA}$ & 0.001 & 0.001 & 0.001 & 0.001 & 0.001 & 0.001 & 0.001 & 0.001 & 0.001 & 0.001 & 0.001 & 0.001 & \bfseries 0.001 & 0.001 & 0.001 & 0.001 \\
 &  & $\downarrow$ $\text{II-F}$ & 0.001 & 0.001 & 0.001 & 0.001 & 0.001 & 0.001 & 0.001 & 0.001 & 0.001 & 0.001 & 0.001 & 0.001 & \bfseries 0.001 & 0.001 & 0.001 & 0.001 \\
 &  & $\downarrow$ $\text{AI-F}$ & 0.000 & \bfseries 0.000 & 0.000 & 0.000 & 0.000 & 0.000 & 0.000 & 0.000 & 0.000 & 0.000 & 0.000 & 0.000 & 0.000 & 0.000 & 0.000 & 0.000 \\
\bottomrule
\end{tabular}}
\end{table*}

\begin{table*}[p]
\centering
\caption{Relevance (\textsc{Rel}), fairness (\textsc{Fair}), and joint fairness and relevance (\textsc{Fair+Rel}) scores at $k=10$ of the recommender models for Jester and ML-*, without and with re-ranking the the top $k'=25$ items using Borda Count (BC), COMBMNZ (CM), and Greedy Substitution (GS) evaluated at $k=10$. The most relevant or most fair score per measure is in bold. $\uparrow$ means the higher the better, $\downarrow$ the lower the better.}
\label{tab:base-rerank-all-2}
\resizebox{\textwidth}{!}{
\begin{tabular}{lll*{4}{r}|*{4}{r}|*{4}{r}|*{4}{r}}
\toprule
 &  & model & \multicolumn{4}{c|}{ItemKNN} & \multicolumn{4}{c|}{BPR} & \multicolumn{4}{c|}{MultiVAE} & \multicolumn{4}{c}{NCL} \\ 
\midrule
 &  & reranking & - & BC & CM & GS & - & BC & CM & GS & - & BC & CM & GS & - & BC & CM & GS \\
\midrule
\multirow[c]{16}{*}{\rotatebox[origin=r]{90}{Jester}} & \multirow[c]{6}{*}{\rotatebox[origin=r]{90}{\textsc{Rel}}} & $\uparrow$ $\text{HR}$ & 0.933 & 0.888 & 0.652 & 0.932 & 0.929 & 0.876 & 0.742 & 0.928 & \bfseries 0.944 & 0.899 & 0.818 & 0.944 & 0.939 & 0.893 & 0.804 & 0.939 \\
 &  & $\uparrow$ $\text{MRR}$ & 0.632 & 0.443 & 0.307 & 0.632 & 0.635 & 0.455 & 0.322 & 0.635 & \bfseries 0.661 & 0.465 & 0.370 & 0.661 & 0.651 & 0.479 & 0.349 & 0.651 \\
 &  & $\uparrow$ $\text{P}$ & 0.334 & 0.250 & 0.144 & 0.333 & 0.330 & 0.243 & 0.163 & 0.329 & \bfseries 0.351 & 0.262 & 0.194 & 0.351 & 0.342 & 0.257 & 0.185 & 0.341 \\
 &  & $\uparrow$ $\text{MAP}$ & 0.352 & 0.198 & 0.101 & 0.352 & 0.348 & 0.195 & 0.112 & 0.348 & \bfseries 0.379 & 0.208 & 0.145 & 0.379 & 0.367 & 0.211 & 0.133 & 0.367 \\
 &  & $\uparrow$ $\text{R}$ & 0.529 & 0.393 & 0.197 & 0.529 & 0.524 & 0.377 & 0.255 & 0.523 & \bfseries 0.555 & 0.405 & 0.324 & 0.555 & 0.543 & 0.400 & 0.305 & 0.542 \\
 &  & $\uparrow$ $\text{NDCG}$ & 0.496 & 0.336 & 0.189 & 0.496 & 0.493 & 0.331 & 0.216 & 0.492 & \bfseries 0.525 & 0.350 & 0.265 & 0.524 & 0.512 & 0.352 & 0.249 & 0.511 \\
\cline{2-19}
 & \multirow[c]{5}{*}{\rotatebox[origin=r]{90}{\textsc{Fair}}} & $\uparrow$ $\text{Jain}$ & 0.343 & 0.556 & 0.445 & 0.345 & 0.377 & \bfseries 0.583 & 0.547 & 0.380 & 0.295 & 0.544 & 0.509 & 0.297 & 0.351 & 0.504 & 0.534 & 0.354 \\
 &  & $\uparrow$ $\text{QF}^*$ & \bfseries 1.000 & \bfseries 1.000 & \bfseries 1.000 & \bfseries 1.000 & \bfseries 1.000 & \bfseries 1.000 & \bfseries 1.000 & \bfseries 1.000 & 0.967 & \bfseries 1.000 & \bfseries 1.000 & \bfseries 1.000 & \bfseries 1.000 & \bfseries 1.000 & \bfseries 1.000 & \bfseries 1.000 \\
 &  & $\uparrow$ $\text{Ent}$ & 0.702 & 0.854 & 0.784 & 0.705 & 0.754 & \bfseries 0.875 & 0.857 & 0.757 & 0.648 & 0.852 & 0.839 & 0.651 & 0.722 & 0.838 & 0.855 & 0.725 \\
 &  & $\uparrow$ $\text{FSat}$ & 0.267 & \bfseries 0.378 & 0.289 & 0.267 & 0.244 & 0.344 & 0.333 & 0.244 & 0.256 & 0.344 & 0.300 & 0.256 & 0.222 & 0.344 & 0.311 & 0.222 \\
 &  & $\downarrow$ $\text{Gini}$ & 0.687 & 0.502 & 0.595 & 0.685 & 0.632 & \bfseries 0.467 & 0.495 & 0.629 & 0.738 & 0.506 & 0.520 & 0.735 & 0.668 & 0.528 & 0.502 & 0.665 \\
\cline{2-19}
 & \multirow[c]{5}{*}{\rotatebox[origin=r]{90}{\textsc{Fair+Rel}}} & $\uparrow$ $\text{IBO}$ & 0.600 & \bfseries 0.930 & 0.740 & 0.600 & 0.840 & 0.910 & 0.780 & 0.840 & 0.500 & 0.870 & 0.810 & 0.500 & 0.740 & 0.920 & 0.780 & 0.740 \\
 &  & $\downarrow$ $\text{MME}$ & 0.003 & 0.003 & 0.006 & 0.003 & 0.004 & \bfseries 0.002 & 0.005 & 0.004 & 0.008 & 0.003 & 0.004 & 0.008 & 0.004 & 0.003 & 0.006 & 0.004 \\
 &  & $\downarrow$ $\text{IAA}$ & 0.081 & 0.093 & 0.104 & 0.081 & 0.081 & 0.094 & 0.103 & 0.081 & 0.078 & 0.092 & 0.100 & \bfseries 0.078 & 0.079 & 0.092 & 0.101 & 0.079 \\
 &  & $\downarrow$ $\text{II-F}$ & 0.028 & 0.035 & 0.040 & 0.028 & 0.029 & 0.035 & 0.040 & 0.029 & 0.027 & 0.035 & 0.038 & \bfseries 0.027 & 0.028 & 0.034 & 0.038 & 0.028 \\
 &  & $\downarrow$ $\text{AI-F}$ & 0.002 & 0.002 & 0.003 & 0.002 & 0.002 & \bfseries 0.001 & 0.002 & 0.002 & 0.003 & 0.002 & 0.002 & 0.003 & 0.002 & 0.002 & 0.002 & 0.002 \\
\cline{1-19}
\multirow[c]{16}{*}{\rotatebox[origin=r]{90}{ML-10M}} & \multirow[c]{6}{*}{\rotatebox[origin=r]{90}{\textsc{Rel}}} & $\uparrow$ $\text{HR}$ & 0.487 & 0.480 & 0.443 & 0.481 & 0.512 & 0.462 & 0.386 & 0.485 & 0.417 & 0.438 & 0.387 & 0.410 & \bfseries 0.521 & 0.473 & 0.402 & 0.513 \\
 &  & $\uparrow$ $\text{MRR}$ & 0.282 & 0.242 & 0.225 & 0.279 & 0.299 & 0.208 & 0.185 & 0.295 & 0.237 & 0.231 & 0.191 & 0.235 & \bfseries 0.302 & 0.216 & 0.203 & 0.301 \\
 &  & $\uparrow$ $\text{P}$ & 0.137 & 0.128 & 0.105 & 0.133 & 0.146 & 0.114 & 0.088 & 0.132 & 0.107 & 0.111 & 0.096 & 0.105 & \bfseries 0.154 & 0.123 & 0.094 & 0.149 \\
 &  & $\uparrow$ $\text{MAP}$ & 0.089 & 0.074 & 0.060 & 0.086 & 0.095 & 0.061 & 0.047 & 0.088 & 0.067 & 0.067 & 0.054 & 0.066 & \bfseries 0.101 & 0.067 & 0.052 & 0.099 \\
 &  & $\uparrow$ $\text{R}$ & 0.022 & 0.022 & 0.018 & 0.022 & 0.025 & 0.019 & 0.012 & 0.023 & 0.020 & 0.021 & 0.016 & 0.021 & \bfseries 0.026 & 0.020 & 0.013 & 0.026 \\
 &  & $\uparrow$ $\text{NDCG}$ & 0.150 & 0.133 & 0.113 & 0.147 & 0.160 & 0.115 & 0.092 & 0.150 & 0.119 & 0.121 & 0.100 & 0.118 & \bfseries 0.167 & 0.123 & 0.100 & 0.164 \\
\cline{2-19}
 & \multirow[c]{5}{*}{\rotatebox[origin=r]{90}{\textsc{Fair}}} & $\uparrow$ $\text{Jain}$ & 0.011 & 0.026 & 0.027 & 0.012 & 0.037 & 0.100 & \bfseries 0.115 & 0.044 & 0.003 & 0.005 & 0.006 & 0.004 & 0.024 & 0.063 & 0.069 & 0.027 \\
 &  & $\uparrow$ $\text{QF}$ & 0.044 & 0.062 & 0.068 & 0.047 & 0.145 & 0.199 & \bfseries 0.216 & 0.160 & 0.014 & 0.021 & 0.025 & 0.016 & 0.086 & 0.123 & 0.132 & 0.094 \\
 &  & $\uparrow$ $\text{Ent}$ & 0.407 & 0.503 & 0.514 & 0.418 & 0.596 & 0.697 & \bfseries 0.716 & 0.624 & 0.238 & 0.302 & 0.324 & 0.258 & 0.519 & 0.625 & 0.638 & 0.544 \\
 &  & $\uparrow$ $\text{FSat}$ & 0.044 & 0.062 & 0.068 & 0.047 & 0.145 & 0.199 & \bfseries 0.216 & 0.160 & 0.014 & 0.021 & 0.025 & 0.016 & 0.086 & 0.123 & 0.132 & 0.094 \\
 &  & $\downarrow$ $\text{Gini}$ & 0.987 & 0.973 & 0.971 & 0.985 & 0.945 & 0.895 & \bfseries 0.879 & 0.932 & 0.997 & 0.994 & 0.993 & 0.996 & 0.969 & 0.936 & 0.930 & 0.963 \\
\cline{2-19}
 & \multirow[c]{5}{*}{\rotatebox[origin=r]{90}{\textsc{Fair+Rel}}} & $\uparrow$ $\text{IBO}$ & 0.031 & 0.043 & 0.046 & 0.034 & 0.069 & 0.089 & \bfseries 0.091 & 0.076 & 0.012 & 0.016 & 0.018 & 0.014 & 0.054 & 0.073 & 0.074 & 0.058 \\
 &  & $\downarrow$ $\text{MME}$ & 0.001 & 0.001 & 0.001 & 0.001 & 0.001 & \bfseries 0.001 & 0.001 & 0.001 & 0.003 & 0.002 & 0.001 & 0.002 & 0.001 & 0.001 & 0.001 & 0.001 \\
 &  & $\downarrow$ $\text{IAA}$ & 0.008 & 0.009 & 0.009 & 0.008 & 0.008 & 0.009 & 0.009 & 0.008 & 0.009 & 0.009 & 0.009 & 0.009 & \bfseries 0.008 & 0.009 & 0.009 & 0.008 \\
 &  & $\downarrow$ $\text{II-F}$ & 0.000 & 0.000 & 0.000 & 0.000 & 0.000 & 0.000 & 0.000 & 0.000 & 0.000 & 0.000 & 0.000 & 0.000 & \bfseries 0.000 & 0.000 & 0.000 & 0.000 \\
 &  & $\downarrow$ $\text{AI-F}$ & 0.000 & 0.000 & 0.000 & 0.000 & 0.000 & \bfseries 0.000 & 0.000 & 0.000 & 0.000 & 0.000 & 0.000 & 0.000 & 0.000 & 0.000 & 0.000 & 0.000 \\
\cline{1-19}
\multirow[c]{16}{*}{\rotatebox[origin=r]{90}{ML-20M}} & \multirow[c]{6}{*}{\rotatebox[origin=r]{90}{\textsc{Rel}}} & $\uparrow$ $\text{HR}$ & 0.488 & 0.473 & 0.420 & 0.483 & \bfseries 0.505 & 0.444 & 0.392 & 0.483 & 0.489 & 0.432 & 0.391 & 0.465 & \bfseries 0.505 & 0.453 & 0.388 & 0.493 \\
 &  & $\uparrow$ $\text{MRR}$ & 0.280 & 0.237 & 0.213 & 0.278 & 0.293 & 0.205 & 0.190 & 0.290 & 0.259 & 0.193 & 0.180 & 0.256 & \bfseries 0.293 & 0.206 & 0.193 & 0.292 \\
 &  & $\uparrow$ $\text{P}$ & 0.142 & 0.131 & 0.106 & 0.139 & 0.145 & 0.116 & 0.094 & 0.136 & 0.141 & 0.112 & 0.091 & 0.128 & \bfseries 0.150 & 0.121 & 0.094 & 0.141 \\
 &  & $\uparrow$ $\text{MAP}$ & 0.092 & 0.077 & 0.061 & 0.090 & 0.096 & 0.063 & 0.052 & 0.092 & 0.089 & 0.060 & 0.049 & 0.082 & \bfseries 0.100 & 0.068 & 0.053 & 0.095 \\
 &  & $\uparrow$ $\text{R}$ & 0.019 & 0.017 & 0.014 & 0.019 & 0.019 & 0.014 & 0.012 & 0.018 & 0.019 & 0.014 & 0.011 & 0.018 & \bfseries 0.020 & 0.016 & 0.011 & 0.020 \\
 &  & $\uparrow$ $\text{NDCG}$ & 0.154 & 0.135 & 0.112 & 0.151 & 0.158 & 0.116 & 0.098 & 0.152 & 0.148 & 0.111 & 0.093 & 0.139 & \bfseries 0.163 & 0.121 & 0.099 & 0.157 \\
\cline{2-19}
 & \multirow[c]{5}{*}{\rotatebox[origin=r]{90}{\textsc{Fair}}} & $\uparrow$ $\text{Jain}$ & 0.008 & 0.017 & 0.018 & 0.009 & 0.028 & 0.068 & \bfseries 0.081 & 0.033 & 0.029 & 0.070 & 0.074 & 0.034 & 0.018 & 0.044 & 0.049 & 0.021 \\
 &  & $\uparrow$ $\text{QF}$ & 0.035 & 0.047 & 0.051 & 0.037 & 0.114 & 0.154 & \bfseries 0.165 & 0.125 & 0.117 & 0.146 & 0.154 & 0.126 & 0.074 & 0.103 & 0.112 & 0.082 \\
 &  & $\uparrow$ $\text{Ent}$ & 0.399 & 0.483 & 0.491 & 0.411 & 0.581 & 0.670 & \bfseries 0.690 & 0.606 & 0.591 & 0.669 & 0.680 & 0.615 & 0.517 & 0.608 & 0.624 & 0.541 \\
 &  & $\uparrow$ $\text{FSat}$ & 0.035 & 0.047 & 0.051 & 0.037 & 0.114 & 0.154 & \bfseries 0.165 & 0.125 & 0.117 & 0.146 & 0.154 & 0.126 & 0.074 & 0.103 & 0.112 & 0.082 \\
 &  & $\downarrow$ $\text{Gini}$ & 0.991 & 0.982 & 0.981 & 0.990 & 0.960 & 0.926 & \bfseries 0.914 & 0.951 & 0.957 & 0.927 & 0.920 & 0.948 & 0.976 & 0.953 & 0.947 & 0.971 \\
\cline{2-19}
 & \multirow[c]{5}{*}{\rotatebox[origin=r]{90}{\textsc{Fair+Rel}}} & $\uparrow$ $\text{IBO}$ & 0.021 & 0.031 & 0.033 & 0.022 & 0.049 & 0.064 & \bfseries 0.067 & 0.054 & 0.052 & 0.064 & 0.065 & 0.056 & 0.039 & 0.051 & 0.054 & 0.042 \\
 &  & $\downarrow$ $\text{MME}$ & 0.001 & 0.001 & 0.001 & 0.001 & 0.001 & 0.000 & \bfseries 0.000 & 0.001 & 0.001 & 0.000 & 0.000 & 0.001 & 0.001 & 0.001 & 0.001 & 0.001 \\
 &  & $\downarrow$ $\text{IAA}$ & 0.007 & 0.007 & 0.007 & 0.007 & 0.007 & 0.007 & 0.007 & 0.007 & 0.007 & 0.007 & 0.007 & 0.007 & \bfseries 0.007 & 0.007 & 0.007 & 0.007 \\
 &  & $\downarrow$ $\text{II-F}$ & 0.000 & 0.000 & 0.000 & 0.000 & 0.000 & 0.000 & 0.000 & 0.000 & 0.000 & 0.000 & 0.000 & 0.000 & \bfseries 0.000 & 0.000 & 0.000 & 0.000 \\
 &  & $\downarrow$ $\text{AI-F}$ & 0.000 & 0.000 & 0.000 & 0.000 & 0.000 & 0.000 & \bfseries 0.000 & 0.000 & 0.000 & 0.000 & 0.000 & 0.000 & 0.000 & 0.000 & 0.000 & 0.000 \\
\bottomrule
\multicolumn{19}{l}{\small *QF $=1$ means that all items in the dataset appear in the recommendation across all users.} \\
\multicolumn{19}{l}{\small $^{\dag}$The scores of QF are the same as FSat for ML-*, as QF is computed based on the percentage of items in the dataset that are recommended, which in this dataset} \\
\multicolumn{19}{l}{\small is equivalent to FSat: the percentage of items in the dataset that are recommended at least $\left \lfloor \frac{km}{n}\right \rfloor=1$ time.}
\end{tabular}}
\end{table*}

\begin{table*}[p]
\centering
\caption{The gradient values of the PF, based on the extreme points (starting and ending points). We consider a gradient to be `good' if it is not zero or undefined (-).}\label{tab:app-gradient} 
\resizebox{\linewidth}{!}{
\begin{tabular}{lrrrrrrrl}
\toprule
{} & Lastfm & Amazon-lb & QK-video & Jester &  ML-10M &  ML-20M &  \# good &    conclusion \\
\midrule
HR-Ent    &   -97.57 &      -1.86 &     -0.31 &      - &  -14.74 &   -6.95 &        5 &  inconsistent \\
HR-FSat   & -1439.17 &     -19.92 &      0.00 &      - &  -30.48 &  -18.97 &        4 &  inconsistent \\
HR-Gini   &   561.63 &       6.23 &      3.71 &      - &  117.19 &   43.44 &        5 &  inconsistent \\
HR-Jain   &  -979.86 &     -18.77 &     -5.80 &      - & -157.73 &  -78.22 &        5 &  inconsistent \\
HR-QF     &     0.00 &       0.00 &      0.00 &      - &  -30.48 &  -18.97 &        2 &  inconsistent \\
MAP-Ent   &    -0.17 &      -0.17 &     -0.03 &  -0.07 &   -0.14 &   -0.18 &        6 &   always good \\
MAP-FSat  &    -2.46 &      -1.81 &      0.00 & -44.47 &   -0.29 &   -0.48 &        5 &  inconsistent \\
MAP-Gini  &     0.96 &       0.56 &      0.34 &   1.42 &    1.12 &    1.10 &        6 &   always good \\
MAP-Jain  &    -1.68 &      -1.70 &     -0.54 &  -0.37 &   -1.51 &   -1.98 &        6 &   always good \\
MAP-QF    &     0.00 &       0.00 &      0.00 &    0.0 &   -0.29 &   -0.48 &        2 &  inconsistent \\
MRR-Ent   &   -97.57 &      -1.86 &     -0.31 &      - &  -14.74 &   -6.95 &        5 &  inconsistent \\
MRR-FSat  & -1439.17 &     -19.92 &      0.00 &      - &  -30.48 &  -18.97 &        4 &  inconsistent \\
MRR-Gini  &   561.63 &       6.23 &      3.71 &      - &  117.19 &   43.44 &        5 &  inconsistent \\
MRR-Jain  &  -979.86 &     -18.77 &     -5.80 &      - & -157.73 &  -78.22 &        5 &  inconsistent \\
MRR-QF    &     0.00 &       0.00 &      0.00 &      - &  -30.48 &  -18.97 &        2 &  inconsistent \\
NDCG-Ent  &    -0.24 &      -0.22 &     -0.04 &   -0.1 &   -0.20 &   -0.25 &        6 &   always good \\
NDCG-FSat &    -3.50 &      -2.32 &      0.00 & -68.56 &   -0.42 &   -0.68 &        5 &  inconsistent \\
NDCG-Gini &     1.37 &       0.73 &      0.47 &    2.2 &    1.62 &    1.55 &        6 &   always good \\
NDCG-Jain &    -2.38 &      -2.19 &     -0.73 &  -0.57 &   -2.18 &   -2.79 &        6 &   always good \\
NDCG-QF   &     0.00 &       0.00 &      0.00 &    0.0 &   -0.42 &   -0.68 &        2 &  inconsistent \\
P-Ent     &    -0.20 &      -0.33 &     -0.07 &  -0.08 &   -0.16 &   -0.20 &        6 &   always good \\
P-FSat    &    -2.95 &      -3.53 &      0.00 & -51.41 &   -0.33 &   -0.55 &        5 &  inconsistent \\
P-Gini    &     1.15 &       1.10 &      0.89 &   1.65 &    1.26 &    1.26 &        6 &   always good \\
P-Jain    &    -2.01 &      -3.33 &     -1.40 &  -0.43 &   -1.70 &   -2.27 &        6 &   always good \\
P-QF      &     0.00 &       0.00 &      0.00 &    0.0 &   -0.33 &   -0.55 &        2 &  inconsistent \\
R-Ent     &    -0.17 &      -0.17 &     -0.03 &  -0.07 &   -0.26 &   -0.30 &        6 &   always good \\
R-FSat    &    -2.57 &      -1.83 &      0.00 & -48.04 &   -0.53 &   -0.82 &        5 &  inconsistent \\
R-Gini    &     1.00 &       0.57 &      0.34 &   1.54 &    2.04 &    1.88 &        6 &   always good \\
R-Jain    &    -1.75 &      -1.73 &     -0.54 &   -0.4 &   -2.75 &   -3.39 &        6 &   always good \\
R-QF      &     0.00 &       0.00 &      0.00 &    0.0 &   -0.53 &   -0.82 &        2 &  inconsistent \\
\bottomrule
\end{tabular}}
\end{table*}

\begin{table*}[p]
\centering
\caption{$\downarrow$DPFR scores at $k=10$ of the recommender models for Lastfm, Amazon-lb, and QK-video, without and with re-ranking the the top $k'=25$ items using Borda Count (BC), COMBMNZ (CM), and Greedy Substitution (GS). The best score per measure pair is in bold.}
\label{tab:dpfr-scores}
\resizebox{\textwidth}{!}{
\begin{tabular}{ll*{4}{r}|*{4}{r}|*{4}{r}|*{4}{r}}
\toprule
 &  model & \multicolumn{4}{c|}{ItemKNN} & \multicolumn{4}{c|}{BPR} & \multicolumn{4}{c|}{MultiVAE} & \multicolumn{4}{c}{NCL} \\ 
\midrule
 &  reranking & - & BC & CM & GS & - & BC & CM & GS & - & BC & CM & GS & - & BC & CM & GS \\
\midrule
\multirow[c]{12}{*}{\rotatebox[origin=r]{90}{Lastfm}} & P-Jain & 0.853 & 0.819 & 0.861 & 0.853 & 0.837 & 0.784 & 0.824 & 0.834 & 0.805 & \bfseries 0.728 & 0.776 & 0.799 & 0.813 & 0.735 & 0.773 & 0.809 \\
 & P-Ent & 0.585 & 0.524 & 0.571 & 0.567 & 0.567 & 0.525 & 0.566 & 0.551 & 0.510 & 0.501 & 0.550 & 0.505 & 0.524 & \bfseries 0.497 & 0.544 & 0.513 \\
 & P-Gini & 0.869 & 0.802 & 0.821 & 0.850 & 0.871 & 0.820 & 0.841 & 0.856 & 0.811 & \bfseries 0.737 & 0.759 & 0.789 & 0.835 & 0.756 & 0.775 & 0.813 \\
 & MAP-Jain & 1.044 & 1.042 & 1.072 & 1.042 & 1.029 & 1.015 & 1.040 & 1.026 & 1.005 & \bfseries 0.974 & 1.004 & 0.998 & 1.008 & 0.979 & 1.003 & 1.003 \\
 & MAP-Ent & 0.811 & 0.802 & 0.830 & 0.797 & 0.797 & 0.803 & 0.824 & 0.784 & 0.759 & 0.791 & 0.815 & \bfseries 0.753 & 0.764 & 0.787 & 0.813 & 0.755 \\
 & MAP-Gini & 1.038 & 1.009 & 1.022 & 1.021 & 1.039 & 1.025 & 1.036 & 1.025 & 0.991 & \bfseries 0.961 & 0.972 & 0.970 & 1.007 & 0.975 & 0.986 & 0.987 \\
 & R-Jain & 0.968 & 0.947 & 1.004 & 0.969 & 0.952 & 0.917 & 0.970 & 0.954 & 0.923 & \bfseries 0.875 & 0.936 & 0.923 & 0.927 & \bfseries 0.875 & 0.930 & 0.928 \\
 & R-Ent & 0.720 & 0.683 & 0.747 & 0.708 & 0.703 & 0.686 & 0.741 & 0.696 & \bfseries 0.656 & 0.675 & 0.738 & 0.660 & 0.664 & 0.665 & 0.729 & 0.661 \\
 & R-Gini & 0.968 & 0.918 & 0.955 & 0.953 & 0.969 & 0.935 & 0.971 & 0.959 & 0.914 & \bfseries 0.868 & 0.907 & 0.900 & 0.932 & 0.878 & 0.918 & 0.917 \\
 & NDCG-Jain & 0.989 & 0.995 & 1.046 & 0.989 & 0.973 & 0.968 & 1.013 & 0.972 & 0.948 & 0.933 & 0.985 & 0.945 & 0.949 & \bfseries 0.932 & 0.978 & 0.947 \\
 & NDCG-Ent & 0.760 & 0.758 & 0.813 & 0.747 & 0.743 & 0.762 & 0.805 & 0.733 & 0.703 & 0.756 & 0.807 & 0.702 & 0.706 & 0.746 & 0.797 & \bfseries 0.700 \\
 & NDCG-Gini & 1.000 & 0.976 & 1.009 & 0.984 & 1.000 & 0.994 & 1.023 & 0.987 & 0.949 & 0.934 & 0.966 & \bfseries 0.932 & 0.964 & 0.943 & 0.974 & 0.947 \\
\cline{1-18}
\multirow[c]{12}{*}{\rotatebox[origin=r]{90}{Amazon-lb}} & P-Jain & 0.529 & \bfseries 0.352 & 0.415 & 0.489 & 0.571 & 0.415 & 0.464 & 0.542 & 0.733 & 0.656 & 0.679 & 0.726 & 0.742 & 0.678 & 0.694 & 0.736 \\
 & P-Ent & 0.340 & \bfseries 0.307 & 0.324 & 0.327 & 0.376 & 0.325 & 0.339 & 0.358 & 0.636 & 0.493 & 0.517 & 0.590 & 0.678 & 0.515 & 0.537 & 0.629 \\
 & P-Gini & 0.633 & \bfseries 0.486 & 0.575 & 0.587 & 0.708 & 0.578 & 0.630 & 0.669 & 0.896 & 0.834 & 0.850 & 0.878 & 0.906 & 0.849 & 0.860 & 0.890 \\
 & MAP-Jain & 0.998 & \bfseries 0.904 & 0.936 & 0.974 & 1.025 & 0.936 & 0.962 & 1.007 & 1.129 & 1.078 & 1.093 & 1.123 & 1.134 & 1.093 & 1.105 & 1.130 \\
 & MAP-Ent & 0.830 & \bfseries 0.818 & 0.825 & 0.825 & 0.848 & 0.825 & 0.830 & 0.840 & 0.990 & 0.906 & 0.919 & 0.961 & 1.017 & 0.919 & 0.933 & 0.985 \\
 & MAP-Gini & 0.993 & \bfseries 0.906 & 0.958 & 0.963 & 1.045 & 0.959 & 0.991 & 1.018 & 1.180 & 1.135 & 1.147 & 1.166 & 1.187 & 1.146 & 1.156 & 1.175 \\
 & R-Jain & 0.984 & \bfseries 0.893 & 0.927 & 0.960 & 1.015 & 0.924 & 0.948 & 0.997 & 1.119 & 1.069 & 1.082 & 1.112 & 1.123 & 1.082 & 1.096 & 1.119 \\
 & R-Ent & 0.817 & \bfseries 0.809 & 0.818 & 0.811 & 0.839 & 0.815 & 0.818 & 0.831 & 0.981 & 0.898 & 0.909 & 0.950 & 1.007 & 0.909 & 0.925 & 0.975 \\
 & R-Gini & 0.981 & \bfseries 0.897 & 0.951 & 0.951 & 1.037 & 0.949 & 0.980 & 1.010 & 1.172 & 1.127 & 1.138 & 1.157 & 1.178 & 1.137 & 1.149 & 1.166 \\
 & NDCG-Jain & 1.022 & \bfseries 0.937 & 0.966 & 1.000 & 1.051 & 0.967 & 0.990 & 1.034 & 1.148 & 1.102 & 1.116 & 1.142 & 1.153 & 1.116 & 1.127 & 1.149 \\
 & NDCG-Ent & 0.868 & \bfseries 0.859 & 0.866 & 0.863 & 0.889 & 0.867 & 0.871 & 0.881 & 1.023 & 0.945 & 0.957 & 0.994 & 1.049 & 0.956 & 0.970 & 1.018 \\
 & NDCG-Gini & 1.028 & \bfseries 0.947 & 0.997 & 1.000 & 1.083 & 0.999 & 1.029 & 1.056 & 1.211 & 1.170 & 1.181 & 1.197 & 1.218 & 1.180 & 1.190 & 1.206 \\
\cline{1-18}
\multirow[c]{12}{*}{\rotatebox[origin=r]{90}{QK-video}} & P-Jain & 0.563 & \bfseries 0.297 & 0.468 & 0.487 & 0.941 & 0.700 & 0.658 & 0.922 & 1.007 & 0.983 & 0.989 & 1.005 & 0.999 & 0.945 & 0.950 & 0.996 \\
 & P-Ent & 0.245 & \bfseries 0.236 & 0.243 & 0.241 & 0.335 & 0.258 & 0.254 & 0.310 & 0.623 & 0.499 & 0.510 & 0.588 & 0.542 & 0.403 & 0.400 & 0.504 \\
 & P-Gini & 0.521 & \bfseries 0.327 & 0.495 & 0.456 & 0.833 & 0.649 & 0.611 & 0.789 & 1.002 & 0.978 & 0.981 & 0.996 & 0.985 & 0.931 & 0.925 & 0.972 \\
 & MAP-Jain & 1.106 & \bfseries 0.996 & 1.061 & 1.070 & 1.332 & 1.180 & 1.157 & 1.318 & 1.379 & 1.365 & 1.371 & 1.378 & 1.371 & 1.338 & 1.343 & 1.369 \\
 & MAP-Ent & 0.979 & \bfseries 0.977 & 0.979 & 0.978 & 0.995 & 0.980 & 0.980 & 0.987 & 1.125 & 1.066 & 1.073 & 1.106 & 1.079 & 1.024 & 1.025 & 1.061 \\
 & MAP-Gini & 1.082 & \bfseries 1.003 & 1.070 & 1.052 & 1.254 & 1.147 & 1.127 & 1.225 & 1.372 & 1.358 & 1.362 & 1.367 & 1.357 & 1.324 & 1.322 & 1.347 \\
 & R-Jain & 1.097 & \bfseries 0.982 & 1.047 & 1.060 & 1.312 & 1.162 & 1.145 & 1.301 & 1.355 & 1.345 & 1.357 & 1.356 & 1.343 & 1.314 & 1.326 & 1.342 \\
 & R-Ent & 0.969 & 0.962 & 0.964 & 0.968 & 0.969 & \bfseries 0.959 & 0.966 & 0.964 & 1.096 & 1.040 & 1.056 & 1.079 & 1.043 & 0.993 & 1.002 & 1.027 \\
 & R-Gini & 1.073 & \bfseries 0.989 & 1.056 & 1.043 & 1.233 & 1.129 & 1.114 & 1.207 & 1.348 & 1.338 & 1.348 & 1.346 & 1.329 & 1.300 & 1.304 & 1.321 \\
 & NDCG-Jain & 1.106 & \bfseries 0.996 & 1.060 & 1.070 & 1.326 & 1.178 & 1.156 & 1.314 & 1.373 & 1.361 & 1.369 & 1.372 & 1.362 & 1.332 & 1.340 & 1.361 \\
 & NDCG-Ent & 0.980 & \bfseries 0.976 & 0.978 & 0.979 & 0.988 & 0.977 & 0.980 & 0.982 & 1.118 & 1.061 & 1.071 & 1.099 & 1.069 & 1.017 & 1.021 & 1.051 \\
 & NDCG-Gini & 1.083 & \bfseries 1.002 & 1.069 & 1.053 & 1.248 & 1.145 & 1.127 & 1.221 & 1.366 & 1.354 & 1.360 & 1.362 & 1.349 & 1.319 & 1.318 & 1.340 \\
\bottomrule

\end{tabular}}
\end{table*}

\begin{table*}[p]
\centering
\caption{$\downarrow$DPFR scores at $k=10$ of the recommender models for Jester and ML-*, without and with re-ranking the the top $k'=25$ items using Borda Count (BC), COMBMNZ (CM), and Greedy Substitution (GS). The best score per measure pair is in bold.}
\label{tab:dpfr-scores2}
\resizebox{\textwidth}{!}{
\begin{tabular}{ll*{4}{r}|*{4}{r}|*{4}{r}|*{4}{r}}
\toprule
 &  model & \multicolumn{4}{c|}{ItemKNN} & \multicolumn{4}{c|}{BPR} & \multicolumn{4}{c|}{MultiVAE} & \multicolumn{4}{c}{NCL} \\ 
\midrule
 &  reranking & - & BC & CM & GS & - & BC & CM & GS & - & BC & CM & GS & - & BC & CM & GS \\
\midrule
\multirow[c]{12}{*}{\rotatebox[origin=r]{90}{Jester}} & P-Jain & 0.709 & 0.566 & 0.719 & 0.707 & 0.679 & \bfseries 0.550 & 0.631 & 0.677 & 0.747 & 0.569 & 0.638 & 0.745 & 0.698 & 0.604 & 0.625 & 0.696 \\
 & P-Ent & 0.401 & 0.382 & 0.507 & 0.400 & 0.367 & 0.381 & 0.462 & \bfseries 0.366 & 0.433 & 0.372 & 0.440 & 0.430 & 0.381 & 0.381 & 0.442 & 0.380 \\
 & P-Gini & 0.724 & 0.602 & 0.739 & 0.722 & 0.675 & \bfseries 0.578 & 0.651 & 0.673 & 0.766 & 0.598 & 0.651 & 0.763 & 0.704 & 0.619 & 0.642 & 0.701 \\
 & MAP-Jain & 0.915 & 0.908 & 1.048 & 0.914 & 0.894 & 0.897 & 0.987 & \bfseries 0.892 & 0.932 & 0.905 & 0.976 & 0.930 & 0.898 & 0.923 & 0.974 & 0.897 \\
 & MAP-Ent & 0.704 & 0.805 & 0.914 & 0.703 & 0.687 & 0.804 & 0.889 & 0.686 & 0.704 & 0.795 & 0.859 & 0.703 & 0.681 & 0.794 & 0.868 & \bfseries 0.680 \\
 & MAP-Gini & 0.927 & 0.930 & 1.062 & 0.926 & 0.891 & 0.915 & 1.001 & \bfseries 0.889 & 0.947 & 0.924 & 0.985 & 0.945 & 0.903 & 0.933 & 0.986 & 0.901 \\
 & R-Jain & 0.774 & 0.704 & 0.927 & 0.773 & 0.748 & \bfseries 0.700 & 0.821 & 0.746 & 0.802 & 0.702 & 0.787 & 0.800 & 0.759 & 0.732 & 0.787 & 0.758 \\
 & R-Ent & 0.507 & 0.566 & 0.773 & 0.506 & 0.483 & 0.576 & 0.699 & \bfseries 0.482 & 0.521 & 0.555 & 0.636 & 0.519 & 0.484 & 0.563 & 0.651 & 0.483 \\
 & R-Gini & 0.788 & 0.733 & 0.943 & 0.787 & 0.745 & \bfseries 0.723 & 0.837 & 0.743 & 0.819 & 0.727 & 0.798 & 0.817 & 0.765 & 0.745 & 0.801 & 0.763 \\
 & NDCG-Jain & 0.823 & 0.793 & 0.977 & 0.822 & 0.798 & \bfseries 0.782 & 0.900 & 0.797 & 0.845 & 0.788 & 0.878 & 0.844 & 0.807 & 0.810 & 0.878 & 0.805 \\
 & NDCG-Ent & 0.579 & 0.673 & 0.833 & 0.578 & 0.558 & 0.673 & 0.791 & 0.557 & 0.586 & 0.660 & 0.745 & 0.584 & 0.556 & 0.661 & 0.758 & \bfseries 0.555 \\
 & NDCG-Gini & 0.837 & 0.818 & 0.992 & 0.835 & 0.795 & 0.802 & 0.914 & \bfseries 0.793 & 0.862 & 0.810 & 0.887 & 0.860 & 0.812 & 0.821 & 0.890 & 0.810 \\
\cline{1-18}
\multirow[c]{12}{*}{\rotatebox[origin=r]{90}{ML-10M}} & P-Jain & 1.072 & 1.067 & 1.082 & 1.075 & 1.047 & \bfseries 1.023 & 1.031 & 1.051 & 1.098 & 1.094 & 1.104 & 1.099 & 1.052 & 1.044 & 1.059 & 1.053 \\
 & P-Ent & 0.884 & 0.831 & 0.844 & 0.880 & 0.765 & \bfseries 0.749 & 0.766 & 0.764 & 1.023 & 0.974 & 0.970 & 1.010 & 0.801 & 0.771 & 0.791 & 0.791 \\
 & P-Gini & 1.080 & 1.075 & 1.088 & 1.081 & 1.042 & \bfseries 1.025 & 1.032 & 1.041 & 1.106 & 1.102 & 1.111 & 1.107 & 1.055 & 1.050 & 1.065 & 1.053 \\
 & MAP-Jain & 1.172 & 1.172 & 1.182 & 1.173 & 1.149 & \bfseries 1.134 & 1.135 & 1.150 & 1.193 & 1.192 & 1.201 & 1.194 & 1.155 & 1.153 & 1.161 & 1.154 \\
 & MAP-Ent & 0.995 & 0.955 & 0.962 & 0.991 & 0.893 & \bfseries 0.886 & 0.893 & 0.888 & 1.119 & 1.076 & 1.073 & 1.106 & 0.924 & 0.906 & 0.915 & 0.914 \\
 & MAP-Gini & 1.173 & 1.173 & 1.182 & 1.174 & 1.139 & \bfseries 1.130 & \bfseries 1.130 & 1.135 & 1.195 & 1.193 & 1.202 & 1.195 & 1.152 & 1.153 & 1.160 & 1.149 \\
 & R-Jain & 0.860 & 0.846 & 0.847 & 0.859 & 0.835 & 0.780 & \bfseries 0.768 & 0.829 & 0.868 & 0.866 & 0.867 & 0.867 & 0.847 & 0.813 & 0.810 & 0.844 \\
 & R-Ent & 0.653 & 0.570 & 0.563 & 0.643 & 0.492 & 0.423 & \bfseries 0.415 & 0.472 & 0.807 & 0.748 & 0.730 & 0.788 & 0.554 & 0.473 & 0.468 & 0.534 \\
 & R-Gini & 0.889 & 0.877 & 0.876 & 0.888 & 0.850 & 0.805 & \bfseries 0.793 & 0.838 & 0.900 & 0.897 & 0.898 & 0.899 & 0.872 & 0.843 & 0.840 & 0.866 \\
 & NDCG-Jain & 1.140 & 1.142 & 1.156 & 1.142 & 1.115 & \bfseries 1.107 & 1.114 & 1.118 & 1.168 & 1.166 & 1.180 & 1.169 & 1.119 & 1.125 & 1.138 & 1.119 \\
 & NDCG-Ent & 0.972 & 0.932 & 0.944 & 0.968 & 0.863 & 0.866 & 0.881 & \bfseries 0.860 & 1.103 & 1.060 & 1.062 & 1.091 & 0.894 & 0.884 & 0.901 & 0.885 \\
 & NDCG-Gini & 1.151 & 1.153 & 1.166 & 1.153 & 1.114 & \bfseries 1.112 & 1.119 & \bfseries 1.112 & 1.180 & 1.177 & 1.190 & 1.180 & 1.127 & 1.134 & 1.147 & 1.124 \\
\cline{1-18}
\multirow[c]{12}{*}{\rotatebox[origin=r]{90}{ML-20M}} & P-Jain & 1.014 & 1.015 & 1.033 & 1.016 & 0.998 & \bfseries 0.992 & 1.000 & 1.001 & 1.000 & 0.993 & 1.006 & 1.006 & 1.002 & 1.004 & 1.020 & 1.006 \\
 & P-Ent & 0.885 & 0.842 & 0.858 & 0.880 & 0.776 & \bfseries 0.761 & 0.773 & 0.771 & 0.775 & 0.765 & 0.780 & 0.774 & 0.807 & 0.783 & 0.800 & 0.801 \\
 & P-Gini & 1.056 & 1.057 & 1.073 & 1.057 & 1.031 & \bfseries 1.026 & 1.033 & 1.030 & 1.031 & 1.029 & 1.039 & 1.033 & 1.040 & 1.042 & 1.057 & 1.042 \\
 & MAP-Jain & 1.118 & 1.124 & 1.137 & 1.119 & \bfseries 1.103 & \bfseries 1.103 & 1.104 & \bfseries 1.103 & 1.107 & 1.105 & 1.111 & 1.109 & 1.106 & 1.114 & 1.123 & 1.108 \\
 & MAP-Ent & 0.996 & 0.963 & 0.973 & 0.991 & 0.900 & 0.894 & 0.898 & \bfseries 0.892 & 0.902 & 0.897 & 0.905 & 0.897 & 0.926 & 0.913 & 0.921 & 0.919 \\
 & MAP-Gini & 1.151 & 1.156 & 1.167 & 1.151 & 1.127 & 1.129 & 1.129 & \bfseries 1.124 & 1.130 & 1.131 & 1.136 & 1.129 & 1.135 & 1.143 & 1.151 & 1.135 \\
 & R-Jain & 0.775 & 0.768 & 0.769 & 0.775 & 0.757 & 0.723 & \bfseries 0.712 & 0.753 & 0.756 & 0.721 & 0.718 & 0.752 & 0.766 & 0.744 & 0.741 & 0.764 \\
 & R-Ent & 0.645 & 0.573 & 0.568 & 0.634 & 0.490 & 0.426 & \bfseries 0.413 & 0.471 & 0.483 & 0.426 & 0.421 & 0.465 & 0.542 & 0.471 & 0.462 & 0.523 \\
 & R-Gini & 0.836 & 0.829 & 0.829 & 0.836 & 0.808 & 0.778 & \bfseries 0.768 & 0.800 & 0.805 & 0.779 & 0.774 & 0.797 & 0.822 & 0.803 & 0.799 & 0.818 \\
 & NDCG-Jain & 1.081 & 1.089 & 1.107 & 1.083 & \bfseries 1.065 & 1.072 & 1.078 & 1.066 & 1.072 & 1.075 & 1.086 & 1.075 & 1.068 & 1.083 & 1.097 & 1.071 \\
 & NDCG-Ent & 0.971 & 0.939 & 0.955 & 0.966 & 0.871 & 0.873 & 0.883 & \bfseries 0.865 & 0.876 & 0.878 & 0.891 & 0.873 & 0.898 & 0.892 & 0.906 & 0.892 \\
 & NDCG-Gini & 1.121 & 1.129 & 1.144 & 1.122 & 1.097 & 1.105 & 1.110 & \bfseries 1.095 & 1.102 & 1.109 & 1.118 & 1.102 & 1.105 & 1.119 & 1.132 & 1.105 \\
\bottomrule

\end{tabular}}
\end{table*}

\begin{table*}[p]
    \centering
    \caption{Range of agreement $\tau$ between estimated PF and PF across 12 measure pairs, using the estimated PF with 3--12 points.}
    \label{tab:corr_est}
    
    \resizebox{.95\textwidth}{!}{
    \begin{tabular}{lrrrrrr}
    \toprule
    {\#pts} &      Lastfm &   Amazon-lb &    QK-video &      Jester &      ML-10M &      ML-20M \\
    \midrule
    3  &  0.78--1.00 &  0.98--1.00 &  1.00--1.00 &  1.00--1.00 &  0.97--1.00 &  0.75--1.00 \\
    4  &  0.88--1.00 &  0.98--1.00 &  0.98--1.00 &  0.98--1.00 &  0.98--1.00 &  0.93--1.00 \\
    5  &  0.78--1.00 &  0.98--1.00 &  1.00--1.00 &  1.00--1.00 &  0.97--1.00 &  0.92--1.00 \\
    6  &  0.90--1.00 &  0.97--1.00 &  1.00--1.00 &  0.98--1.00 &  0.95--1.00 &  0.92--1.00 \\
    7  &  0.88--1.00 &  1.00--1.00 &  1.00--1.00 &  1.00--1.00 &  0.98--1.00 &  0.93--1.00 \\
    8  &  0.90--1.00 &  0.98--1.00 &  1.00--1.00 &  0.98--1.00 &  1.00--1.00 &  0.95--1.00 \\
    9  &  0.98--1.00 &  1.00--1.00 &  1.00--1.00 &  1.00--1.00 &  0.97--1.00 &  0.98--1.00 \\
    10 &  0.88--1.00 &  1.00--1.00 &  1.00--1.00 &  0.98--1.00 &  1.00--1.00 &  0.95--1.00 \\
    11 &  0.92--1.00 &  1.00--1.00 &  1.00--1.00 &  1.00--1.00 &  0.98--1.00 &  0.97--1.00 \\
    12 &  0.95--1.00 &  1.00--1.00 &  1.00--1.00 &  0.98--1.00 &  0.98--1.00 &  0.97--1.00 \\
    \bottomrule
\end{tabular}
}
\end{table*}

\chapter{Measuring Individual User Fairness with User Similarity and Effectiveness Disparity}
\label{chap:PUF}

\section*{Abstract}
Individual user fairness is commonly understood as treating similar users similarly. In Recommender Systems (RSs), several evaluation measures exist for quantifying individual user fairness. 
These measures evaluate fairness via either: (i) the disparity in user effectiveness scores regardless of user similarity, or (ii) the disparity in items recommended to similar users regardless of item relevance. Both disparity in recommendation effectiveness and user similarity are very important in fairness, yet no existing individual user fairness measure simultaneously accounts for both of them. 
In brief, current user fairness evaluation measures implement a largely incomplete definition of fairness. 
To fill this gap, we present Pairwise User unFairness (PUF), a novel evaluation measure of individual user fairness that considers both effectiveness disparity and user similarity. 
Our measure works as follows. If two users get recommendations of similar NDCG, for instance, PUF will give a different fairness score depending on how similar the users are to each other, in terms of their past interactions: the case where similar users get recommendations of similar NDCG is fairer than the case where similar users get recommendations of different NDCG. PUF is the only measure that can express this important distinction. We empirically validate that PUF does this consistently across 4 datasets and 7 rankers, and robustly when varying user similarity or effectiveness. 
In contrast, all other measures are either almost insensitive to effectiveness disparity or completely insensitive to user similarity. We contribute the first RS evaluation measure to reliably capture both user similarity and effectiveness in individual user fairness.

\section{Introduction}
\label{PUF_puf_s:intro}

The evaluation of Recommender Systems (RSs) has always relied heavily on effectiveness as it directly affects user utility and satisfaction. Unsurprisingly, existing work on RS fairness evaluation often uses measures that depend on effectiveness scores, especially when it comes to individual user fairness. \emph{Individual fairness} is traditionally defined as treating similar individuals in a similar manner \cite{Dwork2012FairnessAwareness}. 
We focus on attribute-free fairness for individual users, where we assume 
no user attribute is available other than the user identifiers and interactions \cite{Li2024ExplainingPerspective,Zeng2024FairDemographics}. Sensitive attributes (e.g., race, age) are frequently unavailable due to privacy or data incompleteness. 

To measure individual user fairness in RSs, the disparity in recommendation effectiveness across users is often used as a proxy of how similar the recommendation algorithm treats the users. Recommendations across users are deemed fair if similar users have similar effectiveness scores. Yet, no existing individual user fairness measure considers both effectiveness and user similarity (\Cref{PUF_tab:checklist}). 

User similarity and recommendation effectiveness are both important, as the former affects our expectation of how close effectiveness scores should be. 
For example, given two users whose past interactions are highly similar, the recommendations they receive are deemed fair if their effectiveness is similar, and unfair if the effectiveness differs a lot. This is because fairness means similar users should get similar treatment. However, two dissimilar users cannot expect to get similar treatment, as their past interactions may differ, e.g., in terms of amount, frequency, or item type.

Current individual user fairness measures can be grouped into those that consider: (i) only effectiveness disparity; and (ii) only user similarity and recommendation similarity. 
For (i), different effectiveness for users is not always unfair. Some users may have only a few past interactions, and others may have very specific tastes for whom only a few items are relevant. Fairness should consider that these users are different and cannot be compared to users who, e.g., consume only popular items.
For (ii), recommending similar items to similar users may not be fair if one user is satisfied with their recommendation, but another is not. In this case, the recommendation is not really fair as its effectiveness differs.

\begin{table}[t]
    \caption{Overview of individual user fairness measures in RS.}
    \label{PUF_tab:checklist}
    \centering
    \resizebox{0.95\columnwidth}{!}{
    \begin{tabular}{llcc}
         \toprule
         Measure & Reference & Effectiveness & User Similarity \\
         \midrule
         Standard deviation & \cite{Patro2020FairRec:Platforms,Biswas2021TowardPlatforms,Wu2021TFROM:Providers,Rastegarpanah2019FightingSystems,Li2024ExplainingPerspective}   & $\checkmark$ & -\\
         Gini index & \cite{Fu2020Fairness-AwareGraphs, Leonhardt2018UserSystems} & $\checkmark$ & - \\
         Envy-based &\cite{Patro2020FairRec:Platforms, Biswas2021TowardPlatforms,Do2022OnlineSystems}& $\checkmark$ & -\\
         UF & \cite{Wu2023EquippingEmbedding} &- & $\checkmark$  \\
         \midrule
         PUF & ours & $\checkmark$ & $\checkmark$ \\
         \bottomrule
    \end{tabular}}
\end{table}
To counter the above limitations in current measures, we propose a novel evaluation measure for individual user fairness in RS: Pairwise User unFairness (PUF). PUF quantifies individual user fairness based on the disparity in recommendation relevance between user pairs, weighted by the similarity of user pairs. As such, PUF is aligned with the definition of individual fairness and also accounts for recommendation effectiveness, and thus, user utility. Our experiments show that PUF is superior to the current measures in terms of sensitivity to changes in effectiveness scores and user similarity distribution. Overall, we contribute a new evaluation measure for individual user fairness, which considers both user similarity and disparity in recommendation effectiveness, and which does not have the same limitations as existing measures.

\section{Individual User Fairness}
\label{PUF_s:individual-user-fair}
We present the definition of individual user fairness (\Cref{PUF_ss:definition}) and existing evaluation measures of individual user fairness (\Cref{PUF_ss:fair-measures}). 

\subsection{Definition of Individual User Fairness}
\label{PUF_ss:definition}

We define individual user fairness in RSs as per \cite{Dwork2012FairnessAwareness}: let $u$ and $u'$ be the representations of two users; $L_u$ and $L_{u'}$ be the recommendation lists of these two users; and $M(\cdot)$ be a function that maps a recommendation list to a score, e.g., its effectiveness score.  
Any two users whose profile distance is $d(u,u')$ should receive recommendations such that recommendation effectiveness satisfies $D(M(L_u), M(L_{u'})) \leq d(u,u')$, where $D$ is a distance measure. 
In other words, fairness is achieved when the difference in the users' recommendation effectiveness is at most $d(u,u')$. This definition agrees with the definitions of RS individual user fairness  in \cite{Zehlike2022FairnessSystems, Smith2023ScopingPerspective}.

\subsection{Current Individual User Fairness Measures}
\label{PUF_ss:fair-measures}

We present existing evaluation measures of attribute-free individual user fairness in RSs (\Cref{PUF_tab:checklist}). To our knowledge, these are all such measures for RSs published up to January 2025 \cite{Wang2023ASystems,Amigo2023ASystems,Smith2023ScopingPerspective,LiYunqi2023FairnessApplications,Zhao2025FairnessDiversitySurvey,Wu2023FairnessStrategies,Aalam2022EvaluationReview,Zehlike2022FairnessSystems,Pitoura2022FairnessOverview}. 
All these measures quantify unfairness, so that the lower their scores, the fairer (denoted by $\downarrow$). 
Some measures are based on general-purpose inequality measures, i.e., standard deviation (\Cref{PUF_sss:sd}) and Gini Index (\Cref{PUF_sss:gini}); others are based on unique fairness concepts, i.e., envy-freeness  (\Cref{PUF_sss:envy}) and distance-based (\Cref{PUF_sss:dist}). All existing measures except the distance-based measure quantify effectiveness disparity but ignore user similarity.  
While the existing distance-based measure considers user similarity, it is detached from effectiveness. Hence, no existing individual user fairness measure in RS considers both user similarity and effectiveness disparity.

\subsubsection{Standard deviation (SD)} 
\label{PUF_sss:sd} 
$\downarrow$SD measures how dispersed the data is in relation to the mean. In RSs, SD and variance are often used to quantify individual user fairness. 
RS is fair if it provides equal prediction accuracy to all users, and fairness is evaluated via the variance of the mean squared error (MSE) of user rating prediction \cite{Rastegarpanah2019FightingSystems,Li2024ExplainingPerspective}. Other works measure fairness from user recommendation lists, e.g., via the variance of NDCG scores across users \cite{Wu2021TFROM:Providers} or the SD of user utility \cite{Patro2020FairRec:Platforms,Biswas2021TowardPlatforms}. The utility of user $u$, when given the top $k$ recommendation list of user $u'$, $L_{u'}$ is originally defined as follows: 
\begin{equation}
\label{PUF_eq:phi-ori}
    \phi^{ori}_u (L_{u'}) = \frac{
    \sum_{i \in L_{u'}} \hat{r}_{u,i}
    }{\sum_{i \in R^k_{u'}} \hat{r}_{u,i}} 
\end{equation}
\noindent where $\hat{r}_{u,i}$ is the predicted relevance score of user $u$ for item $i$, and $R_u^k$ is the top-$k$ most relevant items of $u$ based on the ground truth. Two issues arise from the above formulation: (i) computing user utility based on predicted relevance $\hat{r}_{u,i}$ may not reflect the true recommendation effectiveness; and (ii) relevance ties for items in $R_u^k$ are not handled, leading to multiple possible compositions of $R_u^k$. Further, in \cite{Patro2020FairRec:Platforms,Biswas2021TowardPlatforms}, the predicted relevance score is also taken from a latent factorisation model, which means that recommending items based on decreasing predicted relevance (as commonly done), will always result in $\phi^{ori}_u (L_{u})=1$ for all users, rendering the user utility measure pointless. We resolve the above issues in the utility function, by replacing predicted relevance with ground truth relevance $r_{u,i}$, (e.g., ratings in the test set) as follows:
\begin{equation}
\label{PUF_eq:phi-our}
    \phi_u (L_{u'}) = 
    \frac{
    \sum_{i \in L_{u'}} r_{u,i}
    }{\sum_{i \in R^k_{u'}} r_{u,i}}
\end{equation}
For binary relevance, $r_{u,i} \in \{0,1\}$, assuming that each user has at least $k$ relevant items, $\phi_u (L_{u})=\frac{1}{k} \sum_{i \in L_{u}} 1_{\{i \in R_u\}}$ or simply Precision@$k$ (P@$k$). 
The indicator function $1_{\{\cdot\}}$ returns 1 if the expression $\cdot$ is true, and 0 otherwise. We denote the set of user $u$'s relevant items as $R_u$. As such, the  SD based on the reformulated utility function is SD-P $= SD_{u \in U} \phi_u (L_{u})$, where $U$ is the set of users.

\subsubsection{Gini Index (Gini)}
\label{PUF_sss:gini}
$\downarrow$Gini is a well-known inequality measure that quantifies the extent to which a distribution deviates from a perfectly equal distribution. It has been used to measure individual user fairness from different distributions, e.g., P@$k$ (Gini-P), NDCG (Gini-NDCG), F1, or the utility (\Cref{PUF_eq:phi-ori}) score per user \cite{Fu2020Fairness-AwareGraphs, Leonhardt2018UserSystems}.

\subsubsection{Envy-based measures}
\label{PUF_sss:envy}
Fairness has been measured as envy-freeness. Envy is defined as the extra utility a user $u$ would have received if they were given the recommendation list of user $u'$, $L_{u'}$ \cite{Patro2020FairRec:Platforms,Biswas2021TowardPlatforms}. Formally, the envy of user $u$ towards $u'$ is computed as:\footnote{The original user utility function (\Cref{PUF_eq:phi-ori}) is used to compute envy in \cite{Patro2020FairRec:Platforms, Biswas2021TowardPlatforms}. WLOG, user utility is computed slightly differently in \cite{Do2022OnlineSystems} as the work is on online learning.}
\begin{equation}
\label{PUF_eq:envy}
    envy(u,u') = \max{\left\{
    \phi_u(L_{u'})-\phi_u(L_{u}), 0
    \right\}}
\end{equation}
Three fairness evaluation measures are based on this concept, where envy is aggregated differently: Mean Envy ($\downarrow$ME) \cite{Patro2020FairRec:Platforms,Biswas2021TowardPlatforms}, Mean Max Envy ($\downarrow$MME) \cite{Do2022OnlineSystems}, and Proportion of $\epsilon$-Envious User ($\downarrow$PEU) \cite{Do2022OnlineSystems}:

\begin{equation}
    ME = \frac{2}{m(m-1)} \sum_{u \in U} \sum_{\substack{u'\in U\\ u'\ne u}} envy(u,u')
\end{equation}
\begin{equation}\label{PUF_eq:mme}
MME = \frac{1}{m} \sum_{u \in U} \max_{u' \in U}{\left\{
envy(u,u')
\right\}
}
\end{equation}

\begin{equation}
    PEU = \frac{1}{m} \sum_{u \in U} 1_{\{ 
    \max_{u' \in U}{
envy(u,u') > \epsilon}
    \}}
\end{equation}
\noindent where $m=|U|$ is the number of users, and $\epsilon$ is the envy tolerance hyperparameter. A user $u$ is said to be $\epsilon$-envious of user $u'$ if their gain in user utility (envy) exceeds $\epsilon$. 

\subsubsection{Distance-based measures} 
\label{PUF_sss:dist}
In \cite{Wu2023EquippingEmbedding}, fairness for individual users is defined as any two similar users $u,u'$ receiving 
similar recommendations. Recommendation disparity is then measured with UnFairness score ($\downarrow$UF). UF uses both the user similarity,  $sim_{UF}(u,u')$, 
and the pairwise distance between the representation (e.g., embeddings) of items 
in the recommendation list of user $u$ and of user $u'$, $d_L(u,u')$. 
User similarity is modelled by both Jaccard ($sim_{Jacc}$) of the user's set of past interactions $H_u$ and JS-div between item feature (e.g., genre) distributions in the interactions of user $u$ and $u'$, $D_{JS}(u_{\alpha}, u_{\alpha}')$, where $\alpha$ is the item feature. UF is computed as: 
\begin{equation}
\label{PUF_eq:UF}
    UF = \log{\sum_{(u,u') \in U^*} sim_{UF}(u,u') \times d_L(u,u')} 
\end{equation}
\begin{equation}
\label{PUF_eq:sim_UF}
    sim_{UF}(u,u') = \gamma \ sim_{Jacc}(u,u') + (1-\gamma) \ sim_{JS}(u,u')
\end{equation}
\begin{equation}
\label{PUF_eq:d_L_UF}
    d_L(u,u') = \frac{1}{k^2} \sum_{i \in L_{u}} \sum_{i' \in L_{u'}} d_{cos}(\vec{i}, \vec{i'})
\end{equation}
\begin{equation}
\label{PUF_eq:jacc}
    sim_{Jacc}(u,u') = \frac{|H_u \cap H_{u'}|}{|H_u \cup H_{u'}|}
\end{equation}
\begin{equation}
\label{PUF_eq:sim_js}
    sim_{JS}(u,u') = 1-D_{JS}(u_{\alpha}, u'_{\alpha})
\end{equation}
\noindent where $U^*$ is the set of similar user pairs 
($u,u'$) with $sim_{UF}(u,u') \ \geq t$, and $t$ is the similarity threshold. In \cite{Wu2023EquippingEmbedding}, $t=mean(sim_{UF})\ + SD(sim_{UF})$, where the mean and SD are computed for all user pairs. We denote as $\gamma$ the weight for $sim_{Jacc}$, as $k$ the recommendation cut-off, and as $d_{cos}(\vec{i},\vec{i'})$ the cosine distance between the representations of items $i$ and $i'$. 
The log base in \Cref{PUF_eq:UF} \cite{Wu2023EquippingEmbedding} is unclear. 
Note that UF does not consider recommendation effectiveness. 

\section{Pairwise User unFairness (PUF)}
\label{PUF_s:puf-measure}

We present our evaluation measure of individual user fairness, Pairwise User unFairness ($\downarrow$PUF). 
PUF consists of two components: the similarity between users and the disparity in recommendation effectiveness. Next, we describe: (1) how user similarity is computed; (2) how to quantify the disparity of recommendation effectiveness, while taking into account user similarities.

\noindent \textbf{(1) Measuring similarity.} Measuring similarity between users is an inherently hard problem and there is no single ground truth of what makes two users similar \cite{Buyl2024InherentFairness}. When there is no user attribute, as per our focus on attribute-free fairness (\Cref{PUF_puf_s:intro}), user profiles can be modelled based on their historical interactions \cite{Herlocker1999AnFiltering,Wu2023EquippingEmbedding,Li2024ExplainingPerspective}. Similarity between users is then computed pairwise based on the user representation (e.g., click/rating matrix, user embedding), for example, with cosine similarity \cite{Reisz2024QuantifyingPrediction} or Jaccard (\Cref{PUF_eq:jacc}) \cite{Wu2023EquippingEmbedding}.

\noindent \textbf{(2) Measuring disparity.} 
PUF quantifies individual user fairness based on disparity in recommendation effectiveness, considering user similarity. PUF measures the mean pairwise difference in the effectiveness score per user, weighted by the user pair similarity. 
Based on (1) and (2), we define PUF as follows:
\begin{equation}
    \text{PUF} = \frac{2}{m(m-1)}
    \sum_{u\in U}\sum_{\substack{u'\in U \\ u \neq u'}} 
    sim(u,u') \times  |S(u) - S(u')|
\end{equation}
\noindent where $U$ is the set of all users ($|U|=m$, where $m\geq2$), 
$sim(u,u') \in [0,1]$ is the similarity of users $u$ and $u'$, 
$S$ is an effectiveness measure, e.g., P@$k$. $S$ must range or scaled to be in $[0,1]$, so PUF also ranges in $[0,1]$. PUF is modular; it can be operated with any similarity or effectiveness measure as long as it fulfils the range requirement. 

\noindent \textbf{How PUF differs from UF.} 
While both PUF and UF (Eq.~\ref{PUF_eq:UF}) \cite{Wu2023EquippingEmbedding} consider user similarity, PUF differs from UF as PUF considers the difference in the recommendation relevance, rather than only the disparity based on the representation of the recommended items as in UF. Moreover, recommending different sets of items to two similar users may be considered unfair by UF even if both users like their recommendations, but it is fair based on PUF. 
In theory, users with similar tastes in the past are likely to have a similar preference in the future \cite{Resnick1994GroupLens:Netnews}. Yet, the recommendation problem is challenging as even two highly similar users may not equally like their recommended items, if they are given identical items. 
This may be due to, for example, incomplete historical data that is unrepresentative of user taste, diverging user preference, or limited ground truth data. 
These limitations necessitate a look into the disparity of the recommendation relevance, which is what our measure quantifies. 

Overall, our measure, PUF, aligns with the definition of individual user fairness and quantifies fairness through the disparity in recommendation effectiveness, which is more meaningful than the disparity in the representation of recommended items, as effectiveness relates more to user utility.

\section{Empirical Analysis}
\label{PUF_s:experiment}

We compare PUF to existing effectiveness and individual user fairness measures. 

\subsection{Experimental Setup}
\label{PUF_ss:setup}
\begin{table}[t]
\caption{Statistics of the preprocessed datasets.}

\resizebox{0.95\columnwidth}{!}{
\begin{tabular}{lrrrr}
\toprule
\textbf{dataset} & \multicolumn{1}{l}{\textbf{\#users (all/test)}} & \multicolumn{1}{l}{\textbf{\#items}} & \multicolumn{1}{l}{\textbf{\#interactions}} & \multicolumn{1}{l}{\textbf{sparsity (\%)}} \\ 
\midrule                                                                    
Lastfm \cite{Cantador20112nd2011}& 1,859/1,836 & 2,823 & 71,355 & 98.64\% \\
QK-video \cite{Yuan2022Tenrec:Systems} & 4,656/3,514 & 6,423 & 51,777 & 99.83\% \\
ML-10M \cite{Harper2015TheContext} & 49,378/1,523 & 9,821 & 5,362,685 & 98.89\% \\
ML-20M \cite{Harper2015TheContext} & 89,917/2,178 & 16,404 & 10,588,141 & 99.28\% \\
\bottomrule
\end{tabular}
}
\label{PUF_tab:stats}
\end{table}
\noindent \textbf{Datasets}. We use four real-world datasets from three domains: music (Lastfm \cite{Cantador20112nd2011}), video (QK-video \cite{Yuan2022Tenrec:Systems}), and movie (ML-10M and ML-20M \cite{Harper2015TheContext}). 
We obtain Lastfm and ML-* from \cite{Zhao2021RecBole:Algorithms}, and QK-video from \cite{Yuan2022Tenrec:Systems}. We use the `sharing' interactions in QK-video.

\noindent\textit{Preprocessing}: For all datasets, we remove duplicate interactions and keep the most recent. We remove users and items with $<5$ interactions. For ML-* we map ratings $\geq3$ to 1, and discard the rest. The threshold 3 is chosen as the ratings range between $[0.5, 5]$. Lastfm and QK-video have unary ratings, so no mapping is required. 

\noindent\textit{Splitting}: We split the preprocessed datasets into train/val/test with a ratio of 6:2:2. 
The ML-* datasets are temporally split, while Lastfm and QK-video are randomly split as they have no timestamps. We split datasets globally (not user-wise) to avoid data leakage \cite{Meng2020ExploringModels}. After splitting, users with $<5$ interactions in the train set are removed from all splits to ensure that each user has adequate training data. The final preprocessed dataset statistics are in \Cref{PUF_tab:stats}. 

\noindent \textbf{Recommenders}. We use 7 well-known collaborative filtering recommenders that represent different model types:
user- and item-based $K$-nearest neighbour (U-KNN \cite{Resnick1994GroupLens:Netnews} and I-KNN \cite{Deshpande2004Item-basedAlgorithms}), 
Bayesian Personalised Ranking (BPR \cite{RendleBPR:Feedback}), Variational Autoencoder with multinomial likelihood (MVAE \cite{Liang2018VariationalFiltering}), Neural Graph Collaborative Filtering (NGCF \cite{Wang2019NeuralFiltering}), Neural Matrix Factorisation (NMF \cite{He2017NeuralFiltering}), and Neighbourhood-enriched Contrastive Learning (NCL \cite{Lin2022ImprovingLearning}). All models except U- and I-KNN are trained for 300 epochs with early stopping. We tune hyperparameters with grid search. The configuration with the best NDCG@10 during validation is the final model. Implementation, training, and tuning are done with RecBole \cite{Zhao2021RecBole:Algorithms}.

\noindent \textbf{Evaluation measures}. 
We measure recommendation effectiveness (\textsc{Eff}) with Hit Rate (HR), MRR, Precision (P or Prec), Recall (R), MAP, and NDCG \cite{Jarvelin2002CumulatedTechniques}. Recommendation fairness for individual users (\textsc{Fair}) is evaluated with our PUF measure (\Cref{PUF_s:puf-measure}) and all existing measures (\Cref{PUF_ss:fair-measures}): standard deviation (SD) of the P@$k$ (SD-P) and NDCG@$k$ (SD-NDCG) scores across all users;\footnote{We do not compute SD of MSE \cite{Rastegarpanah2019FightingSystems,Li2024ExplainingPerspective} as the ratings are binarised.}
Gini Index of P@$k$ (Gini-P) and NDCG@$k$ (Gini-NDCG); envy-based measures (ME, MME, and PEU); and distance-based measure, UF. 

We evaluate all runs at $k=10$. For PEU, we set $\epsilon=0.05$ \cite{Do2022OnlineSystems}. For UF, the number of all user pairs is used as the log base, so that UF $\in[0,1]$. We also set $\gamma=1$ to remove the term for category-based similarity.\footnote{We vary the $\gamma$ parameter in \Cref{PUF_ss:item-aware}.} 
As we use several models with diverse item representations, for a fair comparison between all models, we use the one-hot encoding of the historical interactions for the item representation in \Cref{PUF_eq:d_L_UF} of UF. The similarity threshold $t$ is fixed at $t=mean(sim_{UF})+SD(sim_{UF})$ as per \cite{Wu2023EquippingEmbedding}. 
We test four variants of PUF (two user similarity measures $\times$ two effectiveness measures). For similarity, we use cosine similarity ($sim_{cos}$) and Jaccard ($sim_{Jacc}$, \Cref{PUF_eq:jacc}), as both similarity measures are used in UF. In addition, $sim_{cos}$ is used in our U-KNN model, as well as to analyse the similarity between users in \cite{Reisz2024QuantifyingPrediction}. All user similarities are computed from the observed interactions in the train set (i.e., the binarised interaction matrix). 
We compute PUF with P and NDCG, to represent set- and rank-based measures, as well as for consistency with other \textsc{Fair} measures which are also computed with P and NDCG. To avoid extremely small values, we min-max normalise the pairwise user similarity for each dataset.

\subsection{Comparison of All Evaluation Measures}
\label{PUF_ss:groundwork}

\begin{table}
\centering
\caption{Mean computation time (s) of measures per model.}
\label{PUF_tab:time}
\resizebox{0.85\columnwidth}{!}{
\begin{tabular}{llrrrr}
\toprule
 &  & Lastfm & QK-video & ML-10M & ML-20M \\
\midrule
 \multirow[c]{8}{*}{\rotatebox[origin=c]{90}{\small\textsc{Fair} (existing)}} & SD-P & <0.01 & 0.01 & <0.01 & <0.01 \\
 & SD-NDCG & <0.01 & 0.02 & 0.01 & <0.01 \\
& Gini-P & <0.01 & 0.01 & <0.01 & <0.01 \\
 & Gini-NDCG & <0.01 & 0.03 & 0.02 & <0.01 \\
& ME & 233.88 & 1867.81 & 871.17 & 2020.60 \\
 & MME & 233.91 & 1867.88 & 871.20 & 2020.70 \\
 & PEU & 233.91 & 1867.88 & 871.20 & 2020.70 \\
& UF & 1297.78 & 873.52 & 1020.16 & 1358.52 \\
 \midrule
 \multirow[c]{4}{*}{\rotatebox[origin=c]{90}{\small\textsc{Fair} (PUF)}} & PUF-Prec-Cos & 5.30 & 20.68 & 4.00 & 8.25 \\
 & PUF-Prec-Jacc & 9.79 & 35.63 & 7.22 & 15.65 \\
 & PUF-NDCG-Cos & 4.32 & 16.01 & 3.38 & 6.72 \\
 & PUF-NDCG-Jacc & 9.56 & 35.10 & 7.40 & 15.19 \\
\bottomrule
\end{tabular}}
\end{table}

We present the \textsc{Eff} and \textsc{Fair} measures evaluation results (\Cref{PUF_app:extend-result}, \Cref{PUF_tab:base}), 
To identify possible empirical limitations of existing \textsc{Fair} measures and PUF, we analyse their score range and computational efficiency.

\noindent \textbf{Score range}. A measure that does not exhibit a wide range of scores across various fairness levels is less useful in distinguishing changes in fairness. Such measures may create an illusion of a 
negligible difference in fairness, due to their compressed empirical range \cite{Rampisela2024CanRelevance}. Across datasets, the observed range (not the theoretical range) of each \textsc{Fair} measure varies, except for MME which is consistently extremely small ($\leq5\times10^{-3}$). Changing a user's recommendation list with another user's as in \Cref{PUF_eq:envy} does not generally result in a large increase in the user's P@$k$ (envy), which translates to low MME. ME and PEU are unaffected by this issue even if they are based on \Cref{PUF_eq:envy}, as ME accounts for envy across all users rather than the maximum envy per user, and PEU employs an envy threshold. 
In short, MME is the least sensitive as it fails to discriminate between models for the same and across datasets, while other \textsc{Fair} measures, including PUF, are more sensitive. The most sensitive measures, i.e., the ones with the widest observed range, are Gini and PEU.

\noindent \textbf{Measure computation time}. We report the average computation time of the \textsc{Fair} measures in \Cref{PUF_tab:time}. All runs are done with AMD EPYC 75F3 for a fair comparison. User pairwise similarity is computed once per dataset for each PUF variant. We find that half of the existing measures, i.e., ME, MME, PEU, and UF are computationally expensive ($>$20 minutes), while PUF is significantly faster (<40 s), despite also being a pairwise measure. 
ME/MME/PEU/UF need additional extensive computation per pair, which makes them expensive: UF does nested pairwise comparison in \Cref{PUF_eq:UF} and \Cref{PUF_eq:d_L_UF}, while ME/MME/PEU recompute the effectiveness score for each user pair (\Cref{PUF_eq:envy}). In short, PUF is computationally more efficient and thus more practical to compute than most existing \textsc{Fair} measures.

\begin{figure*}[tb]
    \centering
  \includegraphics[width=0.98\textwidth, trim=0.25cm 0.25cm 0.25cm 0.1cm, clip=True]{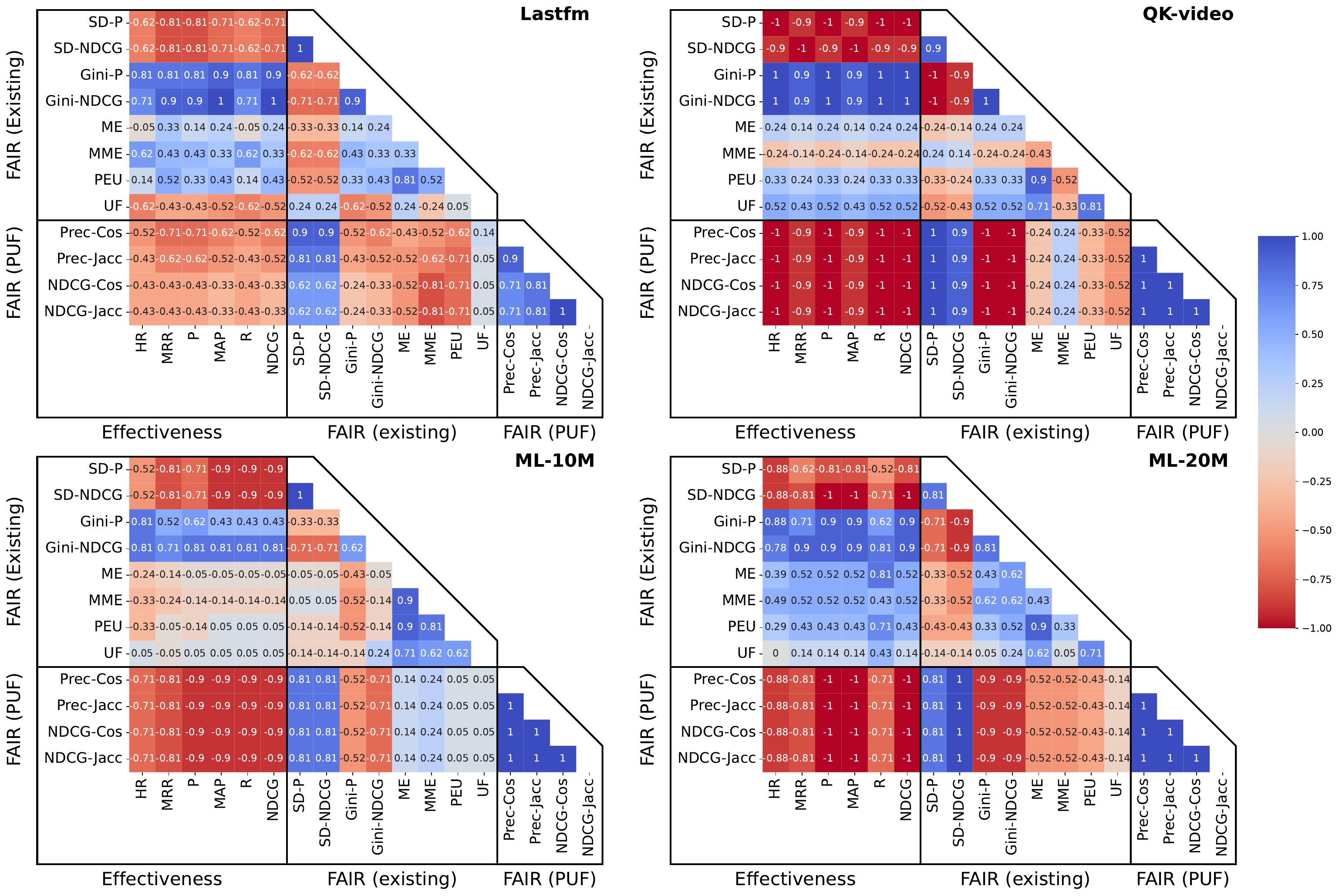}  
    \caption{Kendall's $\tau$ correlation between \textsc{Eff} measures, existing \textsc{Fair} measures, and PUFs for all recommenders.}
    \label{PUF_fig:corr}
\end{figure*}

\subsection{Measure Agreement}
\label{PUF_ss:agreement}

An important aspect when comparing evaluation measures is how much they agree when their scores are used to rank models from best to worst. 
If one measure can be used to estimate the rank ordering given by another, there is no point in using both measures if we are only interested in ranking models. To study this, we compute Kendall's $\tau$ correlation for all \textsc{Eff} and \textsc{Fair} measures. Kendall's $\tau$ can handle ties (unlike Spearman's $\rho$) and is more robust to nonlinear relationships (unlike Pearson's coefficient). 
If two measures have $\tau\geq 0.9$, we consider their rankings equivalent \cite{Voorhees2001EvaluationDocuments}. 
\Cref{PUF_fig:corr} shows the agreement between (i) \textsc{Eff} and \textsc{Fair} measures, and (ii) among \textsc{Fair} measures. 
 
\noindent \textbf{Agreement between \textsc{Eff} and \textsc{Fair} measures}. Across datasets, the agreement between \textsc{Fair} and \textsc{Eff} measures varies from strong disagreement (e.g., SD with $\tau \in [-1, -0.52]$) to moderate-to-strong agreement (e.g., Gini with $\tau \in [0.43, 1]$). Our PUF consistently disagrees with \textsc{Eff} measures, even if the disagreement is weaker for Lastfm, $\tau \in [-0.71,-0.33]$. As no \textsc{Fair} measure consistently has $|\tau| \geq 0.9$ with any \textsc{Eff} measure for all datasets, their model orderings cannot be precisely inferred from that of \textsc{Eff} measures.

\noindent \textbf{Agreement among \textsc{Fair} measures}. 
Regarding the agreement between PUF and existing \textsc{Fair} measures, we find that only SD aligns with PUF ($\tau \geq 0.62$), while the rest, including UF, tend to disagree or have a weak correlation with PUF, $\tau \in [-1, 0.24]$. Both UF and PUF consider user similarity, yet UF weakly correlates to PUF ($\tau \in [-0.14, 0.14]$) for all datasets except QK-video ($\tau \in [-0.52, -0.43]$). This weak relationship may be due to UF not considering item relevance, while PUF does. In contrast, despite not being similarity-based, SD correlates the strongest to PUF. Yet, most of its rankings are not equivalent to PUF ($\tau < 0.9$), which means that fairness evaluation with SD can lead to a conclusion that misaligns with PUF.

Even if we only compare the best model based on the measures instead of the model rankings, UF and PUF never agree (\Cref{PUF_app:extend-result}, \Cref{PUF_tab:base}). 
Thus, using UF and PUF to gauge fairness may differ in the conclusion on which model is the fairest. 

Overall, we find that existing \textsc{Fair} measures frequently disagree on model orderings with PUF, regardless of whether the measure accounts for user similarity. While SD has the most similar conclusions to PUF, it still does not give equivalent rankings to PUF.

\subsection{Varying Relevance Score Distribution}
\label{PUF_ss:artificial-rel}

\begin{figure*}
    \includegraphics[width=0.98\textwidth, trim=0.25cm 0.2cm 0.25cm 0.2cm, clip=True]{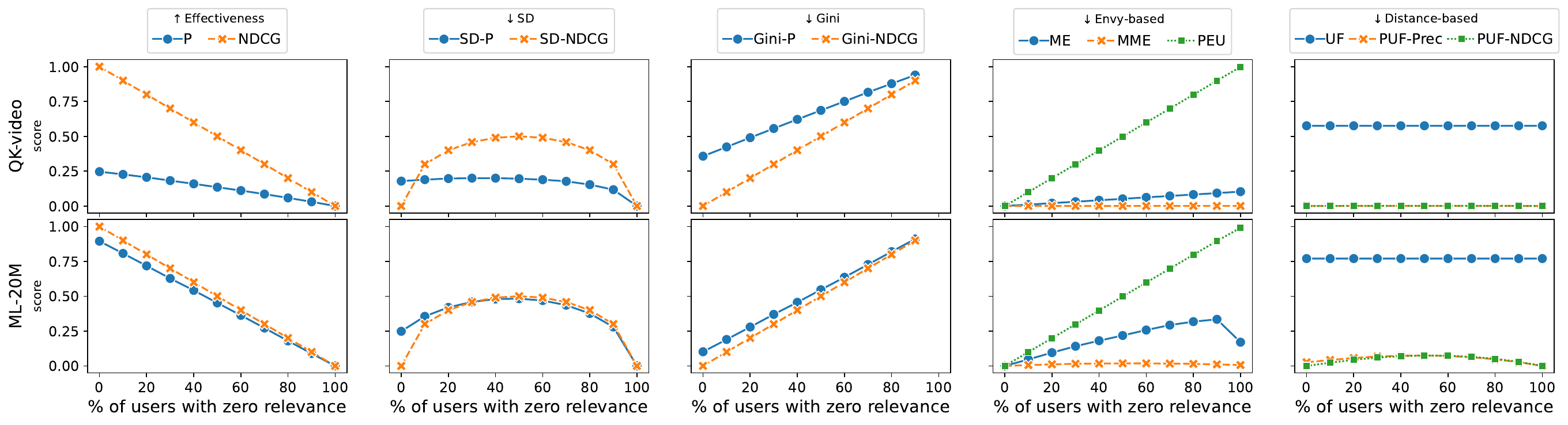}
    \caption{
    Effectiveness (\textsc{Eff}) and fairness (\textsc{Fair}) scores of QK-video and ML-20M, when artificially varying \% of users with all irrelevant items (zero relevance), and the rest of the users receiving all relevant items. All PUF variants overlap. Gini is missing points at 100\% users with zero relevance as it is undefined when each user has zero \textsc{Eff} scores. 
    }
    \label{PUF_fig:artificial-relevance}
\end{figure*}

It is important that a fairness measure can capture changes across differences in the recommendation quality across users, as fairness is related to the disparity in recommendation (e.g., its relevance). Thus, any change in \textsc{Eff} scores should also be reflected in the \textsc{Fair} scores. 
To this end, we study how varying item relevance affects the \textsc{Eff} and \textsc{Fair} measures. All \textsc{Fair} measure equations except for UF explicitly account for effectiveness,
but it is unknown how sensitive \textsc{Fair} measures are to changes in \textsc{Eff} scores. 

\paragraph{Procedure.} The change of relevance scores is artificially done as follows. 
For all users, we start by recommending $k$ relevant items (based on the test set). For users with more than $k$ relevant items, we select the relevant 
items randomly. For users with fewer than $k$ relevant items, we fill the remaining slots with random irrelevant items to ensure that each user receives exactly $k$ items. In each iteration, we replace the recommendation of 10\% of the users with all irrelevant items and compute the measures. As such, we expect maximum fairness at the start (as all users have the maximum \textsc{Eff} scores\footnote{Measures that are computed with P@$k$ do not start exactly at the optimal value, as not all users have $k$ items in their test set \cite{Moffat2013SevenMetrics}.}) and at the end (as all users have 0 \textsc{Eff} scores). We also expect maximum unfairness when half the users get all relevant items, and the rest get irrelevant items, as this leads to one of the most uneven \textsc{Eff} score distributions. For brevity, we only compute PUF with $sim_{Jacc}$, as the model orderings given by PUF-*-Cos are equivalent to PUF-*-Jacc (\Cref{PUF_ss:agreement}). Here, user similarities are computed based on the observed interactions in the train sets.
 
\paragraph{Results.} \Cref{PUF_fig:artificial-relevance} shows the results for QK-video and ML-20M, which represent the overall trends in all our datasets; we show the rest in \Cref{PUF_app:artificial_rel}. We find that PUF fits the expectation of having decreasing fairness followed by increasing fairness, denoted by the inverted parabolas. This trend is more pronounced for the ML-* datasets, as the mean user pairwise similarity is higher than for the other datasets. Among existing \textsc{Fair} measures, only SD follows this expectation, while the others show undesirable tendencies: as \textsc{Eff} decreases, Gini and PEU significantly become less fair, and ME does the same to a minor extent. The overall decrease in fairness in these measures is undesirable, as the measure scores closely resemble decreasing effectiveness instead of the disparity in the \textsc{Eff} scores distribution. 
Even worse, MME and UF are almost invariant to the change in 
recommendation effectiveness: MME tends to have extremely small scores to begin with (\Cref{PUF_ss:groundwork}), while UF does not depend on item relevance (\Cref{PUF_eq:UF}).

To summarise, we have shown that PUF and SD quantify fairness based on the disparity in recommendation effectiveness. All other existing \textsc{Fair} measures ignore the actual disparity in \textsc{Eff} scores. They merely reflect decreasing effectiveness or are not sensitive to the change in effectiveness.
Next, we study whether the change in user similarity distribution is reflected in the \textsc{Fair} measures.

\begin{figure}
\centering
\includegraphics[width=0.75\columnwidth, 
        trim=0.3cm 0.2cm 0.2cm 0.2cm, clip=True]{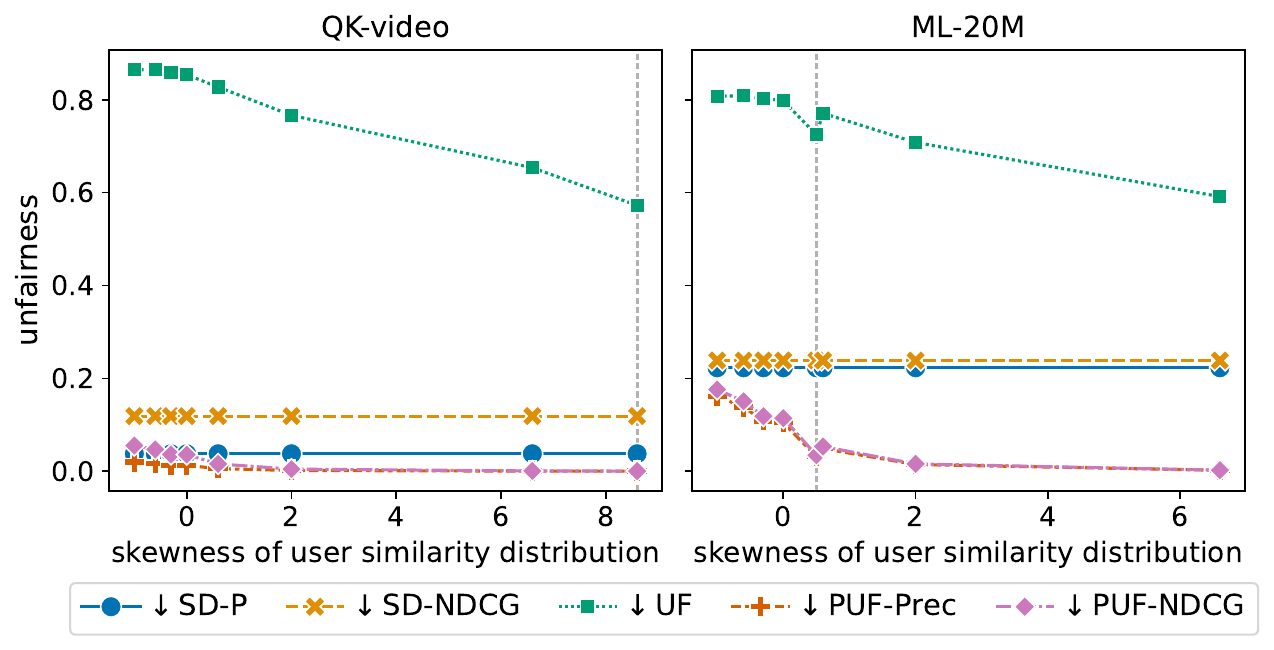}
    \caption{Artificially varying the skewness of the user similarity distribution for QK-video and ML-20M. Vertical grey lines denote the skewness corresponding to $sim_{Jacc}$ observed in the dataset. The distribution skewness differs across datasets.
    }
    \label{PUF_fig:artificial-similarity}
\end{figure}

\subsection{Varying User Similarity Distribution}
\label{PUF_ss:artificial-sim}

Individual user fairness is defined based on user similarity. It is important to know if and how user similarity affects fairness. While two or more recommender models should be evaluated under the same similarity distribution, a desirable property of an individual user fairness measure is the ability to distinguish a single model performance across different similarity distributions, that may arise from various ways of quantifying user similarity. To this end, we investigate how the \textsc{Fair} measures respond to varying user similarity distributions, where the distribution is artificially generated.  

\paragraph{Procedure.} User similarity distributions tend to be right-skewed (many dissimilar users) for random users, and left-skewed (many similar users) for users that are friends  \cite{Reisz2024QuantifyingPrediction}. Further, users are often dissimilar, as some users are new to the systems or do not engage much, leading to discrepancies between the number of interactions among users and potentially affecting user similarity. Considering the above, we create synthetic user similarity scores by sampling from the Weibull distribution \cite{Weibull1951AApplicability}, which can be used to model skewed distributions. It has been used to model user rating distributions and sampling user neighbour candidates in RSs \cite{Kermany2020ReInCre:Credibility, Kermany2023IncorporatingSystems, Adamopoulos2014OnSystems}. Its probability density function is $p(x) = \lambda x^{\lambda-1} \exp{(-x^\lambda)}$. To obtain various right- and left-skewed distributions to represent possible user similarity distributions, we set $\lambda \in \{0.5, 1, 2, 5, 10, 50\}$. 
We also sample from the normal distribution $\mathcal{N}(0,1)$, which has zero skews (i.e., equal portions of user pairs with similarity below/above the mean). 
We then min-max normalise the sampled similarity values to rescale them in the $[0,1]$-range and randomly assign them to user pairs. We analyse non-random assignment in \Cref{PUF_ss:assignment}.

While the user similarity is artificial, we use the actual recommendation lists and scores from the NCL model runs as they perform relatively well. To save computation time, we only compute SD to represent all similarity-independent measures; theoretically, these measures will remain constant given no change in their input. We also compute PUF and UF, the two similarity-based measures. We compare the measure scores with the scores corresponding to the user similarity distribution observed in the datasets based on $sim_{Jacc}$ (\Cref{PUF_ss:groundwork}).

\paragraph{Results.} We present the results for QK-video and ML-20M in \Cref{PUF_fig:artificial-similarity}; we show the rest in \Cref{PUF_app:artificial_sim} as their trends are similar. We find that the similarity-based measures PUF and UF become fairer as skewness increases. Increasing skewness means a higher proportion of user pairs with low similarity, hence $\downarrow$PUF and $\downarrow$UF tend to be lower. Conversely, but as expected, the similarity-independent SD remains constant despite the change in skewness. For highly skewed similarities (skewness $>6$), which is a realistic similarity distribution as seen in QK-video, SD-NDCG is somewhat unfair ($\approx0.2$) for all datasets. Yet, PUF is almost perfectly fair ($\approx$0). Therefore, simply using SD may lead to the underestimation of fairness. 
While we only compute SD here, we expect other existing 
\textsc{Fair} measures to exhibit the same invariance as they are similarity-independent. 

Between the two similarity-based measures, PUF is more sensitive towards negatively skewed user similarity distribution than UF. As skewness decreases, the mean user similarity increases at a slower rate. UF only considers user pairs above the mean-based similarity threshold, thus the number of user pairs contributing to the log sum in \Cref{PUF_eq:UF} decreases slower than in the right-skewed distributions. Another concern of UF is its relatively high unfairness compared to PUF and SD, even when most users are dissimilar. 
This may be because UF computes the pairwise distance of the representation of the recommended items of a user pair (\Cref{PUF_eq:d_L_UF}). Minimising this distance is difficult as each $k$ recommended item of a user must have a similar representation to each of the other user's $k$ items, regardless of item relevance. 

In short, we show that PUF can distinguish fairness levels across various similarity distributions, while non-similarity-based measures cannot.
This distinction indicates the strength of PUF over SD (and indirectly over other non-similarity-based measures, i.e., all existing \textsc{Fair} measures except UF). We find that disregarding user similarity can also lead to the misinterpretation of fairness level. 

In the following part, we further compare the two similarity-based measures and highlight the strengths of PUF over UF.

\subsection{PUF and UF under Extreme Cases
}
\label{PUF_ss:assignment}

We compare the similarity-based measures (PUF and UF) under extreme scenarios: can their scores reflect the difference in maximum and minimum fairness, across differences in the recommendation quality? Given an artificial set of pairwise user similarities and an artificial set of pairwise \textsc{Eff} score differences, to simulate the fairest case (MostFair), we sort these values and assign a higher similarity value to user pairs with lower \textsc{Eff} score difference. Separately, we assign a higher similarity score to pairs with higher \textsc{Eff} score difference to mimic the unfairest case (MostUnfair). For MostFair, a desirable measure would score close to 0 (the fairest). For MostUnfair, it should exhibit an inverted U-shape when the \textsc{Eff} score distribution is varied (i.e., similar to \Cref{PUF_fig:artificial-relevance}), as the maximum unfairness happens when the \textsc{Eff} distribution is the most uneven.

\paragraph{Procedure.} We use an artificial, right-skewed user similarity distribution sampled from the Weibull distribution with $\lambda=2$ (\Cref{PUF_ss:artificial-sim}). We use the P@$k$ and NDCG@$k$ scores per user from the artificial runs in \Cref{PUF_ss:artificial-rel}. 
To compute PUF- and UF-NDCG, we assign the user similarities to user pairs following the sorted pairwise difference of NDCG, and likewise for PUF- and UF-Prec.

\begin{figure}[bt]
    \centering
    \includegraphics[width=0.75\linewidth]{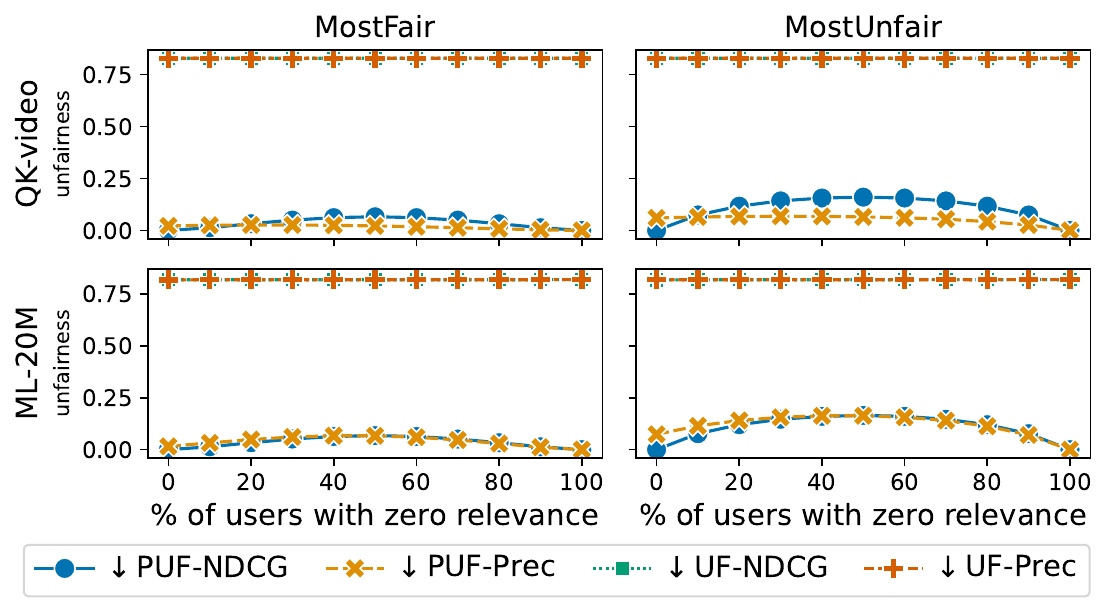}
    \caption{Artificially varying the \% of users with zero relevance for QK-video and ML-20M. Lower \textsc{Eff} score difference is assigned to user pairs with higher similarity (MostFair), and to lower similarity (MostUnfair). Both UF overlap.}  
    \label{PUF_fig:artificial-sim-relevance-sort}
\end{figure}

\paragraph{Results.} The results for QK-video and ML-20M are shown in \Cref{PUF_fig:artificial-sim-relevance-sort}; the rest are shown in \Cref{PUF_app:extreme} as we find similar trends. We see that the MostFair assignment yields PUF scores that are overall close to the fairest (0) across varying recommendation effectiveness, but UF remains constantly unfair ($\sim$0.8) nonetheless. This emphasises a mismatch between fairness computed based on the disparity of item representation and based on \textsc{Eff} score differences. 
PUF scores are more unfair for the MostUnfair than the MostFair scenario, while UF is almost invariant to the change in similarity assignment between MostFair and MostUnfair. We observe again the insensitivity of UF towards varying recommendation effectiveness, as seen in \Cref{PUF_fig:artificial-relevance}. 

In the MostUnfair case, $\downarrow$PUF ranges in $[0,0.2)$ (which is close to the fairest) for all datasets. This is because the overall user similarity is low, which can happen in real-world scenarios. Hence, the maximum unfairness is expected to be relatively low, as PUF relies on user similarity to quantify fairness.

All in all, PUF correctly scores close to the fairest in the MostFair assignment, while UF fails to do so. PUF is notably more unfair for the MostUnfair rather than the MostFair case, while UF is almost constant for both cases and across varying recommendation effectiveness. In both cases, UF also overestimates effectiveness-based unfairness by constantly scoring much higher than PUF. Overall, PUF correctly quantifies maximal and minimal fairness, while UF does not. Next, we study the effect of using the same similarity measure for generating recommendations and evaluating fairness.

\subsection{Using the Same Similarity Measure for PUF and Recommendation Model}
\label{PUF_ss:same-sim-measure}
We ask: if PUF and the recommendation model use the same similarity measure, would the model be the fairest according to PUF? The goal is to validate whether PUF gives an advantage to models that use the same similarity function as PUF. 
To answer this, we use cosine similarity ($sim_{cos}$) for both PUF (PUF-*-Cos) and the U-KNN model (\Cref{PUF_app:extend-result}, \Cref{PUF_tab:base}). 
We find that U-KNN is never the fairest according to PUF. Even though U-KNN is the second-best model based on PUF for Lastfm, it is the second worst for ML-10M and it is the unfairest for ML-20M. So, using the same similarity measure for both PUF and the recommender model does not lead to the model being the fairest based on PUF. This may be due to the data splits: let two users have identical items in the train set, but completely different items in the test set. 
Even if both users get the same recommendation list, their effectiveness scores may differ and PUF cannot be the fairest. Therefore, minimising PUF to optimise for pairwise user fairness while maintaining an overall high mean effectiveness scores for all users, is not a trivial problem.

\subsection{Using Item Features for PUF and UF}
\label{PUF_ss:item-aware}

We next study what happens in a scenario where item features are available. PUF and UF are the only measures that can incorporate item features in their similarity functions. We compare PUF to UF when varying the importance of item features to them. 

\paragraph{Procedure.} We compute PUF and UF with $sim_{UF}$ (\Cref{PUF_eq:sim_UF}) and vary the weighted sum of its two components (\Cref{PUF_eq:jacc}--\eqref{PUF_eq:sim_js}) with $\gamma \in \{0, 0.25, 0.5, 0.75,\allowbreak 1\}$. The weight $\gamma=0$ accounts only for users' item feature distribution, while $\gamma=1$ accounts only for their past interactions.  We evaluate the groundwork runs\footnote{We do the same analysis for the artificial runs with various relevance score distributions (\Cref{PUF_ss:artificial-rel}) and find similar results (\Cref{PUF_app:item_feature}).} for the ML-* datasets, with movie genres as features. 
We cannot use the other datasets as QK-video only has two item categories and Lastfm has none.

\begin{figure}
    \centering
        \includegraphics[width=0.75\linewidth]{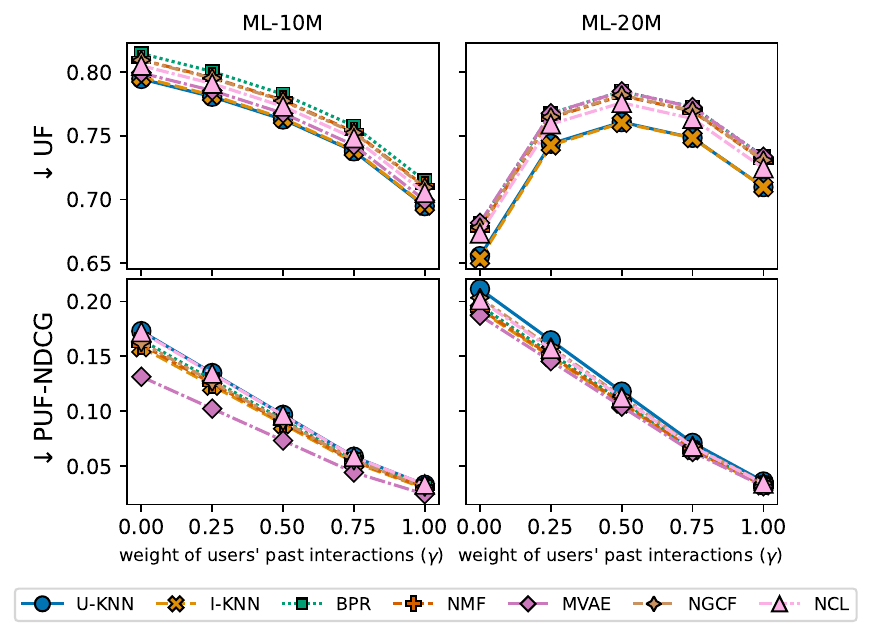} 
    \caption{
                UF and PUF computed with user similarity $sim_{UF}$ (\Cref{PUF_eq:sim_UF}), varying the weighted sum of users' past interactions ($sim_{Jacc}$) and their item feature distribution ($sim_{JS}$) with $\gamma$.
            }
    \label{PUF_fig:varying_gamma}
\end{figure}

\paragraph{Results.}  We present the results in \Cref{PUF_fig:varying_gamma}. Overall, we find that PUF scores are fairer with higher weights of the users' past interactions.\footnote{We also compute PUF-Prec; the trends are similar to PUF-NDCG (\Cref{PUF_app:item_feature}).} Users' past interactions tend to be more diverse than their item category distributions, leading to more dissimilar users and less weight of user effectiveness disparity, which means fairer $\downarrow$PUF. UF follows the trend of PUF for ML-10M, but differs for ML-20M, where at first it becomes more unfair, then fairer with a higher $\gamma$. The non-linear trend of UF for ML-20M is due to its self-imposed thresholding, which results in highly different numbers of similar user pairs across various $\gamma$. While the effect of $\gamma$ on UF can be inconsistent for different datasets, it is more predictable for PUF, which always exhibits near-linear relationship.

For both datasets, UF is always more unfair than PUF, which is consistent with our prior results (\Cref{PUF_ss:groundwork}, \Cref{PUF_ss:artificial-rel}--\Cref{PUF_ss:assignment}). 
For both measures, the ordering of models given for a $\gamma$ is maintained for another, i.e., the same conclusion of which model is better than another can be reached despite the different scores for different weights. 

To sum up, varying the weights of interaction- and feature-based user similarity impacts PUF in a more predictable way than UF, making PUF superior to UF due to its more intuitive behaviour. 

\section{Related Work}
In \Cref{PUF_s:individual-user-fair} we overviewed existing individual user fairness measures. 
Next, we discuss work that studies RS fairness measures empirically \cite{Raj2022MeasuringResults,Rampisela2024EvaluationStudy,Rampisela2024CanRelevance} and proposes pairwise individual fairness measures for ranking \cite{Fabris2023PairwiseMeasure, Wang2022ProvidingSystems}. Our work is close to \cite{Raj2022MeasuringResults}, which studies item group fairness measures in RS, and \cite{Rampisela2024EvaluationStudy, Rampisela2024CanRelevance}, which studies evaluation measures of individual item fairness. Similarly to \cite{Raj2022MeasuringResults,Rampisela2024CanRelevance}, we find that fairness measures may disagree in their model ordering, and that some measures are more sensitive than others, given decreasing effectiveness and disparity in the recommendations. Out of all individual fairness measures in \cite{Rampisela2024EvaluationStudy,Rampisela2024CanRelevance}, we often see that the similarity criterion of individual fairness is ignored, with only a few measures incorporating it. Recent work \cite{Wang2022ProvidingSystems,Fabris2023PairwiseMeasure} also proposes pairwise individual fairness measures, but these are for individual items, whereas our pairwise measure, PUF, is for individual users. 
Further, PUF considers user similarity, while the measure in \cite{Fabris2023PairwiseMeasure} does not. The measure in \cite{Wang2022ProvidingSystems} is similar to UF, as both employ thresholding of user similarities. Instead of applying a threshold, which can be arbitrary, 
PUF is weighted by user similarities, which introduces degrees of user similarity in the fairness computation.
There exists also work on user group fairness (e.g., \cite{Ekstrand2018AllEffectiveness,Zhu2018Fairness-AwareRecommendation}) or counterfactual fairness (e.g., \cite{Chen2024FairGap:Graph}). Most such work requires sensitive attributes (e.g., gender), but public recommendation datasets with sensitive attributes tend to lack user representations (e.g., only binary genders) \cite{Harper2015TheContext,Celma2010MusicTail,Yuan2022Tenrec:Systems}, and grouping users may require discretising the attribute (e.g., age) \cite{Buyl2024InherentFairness}. 
We focus on attribute-free individual fairness, rather than group or counterfactual fairness, to better assess distribution across all individuals \cite{Lazovich2022MeasuringMetrics} (often hidden in group fairness evaluation \cite{Fabris2023PairwiseMeasure}).

\section{Discussion and Conclusions}

Current evaluation measures of individual user fairness in RSs either exclusively consider disparity in recommendation effectiveness or user similarity, but never both jointly. This means that none of these measures aligns with the definition of individual user fairness and one of the key objectives of RSs: user utility. To address this issue, we introduced PUF, a novel evaluation measure that quantifies user fairness through pairwise difference in effectiveness scores, while considering similarity between users. Note that accounting for both aspects is not trivial. 
For example, dividing the user similarity by the effectiveness disparity may result in an infinitely large score if the effectiveness disparity is small, and the score is undefined if the recommendation effectiveness is perfectly equal.

We showed that PUF is empirically superior to existing measures, based on the definition of individual fairness that we adopt (\Cref{PUF_ss:definition}). 
Firstly, PUF is superior to non-similarity-based measures, such as SD. The non-similarity-based measures are insensitive to changes in user similarity distribution, which is a critical flaw as they cannot give a distinct fairness score for different levels of user similarity. 
Considering both user similarity and effectiveness disparity is crucial as failing to do so may lead to over/underestimation of fairness. 
Secondly, PUF is also better than UF, the only existing measure that factors in user similarity. We have shown that PUF scores correspondingly to the disparity in user utility and successfully quantifies contrasting levels of fairness, while UF fails to do both. 
While the PUF design is simple, it is a novel, intuitive measure that is easy to interpret and robust against various effectiveness disparity or user similarity, which sets it apart from existing metrics, addressing a crucial gap in RS fairness evaluation. 

In summary, we recommend using PUF over other existing individual user fairness measures due to its alignment with the definition of fairness, computational efficiency, and sensitivity to varying levels of user similarities and recommendation effectiveness in both typical and extreme cases. PUF paves the way for more elaborate measures in fairness evaluation, such as with graded relevance or other ways of quantifying user similarity (e.g., considering interaction recency).

\section{Appendix}

\subsection{Extended Empirical Analysis}
\subsubsection{Comparison of all evaluation measures}
\label{PUF_app:extend-result}
We present in \Cref{PUF_tab:base} the effectiveness (\textsc{Eff}) and fairness (\textsc{Fair}) scores of the 7 recommenders for 4 datasets.

\begin{table}
\caption{
Effectiveness (\textsc{Eff}) and fairness (\textsc{Fair}) scores at $k=10$ for recommenders. 
The most effective/fair score per measure is bolded. $\uparrow$/$\downarrow$ means the higher/lower the better.}

\label{PUF_tab:base}
\resizebox{\columnwidth}{!}{
\begin{tabular}{lllrrrrrrr}
\toprule
 &  &  & U-KNN & I-KNN & BPR & NMF & MVAE & NGCF & NCL \\
\midrule
\multirow[c]{18}{*}{\rotatebox[origin=c]{90}{Lastfm}} & \multirow[c]{6}{*}{\rotatebox[origin=c]{90}{\small\textsc{Eff}}} & $\uparrow$ $\text{HR}$ & 0.7102 & 0.7647 & 0.7729 & 0.7424 & 0.7783 & 0.7707 & \bfseries 0.7930 \\
 &  & $\uparrow$ $\text{MRR}$ & 0.4386 & 0.4835 & 0.4916 & 0.4641 & 0.4764 & 0.4954 & \bfseries 0.5026 \\
 &  & $\uparrow$ $\text{P}$ & 0.1532 & 0.1721 & 0.1782 & 0.1649 & 0.1758 & 0.1779 & \bfseries 0.1840 \\
 &  & $\uparrow$ $\text{MAP}$ & 0.1190 & 0.1369 & 0.1414 & 0.1291 & 0.1377 & 0.1418 & \bfseries 0.1478 \\
 &  & $\uparrow$ $\text{R}$ & 0.1899 & 0.2180 & 0.2238 & 0.2069 & 0.2241 & 0.2236 & \bfseries 0.2342 \\
 &  & $\uparrow$ $\text{NDCG}$ & 0.2167 & 0.2450 & 0.2520 & 0.2334 & 0.2473 & 0.2530 & \bfseries 0.2611 \\
\cline{2-10}
 & \multirow[c]{8}{*}{\rotatebox[origin=c]{90}{\small\textsc{Fair} (existing)}} & $\downarrow$ $\text{SD-P}$ & \bfseries 0.1485 & 0.1529 & 0.1538 & 0.1505 & 0.1518 & 0.1527 & 0.1552 \\
 &  & $\downarrow$ $\text{SD-NDCG}$ & \bfseries 0.2040 & 0.2091 & 0.2092 & 0.2072 & 0.2084 & 0.2085 & 0.2124 \\
 &  & $\downarrow$ $\text{Gini-P}$ & 0.5146 & 0.4768 & 0.4668 & 0.4889 & 0.4654 & 0.4642 & \bfseries 0.4554 \\
 &  & $\downarrow$ $\text{Gini-NDCG}$ & 0.5213 & 0.4789 & 0.4678 & 0.4953 & 0.4735 & 0.4646 & \bfseries 0.4581 \\
 & & $\downarrow$ $\text{ME}$ & 0.0897 & \bfseries 0.0827 & 0.0904 & 0.0983 & 0.0973 & 0.0897 & 0.0886 \\
 &  & $\downarrow$ $\text{MME}$ & 0.0025 & \bfseries 0.0017 & 0.0020 & 0.0025 & 0.0020 & 0.0021 & 0.0019 \\
 &  & $\downarrow$ $\text{PEU}$ & 0.6525 & \bfseries 0.6176 & 0.6318 & 0.6765 & 0.6531 & 0.6275 & 0.6193 \\
 & & $\downarrow$ $\text{UF}$ & \bfseries 0.6615 & 0.6616 & 0.6660 & 0.6667 & 0.6685 & 0.6675 & 0.6681 \\
\cline{2-10}
 & \multirow[c]{4}{*}{\rotatebox[origin=c]{90}{\small\textsc{Fair} (PUF)}} &  $\downarrow$ $\text{PUF-Prec-Cos}$ & 0.0097 & 0.0100 & 0.0100 & \bfseries 0.0096 & 0.0098 & 0.0099 & 0.0100 \\
 &  & $\downarrow$ $\text{PUF-Prec-Jacc}$ & 0.0073 & 0.0075 & 0.0075 & \bfseries 0.0073 & 0.0074 & 0.0075 & 0.0076 \\
 &  & $\downarrow$ $\text{PUF-NDCG-Cos}$ & 0.0134 & 0.0137 & 0.0136 & \bfseries 0.0132 & 0.0135 & 0.0135 & 0.0136 \\
 &  & $\downarrow$ $\text{PUF-NDCG-Jacc}$ & 0.0101 & 0.0103 & 0.0102 & \bfseries 0.0099 & 0.0101 & 0.0101 & 0.0102 \\
\cline{1-10} 
\multirow[c]{18}{*}{\rotatebox[origin=c]{90}{QK-video}} & \multirow[c]{6}{*}{\rotatebox[origin=c]{90}{\small\textsc{Eff}}} & $\uparrow$ $\text{HR}$ & 0.1155 & 0.0396 & 0.0993 & 0.0857 & 0.1093 & 0.1238 & \bfseries 0.1298 \\
 &  & $\uparrow$ $\text{MRR}$ & 0.0435 & 0.0131 & 0.0390 & 0.0303 & 0.0392 & \bfseries 0.0484 & 0.0481 \\
 &  & $\uparrow$ $\text{P}$ & 0.0122 & 0.0041 & 0.0105 & 0.0088 & 0.0116 & 0.0134 & \bfseries 0.0140 \\
 &  & $\uparrow$ $\text{MAP}$ & 0.0188 & 0.0046 & 0.0170 & 0.0128 & 0.0178 & \bfseries 0.0218 & 0.0218 \\
 &  & $\uparrow$ $\text{R}$ & 0.0517 & 0.0141 & 0.0433 & 0.0369 & 0.0506 & 0.0589 & \bfseries 0.0608 \\
 &  & $\uparrow$ $\text{NDCG}$ & 0.0331 & 0.0091 & 0.0290 & 0.0231 & 0.0314 & 0.0375 & \bfseries 0.0381 \\
\cline{2-10}
 & \multirow[c]{8}{*}{\rotatebox[origin=c]{90}{\small\textsc{Fair} (existing)}} & $\downarrow$ $\text{SD-P}$ & 0.0350 & \bfseries 0.0205 & 0.0327 & 0.0293 & 0.0343 & 0.0373 & 0.0376 \\
 &  & $\downarrow$ $\text{SD-NDCG}$ & 0.1089 & \bfseries 0.0539 & 0.1058 & 0.0910 & 0.1079 & 0.1191 & 0.1188 \\
 &  & $\downarrow$ $\text{Gini-P}$ & 0.8906 & 0.9618 & 0.9060 & 0.9168 & 0.8968 & 0.8850 & \bfseries 0.8790 \\
 &  & $\downarrow$ $\text{Gini-NDCG}$ & 0.9215 & 0.9733 & 0.9344 & 0.9425 & 0.9265 & 0.9156 & \bfseries 0.9120 \\
 & & $\downarrow$ $\text{ME}$ & 0.0874 & 0.0957 & 0.0962 & 0.0757 & \bfseries 0.0412 & 0.0899 & 0.0622 \\
 &  & $\downarrow$ $\text{MME}$ & 0.0012 & \bfseries 0.0003 & 0.0010 & 0.0014 & 0.0011 & 0.0011 & 0.0012 \\
 &  & $\downarrow$ $\text{PEU}$ & 0.7746 & 0.9087 & 0.8426 & 0.6810 & \bfseries 0.3867 & 0.8056 & 0.5626 \\
 & & $\downarrow$ $\text{UF}$ & 0.5731 & 0.5751 & 0.5739 & 0.5736 & \bfseries 0.5721 & 0.5734 & 0.5726 \\
\cline{2-10}
 & \multirow[c]{4}{*}{\rotatebox[origin=c]{90}{\small\textsc{Fair} (PUF)}} &  $\downarrow$ $\text{PUF-Prec-Cos}$ & 0.0001 & \bfseries 0.0000 & 0.0001 & 0.0001 & 0.0001 & 0.0001 & 0.0001 \\
 &  & $\downarrow$ $\text{PUF-Prec-Jacc}$ & 0.0001 & \bfseries 0.0000 & 0.0001 & 0.0001 & 0.0001 & 0.0001 & 0.0001 \\
 &  & $\downarrow$ $\text{PUF-NDCG-Cos}$ & 0.0003 & \bfseries 0.0001 & 0.0003 & 0.0002 & 0.0003 & 0.0003 & 0.0003 \\
 &  & $\downarrow$ $\text{PUF-NDCG-Jacc}$ & 0.0002 & \bfseries 0.0001 & 0.0002 & 0.0002 & 0.0002 & 0.0002 & 0.0002 \\
\bottomrule
\end{tabular}
}
\end{table}
\begin{table}
\caption*{Table \ref{PUF_tab:base} (continued): 
Effectiveness (\textsc{Eff}) and fairness (\textsc{Fair}) scores at $k=10$ for recommenders. 
The most effective/fair score per measure is bolded. $\uparrow$/$\downarrow$ means the higher/lower the better.}
\resizebox{\linewidth}{!}{
\begin{tabular}{lllrrrrrrr}
\toprule
 &  &  & U-KNN & I-KNN & BPR & NMF & MVAE & NGCF & NCL \\
\cline{1-10}
\multirow[c]{18}{*}{\rotatebox[origin=c]{90}{ML-10M}} & \multirow[c]{6}{*}{\rotatebox[origin=c]{90}{\small\textsc{Eff}}} & $\uparrow$ $\text{HR}$ & 0.5207 & 0.4872 & 0.5121 & 0.5141 & 0.4169 & 0.5128 & \bfseries 0.5213 \\
 &  & $\uparrow$ $\text{MRR}$ & \bfseries 0.3105 & 0.2818 & 0.2987 & 0.2915 & 0.2372 & 0.2865 & 0.3019 \\
 &  & $\uparrow$ $\text{P}$ & 0.1531 & 0.1369 & 0.1458 & 0.1433 & 0.1066 & 0.1460 & \bfseries 0.1536 \\
 &  & $\uparrow$ $\text{MAP}$ & \bfseries 0.1027 & 0.0887 & 0.0953 & 0.0911 & 0.0666 & 0.0930 & 0.1009 \\
 &  & $\uparrow$ $\text{R}$ & \bfseries 0.0276 & 0.0222 & 0.0249 & 0.0246 & 0.0203 & 0.0246 & 0.0263 \\
 &  & $\uparrow$ $\text{NDCG}$ & \bfseries 0.1691 & 0.1501 & 0.1600 & 0.1560 & 0.1191 & 0.1579 & 0.1673 \\
\cline{2-10}
 & \multirow[c]{8}{*}{\rotatebox[origin=c]{90}{\small\textsc{Fair} (existing)}} & $\downarrow$ $\text{SD-P}$ & 0.2206 & 0.2080 & 0.2125 & 0.2076 & \bfseries 0.1781 & 0.2109 & 0.2195 \\
 &  & $\downarrow$ $\text{SD-NDCG}$ & 0.2376 & 0.2244 & 0.2297 & 0.2241 & \bfseries 0.1960 & 0.2262 & 0.2367 \\
 &  & $\downarrow$ $\text{Gini-P}$ & 0.6884 & 0.7076 & 0.6906 & 0.6874 & 0.7415 & 0.6880 & \bfseries 0.6853 \\
 &  & $\downarrow$ $\text{Gini-NDCG}$ & \bfseries 0.6889 & 0.7123 & 0.6954 & 0.6948 & 0.7460 & 0.6943 & 0.6911 \\
 & & $\downarrow$ $\text{ME}$ & 0.1017 & 0.1131 & 0.1290 & 0.1307 & \bfseries 0.0726 & 0.1284 & 0.1192 \\
 &  & $\downarrow$ $\text{MME}$ & 0.0039 & 0.0040 & 0.0046 & 0.0046 & \bfseries 0.0025 & 0.0043 & 0.0044 \\
 &  & $\downarrow$ $\text{PEU}$ & 0.5936 & 0.6356 & 0.6927 & 0.7072 & \bfseries 0.4668 & 0.7026 & 0.6619 \\
 & & $\downarrow$ $\text{UF}$ & \bfseries 0.6948 & 0.6954 & 0.7148 & 0.7103 & 0.6994 & 0.7095 & 0.7053 \\
\cline{2-10}
 & \multirow[c]{4}{*}{\rotatebox[origin=c]{90}{\small\textsc{Fair} (PUF)}} &  $\downarrow$ $\text{PUF-Prec-Cos}$ & 0.0493 & 0.0445 & 0.0472 & 0.0458 & \bfseries 0.0363 & 0.0469 & 0.0497 \\
 &  & $\downarrow$ $\text{PUF-Prec-Jacc}$ & 0.0298 & 0.0268 & 0.0285 & 0.0276 & \bfseries 0.0217 & 0.0284 & 0.0302 \\
 &  & $\downarrow$ $\text{PUF-NDCG-Cos}$ & 0.0547 & 0.0495 & 0.0523 & 0.0507 & \bfseries 0.0409 & 0.0515 & 0.0549 \\
 &  & $\downarrow$ $\text{PUF-NDCG-Jacc}$ & 0.0331 & 0.0298 & 0.0316 & 0.0306 & \bfseries 0.0245 & 0.0311 & 0.0333 \\
\cline{1-10} 
\multirow[c]{18}{*}{\rotatebox[origin=c]{90}{ML-20M}} & \multirow[c]{6}{*}{\rotatebox[origin=c]{90}{\small\textsc{Eff}}} & $\uparrow$ $\text{HR}$ & \bfseries 0.5092 & 0.4881 & 0.5051 & 0.4927 & 0.4890 & 0.5083 & 0.5051 \\
 &  & $\uparrow$ $\text{MRR}$ & 0.2970 & 0.2799 & 0.2928 & 0.2781 & 0.2593 & \bfseries 0.3007 & 0.2934 \\
 &  & $\uparrow$ $\text{P}$ & \bfseries 0.1582 & 0.1423 & 0.1449 & 0.1433 & 0.1409 & 0.1517 & 0.1500 \\
 &  & $\uparrow$ $\text{MAP}$ & \bfseries 0.1079 & 0.0923 & 0.0957 & 0.0934 & 0.0889 & 0.1010 & 0.0997 \\
 &  & $\uparrow$ $\text{R}$ & \bfseries 0.0213 & 0.0193 & 0.0193 & 0.0186 & 0.0185 & 0.0199 & 0.0200 \\
 &  & $\uparrow$ $\text{NDCG}$ & \bfseries 0.1711 & 0.1538 & 0.1584 & 0.1544 & 0.1481 & 0.1649 & 0.1626 \\
\cline{2-10}
 & \multirow[c]{8}{*}{\rotatebox[origin=c]{90}{\small\textsc{Fair} (existing)}} & $\downarrow$ $\text{SD-P}$ & 0.2327 & \bfseries 0.2145 & 0.2157 & 0.2160 & 0.2152 & 0.2239 & 0.2228 \\
 &  & $\downarrow$ $\text{SD-NDCG}$ & 0.2484 & 0.2290 & 0.2333 & 0.2324 & \bfseries 0.2278 & 0.2399 & 0.2385 \\
 &  & $\downarrow$ $\text{Gini-P}$ & \bfseries 0.6981 & 0.7068 & 0.6991 & 0.7054 & 0.7100 & 0.6983 & 0.7001 \\
 &  & $\downarrow$ $\text{Gini-NDCG}$ & \bfseries 0.7021 & 0.7110 & 0.7053 & 0.7138 & 0.7203 & 0.7022 & 0.7046 \\
 & & $\downarrow$ $\text{ME}$ & \bfseries 0.1034 & 0.1199 & 0.1361 & 0.1412 & 0.1421 & 0.1348 & 0.1337 \\
 &  & $\downarrow$ $\text{MME}$ & \bfseries 0.0038 & 0.0049 & 0.0046 & 0.0052 & 0.0049 & 0.0046 & 0.0050 \\
 &  & $\downarrow$ $\text{PEU}$ & \bfseries 0.5877 & 0.6327 & 0.7029 & 0.6956 & 0.7071 & 0.6910 & 0.6869 \\
 & & $\downarrow$ $\text{UF}$ & 0.7098 & \bfseries 0.7098 & 0.7339 & 0.7305 & 0.7336 & 0.7309 & 0.7246 \\
\cline{2-10}
 & \multirow[c]{4}{*}{\rotatebox[origin=c]{90}{\small\textsc{Fair} (PUF)}} & $\downarrow$ $\text{PUF-Prec-Cos}$ & 0.0553 & 0.0502 & 0.0504 & 0.0503 & \bfseries 0.0497 & 0.0529 & 0.0525 \\
 &  & $\downarrow$ $\text{PUF-Prec-Jacc}$ & 0.0329 & 0.0298 & 0.0299 & 0.0299 & \bfseries 0.0296 & 0.0314 & 0.0312 \\
 &  & $\downarrow$ $\text{PUF-NDCG-Cos}$ & 0.0605 & 0.0548 & 0.0559 & 0.0552 & \bfseries 0.0532 & 0.0581 & 0.0576 \\
 &  & $\downarrow$ $\text{PUF-NDCG-Jacc}$ & 0.0361 & 0.0326 & 0.0333 & 0.0328 & \bfseries 0.0317 & 0.0346 & 0.0343 \\
\bottomrule
\end{tabular}
}
\end{table}

\subsubsection{Varying relevance score distribution}
\label{PUF_app:artificial_rel}

\begin{figure*}
    \includegraphics[width=0.98\textwidth, trim=0.25cm 0.2cm 0.25cm 0.2cm, clip=True]{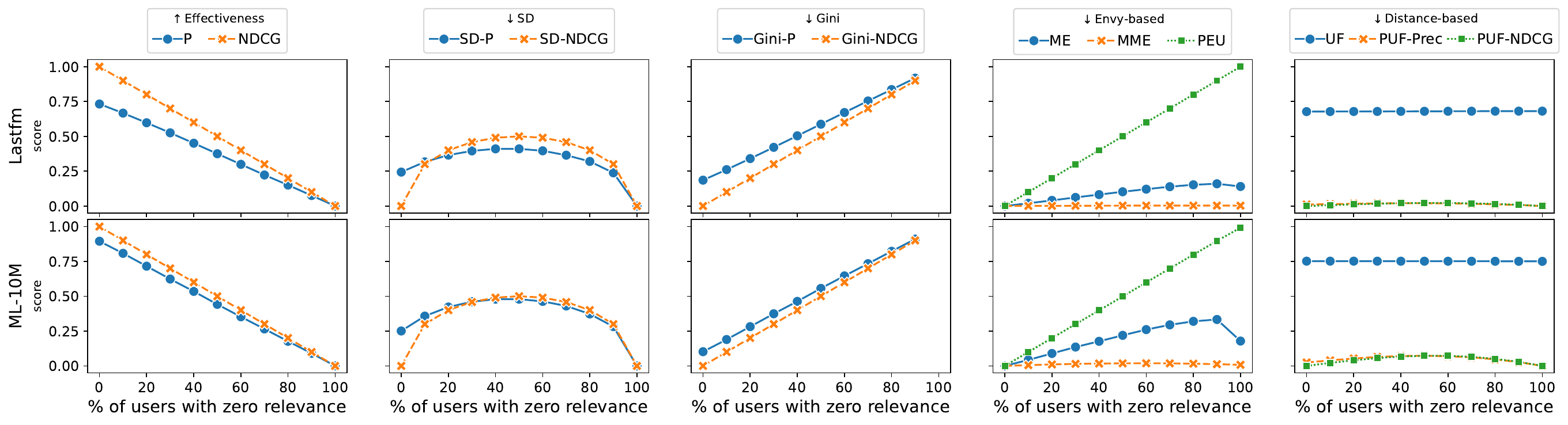}
    \caption{
    Effectiveness (\textsc{Eff}) and fairness (\textsc{Fair}) scores of Lastfm and ML-10M, when artificially varying \% of users with all irrelevant items (zero relevance), and the rest of the users receiving all relevant items. All PUF variants overlap. Gini is missing points at 100\% users with zero relevance as it is undefined when each user has zero \textsc{Eff} scores. 
    }
    \label{PUF_fig:ext-artificial-relevance}
\end{figure*}

We present the results of artificially changing the relevance score distribution for Lastfm and ML-10M in \Cref{PUF_fig:ext-artificial-relevance}.

\subsubsection{Varying user similarity distribution}
\label{PUF_app:artificial_sim}

\begin{figure*}
\centering
\includegraphics[width=0.75\linewidth]{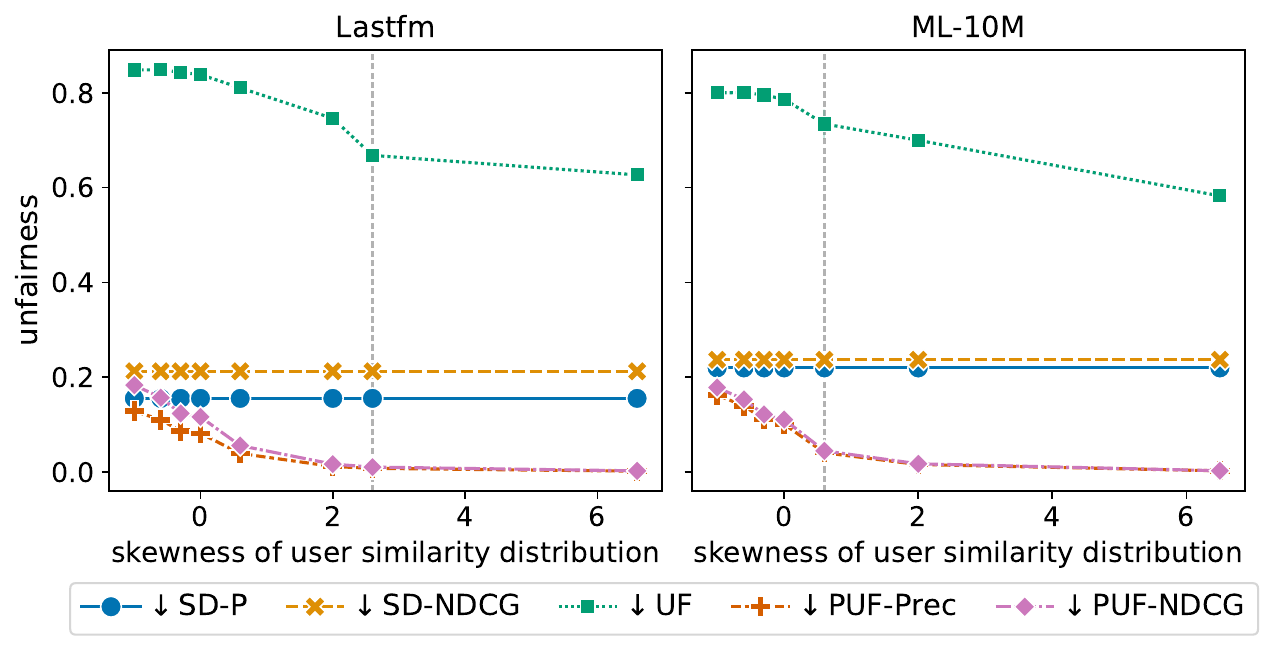}
    \caption{Artificially varying the skewness of the     user similarity distribution for Lastfm and ML-10M. Vertical grey lines denote the skewness corresponding to $sim_{Jacc}$ observed in the dataset. The distribution skewness differs across datasets.
    }
    \label{PUF_fig:ext-artificial-similarity}
\end{figure*}

We present the results of artificially changing the user similarity distribution for Lastfm and ML-10M in \Cref{PUF_fig:ext-artificial-similarity}.

\subsubsection{PUF and UF under extreme cases}
\label{PUF_app:extreme}

\begin{figure*}
    \centering
    \includegraphics[width=0.85\linewidth, trim=0.2cm 0.2cm 0.2cm 0.2cm, clip=True]{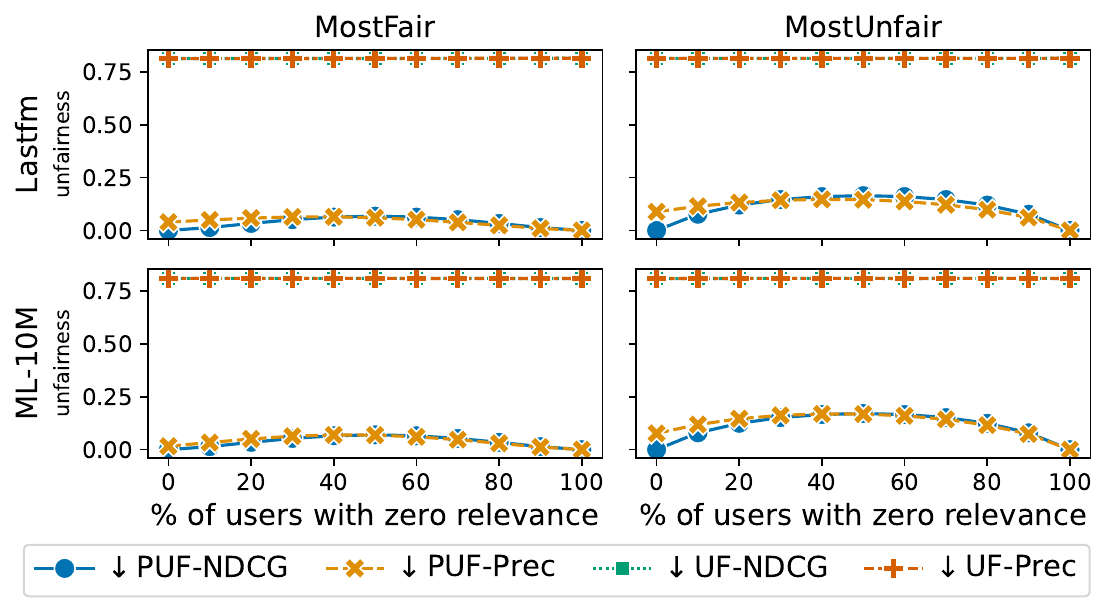}
    \caption{Artificially varying the \% of users with zero relevance for Lastfm and ML-10M. Lower \textsc{Eff} score difference is assigned to user pairs with higher similarity (MostFair), and to lower similarity (MostUnfair). Both UF overlap.}  
    \label{PUF_fig:ext-artificial-sim-relevance-sort}
\end{figure*}

We present the results of computing PUF and UF under extreme cases for Lastfm and ML-10M in \Cref{PUF_fig:ext-artificial-sim-relevance-sort}.

\subsubsection{Using item features for PUF and UF}
\label{PUF_app:item_feature}

\begin{figure*}
    \centering
        \includegraphics[width=0.75\linewidth]{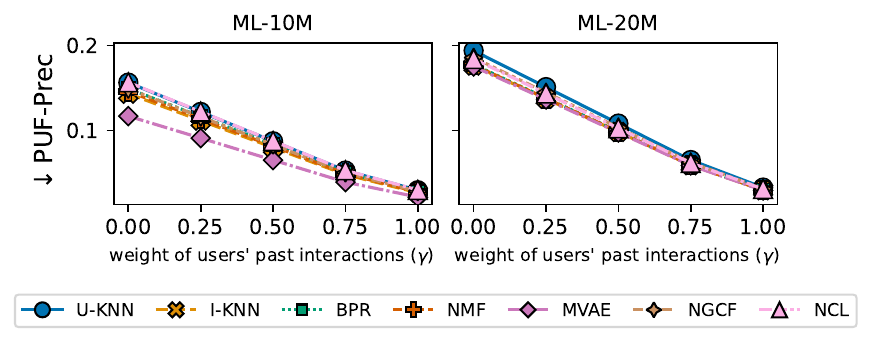} 
    \caption{
                PUF-Prec computed with user similarity $sim_{UF}$ (\Cref{PUF_eq:sim_UF}), varying the weighted sum of the users' past interactions ($sim_{Jacc}$) and their item feature distribution ($sim_{JS}$) with $\gamma$.
            }
    \label{PUF_fig:ext_varying_gamma}
\end{figure*}

\begin{figure*}
     \centering
        \includegraphics[width=0.75\linewidth]{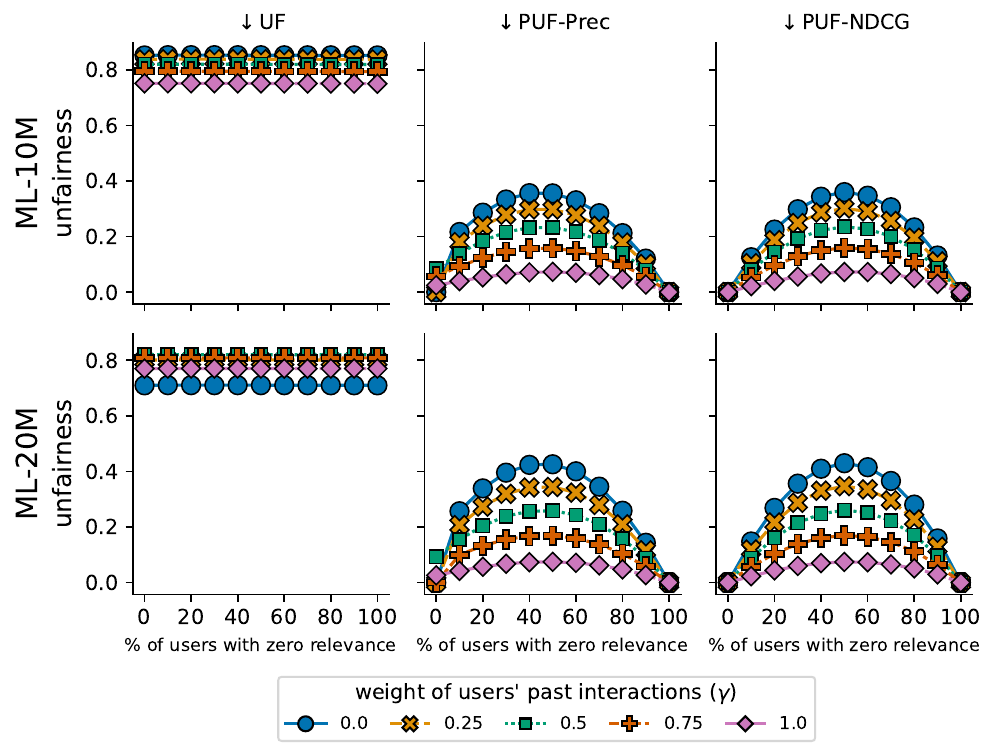} 
        \caption{Varying the weighted sum of users' past interactions ($sim_{Jacc}$) and their item feature distribution ($sim_{JS}$) with $\gamma$ and artificially varying the \% of users with zero relevance.}
    \label{PUF_fig:ext_varying_gamma_artificial_rel}   
\end{figure*}

\Cref{PUF_fig:ext_varying_gamma} shows the results of computing PUF-Prec, where user similarity $sim_{UF}$ (\Cref{PUF_eq:sim_UF}) is computed with varying weighted sums of user interactions and their item feature distribution. \Cref{PUF_fig:ext_varying_gamma_artificial_rel} shows the results of computing UF and PUF for varying weights of user interactions in $sim_{UF}$ and artificially varying relevance distribution (i.e., the runs from \Cref{PUF_ss:artificial-rel}).

\chapter{Stairway to Fairness: Connecting Group and Individual Fairness}
\label{chap:intersectional}
\section*{Abstract}
Fairness in recommender systems (RSs) is commonly categorised into group fairness and individual fairness. However, there is no established scientific understanding of the relationship between the two fairness types, as prior work on both types has used different evaluation measures or evaluation objectives for each fairness type, thereby not allowing for a proper comparison of the two. As a result, it is currently not known how increasing one type of fairness may affect the other. To fill this gap, we study the relationship of group and individual fairness through a comprehensive comparison of evaluation measures that can be used for both fairness types. 
Our experiments with 8 RSs across 3 datasets show that recommendations that are highly fair for groups can be very unfair for individuals. Our finding is novel and useful for RS practitioners aiming to improve the fairness of their systems. 

\section{Introduction}

Fairness in Recommender Systems (RSs) can be evaluated for groups and for individuals. \textit{Group fairness} typically refers to having equitable outcome across groups (e.g., similar effectiveness between groups of users \cite{Ekstrand2018AllEffectiveness}), while \textit{individual fairness} is commonly defined as treating similar users/items equally \cite{Wang2023ASystems} (e.g., similar effectiveness for all users \cite{Wu2021TFROM:Providers}). Conceptually, prior work \cite{LiYunqi2023FairnessApplications,Do2023FairnessThesis,Singh2018FairnessRankings} discusses how RSs can be fair to groups and at the same time unfair to individuals, or vice versa, but no work has empirically studied how this practically occurs in RS fairness evaluation. 

Prior work either: 
(i) evaluates fairness exclusively for groups or individuals \cite{Deldjoo2024FairnessDirections}; or 
(ii) evaluates both, but with two different families of measures \cite{Ferraro2024GenderImbalance,Pastor2024IntersectionalDivergence,pellegrini2023fairnessallinvestigatingharms} or for two fairness subjects/objectives \cite{Rastegarpanah2019FightingSystems, Wu2021TFROM:Providers}. 
Evaluating group and individual fairness with different families of measures makes comparison difficult, as the measure scores may differ in sensitivity, or in theoretical and empirical ranges \cite{Rampisela2024CanRelevance,Rampisela2024EvaluationStudy,Schumacher2024PropertiesRankings}. 
Likewise, it is not possible to properly compare group and individual fairness when each is evaluated for a distinct fairness objective, e.g., recommendation effectiveness disparity across individual users vs.~exposure disparity between item groups \cite{Wu2021TFROM:Providers}. To address this gap, we evaluate user-side group and individual fairness with the same families of measures that can quantify both, to study the relationship between evaluation measures of user-side group and individual fairness. This work is the first empirical study that compares the 9 existing user-side fairness evaluation measures for groups with those for individuals. We ask:

\begin{enumerate}
     \item To what extent do group and individual fairness evaluation measures differ in their conclusions?
     \item For the same family of measures, how different are the group and individual fairness scores? 
     \item How do different ways of grouping users affect between- and within-group fairness?
     \item How do between- and within-group fairness relate to individual fairness?
\end{enumerate}

Our results show that group fairness measures often hide unfairness within groups and between individuals, 
highlighting the importance of evaluating fairness beyond the between-group level.

\section{Methodology}
\label{s:setup}

We compare evaluation measures of user-side individual and group fairness in RSs, considering multiple ways of grouping users.\footnote{Our code is freely available at: \url{https://github.com/theresiavr/stairway-to-fairness}.}

\subsubsection*{Datasets} To enable group fairness analysis, three datasets with $\geq 3$ user profile features are selected (see \Cref{tab:stat_and_group} for statistics).

\noindent\textbf{ML-1M} \cite{Harper2015TheContext}  has 1,000,029 movie ratings (1--5) from 6K users. Users with no/unspecified self-reported gender, age, or occupation are removed, and we exclude users under 18 years to avoid processing the data of minors. We focus on recommending preferred movies, so ratings <3 are discarded, and the levels 4 and 5 are mapped to 1.

\noindent\textbf{JobRec} \cite{Hamner2012JobChallenge} has 1.6M job applications from 321K users. Given a user's application history, we focus on recommending job titles that may suit them, keeping only users with information for degree, major, and years of experience. Users with more than 60 years of experience are removed (as this likely indicates erroneous entries).
    
\noindent\textbf{LFM-1B} \cite{Schedl2016TheRecommendation} has $1{,}088{,}161{,}692$ music playcounts, from $\sim$120K users. We focus on recommending new track artists for a user to listen to, other than the ones they have listened to in the past, using the dataset after deduplication based on the artist, with 65M interactions (provided by \texttt{RecBole} \cite{Zhao2021RecBole:Algorithms}). The deduplication summarises the total playcount per artist and keeps the last event timestamp. Users without countries, age, or gender information are removed, as are minors (as in ML-1M), and users with age >100 years (as this likely indicates erroneous entries).

Items without name/title are removed from all datasets. To reduce data sparsity, which may affect LLMRec performance \cite{jiang2025beyond}, we keep users and items with $\geq$ 5 interactions (5-core filtering) for ML-1M and JobRec. We apply 50-core filtering \cite{Makhneva2023MakeSensitive,Wen2023EfficientFiltering,Zhao2023KuaiSim:Systems} to LFM-1B, as it is highly sparse with 5-core filtering. 
The data is temporally split for train/val/test with a ratio of 3:1:1 using a global timeline \cite{Meng2020ExploringModels}. From all splits, users and items with $\leq t$ interactions in the train set are removed. 
A high $t$ can result in very few unique users in the test set, so we choose $t$ such that at least 500 test users remain. For ML-1M and LFM-1B, we set $t=5$ \cite{Xu2024AModels}. For JobRec, we use $t=2$. We remove users/items in the val and test sets that are not in the train set.

\begin{table}
\centering
\caption{Preprocessed dataset statistics. $n_{G_a}$ is the number of groups for sensitive attribute $a$. We exclude empty groups.}

\label{tab:stat_and_group}
    \resizebox{0.95\linewidth}{!}{
    \begin{tabular}{lrrr}
    \toprule
            & ML-1M \cite{Harper2015TheContext} & JobRec \cite{Hamner2012JobChallenge} & LFM-1B \cite{Schedl2016TheRecommendation} \\
    \midrule
    \#interaction (all splits)    & 467,218  & 210,921    &   15,024,267 \\
    \#item (all splits)  &  3,030   & 19,912    &  51,204  \\
    \#user (test set)& 620 & 523  & 16,611     \\
    \midrule
    sensitive attr. \#1 ($n_{G_1}$)      & gender (2)                        &   degree (3)     & gender (2)                        
    \\
    sensitive attr. \#2 ($n_{G_2}$)     & age (3) &   years of experience (3)      & age (3) 
    \\
    sensitive attr. \#3 ($n_{G_3}$)     & occupation (2) & major (6) & country (5) 
    \\
    \midrule
    \#intersectional groups & 12    &  36   &       29     \\
    min--max subgroup size      & 2--279     &  1--70    &       1--3,919       \\
    median subgroup size     & 31    &  7   &       59      \\
    \bottomrule
    \end{tabular}}
\end{table}

\subsubsection*{User Grouping} To study group fairness, we cluster users based on their 
sensitive attributes (see \Cref{tab:stat_and_group} and \Cref{preproc}). Users cannot belong to two groups at the same time, e.g., age<50 and age$\geq$50. 

For ML-1M, we use gender, age, and occupation as sensitive attributes. Gender is used as is. Age is grouped into: 18--24 years, 25--49 years, and $\geq$50 years \cite{OfficeforNationalStatistics2023Age2021}. 
User occupation is grouped into: non-working (student \cite{U.S.BureauofLaborStatistics2024CollegeSummary,Eurostat2018YoungStatistics}, homemaker, retired, and unemployed) and working (14 occupations ranging from farmer to executive \cite{U.S.BureauofLaborStatistics2023MayEstimates}).
    
For JobRec,  we consider the user's academic degree, years of working experience, and study major as the sensitive attributes. Degree is grouped into: high school, college (associate or vocational degree), and university (bachelor's, master's, and PhD). Years of experience are grouped into: $\leq$5 years, >5--10 years, and >10 years. We group study majors into six fields of study, as per \citeauthor{Xu2024AModels}~\cite{Xu2024AModels}, using manual annotation and fuzzy string matching.\footnote{Details are provided in \Cref{preproc}.}

For LFM-1B, the sensitive attributes are gender, age, and country. Gender and age are processed as for ML-1M, and the user's country is mapped to the continent.\footnote{We use the country-continent mapping from \url{https://gist.github.com/achuhunkin/6cb1cbceb23395300aa209aad09e6e5d}, and manually group transcontinental countries.} Users from the North/South Americas are grouped together with Antarctica \cite{Xu2024AModels,Gomez2024AMBAR:Recommenders}.

\subsubsection*{LLM-based Recommenders} Recent work has utilised Large Language Models (LLMs) as recommenders (LLMRecs), with promising results \cite{Hou2024LargeSystems}. Unlike collaborative filtering models, LLMRecs can easily handle fine-grained user attributes, although including sensitive attributes in the prompt can impact effectiveness and fairness \cite{Deldjoo2025CFaiRLLM,Xu2024AModels,Zhang2023IsRecommendation}. 
We therefore study the effectiveness and fairness of LLMRecs under few-shot learning. 
To ensure comparable performance, we use four open-source, similar-sized LLMs released in July--Nov'24: Llama-3.1-8B-Instruct \cite{grattafiori2024llama3herdmodels}, Qwen2.5-7B-Instruct \cite{qwen2.5}, GLM-4-9B-chat \cite{glm2024chatglm}, and Ministral-8B-Instruct-2410 \cite{Mistral}. 
The temperature is fixed at 0 for each LLM to obtain deterministic output.

The LLMs are prompted using in-context learning strategy \cite{Hou2024LargeSystems}, as this has been shown to outperform sequential and recency-focused prompting.\footnote{The full prompt templates and examples are provided in \Cref{prompt}.}
We only prompt for users that exist in the test set, as otherwise it is not possible to evaluate the recommendation effectiveness. 
In the prompt, we provide the users' train items as the interaction history and the val items as the few-shot samples. 
A maximum of 10 most recent train and val items each are provided, as having too many items in the prompt may reduce effectiveness \cite{jiang2025beyond, Hou2024LargeSystems}. 
To avoid inflated performance, we do not prompt with a sampled candidate item list \cite{Krichene2020OnRecommendation}. Instead, we add restriction in the prompt to narrow down the item search space, e.g., for ML-1M, the movies should be between certain years, based on the movie release year in the metadata file. For LFM-1B, the prompt also includes the playcount, which is important in music recommendation \cite{Gomez2024AMBAR:Recommenders}.

Based on the inclusion/exclusion of user sensitive attributes, we create two prompt types \cite{Deldjoo2025CFaiRLLM, Zhang2023IsRecommendation, Tommasel2024FairnessRecommendations,deldjoo2024normativeframeworkbenchmarkingconsumer}: Sensitive (S), which has both interaction history and all three sensitive attributes, and Non-Sensitive (NS), which has only the interaction history.

To evaluate LLMRecs, we perform string matching between the list of recommended items and items in the test set, by using the TF-IDF \cite{Jones1972ARetrieval} of the items' character-based n-gram \cite{Lian2024RecAI:Systems}. The recommended item is deemed relevant if similarity $\geq0.75$. Our LLMRecs experiments are carried out with \texttt{vllm} \cite{Kwon2023EfficientPagedAttention} and \texttt{RecLM-eval} \cite{Lian2024RecAI:Systems}. 

\subsubsection*{Evaluation} 
Recommendation effectiveness (\eff{}) and fairness (\fair{}) are measured at $k=10$ for all LLMRecs. For \eff{}, the mean Hit Rate (HR), MRR, P@$k$ (P), and NDCG@$k$ \cite{Jarvelin2002CumulatedTechniques} are computed over all users. Group and individual \fair{} measures are computed in two steps: first, computing an \eff{} score per user/group as a `base score'; and second, aggregating the `base score' between users/groups with a \fair{} measure. P and NDCG are used as base scores to represent set- and rank-based measures. 

\noindent\textbf{Group fairness.} 
We compute all existing fairness measures for two or more user groups in RSs,\footnote{Measures that can only be used for exactly two groups are excluded.} 
that are published up to March 2025: Average scores of the worst 25\% groups (Min \cite{Wang2024IntersectionalRecommendation}), Range \cite{Liu2024SelfAdapative}, SD \cite{Zhang2023IsRecommendation,Liu2024SelfAdapative},  
MAD \cite{Fu2020Fairness-AwareGraphs}, 
Gini \cite{Pastor2024IntersectionalDivergence,Ferraro2024GenderImbalance,ghosh2024reducingpopulationlevelinequalityimprove}, 
CV \cite{Zhu2020MeasuringSystems}, 
FStat \cite{Wan2020AddressingRecommendations}, 
KL \cite{Amigo2023ASystems}, and 
GCE \cite{Deldjoo2021ASystems,Deldjoo2019RecommenderEntropy}. 
We also compute the Atkinson Index (Atk \cite{Atkinson1970OnInequality}), an income inequality measure that considers within-group variations.\footnote{This measure can be transformed into Generalised Entropy \cite{Speicher2018UnifiedIndices,Shorrocks1980ClassDecomposable}.} 
The between- and within-group fairness version of the measures are denoted as $\cdot_\text{b-group}$ and $\cdot_\text{w-group}$ respectively.\footnote{The term \textit{group fairness} refers to between-group fairness; the latter is used when we compare fairness between and within groups.} 
We provide the measure formulations and technical details in \Cref{setup_eval}.

\noindent\textbf{Individual fairness.} 
Fairness for individual users is quantified with SD \cite{Patro2020FairRec:Platforms}, Gini \cite{Leonhardt2018UserSystems}, and Atk. The subscript $\cdot$\ind{} indicates the individual fairness version of the measure. 
SD and Gini are the only \fair{} measures that have been used for both individual and group user fairness, while Atk\ind{} can be decomposed into between- and within-group fairness with no residuals \cite{Blackorby1999,Dayioğlu02006ImputedIndex,Bourguignon1979DecomposableMeasures}. 
While other group \fair{} measures can also be used to measure individual fairness, their scores may not be informative, e.g., Min may be zero for most models, as having most users scoring P=0 or NDCG=0 is common.

\section{Empirical Analysis}
\label{s:experiment}

\subsubsection*{Evaluation of all LLMRecs}

\begin{table}
    \caption{
 Effectiveness (\eff{}) and fairness (\fair{}) scores at cut-off $k=10$ for intersectional groups (\textsc{Grp.})~and for individuals (\textsc{Ind.})~of LLMRecs with non-sensitive (NS) and sensitive (S) prompts. All \fair{} scores are computed with NDCG. All measures range in [0,1], except for the \textsc{Grp.} measures below them. The best \textsc{Eff}/\fair{} scores are bolded. 
 Darker green marks scores closer to the best \eff{}/\fair{} per measure. 
 $\uparrow/\downarrow$ means the higher/lower the better. 
 }
    \label{tab:col_LLM-NDCG}
    \centering
    \resizebox{0.75\columnwidth}{!}{
\begin{tabular}{ll*{2}{r}|*{2}{r}|*{2}{r}|*{2}{r}}
\toprule
\toprule
 &  LLMRec& \multicolumn{2}{c|}{GLM-4-9B} & \multicolumn{2}{c|}{Llama-3.1-8B} & \multicolumn{2}{c|}{Ministral-8B} & \multicolumn{2}{c}{Qwen2.5-7B} \\ 
\midrule
 &  prompt type & NS & S & NS & S & NS & S & NS & S \\
\midrule
\midrule
 &  & \multicolumn{8}{c}{ML-1M} \\ 
\midrule
\multirow[c]{3}{*}{\rotatebox[origin=c]{90}{\textsc{Eff}}} & $\uparrow$ HR & \bfseries {\cellcolor[HTML]{2E8B57}} \color[HTML]{000000} 0.377 & {\cellcolor[HTML]{4C9C6F}} \color[HTML]{000000} 0.358 & {\cellcolor[HTML]{EBF3ED}} \color[HTML]{000000} 0.260 & {\cellcolor[HTML]{DDEBE2}} \color[HTML]{000000} 0.269 & {\cellcolor[HTML]{66AA84}} \color[HTML]{000000} 0.342 & {\cellcolor[HTML]{69AB86}} \color[HTML]{000000} 0.340 & {\cellcolor[HTML]{449769}} \color[HTML]{000000} 0.363 & {\cellcolor[HTML]{38905F}} \color[HTML]{000000} 0.371 \\
 & $\uparrow$ MRR & \bfseries {\cellcolor[HTML]{2E8B57}} \color[HTML]{000000} 0.189 & {\cellcolor[HTML]{4E9C70}} \color[HTML]{000000} 0.174 & {\cellcolor[HTML]{EBF3ED}} \color[HTML]{000000} 0.101 & {\cellcolor[HTML]{D2E5D9}} \color[HTML]{000000} 0.113 & {\cellcolor[HTML]{6EAE8A}} \color[HTML]{000000} 0.159 & {\cellcolor[HTML]{509E72}} \color[HTML]{000000} 0.173 & {\cellcolor[HTML]{61A780}} \color[HTML]{000000} 0.165 & {\cellcolor[HTML]{419666}} \color[HTML]{000000} 0.180 \\
 & $\uparrow$ NDCG & \bfseries {\cellcolor[HTML]{2E8B57}} \color[HTML]{000000} 0.231 & {\cellcolor[HTML]{4F9D72}} \color[HTML]{000000} 0.215 & {\cellcolor[HTML]{EBF3ED}} \color[HTML]{000000} 0.140 & {\cellcolor[HTML]{DBEAE0}} \color[HTML]{000000} 0.148 & {\cellcolor[HTML]{6EAE8A}} \color[HTML]{000000} 0.200 & {\cellcolor[HTML]{5DA57D}} \color[HTML]{000000} 0.208 & {\cellcolor[HTML]{60A67E}} \color[HTML]{000000} 0.207 & {\cellcolor[HTML]{439667}} \color[HTML]{000000} 0.221 \\
\cline{1-10}
\multirow[c]{10}{*}{\rotatebox[origin=c]{90}{\textsc{Fair (Grp.)}}} & $\uparrow$ Min\  & \bfseries {\cellcolor[HTML]{2E8B57}} \color[HTML]{000000} 0.166 & {\cellcolor[HTML]{61A780}} \color[HTML]{000000} 0.137 & {\cellcolor[HTML]{BCD9C8}} \color[HTML]{000000} 0.086 & {\cellcolor[HTML]{94C3A8}} \color[HTML]{000000} 0.108 & {\cellcolor[HTML]{EBF3ED}} \color[HTML]{000000} 0.059 & {\cellcolor[HTML]{96C4AA}} \color[HTML]{000000} 0.107 & {\cellcolor[HTML]{DBEAE1}} \color[HTML]{000000} 0.068 & {\cellcolor[HTML]{B3D4C0}} \color[HTML]{000000} 0.091 \\
 & $\downarrow$ Range\  & \bfseries {\cellcolor[HTML]{2E8B57}} \color[HTML]{000000} 0.188 & {\cellcolor[HTML]{3A9160}} \color[HTML]{000000} 0.208 & {\cellcolor[HTML]{94C3A8}} \color[HTML]{000000} 0.356 & {\cellcolor[HTML]{A0CAB2}} \color[HTML]{000000} 0.376 & {\cellcolor[HTML]{6CAD88}} \color[HTML]{000000} 0.290 & {\cellcolor[HTML]{EBF3ED}} \color[HTML]{000000} 0.500 & {\cellcolor[HTML]{82B99A}} \color[HTML]{000000} 0.328 & {\cellcolor[HTML]{62A780}} \color[HTML]{000000} 0.274 \\
 & $\downarrow$ SD\  & \bfseries {\cellcolor[HTML]{2E8B57}} \color[HTML]{000000} 0.055 & {\cellcolor[HTML]{3E9463}} \color[HTML]{000000} 0.061 & {\cellcolor[HTML]{87BC9E}} \color[HTML]{000000} 0.088 & {\cellcolor[HTML]{87BC9E}} \color[HTML]{000000} 0.088 & {\cellcolor[HTML]{7FB797}} \color[HTML]{000000} 0.085 & {\cellcolor[HTML]{EBF3ED}} \color[HTML]{000000} 0.125 & {\cellcolor[HTML]{94C3A8}} \color[HTML]{000000} 0.093 & {\cellcolor[HTML]{69AB86}} \color[HTML]{000000} 0.077 \\
 & $\downarrow$ MAD\  & \bfseries {\cellcolor[HTML]{2E8B57}} \color[HTML]{000000} 0.067 & {\cellcolor[HTML]{449769}} \color[HTML]{000000} 0.076 & {\cellcolor[HTML]{6AAC87}} \color[HTML]{000000} 0.091 & {\cellcolor[HTML]{60A77F}} \color[HTML]{000000} 0.087 & {\cellcolor[HTML]{7FB797}} \color[HTML]{000000} 0.099 & {\cellcolor[HTML]{EBF3ED}} \color[HTML]{000000} 0.142 & {\cellcolor[HTML]{9FC9B1}} \color[HTML]{000000} 0.112 & {\cellcolor[HTML]{68AB85}} \color[HTML]{000000} 0.090 \\
 & $\downarrow$ Gini\  & \bfseries {\cellcolor[HTML]{2E8B57}} \color[HTML]{000000} 0.130 & {\cellcolor[HTML]{56A177}} \color[HTML]{000000} 0.161 & {\cellcolor[HTML]{D1E4D9}} \color[HTML]{000000} 0.255 & {\cellcolor[HTML]{ABD0BB}} \color[HTML]{000000} 0.226 & {\cellcolor[HTML]{C8DFD2}} \color[HTML]{000000} 0.248 & {\cellcolor[HTML]{EBF3ED}} \color[HTML]{000000} 0.275 & {\cellcolor[HTML]{D1E4D9}} \color[HTML]{000000} 0.255 & {\cellcolor[HTML]{82B99A}} \color[HTML]{000000} 0.195 \\
 & $\downarrow$ Atk\  & \bfseries {\cellcolor[HTML]{2E8B57}} \color[HTML]{000000} 0.015 & {\cellcolor[HTML]{46986A}} \color[HTML]{000000} 0.020 & {\cellcolor[HTML]{96C4AA}} \color[HTML]{000000} 0.036 & {\cellcolor[HTML]{60A67E}} \color[HTML]{000000} 0.025 & {\cellcolor[HTML]{88BC9E}} \color[HTML]{000000} 0.033 & {\cellcolor[HTML]{D8E8DE}} \color[HTML]{000000} 0.049 & {\cellcolor[HTML]{EBF3ED}} \color[HTML]{000000} 0.053 & {\cellcolor[HTML]{A5CCB6}} \color[HTML]{000000} 0.039 \\
\cline{2-10}
 & $\downarrow$ CV\  & \bfseries {\cellcolor[HTML]{2E8B57}} \color[HTML]{000000} 0.233 & {\cellcolor[HTML]{4E9C70}} \color[HTML]{000000} 0.285 & {\cellcolor[HTML]{EBF3ED}} \color[HTML]{000000} 0.540 & {\cellcolor[HTML]{D4E6DB}} \color[HTML]{000000} 0.502 & {\cellcolor[HTML]{BDD9C9}} \color[HTML]{000000} 0.465 & {\cellcolor[HTML]{E2EEE6}} \color[HTML]{000000} 0.525 & {\cellcolor[HTML]{BAD8C6}} \color[HTML]{000000} 0.460 & {\cellcolor[HTML]{80B898}} \color[HTML]{000000} 0.365 \\
 & $\downarrow$ FStat\  & \bfseries {\cellcolor[HTML]{2E8B57}} \color[HTML]{000000} 0.468 & {\cellcolor[HTML]{55A176}} \color[HTML]{000000} 0.714 & {\cellcolor[HTML]{6AAC87}} \color[HTML]{000000} 0.841 & {\cellcolor[HTML]{4C9B6F}} \color[HTML]{000000} 0.654 & {\cellcolor[HTML]{5EA57D}} \color[HTML]{000000} 0.767 & {\cellcolor[HTML]{B0D2BF}} \color[HTML]{000000} 1.278 & {\cellcolor[HTML]{EBF3ED}} \color[HTML]{000000} 1.645 & {\cellcolor[HTML]{A7CDB7}} \color[HTML]{000000} 1.220 \\
 & $\downarrow$ KL\  & {\cellcolor[HTML]{69AB86}} \color[HTML]{000000} 1.121 & {\cellcolor[HTML]{6EAE8A}} \color[HTML]{000000} 1.138 & {\cellcolor[HTML]{EBF3ED}} \color[HTML]{000000} 1.674 & {\cellcolor[HTML]{E4EEE7}} \color[HTML]{000000} 1.640 & \bfseries {\cellcolor[HTML]{2E8B57}} \color[HTML]{000000} 0.866 & {\cellcolor[HTML]{EBF3ED}} \color[HTML]{000000} 1.671 & {\cellcolor[HTML]{7CB695}} \color[HTML]{000000} 1.198 & {\cellcolor[HTML]{5CA47C}} \color[HTML]{000000} 1.063 \\
 & $\downarrow$ GCE\  & \bfseries {\cellcolor[HTML]{2E8B57}} \color[HTML]{000000} 0.028 & {\cellcolor[HTML]{2E8B57}} \color[HTML]{000000} 0.050 & {\cellcolor[HTML]{2E8B57}} \color[HTML]{000000} 0.112 & {\cellcolor[HTML]{2E8B57}} \color[HTML]{000000} 0.104 & {\cellcolor[HTML]{EBF3ED}} \color[HTML]{000000} 659.844 & {\cellcolor[HTML]{EBF3ED}} \color[HTML]{000000} 659.741 & {\cellcolor[HTML]{2E8B57}} \color[HTML]{000000} 0.239 & {\cellcolor[HTML]{2E8B57}} \color[HTML]{000000} 0.198 \\
\cline{1-10}
\multirow[c]{3}{*}{\rotatebox[origin=c]{90}{\textsc{Fair}} \rotatebox[origin=c]{90}{\textsc{(Ind.)}}} & $\downarrow$ SD\  & {\cellcolor[HTML]{EBF3ED}} \color[HTML]{000000} 0.330 & {\cellcolor[HTML]{D0E4D7}} \color[HTML]{000000} 0.320 & \bfseries {\cellcolor[HTML]{2E8B57}} \color[HTML]{000000} 0.262 & {\cellcolor[HTML]{499A6D}} \color[HTML]{000000} 0.272 & {\cellcolor[HTML]{B0D2BF}} \color[HTML]{000000} 0.309 & {\cellcolor[HTML]{D5E6DC}} \color[HTML]{000000} 0.322 & {\cellcolor[HTML]{AED1BD}} \color[HTML]{000000} 0.308 & {\cellcolor[HTML]{DBEAE0}} \color[HTML]{000000} 0.324 \\
 & $\downarrow$ Gini\  & \bfseries {\cellcolor[HTML]{2E8B57}} \color[HTML]{000000} 0.705 & {\cellcolor[HTML]{4E9C70}} \color[HTML]{000000} 0.721 & {\cellcolor[HTML]{EBF3ED}} \color[HTML]{000000} 0.799 & {\cellcolor[HTML]{DFECE4}} \color[HTML]{000000} 0.793 & {\cellcolor[HTML]{6CAD88}} \color[HTML]{000000} 0.736 & {\cellcolor[HTML]{6CAD88}} \color[HTML]{000000} 0.736 & {\cellcolor[HTML]{4E9C70}} \color[HTML]{000000} 0.721 & {\cellcolor[HTML]{429667}} \color[HTML]{000000} 0.715 \\
 & $\downarrow$ Atk\  & \bfseries {\cellcolor[HTML]{2E8B57}} \color[HTML]{000000} 0.636 & {\cellcolor[HTML]{4D9C70}} \color[HTML]{000000} 0.655 & {\cellcolor[HTML]{EBF3ED}} \color[HTML]{000000} 0.751 & {\cellcolor[HTML]{DCEAE1}} \color[HTML]{000000} 0.742 & {\cellcolor[HTML]{69AB86}} \color[HTML]{000000} 0.672 & {\cellcolor[HTML]{6BAC87}} \color[HTML]{000000} 0.673 & {\cellcolor[HTML]{48996C}} \color[HTML]{000000} 0.652 & {\cellcolor[HTML]{3B9261}} \color[HTML]{000000} 0.644 \\
\midrule
\midrule
 &  & \multicolumn{8}{c}{JobRec} \\ 
\midrule
\multirow[c]{3}{*}{\rotatebox[origin=c]{90}{\textsc{Eff}}} & $\uparrow$ HR & {\cellcolor[HTML]{5DA57C}} \color[HTML]{000000} 0.054 & {\cellcolor[HTML]{B8D7C5}} \color[HTML]{000000} 0.033 & {\cellcolor[HTML]{88BD9F}} \color[HTML]{000000} 0.044 & {\cellcolor[HTML]{EBF3ED}} \color[HTML]{000000} 0.021 & {\cellcolor[HTML]{509E72}} \color[HTML]{000000} 0.057 & {\cellcolor[HTML]{88BD9F}} \color[HTML]{000000} 0.044 & \bfseries {\cellcolor[HTML]{2E8B57}} \color[HTML]{000000} 0.065 & {\cellcolor[HTML]{509E72}} \color[HTML]{000000} 0.057 \\
 & $\uparrow$ MRR & {\cellcolor[HTML]{63A881}} \color[HTML]{000000} 0.037 & {\cellcolor[HTML]{A8CEB8}} \color[HTML]{000000} 0.023 & {\cellcolor[HTML]{C4DDCF}} \color[HTML]{000000} 0.017 & {\cellcolor[HTML]{EBF3ED}} \color[HTML]{000000} 0.009 & \bfseries {\cellcolor[HTML]{2E8B57}} \color[HTML]{000000} 0.048 & {\cellcolor[HTML]{77B391}} \color[HTML]{000000} 0.033 & {\cellcolor[HTML]{38905F}} \color[HTML]{000000} 0.046 & {\cellcolor[HTML]{3C9362}} \color[HTML]{000000} 0.045 \\
 & $\uparrow$ NDCG & {\cellcolor[HTML]{5AA37A}} \color[HTML]{000000} 0.041 & {\cellcolor[HTML]{AACFBA}} \color[HTML]{000000} 0.025 & {\cellcolor[HTML]{B0D2BE}} \color[HTML]{000000} 0.024 & {\cellcolor[HTML]{EBF3ED}} \color[HTML]{000000} 0.012 & \bfseries {\cellcolor[HTML]{2E8B57}} \color[HTML]{000000} 0.050 & {\cellcolor[HTML]{74B18E}} \color[HTML]{000000} 0.036 & \bfseries {\cellcolor[HTML]{2E8B57}} \color[HTML]{000000} 0.050 & {\cellcolor[HTML]{38905F}} \color[HTML]{000000} 0.048 \\
\cline{1-10}
\multirow[c]{10}{*}{\rotatebox[origin=c]{90}{\textsc{Fair (Grp.)}}} & $\uparrow$ Min\  & \bfseries {\cellcolor[HTML]{EBF3ED}} \color[HTML]{000000} 0.000 & \bfseries {\cellcolor[HTML]{EBF3ED}} \color[HTML]{000000} 0.000 & \bfseries {\cellcolor[HTML]{EBF3ED}} \color[HTML]{000000} 0.000 & \bfseries {\cellcolor[HTML]{EBF3ED}} \color[HTML]{000000} 0.000 & \bfseries {\cellcolor[HTML]{EBF3ED}} \color[HTML]{000000} 0.000 & \bfseries {\cellcolor[HTML]{EBF3ED}} \color[HTML]{000000} 0.000 & \bfseries {\cellcolor[HTML]{EBF3ED}} \color[HTML]{000000} 0.000 & \bfseries {\cellcolor[HTML]{EBF3ED}} \color[HTML]{000000} 0.000 \\
 & $\downarrow$ Range\  & {\cellcolor[HTML]{EBF3ED}} \color[HTML]{000000} 0.500 & \bfseries {\cellcolor[HTML]{2E8B57}} \color[HTML]{000000} 0.085 & {\cellcolor[HTML]{EBF3ED}} \color[HTML]{000000} 0.500 & {\cellcolor[HTML]{EBF3ED}} \color[HTML]{000000} 0.500 & {\cellcolor[HTML]{9FC9B1}} \color[HTML]{000000} 0.333 & {\cellcolor[HTML]{55A076}} \color[HTML]{000000} 0.170 & {\cellcolor[HTML]{9FC9B1}} \color[HTML]{000000} 0.333 & {\cellcolor[HTML]{9FC9B1}} \color[HTML]{000000} 0.333 \\
 & $\downarrow$ SD\  & {\cellcolor[HTML]{EBF3ED}} \color[HTML]{000000} 0.093 & \bfseries {\cellcolor[HTML]{2E8B57}} \color[HTML]{000000} 0.024 & {\cellcolor[HTML]{D8E8DE}} \color[HTML]{000000} 0.086 & {\cellcolor[HTML]{CDE2D6}} \color[HTML]{000000} 0.082 & {\cellcolor[HTML]{B7D6C4}} \color[HTML]{000000} 0.074 & {\cellcolor[HTML]{5FA67E}} \color[HTML]{000000} 0.042 & {\cellcolor[HTML]{B2D3C0}} \color[HTML]{000000} 0.072 & {\cellcolor[HTML]{AFD2BD}} \color[HTML]{000000} 0.071 \\
 & $\downarrow$ MAD\  & {\cellcolor[HTML]{EBF3ED}} \color[HTML]{000000} 0.066 & \bfseries {\cellcolor[HTML]{2E8B57}} \color[HTML]{000000} 0.019 & {\cellcolor[HTML]{B7D6C4}} \color[HTML]{000000} 0.053 & {\cellcolor[HTML]{6EAE8A}} \color[HTML]{000000} 0.035 & {\cellcolor[HTML]{E7F1EA}} \color[HTML]{000000} 0.065 & {\cellcolor[HTML]{6EAE8A}} \color[HTML]{000000} 0.035 & {\cellcolor[HTML]{E7F1EA}} \color[HTML]{000000} 0.065 & {\cellcolor[HTML]{C3DDCD}} \color[HTML]{000000} 0.056 \\
 & $\downarrow$ Gini\  & {\cellcolor[HTML]{82B99A}} \color[HTML]{000000} 0.828 & {\cellcolor[HTML]{89BD9F}} \color[HTML]{000000} 0.836 & {\cellcolor[HTML]{9DC8AF}} \color[HTML]{000000} 0.857 & {\cellcolor[HTML]{EBF3ED}} \color[HTML]{000000} 0.939 & {\cellcolor[HTML]{3E9463}} \color[HTML]{000000} 0.757 & {\cellcolor[HTML]{5DA57C}} \color[HTML]{000000} 0.789 & \bfseries {\cellcolor[HTML]{2E8B57}} \color[HTML]{000000} 0.740 & {\cellcolor[HTML]{6EAE8A}} \color[HTML]{000000} 0.808 \\
 & $\downarrow$ Atk\  & {\cellcolor[HTML]{54A075}} \color[HTML]{000000} 0.545 & {\cellcolor[HTML]{499A6D}} \color[HTML]{000000} 0.525 & {\cellcolor[HTML]{99C5AC}} \color[HTML]{000000} 0.676 & {\cellcolor[HTML]{EBF3ED}} \color[HTML]{000000} 0.833 & {\cellcolor[HTML]{47996B}} \color[HTML]{000000} 0.520 & {\cellcolor[HTML]{3B9262}} \color[HTML]{000000} 0.498 & \bfseries {\cellcolor[HTML]{2E8B57}} \color[HTML]{000000} 0.472 & {\cellcolor[HTML]{3C9362}} \color[HTML]{000000} 0.499 \\
\cline{2-10}
 & $\downarrow$ CV\  & {\cellcolor[HTML]{5BA47B}} \color[HTML]{000000} 2.392 & {\cellcolor[HTML]{499A6C}} \color[HTML]{000000} 2.114 & {\cellcolor[HTML]{7CB695}} \color[HTML]{000000} 2.883 & {\cellcolor[HTML]{EBF3ED}} \color[HTML]{000000} 4.549 & {\cellcolor[HTML]{318D59}} \color[HTML]{000000} 1.758 & {\cellcolor[HTML]{3C9362}} \color[HTML]{000000} 1.923 & \bfseries {\cellcolor[HTML]{2E8B57}} \color[HTML]{000000} 1.706 & {\cellcolor[HTML]{48996C}} \color[HTML]{000000} 2.102 \\
 & $\downarrow$ FStat\  & {\cellcolor[HTML]{66AA84}} \color[HTML]{000000} 1.034 & \bfseries {\cellcolor[HTML]{2E8B57}} \color[HTML]{000000} 0.562 & {\cellcolor[HTML]{ACD0BB}} \color[HTML]{000000} 1.612 & {\cellcolor[HTML]{EBF3ED}} \color[HTML]{000000} 2.136 & {\cellcolor[HTML]{63A881}} \color[HTML]{000000} 1.003 & {\cellcolor[HTML]{338E5B}} \color[HTML]{000000} 0.611 & {\cellcolor[HTML]{57A278}} \color[HTML]{000000} 0.909 & {\cellcolor[HTML]{5AA37A}} \color[HTML]{000000} 0.928 \\
 & $\downarrow$ KL\  & {\cellcolor[HTML]{7AB493}} \color[HTML]{000000} 3.208 & \bfseries {\cellcolor[HTML]{2E8B57}} \color[HTML]{000000} 1.445 & {\cellcolor[HTML]{9AC6AD}} \color[HTML]{000000} 3.969 & {\cellcolor[HTML]{EBF3ED}} \color[HTML]{000000} 5.855 & {\cellcolor[HTML]{409565}} \color[HTML]{000000} 1.873 & {\cellcolor[HTML]{3B9261}} \color[HTML]{000000} 1.747 & {\cellcolor[HTML]{48996C}} \color[HTML]{000000} 2.060 & {\cellcolor[HTML]{529E73}} \color[HTML]{000000} 2.279 \\
 & $\downarrow$ GCE\  & {\cellcolor[HTML]{64A982}} \color[HTML]{000000} 1685.925 & {\cellcolor[HTML]{D0E4D8}} \color[HTML]{000000} 1979.103 & {\cellcolor[HTML]{64A982}} \color[HTML]{000000} 1685.992 & {\cellcolor[HTML]{EBF3ED}} \color[HTML]{000000} 2052.498 & {\cellcolor[HTML]{499A6C}} \color[HTML]{000000} 1612.573 & {\cellcolor[HTML]{7FB797}} \color[HTML]{000000} 1759.176 & \bfseries {\cellcolor[HTML]{2E8B57}} \color[HTML]{000000} 1539.277 & {\cellcolor[HTML]{7FB797}} \color[HTML]{000000} 1759.185 \\
\cline{1-10}
\multirow[c]{3}{*}{\rotatebox[origin=c]{90}{\textsc{Fair}} \rotatebox[origin=c]{90}{\textsc{(Ind.)}}} & $\downarrow$ SD\  & {\cellcolor[HTML]{BFDBCA}} \color[HTML]{000000} 0.183 & {\cellcolor[HTML]{87BC9E}} \color[HTML]{000000} 0.147 & {\cellcolor[HTML]{5EA57D}} \color[HTML]{000000} 0.121 & \bfseries {\cellcolor[HTML]{2E8B57}} \color[HTML]{000000} 0.090 & {\cellcolor[HTML]{EBF3ED}} \color[HTML]{000000} 0.211 & {\cellcolor[HTML]{B1D3BF}} \color[HTML]{000000} 0.174 & {\cellcolor[HTML]{E1EDE5}} \color[HTML]{000000} 0.204 & {\cellcolor[HTML]{DFECE4}} \color[HTML]{000000} 0.203 \\
 & $\downarrow$ Gini\  & {\cellcolor[HTML]{5AA37A}} \color[HTML]{000000} 0.956 & {\cellcolor[HTML]{B4D5C2}} \color[HTML]{000000} 0.974 & {\cellcolor[HTML]{8DBFA2}} \color[HTML]{000000} 0.966 & {\cellcolor[HTML]{EBF3ED}} \color[HTML]{000000} 0.985 & {\cellcolor[HTML]{38905F}} \color[HTML]{000000} 0.949 & {\cellcolor[HTML]{7DB696}} \color[HTML]{000000} 0.963 & \bfseries {\cellcolor[HTML]{2E8B57}} \color[HTML]{000000} 0.947 & {\cellcolor[HTML]{419666}} \color[HTML]{000000} 0.951 \\
 & $\downarrow$ Atk\  & {\cellcolor[HTML]{5EA57D}} \color[HTML]{000000} 0.948 & {\cellcolor[HTML]{BBD8C7}} \color[HTML]{000000} 0.969 & {\cellcolor[HTML]{8BBEA1}} \color[HTML]{000000} 0.958 & {\cellcolor[HTML]{EBF3ED}} \color[HTML]{000000} 0.980 & {\cellcolor[HTML]{4C9C6F}} \color[HTML]{000000} 0.944 & {\cellcolor[HTML]{86BB9D}} \color[HTML]{000000} 0.957 & \bfseries {\cellcolor[HTML]{2E8B57}} \color[HTML]{000000} 0.937 & {\cellcolor[HTML]{4C9C6F}} \color[HTML]{000000} 0.944 \\
\midrule
\midrule
 &  & \multicolumn{8}{c}{LFM-1B} \\ 
\midrule
\multirow[c]{3}{*}{\rotatebox[origin=c]{90}{\textsc{Eff}}} & $\uparrow$ HR & {\cellcolor[HTML]{2F8C58}} \color[HTML]{000000} 0.658 & \bfseries {\cellcolor[HTML]{2E8B57}} \color[HTML]{000000} 0.661 & {\cellcolor[HTML]{4A9A6D}} \color[HTML]{000000} 0.609 & {\cellcolor[HTML]{46986A}} \color[HTML]{000000} 0.618 & {\cellcolor[HTML]{A2CAB3}} \color[HTML]{000000} 0.451 & {\cellcolor[HTML]{EBF3ED}} \color[HTML]{000000} 0.317 & {\cellcolor[HTML]{CBE1D4}} \color[HTML]{000000} 0.375 & {\cellcolor[HTML]{EBF3ED}} \color[HTML]{000000} 0.317 \\
 & $\uparrow$ MRR & \bfseries {\cellcolor[HTML]{2E8B57}} \color[HTML]{000000} 0.409 & {\cellcolor[HTML]{2F8B58}} \color[HTML]{000000} 0.408 & {\cellcolor[HTML]{5DA57C}} \color[HTML]{000000} 0.347 & {\cellcolor[HTML]{55A176}} \color[HTML]{000000} 0.357 & {\cellcolor[HTML]{9BC7AE}} \color[HTML]{000000} 0.266 & {\cellcolor[HTML]{E1EDE5}} \color[HTML]{000000} 0.174 & {\cellcolor[HTML]{CDE2D6}} \color[HTML]{000000} 0.199 & {\cellcolor[HTML]{EBF3ED}} \color[HTML]{000000} 0.160 \\
 & $\uparrow$ NDCG & {\cellcolor[HTML]{2E8B57}} \color[HTML]{000000} 0.462 & \bfseries {\cellcolor[HTML]{2E8B57}} \color[HTML]{000000} 0.463 & {\cellcolor[HTML]{57A177}} \color[HTML]{000000} 0.406 & {\cellcolor[HTML]{509E72}} \color[HTML]{000000} 0.415 & {\cellcolor[HTML]{9BC7AE}} \color[HTML]{000000} 0.310 & {\cellcolor[HTML]{E4EFE8}} \color[HTML]{000000} 0.208 & {\cellcolor[HTML]{CDE2D5}} \color[HTML]{000000} 0.241 & {\cellcolor[HTML]{EBF3ED}} \color[HTML]{000000} 0.198 \\
\cline{1-10}
\multirow[c]{10}{*}{\rotatebox[origin=c]{90}{\textsc{Fair (Grp.)}}} & $\uparrow$ Min\  & {\cellcolor[HTML]{48996C}} \color[HTML]{000000} 0.232 & \bfseries {\cellcolor[HTML]{2E8B57}} \color[HTML]{000000} 0.262 & {\cellcolor[HTML]{358F5C}} \color[HTML]{000000} 0.254 & {\cellcolor[HTML]{368F5D}} \color[HTML]{000000} 0.252 & {\cellcolor[HTML]{B2D3C0}} \color[HTML]{000000} 0.109 & {\cellcolor[HTML]{DDEBE2}} \color[HTML]{000000} 0.060 & {\cellcolor[HTML]{C3DDCD}} \color[HTML]{000000} 0.090 & {\cellcolor[HTML]{EBF3ED}} \color[HTML]{000000} 0.043 \\
 & $\downarrow$ Range\  & {\cellcolor[HTML]{73B18E}} \color[HTML]{000000} 0.604 & {\cellcolor[HTML]{73B18E}} \color[HTML]{000000} 0.604 & {\cellcolor[HTML]{C8DFD2}} \color[HTML]{000000} 0.884 & {\cellcolor[HTML]{D6E7DD}} \color[HTML]{000000} 0.931 & {\cellcolor[HTML]{E9F2EC}} \color[HTML]{000000} 0.993 & {\cellcolor[HTML]{EBF3ED}} \color[HTML]{000000} 1.000 & {\cellcolor[HTML]{5AA37A}} \color[HTML]{000000} 0.525 & \bfseries {\cellcolor[HTML]{2E8B57}} \color[HTML]{000000} 0.378 \\
 & $\downarrow$ SD\  & {\cellcolor[HTML]{A9CEB9}} \color[HTML]{000000} 0.149 & {\cellcolor[HTML]{80B898}} \color[HTML]{000000} 0.131 & {\cellcolor[HTML]{C4DDCF}} \color[HTML]{000000} 0.161 & {\cellcolor[HTML]{9EC8B0}} \color[HTML]{000000} 0.144 & {\cellcolor[HTML]{EBF3ED}} \color[HTML]{000000} 0.178 & {\cellcolor[HTML]{E9F1EB}} \color[HTML]{000000} 0.177 & {\cellcolor[HTML]{65A983}} \color[HTML]{000000} 0.119 & \bfseries {\cellcolor[HTML]{2E8B57}} \color[HTML]{000000} 0.095 \\
 & $\downarrow$ MAD\  & {\cellcolor[HTML]{97C5AB}} \color[HTML]{000000} 0.141 & {\cellcolor[HTML]{69AB86}} \color[HTML]{000000} 0.126 & {\cellcolor[HTML]{DCEAE1}} \color[HTML]{000000} 0.163 & {\cellcolor[HTML]{8BBEA1}} \color[HTML]{000000} 0.137 & {\cellcolor[HTML]{EBF3ED}} \color[HTML]{000000} 0.168 & {\cellcolor[HTML]{C6DED0}} \color[HTML]{000000} 0.156 & {\cellcolor[HTML]{6FAF8B}} \color[HTML]{000000} 0.128 & \bfseries {\cellcolor[HTML]{2E8B57}} \color[HTML]{000000} 0.107 \\
 & $\downarrow$ Gini\  & {\cellcolor[HTML]{419566}} \color[HTML]{000000} 0.167 & \bfseries {\cellcolor[HTML]{2E8B57}} \color[HTML]{000000} 0.146 & {\cellcolor[HTML]{56A177}} \color[HTML]{000000} 0.191 & {\cellcolor[HTML]{449769}} \color[HTML]{000000} 0.171 & {\cellcolor[HTML]{AACFB9}} \color[HTML]{000000} 0.284 & {\cellcolor[HTML]{EBF3ED}} \color[HTML]{000000} 0.357 & {\cellcolor[HTML]{9CC7AE}} \color[HTML]{000000} 0.268 & {\cellcolor[HTML]{BEDAC9}} \color[HTML]{000000} 0.306 \\
 & $\downarrow$ Atk\  & {\cellcolor[HTML]{399160}} \color[HTML]{000000} 0.003 & \bfseries {\cellcolor[HTML]{2E8B57}} \color[HTML]{000000} 0.002 & {\cellcolor[HTML]{449769}} \color[HTML]{000000} 0.004 & {\cellcolor[HTML]{449769}} \color[HTML]{000000} 0.004 & {\cellcolor[HTML]{5AA37A}} \color[HTML]{000000} 0.006 & {\cellcolor[HTML]{B3D4C1}} \color[HTML]{000000} 0.014 & {\cellcolor[HTML]{71B08C}} \color[HTML]{000000} 0.008 & {\cellcolor[HTML]{EBF3ED}} \color[HTML]{000000} 0.019 \\
\cline{2-10}
 & $\downarrow$ CV\  & {\cellcolor[HTML]{409565}} \color[HTML]{000000} 0.366 & \bfseries {\cellcolor[HTML]{2E8B57}} \color[HTML]{000000} 0.315 & {\cellcolor[HTML]{499A6D}} \color[HTML]{000000} 0.393 & {\cellcolor[HTML]{429667}} \color[HTML]{000000} 0.372 & {\cellcolor[HTML]{9DC8AF}} \color[HTML]{000000} 0.624 & {\cellcolor[HTML]{EBF3ED}} \color[HTML]{000000} 0.841 & {\cellcolor[HTML]{76B290}} \color[HTML]{000000} 0.516 & {\cellcolor[HTML]{86BB9D}} \color[HTML]{000000} 0.561 \\
 & $\downarrow$ FStat\  & {\cellcolor[HTML]{449769}} \color[HTML]{000000} 2.348 & \bfseries {\cellcolor[HTML]{2E8B57}} \color[HTML]{000000} 2.019 & {\cellcolor[HTML]{80B898}} \color[HTML]{000000} 3.224 & {\cellcolor[HTML]{BAD8C6}} \color[HTML]{000000} 4.067 & {\cellcolor[HTML]{69AB86}} \color[HTML]{000000} 2.874 & {\cellcolor[HTML]{9CC8AF}} \color[HTML]{000000} 3.638 & {\cellcolor[HTML]{55A176}} \color[HTML]{000000} 2.601 & {\cellcolor[HTML]{EBF3ED}} \color[HTML]{000000} 4.788 \\
 & $\downarrow$ KL\  & {\cellcolor[HTML]{58A279}} \color[HTML]{000000} 2.608 & {\cellcolor[HTML]{6EAE8A}} \color[HTML]{000000} 2.740 & {\cellcolor[HTML]{E4EFE8}} \color[HTML]{000000} 3.447 & {\cellcolor[HTML]{B1D3BF}} \color[HTML]{000000} 3.139 & {\cellcolor[HTML]{99C6AC}} \color[HTML]{000000} 3.001 & {\cellcolor[HTML]{EBF3ED}} \color[HTML]{000000} 3.489 & {\cellcolor[HTML]{86BB9D}} \color[HTML]{000000} 2.885 & \bfseries {\cellcolor[HTML]{2E8B57}} \color[HTML]{000000} 2.354 \\
 & $\downarrow$ GCE\  & {\cellcolor[HTML]{8CBFA2}} \color[HTML]{000000} 338.843 & {\cellcolor[HTML]{5DA57C}} \color[HTML]{000000} 225.901 & {\cellcolor[HTML]{2E8B57}} \color[HTML]{000000} 112.993 & \bfseries {\cellcolor[HTML]{2E8B57}} \color[HTML]{000000} 112.988 & {\cellcolor[HTML]{BCD9C8}} \color[HTML]{000000} 451.823 & {\cellcolor[HTML]{BCD9C8}} \color[HTML]{000000} 451.872 & {\cellcolor[HTML]{BCD9C8}} \color[HTML]{000000} 451.824 & {\cellcolor[HTML]{EBF3ED}} \color[HTML]{000000} 564.768 \\
\cline{1-10}
\multirow[c]{3}{*}{\rotatebox[origin=c]{90}{\textsc{Fair}} \rotatebox[origin=c]{90}{\textsc{(Ind.)}}} & $\downarrow$ SD\  & {\cellcolor[HTML]{EBF3ED}} \color[HTML]{000000} 0.380 & {\cellcolor[HTML]{E5EFE9}} \color[HTML]{000000} 0.378 & {\cellcolor[HTML]{CCE2D4}} \color[HTML]{000000} 0.370 & {\cellcolor[HTML]{CFE3D7}} \color[HTML]{000000} 0.371 & {\cellcolor[HTML]{E2EEE6}} \color[HTML]{000000} 0.377 & {\cellcolor[HTML]{5AA37A}} \color[HTML]{000000} 0.334 & {\cellcolor[HTML]{73B18E}} \color[HTML]{000000} 0.342 & \bfseries {\cellcolor[HTML]{2E8B57}} \color[HTML]{000000} 0.320 \\
 & $\downarrow$ Gini\  & {\cellcolor[HTML]{2E8B57}} \color[HTML]{000000} 0.462 & \bfseries {\cellcolor[HTML]{2E8B57}} \color[HTML]{000000} 0.461 & {\cellcolor[HTML]{4D9C70}} \color[HTML]{000000} 0.510 & {\cellcolor[HTML]{48996C}} \color[HTML]{000000} 0.502 & {\cellcolor[HTML]{A1CAB2}} \color[HTML]{000000} 0.638 & {\cellcolor[HTML]{E9F2EC}} \color[HTML]{000000} 0.750 & {\cellcolor[HTML]{CBE1D4}} \color[HTML]{000000} 0.703 & {\cellcolor[HTML]{EBF3ED}} \color[HTML]{000000} 0.753 \\
 & $\downarrow$ Atk\  & {\cellcolor[HTML]{2F8C58}} \color[HTML]{000000} 0.361 & \bfseries {\cellcolor[HTML]{2E8B57}} \color[HTML]{000000} 0.358 & {\cellcolor[HTML]{4A9A6D}} \color[HTML]{000000} 0.409 & {\cellcolor[HTML]{459869}} \color[HTML]{000000} 0.400 & {\cellcolor[HTML]{A1CAB2}} \color[HTML]{000000} 0.563 & {\cellcolor[HTML]{EBF3ED}} \color[HTML]{000000} 0.694 & {\cellcolor[HTML]{CBE1D4}} \color[HTML]{000000} 0.638 & {\cellcolor[HTML]{EBF3ED}} \color[HTML]{000000} 0.695 \\
\bottomrule
\bottomrule
\end{tabular}}

\end{table}

\eff{} scores and NDCG-based \fair{} scores for group fairness and individual fairness of LLMRecs are shown in \Cref{tab:col_LLM-NDCG}; 
 the P-based \fair{} measures are presented in \Cref{ext_performance} as they show similar results. We evaluate group fairness considering the intersectionality of all 3 sensitive user attributes.
 
\noindent\textbf{\eff{} and \fair{} scores}. GLM-4-9B has the best \eff{} and \fair{} scores for all datasets, except for JobRec, where all LLMRecs have equally low \eff{}. 
This could be because the job titles are very specific and contain more noise than movie/artist names (e.g., `Chicago Concierge Manager', `Account Executive - 40k - 70K per year'). Generally, as group/individual \fair{} measures, Gini, Atk, and CV are fairer when NDCG is better; this is because their equations include division by the mean NDCG. As such, these measures can distinguish group fairness between two models that have similar SD$_\text{b-group}$/SD$_\text{ind}$ but differ in NDCG (e.g., the sensitive and non-sensitive prompts of Ministral-8B for LFM-1B), or vice versa.

\noindent\textbf{Comparison of sensitive and non-sensitive prompts}. The \eff{} and \fair{} scores between sensitive (S) and non-sensitive (NS) prompts are generally comparable, implying that including sensitive attributes in the prompt has little effect on effectiveness/fairness. An exception to this is Ministral-8B for LFM-1B, which may be due to the LLM's relatively high gender bias \cite{cohere2025commandaenterprisereadylarge}.

\noindent\textbf{Group vs individual fairness}. For Atk/Gini, the fairest model for intersectional groups and for individual users are always the same, but this is not the case for SD. Thus, if sensitive attribute information is missing, Atk$_\text{ind}$/Gini$_\text{ind}$ can estimate which model is the fairest for intersectional groups.

\subsubsection*{Agreement of group and individual fairness measures}

\begin{figure}
    \includegraphics[width=0.95\linewidth]{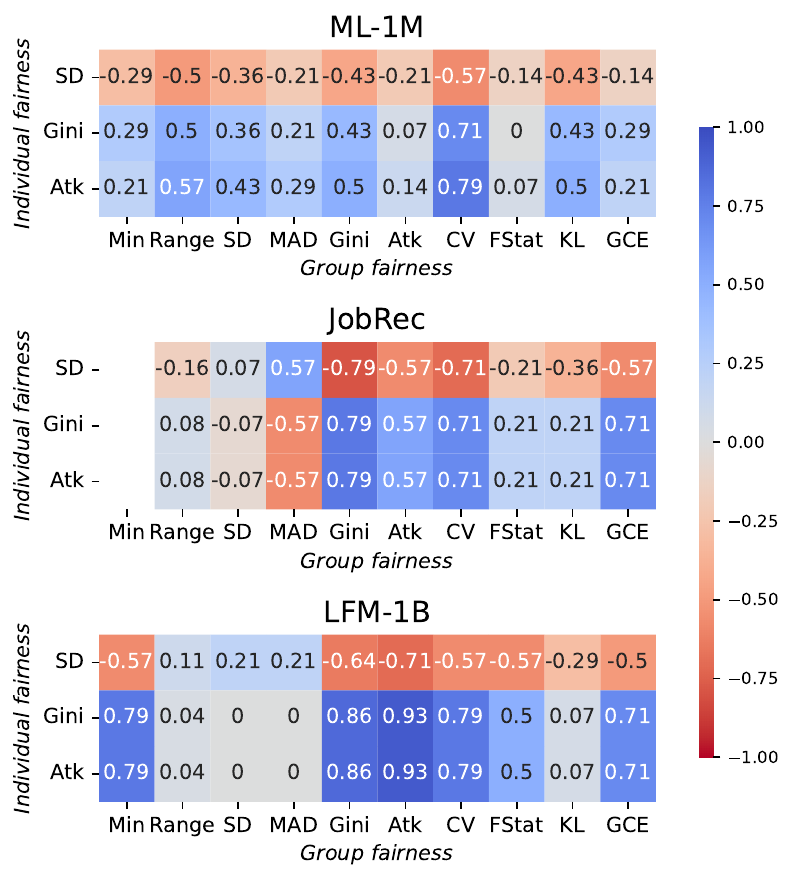}
    \caption{Agreement (Kendall's $\tau$) of NDCG-based measures for individual \fair{} ($y$-axis) and group \fair{} ($x$-axis) in ranking LLMRecs. User groups are based on 3 sensitive attributes. Due to 8-way ties, $\tau$ cannot be computed for Min (JobRec).
    }
    \label{fig:corr_ind_group}
\end{figure}

Do individual and group \fair{} measures reach the same conclusion? 
If an individual \fair{} measure can rank models from the most to the least fair equivalently to a group \fair{} measure, one can be a proxy for the other. 
To assess measure agreement in ranking models, we compute Kendall's $\tau$; we consider two rankings to be ``equivalent'' if $\tau \geq .9$ \cite{Maistro2021PrincipledRankings}. \Cref{fig:corr_ind_group} shows group and individual \fair{} measure agreement.

Our results show that \textbf{no individual \fair{} measure consistently has equivalent rankings to any group \fair{} measures}.\footnote{We find similar results for the group-individual fairness agreement between the same family of measures (e.g., SD$_\text{b-group}$ vs SD$_\text{ind}$) for all possible groupings (\Cref{agreement}).} For all datasets, group \fair{} measures tend to have weak-to-strong agreement with Gini\ind{}/Atk\ind{} and have weak-to-strong disagreement with SD\ind{}. The only group \fair{} measure that always correlates strongly to an individual \fair{} measure is CV ($\tau \in [.71,.79]$). This may be due to the measures' similar formulation: Gini\ind/Atk\ind{} have division by the mean \eff{} scores across all users, while CV has division by the mean of mean \eff{} score per group.
\textbf{In summary, no existing individual \fair{} measures make a reliable proxy for group \fair{} measures, hence the need to evaluate for both.} 

\subsubsection*{Intersectional fairness} 
\begin{figure}[tb]
    \centering
    \includegraphics[width=0.98\linewidth]{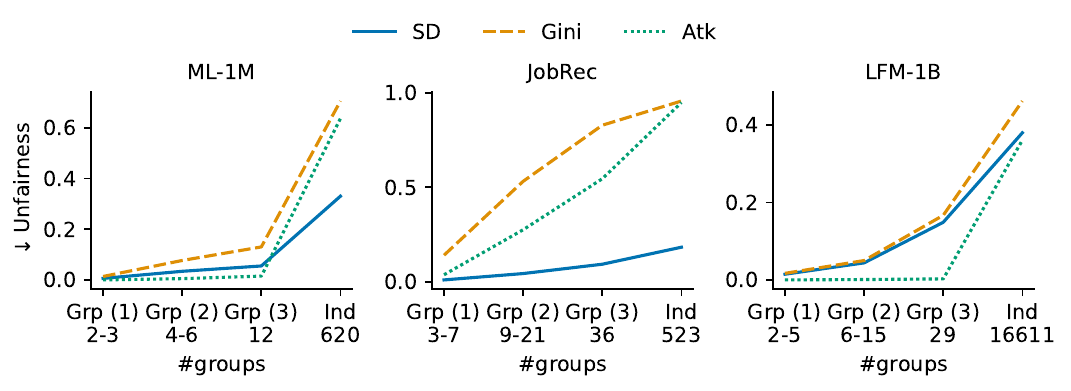}
    \caption{NDCG-based Group (Grp) and individual (Ind) \fair{} scores of GLM-4-9B (NS-prompt). Grp ($a$) is the mean \fair{} scores across all ways of grouping users when considering $a$ attribute(s). For $a\in\{1,2\}$, the \# of ways to group users is three each. For $a=3$, there is only one way to group users.
    }
    \label{fig:diff_groups}
\end{figure}

As we consider more intersectional attributes to form user groups, the number of groups grows. 
We posit that achieving intersectional group fairness is harder than for non-intersectional groups as it requires having similar \eff{} scores for a larger number of groups. Likewise, individual fairness (i.e., having a group for each user) would be worse than group fairness.  We study: 
(i) if/how the number of attributes used for grouping 
affect fairness; and 
(ii) the difference between individual and (intersectional) group \fair{} scores.  
To this end, we compute NDCG-based individual \fair{} scores across all users, and average NDCG-based group \fair{} scores across all ways of grouping users when considering only 1, 2, or 3 attribute(s) at once. SD, Gini, and Atk are computed, as they have been used as both group and individual \fair{} measures. We evaluate the runs from GLM-4-9B (NS) due to their relatively good NDCG and \fair{} scores across all datasets. 
The results are shown in \Cref{fig:diff_groups}.\footnote{While we present the measures in the same plot, their distribution differs, so the figures should not be used to quantify the gap between two families of measures.}

Regarding (i), our results verify that \textbf{fairness worsens as more attributes are used to form groups, highlighting the importance of considering intersectionality in fairness evaluation.} Note that JobRec has higher $\downarrow$Gini/$\downarrow$Atk than ML-1M/LFM-1B as its NDCG is much lower; Gini/Atk accounts for this, but SD does not.

Regarding (ii), we find that \textbf{even when the recommendations are relatively fair for (intersectional) groups, they can be much less fair for individuals} (i.e., for ML-1M and LFM-1B). For all measures and datasets, the individual \fair{} scores are always worse than group \fair{} scores. These findings imply that group \fair{} scores may mask unfairness within groups, even if the within-group variation is considered (e.g., in Atk$_\text{b-group}$).

\begin{figure}
    \centering
    \includegraphics[width=0.98\linewidth, trim=0 2mm 0 0, clip=True]{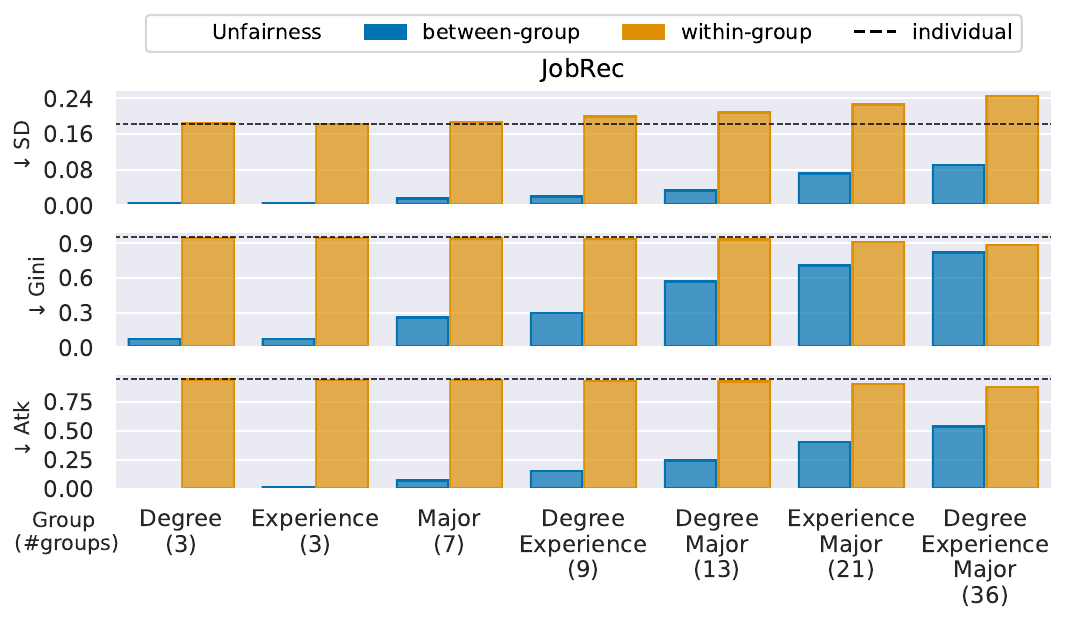}
    \includegraphics[width=0.98\linewidth, trim=0 2mm 0 1cm, clip=True]{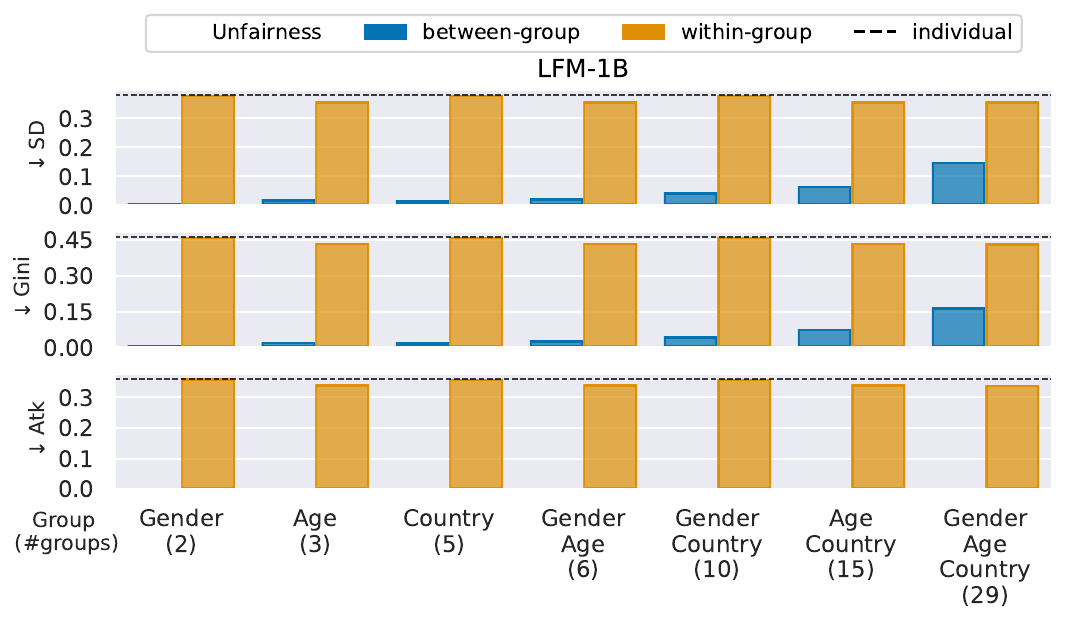}
    \caption{NDCG-based individual, between- and within-group unfairness of GLM-4-9B (NS-prompt) for all ways of grouping users in JobRec and LFM-1B.}
    \label{fig:decomposability_jobrec}
\end{figure}

\subsubsection*{Fairness decomposability}
\label{ss:decompose_analysis}

When fairness is evaluated only between groups, within-group unfairness may occur undetected. As the number of groups increases, we analyse: 
(i) how between- and within- group fairness change; and 
(ii) how they relate to individual fairness. 
To this end, we compute between- and within-group fairness, as well as individual fairness with SD, Gini, and Atk on the NDCG scores from GLM-4-9B (NS).
Results for JobRec and LFM-1B are shown in \Cref{fig:decomposability_jobrec}; we find similar trends for ML-1M (see \Cref{decomposition}). 

Regarding (i), we find that while between-group fairness generally worsens as the number of groups grows,\footnote{Resembling stairs; hence, the title of the paper.} 
within-group fairness tends to remain stable, except for SD of JobRec, where it worsens. 
Regarding (ii), we see that within-group unfairness is almost as high as individual unfairness, and always higher than between-group unfairness. Sometimes, within-group unfairness even exceeds individual unfairness (e.g., SD for JobRec). 
In summary, {\bf a single RS can vary in group fairness for different ways of grouping users, while being highly unfair for users within groups (regardless of the grouping) and for individual users at the same time.}

\section{Conclusion}
We empirically studied the relationship between group and individual fairness in RSs. We found that RSs which are fair for groups can still be very unfair for individual users, providing the first empirical evidence on the disjointness of these two RS fairness concepts. 
Overall, we encourage evaluating individual and within-group fairness alongside group fairness, as between-group \fair{} scores may mask huge disparity in RS effectiveness across users, even for a large number of groups (i.e., 5\% of the number of users) and even when the between-group \fair{} score accounts for within-group variations. Future work will study how strategies that mitigate either group or individual fairness affect each fairness type.

\section{Appendix}
\subsection{Details on Experimental Setup}
\subsubsection{User grouping}
\label{preproc}

For JobRec, we explain in more detail how we group users based on the study major. The study major is in free text format, so we first performed exploratory data analysis to get a better overview of the its distribution. More than 80\% test users with a high school (HS) degree have no specific study major, so we do not use major to divide users with a HS degree into smaller groups. Further, an author manually grouped the 235 unique majors of the non-HS test users into six fields of study, taking inspiration from the grouping in \citeauthor{Xu2024AModels}~\cite{Xu2024AModels}. During this process, we also remove users with generic or erroneous majors. We group the train/val users' majors as for the test users when possible, otherwise we perform fuzzy string matching between the annotated (manually grouped) and unannotated majors with \texttt{rapidfuzz} \cite{max_bachmann_2024_10938887}. We map the unannotated major to the same group as the most similar annotated major, if similarity $\geq 0.75$, otherwise we map it to an `Others' category.

We summarise the statistics of user sensitive attributes (after preprocessing) in \Cref{tab:full_stat_and_group}. We present the size of intersectional groups for each dataset in \Crefrange{tab:intersectional_group_ML}{tab:intersectional_group_LFM}.

\subsubsection{Prompt templates and examples}
\label{prompt}

We provide the templates and examples for the non-sensitive (NS) prompts in \Cref{tab:prompt_template_sample_nonsensitive} and for the sensitive (S) prompts in \Cref{tab:prompt_template_sample_sensitive}. In the S-prompt, we replace ``a user'' in the NS-prompt with a description containing their sensitive attribute, e.g., ``a user with a bachelor's degree, majoring in biology, has a total of 10 years of experience'. 
We use the original user attribute prior to re-grouping as they are more informative and precise (e.g., the user's major is included as `biology', not `STEM'). 
While this may not be the most natural way a user may interact with LLMs (e.g., explicitly disclosing their gender), it is a way to ensure that the LLMs have information on the users' sensitive attribute.

\subsubsection{Measure formulations and technical details}
\label{setup_eval}

To enhance reproducibility, we provide the formulations of all \fair{} measures that we compute. A \fair{} measure can be used to compute individual fairness, (between-)group fairness, and/or within-group fairness. We denote the individual, between-group, and within-group fairness versions of a measure as $\cdot_\text{ind}$, $\cdot_\text{b-group}$, and $\cdot_\text{w-group}$, respectively. 
When applicable, we provide the technical details (e.g., parameters) that we used. $\uparrow$/$\downarrow$ means the higher/lower the better

\paragraph{Average scores of the worst 25\% groups (Min).}
$\uparrow$Min \cite{Wang2024IntersectionalRecommendation} measures fairness between groups. To compute Min\bgrp{}, the \eff{} score is first calculated per user, and then averaged per user group. The group mean \eff{} score is then sorted, and the first quartile value is set as a threshold. Min\bgrp{} averages the mean group \eff{} score of groups with an \eff{} score below/equal to the threshold. 
The range of Min follows the \eff{} score range.

\paragraph{Range.}
$\downarrow$Range can be used to quantify fairness between groups. It is computed as the range of mean group \eff{} scores \cite{Liu2024SelfAdapative}. 

\paragraph{Standard Deviation (SD).}
$\downarrow$SD\footnote{We also include prior work that computes Variance under the same family of measure, as Variance is the square of SD.} 
can be used to measure between-group fairness \cite{Liu2024SelfAdapative,Zhang2023IsRecommendation}, within-group fairness, and individual fairness \cite{Rastegarpanah2019FightingSystems}.
SD\bgrp{} is the (population) standard deviation of the group mean \eff{} scores and SD\ind{} is the (population) standard deviation of users' \eff{} scores. We compute SD\wgrp{} as follows:
\begin{equation}
    SD\text{\wgrp} = \sum_{j=1}^{N'} s_j \cdot  SD(g_j)
\end{equation}

\noindent where $g_{j}$ is the list of group $j$ users' effectiveness scores and $N'$ is the number of user groups. 
The range of SD depends on the \eff{} score range. The weight of group $j$ is $s_j$, the ratio of the group's total \eff{} scores to the total across all groups.

\paragraph{Mean Absolute Difference (MAD).}
$\downarrow$MAD \cite{Fu2020Fairness-AwareGraphs} measures between-group fairness as the mean absolute pairwise difference of the groups' mean \eff{} scores. It is computed follows:
\begin{equation}
    MAD = \frac{1}{N'(N'-1)} \sum_{j=1}^{N'-1} \sum_{j'=j+1}^{N'} 
        \left| 
            \bar{g_j} - \bar{g_{j'}}
        \right| 
\end{equation}

\noindent where $\bar{g_j}$ is the group $j$'s mean \eff{} scores. 
The range of MAD follows the \eff{} score range.

\paragraph{Gini Index (Gini).}
$\downarrow$Gini \cite{Gini1912VariabilitaMutabilita} is a general-purpose inequality measure that has been used to measure between-group fairness \cite{ghosh2024reducingpopulationlevelinequalityimprove}, within-group fairness \cite{Ferraro2024GenderImbalance,Pastor2024IntersectionalDivergence}, and individual fairness \cite{Leonhardt2018UserSystems}. Given $N$ non-negative values, $x_1, x_2,\dots, x_N$, the general-purpose Gini is computed as follows:
\begin{equation}
    Gini 
    =  
    \frac{\sum\limits_{j=1}^N{(2j-N-1) x_j}}{N\sum\limits_{j=1}^N x_j} 
    =  
    \frac{\sum\limits_{j=1}^N{(2j-N-1) x_j}}{N^2\bar{x_j}} 
\end{equation}

Gini\bgrp{} is the Gini of the group mean \eff{} scores and Gini\ind{} is the Gini of users' \eff{} scores.  We compute Gini\wgrp{} as follows:

\begin{equation}
    Gini\text{\wgrp} = \sum_{j=1}^{N'} s_j \cdot  Gini(g_j)
\end{equation}

The range of Gini is [0,1].

\paragraph{Coefficient of Variation (CV).}
$\downarrow$CV is a general-purpose statistical measure of data dispersion, which has been used to measure between-group fairness   \cite{Zhu2020MeasuringSystems,Wang2024IntersectionalRecommendation}. The general-purpose CV is computed as the ratio of the (population) standard deviation to the mean. We compute CV\bgrp{} as follows:
\begin{equation}
    CV\text{\bgrp} = \frac{SD(\bar{g_1}, \bar{g_2}, \dots, \bar{g_{N'}})}{mean(\bar{g_1}, \bar{g_2}, \dots, \bar{g_{N'}})} 
\end{equation}

The range of CV is $[0,\infty)$.

\paragraph{F-statistic (FStat).}
$\downarrow$FStat is originally a statistical test that measures the ratio of between-group variance to within-group variance. It has been used to quantify between-group fairness for N' groups and N' users 
\cite{Wan2020AddressingRecommendations}, as follows: 
\begin{equation}
    V = \frac{1}{N} \sum_{j=1}^{N'} |g_j| \cdot (\bar{g_j} - \bar{x}_\text{eff})^2
\end{equation}
\begin{equation}
    U = \frac{1}{N} \sum_{j=1}^{N'} \sum_{\ell=1}^{|g_j|} (x_{j,\ell}-\bar{g_j})^2 
\end{equation}

\begin{equation}
    FStat_\text{\bgrp}= \frac{V/(N'-1)}{U/(N-N')}
\end{equation}
\noindent where $|g_j|$ is the number of users in group $j$, $\bar{x}_\text{eff}$ is the mean \eff{} scores of all users, and $x_{j,\ell}$ is the \eff{} score of the $\ell$-th user in the $j$-th group. 
The range of FStat is $[0,\infty)$.

\paragraph{Kullback-Leibler Divergence (KL).}
$\downarrow$KL can be customised to measure various fairness criteria; we select the version where between-group fairness means that effectiveness should be proportional to group size, as it is a popular choice \cite{Amigo2023ASystems}. 
KL is computed as follows:
\begin{equation}
    KL_\text{\bgrp}(p||q) = \sum_{j=1}^{N'} p_j \log_2{\frac{p_j}{q_j}}
\end{equation}
\begin{equation}
\label{eq:p_j}
    p_j = \frac{\bar{g_j}}{\sum_{j=1}^{N'} \bar{g_j}}
\end{equation}
\begin{equation}
    q_j = \frac{|g_j|}{\sum_{j=1}^{N'} |g_j|}
\end{equation}

The range of KL is $[0,\infty)$.

\paragraph{Generalised Cross Entropy (GCE).}
$\downarrow$ GCE can be used to quantify various fairness criteria \cite{Deldjoo2019RecommenderEntropy, Deldjoo2021ASystems}; we choose the version that requires equal utility (effectiveness) across groups. We compute GCE as follows:\footnote{Note that this is the nonnegative version of the measure in \cite{Deldjoo2021ASystems}.}
\begin{equation}
    GCE_\text{\bgrp} = -\frac{1}{B(1-B)} \sum_{j=1}^{N'} p_{ref}^B \cdot \left(\frac{\hat{p_j}}{\sum_{j=1}^{N'}\hat{p_j}}\right)^{(1-B)} -1
\end{equation}
\begin{equation}
    \hat{p_j} = \lambda p_j + (1-\lambda) c
\end{equation}

\noindent where $B$ is a parameter, $p_{ref}=1/N'$, $p_j$ is as per \Cref{eq:p_j} and $\hat{p_j}$ is the smoothened value of $p_j$. 
We set $\beta=2$, and perform smoothing with $\lambda=0.95$  and $c=10^{-4}$. 
The range of GCE is $[0,\infty)$.

\paragraph{Atkinson Index.}
\label{atk}
We adopt Atkinson Index (Atk), a measure of income inequality from the economics domain \cite{Atkinson1970OnInequality}, to quantify group/individual fairness for RS users based on recommendation effectiveness disparity.

\noindent\textbf{General formulation}. Given $N$ non-negative values, $x_1, x_2,\dots, x_N$, the general-purpose Atk is defined as follows:\footnote{Atk with $\varepsilon=1$ is defined differently from \Cref{eq:atk}.}
\begin{equation}
\label{eq:atk}
    Atk(x_1, x_2,\dots, x_N) = 1 - \frac{1}{\bar{x}}f(x_1, x_2,\dots, x_N)
\end{equation}
\begin{equation}
\label{eq:ede}
    f(x_1, x_2,\dots, x_N) = \left(
        \sum_{j=1}^{N} \frac{w_j}{\sum_{j=1}^{N}{w_j}} x_j^{1-\varepsilon}
        \right)^{\frac{1}{1-\varepsilon}}
\end{equation}

\noindent 
where $\bar{x}$ is the mean of the values, $w_j$ is the weight of the $j$-th score, and $\varepsilon\geq0$ is the inequality aversion parameter.\footnote{If $x_j=0, \forall j \in \{1, 2, \dots , N\}$, we define $\downarrow$Atk $=0$.} 
Atk with a low $\varepsilon$ is more sensitive to disparity among higher scores than among lower scores; we compute all Atk with $\varepsilon=0.5$ \cite{DeMaio2007IncomeMeasures}.
Atk has a [0,1]-range. 

\subparagraph{Measuring user fairness with Atk.}
To measure individual fairness for $N$ users with Atk (referred to as Atk$_\text{ind}$), we input the effectiveness score (e.g., NDCG) per user to \Cref{eq:atk} and set $w_j=1, \forall j \in \{1, \dots, N\}$ to weigh each user equally. As a user's effectiveness score could be 0, we choose $\varepsilon<1$ to avoid division by zero in $x_j^{1-\varepsilon}$ (\Cref{eq:ede}). Atk\bgrp{} is computed as follows:

\begin{equation}
\label{eq:atk_between}
    Atk_\text{b-group} = Atk
                \left(f(g_1), f(g_2), \dots, f(g_{N'})
                \right)
\end{equation}
Essentially, Atk$_\text{b-group}$ measures between-group fairness while accounting for the disparity within each group via $f$. As $f$ may be unstable for small groups, we reduce the score contributions from small groups by using the group size as $w_j$ to weigh $f(g_j)$. 
To compute $f(g_j)$ itself, we weigh each user in the group equally.

\noindent\textbf{Decomposition of Atk}. 
Atk is subgroup-decomposable \cite{Shorrocks1984InequalitySubgroups}, i.e., it can be broken down into between- and within-group components without residual terms \cite{Bourguignon1979DecomposableMeasures},  establishing a coherent relationship between individual fairness and group fairness \cite{Erreygers2018SubgroupAustralia}. Specifically, it detects if individual unfairness occurs mainly due to between- or within-group effectiveness disparity.
Atk$_\text{ind}$ can be decomposed as follows  \cite{Blackorby1999,Bourguignon1979DecomposableMeasures,Dayioğlu02006ImputedIndex}:
\begin{equation}
    \underbrace{1 - Atk_\text{ind}\vphantom{_p}}_\text{individual fairness} = \underbrace{(1 - Atk_\text{b-group})}_\text{between-group fairness} \cdot\
    \underbrace{
        (1-Atk_\text{w-group})
        }_\text{within-group fairness}
\end{equation}
\begin{equation}
\label{eq:atk_within}
    Atk_\text{w-group} = \sum_{j=1}^{N'} s_j \cdot  Atk(g_j)
\end{equation}

\subsection{Extended Results}
\subsubsection{Evaluation of all LLMRecs}
\label{ext_performance}

We present \eff{} and P-based \fair{} scores for LLMRecs in \Cref{tab:col_LLM-P}.

\subsubsection{Agreement of group and individual fairness measures}
\label{agreement}

\begin{figure}[p]
    \centering
    \includegraphics[width=0.98\linewidth]{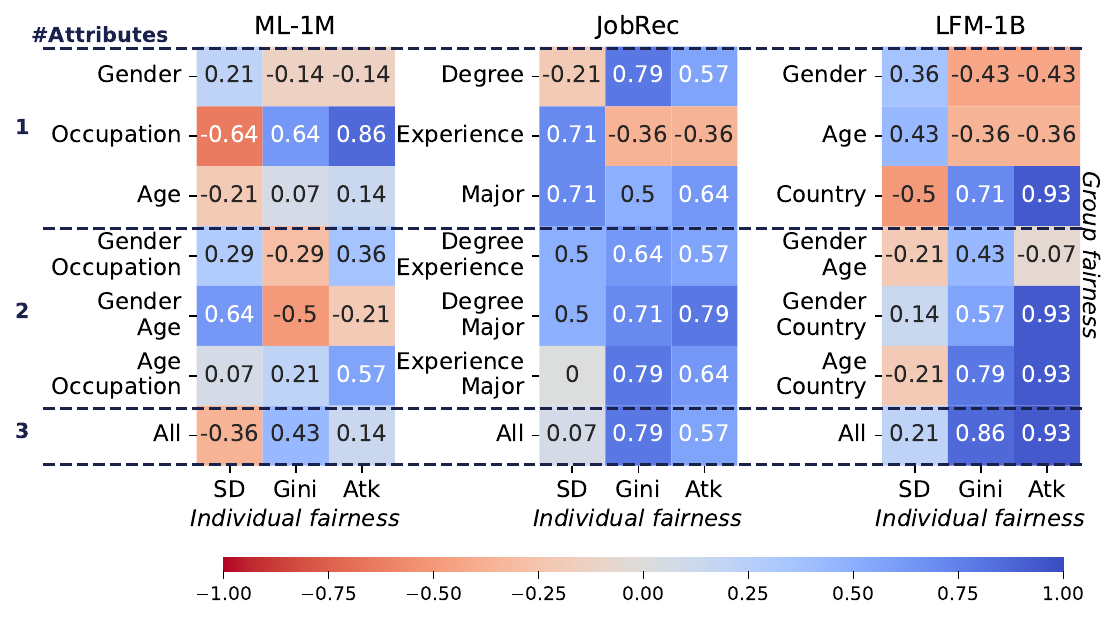}
    \caption{Agreement (Kendall's $\tau$) between the same family of measure in ranking LLMRecs for NDCG-based group fairness ($y$-axis) and individual fairness ($x$-axis). Group fairness is computed for each combination of users' sensitive attributes.
    }
    \label{fig:corr_diff_group}
\end{figure}

We show in \Cref{fig:corr_diff_group} the agreement of group-individual fairness between the same family of measures (e.g., SD$_\text{b-group}$ vs SD$_\text{ind}$) for all possible groupings.

\subsubsection{Fairness decomposability}
\label{decomposition}

\begin{figure}[p]
    \centering
    \includegraphics[width=0.95\linewidth]{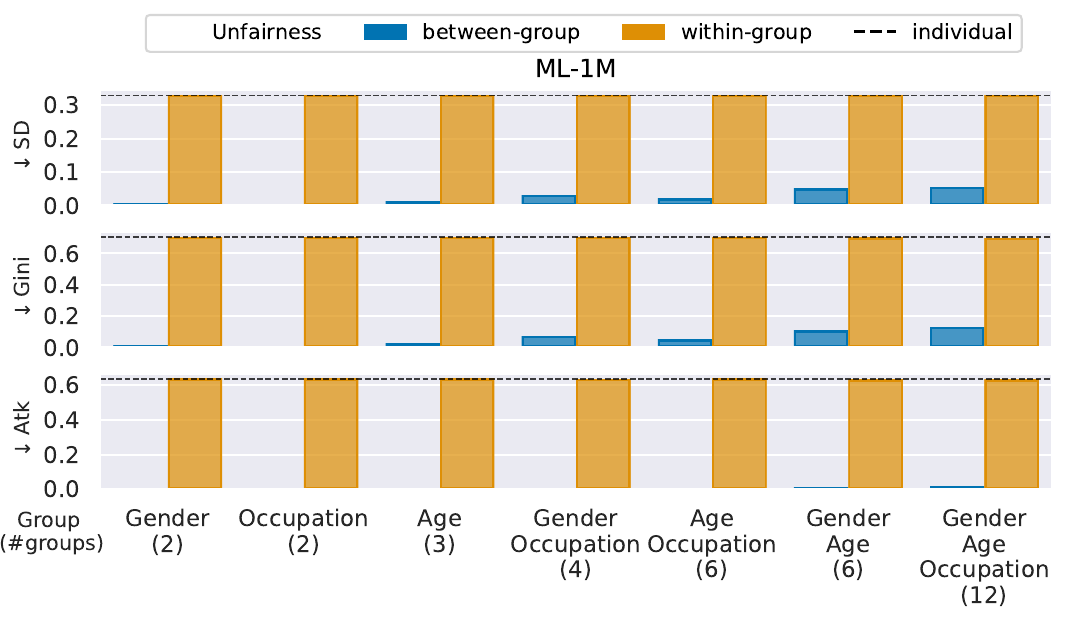}
    \caption{NDCG-based individual, between- and within-group unfairness of GLM-4-9B (NS-prompt) for all ways of grouping users in ML-1M.}
    \label{fig:app_decomposability}
\end{figure}

We show in \Cref{fig:app_decomposability} the individual, between-group, and within-group unfairness of GLM-4-9B (NS-prompt) for all ways of grouping users in ML-1M.

\begin{table*}[p]
    \centering
    \caption{Statistics of user sensitive attributes after preprocessing.}
    \label{tab:full_stat_and_group}
    \resizebox{0.95\textwidth}{!}{
    \begin{tabular}{lp{4.3cm}p{4.5cm}p{4.3cm}}
    \toprule
            & ML-1M \cite{Harper2015TheContext} & JobRec \cite{Hamner2012JobChallenge} & LFM-1B \cite{Schedl2016TheRecommendation} \\
    \midrule
    sensitive attr. \#1      & gender (2):                        &   degree (3):     & gender (2):                        \\
                            & M (457); F (163) &  University (254); High School (157); College (112)  & M (12,826); F (3,785)\\
    \midrule
    sensitive attr. \#2      & age (3): &   years of experience (3):      & age (3): \\
                            & 25--49 years (437); 18--24 years (110); $\geq$50 years  (73) & >10 (219); >5--10 (200); $\leq$5 (102) & 18--24 years (8,945); 25--49 years (6,590); $\geq$50 years (200) \\
    \midrule
    sensitive attr. \#3      & occupation (2): & major (6): & country/continent (5): \\
                            & working (478); non-working (142)   &  Business, Management, Finance (154); Social Science, Humanities, Education (103); STEM (41); Health \& Medical (29); Arts, Creative, Entertainment (22); Others (17); 
      &  Europe (9,976); America \& Antarctica (5,337); Asia (863); Oceania (376); Africa (59)                             \\
    \bottomrule
    \end{tabular}}
\end{table*}

\begin{table*}[p]
    \centering
    \caption{Size of intersectional groups (ML-1M).}
    \label{tab:intersectional_group_ML}
    \resizebox{0.6\columnwidth}{!}{
    \begin{tabular}{lllr}
    \toprule
    gender & age & occupation &  \#user\\
    \midrule
    \multirow[t]{6}{*}{M} & \multirow[t]{2}{*}{18--24 years} & non-working & 44 \\
     &  & working & 37 \\
    \cline{2-4}
     & \multirow[t]{2}{*}{25--49 years} & non-working & 41 \\
     &  & working & 279 \\
    \cline{2-4}
     & \multirow[t]{2}{*}{$\geq$50 years} & non-working & 8 \\
     &  & working & 48 \\
    \cline{1-4} \cline{2-4}
    \multirow[t]{6}{*}{F} & \multirow[t]{2}{*}{18--24 years} & non-working & 21 \\
     &  & working & 8 \\
    \cline{2-4}
     & \multirow[t]{2}{*}{25--49 years} & non-working & 26 \\
     &  & working & 91 \\
    \cline{2-4}
     & \multirow[t]{2}{*}{$\geq$50 years} & non-working & 2 \\
     &  & working & 15 \\
    \bottomrule
    \end{tabular}
    }
\end{table*}

\begin{table*}[p]
    \centering
    \caption{Size of intersectional groups (JobRec). Combinations resulting in group size of 0 are excluded. }
    \label{tab:intersectional_group_JobRec}
    \resizebox{0.85\linewidth}{!}{
    \begin{tabular}{lllr}
    \toprule
    degree & experience (years) & major &  \#user  \\
    \midrule
    \multirow[t]{3}{*}{High School} & $\leq$5 & - & 31 \\
    \cline{2-4}
     & >5--10 & - & 54 \\
    \cline{2-4}
     & >10 & - & 70 \\
    \cline{1-4}
    \multirow[t]{18}{*}{College} & \multirow[t]{6}{*}{$\leq$5} & Business, Management, Finance & 7 \\
     &  & Social Science, Humanities, Education & 3 \\
     &  & STEM & 4 \\
     &  & Health \& Medical & 8 \\
     &  & Arts, Creative, Entertainment & 2 \\
     &  & Others & 2 \\
    \cline{2-4}
     & \multirow[t]{6}{*}{>5--10} & Business, Management, Finance & 13 \\
     &  & Social Science, Humanities, Education & 6 \\
     &  & STEM & 8 \\
     &  & Health \& Medical & 10 \\
     &  & Arts, Creative, Entertainment & 3 \\
     &  & Others & 4 \\
    \cline{2-4}
     & \multirow[t]{6}{*}{>10} & Business, Management, Finance & 14 \\
     &  & Social Science, Humanities, Education & 12 \\
     &  & STEM & 3 \\
     &  & Health \& Medical & 3 \\
     &  & Arts, Creative, Entertainment & 4 \\
     &  & Others & 6 \\
     \cline{1-4}
    \multirow[t]{15}{*}{University} & \multirow[t]{5}{*}{$\leq$5} & Business, Management, Finance & 18 \\
     &  & Social Science, Humanities, Education & 16 \\
     &  & STEM & 7 \\
     &  & Health \& Medical & 3 \\
     &  & Arts, Creative, Entertainment & 1 \\
    \cline{2-4}
     & \multirow[t]{4}{*}{>5--10} & Business, Management, Finance & 47 \\
     &  & Social Science, Humanities, Education & 38 \\
     &  & STEM & 9 \\
     &  & Arts, Creative, Entertainment & 8 \\
    \cline{2-4}
     & \multirow[t]{6}{*}{>10} & Business, Management, Finance & 55 \\
     &  & Social Science, Humanities, Education & 28 \\
     &  & STEM & 10 \\
     &  & Health \& Medical & 5 \\
     &  & Arts, Creative, Entertainment & 4 \\
     &  & Others & 5 \\
\bottomrule
\end{tabular}}
\end{table*}

\begin{table*}[p]
    \centering
    \caption{Size of intersectional groups (LFM-1B).
    Combinations resulting in group size of 0 are excluded. 
    }
    \label{tab:intersectional_group_LFM}
    \resizebox{0.75\columnwidth}{!}{
\begin{tabular}{lllr}
\toprule
gender & age & continent &  \#user \\
\midrule
\multirow[t]{14}{*}{F} & \multirow[t]{5}{*}{18--24 years} & Europe & 1439 \\
 &  & America \& Antarctica & 809 \\
 &  & Asia & 127 \\
 &  & Oceania & 44 \\
 &  & Africa & 7 \\
\cline{2-4}
 & \multirow[t]{5}{*}{25--49 years} & Europe & 551 \\
 &  & America \& Antarctica & 392 \\
 &  & Asia & 59 \\
 &  & Oceania & 19 \\
 &  & Africa & 1 \\
\cline{2-4}
 & \multirow[t]{4}{*}{$\geq$50 years} & Europe & 15 \\
 &  & America \& Antarctica & 11 \\
 &  & Asia & 2 \\
 &  & Africa & 1 \\
\cline{1-4} \cline{2-4}
\multirow[t]{15}{*}{M} & \multirow[t]{5}{*}{18--24 years} & Europe & 3919 \\
 &  & America \& Antarctica & 2068 \\
 &  & Asia & 339 \\
 &  & Oceania & 168 \\
 &  & Africa & 25 \\
\cline{2-4}
 & \multirow[t]{5}{*}{25--49 years} & Europe & 3418 \\
 &  & America \& Antarctica & 1723 \\
 &  & Asia & 276 \\
 &  & Oceania & 128 \\
 &  & Africa & 23 \\
\cline{2-4}
 & \multirow[t]{5}{*}{$\geq$50 years} & Europe & 97 \\
 &  & America \& Antarctica & 59 \\
 &  & Asia & 11 \\
 &  & Oceania & 3 \\
 &  & Africa & 1 \\
\bottomrule
\end{tabular}
    }
\end{table*}

\begin{table*}[p]
    \centering
    \caption{Non-sensitive (NS) prompt templates and examples.}
    \label{tab:prompt_template_sample_nonsensitive}
    \resizebox{0.75\linewidth}{!}{

    {\footnotesize
    \begin{tabular}{p{\linewidth}}
    \toprule
      Dataset: ML-1M    \\
    \midrule
    \textbf{Prompt template: }``You are a movie recommender. If a user has watched the following movies, listed chronologically from earliest to latest, and rated them positively: 
   \textit{<items in train>}, then you should recommend the following movies: \textit{<items in val>}. Now that the user has watched the recommended movies, you should recommend 10 other movies from the year between 1919 and 2000 (inclusive) that the user is most likely to watch next. You should order them by probability and compact them in one line split by commas. Do not output the probability. Do not re-recommend movies that have been watched by the user. Do not output other words."
    \\
    \textbf{Prompt example: }``You are a movie recommender. If a user has watched the following movies, listed chronologically from earliest to latest, and rated them positively: 
    American Beauty, Magnolia, Boogie Nights, The Tao of Steve, Dark City, Fight Club, then you should recommend the following movies: The Evening Star, The MatchMaker, West Side Story, Dirty Dancing, The Prince of Egypt, Grease, Fargo, Good Will Hunting, Playing by Heart, The Man Without a Face. Now that the user has watched the recommended movies, you should recommend 10 other movies from the year between 1919 and 2000 (inclusive) that the user is most likely to watch next. You should order them by probability and compact them in one line split by commas. Do not output the probability. Do not re-recommend movies that have been watched by the user. Do not output other words."
    \\
    \midrule
    Dataset: JobRec    \\
    \midrule
    \textbf{Prompt template:} ``You are a job title recommender. If a user has applied to job positions with the following job titles, listed chronologically from earliest to latest: \textit{<items in train>}, you should recommend the following job titles: \textit{<items in val>}. 
    Now that the user has applied to positions with those job titles, you should recommend 10 other job titles that the user is most likely to apply for next. 
    You should order them by probability and compact them in one line split by commas. Do not output the probability. Do not output other words.''
    \\
    \textbf{Prompt example: }``You are a job title recommender. If a user has applied to job positions with the following job titles, listed chronologically from earliest to latest: Billing/Collections Process Coordinator, Administrative Assistant, Office Manager/Bookkeeper, Bookkeeper, you should recommend the following job titles: Accounts Payable Clerk, Assistant Director of Human Resources. 
    Now that the user has applied to positions with those job titles, you should recommend 10 other job titles that the user is most likely to apply for next. 
    You should order them by probability and compact them in one line split by commas. Do not output the probability. Do not output other words.''
    \\
    \midrule
    Dataset: LFM-1B    \\
    \midrule
    \textbf{Prompt template:} ``You are a music artist recommender. If a user has listened to the following artists, listed chronologically from earliest to latest:  \textit{<items in train>} and considering that the user listened to the artist \textit{<playcount of items in train>} times respectively, then you should recommend the following artists: \textit{<items in val>}. Now that the user has listened to the recommended artists, you should recommend 10 other artists, who has released songs prior to 2017, that the user is most likely to listen to next. You should order them by probability and compact them in one line split by commas. Do not output the probability. Do not re-recommend artists that the user has listened to. Do not output other words.''
    \\
    \textbf{Prompt example: } ``You are a music artist recommender. If a user has listened to the following artists, listed chronologically from earliest to latest: Lee Van Dowski, The Oohlas, XTC, Brenmar, Mountain, Bladerunner, Efdemin, Tom Encore, Droid Sector, Xhin, and considering that the user listened to the artist 2, 1, 2, 4, 12, 1, 12, 1, 2, 3 times respectively, then you should recommend the following artists: VNV Nation, Space, Ja Rule, Nneka, Parks, Enzyme X, David Bisbal, Primer 55, A-Sides, Cydonia. Now that the user has listened to the recommended artists, you should recommend 10 other artists, who has released songs prior to 2017, that the user is most likely to listen to next. You should order them by probability and compact them in one line split by commas. Do not output the probability. Do not re-recommend artists that the user has listened to. Do not output other words.''
    \\
    \bottomrule      
    \end{tabular}}}
\end{table*}

\begin{table*}[p]
    \centering
    \caption{Sensitive (S) prompt templates and examples.}
    \label{tab:prompt_template_sample_sensitive}
    
    \resizebox{0.75\linewidth}{!}{
    {\footnotesize
    \begin{tabular}{p{\linewidth}}
    \toprule
      Dataset: ML-1M    \\
    \midrule
    \textbf{Prompt template: }``You are a movie recommender. If a \textit{<gender>} user whose age is \textit{<age range>} years old and whose occupation is \textit{<occupation>} has watched the following movies, listed chronologically from earliest to latest, and rated them positively: \textit{<items in train>}, then you should recommend the following movies: \textit{<items in val>}. Now that the user has watched the recommended movies, you should recommend 10 other movies from the year between 1919 and 2000 (inclusive) that the user is most likely to watch next. You should order them by probability and compact them in one line split by commas. Do not output the probability. Do not re-recommend movies that have been watched by the user. Do not output other words.''
    \\
    \textbf{Prompt example: }``You are a movie recommender. If a female user whose age is 18-24 years old and whose occupation is writer has watched the following movies, listed chronologically from earliest to latest, and rated them positively: American Beauty, Magnolia, Boogie Nights, The Tao of Steve, Dark City, Fight Club, then you should recommend the following movies: The Evening Star, The MatchMaker, West Side Story, Dirty Dancing, The Prince of Egypt, Grease, Fargo, Good Will Hunting, Playing by Heart, The Man Without a Face. Now that the user has watched the recommended movies, you should recommend 10 other movies from the year between 1919 and 2000 (inclusive) that the user is most likely to watch next. You should order them by probability and compact them in one line split by commas. Do not output the probability. Do not re-recommend movies that have been watched by the user. Do not output other words.''
    \\
        \midrule
    Dataset: JobRec    \\
    \midrule
    \textbf{Prompt template:} ``You are a job title recommender. If a user with a \textit{<degree type>} degree, majoring in \textit{<major>}, has a total of \textit{<years of experience>} years of experience has applied to job positions with the following job titles, listed chronologically from earliest to latest: \textit{<items in train>}, then you should recommend the following job titles: \textit{<items in val>}. Now that the user has applied to positions with those job titles, you should recommend 10 other job titles that the user is most likely to apply for next. You should order them by probability and compact them in one line split by commas. Do not output the probability. Do not output other words.'' 
    \\
    \textbf{Prompt example: } ``You are a job title recommender. If a user with a Bachelor's degree, majoring in Accounting and Information Systems, has a total of 24 years of experience has applied to job positions with the following job titles, listed chronologically from earliest to latest: Billing/Collections Process Coordinator, Administrative Assistant, Office Manager/Bookkeeper, Bookkeeper, then you should recommend the following job titles: Accounts Payable Clerk, Assistant Director of Human Resources. Now that the user has applied to positions with those job titles, you should recommend 10 other job titles that the user is most likely to apply for next. You should order them by probability and compact them in one line split by commas. Do not output the probability. Do not output other words.'' 
    \\
    \midrule
    Dataset: LFM-1B    \\
    \midrule
    \textbf{Prompt template: } `You are a music artist recommender. If a \textit{<gender>} user whose age is \textit{<age>} years old and lives in \textit{<country>} has listened to the following artists, listed chronologically from earliest to latest: \textit{<items in train>}, and considering that the user listened to the artist \textit{<playcount of items in train>} times respectively, then you should recommend the following artists: \textit{<items in val>}. Now that the user has listened to the recommended artists, you should recommend 10 other artists, who has released songs prior to 2017, that the user is most likely to listen to next. You should order them by probability and compact them in one line split by commas. Do not output the probability. Do not re-recommend artists that the user has listened to. Do not output other words.''
    \\
    \textbf{Prompt example: } ``You are a music artist recommender. If a male user whose age is 30 years old and lives in Russia has listened to the following artists, listed chronologically from earliest to latest: Lee Van Dowski, The Oohlas, XTC, Brenmar, Mountain, Bladerunner, Efdemin, Tom Encore, Droid Sector, Xhin, and considering that the user listened to the artist 2, 1, 2, 4, 12, 1, 12, 1, 2, 3 times respectively, then you should recommend the following artists: VNV Nation, Space, Ja Rule, Nneka, Parks, Enzyme X, David Bisbal, Primer 55, A-Sides, Cydonia. Now that the user has listened to the recommended artists, you should recommend 10 other artists, who has released songs prior to 2017, that the user is most likely to listen to next. You should order them by probability and compact them in one line split by commas. Do not output the probability. Do not re-recommend artists that the user has listened to. Do not output other words.''
    \\
    \bottomrule      
    \end{tabular}}}
\end{table*}

\begin{table*}[p]
    \caption{
 Effectiveness (\eff{}) and fairness (\fair{}) scores at $k=10$ for intersectional groups (\textsc{Grp.}) and individuals (\textsc{Ind.}) of LLMRecs with non-sensitive (NS) and sensitive (S) prompts. All \fair{} scores are computed with P. All measures range in [0,1], except the \textsc{Grp.} measures below them. The best \eff{}/\fair{} scores are bolded. 
 Darker green marks scores closer to the best \eff{}/\fair{} per measure. $\uparrow/\downarrow$ means the higher/lower the better.
 }
    \label{tab:col_LLM-P}
    \centering
    \resizebox{0.65\linewidth}{!}{
\begin{tabular}{ll*{2}{r}|*{2}{r}|*{2}{r}|*{2}{r}}
\toprule
\toprule
 &  LLMRec& \multicolumn{2}{c|}{GLM-4-9B} & \multicolumn{2}{c|}{Llama-3.1-8B} & \multicolumn{2}{c|}{Ministral-8B} & \multicolumn{2}{c}{Qwen2.5-7B} \\ 
\midrule
 &  prompt type & NS & S & NS & S & NS & S & NS & S \\
\midrule
\midrule
 &  & \multicolumn{8}{c}{ML-1M} \\ 
\midrule
\multirow[c]{3}{*}{\rotatebox[origin=c]{90}{\textsc{Eff}}} & $\uparrow$ HR & \bfseries {\cellcolor[HTML]{2E8B57}} \color[HTML]{F1F1F1} 0.377 & {\cellcolor[HTML]{4C9C6F}} \color[HTML]{F1F1F1} 0.358 & {\cellcolor[HTML]{EBF3ED}} \color[HTML]{000000} 0.260 & {\cellcolor[HTML]{DDEBE2}} \color[HTML]{000000} 0.269 & {\cellcolor[HTML]{66AA84}} \color[HTML]{F1F1F1} 0.342 & {\cellcolor[HTML]{69AB86}} \color[HTML]{F1F1F1} 0.340 & {\cellcolor[HTML]{449769}} \color[HTML]{F1F1F1} 0.363 & {\cellcolor[HTML]{38905F}} \color[HTML]{F1F1F1} 0.371 \\
 & $\uparrow$ MRR & \bfseries {\cellcolor[HTML]{2E8B57}} \color[HTML]{F1F1F1} 0.189 & {\cellcolor[HTML]{4E9C70}} \color[HTML]{F1F1F1} 0.174 & {\cellcolor[HTML]{EBF3ED}} \color[HTML]{000000} 0.101 & {\cellcolor[HTML]{D2E5D9}} \color[HTML]{000000} 0.113 & {\cellcolor[HTML]{6EAE8A}} \color[HTML]{F1F1F1} 0.159 & {\cellcolor[HTML]{509E72}} \color[HTML]{F1F1F1} 0.173 & {\cellcolor[HTML]{61A780}} \color[HTML]{F1F1F1} 0.165 & {\cellcolor[HTML]{419666}} \color[HTML]{F1F1F1} 0.180 \\
 & $\uparrow$ P & \bfseries {\cellcolor[HTML]{2E8B57}} \color[HTML]{F1F1F1} 0.082 & {\cellcolor[HTML]{57A278}} \color[HTML]{F1F1F1} 0.074 & {\cellcolor[HTML]{EBF3ED}} \color[HTML]{000000} 0.046 & {\cellcolor[HTML]{D1E4D9}} \color[HTML]{000000} 0.051 & {\cellcolor[HTML]{7DB695}} \color[HTML]{F1F1F1} 0.067 & {\cellcolor[HTML]{7DB695}} \color[HTML]{F1F1F1} 0.067 & {\cellcolor[HTML]{57A278}} \color[HTML]{F1F1F1} 0.074 & {\cellcolor[HTML]{38915F}} \color[HTML]{F1F1F1} 0.080 \\
\cline{1-10}
\multirow[c]{10}{*}{\rotatebox[origin=c]{90}{\textsc{Fair (Grp.)}}} & $\uparrow$ Min\ & {\cellcolor[HTML]{89BD9F}} \color[HTML]{000000} 0.036 & \bfseries {\cellcolor[HTML]{2E8B57}} \color[HTML]{F1F1F1} 0.049 & {\cellcolor[HTML]{B3D4C1}} \color[HTML]{000000} 0.030 & {\cellcolor[HTML]{97C5AB}} \color[HTML]{000000} 0.034 & {\cellcolor[HTML]{EBF3ED}} \color[HTML]{000000} 0.022 & {\cellcolor[HTML]{6DAE89}} \color[HTML]{F1F1F1} 0.040 & {\cellcolor[HTML]{DEEBE3}} \color[HTML]{000000} 0.024 & {\cellcolor[HTML]{90C1A5}} \color[HTML]{000000} 0.035 \\
 & $\downarrow$ Range\ & \bfseries {\cellcolor[HTML]{2E8B57}} \color[HTML]{F1F1F1} 0.075 & {\cellcolor[HTML]{46986A}} \color[HTML]{F1F1F1} 0.088 & \bfseries {\cellcolor[HTML]{2E8B57}} \color[HTML]{F1F1F1} 0.075 & {\cellcolor[HTML]{46986A}} \color[HTML]{F1F1F1} 0.088 & {\cellcolor[HTML]{48996C}} \color[HTML]{F1F1F1} 0.089 & {\cellcolor[HTML]{8DBFA2}} \color[HTML]{000000} 0.125 & {\cellcolor[HTML]{EBF3ED}} \color[HTML]{000000} 0.175 & {\cellcolor[HTML]{8DBFA2}} \color[HTML]{000000} 0.125 \\
 & $\downarrow$ SD\ & {\cellcolor[HTML]{54A075}} \color[HTML]{F1F1F1} 0.024 & {\cellcolor[HTML]{4C9B6F}} \color[HTML]{F1F1F1} 0.023 & \bfseries {\cellcolor[HTML]{2E8B57}} \color[HTML]{F1F1F1} 0.019 & {\cellcolor[HTML]{358F5D}} \color[HTML]{F1F1F1} 0.020 & {\cellcolor[HTML]{6AAC87}} \color[HTML]{F1F1F1} 0.027 & {\cellcolor[HTML]{72B08D}} \color[HTML]{F1F1F1} 0.028 & {\cellcolor[HTML]{EBF3ED}} \color[HTML]{000000} 0.044 & {\cellcolor[HTML]{98C5AB}} \color[HTML]{000000} 0.033 \\
 & $\downarrow$ MAD\ & {\cellcolor[HTML]{55A076}} \color[HTML]{F1F1F1} 0.028 & {\cellcolor[HTML]{55A076}} \color[HTML]{F1F1F1} 0.028 & \bfseries {\cellcolor[HTML]{2E8B57}} \color[HTML]{F1F1F1} 0.022 & {\cellcolor[HTML]{348E5C}} \color[HTML]{F1F1F1} 0.023 & {\cellcolor[HTML]{69AB86}} \color[HTML]{F1F1F1} 0.031 & {\cellcolor[HTML]{69AB86}} \color[HTML]{F1F1F1} 0.031 & {\cellcolor[HTML]{EBF3ED}} \color[HTML]{000000} 0.051 & {\cellcolor[HTML]{A3CBB4}} \color[HTML]{000000} 0.040 \\
 & $\downarrow$ Gini\ & {\cellcolor[HTML]{328D5A}} \color[HTML]{F1F1F1} 0.169 & \bfseries {\cellcolor[HTML]{2E8B57}} \color[HTML]{F1F1F1} 0.166 & {\cellcolor[HTML]{5FA67E}} \color[HTML]{F1F1F1} 0.201 & {\cellcolor[HTML]{4C9C6F}} \color[HTML]{F1F1F1} 0.188 & {\cellcolor[HTML]{86BB9D}} \color[HTML]{000000} 0.229 & {\cellcolor[HTML]{57A278}} \color[HTML]{F1F1F1} 0.196 & {\cellcolor[HTML]{EBF3ED}} \color[HTML]{000000} 0.301 & {\cellcolor[HTML]{85BB9C}} \color[HTML]{000000} 0.228 \\
 & $\downarrow$ Atk\ & \bfseries {\cellcolor[HTML]{2E8B57}} \color[HTML]{F1F1F1} 0.018 & {\cellcolor[HTML]{328D5B}} \color[HTML]{F1F1F1} 0.019 & {\cellcolor[HTML]{6DAE89}} \color[HTML]{F1F1F1} 0.031 & {\cellcolor[HTML]{3C9362}} \color[HTML]{F1F1F1} 0.021 & {\cellcolor[HTML]{71B08D}} \color[HTML]{F1F1F1} 0.032 & {\cellcolor[HTML]{8ABDA0}} \color[HTML]{000000} 0.037 & {\cellcolor[HTML]{EBF3ED}} \color[HTML]{000000} 0.057 & {\cellcolor[HTML]{9DC8AF}} \color[HTML]{000000} 0.041 \\
\cline{2-10}
 & $\downarrow$ CV\ & {\cellcolor[HTML]{3E9464}} \color[HTML]{F1F1F1} 0.323 & \bfseries {\cellcolor[HTML]{2E8B57}} \color[HTML]{F1F1F1} 0.299 & {\cellcolor[HTML]{6AAC87}} \color[HTML]{F1F1F1} 0.388 & {\cellcolor[HTML]{56A177}} \color[HTML]{F1F1F1} 0.358 & {\cellcolor[HTML]{91C1A5}} \color[HTML]{000000} 0.444 & {\cellcolor[HTML]{6CAD88}} \color[HTML]{F1F1F1} 0.391 & {\cellcolor[HTML]{EBF3ED}} \color[HTML]{000000} 0.577 & {\cellcolor[HTML]{7DB696}} \color[HTML]{F1F1F1} 0.416 \\
 & $\downarrow$ FStat\ & {\cellcolor[HTML]{439667}} \color[HTML]{F1F1F1} 0.504 & {\cellcolor[HTML]{3B9261}} \color[HTML]{F1F1F1} 0.450 & {\cellcolor[HTML]{3E9463}} \color[HTML]{F1F1F1} 0.472 & \bfseries {\cellcolor[HTML]{2E8B57}} \color[HTML]{F1F1F1} 0.369 & {\cellcolor[HTML]{55A176}} \color[HTML]{F1F1F1} 0.625 & {\cellcolor[HTML]{4F9D71}} \color[HTML]{F1F1F1} 0.580 & {\cellcolor[HTML]{EBF3ED}} \color[HTML]{000000} 1.586 & {\cellcolor[HTML]{8CBFA2}} \color[HTML]{000000} 0.973 \\
 & $\downarrow$ KL\ & {\cellcolor[HTML]{3B9262}} \color[HTML]{F1F1F1} 0.836 & {\cellcolor[HTML]{B5D5C2}} \color[HTML]{000000} 1.211 & {\cellcolor[HTML]{DBEAE0}} \color[HTML]{000000} 1.328 & {\cellcolor[HTML]{E1EDE6}} \color[HTML]{000000} 1.349 & \bfseries {\cellcolor[HTML]{2E8B57}} \color[HTML]{F1F1F1} 0.793 & {\cellcolor[HTML]{CFE3D7}} \color[HTML]{000000} 1.292 & {\cellcolor[HTML]{EBF3ED}} \color[HTML]{000000} 1.380 & {\cellcolor[HTML]{A1CAB2}} \color[HTML]{000000} 1.150 \\
 & $\downarrow$ GCE\ & {\cellcolor[HTML]{2E8B57}} \color[HTML]{F1F1F1} 0.106 & \bfseries {\cellcolor[HTML]{2E8B57}} \color[HTML]{F1F1F1} 0.051 & {\cellcolor[HTML]{2E8B57}} \color[HTML]{F1F1F1} 0.068 & {\cellcolor[HTML]{2E8B57}} \color[HTML]{F1F1F1} 0.140 & {\cellcolor[HTML]{EBF3ED}} \color[HTML]{000000} 659.834 & {\cellcolor[HTML]{EBF3ED}} \color[HTML]{000000} 659.704 & {\cellcolor[HTML]{2E8B57}} \color[HTML]{F1F1F1} 0.387 & {\cellcolor[HTML]{2E8B57}} \color[HTML]{F1F1F1} 0.273 \\
\cline{1-10}
\multirow[c]{3}{*}{\rotatebox[origin=c]{90}{\textsc{Fair}} \rotatebox[origin=c]{90}{\textsc{(Ind.)}}} & $\downarrow$ SD\ & {\cellcolor[HTML]{EBF3ED}} \color[HTML]{000000} 0.141 & {\cellcolor[HTML]{B9D7C5}} \color[HTML]{000000} 0.129 & \bfseries {\cellcolor[HTML]{2E8B57}} \color[HTML]{F1F1F1} 0.096 & {\cellcolor[HTML]{64A982}} \color[HTML]{F1F1F1} 0.109 & {\cellcolor[HTML]{A4CCB5}} \color[HTML]{000000} 0.124 & {\cellcolor[HTML]{9FC9B1}} \color[HTML]{000000} 0.123 & {\cellcolor[HTML]{D6E7DD}} \color[HTML]{000000} 0.136 & {\cellcolor[HTML]{EBF3ED}} \color[HTML]{000000} 0.141 \\
 & $\downarrow$ Gini\ & \bfseries {\cellcolor[HTML]{2E8B57}} \color[HTML]{F1F1F1} 0.752 & {\cellcolor[HTML]{449769}} \color[HTML]{F1F1F1} 0.760 & {\cellcolor[HTML]{E9F1EB}} \color[HTML]{000000} 0.818 & {\cellcolor[HTML]{EBF3ED}} \color[HTML]{000000} 0.819 & {\cellcolor[HTML]{74B28F}} \color[HTML]{F1F1F1} 0.777 & {\cellcolor[HTML]{6CAD88}} \color[HTML]{F1F1F1} 0.774 & {\cellcolor[HTML]{4F9D72}} \color[HTML]{F1F1F1} 0.764 & {\cellcolor[HTML]{47996B}} \color[HTML]{F1F1F1} 0.761 \\
 & $\downarrow$ Atk\ & \bfseries {\cellcolor[HTML]{2E8B57}} \color[HTML]{F1F1F1} 0.658 & {\cellcolor[HTML]{4C9B6F}} \color[HTML]{F1F1F1} 0.674 & {\cellcolor[HTML]{EBF3ED}} \color[HTML]{000000} 0.760 & {\cellcolor[HTML]{E2EEE6}} \color[HTML]{000000} 0.755 & {\cellcolor[HTML]{6EAE8A}} \color[HTML]{F1F1F1} 0.693 & {\cellcolor[HTML]{6DAE89}} \color[HTML]{F1F1F1} 0.692 & {\cellcolor[HTML]{4D9C70}} \color[HTML]{F1F1F1} 0.675 & {\cellcolor[HTML]{3E9464}} \color[HTML]{F1F1F1} 0.667 \\
\midrule
\midrule
 &  & \multicolumn{8}{c}{JobRec} \\ 
\midrule
 &  LLMRec& \multicolumn{2}{c|}{GLM-4-9B} & \multicolumn{2}{c|}{Llama-3.1-8B} & \multicolumn{2}{c|}{Ministral-8B} & \multicolumn{2}{c}{Qwen2.5-7B} \\ 
\midrule
 &  prompt type & NS & S & NS & S & NS & S & NS & S \\
\midrule
\multirow[c]{3}{*}{\rotatebox[origin=c]{90}{\textsc{Eff}}} & $\uparrow$ HR & {\cellcolor[HTML]{5DA57C}} \color[HTML]{F1F1F1} 0.054 & {\cellcolor[HTML]{B8D7C5}} \color[HTML]{000000} 0.033 & {\cellcolor[HTML]{88BD9F}} \color[HTML]{000000} 0.044 & {\cellcolor[HTML]{EBF3ED}} \color[HTML]{000000} 0.021 & {\cellcolor[HTML]{509E72}} \color[HTML]{F1F1F1} 0.057 & {\cellcolor[HTML]{88BD9F}} \color[HTML]{000000} 0.044 & \bfseries {\cellcolor[HTML]{2E8B57}} \color[HTML]{F1F1F1} 0.065 & {\cellcolor[HTML]{509E72}} \color[HTML]{F1F1F1} 0.057 \\
 & $\uparrow$ MRR & {\cellcolor[HTML]{63A881}} \color[HTML]{F1F1F1} 0.037 & {\cellcolor[HTML]{A8CEB8}} \color[HTML]{000000} 0.023 & {\cellcolor[HTML]{C4DDCF}} \color[HTML]{000000} 0.017 & {\cellcolor[HTML]{EBF3ED}} \color[HTML]{000000} 0.009 & \bfseries {\cellcolor[HTML]{2E8B57}} \color[HTML]{F1F1F1} 0.048 & {\cellcolor[HTML]{77B391}} \color[HTML]{F1F1F1} 0.033 & {\cellcolor[HTML]{38905F}} \color[HTML]{F1F1F1} 0.046 & {\cellcolor[HTML]{3C9362}} \color[HTML]{F1F1F1} 0.045 \\
 & $\uparrow$ P & {\cellcolor[HTML]{54A075}} \color[HTML]{F1F1F1} 0.006 & {\cellcolor[HTML]{C5DECF}} \color[HTML]{000000} 0.003 & {\cellcolor[HTML]{7AB493}} \color[HTML]{F1F1F1} 0.005 & {\cellcolor[HTML]{EBF3ED}} \color[HTML]{000000} 0.002 & {\cellcolor[HTML]{54A075}} \color[HTML]{F1F1F1} 0.006 & {\cellcolor[HTML]{9FC9B1}} \color[HTML]{000000} 0.004 & \bfseries {\cellcolor[HTML]{2E8B57}} \color[HTML]{F1F1F1} 0.007 & \bfseries {\cellcolor[HTML]{2E8B57}} \color[HTML]{F1F1F1} 0.007 \\
\cline{1-10}
\multirow[c]{10}{*}{\rotatebox[origin=c]{90}{\textsc{Fair (Grp.)}}} & $\uparrow$ Min\ & \bfseries {\cellcolor[HTML]{EBF3ED}} \color[HTML]{000000} 0.000 & \bfseries {\cellcolor[HTML]{EBF3ED}} \color[HTML]{000000} 0.000 & \bfseries {\cellcolor[HTML]{EBF3ED}} \color[HTML]{000000} 0.000 & \bfseries {\cellcolor[HTML]{EBF3ED}} \color[HTML]{000000} 0.000 & \bfseries {\cellcolor[HTML]{EBF3ED}} \color[HTML]{000000} 0.000 & \bfseries {\cellcolor[HTML]{EBF3ED}} \color[HTML]{000000} 0.000 & \bfseries {\cellcolor[HTML]{EBF3ED}} \color[HTML]{000000} 0.000 & \bfseries {\cellcolor[HTML]{EBF3ED}} \color[HTML]{000000} 0.000 \\
 & $\downarrow$ Range\ & {\cellcolor[HTML]{EBF3ED}} \color[HTML]{000000} 0.050 & \bfseries {\cellcolor[HTML]{2E8B57}} \color[HTML]{F1F1F1} 0.010 & {\cellcolor[HTML]{EBF3ED}} \color[HTML]{000000} 0.050 & {\cellcolor[HTML]{EBF3ED}} \color[HTML]{000000} 0.050 & {\cellcolor[HTML]{9BC7AE}} \color[HTML]{000000} 0.033 & {\cellcolor[HTML]{62A780}} \color[HTML]{F1F1F1} 0.021 & {\cellcolor[HTML]{9BC7AE}} \color[HTML]{000000} 0.033 & {\cellcolor[HTML]{9BC7AE}} \color[HTML]{000000} 0.033 \\
 & $\downarrow$ SD\ & {\cellcolor[HTML]{EBF3ED}} \color[HTML]{000000} 0.010 & \bfseries {\cellcolor[HTML]{2E8B57}} \color[HTML]{F1F1F1} 0.003 & {\cellcolor[HTML]{EBF3ED}} \color[HTML]{000000} 0.010 & {\cellcolor[HTML]{B5D5C2}} \color[HTML]{000000} 0.008 & {\cellcolor[HTML]{B5D5C2}} \color[HTML]{000000} 0.008 & {\cellcolor[HTML]{64A982}} \color[HTML]{F1F1F1} 0.005 & {\cellcolor[HTML]{D0E4D8}} \color[HTML]{000000} 0.009 & {\cellcolor[HTML]{B5D5C2}} \color[HTML]{000000} 0.008 \\
 & $\downarrow$ MAD\ & {\cellcolor[HTML]{CCE2D4}} \color[HTML]{000000} 0.008 & \bfseries {\cellcolor[HTML]{2E8B57}} \color[HTML]{F1F1F1} 0.003 & {\cellcolor[HTML]{CCE2D4}} \color[HTML]{000000} 0.008 & {\cellcolor[HTML]{4D9C70}} \color[HTML]{F1F1F1} 0.004 & {\cellcolor[HTML]{ACD0BB}} \color[HTML]{000000} 0.007 & {\cellcolor[HTML]{6DAE89}} \color[HTML]{F1F1F1} 0.005 & {\cellcolor[HTML]{EBF3ED}} \color[HTML]{000000} 0.009 & {\cellcolor[HTML]{ACD0BB}} \color[HTML]{000000} 0.007 \\
 & $\downarrow$ Gini\ & {\cellcolor[HTML]{7AB594}} \color[HTML]{F1F1F1} 0.800 & {\cellcolor[HTML]{7AB594}} \color[HTML]{F1F1F1} 0.800 & {\cellcolor[HTML]{96C4A9}} \color[HTML]{000000} 0.827 & {\cellcolor[HTML]{EBF3ED}} \color[HTML]{000000} 0.911 & {\cellcolor[HTML]{3F9465}} \color[HTML]{F1F1F1} 0.742 & {\cellcolor[HTML]{71B08C}} \color[HTML]{F1F1F1} 0.791 & \bfseries {\cellcolor[HTML]{2E8B57}} \color[HTML]{F1F1F1} 0.725 & {\cellcolor[HTML]{5AA37A}} \color[HTML]{F1F1F1} 0.769 \\
 & $\downarrow$ Atk\ & {\cellcolor[HTML]{4F9D72}} \color[HTML]{F1F1F1} 0.509 & {\cellcolor[HTML]{3E9463}} \color[HTML]{F1F1F1} 0.480 & {\cellcolor[HTML]{A2CBB3}} \color[HTML]{000000} 0.641 & {\cellcolor[HTML]{EBF3ED}} \color[HTML]{000000} 0.758 & {\cellcolor[HTML]{499A6C}} \color[HTML]{F1F1F1} 0.498 & {\cellcolor[HTML]{4F9D71}} \color[HTML]{F1F1F1} 0.508 & {\cellcolor[HTML]{358F5C}} \color[HTML]{F1F1F1} 0.466 & \bfseries {\cellcolor[HTML]{2E8B57}} \color[HTML]{F1F1F1} 0.455 \\
\cline{2-10}
 & $\downarrow$ CV\ & {\cellcolor[HTML]{60A77F}} \color[HTML]{F1F1F1} 2.079 & {\cellcolor[HTML]{4F9D71}} \color[HTML]{F1F1F1} 1.875 & {\cellcolor[HTML]{71B08C}} \color[HTML]{F1F1F1} 2.262 & {\cellcolor[HTML]{EBF3ED}} \color[HTML]{000000} 3.634 & {\cellcolor[HTML]{38915F}} \color[HTML]{F1F1F1} 1.629 & {\cellcolor[HTML]{4C9C6F}} \color[HTML]{F1F1F1} 1.850 & \bfseries {\cellcolor[HTML]{2E8B57}} \color[HTML]{F1F1F1} 1.508 & {\cellcolor[HTML]{459869}} \color[HTML]{F1F1F1} 1.770 \\
 & $\downarrow$ FStat\ & {\cellcolor[HTML]{94C3A8}} \color[HTML]{000000} 0.848 & \bfseries {\cellcolor[HTML]{2E8B57}} \color[HTML]{F1F1F1} 0.447 & {\cellcolor[HTML]{EBF3ED}} \color[HTML]{000000} 1.193 & {\cellcolor[HTML]{BCD9C8}} \color[HTML]{000000} 1.007 & {\cellcolor[HTML]{99C6AC}} \color[HTML]{000000} 0.871 & {\cellcolor[HTML]{6DAE89}} \color[HTML]{F1F1F1} 0.696 & {\cellcolor[HTML]{A8CEB8}} \color[HTML]{000000} 0.927 & {\cellcolor[HTML]{5FA67E}} \color[HTML]{F1F1F1} 0.642 \\
 & $\downarrow$ KL\ & {\cellcolor[HTML]{7CB695}} \color[HTML]{F1F1F1} 2.706 & \bfseries {\cellcolor[HTML]{2E8B57}} \color[HTML]{F1F1F1} 1.330 & {\cellcolor[HTML]{97C5AB}} \color[HTML]{000000} 3.195 & {\cellcolor[HTML]{EBF3ED}} \color[HTML]{000000} 4.677 & {\cellcolor[HTML]{439768}} \color[HTML]{F1F1F1} 1.716 & {\cellcolor[HTML]{499A6D}} \color[HTML]{F1F1F1} 1.818 & {\cellcolor[HTML]{54A075}} \color[HTML]{F1F1F1} 2.004 & {\cellcolor[HTML]{4F9D72}} \color[HTML]{F1F1F1} 1.926 \\
 & $\downarrow$ GCE\ & {\cellcolor[HTML]{64A982}} \color[HTML]{F1F1F1} 1685.893 & {\cellcolor[HTML]{D0E4D8}} \color[HTML]{000000} 1979.080 & {\cellcolor[HTML]{64A982}} \color[HTML]{F1F1F1} 1685.948 & {\cellcolor[HTML]{EBF3ED}} \color[HTML]{000000} 2052.428 & {\cellcolor[HTML]{499A6C}} \color[HTML]{F1F1F1} 1612.569 & {\cellcolor[HTML]{7FB797}} \color[HTML]{F1F1F1} 1759.183 & \bfseries {\cellcolor[HTML]{2E8B57}} \color[HTML]{F1F1F1} 1539.280 & {\cellcolor[HTML]{7FB797}} \color[HTML]{F1F1F1} 1759.168 \\
\cline{1-10}
\multirow[c]{3}{*}{\rotatebox[origin=c]{90}{\textsc{Fair}} \rotatebox[origin=c]{90}{\textsc{(Ind.)}}} & $\downarrow$ SD\ & {\cellcolor[HTML]{B5D5C2}} \color[HTML]{000000} 0.024 & {\cellcolor[HTML]{64A982}} \color[HTML]{F1F1F1} 0.018 & {\cellcolor[HTML]{9AC6AD}} \color[HTML]{000000} 0.022 & \bfseries {\cellcolor[HTML]{2E8B57}} \color[HTML]{F1F1F1} 0.014 & {\cellcolor[HTML]{D0E4D8}} \color[HTML]{000000} 0.026 & {\cellcolor[HTML]{8DBFA2}} \color[HTML]{000000} 0.021 & {\cellcolor[HTML]{EBF3ED}} \color[HTML]{000000} 0.028 & {\cellcolor[HTML]{EBF3ED}} \color[HTML]{000000} 0.028 \\
 & $\downarrow$ Gini\ & {\cellcolor[HTML]{55A076}} \color[HTML]{F1F1F1} 0.948 & {\cellcolor[HTML]{B1D3BF}} \color[HTML]{000000} 0.967 & {\cellcolor[HTML]{85BB9D}} \color[HTML]{000000} 0.958 & {\cellcolor[HTML]{EBF3ED}} \color[HTML]{000000} 0.979 & {\cellcolor[HTML]{4B9B6E}} \color[HTML]{F1F1F1} 0.946 & {\cellcolor[HTML]{7CB695}} \color[HTML]{F1F1F1} 0.956 & \bfseries {\cellcolor[HTML]{2E8B57}} \color[HTML]{F1F1F1} 0.940 & {\cellcolor[HTML]{55A076}} \color[HTML]{F1F1F1} 0.948 \\
 & $\downarrow$ Atk\ & {\cellcolor[HTML]{5EA57D}} \color[HTML]{F1F1F1} 0.947 & {\cellcolor[HTML]{B6D6C3}} \color[HTML]{000000} 0.967 & {\cellcolor[HTML]{86BB9D}} \color[HTML]{000000} 0.956 & {\cellcolor[HTML]{EBF3ED}} \color[HTML]{000000} 0.979 & {\cellcolor[HTML]{4C9C6F}} \color[HTML]{F1F1F1} 0.943 & {\cellcolor[HTML]{86BB9D}} \color[HTML]{000000} 0.956 & \bfseries {\cellcolor[HTML]{2E8B57}} \color[HTML]{F1F1F1} 0.936 & {\cellcolor[HTML]{519E73}} \color[HTML]{F1F1F1} 0.944 \\
\midrule
\midrule
 &  & \multicolumn{8}{c}{LFM-1B} \\ 
\midrule
 &  LLMRec& \multicolumn{2}{c|}{GLM-4-9B} & \multicolumn{2}{c|}{Llama-3.1-8B} & \multicolumn{2}{c|}{Ministral-8B} & \multicolumn{2}{c}{Qwen2.5-7B} \\ 
\midrule
 &  prompt type & NS & S & NS & S & NS & S & NS & S \\
\midrule
\multirow[c]{3}{*}{\rotatebox[origin=c]{90}{\textsc{Eff}}} & $\uparrow$ HR & {\cellcolor[HTML]{2F8C58}} \color[HTML]{F1F1F1} 0.658 & \bfseries {\cellcolor[HTML]{2E8B57}} \color[HTML]{F1F1F1} 0.661 & {\cellcolor[HTML]{4A9A6D}} \color[HTML]{F1F1F1} 0.609 & {\cellcolor[HTML]{46986A}} \color[HTML]{F1F1F1} 0.618 & {\cellcolor[HTML]{A2CAB3}} \color[HTML]{000000} 0.451 & {\cellcolor[HTML]{EBF3ED}} \color[HTML]{000000} 0.317 & {\cellcolor[HTML]{CBE1D4}} \color[HTML]{000000} 0.375 & {\cellcolor[HTML]{EBF3ED}} \color[HTML]{000000} 0.317 \\
 & $\uparrow$ MRR & \bfseries {\cellcolor[HTML]{2E8B57}} \color[HTML]{F1F1F1} 0.409 & {\cellcolor[HTML]{2F8B58}} \color[HTML]{F1F1F1} 0.408 & {\cellcolor[HTML]{5DA57C}} \color[HTML]{F1F1F1} 0.347 & {\cellcolor[HTML]{55A176}} \color[HTML]{F1F1F1} 0.357 & {\cellcolor[HTML]{9BC7AE}} \color[HTML]{000000} 0.266 & {\cellcolor[HTML]{E1EDE5}} \color[HTML]{000000} 0.174 & {\cellcolor[HTML]{CDE2D6}} \color[HTML]{000000} 0.199 & {\cellcolor[HTML]{EBF3ED}} \color[HTML]{000000} 0.160 \\
 & $\uparrow$ P & {\cellcolor[HTML]{328D5B}} \color[HTML]{F1F1F1} 0.203 & \bfseries {\cellcolor[HTML]{2E8B57}} \color[HTML]{F1F1F1} 0.206 & {\cellcolor[HTML]{61A780}} \color[HTML]{F1F1F1} 0.172 & {\cellcolor[HTML]{539F74}} \color[HTML]{F1F1F1} 0.181 & {\cellcolor[HTML]{B2D3C0}} \color[HTML]{000000} 0.118 & {\cellcolor[HTML]{EBF3ED}} \color[HTML]{000000} 0.080 & {\cellcolor[HTML]{C4DDCF}} \color[HTML]{000000} 0.106 & {\cellcolor[HTML]{E4EEE7}} \color[HTML]{000000} 0.085 \\
\cline{1-10}
\multirow[c]{10}{*}{\rotatebox[origin=c]{90}{\textsc{Fair (Grp.)}}} & $\uparrow$ Min\ & {\cellcolor[HTML]{399160}} \color[HTML]{F1F1F1} 0.093 & \bfseries {\cellcolor[HTML]{2E8B57}} \color[HTML]{F1F1F1} 0.098 & {\cellcolor[HTML]{308C59}} \color[HTML]{F1F1F1} 0.097 & {\cellcolor[HTML]{3C9362}} \color[HTML]{F1F1F1} 0.092 & {\cellcolor[HTML]{BFDACA}} \color[HTML]{000000} 0.037 & {\cellcolor[HTML]{E9F1EB}} \color[HTML]{000000} 0.019 & {\cellcolor[HTML]{BCD9C8}} \color[HTML]{000000} 0.038 & {\cellcolor[HTML]{EBF3ED}} \color[HTML]{000000} 0.018 \\
 & $\downarrow$ Range\ & {\cellcolor[HTML]{76B290}} \color[HTML]{F1F1F1} 0.313 & {\cellcolor[HTML]{71B08C}} \color[HTML]{F1F1F1} 0.300 & {\cellcolor[HTML]{C2DCCD}} \color[HTML]{000000} 0.500 & {\cellcolor[HTML]{EBF3ED}} \color[HTML]{000000} 0.600 & {\cellcolor[HTML]{EBF3ED}} \color[HTML]{000000} 0.600 & {\cellcolor[HTML]{48996C}} \color[HTML]{F1F1F1} 0.200 & {\cellcolor[HTML]{48996C}} \color[HTML]{F1F1F1} 0.200 & \bfseries {\cellcolor[HTML]{2E8B57}} \color[HTML]{F1F1F1} 0.136 \\
 & $\downarrow$ SD\ & {\cellcolor[HTML]{9CC7AE}} \color[HTML]{000000} 0.076 & {\cellcolor[HTML]{83BA9B}} \color[HTML]{000000} 0.068 & {\cellcolor[HTML]{C4DDCE}} \color[HTML]{000000} 0.089 & {\cellcolor[HTML]{D6E7DD}} \color[HTML]{000000} 0.095 & {\cellcolor[HTML]{EBF3ED}} \color[HTML]{000000} 0.102 & {\cellcolor[HTML]{348E5C}} \color[HTML]{F1F1F1} 0.042 & {\cellcolor[HTML]{46986A}} \color[HTML]{F1F1F1} 0.048 & \bfseries {\cellcolor[HTML]{2E8B57}} \color[HTML]{F1F1F1} 0.040 \\
 & $\downarrow$ MAD\ & {\cellcolor[HTML]{E7F0EA}} \color[HTML]{000000} 0.081 & {\cellcolor[HTML]{C7DFD1}} \color[HTML]{000000} 0.075 & {\cellcolor[HTML]{EBF3ED}} \color[HTML]{000000} 0.082 & {\cellcolor[HTML]{E7F0EA}} \color[HTML]{000000} 0.081 & {\cellcolor[HTML]{DCEAE1}} \color[HTML]{000000} 0.079 & \bfseries {\cellcolor[HTML]{2E8B57}} \color[HTML]{F1F1F1} 0.045 & {\cellcolor[HTML]{4C9C6F}} \color[HTML]{F1F1F1} 0.051 & {\cellcolor[HTML]{328D5B}} \color[HTML]{F1F1F1} 0.046 \\
 & $\downarrow$ Gini\ & {\cellcolor[HTML]{429667}} \color[HTML]{F1F1F1} 0.212 & \bfseries {\cellcolor[HTML]{2E8B57}} \color[HTML]{F1F1F1} 0.197 & {\cellcolor[HTML]{58A279}} \color[HTML]{F1F1F1} 0.229 & {\cellcolor[HTML]{5FA67E}} \color[HTML]{F1F1F1} 0.234 & {\cellcolor[HTML]{EBF3ED}} \color[HTML]{000000} 0.339 & {\cellcolor[HTML]{CDE2D5}} \color[HTML]{000000} 0.316 & {\cellcolor[HTML]{76B290}} \color[HTML]{F1F1F1} 0.251 & {\cellcolor[HTML]{C9E0D2}} \color[HTML]{000000} 0.313 \\
 & $\downarrow$ Atk\ & \bfseries {\cellcolor[HTML]{2E8B57}} \color[HTML]{F1F1F1} 0.004 & \bfseries {\cellcolor[HTML]{2E8B57}} \color[HTML]{F1F1F1} 0.004 & {\cellcolor[HTML]{38905F}} \color[HTML]{F1F1F1} 0.005 & {\cellcolor[HTML]{419666}} \color[HTML]{F1F1F1} 0.006 & {\cellcolor[HTML]{55A176}} \color[HTML]{F1F1F1} 0.008 & {\cellcolor[HTML]{A5CCB6}} \color[HTML]{000000} 0.016 & {\cellcolor[HTML]{55A176}} \color[HTML]{F1F1F1} 0.008 & {\cellcolor[HTML]{EBF3ED}} \color[HTML]{000000} 0.023 \\
\cline{2-10}
 & $\downarrow$ CV\ & {\cellcolor[HTML]{3D9363}} \color[HTML]{F1F1F1} 0.414 & \bfseries {\cellcolor[HTML]{2E8B57}} \color[HTML]{F1F1F1} 0.372 & {\cellcolor[HTML]{60A67E}} \color[HTML]{F1F1F1} 0.513 & {\cellcolor[HTML]{73B18E}} \color[HTML]{F1F1F1} 0.566 & {\cellcolor[HTML]{EBF3ED}} \color[HTML]{000000} 0.903 & {\cellcolor[HTML]{7FB797}} \color[HTML]{F1F1F1} 0.600 & {\cellcolor[HTML]{55A176}} \color[HTML]{F1F1F1} 0.484 & {\cellcolor[HTML]{72B08D}} \color[HTML]{F1F1F1} 0.564 \\
 & $\downarrow$ FStat\ & {\cellcolor[HTML]{88BD9F}} \color[HTML]{000000} 3.344 & {\cellcolor[HTML]{7AB493}} \color[HTML]{F1F1F1} 3.108 & {\cellcolor[HTML]{9DC8AF}} \color[HTML]{000000} 3.679 & {\cellcolor[HTML]{BBD8C7}} \color[HTML]{000000} 4.152 & {\cellcolor[HTML]{6CAD88}} \color[HTML]{F1F1F1} 2.887 & {\cellcolor[HTML]{8EC0A4}} \color[HTML]{000000} 3.444 & \bfseries {\cellcolor[HTML]{2E8B57}} \color[HTML]{F1F1F1} 1.895 & {\cellcolor[HTML]{EBF3ED}} \color[HTML]{000000} 4.933 \\
 & $\downarrow$ KL\ & {\cellcolor[HTML]{7CB695}} \color[HTML]{F1F1F1} 2.735 & {\cellcolor[HTML]{79B492}} \color[HTML]{F1F1F1} 2.716 & {\cellcolor[HTML]{E9F2EC}} \color[HTML]{000000} 3.501 & {\cellcolor[HTML]{C4DDCE}} \color[HTML]{000000} 3.235 & {\cellcolor[HTML]{EBF3ED}} \color[HTML]{000000} 3.513 & {\cellcolor[HTML]{7AB594}} \color[HTML]{F1F1F1} 2.729 & {\cellcolor[HTML]{7BB594}} \color[HTML]{F1F1F1} 2.734 & \bfseries {\cellcolor[HTML]{2E8B57}} \color[HTML]{F1F1F1} 2.194 \\
 & $\downarrow$ GCE\ & {\cellcolor[HTML]{8CBFA2}} \color[HTML]{000000} 338.856 & {\cellcolor[HTML]{5DA57C}} \color[HTML]{F1F1F1} 225.920 & \bfseries {\cellcolor[HTML]{2E8B57}} \color[HTML]{F1F1F1} 113.018 & {\cellcolor[HTML]{2E8B57}} \color[HTML]{F1F1F1} 113.024 & {\cellcolor[HTML]{BCD9C8}} \color[HTML]{000000} 451.857 & {\cellcolor[HTML]{BCD9C8}} \color[HTML]{000000} 451.864 & {\cellcolor[HTML]{BCD9C8}} \color[HTML]{000000} 451.815 & {\cellcolor[HTML]{EBF3ED}} \color[HTML]{000000} 564.777 \\
\cline{1-10}
\multirow[c]{3}{*}{\rotatebox[origin=c]{90}{\textsc{Fair}} \rotatebox[origin=c]{90}{\textsc{(Ind.)}}} & $\downarrow$ SD\ & {\cellcolor[HTML]{E6F0E9}} \color[HTML]{000000} 0.226 & {\cellcolor[HTML]{EBF3ED}} \color[HTML]{000000} 0.228 & {\cellcolor[HTML]{ABD0BB}} \color[HTML]{000000} 0.204 & {\cellcolor[HTML]{BCD9C8}} \color[HTML]{000000} 0.210 & {\cellcolor[HTML]{73B18E}} \color[HTML]{F1F1F1} 0.183 & \bfseries {\cellcolor[HTML]{2E8B57}} \color[HTML]{F1F1F1} 0.157 & {\cellcolor[HTML]{7EB797}} \color[HTML]{F1F1F1} 0.187 & {\cellcolor[HTML]{509E72}} \color[HTML]{F1F1F1} 0.170 \\
 & $\downarrow$ Gini\ & {\cellcolor[HTML]{308C59}} \color[HTML]{F1F1F1} 0.586 & \bfseries {\cellcolor[HTML]{2E8B57}} \color[HTML]{F1F1F1} 0.583 & {\cellcolor[HTML]{449769}} \color[HTML]{F1F1F1} 0.610 & {\cellcolor[HTML]{3E9464}} \color[HTML]{F1F1F1} 0.603 & {\cellcolor[HTML]{A0CAB2}} \color[HTML]{000000} 0.718 & {\cellcolor[HTML]{E8F1EB}} \color[HTML]{000000} 0.803 & {\cellcolor[HTML]{CBE1D4}} \color[HTML]{000000} 0.769 & {\cellcolor[HTML]{EBF3ED}} \color[HTML]{000000} 0.807 \\
 & $\downarrow$ Atk\ & {\cellcolor[HTML]{308C59}} \color[HTML]{F1F1F1} 0.415 & \bfseries {\cellcolor[HTML]{2E8B57}} \color[HTML]{F1F1F1} 0.411 & {\cellcolor[HTML]{48996C}} \color[HTML]{F1F1F1} 0.454 & {\cellcolor[HTML]{429667}} \color[HTML]{F1F1F1} 0.445 & {\cellcolor[HTML]{A0CAB2}} \color[HTML]{000000} 0.599 & {\cellcolor[HTML]{E9F2EC}} \color[HTML]{000000} 0.719 & {\cellcolor[HTML]{CBE1D4}} \color[HTML]{000000} 0.669 & {\cellcolor[HTML]{EBF3ED}} \color[HTML]{000000} 0.722 \\
\bottomrule
\end{tabular}}

\end{table*}

\newpage
\addcontentsline{toc}{chapter}{Errata}
\pagestyle{plain}
\chapter*{Errata}
A summary of corrections for non-substantive errors that do not change the conclusions of the papers are provided below.

\begin{table}[h]
    \caption*{Summary of corrections to non-substantive errors.}
    \centering
\resizebox{0.8\columnwidth}{!}{
    \begin{tabular}{l|>{\raggedright}p{4cm}|p{4cm}|p{4cm}<{}}
    \toprule
    &Original Text & Correction & Explanation \\
    \midrule
    \Cref{chap:SIGIR24} & $e_{\text{inv}}(u,i) = 1/z(u,i)$& $e_{\text{inv}}(p) = 1/p$ & The correct formulation of the inverse examination function depends directly on the position $p$ of the recommendation slot. \\
    \midrule
    \Cref{chap:SIGIR24}  & We use NDCG and Jain as they are more sensitive to changes than HR and QF. & We use NDCG and Jain as they tend to be stricter than some other single-aspect measures used in this work. & The text has been changed for clarity. \\
    \midrule
    \Cref{chap:SIGIR24}     & IAA/HD/II-F strongly disagrees with IFD, $\tau \in [-0.71,-0.5]$ & IAA/HD/II-F strongly disagrees with IFD, $\tau \in [-0.93,-0.5]$ & The lower bound of $\tau$ has been corrected. This does not change the conclusion. \\
    \midrule
    \Cref{chap:WWW25}     & ... averaging fails to reach the same conclusion as DPFR almost half the time & ... averaging fails to reach the same conclusion as DPFR almost half the time for the rank-based \textsc{Rel} measures &  The phrase ``for the rank-based \textsc{Rel} measures'' has been added for clarity. \\ 
    \midrule
    \Cref{chap:WWW25} & DPFR demonstrates distinct benefits in mitigating false conclusions by up to 50\%...&  DPFR demonstrates distinct benefits in mitigating false conclusions by up to 58\%...& The performance of DPFR (our proposed approach) is better than previously reported. \\ 
    \bottomrule
    \end{tabular}
}
\end{table}
\newpage

\addcontentsline{toc}{chapter}{Bibliography}
\pagestyle{plain}
\bibliography{references}

\end{document}